%% file: main.tex
\documentclass{article}

\usepackage[a4paper, left=2.7cm, right=2.7cm, top=1in, bottom=1in]{geometry}
\usepackage{authblk}
\usepackage[numbers]{natbib}
\usepackage{graphicx} % needed for \textsuperscript
\usepackage{amsmath}
\usepackage{amssymb} % for checkmark symbol
\usepackage{array}
\usepackage{bm} % for printing math symbols in bold face
\usepackage{nicefrac} % for nice in-line fractions in figures
\usepackage{booktabs} % for more fancy tables
\usepackage{float} % for "H" positioning option
\usepackage{multirow} % for combining rows/columns in tables
\usepackage{pgf} % for pgf figures
\usepackage{siunitx}
\usepackage[acronym]{glossaries} % for acronyms
\usepackage{mathtools} % for \coloneqq
\def\mathdefault#1{\ifmmode#1\else\text{#1}\fi}
\usepackage{todonotes}
\usepackage{subcaption} % for subfigure environment
\usetikzlibrary{arrows,patterns,patterns.meta}   % Include more stuff from tikz

\input{custom-colors}

\renewcommand{\cite}[1]{\textsuperscript{\citenum{#1}}}

\newacronym{bc}{BC}{boundary condition}
\newacronym{binn}{BINN}{boundary-informed neural network}
\newacronym{cfd}{CFD}{computational fluid dynamics}
\newacronym{ibc}{IBC}{initial and boundary condition}
\newacronym{fem}{FEM}{finite element method}
\newacronym{ml}{ML}{machine learning}
\newacronym{nn}{NN}{neural network}
\newacronym{pde}{PDE}{partial differential equation}
\newacronym{pinn}{PINN}{physics-informed neural network}
\newacronym{rans}{RANS}{Reynolds-averaged Navier-Stokes}
\newacronym{rom}{ROM}{reduced order model}

\begin{document}

\title{Quantifying data needs in surrogate modeling for flow fields in two-dimensional stirred tanks with physics-informed neural networks}

\author[1]{Veronika Trávníková}       
\author[2,3]{Eric von Lieres}          
\author[1]{Marek Behr}        

\affil[1]{Chair for Computational Analysis of Technical Systems, RWTH Aachen University, Germany}
\affil[2]{Institute of Bio- and Geosciences 1: Biotechnology, Forschungszentrum Jülich, Germany}
\affil[3]{Computational Systems Biotechnology, RWTH Aachen University, Germany}

\maketitle 
\begin{abstract}
Stirred tanks are vital in chemical and biotechnological processes, particularly as bioreactors. Although \acrfull{cfd} is widely used to model the flow in stirred tanks, its high computational cost—especially in multi-query scenarios for process design and optimization—drives the need for efficient data-driven surrogate models. However, acquiring sufficiently large datasets can be costly. \Acrfullpl{pinn} offer a promising solution to reduce data requirements while maintaining accuracy by embedding underlying physics into \acrfull{nn} training. This study quantifies the data requirements of vanilla \glspl{pinn} for developing surrogate models of a flow field in a 2D stirred tank. We compare these requirements with classical supervised neural networks and \acrfullpl{binn}. Our findings demonstrate that surrogate models can achieve prediction errors around \qty{3}{\percent} across Reynolds numbers from \num{50} to \num{5000} using as few as six datapoints. Moreover, employing an approximation of the velocity profile in place of real data labels leads to prediction errors of around \qty{2.5}{\percent}. These results indicate that even with limited or approximate datasets, \glspl{pinn} can be effectively trained to deliver high accuracy comparable to high-fidelity data.

\end{abstract}

\section{Introduction}
\label{sec:introduction}
Stirred tanks are widely used for mixing and fermentation in chemical and biotechnological processes. A typical configuration of a stirred tank consists of a cylindrical vessel with one or more impellers attached to a central shaft, along with baffles fixed to the inner walls of the vessel. The flow field within these systems is influenced by numerous geometrical parameters, such as impeller diameter, tank radius, and number of baffles, as well as fluid properties and operational conditions like stirring rate. These variables allow for flexibility in process design while simultaneously presenting challenges in the design and scaling of processes that incorporate stirred tanks.\cite{Jaszczur2020, Garcia-Ochoa2011} We are particularly focused on stirred tanks in their application as bioreactors, where they play an integral role in bioprocesses, such as the production of antibiotics or specific antibodies. In bioprocesses, the hydrodynamics of mixing are closely coupled to the kinetics of biochemical reactions; heterogeneities in nutrient concentration result in variations in intracellular composition among cells, thereby affecting production.\cite{Haringa2017} Consequently, there is a need for models that can quickly evaluate flow fields and be integrated into optimization loops for bioprocess design.
\Gls{cfd} is commonly used to simulate stirred tanks,\cite{Ebrahimi2019, Reid2024} addressing the limitations of experimental methods where measurements can be prohibitively complex or expensive for gaining insights into internal conditions. However, high-fidelity simulations entail substantial computational costs, particularly when repeated evaluations of the model under different parameters are required. 
This drives the development of \glspl{rom} or surrogate models that could provide fast approximate solutions at low computational cost. Data-driven approaches have been proposed to address this need, leveraging the advantages of offline pretraining and swift model evaluation offered by \gls{ml}. One method for constructing \glspl{rom} using \gls{ml} involves autoencoders---artificial neural networks designed to map high-dimensional inputs onto low-dimensional spaces.\cite{yuNonintrusiveReducedorderModeling2019} Additionally, methods such as neural operators\cite{luLearningNonlinearOperators2021} have been proposed as promising data-driven surrogates. However, these methods often demand large amounts of data, making it prohibitively expensive to gather a sufficiently extensive training dataset for our specific application. Therefore, we aim to harness insights into the underlying physics of the problem to reduce or potentially eliminate reliance on extensive training data.
To this end, we employ \glspl{pinn}, which are specifically designed to embed the governing equations of a problem directly into the loss function of the \gls{nn}. The integration of physical constraints into the training process of \glspl{nn} was first introduced by Lagaris et al.\cite{Lagaris1998} in 1998. However, it has only gained substantial attention following the work of Raissi et al.\cite{Raissi2019} Since then, \glspl{pinn} have been successfully employed across a range of fields,\cite{cuomoScientificMachineLearning2022} including solid mechanics,\cite{Haghighat2021} hemodynamics,\cite{chenHemodynamicsModelingPhysicsinformed2025} or reaction kinetics in fed-batch bioreactors.\cite{yangHybridModelingFedBatch2024} Moreover, they have been utilized to tackle various problems in fluid mechanics,\cite{JinNSFnets2021, tillmannShapeoptimizationExtrusiondiesParameterized2023, zhouPhysicsInformedNeuralNetworks2024, lorenzenPotentialPhysicsinformedNeural2024, FaroughiReview2024} including acting as closure for \gls{rans} equations\cite{Eivazi2022} and reconstructing flow fields from sparse measurement\cite{steinfurthPhysicsinformedNeuralNetworks2024, hosseiniFlowFieldReconstruction2024} or visualization\cite{RaissiHFM2020} data.
Numerous strategies have been proposed to address the limitations of \glspl{pinn} in their basic (or vanilla) configuration, including strong imposition of \glspl{bc},\cite{Sheng2021LFF} domain decomposition with non-overlapping\cite{Shukla2021PPINNs} and overlapping\cite{Moseley2023FBPINNs} subdomains, Fourier feature encoding,\cite{zengTrainingDynamicsPhysicsInformed2024} adaptive loss scaling methods,\cite{Wang2021LossScaleAnnealing, McClennySAPINN2023} alternative optimizers\cite{wangGradientAlignmentPhysicsinformed2025, yaoMultiAdamParameterwiseScaleinvariant2023a} and adaptive sampling techniques.\cite{visserPACMANNPointAdaptive2024, wuComprehensiveStudyNonadaptive2023} In our previous work, we developed an accurate surrogate model for a 2D stirred tank using some of these methods---specifically strong \gls{bc} imposition and domain decomposition---without relying on labeled data during training.\cite{TravnikovaWolff2024} However, constructing and evaluating the interpolation function required for strong \gls{bc} imposition in complex geometries---and performing backpropagation through this function during training---can be computationally demanding and memory-intensive, particularly when increasing the spatial dimensionality. With the ultimate goal of developing surrogate models for flow fields in 3D stirred tanks, we aim to assess the capabilities of vanilla \glspl{pinn} in their most basic configuration, enhanced by some labeled simulation or measurement data. Previous research\cite{bhatnagarPhysicsInformedNeural2024, ghoshGeometryawarePINNsTurbulent2024} indicates that \glspl{pinn} require significantly smaller labeled datasets compared to traditional \glspl{nn}. Our objective is to conduct a quantitative study comparing the data requirements of a \gls{pinn} with those of a classical \gls{nn} model, as well as a configuration where only \gls{bc} residuals are incorporated into the \gls{nn} loss function, which we refer to as \gls{binn}. Furthermore, we investigate whether lower fidelity or approximate data can be effectively utilized for training to achieve results comparable to that obtained through high-fidelity simulation.

\section{Methods}
\label{sec:methods}
This section provides an overview of the methodologies employed in this paper. First, it reviews the fundamental concepts of fully-connected feed-forward \glspl{nn}, the architecture used in this work, before introducing our \gls{pinn}-related notation. 

% ----

\subsection{Fully-connected feed-forward \glspl{nn}}
\label{subsec:methods:fcffnn}
The models presented in this paper exclusively use fully-connected feed-forward \glspl{nn}, where each neuron in one layer is linked to every neuron in the next layer, with information flowing only forward. Mathematically, a \gls{nn} mapping input space $\mathcal{X}$ to output space $\mathcal{U}$ with $N_{L}$ layers can be defined as:
\begin{eqnarray}
    \label{eq:methods:fcffnn:network}
    \bm{\mathcal{NN}} : \mathcal{X} \to \mathcal{U} \;, \nonumber \\
    \bm{x}\mapsto\bm{\mathcal{NN}}(\bm{x};\bm{\theta})&=\left(\bm{f}^{(N_{L})} \circ \hdots \circ \bm{f}^{(1)}\right)(\bm{x};\bm{\theta}) \; ,
\end{eqnarray}
where each layer is defined by a function:

\begin{align}
    \label{eq:methodology:fcffnn:layer}
    \bm{f}^{(l)}\left(\bm{x};\bm{\theta}^{(l)}\right) &= \sigma^{(l)}\left(\bm{W}^{(l)}\bm{x}+\bm{b}^{(l)}\right) \quad \forall \; l\in\{1,\hdots,N_{L}\} \; .
\end{align}
Here, $\bm{x}\in\mathcal{X}$ denotes the input vector, $\bm{\theta}^{(l)}=\{\bm{W}^{(l)},\bm{b}^{(l)}\}$ are the trainable parameters (weights and biases) of the \gls{nn}, and $\sigma$ is the activation function. In this paper, all hidden layers employ the same activation function, expressed as $\sigma^{(l)}(\bm{z})=\sigma(\bm{z}) \;\forall\; l \in \{1,\hdots,N_{L-1}\}$, with the exception of the final layer, which utilizes a linear activation function: $\sigma^{(N_L)}(\bm{z})=\bm{z}$.

% ----

\subsection{\Acrfullpl{pinn}}
\label{subsec:methods:pinn}

\Glspl{pinn} utilize neural networks to approximate the unknown solution fields of (partial) differential equations.
We approximate the unknown solution $\bm{u}$ with a neural network, expressed as:
\begin{align}
    \label{eq:methods:pinn:approximation}
    \bm{u}(\bm{x}) \approx \bm{u}_{\bm{\theta}}(\bm{x}) \coloneqq \bm{\mathcal{NN}}(\bm{x};\bm{\theta}) \; .
\end{align}
The neural network's trainable parameters, $\bm{\theta}$, are optimized by minimizing the loss function $\mathcal{L}(\bm{\theta})$, defined by:
\begin{align}
    \label{eq:methods:pinn:minimization}
    \bm{\theta}^* = \arg\min_{\bm{\theta}} \mathcal{L}(\bm{\theta};\bm{X}).
\end{align}
The loss function comprises several components:
\begin{eqnarray}
    \label{eq:methods:pinn:loss}
    \mathcal{L}(\bm{\theta};\bm{X}) &= \bm{\alpha}_{\text{PDE}} \cdot \bm{\mathcal{L}}_{\text{PDE}}(\bm{\theta}; \bm{X}_{\text{PDE}}) \nonumber \\
     & + \bm{\alpha}_{\text{IBC}} \cdot \bm{\mathcal{L}}_{\text{IBC}}(\bm{\theta}; \bm{X}_{\text{IBC}}) \nonumber \\
     & + \bm{\alpha}_{\text{data}} \cdot \bm{\mathcal{L}}_{\text{data}}(\bm{\theta}; \bm{X}_{\text{data}})\; ,
\end{eqnarray}
where:
\begin{align}
    \label{eq:methods:pinn:physicslosses:pdes}
    \bm{\mathcal{L}}_{\text{PDE}}(\bm{\theta}; \bm{X}_{\text{PDE}}) 
    &= \frac{1}{N_{\text{PDE}}} \sum_{i=1}^{N_{\text{PDE}}} \left(\bm{\mathcal{R}}\left[\bm{u}_{\bm{\theta}}\left(\bm{x}^{(i)}_{\text{PDE}}\right)\right]\right)^2, \\
    \label{eq:methods:pinn:physicslosses:bcs}
    \bm{\mathcal{L}}_{\text{IBC}}(\bm{\theta}; \bm{X}_{\text{IBC}})
    &= \frac{1}{N_{\text{IBC}}} \sum_{i=1}^{N_{\text{IBC}}} \left(\bm{\mathcal{B}}\left[\bm{u}_{\bm{\theta}}\left(\bm{x}^{(i)}_{\text{IBC}}\right)\right]\right)^2.
\end{align}
These terms represent the physics-informed losses, quantifying the deviation from the true solution by evaluating the residuals of the \gls{pde} denoted by $\bm{\mathcal{R}}[\,\cdot\,]$ and the \acrfull{ibc} denoted by $\bm{\mathcal{B}}[\,\cdot\,]$  at the predicted values:
\begin{subequations}
\label{eq:methods:pinn:pdeandbc}
\begin{align}
    \bm{\mathcal{R}}[\bm{u}] &= \bm{0} \quad \text{in} \; \Omega \; ,\\
    \bm{\mathcal{B}}[\bm{u}] &= \bm{0} \quad \text{on} \; \Gamma \; .
\end{align}
\end{subequations}
Additionally, the loss function can include a data-driven component:
\begin{align}
    \bm{\mathcal{L}}_{\text{data}}(\bm{\theta}; \bm{X}_{\text{data}}) &= \frac{1}{N_{\text{data}}} \sum_{i=1}^{N_{\text{data}}} \left( \bm{u}_{\bm{\theta}}\left(\bm{x}^{(i)}_{\text{data}}\right)-\bm{u}^{(i)}_{\text{data}} \right)^2 \; ,
\end{align}
which accounts for the discrepancy with labeled data obtained from high-fidelity simulations or measurements.

In these equations, $\bm{X}=\{\bm{X}_{\text{PDE}},\bm{X}_{\text{IBC}}\}$ denotes the collection of evaluation points, also known as collocation points, at which the \gls{pde} and the (initial and) boundary conditions are evaluated. Specifically, $\bm{X}_{\text{PDE}}=\left\{\bm{x}^{(1)}_{\text{PDE}},\hdots,\bm{x}^{(N_{\text{PDE}})}_{\text{PDE}}\right\}$ and $\bm{X}_{\text{IBC}}=\left\{\bm{x}^{(1)}_{\text{IBC}},\hdots,\bm{x}^{(N_{\text{IBC}})}_{\text{IBC}}\right\}$ represent the collocation points within the domain and on its boundary, respectively. The set $\bm{X}_{\text{data}}=\left\{\left(\bm{x}^{(1)}_{\text{data}},\bm{u}^{(1)}_{\text{data}}\right),\hdots,\left(\bm{x}^{(N_{\text{data}})}_{\text{data}},\bm{u}^{(N_{\text{data}})}_{\text{data}}\right)\right\}$ comprises the training data with information on the solution $\bm{u}^{(i)}_{\text{data}}$ at specific input locations $\bm{x}^{(i)}_{\text{data}}$. The scaling factors $\bm{\alpha}_{\text{PDE}}$, $\bm{\alpha}_{\text{IBC}}$, and $\bm{\alpha}_{\text{data}}$ enable weighting the individual loss components appropriately.
%Various strategies have been developed to effectively determine loss scaling factors, as their selection is crucial for both the convergence of the training process and the accuracy of PINN predictions \cite{Wang2021LossScaleAnnealing}. Most of these approaches involve dynamic adjustments during training. These methods include a bi-level optimization algorithm that updates loss scales through gradient descent steps, as proposed by Li et al. (2022) \cite{Li2022BiLevelOptiWeights}; modifying loss scales based on the magnitudes of each loss component's gradients \cite{Wang2021LossScaleAnnealing}; and employing updates informed by neural tangent kernel theory \cite{Wang2022NTK}. Additionally, McClenny and Braga-Neto \cite{McClennySAPINN2023} developed a technique for adjusting loss scaling factors by evaluating the gradients of loss components on a point-wise basis.

\subsubsection*{\emph{Input and output scaling}}
In the models outlined in this paper, we apply a pre-processing function to non-dimensionalize the inputs of the network and a post-processing function to adjust the \gls{nn} outputs to their original scale, as recommended in the literature.\cite{Wang2023ExpertGuide} The pre-processing function serves as a fixed transformation layer, converting input features to fall within the range $[-1,1]^{N_\text{in}}$. Meanwhile, the post-processing function is another fixed transformation layer that restores outputs from $[-1,1]^{N_\text{out}}$ to their actual dimensional values. This approach can be expressed by modifying Eq. ~\ref{eq:methods:pinn:approximation} as follows:
\begin{align}
    \label{eq:methods:inputoutputscaling:ansatz}
    \bm{u}_{\bm{\theta}}(\bm{x}) &= \bm{f}_\text{post}(\bm{x}) \circ \bm{\mathcal{NN}}(\bm{x};\bm{\theta}) \circ \bm{f}_\text{pre}(\bm{x}) \; .
\end{align}

\section{Model}
\label{sec:model}
In this section, we introduce the benchmark case used in this paper, detailing the geometry of the computational domain, governing equations, and boundary conditions. We also provide a concise overview of how the \gls{pinn} models were validated against a high-fidelity solution generated using the \acrfull{fem} and how this solution was obtained.

In the modeling of stirred tanks, it is common to introduce simplifying assumptions to make the analysis more tractable.\cite{JoshiCFDSimulationSTR2011, Reid2024} One such assumption is that the flow is steady, aiming to find the solution for the flow at a specific moment in time without considering transient effects. Moreover, treating the fluid as Newtonian and incompressible—with constant density and viscosity—simplifies the governing equations. Additionally, initial studies usually concentrate on single-phase flow models before progressing to include the dynamics of multiphase flows.

 Assuming steady flow enables us to sidestep complexities associated with a moving domain, which would arise from the rotating stirrer.

\subsection{Problem domain}
\begin{figure}
    \begin{subfigure}[b]{0.47\textwidth}
        \input{figures/3D_domain}
        \caption{Simplified representation of a three-dimensional stirred tank. The yellow circle highlights the two-dimensional cross-sectional plane through the stirrer, which is detailed in Figure~\ref{fig:model:domain:overview:2d}.}
        \label{fig:model:domain:overview:3d}
    \end{subfigure}
    \hfill
    \begin{subfigure}[b]{0.47\textwidth}
        \input{figures/2D_domain}
        \caption{Schematic representation of the problem domain $\Omega$ inspired by a stirred tank configuration and the problem domain $\Omega_\text{sym} \subset \Omega$ for models which exploit symmetry. The stirrer rotates at a constant angular velocity $\omega$, which leads to a linear velocity profile on the stirrer blades $\Gamma_\text{stirrer}$.}
        \label{fig:model:domain:overview:2d}
    \end{subfigure}
    \label{fig:model:domain:overview}
    \caption{Depiction of a simplified three-dimensional stirred tank geometry (A) and the simplified two-dimensional problem domain that is the focus of this study (B).}
\end{figure}
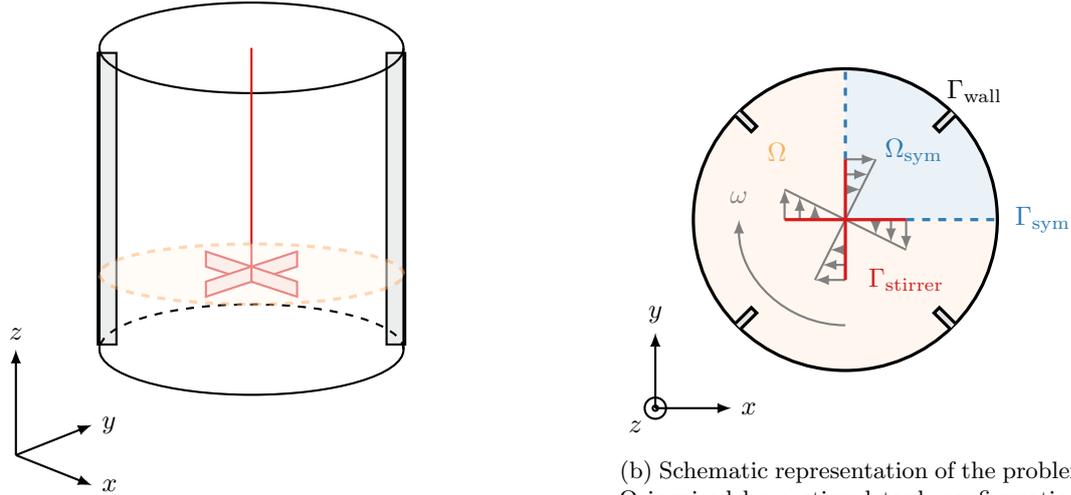

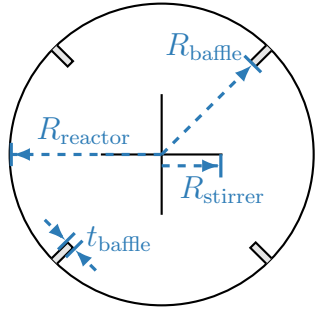
\begin{figure}[ht]
    \centering
    \input{figures/2D_dimensions}
    \caption{Geometrical dimensions of the 2D problem domain.}
    \label{fig:model:domain:dimensions}
\end{figure}

Due to the complexities of modeling flow within realistic three-dimensional stirred tank geometries, we simplify our approach in this case study by focusing on a two-dimensional scenario. As shown in Figure~\ref{fig:model:domain:overview:3d}, we examine the planar flow in a cross-sectional slice through the stirrer plane. This approach simplifies the three-dimensional problem to a two-dimensional one, leading to the computational domain presented in Figure~\ref{fig:model:domain:overview:2d}.
Our reference frame is fixed and aligned with the Cartesian coordinate system centered at the origin of the domain. The stirrer blades are aligned with the $x$- and $y$-axes. The model is trained to predict the flow field in this specific instance. Given the apparent symmetry of the problem, we can reduce the computational domain to $\Omega_\text{sym} \subset \Omega$. The dimensions of the geometry are illustrated in Figure~\ref{fig:model:domain:dimensions}, and their numerical values, along with material properties of the fluid, are listed in Table~\ref{tab:model:domain:data}.

\begin{table}[ht]
    \centering
    \caption{Geometrical dimensions of the 2D problem domain and material properties that are considered constant throughout the paper.}
    \label{tab:model:domain:data}
    \begin{tabular}{lcl}
        %\toprule
        \hline
        Quantity & Value & Unit \\
        %\midrule
        \hline
        $R_\text{stirrer}$ & \num{0.040} & \unit{m} \\ 
        $R_\text{baffle}$ & \num{0.085} & \unit{\meter} \\
        $R_\text{reactor}$ & \num{0.100} & \unit{\meter} \\
        $t_\text{baffle}$ & \num{0.005} & \unit{\meter} \\
        $\rho$ & \num{1000}  & \unit{\kilogram\per\meter\cubed} \\
        $\mu$ & \num{0.001} & \unit{\kilogram\per\meter\per\second} \\
        \hline
        %\bottomrule
    \end{tabular}
\end{table}

\subsection{Governing equations and \glspl{bc}}
Applying the simplifying assumptions outlined above, the system reduces to the stationary, incompressible Navier-Stokes equations for the momentum and mass conservation laws:
\begin{subequations} 
\label{eq:model:governingequations} 
    \begin{align}
    \bm{\mathcal{R}}_{\text{momentum}}[\bm{u}] &= \rho (\bm{v}\cdot\bm{\nabla}) \bm{v} + \bm{\nabla} p - \mu \Delta \bm{v} \;, \\
    \mathcal{R}_{\text{mass}}[\bm{u}] &= \bm{\nabla}\cdot\bm{v} \;.
\end{align}
\end{subequations}
Accordingly, the total PDE residual in Eq.~\ref{eq:methods:pinn:physicslosses:pdes} can be written as $\bm{\mathcal{R}}[\bm{u}] = (\bm{\mathcal{R}}_\text{momentum}[\bm{u}]^T, \mathcal{R}_\text{mass}[\bm{u}])^T$. The outputs of the model---the variables we are solving for---are the velocity vector components and pressure, expressed as $\bm{u} = (v_x, v_y, p)^T$, or  $\bm{u} = (\bm{v}^T, p)^T$.

We apply a no-slip \gls{bc} on the outer tank wall, assuming the wall velocity to be zero. To model the rotation of the stirrer, we impose the \gls{bc} $\bm{v} = \bm{v}_\text{stirrer} \;\text{on} \; \Gamma_\text{stirrer}$, where the stirrer velocity is given by
\begin{align}
    \label{eq:model:stirrerbc}
    \bm{v} = \bm{\omega} \times \bm{r} \;.
\end{align}
In this expression, $\bm{\omega}$ is the angular velocity vector (aligned with the $z$-axis) and $\bm{r}$ is the position vector from the center of rotation. Since $\bm{\omega}$ has non-zero component only in the $z$-direction, the cross product simplifies to 
\begin{align}
    \label{eq:model:domain:2Dstirrervelocity}
    \bm{v}_\text{stirrer} &= \begin{pmatrix} \phantom{-}\omega y \\ -\omega x \end{pmatrix} \; .
\end{align}

Consequently, the boundary residuals on boundaries depicted in Figure~\ref{fig:model:domain:overview:2d} are defined as follows:
\begin{subequations}
\label{eq:model:domain:bcs}
\begin{align}
    \label{eq:model:bcresidualss:wall}
   \bm{\mathcal{B}}_\text{wall}[\bm{u}]  &= \begin{pmatrix} v_x \\ v_y \end{pmatrix} \; , \\
    \label{eq:model:bcresiduals:stirrer}
   \bm{\mathcal{B}}_\text{stirrer}[\bm{u}]  &= \begin{pmatrix} v_x - \omega y \\ v_y + \omega x \end{pmatrix} \; ,
\end{align}
\label{eq:model:bcresiduals}
\end{subequations}
and the \gls{bc} residual in Eq.~\ref{eq:methods:pinn:physicslosses:bcs} is composed as $\bm{\mathcal{B}}[\bm{u}]=(\bm{\mathcal{B}}_\text{wall}[\bm{u}]^T,\bm{\mathcal{B}}_\text{stirrer}[\bm{u}]^T)^T$.

We include the angular velocity as an extra input to the \gls{nn}, training it to predict for a range of impeller speeds. The flow regime is characterized by the Reynolds number, defined as:\cite{Rosseburg2018}
\begin{subequations}
\label{eq:model:bcs:reynolds}
\begin{align}
    \mathrm{Re} &= \frac{N\rho(2R_\text{stirrer})^2}{\mu} \; ,
\end{align}
where
\begin{align}
    N &= \frac{\omega}{2\pi} \; .
\end{align}
\end{subequations}
The models presented in this study are trained for $\mathrm{Re} \in [50, 5000]$. We acknowledge that restricting our approach to the steady Navier-Stokes equations constrains our ability to extend the model to higher $\mathrm{Re}$ values. For these values, directly resolving turbulent fluctuations---an inherently transient phenomenon---is not feasible. In future work, we intend to incorporate a turbulence model into the steady formulation.

If the model leverages the symmetry of the problem, we introduce a \gls{bc} to maintain continuity of the solution over $\Gamma_\text{sym}$, resulting in an additional \gls{bc} residual defined as:
%\begin{align}
%    \label{eq:model:bcs:symmetry}
%    \bm{u}\rvert_{\bm{x}^{(+)}} = \bm{u}\rvert_{\bm{x}^{(-)}} \quad \text{on} \; \Gamma_\text{sym} \; ,
%\end{align}
%resulting in an additional \gls{bc} residual
\begin{align}
    \label{eq:model:bcresiduals:symmetry}
   \bm{\mathcal{B}}_\text{sym}[\bm{u}]  &= \begin{pmatrix} v_x\rvert_{\bm{x}^{(+)}} + v_y\rvert_{\bm{x}^{(-)}}\\ v_x\rvert_{\bm{x}^{(-)}} - v_y\rvert_{\bm{x}^{(+)}} \\ p\rvert_{\bm{x}^{(+)}} - p\rvert_{\bm{x}^{(-)}}
   \end{pmatrix} \; ,
\end{align}
where $\bm{x}^{(+)}$ and $\bm{x}^{(-)}$ are the limiting values approaching approaching $\Gamma_\text{sym}$ from left and right, respectively.

In Section~\ref{sec:results}, we introduce a configuration where only the Dirichlet boundary residuals, as defined in Eq.~\ref{eq:model:bcresiduals}, are added to the classical supervised \gls{nn} loss function, without including \gls{pde} residuals. For simplicity, we will refer to this setup as \acrfull{binn}.

\subsubsection*{\emph{Input and output scaling}}
\label{subsubsec:model:governingequations:inputoutputscaling}
For the basic configuration, when training the model on the computational domain 
$\Omega$ shown in Figure~\ref{fig:model:domain:overview:2d}, the pre-processing function $\bm{f}_\text{pre}$ from Eq.~\ref{eq:methods:inputoutputscaling:ansatz} is defined as follows:
\begin{align}
    \label{eq:model:governingequations:inputoutputscaling:input}
    \tilde{\bm{x}} &= \bm{f}_\text{pre}(\bm{x}) = \begin{pmatrix} \nicefrac{x}{R_\text{reactor}} \\ \nicefrac{y}{R_\text{reactor}} \\ \nicefrac{(\omega-\omega_\text{avg})}{(\omega_\text{max}-\omega_\text{avg})} \end{pmatrix} \; .
\end{align}
Here, $\bm{x}$ is the vector of spatial and parametric inputs, $\bm{x}=(x, y, \omega)^T$ and $\omega_\text{avg}$ is defined as $\omega_\text{avg}=\frac{(\omega_\text{max}-\omega_\text{min})}{2}$.
The network's non-dimensional outputs are then scaled according to:
\begin{align}
    \label{eq:model:governingequations:inputoutputscaling:output}
    \bm{u} &= \bm{f}_\text{post}(\tilde{\bm{u}}) = \begin{pmatrix} \phantom{\rho}v_\text{tip} \tilde{\bm{v}} \\ \rho v_\text{tip}^2 \tilde{p} \end{pmatrix} \; .
\end{align}
In Sections~\ref{subsec:results:targetedlabels} and \ref{subsec:results:labelapproximations}, we employ configurations that leverage symmetry of the problem, training the models on $\Omega_\text{sym}$. In these configurations, spatial inputs are projected onto $\Omega_\text{sym}$ prior to applying $\bm{f}_\text{pre}$. Consequently, the network outputs undergo inverse transformation from $\Omega_\text{sym}$ back to the full domain $\Omega$. The transformations are detailed in Appendix~\ref{app:transformations}.

\subsection{Data \& validation}
\label{subsec:model:datavalidation}
\subsubsection{Reference solutions}
\label{subsubsec:model:datavalidation:refsol}
The reference solutions used as ground truth in this paper were generated through high-fidelity simulations conducted with our in-house solver XNS. \cite{Dirkes2024} Given the relative complexity of the computational domain---which will increase further in three-dimensional scenarios---we represent the geometry using nodes (vertices) of a finite element mesh. Two meshes of varying coarseness are employed---one for the training dataset and another for testing purposes. The training mesh consists of \num{73164} triangular elements and \num{37117} nodes. The testing mesh was refined around the impeller tips and along $\Gamma_\text{wall}$ to accurately capture the Kolmogorov scale for $\mathrm{Re} = 5000$, which was estimated to be approximately $\num{5.4e-5}$. It contains \num{2464018} nodes and \num{4914370} elements. The parametric coordinate $\omega$ of the collocation points was sampled from a random uniform distribution corresponding to $\mathrm{Re} \sim U(50, 5000)$.

\begin{figure}
    \begin{subfigure}[b]{0.47\textwidth}
    \centering
    \input{figures/reference_sol/re_1000/vel_mag.pgf}
    \caption{Magnitude of the velocity field $\|\bm{v}\|_2$.}
    \label{fig:model:datavalidation:refsolution:velocity}
    \end{subfigure}
    \begin{subfigure}[b]{0.47\textwidth}
    \centering
    \input{figures/reference_sol/re_1000/p.pgf}
    \caption{Pressure field $p$.}
    \label{fig:model:datavalidation:refsolution:pressure}
    \end{subfigure}
    \caption{High-fidelity solution computed on a finite element mesh with  \num{4914370} elements for velocity (A) and pressure (B) at Re = 1000 that was used to validate predictions of the PINN models.}
    \label{fig:model:datavalidation:refsolution}
\end{figure}
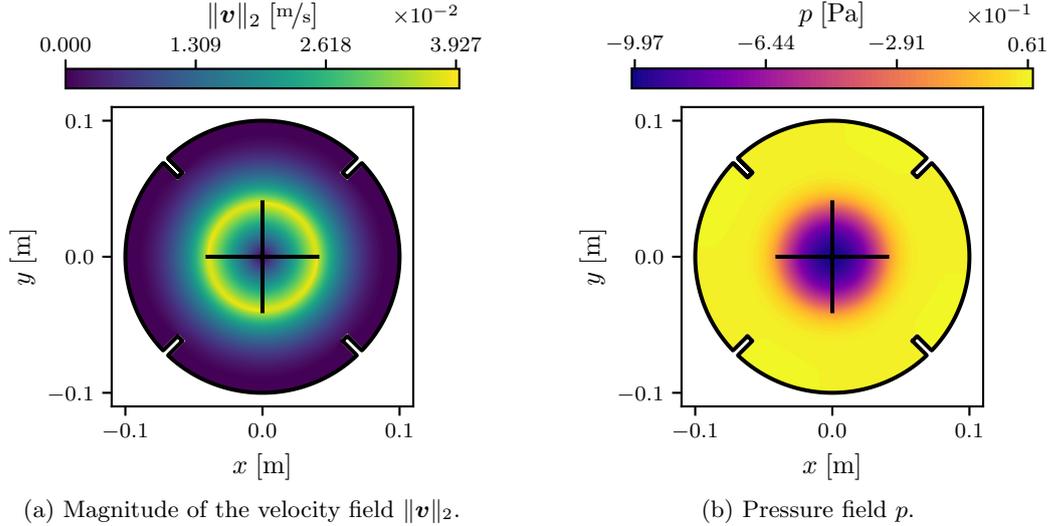

The training data labels were obtained from simulations using the coarser mesh at Reynolds numbers $\mathrm{Re}_\text{train} \in \{\num{50}, \num{500}, \num{2000}, \num{5000}\}$. Subsequently, the trained \gls{nn} and \gls{pinn} models were validated against high-fidelity simulation results obtained on the very fine mesh for Reynolds numbers $\mathrm{Re}_\text{test} \in \{\num{30}, \num{50}, \num{100}, \num{500}, \num{1000}, \num{2000}, \num{3000}, \num{4000}, \num{5000}, \num{6000}\}$. The values of $\mathrm{Re}_\text{test}=\num{30}$ and $\mathrm{Re}_\text{test}=\num{6000}$ fall outside the parameter range for which the \gls{pinn} was trained and serve as tests for its extrapolation ability. It is important to note that the testing mesh may be under-resolved for $\mathrm{Re}=\num{6000}$, given that its minimal mesh width is $\num{5.0e-5}$ while the estimated Kolmogorov scale for this case is $\num{4.4e-5}$; hence some accuracy in the reference solution might be compromised. Figure~\ref{fig:model:datavalidation:refsolution} shows reference solutions for velocity and pressure at $\mathrm{Re}=1000$ on the test mesh composed of \num{4914370} elements.
\subsubsection*{\emph{Remark on reference solution simulation timings}}
All simulations were executed in parallel using \num{256} CPUs. The simulations on the coarser mesh for obtaining training data required between \SIrange{200}{300}{\minute} of CPU time, corresponding to \SIrange{1}{2}{\minute} of wall-clock time. For $\mathrm{Re} = 500$ and higher, simulations needed to be restarted from a previous solution at a lower $\mathrm{Re}$ to ensure an accurate initial guess for convergence. Consequently, for $\mathrm{Re} \geq 500$, the simulation runtime must be calculated as the accumulated time for all preceding Reynolds numbers. For $\mathrm{Re} = 5000$, this results in an accumulated CPU time of \SI{1698}{\minute}, equating to \SI{6.6}{\minute} on \num{256} cores. Test simulations on the finely refined mesh took between \SIrange{800}{2700}{\minute} of CPU time, translating to approximately \SIrange{3}{10}{\minute} of wall-clock time when parallelized across \num{256} cores. These simulations also required restarting from lower $\mathrm{Re}$, resulting in an accumulated wall-clock time of \SI{72}{\minute} on \num{256} cores for $\mathrm{Re} = 6000$.
For more comprehensive information on the simulation execution, please refer to our preceding publication by Trávníková, Wolff, et al.\cite{TravnikovaWolff2024}
\subsubsection{Error metrics}
\label{subsubsec:model:datavalidation:errormetrics}
To assess the performance of the \gls{nn} models, we employed the normalized point-wise $\ell^1$~%and $\ell^2$ 
error metric defined as follows:
\begin{subequations}
\label{eq:model:datavalidation:errormetrics}
\begin{align}
    \delta_{\ell^1}^{(q)} &= \frac{1}{q_\text{norm}}\frac{1}{N_\text{eval}} \sum_{k=1}^{N_\text{eval}}\left|\bm{f}^{(q)}_\text{err}(\bm{x}_k)\right| \; .%, \\
    %\delta_{\ell^2}^{(q)} &= \frac{1}{q_\text{norm}}\sqrt{\frac{1}{N_\text{eval}} \sum_{k=1}^{N_\text{eval}}\left(\bm{f}^{(q)}_\text{err}(\bm{x}_k)\right)^2} \; .
\end{align}
\end{subequations}
In these equations, $N_\text{eval}$ is the number of points at which the \gls{pinn} was evaluated, specifically set to \num{50000}. These points constitute a subset of nodes from the very fine mesh utilized for testing, chosen such that they evenly cover the entire domain. $f_\text{err}^{(q)}$ is a difference between the predicted value and the reference solution defined as
\begin{align}
\label{eq:datavalidation:errorfunction}
    f_\text{err}^{(q)}(\bm{x}) = ||q^\text{PINN}(\bm{x})||_2 - ||q^\text{ref}_{\bm{x}}||_2 \; ,
\end{align}
and $q$ represents the quantity of interest---either the velocity vector $\bm{v}$ or the adjusted pressure $\tilde{p}$. The pressure transformation is performed using
\begin{subequations}
\label{eq:model:datavalidation:pressuretransform}
    \begin{align}
    \tilde{p}^\text{predicted}(\bm{x}) = p^\text{predicted}(\bm{x}) - \max\limits_{\bm{x}\in\bm{X}_\text{eval}}p^\text{predicted} \; ,
    \end{align}
    \begin{align}
    \tilde{p}^\text{ref}_{\bm{x}} = p^\text{ref}_{\bm{x}} - \max\limits_{\bm{x}\in\bm{X}_\text{eval}}p^\text{ref} \; ,
    \end{align}
\end{subequations}
to ensure that the predicted and reference pressure values are on the same scale by setting the maximum pressure value to zero for both.
In Eq.~\ref{eq:model:datavalidation:errormetrics}, $q_\text{norm}$ is the normalizing quantity used to scale the errors. For the pressure error $\delta^{(\tilde{p})}$, we normalize by the difference between the maximum and minimum reference pressures, $\tilde{p}_\text{norm}\equiv p^\text{ref}_\Delta \coloneqq p^\text{ref}_\text{max} - p^\text{ref}_\text{min}$ and for the velocity error metric $\delta^{(\bm{v})}$, we utilize the tip velocity of the stirrer as the normalization factor, $v_\text{norm}\equiv v_\text{tip}\coloneqq \omega R_\text{stirrer}$.

\section{Results \& Discussion}
\label{sec:results}
The following section presents a detailed analysis of the results obtained in this work. Specifically, the effects of label quantity, label placement, and loss formulation on the prediction accuracy of a vanilla \gls{pinn} are investigated and compared to those of a classical supervised \gls{nn}. The study focuses on a surrogate model of a stirred tank, as defined in Section~\ref{sec:model}, trained across a Reynolds number range of $\mathrm{Re} \in [50, 5000]$. We compare various configurations to quantify the number of labeled points required, thereby estimating the amount of measurements or simulations that need to be performed to construct an accurate surrogate model. All experiments were conducted using a single feed-forward architecture with fixed hyperparameters optimized for the vanilla \gls{pinn} via grid search (see Table~~\ref{tab:results:hyperparameters}), allowing for a consistent comparison across configurations. In every configuration, training points are processed in randomly sampled batches, with resampling occurring at a batch-resampling rate of \num{1000}. Configuration-specific hyperparameters, including batch size and loss component scaling, are detailed in Appendix~\ref{app:hyperparameters}. The training times reported in this section were measured on the GPUs of the high-performance computing cluster CLAIX, managed by the IT Center at RWTH Aachen University. It is important to note that training times are influenced by the specific hardware available and may vary across different systems. Therefore, these times should be regarded as a general guideline for comparing the computational expense of training different models.
\begin{table}[ht]
    \centering
    \caption{Shared hyperparameter settings for all configurations.}
    \label{tab:results:hyperparameters}
    \begin{tabular}{lll}
        \hline
        & Hyperparameter & Value \\
        \hline
        \multirow{3}{*}{Architecture} & Number of layers & \num{3} \\
        & Number of neurons per layer & \num{100} \\
        & Activation function & $\tanh$ \\
        \hline
        \multirow{2}{*}{Optimization} & Optimizer & L-BFGS \\
        & Epochs & \num{25000} \\
        \hline
        Sampling & Batch-resampling rate & \num{1000} \\
        \hline
    \end{tabular}
\end{table}

\subsection{Impact of labeled dataset size on prediction accuracy}
\label{subsec:results:datasetsize}
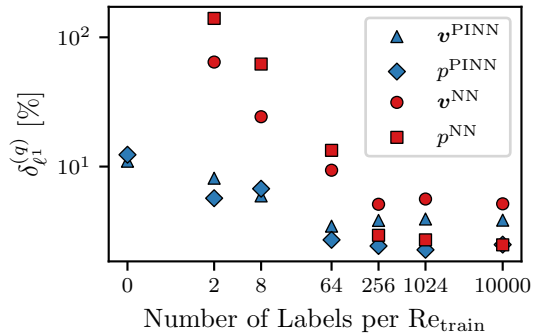
\begin{figure}[ht]
    \centering
    \input{figures/section1/n_points_per_Re.pgf}
    \caption{Effect of the number of labeled points per $\mathrm{Re}_\text{train} \in \{50, 500, 2000, 5000\}$ on the prediction accuracy of the vanilla \gls{pinn} and the classical supervised \gls{nn}. Labels were added at four distinct process parameter values to points randomly distributed across the spatial domain. The models were evaluated at $\mathrm{Re}_\text{test}=1000$.}
    \label{fig:results:datasetsize:comparison}
\end{figure}

First, labels were added at four distinct process parameter values, $\mathrm{Re}_\text{train} \in \{50, 500, 2000, 5000\}$ to randomly sampled points in the spatial domain. Figure~\ref{fig:results:datasetsize:comparison} demonstrates that even a small number of labels can greatly enhance the performance of the vanilla \gls{pinn}, with optimal accuracy achieved using \num{64} labels per $\mathrm{Re}$. At $\mathrm{Re}_\text{test}=1000$, the  $\ell^1$ error on unseen spatial data for the vanilla \gls{pinn} was \qty{11.00}{\percent} for velocity and \qty{12.35}{\percent} for pressure. These errors are reduced to \qty{3.68}{\percent} and \qty{3.52}{\percent}, respectively, upon incorporating \num{64} labels per $\mathrm{Re}_\text{train}$. Increasing the number of labels beyond this point did not lead to further notable improvements in prediction error. 

Conversely, the supervised \gls{nn} only achieved comparable accuracy on unseen data, as measured by the $\ell^1$ error, with errors \qty{5.15}{\percent} for velocity and \qty{2.48}{\percent} for pressure when trained using up to \num{10000} labels per $\mathrm{Re}_\text{train}$. However, Figure~\ref{fig:results:datasetsize:dataonly:velerrordistribution} illustrates that despite being trained on an extensive labeled dataset, the model fails to uphold boundary constraints and maintain problem symmetry without incorporating any information about the underlying physics. Table~\ref{tab:results:datasetsize:timings} indicates that the training duration for the \gls{pinn} model is as much as \num{2.7} times longer than that of the classical supervised \gls{nn}.
\begin{table}[ht]
    \centering
    \caption{Training times of the \gls{pinn} model compared to the training time of the classical supervised \gls{nn} model.}
    \label{tab:results:datasetsize:timings}
    \begin{tabular}{lr}
    %\toprule
      \hline
        Model & Training time [min] \\
    %\midrule
        \hline
        \gls{pinn} & \num{8.12}\\
        \gls{nn} & \num{3.01}\\
        \hline
        %\bottomrule
    \end{tabular}
\end{table}

\begin{figure}[ht]
    \centering
    \input{figures/section1/data/10000p/re_1000/err_vmag.pgf}
    \caption{Normalized velocity magnitude error for predictions by a classical \gls{nn} lacking physics information trained on a dataset with \num{10000} labels per $\mathrm{Re}_\text{train}$ evaluated at $\mathrm{Re}_\text{test}=1000$. Errors are evident near the wall, implying that the model does not adhere to the no-slip boundary condition.}
    \label{fig:results:datasetsize:dataonly:velerrordistribution}
\end{figure}
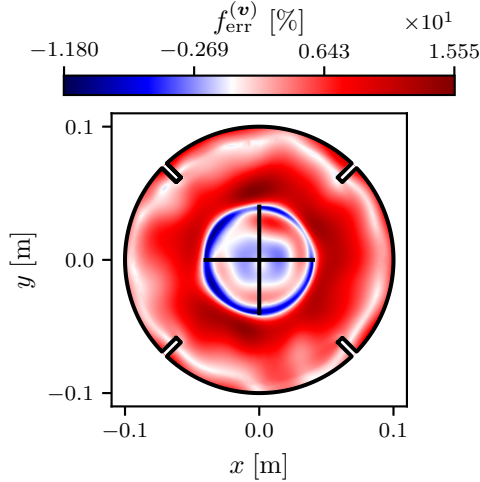
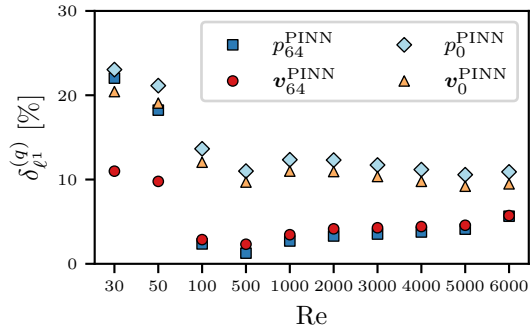
\begin{figure}
    \centering
    \input{figures/section1/errors_all_Re.pgf}
    \caption{Comparison of mean normalized errors in velocity and pressure for physics-based vanilla \gls{pinn} and labeled data-enhanced \gls{pinn} across various Reynolds numbers ($\mathrm{Re}_\text{test}$). The labeled data-enhanced model demonstrates superior performance, particularly within $\mathrm{Re}_\text{test}=100$ and $\mathrm{Re}_\text{test}=5000$, while both models show increased errors at $\mathrm{Re}_\text{test}=50$. Extrapolation errors at $\mathrm{Re}_\text{test}=30$ and $\mathrm{Re}_\text{test}=6000$ are also presented.}
    \label{fig:results:datasetsize:errorsallRe}
\end{figure}

Figure~\ref{fig:results:datasetsize:errorsallRe} shows the comparison of mean normalized errors in velocity and pressure, as obtained from two distinct models: a purely physics-based vanilla \gls{pinn} without labeled data, and a \gls{pinn} enhanced with \num{64} labels per $\mathrm{Re}_\text{train}$ for the four selected $\mathrm{Re}_\text{train}$ values stated above. Both models underwent evaluation across various $\mathrm{Re}_\text{test}$ values spanning the entire training range. Furthermore, the extrapolation capabilities of the model are shown by prediction errors at $\mathrm{Re}_\text{test}=30$ and $\mathrm{Re}_\text{test}=6000$, which lay beyond the established training range.
The \gls{pinn} enhanced with labeled data consistently outperforms across all $\mathrm{Re}_\text{test}$ values, showing stable results particularly between $\mathrm{Re}_\text{test}=100$ and $\mathrm{Re}_\text{test}=5000$, and successfully extrapolating at $\mathrm{Re}_\text{test}=6000$. Both models show reduced accuracy at $\mathrm{Re}_\text{test}=50$, which can be explained by the predicted variables being so small that they lead to trivial solutions. Additionally, the smaller normalizing factor results in relatively large normalized errors from minor absolute differences. We further explore the behavior at lower $\mathrm{Re}$ values in Section~\ref{subsec:results:labelapproximations}. The errors for $\mathrm{Re}_\text{test}=30$, representing extrapolation beyond the training range, aligns with those observed at $\mathrm{Re}_\text{test}=50$.
\begin{figure}[h]
    \centering
    \input{figures/section1/n_points_per_Re_velonly.pgf}
    \caption{Comparison of \gls{pinn} model accuracy with labels for both velocity and pressure versus velocity-only labels at $\mathrm{Re}_\text{test}=1000$ across varying label counts. The figure highlights that mean normalized errors in velocity remain similar between the two models, while the absence of pressure labels results in slightly reduced precision in pressure prediction.}
    \label{fig:results:datasetsize:velocityonly}
\end{figure}
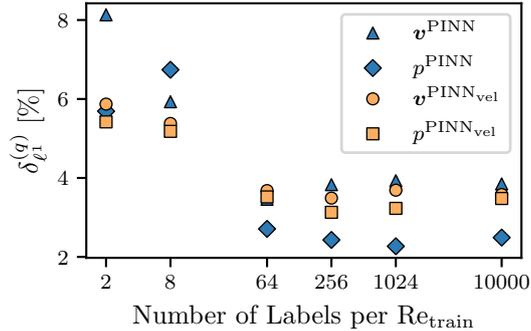

In many instances, particularly when dealing with measurement data rather than high-fidelity simulations, only velocity labels may be available. Figure~\ref{fig:results:datasetsize:velocityonly} evaluates the accuracy of a \gls{pinn} model enhanced with labels for both velocity and pressure against a \gls{pinn} model with an equivalent number of labels added exclusively for velocity, as a function of the label count. The figure demonstrates that mean normalized errors in velocity are comparable between the two models, whereas the model without pressure labels shows slightly decreased accuracy in predicting pressure.

\subsection{Effect of loss components on prediction accuracy}
\label{subsec:results:losscomponents}
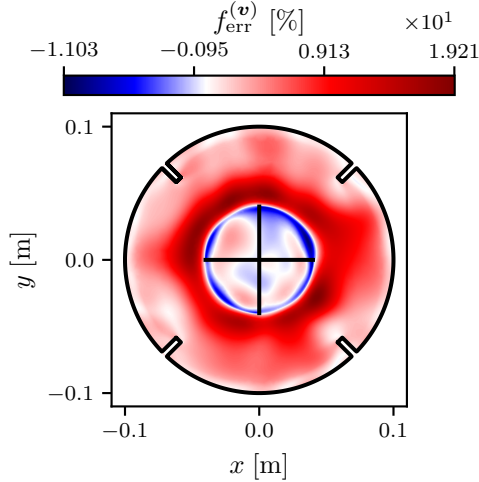
\begin{figure}[h]
    \centering
    \input{figures/section2/dataBC/10000p/re_1000/err_vmag.pgf}
    \caption{Normalized velocity magnitude error at $\mathrm{Re}_\text{test}=1000$ for predictions by a classical \gls{nn} enhanced by boundary residuals as defined in Eq.~\ref{eq:model:bcresiduals} (\gls{binn}) trained on a dataset with \num{10000} labels per $\mathrm{Re}_\text{train}$. Compared to Figure~\ref{fig:results:datasetsize:dataonly:velerrordistribution}, the reduced errors near the walls indicate that the model successfully adheres to the no-slip condition.}
    \label{fig:results:losscomponents:dataBC:velerrordistribution}
\end{figure}

Building on the insights from Section~\ref{subsec:results:datasetsize}, a key question arises: is it essential to incorporate the \gls{pde} residual in training, or would adding only the residuals representing the Dirichlet \gls{bc} suffice for achieving accurate predictions? This approach---omitting the \gls{pde} residual---would enhance efficiency by avoiding the computational complexity associated with gradient calculations through backpropagation. Table~\ref{tab:results:losscomponents} provides a comparative analysis of models with distinct configurations, exploring various combinations of loss components considered in the objective function for cases with \num{64} labels per $\mathrm{Re}$ and \num{10000} labels per $\mathrm{Re}$. The table indicates that when a sufficiently large dataset is available, a classical \gls{nn} augmented with \gls{bc} residuals as defined in Eq.~\ref{eq:model:bcresiduals} (Model III) achieves nearly equivalent performance for both velocity and pressure compared to the \gls{pinn} with incorporated data loss (Model II), according to the selected error metric at $\mathrm{Re}_\text{test}=1000$. For the purposes of this paper, we will refer to the configuration of Model III as \acrfull{binn}. The distribution of the velocity magnitude error of the predictions of Model III is shown in Figure~\ref{fig:results:losscomponents:dataBC:velerrordistribution} and indicates that the model complies with the no-slip \gls{bc} at $\Gamma_\text{wall}$.

\begin{table*}%[ht]
    \centering
    \caption{Comparison of model configurations based on different combinations of loss components in the objective function, evaluated on the test point coordinates at $\mathrm{Re}_\text{test}=1000$ for cases with \num{64} labels per $\mathrm{Re}_\text{train}$ and \num{10000} labels per $\mathrm{Re}_\text{train}$, where $\mathrm{Re}_\text{train} \in \{50, 500, 2000, 5000\}$.}
    \label{tab:results:losscomponents}
    \begin{tabular}{lcccc|rrrrr}
        \hline
        \multirow{3}{*}{Model}& \multirow{3}{*}{$\bm{\mathcal{L}}_{\text{PDE}}$}& \multirow{3}{*}{$\bm{\mathcal{L}}_{\text{BC}}$}& \multirow{3}{*}{$\bm{\mathcal{L}}_{\text{data},\bm{v}}$}& \multirow{3}{*}{$\bm{\mathcal{L}}_{\text{data},p}$}& \multicolumn{4}{c}{$n_\text{labels}$} & \multirow{3}{*}{Training time [min]}\\
        & & & & & \multicolumn{2}{c}{\num{64}}& \multicolumn{2}{c}{\num{10000}}& \\
        & & & & & $\delta_{\ell^1}^{(\bm{v})} \ [\%]$ & $\delta_{\ell^1}^{(p)} \ [\%]$ & $\delta_{\ell^1}^{(\bm{v})} \ [\%]$ & $\delta_{\ell^1}^{(p)} \ [\%]$ & \\
        \hline
        %\midrule
        I. & \checkmark & \checkmark & $\times$ & $\times$ & \num{11.00} & \num{12.35} & - & - & 7.95\\ 
        \textbf{II.} & \textbf{\checkmark} & \textbf{\checkmark} & \textbf{\checkmark} & \textbf{\checkmark}  & \textbf{\num[text-series-to-math]{3.46}} & \textbf{\num[text-series-to-math]{2.71}} & \num{3.85} & \num{2.49} & \textbf{\num[text-series-to-math]8.12}\\
        \textbf{III.} & $\mathbf{\times}$ & \textbf{\checkmark} & \textbf{\checkmark} & \textbf{\checkmark} & \num{4.60} & \num{6.86} & \textbf{\num[text-series-to-math]{4.24}} & \textbf{\num[text-series-to-math]{2.51}} & \textbf{\num[text-series-to-math]3.55}\\
        IV. & \checkmark & $\times$ & \checkmark & \checkmark & \num{6.23} & \num{3.26} & \num{5.11} & \num{2.34} & \num{7.56}\\
        V. & $\times$ & $\times$ & \checkmark & \checkmark & \num{9.37} & \num{13.35} & \num{5.15} & \num{2.48} & \num{3.01}\\
        VI. & \checkmark & \checkmark & \checkmark & $\times$ & \num{3.68} & \num{3.52} & \num{3.58} & \num{3.48} & \num{8.10}\\
        %\bottomrule
        \hline
    \end{tabular}
\end{table*}

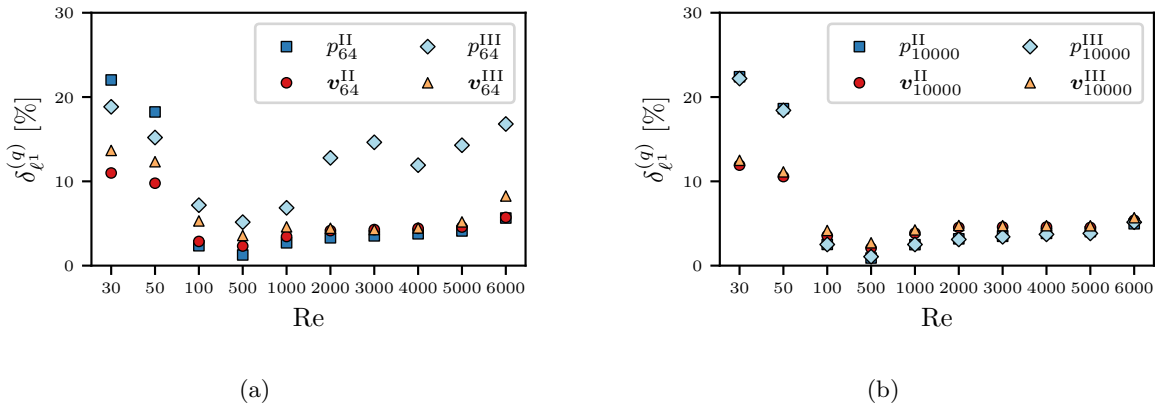
\begin{figure}
    \begin{subfigure}[b]{0.47\textwidth}
        \input{figures/section2/errors_all_Re_64.pgf}
        \caption{}
        \label{fig:results:losscomponents:errorsall:64p}
    \end{subfigure}
    \hfill
    \begin{subfigure}[b]{0.47\textwidth}
        \input{figures/section2/errors_all_Re_10k.pgf}
        \caption{}
        \label{fig:results:losscomponents:errorsall:10kp}
    \end{subfigure}
    \caption{Comparison of mean normalized errors in velocity and pressure for a labeled data-enhanced \gls{pinn} (Model II as defined in Table~\ref{tab:results:losscomponents}) and a \gls{binn} augmented solely by \gls{bc} residuals (Model III) across various $\mathrm{Re}_\text{test}$ values. The comparison is made for two cases with differing labeled dataset sizes: \num{64} labels per $\mathrm{Re}_\text{train}$ (a) and \num{10000} labels per $\mathrm{Re}_\text{train}$ (b).}
    \label{fig:results:losscomponents:errorsall}
\end{figure}

Figure~\ref{fig:results:losscomponents:errorsall} provides a comparison of $\ell^1$-errors for velocity and pressure predictions by Model II and Model III across various $\mathrm{Re}$ values, both within and beyond the training range. Two cases are with different labeled dataset sizes are considered: \num{64} labeled points per $\mathrm{Re}_\text{train}$ and \num{10000} labeled points per $\mathrm{Re}_\text{train}$. In the scenario with the smaller labeled dataset (Figure~\ref{fig:results:losscomponents:errorsall:64p}), Model II outperforms Model III, particularly in pressure prediction and for larger $\mathrm{Re}_\text{test}$. Additionally, Model II demonstrates superior extrapolation to higher $\mathrm{Re}_\text{test}$ values outside the training range compared to Model III. However, when trained on the larger dataset, both models exhibit nearly identical performance (Figure~\ref{fig:results:losscomponents:errorsall:10kp}).

\subsection{Impact of the number of labeled process parameter values on prediction accuracy}
\label{subsec:results:parameterdensity}
\begin{figure}
    \begin{subfigure}[b]{0.47\textwidth}
        \input{figures/section3/n_Re_velocity.pgf}
        \caption{Mean normalized velocity errors}
        \label{fig:results:parameterdensity:comparison:nRevelocity}
    \end{subfigure}
    \hfill
    \begin{subfigure}[b]{0.47\textwidth}
        \input{figures/section3/n_Re_pressure.pgf}
        \caption{Mean normalized pressure errors}
        \label{fig:results:parameterdensity:comparison:nRepressure}
    \end{subfigure}
    \caption{Prediction accuracy of surrogate models with varying $\mathrm{Re}_\text{train}$ values in the labeled training dataset across three configurations: data-enhanced \gls{pinn}, classical supervised \gls{nn}, and \gls{binn}. Subfigure (a) shows velocity errors, while subfigure (b) depicts pressure errors. For \gls{nn} and \gls{binn}, labeled points per $\mathrm{Re}_\text{train}$ were allocated to maintain a constant total number across experiments: \num{10000} labeled points per $\mathrm{Re}_\text{train}$ for \num{4} $\mathrm{Re}_\text{train}$ values ($\mathrm{Re}_\text{train}\in\{50, 500, 2000, 5000\}$), \num{13500} for \num{3} $\mathrm{Re}$ values ($\mathrm{Re}_\text{train}\in\{50, 500, 5000\}$), and \num{20000} for \num{2} $\mathrm{Re}$ values ($\mathrm{Re}_\text{train}\in\{50, 5000\}$). For the \gls{pinn}, \num{64} labeled points per $\mathrm{Re}_\text{train}$ were added. Errors were evaluated at $\mathrm{Re}_\text{test}=1000$.}
    \label{fig:results:parameterdensity:comparison}
\end{figure}
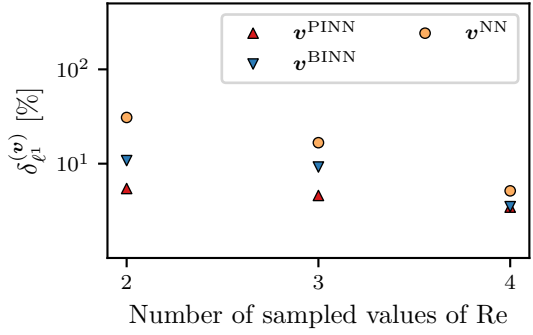
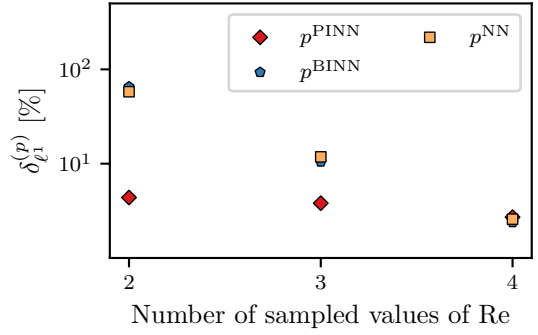

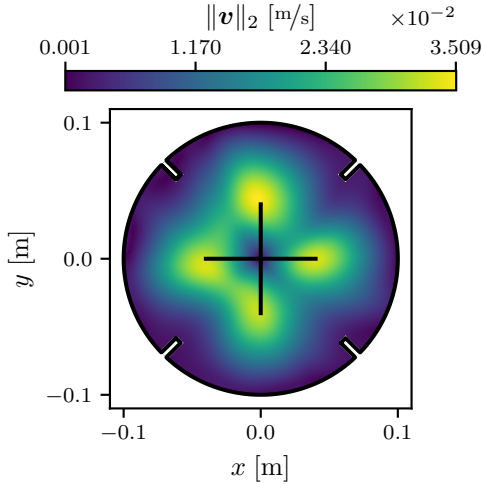
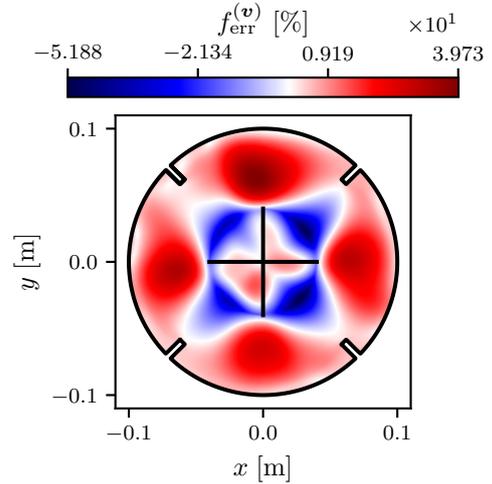
\begin{figure}
    \begin{subfigure}{0.47\textwidth}
        \input{figures/section3/dataBC/3Re/13500p/re_1000/vmag.pgf}
        \caption{Velocity magnitude.}
        \label{fig:results:parameterdensity:3RedataBC:velprediction}
    \end{subfigure}
    \hfill
    \begin{subfigure}{0.47\textwidth}
        \input{figures/section3/dataBC/3Re/13500p/re_1000/err_vmag.pgf}
        \caption{Normalized velocity magnitude error.}
        \label{fig:results:parameterdensity:3RedataBC:velerror}
    \end{subfigure}
    \caption{Velocity magnitude field (a) and normalized velocity magnitude error (b) at $\mathrm{Re}_\text{test}=1000$ of the \gls{binn} model with labeled data added for $\mathrm{Re}_\text{train}\in \{50, 500, 5000\}$. Although the mean normalized velocity error of \qty{9.21}{\percent} does not seem high, the predictions significantly deviate from the reference solution.}
    \label{fig:results:parameterdensity:3RedataBC}
\end{figure}

Thus far, we have focused on how the number of labeled points added per $\mathrm{Re}_\text{train}$ affects model performance, while keeping both the values and count of $\mathrm{Re}_\text{train}$, for which the labels are added, constant. However, in certain scenarios---such as when training a surrogate model with data from high-fidelity simulations rather than measurement data---we face little restrictions on the spatial points available for labeling.  Instead, limitations arise from the number of process parameter (in this case, $\mathrm{Re}$) for which labeled data must be acquired. %--essentially, by the sampling density of this parameter.  
Figure~\ref{fig:results:parameterdensity:comparison} illustrates the prediction accuracy of surrogate models with varying count of $\mathrm{Re}_\text{train}$ values in the labeled training dataset, across three configurations: the data-enhanced \gls{pinn}, classical supervised \gls{nn}, and \gls{binn}. For the \gls{pinn}, only \num{64} labeled points per $\mathrm{Re}_\text{train}$ were used, as Figure~\ref{fig:results:datasetsize:comparison} indicates that increasing the size of the labeled dataset does not further enhance model performance. For both the \gls{nn} and \gls{binn}, labeled points per $\mathrm{Re}_\text{train}$ were allocated to maintain an approximately constant total number of labeled points across experiments: \num{10000} labeled points per $\mathrm{Re}_\text{train}$ for \num{4} $\mathrm{Re}_\text{train}$ values, \num{13500} for \num{3} $\mathrm{Re}_\text{train}$ values, and \num{20000} for \num{2} $\mathrm{Re}_\text{train}$ values. It is evident that with \num{4} $\mathrm{Re}_\text{train}$ values and a sufficiently large labeled dataset, a supervised \gls{nn} can produce an adequately accurate surrogate model, and the \gls{binn} can achieve even greater accuracy (as demonstrated earlier in Section~\ref{subsec:results:losscomponents}). However, when the number of $\mathrm{Re}_\text{train}$ values decreases, only the \gls{pinn} maintains reasonable accuracy for values not included in the dataset. The figure illustrates that an accurate \gls{pinn} surrogate can be achieved by supplying simulation or measurement data exclusively for the boundary values of the process parameter range interval. This model achieves mean normalized errors of \qty{5.45}{\percent} for velocity and \qty{4.38}{\percent} for pressure.
The mean normalized error values of \qty{9.21}{\percent} and \qty{10.44}{\percent} for velocity and pressure, respectively, for the \gls{binn} for \num{3} $\mathrm{Re}_\text{train}$ values might not seem particularly high, especially in comparison to the vanilla \gls{pinn} relying solely on physics information (Model I in Table~\ref{tab:results:losscomponents}). However, the predicted velocity magnitude field and the associated error distribution as presented in Figure~\ref{fig:results:parameterdensity:3RedataBC} show that the results diverge considerably from the reference solution depicted in Figure~\ref{fig:model:datavalidation:refsolution:velocity}.

\subsection{Robustness analysis: Influence of randomized label placement}
\label{subsec:results:robustness}
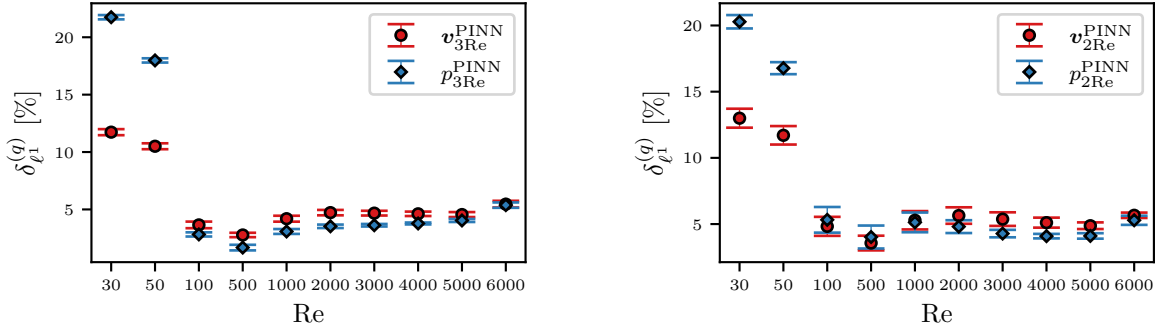
\begin{figure}
    \begin{subfigure}{0.47\textwidth}
        \input{figures/section4/3Re/avg_errors_Re.pgf}
        \caption{Configuration with \num{64} labels added for each $\mathrm{Re}_\text{train} \in \{50, 500, 5000\}$.}
        \label{fig:results:robustness:3Re}
    \end{subfigure}
    \hfill
    \begin{subfigure}{0.47\textwidth}
        \input{figures/section4/2Re/avg_errors_Re.pgf}
        \caption{Configuration with \num{64} labels added for each $\mathrm{Re}_\text{train} \in \{50, 5000\}$.}
        \label{fig:results:robustness:2Re}
    \end{subfigure}
    \caption{Robustness of \gls{pinn} models with respect to random label placement for two different configurations. Figure (a) illustrates the configuration with labels added for \num{3} $\mathrm{Re}_\text{train}$ values, showing minimal fluctuation in errors, with standard deviations less than \qty{0.3}{\percent}, equating to up to \qtyrange{5}{10}{\percent} of the average mean normalized error values. Figure (b) presents the configuration with labels added for \num{2} $\mathrm{Re}_\text{train}$ values, highlighting increased variability, especially for values not seen during training. Standard deviations can reach up to \qty{0.7}{\percent} for velocity and \qty{1}{\percent} for pressure, corresponding to up to \qty{15}{\percent} and \qty{20}{\percent} of their respective average mean normalized errors.}
    \label{fig:results:robustness}
\end{figure}

In Section~\ref{subsec:results:datasetsize}, we demonstrated how the prediction accuracy of the labeled data-enhanced \gls{pinn} is influenced by the size of the labeled dataset, with labels allocated to a random subset of training points. This naturally raises the question: how does the location of these labels affect model performance? In this subsection, we investigate the robustness of two selected \gls{pinn} models: one with \num{64} labeled points added for each $\mathrm{Re}_\text{train} \in \{50, 500, 5000\}$ and another with \num{64} labeled points added for each $\mathrm{Re}_\text{train} \in \{50, 5000\}$. We asses their robustness by re-training each model ten times using different randomly selected subsets. Figure~\ref{fig:results:robustness} presents the average mean normalized errors for velocity and pressure, along with their standard deviations, calculated from \num{10} models trained using the same configuration but with different randomly selected subsets across each $\mathrm{Re}_\text{train}$. Figure~\ref{fig:results:robustness:3Re} corresponds to the configuration with \num{3} $\mathrm{Re}_\text{train}$ values, while Figure~\ref{fig:results:robustness:2Re} corresponds to the configuration with \num{2} $\mathrm{Re}_\text{train}$ values.
Figure~\ref{fig:results:robustness:3Re} demonstrates that the configuration with labels added for \num{3} $\mathrm{Re}_\text{train}$ values exhibits strong robustness regarding the random placement of labeled points. Standard deviations for both mean velocity and pressure prediction errors remain below \qty{0.3}{\percent} in normalized error units, which corresponds to up to \qtyrange{5}{10}{\percent} of the average mean normalized error values. This indicates that despite variations in label locations, the overall performance remains consistent across all $\mathrm{Re}_\text{test}$ values. For the configuration with labels added for \num{2} $\mathrm{Re}_\text{train}$ values, Figure~\ref{fig:results:robustness:2Re} reveals greater variability in errors, particularly for $\mathrm{Re}_\text{test}$ values unseen during training. The standard deviations can reach up to \qty{0.7}{\percent} for velocity and \qty{1}{\percent} for pressure in normalized error units, representing up to \qty{15}{\percent} of the average mean normalized velocity error and up to \qty{20}{\percent} of the average mean normalized pressure error. This indicates a less stable performance when labels for fewer
$\mathrm{Re}_\text{train}$ values are used. However, increasing the number of labeled spatial points could potentially increase robustness of the model.

\subsection{Targeted label allocation: Reducing label count by focusing on high-error areas}
\label{subsec:results:targetedlabels}
Section~\ref{subsec:results:robustness} demonstrated the robustness of models to random label allocations. In this section, we investigate the impact of intentionally adding labeled data to regions with the highest error values. Experience indicates that vanilla \gls{pinn} models exhibit their largest prediction errors immediately after the impeller tips. We strategically introduce labels within this area (with $r$ ranging from \qtyrange{0.04}{0.07}{m}). To achieve optimal prediction accuracy with the least possible number of labels, we leverage the symmetry of the problem by reducing the spatial computational domain to $\Omega_\text{sym}$ as defined in Figure~\ref{fig:model:domain:overview:2d}, and imposing additional \gls{bc} on $\Gamma_\text{sym}$ according to Eq.~\ref{eq:model:bcresiduals:symmetry}. For efficiency, we add only velocity labels, as Section~\ref{subsec:results:datasetsize}, Figure~\ref{fig:results:datasetsize:velocityonly} demonstrated that pressure labels can be omitted without compromising accuracy. This approach is advantageous in situations where only measurement data is accessible, as typically only velocity labels are available.

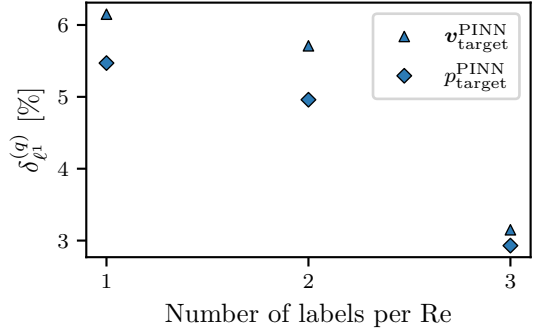
\begin{figure}[t]
    \centering
    \input{figures/section5/n_points_per_Re_targeted.pgf}
    \caption{Mean velocity and pressure errors as a function of the number of labeled points per $\mathrm{Re}_\text{train}$ for a \gls{pinn} utilizing symmetry, with labels strategically placed in critical regions ($0.04 <= r <= 0.07$) for $\mathrm{Re}_\text{train} \in \{50, 5000\}$. The errors were evaluated at $\mathrm{Re}_\text{test}=1000$.}
    \label{fig:results:targetedallocation:datasetsize}
\end{figure}

Figure~\ref{fig:results:targetedallocation:datasetsize} presents the mean velocity and pressure errors for a \gls{pinn} that exploits symmetry and has labels strategically added to points with $r$ between \num{0.04} and \num{0.07} for the boundary values of the process parameter range ($\mathrm{Re}_\text{train} \in \{50, 5000\}$). The figure compares errors across models where \num{3}, \num{2} and \num{1} labeled point per $\mathrm{Re}_\text{train}$ were added.
The figure demonstrates that the model with \num{3} added points within the range $0.04 <= r <= 0.07$  for each $\mathrm{Re}_\text{train}$ achieves mean normalized errors of \qty{3.15}{\percent} for velocity and \qty{2.93}{\percent} for pressure at $\mathrm{Re}_\text{test}=1000$. 

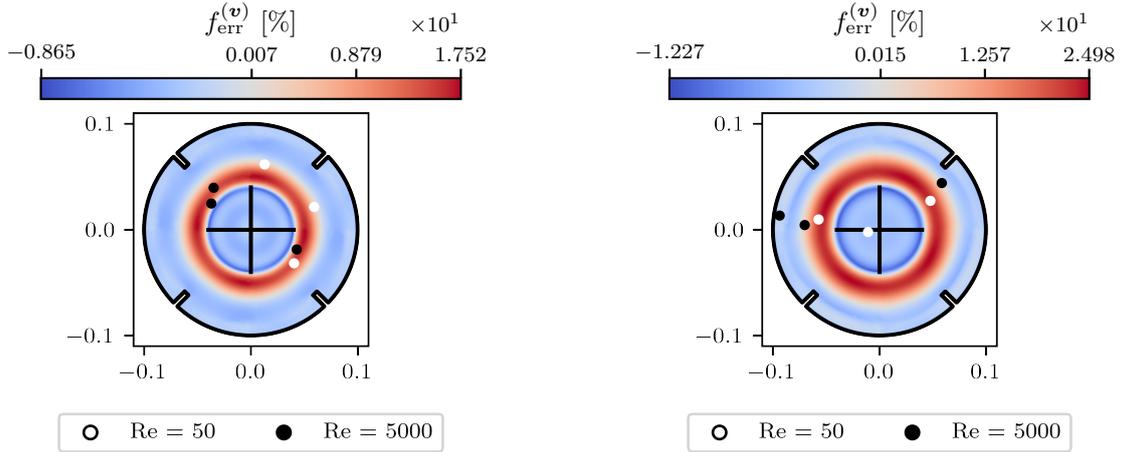
\begin{figure}
    \begin{subfigure}{0.47\textwidth}
        \input{figures/section5/targeted/re_1000/err_vmag.pgf}
        \caption{Targeted label allocation.}
        \label{fig:results:targetedallocation:pointlocations:targeted}
    \end{subfigure}
    \hfill
    \begin{subfigure}{0.47\textwidth}
        \input{figures/section5/random/re_1000/err_vmag.pgf}
        \caption{Random label allocation.}
        \label{fig:results:targetedallocation:pointlocations:random}
    \end{subfigure}
    \caption{Comparison of label allocation strategies in training datasets. Locations of strategically placed labels in the region with the largest error ($0.04 <= r <= 0.07$) (a) and positions of randomly added labels (b) for configurations with \num{3} labels added for each $\mathrm{Re}_\text{train} \in \{50, 5000\}$.}
    \label{fig:results:targetedallocation:pointlocations}
\end{figure}

\begin{figure}
    \begin{subfigure}{0.47\textwidth}
        \input{figures/section5/vel_error_Re_targeted_vs_random.pgf}
        \caption{Mean normalized errors in velocity magnitude prediction.}
        \label{fig:results:targetedallocation:errorsallRe:velocity}
    \end{subfigure}
    \hfill
    \begin{subfigure}{0.47\textwidth}
    \input{figures/section5/press_error_Re_targeted_vs_random.pgf}
        \caption{Mean normalized errors in pressure prediction.}
        \label{fig:results:targetedallocation:errorsallRe:pressure}
    \end{subfigure}
    \caption{Comparison of mean normalized errors for velocity and pressure between models with different label placement strategies across various $\mathrm{Re}_\text{test}$ values.}
    \label{fig:results:targetedallocation:errorsallRe}
\end{figure}
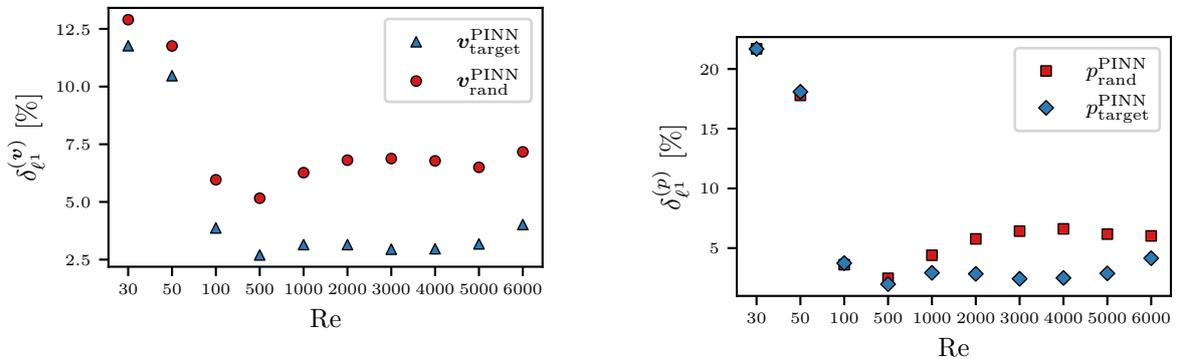

Next, we assess the prediction accuracy of the model trained with strategically added labeled points against a baseline model trained with an equal number of randomly added labeled points. Figure~\ref{fig:results:targetedallocation:pointlocations:targeted} illustrates the locations of strategically placed labels in areas with the largest error. Figure~\ref{fig:results:targetedallocation:pointlocations:random} shows the positions of randomly added labels for a configuration with \num{3} labeled points per $\mathrm{Re}_\text{train}$. Both sets of points and their corresponding labels were projected onto $\Omega_\text{sym}$ using input and output transformations. 
Figure~\ref{fig:results:targetedallocation:errorsallRe} presents a comparison of mean normalized errors for both velocity and pressure between the two models across different 
$\mathrm{Re}_\text{test}$ values. The model with strategically added labels consistently outperforms in velocity predictions across all $\mathrm{Re}_\text{test}$ values, reducing errors by more than half for most $\mathrm{Re}_\text{test}$ values (e.g., from \qty{7}{\percent} to \qty{3}{\percent}). For pressure, both models show comparable performance at lower $\mathrm{Re}_\text{test}$ values; however, for $\mathrm{Re}_\text{test}>500$, the model with targeted labels reduces errors by a factor of two (e.g., from \qty{6}{\percent} to \qty{3}{\percent}).

\subsection{Replacing real labels with approximations: Reducing need for simulation data}
\label{subsec:results:labelapproximations}
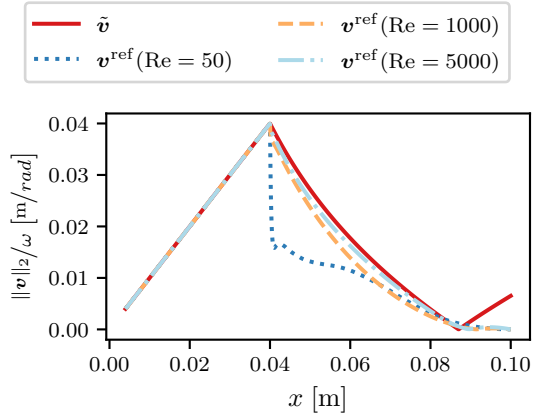
\begin{figure}
    \centering
    \input{figures/section6/vel_profile_phi0.pgf}
    \caption{Velocity magnitude profile along the reactor radius at $y=0$, with magnitudes normalized by angular velocity $\omega$ for uniform scaling. The figure illustrates that the normalized profiles for $\mathrm{Re}=1000$ and $\mathrm{Re}=5000$ are notably similar and their magnitude can be approximated using a polynomial function $\|\tilde{\bm{v}}\|_2$.}
    \label{fig:results:approximation:profiles}
\end{figure}

We have demonstrated that an accurate surrogate model for the 2D sirred tank can be constructed using a vanilla \gls{pinn} enhanced with velocity labels from as few as six datapoints. However, in this section, we will illustrate how the prediction accuracy of our vanilla \gls{pinn} model can be significantly improved without relying on actual simulation or measurement data, but rather by utilizing an approximation of the velocity field. Figure~\ref{fig:results:approximation:profiles} depicts the velocity magnitude profile along the reactor radius $r$ (represented by the $x$-axis at $y=0$), with magnitudes normalized by angular velocity $\omega$ to ensure uniform scaling. The figure reveals that normalized profiles for $\mathrm{Re}=1000$ and $\mathrm{Re}=5000$ are remarkably similar and can be approximated by the following polynomial function:
\begin{align}
    \label{eq:results:approximation:vBC}
    \|\tilde{\bm{v}}\|_2 &= \begin{cases}
        \omega r & r\leq R_\text{stirrer} \\
        \omega R_\text{stirrer}\frac{R_\text{stirrer}(r^2-R_\star^2)}{r(R_\text{stirrer}^2-R_\star^2)} & R_\text{stirrer}<r
    \end{cases} \; ,
\end{align}
where $r=\sqrt{x^2 + y^2}$ and $R_\star$ is a tuning parameter to modify the slope of the velocity profile. Here, $R_\star=0.0875\;\text{m}$ was selected. Further details regarding the choice of this value and the derivation process for $\tilde{\bm{v}}$ are available in our previous paper.\cite{TravnikovaWolff2024} The approximate labels for velocity components $\tilde{v}_x$ and $\tilde{v}_y$ were obtained by projecting $\|\tilde{\bm{v}}\|$ onto their respective directions:
\begin{subequations}
    \label{eq:results:approximation:vxvyprojection}
    \begin{align}
        \tilde{v}_x &= \|\tilde{\bm{v}}\| \cdot \cos{\theta}\;, \\
        \tilde{v}_y &= \|\tilde{\bm{v}}\| \cdot \sin{\theta}\;,
    \end{align}
\end{subequations}
where $\theta$ is the angle between the velocity vector and the $x$-axis.
\\As apparent from Figure~\ref{fig:results:approximation:profiles}, the polynomial function diverges notably from real profiles near $x=0.1$, close to $\Gamma_\text{wall}$; however, our focus is on adding approximate labels in the high-error region just after the impeller tips, where the approximation appears accurate for $\mathrm{Re}=1000$ and $\mathrm{Re}=5000$. It is evident that between $\mathrm{Re}=1000$ and $\mathrm{Re}=50$, the flow character changes significantly, with a discontinuity in the velocity profile at the impeller tip for $\mathrm{Re}=50$ (and presumably all $\mathrm{Re} <= 50$), which $\tilde{\bm{v}}$ fails to capture effectively.

The \gls{pinn} was trained on $\Omega_\text{sym}$ and the data loss was calculated for a dataset with \num{256} points with $0.04 < r < 0.06$ and $\omega$ uniformly sampled from the range corresponding to $\mathrm{Re} \in [50, 5000]$. Labels for these points were obtained by evaluating the approximate polynomial function $\tilde{\bm{v}}$ at these points.

\begin{figure}
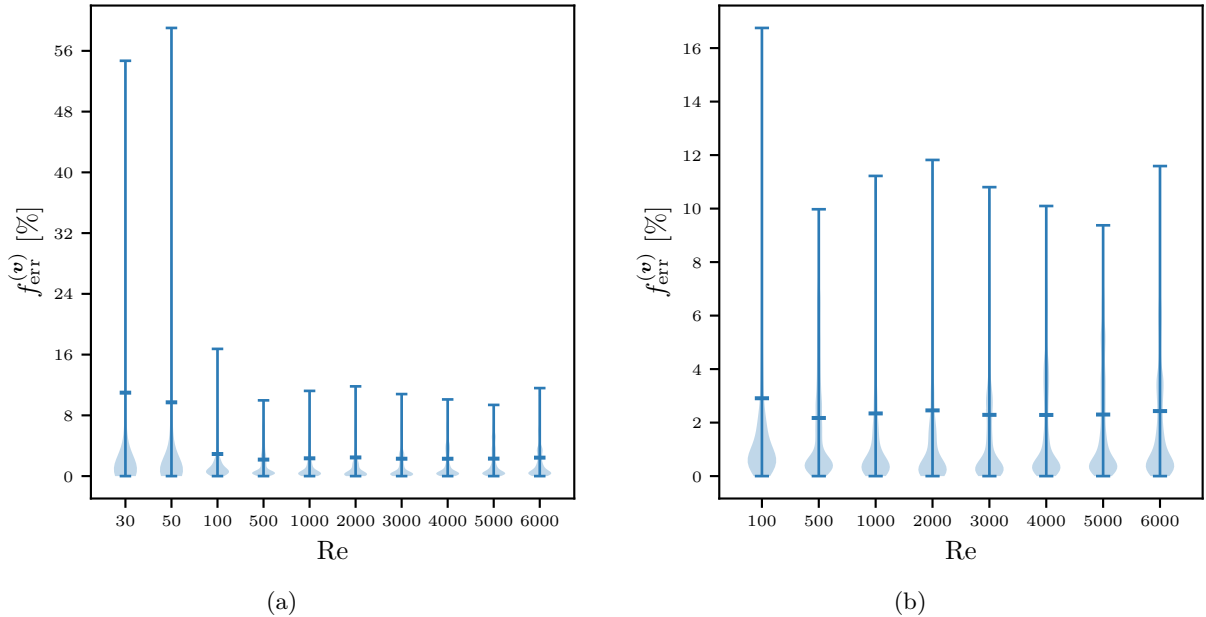

    \begin{subfigure}{0.47\textwidth}
        \input{figures/section6/errors_vmag.pgf}
    \caption{}
    \label{fig:results:approximation:vmag_errors:all}
    \end{subfigure}
    \hfill
    \begin{subfigure}{0.47\textwidth}
    \input{figures/section6/errors_vmag_100-6000.pgf}
    \caption{}
    \label{fig:results:approximation:vmag_errors:100-6000}
    \end{subfigure}
    \caption{(a) Violin plot showing velocity magnitude prediction error distributions across all $\mathrm{Re}_\text{test}$ values, highlighting variations in model performance. (b) Zoomed-in violin plot focusing on velocity prediction error distributions for $\mathrm{Re}_\text{test}$ values between \num{100} and \num{6000}, providing a clearer view of consistent performance across this range.}
    \label{fig:results:approximation:vmag_errors}
\end{figure}

\begin{figure}
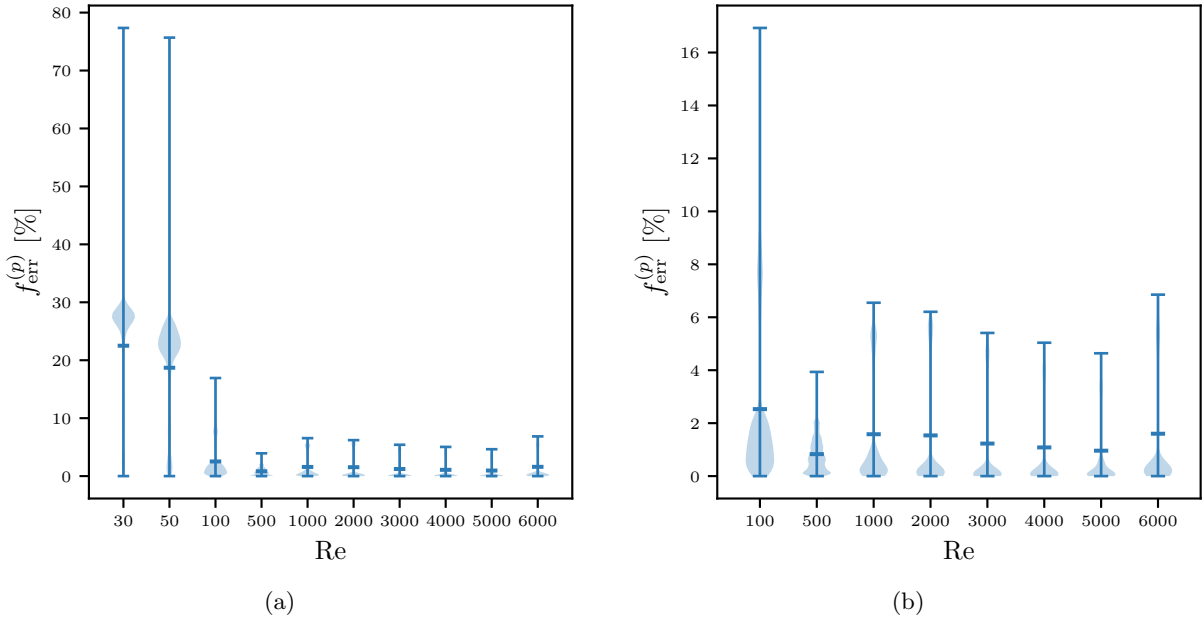

    \begin{subfigure}{0.47\textwidth}
        \input{figures/section6/errors_p.pgf}
    \caption{}
    \label{fig:results:approximation:press_errors:all}
    \end{subfigure}
    \hfill
    \begin{subfigure}{0.47\textwidth}
    \input{figures/section6/errors_p_100-6000.pgf}
    \caption{}
    \label{fig:results:approximation:press_errors:100-6000}
    \end{subfigure}
    \caption{(a) Violin plot showing pressure prediction error distributions across all $\mathrm{Re}_\text{test}$ values, highlighting variations in model performance. (b) Zoomed-in violin plot focusing on velocity prediction error distributions for $\mathrm{Re}_\text{test}$ values between \num{100} and \num{6000} for greater clarity.}
    \label{fig:results:approximation:press_errors}
\end{figure}

Figure~\ref{fig:results:approximation:vmag_errors:all} illustrates the distribution of normalized errors for velocity magnitude fields, while Figure~\ref{fig:results:approximation:press_errors:all} depicts the distribution of normalized errors for pressure fields predicted by the model across various $\mathrm{Re}_\text{test}$ values. The figures clearly indicate that both mean and maximum error values for velocity and pressure at $\mathrm{Re}_\text{test} <= 50$ are substantially larger compared to those at higher $\mathrm{Re}_\text{test}$ values. As previously demonstrated in Figure~\ref{fig:results:approximation:profiles}, this discrepancy arises from the distinct nature of the velocity profile at lower $\mathrm{Re}$ values, which is poorly approximated by $\tilde{\bm{v}}$, leading to training on inaccurate labels for this range. The discontinuity at the stirrer tip poses challenges even for models trained on high-fidelity simulation data at $\mathrm{Re}_\text{test}=50$, as shown for instance in Figure~\ref{fig:results:targetedallocation:errorsallRe}, where errors for $\mathrm{Re}_\text{test}=30$ and $\mathrm{Re}_\text{test}=50$ remain significantly larger than for other $\mathrm{Re}_\text{test}$ values. 

Figures~\ref{fig:results:approximation:vmag_errors:100-6000} and \ref{fig:results:approximation:press_errors:100-6000} offer a zoomed-in perspective on the violin plots showing velocity and pressure prediction error distributions for $\mathrm{Re}_\text{test} \in \{100, 500, 1000, 2000, 3000, 4000, 5000, 6000\}$ for greater clarity. These figures clearly demonstrate that the model maintains consistent performance across these $\mathrm{Re}_\text{test}$ values, exhibiting mean normalized velocity and pressure prediction errors around \qty{2.5}{\percent}. This consistency is evident even for $\mathrm{Re}_\text{test}=6000$, which lies beyond the training range and requires the \gls{pinn} to extrapolate. The slightly elevated errors at $\mathrm{Re}_\text{test}=100$ are likely due to the differing flow characteristics associated with lower $\mathrm{Re}$ values.

\subsection{Summary}
\label{subsec:summary}

In this study, we demonstrated the capability of vanilla \glspl{pinn} to efficiently construct surrogate models for flow fields in stirred tanks with minimal data requirements. We first quantified the number of labeled data points needed at each process parameter value to achieve high prediction accuracy, comparing a vanilla \gls{pinn} to a classical supervised \gls{nn} without physical constraints. Our findings reveal that adding as few as \num{8} labeled datapoints per Reynolds number value to the \gls{pinn} training reduces prediction errors from \qty{11}{\percent} to \qty{5.9}{\percent} for velocity and from \qty{12.4}{\percent} to \qty{6.7}{\percent} for pressure, compared to \glspl{pinn} without data loss.

Moreover, regardless of the number of labeled training points used, \glspl{pinn} consistently outperform purely data-driven \glspl{nn} on unseen data. However, when using \num{256} labeled points per $\mathrm{Re}_\text{train}$ or more, velocity errors for \glspl{pinn} range from \qtyrange{3}{4}{\percent}, while \gls{nn} errors are approximately \qtyrange{5}{6}{\percent}; pressure errors are comparable between both models. This prompts consideration of whether the increased training times—more than double those of classical \glspl{nn}—are warranted if large labeled datasets consisting of hundreds or thousands of points, are accessible.

We also examined the performance differences between the \gls{pinn} model with labels added for all solution variables versus a \gls{pinn} model with velocity labels only, acknowledging that pressure data can be challenging to obtain from measurements. Surprisingly, when limited labeled points are available, models using only velocity labels achieve slightly lower errors than those with complete labeling. This may be due to variability in the magnitude of pressure labels across different $\mathrm{Re}$ values being much higher than that of \gls{pde} residuals, potentially disrupting optimization landscapes--a challenge that could be mitigated by normalizing pressure labels or carefully scaling loss components.

Next, we assessed the impact of individual loss components--specifically the \gls{pde} residual, \glspl{bc}, and data loss--on prediction accuracy. Our findings indicate that when a sufficiently large dataset is available, incorporating only the \gls{bc} residuals into the loss function---a configuration we refer to as \gls{binn}---yields performance comparable to the \gls{pinn} (with prediction errors around \qty{4}{\percent} for velocity and \qty{2.5}{\percent} for pressure at $\mathrm{Re}_\text{test}=1000$). This approach does not increase training time relative to a classical supervised \gls{nn}, as the primary factor contributing to longer training times is the computation of derivatives required by the \gls{pde} residual. However, for smaller labeled datasets, incorporating the \gls{pde} residual becomes essential for achieving accurate predictions, especially for pressure.

Subsequently, we investigated the influence of the number of process parameter values (specifically, $\mathrm{Re}_\text{train}$), for which labeled data is added, on prediction accuracy. We showed that, unlike classical \glspl{nn} or \glspl{binn}, the \gls{pinn} model maintains its accuracy even when labels are provided for fewer process parameter values. Specifically, accurate \gls{binn} or \gls{nn} models require labeled data from at least \num{4} $\mathrm{Re}_\text{train}$ values. Conversely, a \gls{pinn} preserves its accuracy across the entire parameter range with labels for only \num{2} values---the boundaries of the range. This capability is particularly advantageous when constructing surrogate models to interpolate between simulation results, as only \num{2} high-fidelity simulations are required to train an accurate surrogate across the full process parameter range. \\
However, our results indicate that the \gls{pinn} exhibits greater robustness regarding random label placement when labels are added for \num{3} $\mathrm{Re}_\text{train}$ values. In contrast, models with labels for only \num{2} $\mathrm{Re}_\text{train}$ values experience a two- to threefold increase in the standard deviation of prediction errors. This issue may be alleviated by increasing the number of labels added per $\mathrm{Re}_\text{train}$ value.

Next, we showed that by strategically placing velocity labels in areas with high prediction errors, the data requirements for \glspl{pinn} can be reduced. This approach enables the creation of a model with prediction errors around \qty{3}{\percent} for both velocity and pressure across the training process parameter range, and only slightly increased errors when extrapolating to $\mathrm{Re}_\text{test}=6000$, which lies outside the training range (approximately \qty{4}{\percent} for both velocity and pressure) with just \num{6} labeled datapoints in total. Notably, these results are achieved using labels derived from simulations conducted on a mesh with significantly lower resolution than the test mesh used to evaluate prediction accuracy.

Finally, we demonstrated that data requirements can be further reduced if the approximate character of the flow is known, by using approximate velocity labels instead of real simulation or measurement data. This approach results in prediction errors around \qty{2.5}{\percent} and \qty{2}{\percent} for velocity and pressure, respectively, across most values within the process parameter range and enables successful extrapolation to $\mathrm{Re}_\text{test}=6000$. However, the model underperforms at lower $\mathrm{Re}$ values ($\mathrm{Re}_\text{test}<100$), where the flow character approximation does not hold. Other models discussed in this work also face challenges in predicting lower $\mathrm{Re}$ values due to the pronounced kink in the velocity profile at the impeller tip, which becomes more significant at lower $\mathrm{Re}$. Both \glspl{pinn} and classical \glspl{nn} tend to smooth out these discontinuities, leading to inaccuracies. 

\section{Conclusion}
\label{sec:conclusion}
In this work, we provide quantitative insights into the significantly reduced labeled dataset requirements for vanilla \glspl{pinn} compared to classical \glspl{nn}, offering practical guidelines for selecting surrogate model configurations for stirred tanks based on the size and nature of available datasets. 

Our findings reveal that prediction errors of approximately \qty{3}{\percent} for both velocity and pressure across a range of $\mathrm{Re}$ can be achieved using as few as six strategically placed datapoints with velocity labels or by employing an approximation of the velocity profile, in contrast to the \qty{11}{\percent} and \qty{12}{\percent} errors observed in velocity and pressure respectively for a \gls{pinn} trained without labeled data. While our previous work has shown that improvements in vanilla \gls{pinn} accuracy can be achieved through various strategies without data, this paper focuses on evaluating the capabilities of the vanilla \gls{pinn}, which is straightforward to implement and use.
The comparison with classical \glspl{nn}, which require thousands of labeled datapoints to attain similar results, underscores the efficiency of \glspl{pinn} in reducing data requirements while preserving accuracy.

Looking forward, we aim to extend our surrogate models to 3D geometries and explore turbulence modeling to enable progress toward higher Reynolds numbers. This will help address challenges associated with the need for extremely fine meshes for high-fidelity solutions---a limitation not inherent to \glspl{pinn} but relevant to stirred tank models overall, and essential for obtaining meaningful training and reference data.

\section*{Acknowledgements}
This work was performed as part of the Helmholtz School for Data Science in Life, Earth and Energy (HDS-LEE) and received funding from the Helmholtz Association of German Research Centres. This work was also supported by the Deutsche Forschungsgemeinschaft (DFG, German Research Foundation) – 333849990/GRK2379 (IRTG Hierarchical and Hybrid Approaches in Modern Inverse Problems).
The authors gratefully acknowledge the computing time provided to them at the NHR Centers NHR4CES at TU Darmstadt (project number p0020502) and RWTH Aachen University (project number p0024828). This is funded by the Federal Ministry of Education and Research, and the state governments participating on the basis of the resolutions of the GWK for national high performance computing at universities (www.nhr-verein.de/unsere-partner).

\section*{Data availability statement}

The data that support the findings of this study are available from the corresponding author upon reasonable request.

\section*{Use of AI tools declaration}
The authors declare they have used Artificial Intelligence (AI) tools in the creation of this article: In some paragraphs throughout the article, AI tools have been used to rephrase and improve sentences grammatically and language-wise. We emphasize that no content has been generated by AI tools.

\appendix
\section{Appendix}
\subsection{Input and output transformations between $\Omega$ and $\Omega_\text{sym}$}
\label{app:transformations}

For models in Sections~\ref{subsec:results:targetedlabels} and \ref{subsec:results:labelapproximations} that leverage the problem's symmetry, spatial inputs are projected from $\Omega$ to $\Omega_\text{sym}$ prior to being processed by the non-dimensionalizing function $\bm{f}_\text{pre}$ defined in Eq~\ref{eq:model:governingequations:inputoutputscaling:input}. The projection is defined as:

\begin{align}
    \label{eq:appendix:transformations:input}
    \bm{x}_\text{sym} &= \bm{p}(\bm{x}) = 
    \begin{cases} 
        (y, -x, \omega), & \text{if } x < 0 \text{ and } y \geq 0 \\ 
        (-y, x, \omega), & \text{if } x \geq 0 \text{ and } y < 0 \\ 
        (-x, -y, \omega), & \text{if } x < 0 \text{ and } y < 0 \\ 
        (x, y, \omega), & \text{otherwise},
    \end{cases}
\end{align}
where $\bm{x}_\text{cart}=(x,y)$. The transformation $\bm{g}$, detailed below, reverts the components of the velocity vector predicted by the \gls{pinn} to their original positions applying the output scaling function $\bm{f}_\text{post}$, as defined in Eq~\ref{eq:model:governingequations:inputoutputscaling:output}.

\begin{align}
\bm{\tilde{v}} &= \bm{g}(\bm{v}_\text{sym}) = 
\begin{cases} 
(-v_y, v_x), & \text{if } x < 0 \text{ and } y \geq 0 \\ 
(v_y, -v_x), & \text{if } x \geq 0 \text{ and } y < 0 \\ 
(-v_x, -v_y), & \text{if } x < 0 \text{ and } y < 0 \\ 
(v_x, v_y), & \text{otherwise}.
\end{cases}
\end{align}

\subsection{Hyperparameters for models presented in Section~\ref{sec:results}}
\label{app:hyperparameters}

\subsubsection{Hyperparameters of the vanilla \gls{pinn}}
The hyperparameters used for the training of the vanilla \gls{pinn} model in its most basic configuration presented in Sections~\ref{subsec:results:datasetsize},\ref{subsec:results:losscomponents}, \ref{subsec:results:parameterdensity} and \ref{subsec:results:robustness} are shown in Table~\ref{tab:appendix:hyperparamsvanillaPINN}. The training points in the domain and on the boundaries were resampled at a rate specified in Table~\ref{tab:results:hyperparameters}. Due to limitations imposed by the size of the training mesh when sampling its vertices as collocation points at which the \gls{pde} and \gls{bc} residuals are evaluated, both domain and boundary spatial points are sampled with replacement. This approach can theoretically result in some spatial points being presented to the network multiple times across different batches; however, they are always resampled in a new order and paired with a different $\omega$ value from a uniform distribution as the parametric coordinate. When fewer than \num{1024} labeled datapoints per $\mathrm{Re}$ were added to the training, these points were not resampled during training. Conversely, if more than \num{1024} labeled datapoints per $\mathrm{Re}$ were included, they were randomly resampled in batches of \num{1024} points. The batch sizes and scaling factors for individual loss components apply to other models listed in Table~\ref{tab:results:losscomponents}, which feature various combinations of loss components during training (e.g., \gls{binn}), except for the purely supervised \gls{nn}, where the batch size consistently comprised \num{3072} points. 

\begin{table}[H]
    \centering
    \caption{Hyperparameters of the vanilla \gls{pinn} model.}
    \label{tab:appendix:hyperparamsvanillaPINN}
    \begin{tabular}{lcl}
        \hline
         & Hyperparameter & Value  \\
         \hline
         \multirow{7}{*}{Sampling}& Num domain points & \num{2048} \\
         & \multirow{2}{*}{Num boundary points} & \num{512} on $\Gamma_\text{stirrer}$ \\
         & & \num{512} on $\Gamma_\text{wall}$\\
         & \multirow{4}{*}{Num labeled datapoints} & Fixed \\
         & & if $<1024$ per $\mathrm{Re}_\text{train}$\\
         & & Batches of \num{1024} \\
         & & if $>1024$ per $\mathrm{Re}_\text{train}$\\
         \hline
         \multirow{6}{*}{Loss scaling} & $\alpha_{\text{momentum,}x}$ & \num{1}\\
         & $\alpha_{\text{momentum,}y}$ & \num{1}\\
         & $\alpha_\text{mass}$ & \num{1}\\
         & $\alpha_\text{wall}$ & \num{1}\\
         & $\alpha_\text{impeller}$ & \num{1}\\
         & $\alpha_\text{data}$ & \num{1}\\
        \hline
    \end{tabular}
\end{table}

\subsubsection{Hyperparameters of the \gls{pinn} model with targeted label allocation and symmetry}

The hyperparameters employed for training the \gls{pinn} model on $\Omega_\text{sym}$ with targeted label allocation, as detailed in Section~\ref{subsec:results:targetedlabels}, are listed in Table~\ref{tab:appendix:hyperparamstargetedPINN}. It is important to emphasize that neither the points on $\Gamma_\text{sym}$ nor the labeled datapoints were resampled during training. Furthermore, the data loss was calculated solely using velocity labels, with no pressure labels incorporated into the training process. 

\begin{table}[H]
    \centering
    \caption{Hyperparameters of the \gls{pinn} model with targeted label allocation and symmetry. The number of labeled datapoints ranged from \num{1} to \num{3} per $\mathrm{Re}$, depending on the specific configuration.}
    \label{tab:appendix:hyperparamstargetedPINN}
    \begin{tabular}{lcl}
        \hline
         & Hyperparameter & Value  \\
         \hline
         \multirow{5}{*}{Sampling}& Num domain points & \num{2048} \\
         & \multirow{3}{*}{Num boundary points} & \num{512} on $\Gamma_\text{stirrer}$ \\
         & & \num{512} on $\Gamma_\text{wall}$\\
         & & Fixed \num{512} on $\Gamma_\text{sym}$\\
         & {Num labeled datapoints} & Fixed \\
         \hline
         \multirow{6}{*}{Loss scaling} & $\alpha_{\text{momentum,}x}$ & \num{1}\\
         & $\alpha_{\text{momentum,}y}$ & \num{1}\\
         & $\alpha_\text{mass}$ & \num{1}\\
         & $\alpha_\text{wall}$ & \num{1}\\
         & $\alpha_\text{impeller}$ & \num{1}\\
         & $\alpha_\text{data}$ & \num{25}\\
        \hline
    \end{tabular}
\end{table}

\subsubsection{Hyperparameters of the \gls{pinn} model with approximate labels and symmetry}

The hyperparameters employed for training the \gls{pinn} model on $\Omega_\text{sym}$ with approximate velocity labels described in Section~\ref{subsec:results:labelapproximations}, are listed in Table~\ref{tab:appendix:hyperparamsapproxPINN}. The labeled datapoints and points on $\Gamma_\text{sym}$ were fixed during training.

\begin{table}[H]
    \centering
    \caption{Hyperparameters of the \gls{pinn} model with approximate labels and symmetry.}
    \label{tab:appendix:hyperparamsapproxPINN}
    \begin{tabular}{lcl}
        \hline
         & Hyperparameter & Value  \\
         \hline
         \multirow{5}{*}{Sampling}& Num domain points & \num{2048} \\
         & \multirow{3}{*}{Num boundary points} & \num{512} on $\Gamma_\text{stirrer}$ \\
         & & \num{512} on $\Gamma_\text{wall}$\\
         & & Fixed \num{256} on $\Gamma_\text{sym}$\\
         & {Num labeled datapoints} & Fixed \num{256}\\
         \hline
         \multirow{6}{*}{Loss scaling} & $\alpha_{\text{momentum,}x}$ & \num{1}\\
         & $\alpha_{\text{momentum,}y}$ & \num{1}\\
         & $\alpha_\text{mass}$ & \num{1}\\
         & $\alpha_\text{wall}$ & \num{1}\\
         & $\alpha_\text{impeller}$ & \num{1}\\
         & $\alpha_\text{data}$ & \num{9}\\
        \hline
    \end{tabular}
\end{table}

\bibliography{main}

\end{document}

%% file: custom-colors.tex
\definecolor{custom-red}   {RGB}{215  25  28}

\definecolor{custom-blue}   {RGB}{  44  123 182}
\definecolor{custom-blue-10}{RGB}{232 242 249}

\definecolor{custom-black}   {RGB}{  0   0   0}
\definecolor{custom-black-10}{RGB}{236 237 237}

\definecolor{custom-orange}   {RGB}{253 174 97}
\definecolor{custom-orange-10}{RGB}{254 213 173}

%% file: figures/3D_domain.tex
\tikzset{
	%Define standard arrow tip
	>=stealth'
}
\newlength{\gridlength}
\setlength{\gridlength}{20cm}

\begin{tikzpicture}[x=\gridlength, y=\gridlength]]
		% baffles
		\node[thick, draw=custom-black, fill=custom-black-10, shape=rectangle, minimum width=7, minimum height=110, inner sep=0pt, anchor=center] at (-0.095,0.05) {};
		\node[thick, draw=custom-black, fill=custom-black-10, shape=rectangle, minimum width=7, minimum height=110, inner sep=0pt, anchor=center] at (0.095,0.05) {};
		
		% reactor wall
		% Top ellipse (at y = 1.5): front half solid, back half dashed
		\draw[thick] (0.1,0.15) arc (0:180:0.1 and 0.03);
		\draw[thick] (-0.1,0.15) arc (180:360:0.1 and 0.03);
		
		% Vertical sides of the cylinder
		\draw[thick] (-0.1,0.15) -- (-0.1,-0.05);
		\draw[thick] (0.1,0.15) -- (0.1,-0.05);
		
		% Bottom ellipse (at y = -0.5): back half solid, front half dashed
		\draw[thick] (-0.1,-0.05) arc (180:360:0.1 and 0.03);
		\draw[thick, dashed] (0.1,-0.05) arc (0:180:0.1 and 0.03);
		
		% impeller shaft
		\draw[thick, custom-red] (0,0.15) -- (0,0);
		
		% Impeller blades 
		%\foreach \angle in {45,135,225,315} {
			%\draw[custom-red] (0,0) -- %({0.5*cos(\angle)},{0.5*sin(\angle)*0.3});
		%}
		\fill[thick, custom-red!10, draw=custom-red]
		(-0.03,0.015) -- (0,0.005) -- (0,-0.005) -- (-0.03,0.005) -- cycle;
		\fill[thick, custom-red!10, draw=custom-red]
		(0,0.005) -- (0.03,0.015) -- (0.03,0.005) -- (0,-0.005) -- cycle;
		 \fill[thick, custom-red!10, draw=custom-red] (-0.03,-0.005) -- (0,0.005) -- (0,-0.005) -- (-0.03,-0.015) -- cycle;
		 \fill[thick, custom-red!10, draw=custom-red] 
		 (0,0.005) -- (0.03,-0.005) -- (0.03,-0.015) -- (0,-0.005) -- cycle;
		 
		  % section that we are considering
		 \fill[dashed, very thick,custom-orange!10, draw=custom-orange, opacity=0.5] (0,0) ellipse (0.1 and 0.02);
		 
		  % coordinate system
		 \draw[thick, custom-black, -{latex}] (-0.155,-0.12) -- (-0.155,-0.05) node[above] {$z$};
		 \draw[thick, custom-black, -{latex}] (-0.155,-0.12) -- ++(0.05,0.02) node[right] {$y$};
		 \draw[thick, custom-black, -{latex}] (-0.155,-0.12) -- ++(0.05,-0.02) node[right] {$x$};

\end{tikzpicture}

%% file: figures/2D_domain.tex
\tikzset{
	%Define standard arrow tip
	>=stealth'
}
\setlength{\gridlength}{20cm}

\begin{tikzpicture}[ x=\gridlength, y=\gridlength]
  
  % reactor
  \fill[fill=custom-orange!10] (0,0) circle (0.1);
  \node[custom-orange] at (-0.045,0.045) {$\Omega$};

  % symmetry
  \fill[custom-blue!10] (0,0) -- (0.1,0) arc[start angle=0, end angle=90, radius=0.1] -- cycle;

  % outline of reactor
  \draw[very thick, custom-black] (0,0) circle (0.1);
  \node[custom-black] at (0.085,0.085) {$\Gamma_\text{wall}$};

  % symmetry boundary
  \draw[very thick, custom-blue, dashed] (0.0, 0.04) -- (0.0, 0.1);
  \draw[very thick, custom-blue, dashed] (0.04, 0.0) -- (0.1, 0.0);
  \node[custom-blue] at (0.045,0.045) {$\Omega_\text{sym}$};
  \node[custom-blue] at (0.13,0.0) {$\Gamma_\text{sym}$};

  \begin{scope}
  	\clip (0,0) circle (0.1);
  	% baffles
  	\def \sqrttwo {1.41421356237}
  	\def \position {0.1/\sqrttwo}
  	\foreach \x/\y/\d in {\position/\position/-45, \position/-\position/45, -\position/-\position/-45, -\position/\position/45} {
  		\node[very thick, draw=custom-black, fill=custom-black!10, shape=rectangle, minimum width=0.005\gridlength, minimum height=0.03\gridlength, inner sep=0pt, anchor=center, rotate=\d] at (\x,\y) {};
  	}
  \end{scope}
  
  % impeller BC
  \draw[thick, gray] (-0.04,0.02) -- (0.04,-0.02);
  \draw[thick, gray] (-0.02,-0.04) -- (0.02,0.04);
  \foreach \r in {0.02, 0.03, 0.04} {
  	% phi = 0
  	\draw[thick, gray, -{latex}] (\r,0) -- (\r,-0.5*\r);
  	% phi = 90
  	\draw[thick, gray, -{latex}] (0,\r) -- (0.5*\r,\r);
  	% phi = 180
  	\draw[thick, gray, -{latex}] (-\r,0) -- (-\r,0.5*\r);
  	% phi = 270
  	\draw[thick, gray, -{latex}] (0,-\r) -- (-0.5*\r,-\r);
  }
  
  % impeller
  \foreach \r in {-0.04, 0.04} {
  	\draw[very thick, custom-red] (0,0) -- (0,\r);
  	\draw[very thick, custom-red] (0,0) -- (\r,0);
  }
  \node[custom-red] at (0.04,-0.04) {$\Gamma_\text{stirrer}$};
  
  % rotation direction
  \draw[thick, gray, -{latex}] (0,-0.07) to [out=180,in=270, looseness=1] (-0.07,0);
  \node[gray] at (-0.07,0.015) {$\omega$};

  % symmetry

  % coordinate system
  \draw[thick, custom-black, -{latex}] (-0.125,-0.125) -- (-0.125,-0.075) node[above] {$y$};
  \draw[thick, custom-black, -{latex}] (-0.125,-0.125) -- (-0.075,-0.125) node[right] {$x$};
  \draw[thick, custom-black, -{latex}] (-0.125,-0.125) circle (1pt);
  \draw[thick, custom-black, -{latex}] (-0.125,-0.125) circle (4pt) node[below left=0.003] {$z$};

\end{tikzpicture}

%% file: figures/2D_dimensions.tex
\tikzset{
	%Define standard arrow tip
	>=stealth'
}
\setlength{\gridlength}{20cm}

\begin{tikzpicture}[font=\large, x=\gridlength, y=\gridlength]
    
    \def \sqrttwo {1.41421356237}
    \def \position {0.1/\sqrttwo}
    
    % reactor
    \draw[thick, draw=black] (0,0) circle (0.1);
    
    \begin{scope}
        \clip (0,0) circle (0.1);
        % baffles
        \foreach \x/\y/\d in {\position/\position/-45, \position/-\position/45, -\position/-\position/-45, -\position/\position/45} {
            \node[thick, draw=black, fill=black!10, shape=rectangle, minimum width=0.005\gridlength, minimum height=0.03\gridlength, inner sep=0pt, anchor=center, rotate=\d] at (\x,\y) {};
        }
    \end{scope}
    
    % impeller
    \foreach \r in {-0.04, 0.04} {
        \draw[thick] (0,0) -- (0,\r);
        \draw[thick] (0,0) -- (\r,0);
    }
    
    % Annotations
    
    % radius to baffle
    \draw[custom-blue, very thick, dashed, -{latex}|] (0,0) -- (\position-0.01,\position-0.01) node[above left, yshift=-1mm] {$R_\text{baffle}$};
    
    % radius to wall
    \draw[custom-blue, very thick, dashed, -{latex}|] (0,0) -- (-0.1,0.0) node[above, xshift=6mm, text width=20mm, align=right] {$R_\text{reactor}$};
    
    % impeller radius
    \draw[custom-blue, very thick, dashed, -{latex}|] (0,-0.0075) -- (0.04,-0.0075) node[below] {$R_\text{stirrer}$};
    
    % baffle thickness
    \coordinate (AH) at (-0.0769, -0.0433);
    \coordinate (AT) at (-0.0619, -0.0583);
    \coordinate (BH) at (-0.0433, -0.0769);
    \coordinate (BT) at (-0.0583, -0.0619);
    \draw[custom-blue, very thick, dashed, -{latex}|] (AH) -- (AT);
    \draw[custom-blue, very thick, dashed, -{latex}|] (BH) -- (BT);
    \node[custom-blue, right] at (0.015-\position,0.015-\position) {$t_\text{baffle}$};
    
\end{tikzpicture}

%% file: figures/reference_sol/re_1000/vel_mag.pgf
%% Creator: Matplotlib, PGF backend
%%
%% To include the figure in your LaTeX document, write
%%   \input{<filename>.pgf}
%%
%% Make sure the required packages are loaded in your preamble
%%   \usepackage{pgf}
%%
%% Also ensure that all the required font packages are loaded; for instance,
%% the lmodern package is sometimes necessary when using math font.
%%   \usepackage{lmodern}
%%
%% Figures using additional raster images can only be included by \input if
%% they are in the same directory as the main LaTeX file. For loading figures
%% from other directories you can use the `import` package
%% and then include the figures with
%%   \import{<path to file>}{<filename>.pgf}
%%
%% Matplotlib used the following preamble
%%   \def\mathdefault#1{#1}
%%   \everymath=\expandafter{\the\everymath\displaystyle}
%%   \usepackage{amsmath}\usepackage{bm}
%%   \makeatletter\@ifpackageloaded{underscore}{}{\usepackage[strings]{underscore}}\makeatother
%%
\begingroup%
\makeatletter%
\begin{pgfpicture}%
\pgfpathrectangle{\pgfpointorigin}{\pgfqpoint{2.500000in}{2.500000in}}%
\pgfusepath{use as bounding box, clip}%
\begin{pgfscope}%
\pgfsetbuttcap%
\pgfsetmiterjoin%
\definecolor{currentfill}{rgb}{1.000000,1.000000,1.000000}%
\pgfsetfillcolor{currentfill}%
\pgfsetlinewidth{0.000000pt}%
\definecolor{currentstroke}{rgb}{1.000000,1.000000,1.000000}%
\pgfsetstrokecolor{currentstroke}%
\pgfsetdash{}{0pt}%
\pgfpathmoveto{\pgfqpoint{0.000000in}{0.000000in}}%
\pgfpathlineto{\pgfqpoint{2.500000in}{0.000000in}}%
\pgfpathlineto{\pgfqpoint{2.500000in}{2.500000in}}%
\pgfpathlineto{\pgfqpoint{0.000000in}{2.500000in}}%
\pgfpathlineto{\pgfqpoint{0.000000in}{0.000000in}}%
\pgfpathclose%
\pgfusepath{fill}%
\end{pgfscope}%
\begin{pgfscope}%
\pgfsetbuttcap%
\pgfsetmiterjoin%
\definecolor{currentfill}{rgb}{1.000000,1.000000,1.000000}%
\pgfsetfillcolor{currentfill}%
\pgfsetlinewidth{0.000000pt}%
\definecolor{currentstroke}{rgb}{0.000000,0.000000,0.000000}%
\pgfsetstrokecolor{currentstroke}%
\pgfsetstrokeopacity{0.000000}%
\pgfsetdash{}{0pt}%
\pgfpathmoveto{\pgfqpoint{0.575720in}{0.386658in}}%
\pgfpathlineto{\pgfqpoint{2.138239in}{0.386658in}}%
\pgfpathlineto{\pgfqpoint{2.138239in}{1.949178in}}%
\pgfpathlineto{\pgfqpoint{0.575720in}{1.949178in}}%
\pgfpathlineto{\pgfqpoint{0.575720in}{0.386658in}}%
\pgfpathclose%
\pgfusepath{fill}%
\end{pgfscope}%
\begin{pgfscope}%
\pgfsys@transformshift{0.646250in}{0.457500in}%
\pgftext[left,bottom]{\includegraphics[interpolate=true,width=1.421250in,height=1.421250in]{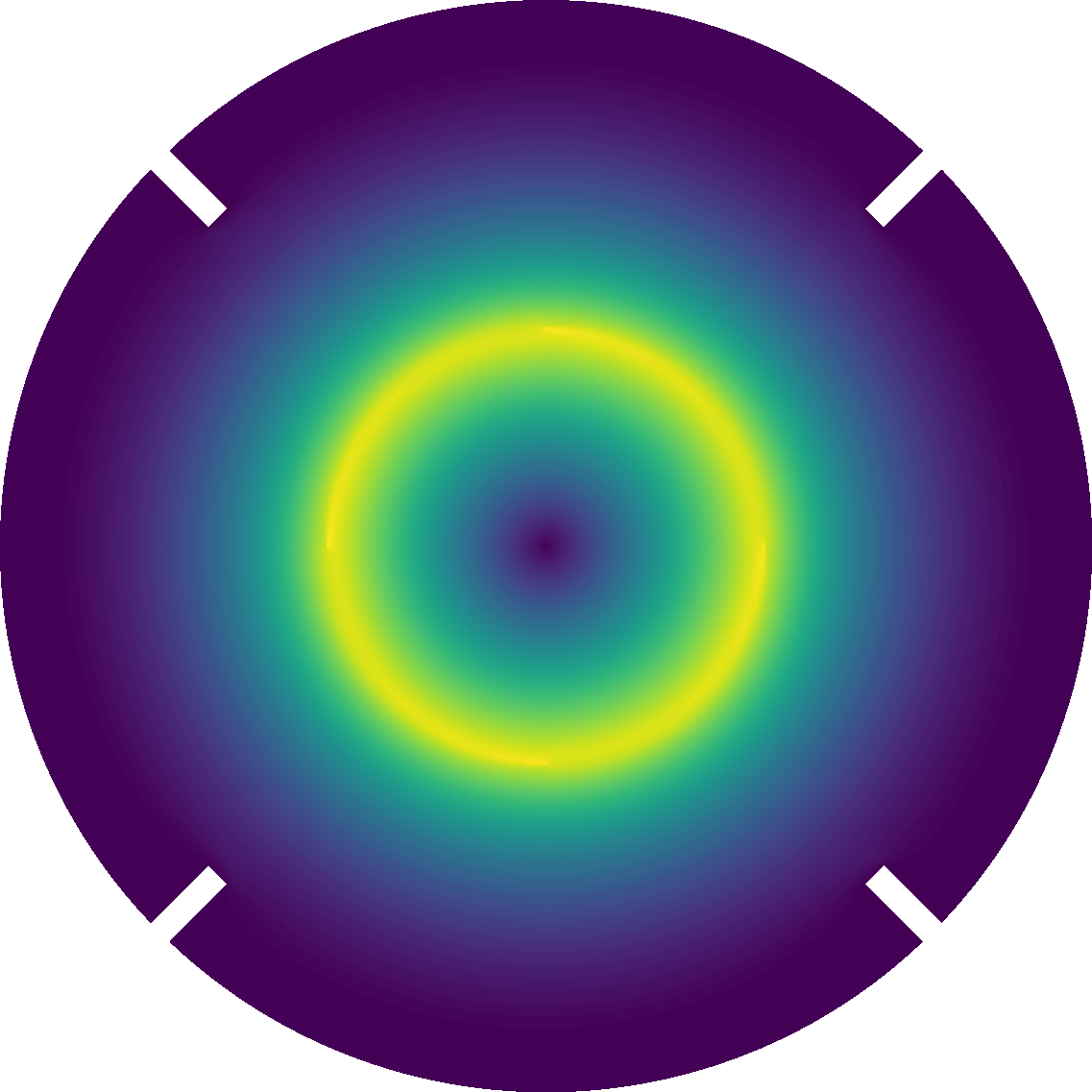}}%
\end{pgfscope}%
\begin{pgfscope}%
\pgfsetbuttcap%
\pgfsetroundjoin%
\definecolor{currentfill}{rgb}{0.000000,0.000000,0.000000}%
\pgfsetfillcolor{currentfill}%
\pgfsetlinewidth{0.803000pt}%
\definecolor{currentstroke}{rgb}{0.000000,0.000000,0.000000}%
\pgfsetstrokecolor{currentstroke}%
\pgfsetdash{}{0pt}%
\pgfsys@defobject{currentmarker}{\pgfqpoint{0.000000in}{-0.048611in}}{\pgfqpoint{0.000000in}{0.000000in}}{%
\pgfpathmoveto{\pgfqpoint{0.000000in}{0.000000in}}%
\pgfpathlineto{\pgfqpoint{0.000000in}{-0.048611in}}%
\pgfusepath{stroke,fill}%
}%
\begin{pgfscope}%
\pgfsys@transformshift{0.646743in}{0.386658in}%
\pgfsys@useobject{currentmarker}{}%
\end{pgfscope}%
\end{pgfscope}%
\begin{pgfscope}%
\definecolor{textcolor}{rgb}{0.000000,0.000000,0.000000}%
\pgfsetstrokecolor{textcolor}%
\pgfsetfillcolor{textcolor}%
\pgftext[x=0.646743in,y=0.296381in,,top]{\color{textcolor}{\rmfamily\fontsize{8.330000}{9.996000}\selectfont\catcode`\^=\active\def^{\ifmmode\sp\else\^{}\fi}\catcode`\%=\active\def%{\%}$\mathdefault{\ensuremath{-}0.1}$}}%
\end{pgfscope}%
\begin{pgfscope}%
\pgfsetbuttcap%
\pgfsetroundjoin%
\definecolor{currentfill}{rgb}{0.000000,0.000000,0.000000}%
\pgfsetfillcolor{currentfill}%
\pgfsetlinewidth{0.803000pt}%
\definecolor{currentstroke}{rgb}{0.000000,0.000000,0.000000}%
\pgfsetstrokecolor{currentstroke}%
\pgfsetdash{}{0pt}%
\pgfsys@defobject{currentmarker}{\pgfqpoint{0.000000in}{-0.048611in}}{\pgfqpoint{0.000000in}{0.000000in}}{%
\pgfpathmoveto{\pgfqpoint{0.000000in}{0.000000in}}%
\pgfpathlineto{\pgfqpoint{0.000000in}{-0.048611in}}%
\pgfusepath{stroke,fill}%
}%
\begin{pgfscope}%
\pgfsys@transformshift{1.356980in}{0.386658in}%
\pgfsys@useobject{currentmarker}{}%
\end{pgfscope}%
\end{pgfscope}%
\begin{pgfscope}%
\definecolor{textcolor}{rgb}{0.000000,0.000000,0.000000}%
\pgfsetstrokecolor{textcolor}%
\pgfsetfillcolor{textcolor}%
\pgftext[x=1.356980in,y=0.296381in,,top]{\color{textcolor}{\rmfamily\fontsize{8.330000}{9.996000}\selectfont\catcode`\^=\active\def^{\ifmmode\sp\else\^{}\fi}\catcode`\%=\active\def%{\%}$\mathdefault{0.0}$}}%
\end{pgfscope}%
\begin{pgfscope}%
\pgfsetbuttcap%
\pgfsetroundjoin%
\definecolor{currentfill}{rgb}{0.000000,0.000000,0.000000}%
\pgfsetfillcolor{currentfill}%
\pgfsetlinewidth{0.803000pt}%
\definecolor{currentstroke}{rgb}{0.000000,0.000000,0.000000}%
\pgfsetstrokecolor{currentstroke}%
\pgfsetdash{}{0pt}%
\pgfsys@defobject{currentmarker}{\pgfqpoint{0.000000in}{-0.048611in}}{\pgfqpoint{0.000000in}{0.000000in}}{%
\pgfpathmoveto{\pgfqpoint{0.000000in}{0.000000in}}%
\pgfpathlineto{\pgfqpoint{0.000000in}{-0.048611in}}%
\pgfusepath{stroke,fill}%
}%
\begin{pgfscope}%
\pgfsys@transformshift{2.067216in}{0.386658in}%
\pgfsys@useobject{currentmarker}{}%
\end{pgfscope}%
\end{pgfscope}%
\begin{pgfscope}%
\definecolor{textcolor}{rgb}{0.000000,0.000000,0.000000}%
\pgfsetstrokecolor{textcolor}%
\pgfsetfillcolor{textcolor}%
\pgftext[x=2.067216in,y=0.296381in,,top]{\color{textcolor}{\rmfamily\fontsize{8.330000}{9.996000}\selectfont\catcode`\^=\active\def^{\ifmmode\sp\else\^{}\fi}\catcode`\%=\active\def%{\%}$\mathdefault{0.1}$}}%
\end{pgfscope}%
\begin{pgfscope}%
\definecolor{textcolor}{rgb}{0.000000,0.000000,0.000000}%
\pgfsetstrokecolor{textcolor}%
\pgfsetfillcolor{textcolor}%
\pgftext[x=1.356980in,y=0.142060in,,top]{\color{textcolor}{\rmfamily\fontsize{10.000000}{12.000000}\selectfont\catcode`\^=\active\def^{\ifmmode\sp\else\^{}\fi}\catcode`\%=\active\def%{\%}$x\;[\text{m}]$}}%
\end{pgfscope}%
\begin{pgfscope}%
\pgfsetbuttcap%
\pgfsetroundjoin%
\definecolor{currentfill}{rgb}{0.000000,0.000000,0.000000}%
\pgfsetfillcolor{currentfill}%
\pgfsetlinewidth{0.803000pt}%
\definecolor{currentstroke}{rgb}{0.000000,0.000000,0.000000}%
\pgfsetstrokecolor{currentstroke}%
\pgfsetdash{}{0pt}%
\pgfsys@defobject{currentmarker}{\pgfqpoint{-0.048611in}{0.000000in}}{\pgfqpoint{-0.000000in}{0.000000in}}{%
\pgfpathmoveto{\pgfqpoint{-0.000000in}{0.000000in}}%
\pgfpathlineto{\pgfqpoint{-0.048611in}{0.000000in}}%
\pgfusepath{stroke,fill}%
}%
\begin{pgfscope}%
\pgfsys@transformshift{0.575720in}{0.457682in}%
\pgfsys@useobject{currentmarker}{}%
\end{pgfscope}%
\end{pgfscope}%
\begin{pgfscope}%
\definecolor{textcolor}{rgb}{0.000000,0.000000,0.000000}%
\pgfsetstrokecolor{textcolor}%
\pgfsetfillcolor{textcolor}%
\pgftext[x=0.242769in, y=0.419102in, left, base]{\color{textcolor}{\rmfamily\fontsize{8.330000}{9.996000}\selectfont\catcode`\^=\active\def^{\ifmmode\sp\else\^{}\fi}\catcode`\%=\active\def%{\%}$\mathdefault{\ensuremath{-}0.1}$}}%
\end{pgfscope}%
\begin{pgfscope}%
\pgfsetbuttcap%
\pgfsetroundjoin%
\definecolor{currentfill}{rgb}{0.000000,0.000000,0.000000}%
\pgfsetfillcolor{currentfill}%
\pgfsetlinewidth{0.803000pt}%
\definecolor{currentstroke}{rgb}{0.000000,0.000000,0.000000}%
\pgfsetstrokecolor{currentstroke}%
\pgfsetdash{}{0pt}%
\pgfsys@defobject{currentmarker}{\pgfqpoint{-0.048611in}{0.000000in}}{\pgfqpoint{-0.000000in}{0.000000in}}{%
\pgfpathmoveto{\pgfqpoint{-0.000000in}{0.000000in}}%
\pgfpathlineto{\pgfqpoint{-0.048611in}{0.000000in}}%
\pgfusepath{stroke,fill}%
}%
\begin{pgfscope}%
\pgfsys@transformshift{0.575720in}{1.167918in}%
\pgfsys@useobject{currentmarker}{}%
\end{pgfscope}%
\end{pgfscope}%
\begin{pgfscope}%
\definecolor{textcolor}{rgb}{0.000000,0.000000,0.000000}%
\pgfsetstrokecolor{textcolor}%
\pgfsetfillcolor{textcolor}%
\pgftext[x=0.334591in, y=1.129338in, left, base]{\color{textcolor}{\rmfamily\fontsize{8.330000}{9.996000}\selectfont\catcode`\^=\active\def^{\ifmmode\sp\else\^{}\fi}\catcode`\%=\active\def%{\%}$\mathdefault{0.0}$}}%
\end{pgfscope}%
\begin{pgfscope}%
\pgfsetbuttcap%
\pgfsetroundjoin%
\definecolor{currentfill}{rgb}{0.000000,0.000000,0.000000}%
\pgfsetfillcolor{currentfill}%
\pgfsetlinewidth{0.803000pt}%
\definecolor{currentstroke}{rgb}{0.000000,0.000000,0.000000}%
\pgfsetstrokecolor{currentstroke}%
\pgfsetdash{}{0pt}%
\pgfsys@defobject{currentmarker}{\pgfqpoint{-0.048611in}{0.000000in}}{\pgfqpoint{-0.000000in}{0.000000in}}{%
\pgfpathmoveto{\pgfqpoint{-0.000000in}{0.000000in}}%
\pgfpathlineto{\pgfqpoint{-0.048611in}{0.000000in}}%
\pgfusepath{stroke,fill}%
}%
\begin{pgfscope}%
\pgfsys@transformshift{0.575720in}{1.878155in}%
\pgfsys@useobject{currentmarker}{}%
\end{pgfscope}%
\end{pgfscope}%
\begin{pgfscope}%
\definecolor{textcolor}{rgb}{0.000000,0.000000,0.000000}%
\pgfsetstrokecolor{textcolor}%
\pgfsetfillcolor{textcolor}%
\pgftext[x=0.334591in, y=1.839574in, left, base]{\color{textcolor}{\rmfamily\fontsize{8.330000}{9.996000}\selectfont\catcode`\^=\active\def^{\ifmmode\sp\else\^{}\fi}\catcode`\%=\active\def%{\%}$\mathdefault{0.1}$}}%
\end{pgfscope}%
\begin{pgfscope}%
\definecolor{textcolor}{rgb}{0.000000,0.000000,0.000000}%
\pgfsetstrokecolor{textcolor}%
\pgfsetfillcolor{textcolor}%
\pgftext[x=0.187213in,y=1.167918in,,bottom,rotate=90.000000]{\color{textcolor}{\rmfamily\fontsize{10.000000}{12.000000}\selectfont\catcode`\^=\active\def^{\ifmmode\sp\else\^{}\fi}\catcode`\%=\active\def%{\%}$y\;[\text{m}]$}}%
\end{pgfscope}%
\begin{pgfscope}%
\pgfpathrectangle{\pgfqpoint{0.575720in}{0.386658in}}{\pgfqpoint{1.562520in}{1.562520in}}%
\pgfusepath{clip}%
\pgfsetbuttcap%
\pgfsetmiterjoin%
\definecolor{currentfill}{rgb}{1.000000,1.000000,1.000000}%
\pgfsetfillcolor{currentfill}%
\pgfsetlinewidth{1.505625pt}%
\definecolor{currentstroke}{rgb}{1.000000,1.000000,1.000000}%
\pgfsetstrokecolor{currentstroke}%
\pgfsetdash{}{0pt}%
\pgfpathmoveto{\pgfqpoint{1.839789in}{1.675838in}}%
\pgfpathlineto{\pgfqpoint{1.832940in}{1.668990in}}%
\pgfpathlineto{\pgfqpoint{1.826092in}{1.662141in}}%
\pgfpathlineto{\pgfqpoint{1.819244in}{1.655293in}}%
\pgfpathlineto{\pgfqpoint{1.812395in}{1.648445in}}%
\pgfpathlineto{\pgfqpoint{1.805547in}{1.641596in}}%
\pgfpathlineto{\pgfqpoint{1.798699in}{1.634748in}}%
\pgfpathlineto{\pgfqpoint{1.791850in}{1.627900in}}%
\pgfpathlineto{\pgfqpoint{1.785002in}{1.621051in}}%
\pgfpathlineto{\pgfqpoint{1.778154in}{1.614203in}}%
\pgfpathlineto{\pgfqpoint{1.771305in}{1.607355in}}%
\pgfpathlineto{\pgfqpoint{1.777583in}{1.601077in}}%
\pgfpathlineto{\pgfqpoint{1.783860in}{1.594799in}}%
\pgfpathlineto{\pgfqpoint{1.790138in}{1.588522in}}%
\pgfpathlineto{\pgfqpoint{1.796416in}{1.582244in}}%
\pgfpathlineto{\pgfqpoint{1.803264in}{1.589092in}}%
\pgfpathlineto{\pgfqpoint{1.810113in}{1.595941in}}%
\pgfpathlineto{\pgfqpoint{1.816961in}{1.602789in}}%
\pgfpathlineto{\pgfqpoint{1.823809in}{1.609637in}}%
\pgfpathlineto{\pgfqpoint{1.830658in}{1.616486in}}%
\pgfpathlineto{\pgfqpoint{1.837506in}{1.623334in}}%
\pgfpathlineto{\pgfqpoint{1.844354in}{1.630182in}}%
\pgfpathlineto{\pgfqpoint{1.851203in}{1.637031in}}%
\pgfpathlineto{\pgfqpoint{1.858051in}{1.643879in}}%
\pgfpathlineto{\pgfqpoint{1.864899in}{1.650728in}}%
\pgfpathlineto{\pgfqpoint{1.839789in}{1.675838in}}%
\pgfpathclose%
\pgfusepath{stroke,fill}%
\end{pgfscope}%
\begin{pgfscope}%
\pgfpathrectangle{\pgfqpoint{0.575720in}{0.386658in}}{\pgfqpoint{1.562520in}{1.562520in}}%
\pgfusepath{clip}%
\pgfsetbuttcap%
\pgfsetmiterjoin%
\definecolor{currentfill}{rgb}{1.000000,1.000000,1.000000}%
\pgfsetfillcolor{currentfill}%
\pgfsetlinewidth{1.505625pt}%
\definecolor{currentstroke}{rgb}{1.000000,1.000000,1.000000}%
\pgfsetstrokecolor{currentstroke}%
\pgfsetdash{}{0pt}%
\pgfpathmoveto{\pgfqpoint{1.864899in}{0.685109in}}%
\pgfpathlineto{\pgfqpoint{1.858051in}{0.691958in}}%
\pgfpathlineto{\pgfqpoint{1.851203in}{0.698806in}}%
\pgfpathlineto{\pgfqpoint{1.844354in}{0.705654in}}%
\pgfpathlineto{\pgfqpoint{1.837506in}{0.712503in}}%
\pgfpathlineto{\pgfqpoint{1.830658in}{0.719351in}}%
\pgfpathlineto{\pgfqpoint{1.823809in}{0.726199in}}%
\pgfpathlineto{\pgfqpoint{1.816961in}{0.733048in}}%
\pgfpathlineto{\pgfqpoint{1.810113in}{0.739896in}}%
\pgfpathlineto{\pgfqpoint{1.803264in}{0.746744in}}%
\pgfpathlineto{\pgfqpoint{1.796416in}{0.753593in}}%
\pgfpathlineto{\pgfqpoint{1.790013in}{0.747190in}}%
\pgfpathlineto{\pgfqpoint{1.783609in}{0.740786in}}%
\pgfpathlineto{\pgfqpoint{1.777206in}{0.734383in}}%
\pgfpathlineto{\pgfqpoint{1.770803in}{0.727980in}}%
\pgfpathlineto{\pgfqpoint{1.771305in}{0.728482in}}%
\pgfpathlineto{\pgfqpoint{1.778154in}{0.721634in}}%
\pgfpathlineto{\pgfqpoint{1.785002in}{0.714785in}}%
\pgfpathlineto{\pgfqpoint{1.791850in}{0.707937in}}%
\pgfpathlineto{\pgfqpoint{1.798699in}{0.701089in}}%
\pgfpathlineto{\pgfqpoint{1.805547in}{0.694240in}}%
\pgfpathlineto{\pgfqpoint{1.812395in}{0.687392in}}%
\pgfpathlineto{\pgfqpoint{1.819244in}{0.680544in}}%
\pgfpathlineto{\pgfqpoint{1.826092in}{0.673695in}}%
\pgfpathlineto{\pgfqpoint{1.832940in}{0.666847in}}%
\pgfpathlineto{\pgfqpoint{1.839789in}{0.659999in}}%
\pgfpathlineto{\pgfqpoint{1.864899in}{0.685109in}}%
\pgfpathclose%
\pgfusepath{stroke,fill}%
\end{pgfscope}%
\begin{pgfscope}%
\pgfpathrectangle{\pgfqpoint{0.575720in}{0.386658in}}{\pgfqpoint{1.562520in}{1.562520in}}%
\pgfusepath{clip}%
\pgfsetbuttcap%
\pgfsetmiterjoin%
\definecolor{currentfill}{rgb}{1.000000,1.000000,1.000000}%
\pgfsetfillcolor{currentfill}%
\pgfsetlinewidth{1.505625pt}%
\definecolor{currentstroke}{rgb}{1.000000,1.000000,1.000000}%
\pgfsetstrokecolor{currentstroke}%
\pgfsetdash{}{0pt}%
\pgfpathmoveto{\pgfqpoint{0.874170in}{0.659999in}}%
\pgfpathlineto{\pgfqpoint{0.881019in}{0.666847in}}%
\pgfpathlineto{\pgfqpoint{0.887867in}{0.673695in}}%
\pgfpathlineto{\pgfqpoint{0.894715in}{0.680544in}}%
\pgfpathlineto{\pgfqpoint{0.901564in}{0.687392in}}%
\pgfpathlineto{\pgfqpoint{0.908412in}{0.694240in}}%
\pgfpathlineto{\pgfqpoint{0.915261in}{0.701089in}}%
\pgfpathlineto{\pgfqpoint{0.922109in}{0.707937in}}%
\pgfpathlineto{\pgfqpoint{0.928957in}{0.714785in}}%
\pgfpathlineto{\pgfqpoint{0.935806in}{0.721634in}}%
\pgfpathlineto{\pgfqpoint{0.942654in}{0.728482in}}%
\pgfpathlineto{\pgfqpoint{0.936251in}{0.734885in}}%
\pgfpathlineto{\pgfqpoint{0.929848in}{0.741289in}}%
\pgfpathlineto{\pgfqpoint{0.923444in}{0.747692in}}%
\pgfpathlineto{\pgfqpoint{0.917041in}{0.754095in}}%
\pgfpathlineto{\pgfqpoint{0.917543in}{0.753593in}}%
\pgfpathlineto{\pgfqpoint{0.910695in}{0.746744in}}%
\pgfpathlineto{\pgfqpoint{0.903847in}{0.739896in}}%
\pgfpathlineto{\pgfqpoint{0.896998in}{0.733048in}}%
\pgfpathlineto{\pgfqpoint{0.890150in}{0.726199in}}%
\pgfpathlineto{\pgfqpoint{0.883302in}{0.719351in}}%
\pgfpathlineto{\pgfqpoint{0.876453in}{0.712503in}}%
\pgfpathlineto{\pgfqpoint{0.869605in}{0.705654in}}%
\pgfpathlineto{\pgfqpoint{0.862756in}{0.698806in}}%
\pgfpathlineto{\pgfqpoint{0.855908in}{0.691958in}}%
\pgfpathlineto{\pgfqpoint{0.849060in}{0.685109in}}%
\pgfpathlineto{\pgfqpoint{0.874170in}{0.659999in}}%
\pgfpathclose%
\pgfusepath{stroke,fill}%
\end{pgfscope}%
\begin{pgfscope}%
\pgfpathrectangle{\pgfqpoint{0.575720in}{0.386658in}}{\pgfqpoint{1.562520in}{1.562520in}}%
\pgfusepath{clip}%
\pgfsetbuttcap%
\pgfsetmiterjoin%
\definecolor{currentfill}{rgb}{1.000000,1.000000,1.000000}%
\pgfsetfillcolor{currentfill}%
\pgfsetlinewidth{1.505625pt}%
\definecolor{currentstroke}{rgb}{1.000000,1.000000,1.000000}%
\pgfsetstrokecolor{currentstroke}%
\pgfsetdash{}{0pt}%
\pgfpathmoveto{\pgfqpoint{0.849060in}{1.650728in}}%
\pgfpathlineto{\pgfqpoint{0.855908in}{1.643879in}}%
\pgfpathlineto{\pgfqpoint{0.862756in}{1.637031in}}%
\pgfpathlineto{\pgfqpoint{0.869605in}{1.630182in}}%
\pgfpathlineto{\pgfqpoint{0.876453in}{1.623334in}}%
\pgfpathlineto{\pgfqpoint{0.883302in}{1.616486in}}%
\pgfpathlineto{\pgfqpoint{0.890150in}{1.609637in}}%
\pgfpathlineto{\pgfqpoint{0.896998in}{1.602789in}}%
\pgfpathlineto{\pgfqpoint{0.903847in}{1.595941in}}%
\pgfpathlineto{\pgfqpoint{0.910695in}{1.589092in}}%
\pgfpathlineto{\pgfqpoint{0.917543in}{1.582244in}}%
\pgfpathlineto{\pgfqpoint{0.923821in}{1.588522in}}%
\pgfpathlineto{\pgfqpoint{0.930099in}{1.594799in}}%
\pgfpathlineto{\pgfqpoint{0.936376in}{1.601077in}}%
\pgfpathlineto{\pgfqpoint{0.942654in}{1.607355in}}%
\pgfpathlineto{\pgfqpoint{0.935806in}{1.614203in}}%
\pgfpathlineto{\pgfqpoint{0.928957in}{1.621051in}}%
\pgfpathlineto{\pgfqpoint{0.922109in}{1.627900in}}%
\pgfpathlineto{\pgfqpoint{0.915261in}{1.634748in}}%
\pgfpathlineto{\pgfqpoint{0.908412in}{1.641596in}}%
\pgfpathlineto{\pgfqpoint{0.901564in}{1.648445in}}%
\pgfpathlineto{\pgfqpoint{0.894715in}{1.655293in}}%
\pgfpathlineto{\pgfqpoint{0.887867in}{1.662141in}}%
\pgfpathlineto{\pgfqpoint{0.881019in}{1.668990in}}%
\pgfpathlineto{\pgfqpoint{0.874170in}{1.675838in}}%
\pgfpathlineto{\pgfqpoint{0.849060in}{1.650728in}}%
\pgfpathclose%
\pgfusepath{stroke,fill}%
\end{pgfscope}%
\begin{pgfscope}%
\pgfpathrectangle{\pgfqpoint{0.575720in}{0.386658in}}{\pgfqpoint{1.562520in}{1.562520in}}%
\pgfusepath{clip}%
\pgfsetrectcap%
\pgfsetroundjoin%
\pgfsetlinewidth{1.505625pt}%
\definecolor{currentstroke}{rgb}{0.000000,0.000000,0.000000}%
\pgfsetstrokecolor{currentstroke}%
\pgfsetdash{}{0pt}%
\pgfpathmoveto{\pgfqpoint{0.849060in}{1.650728in}}%
\pgfpathlineto{\pgfqpoint{0.917543in}{1.582244in}}%
\pgfpathlineto{\pgfqpoint{0.942654in}{1.607355in}}%
\pgfpathlineto{\pgfqpoint{0.874170in}{1.675838in}}%
\pgfpathlineto{\pgfqpoint{0.868085in}{1.683106in}}%
\pgfpathlineto{\pgfqpoint{0.884282in}{1.698006in}}%
\pgfpathlineto{\pgfqpoint{0.900933in}{1.712397in}}%
\pgfpathlineto{\pgfqpoint{0.918022in}{1.726266in}}%
\pgfpathlineto{\pgfqpoint{0.935532in}{1.739598in}}%
\pgfpathlineto{\pgfqpoint{0.953447in}{1.752382in}}%
\pgfpathlineto{\pgfqpoint{0.971750in}{1.764604in}}%
\pgfpathlineto{\pgfqpoint{0.990422in}{1.776253in}}%
\pgfpathlineto{\pgfqpoint{1.009447in}{1.787318in}}%
\pgfpathlineto{\pgfqpoint{1.028805in}{1.797789in}}%
\pgfpathlineto{\pgfqpoint{1.048478in}{1.807654in}}%
\pgfpathlineto{\pgfqpoint{1.068447in}{1.816906in}}%
\pgfpathlineto{\pgfqpoint{1.088694in}{1.825534in}}%
\pgfpathlineto{\pgfqpoint{1.109198in}{1.833531in}}%
\pgfpathlineto{\pgfqpoint{1.129940in}{1.840888in}}%
\pgfpathlineto{\pgfqpoint{1.150900in}{1.847600in}}%
\pgfpathlineto{\pgfqpoint{1.172058in}{1.853658in}}%
\pgfpathlineto{\pgfqpoint{1.193393in}{1.859059in}}%
\pgfpathlineto{\pgfqpoint{1.214885in}{1.863795in}}%
\pgfpathlineto{\pgfqpoint{1.236514in}{1.867864in}}%
\pgfpathlineto{\pgfqpoint{1.258259in}{1.871260in}}%
\pgfpathlineto{\pgfqpoint{1.280098in}{1.873981in}}%
\pgfpathlineto{\pgfqpoint{1.302012in}{1.876024in}}%
\pgfpathlineto{\pgfqpoint{1.323978in}{1.877387in}}%
\pgfpathlineto{\pgfqpoint{1.345975in}{1.878069in}}%
\pgfpathlineto{\pgfqpoint{1.367984in}{1.878069in}}%
\pgfpathlineto{\pgfqpoint{1.389981in}{1.877387in}}%
\pgfpathlineto{\pgfqpoint{1.411947in}{1.876024in}}%
\pgfpathlineto{\pgfqpoint{1.433861in}{1.873981in}}%
\pgfpathlineto{\pgfqpoint{1.455700in}{1.871260in}}%
\pgfpathlineto{\pgfqpoint{1.477445in}{1.867864in}}%
\pgfpathlineto{\pgfqpoint{1.499074in}{1.863795in}}%
\pgfpathlineto{\pgfqpoint{1.520566in}{1.859059in}}%
\pgfpathlineto{\pgfqpoint{1.541902in}{1.853658in}}%
\pgfpathlineto{\pgfqpoint{1.563059in}{1.847600in}}%
\pgfpathlineto{\pgfqpoint{1.584019in}{1.840888in}}%
\pgfpathlineto{\pgfqpoint{1.604761in}{1.833531in}}%
\pgfpathlineto{\pgfqpoint{1.625266in}{1.825534in}}%
\pgfpathlineto{\pgfqpoint{1.645512in}{1.816906in}}%
\pgfpathlineto{\pgfqpoint{1.665481in}{1.807654in}}%
\pgfpathlineto{\pgfqpoint{1.685155in}{1.797789in}}%
\pgfpathlineto{\pgfqpoint{1.704513in}{1.787318in}}%
\pgfpathlineto{\pgfqpoint{1.723537in}{1.776253in}}%
\pgfpathlineto{\pgfqpoint{1.742209in}{1.764604in}}%
\pgfpathlineto{\pgfqpoint{1.760512in}{1.752382in}}%
\pgfpathlineto{\pgfqpoint{1.778427in}{1.739598in}}%
\pgfpathlineto{\pgfqpoint{1.795937in}{1.726266in}}%
\pgfpathlineto{\pgfqpoint{1.813026in}{1.712397in}}%
\pgfpathlineto{\pgfqpoint{1.829677in}{1.698006in}}%
\pgfpathlineto{\pgfqpoint{1.845874in}{1.683106in}}%
\pgfpathlineto{\pgfqpoint{1.839789in}{1.675838in}}%
\pgfpathlineto{\pgfqpoint{1.771305in}{1.607355in}}%
\pgfpathlineto{\pgfqpoint{1.796416in}{1.582244in}}%
\pgfpathlineto{\pgfqpoint{1.864899in}{1.650728in}}%
\pgfpathlineto{\pgfqpoint{1.872167in}{1.656813in}}%
\pgfpathlineto{\pgfqpoint{1.887067in}{1.640616in}}%
\pgfpathlineto{\pgfqpoint{1.901458in}{1.623965in}}%
\pgfpathlineto{\pgfqpoint{1.915327in}{1.606876in}}%
\pgfpathlineto{\pgfqpoint{1.928659in}{1.589366in}}%
\pgfpathlineto{\pgfqpoint{1.941443in}{1.571451in}}%
\pgfpathlineto{\pgfqpoint{1.953665in}{1.553148in}}%
\pgfpathlineto{\pgfqpoint{1.965314in}{1.534476in}}%
\pgfpathlineto{\pgfqpoint{1.976380in}{1.515451in}}%
\pgfpathlineto{\pgfqpoint{1.986850in}{1.496093in}}%
\pgfpathlineto{\pgfqpoint{1.996716in}{1.476420in}}%
\pgfpathlineto{\pgfqpoint{2.005967in}{1.456451in}}%
\pgfpathlineto{\pgfqpoint{2.014595in}{1.436204in}}%
\pgfpathlineto{\pgfqpoint{2.022592in}{1.415700in}}%
\pgfpathlineto{\pgfqpoint{2.029949in}{1.394958in}}%
\pgfpathlineto{\pgfqpoint{2.036661in}{1.373998in}}%
\pgfpathlineto{\pgfqpoint{2.042720in}{1.352840in}}%
\pgfpathlineto{\pgfqpoint{2.048120in}{1.331505in}}%
\pgfpathlineto{\pgfqpoint{2.052857in}{1.310012in}}%
\pgfpathlineto{\pgfqpoint{2.056925in}{1.288384in}}%
\pgfpathlineto{\pgfqpoint{2.060321in}{1.266639in}}%
\pgfpathlineto{\pgfqpoint{2.063042in}{1.244799in}}%
\pgfpathlineto{\pgfqpoint{2.065086in}{1.222886in}}%
\pgfpathlineto{\pgfqpoint{2.066449in}{1.200920in}}%
\pgfpathlineto{\pgfqpoint{2.067131in}{1.178922in}}%
\pgfpathlineto{\pgfqpoint{2.067131in}{1.156914in}}%
\pgfpathlineto{\pgfqpoint{2.066449in}{1.134917in}}%
\pgfpathlineto{\pgfqpoint{2.065086in}{1.112950in}}%
\pgfpathlineto{\pgfqpoint{2.063042in}{1.091037in}}%
\pgfpathlineto{\pgfqpoint{2.060321in}{1.069198in}}%
\pgfpathlineto{\pgfqpoint{2.056925in}{1.047453in}}%
\pgfpathlineto{\pgfqpoint{2.052857in}{1.025824in}}%
\pgfpathlineto{\pgfqpoint{2.048120in}{1.004332in}}%
\pgfpathlineto{\pgfqpoint{2.042720in}{0.982996in}}%
\pgfpathlineto{\pgfqpoint{2.036661in}{0.961838in}}%
\pgfpathlineto{\pgfqpoint{2.029949in}{0.940878in}}%
\pgfpathlineto{\pgfqpoint{2.022592in}{0.920136in}}%
\pgfpathlineto{\pgfqpoint{2.014595in}{0.899632in}}%
\pgfpathlineto{\pgfqpoint{2.005967in}{0.879386in}}%
\pgfpathlineto{\pgfqpoint{1.996716in}{0.859417in}}%
\pgfpathlineto{\pgfqpoint{1.986850in}{0.839743in}}%
\pgfpathlineto{\pgfqpoint{1.976380in}{0.820385in}}%
\pgfpathlineto{\pgfqpoint{1.965314in}{0.801361in}}%
\pgfpathlineto{\pgfqpoint{1.953665in}{0.782689in}}%
\pgfpathlineto{\pgfqpoint{1.941443in}{0.764386in}}%
\pgfpathlineto{\pgfqpoint{1.928659in}{0.746471in}}%
\pgfpathlineto{\pgfqpoint{1.915327in}{0.728961in}}%
\pgfpathlineto{\pgfqpoint{1.901458in}{0.711872in}}%
\pgfpathlineto{\pgfqpoint{1.887067in}{0.695221in}}%
\pgfpathlineto{\pgfqpoint{1.872167in}{0.679024in}}%
\pgfpathlineto{\pgfqpoint{1.864899in}{0.685109in}}%
\pgfpathlineto{\pgfqpoint{1.796416in}{0.753593in}}%
\pgfpathlineto{\pgfqpoint{1.770803in}{0.727980in}}%
\pgfpathlineto{\pgfqpoint{1.771305in}{0.728482in}}%
\pgfpathlineto{\pgfqpoint{1.839789in}{0.659999in}}%
\pgfpathlineto{\pgfqpoint{1.845874in}{0.652731in}}%
\pgfpathlineto{\pgfqpoint{1.829677in}{0.637831in}}%
\pgfpathlineto{\pgfqpoint{1.813026in}{0.623440in}}%
\pgfpathlineto{\pgfqpoint{1.795937in}{0.609571in}}%
\pgfpathlineto{\pgfqpoint{1.778427in}{0.596239in}}%
\pgfpathlineto{\pgfqpoint{1.760512in}{0.583455in}}%
\pgfpathlineto{\pgfqpoint{1.742209in}{0.571233in}}%
\pgfpathlineto{\pgfqpoint{1.723537in}{0.559584in}}%
\pgfpathlineto{\pgfqpoint{1.704513in}{0.548518in}}%
\pgfpathlineto{\pgfqpoint{1.685155in}{0.538048in}}%
\pgfpathlineto{\pgfqpoint{1.665481in}{0.528182in}}%
\pgfpathlineto{\pgfqpoint{1.645512in}{0.518931in}}%
\pgfpathlineto{\pgfqpoint{1.625266in}{0.510303in}}%
\pgfpathlineto{\pgfqpoint{1.604761in}{0.502306in}}%
\pgfpathlineto{\pgfqpoint{1.584019in}{0.494949in}}%
\pgfpathlineto{\pgfqpoint{1.563059in}{0.488237in}}%
\pgfpathlineto{\pgfqpoint{1.541902in}{0.482178in}}%
\pgfpathlineto{\pgfqpoint{1.520566in}{0.476778in}}%
\pgfpathlineto{\pgfqpoint{1.499074in}{0.472041in}}%
\pgfpathlineto{\pgfqpoint{1.477445in}{0.467973in}}%
\pgfpathlineto{\pgfqpoint{1.455700in}{0.464576in}}%
\pgfpathlineto{\pgfqpoint{1.433861in}{0.461855in}}%
\pgfpathlineto{\pgfqpoint{1.411947in}{0.459812in}}%
\pgfpathlineto{\pgfqpoint{1.389981in}{0.458449in}}%
\pgfpathlineto{\pgfqpoint{1.367984in}{0.457767in}}%
\pgfpathlineto{\pgfqpoint{1.345975in}{0.457767in}}%
\pgfpathlineto{\pgfqpoint{1.323978in}{0.458449in}}%
\pgfpathlineto{\pgfqpoint{1.302012in}{0.459812in}}%
\pgfpathlineto{\pgfqpoint{1.280098in}{0.461855in}}%
\pgfpathlineto{\pgfqpoint{1.258259in}{0.464576in}}%
\pgfpathlineto{\pgfqpoint{1.236514in}{0.467973in}}%
\pgfpathlineto{\pgfqpoint{1.214885in}{0.472041in}}%
\pgfpathlineto{\pgfqpoint{1.193393in}{0.476778in}}%
\pgfpathlineto{\pgfqpoint{1.172058in}{0.482178in}}%
\pgfpathlineto{\pgfqpoint{1.150900in}{0.488237in}}%
\pgfpathlineto{\pgfqpoint{1.129940in}{0.494949in}}%
\pgfpathlineto{\pgfqpoint{1.109198in}{0.502306in}}%
\pgfpathlineto{\pgfqpoint{1.088694in}{0.510303in}}%
\pgfpathlineto{\pgfqpoint{1.068447in}{0.518931in}}%
\pgfpathlineto{\pgfqpoint{1.048478in}{0.528182in}}%
\pgfpathlineto{\pgfqpoint{1.028805in}{0.538048in}}%
\pgfpathlineto{\pgfqpoint{1.009447in}{0.548518in}}%
\pgfpathlineto{\pgfqpoint{0.990422in}{0.559584in}}%
\pgfpathlineto{\pgfqpoint{0.971750in}{0.571233in}}%
\pgfpathlineto{\pgfqpoint{0.953447in}{0.583455in}}%
\pgfpathlineto{\pgfqpoint{0.935532in}{0.596239in}}%
\pgfpathlineto{\pgfqpoint{0.918022in}{0.609571in}}%
\pgfpathlineto{\pgfqpoint{0.900933in}{0.623440in}}%
\pgfpathlineto{\pgfqpoint{0.884282in}{0.637831in}}%
\pgfpathlineto{\pgfqpoint{0.868085in}{0.652731in}}%
\pgfpathlineto{\pgfqpoint{0.874170in}{0.659999in}}%
\pgfpathlineto{\pgfqpoint{0.942654in}{0.728482in}}%
\pgfpathlineto{\pgfqpoint{0.917041in}{0.754095in}}%
\pgfpathlineto{\pgfqpoint{0.917543in}{0.753593in}}%
\pgfpathlineto{\pgfqpoint{0.849060in}{0.685109in}}%
\pgfpathlineto{\pgfqpoint{0.841792in}{0.679024in}}%
\pgfpathlineto{\pgfqpoint{0.826892in}{0.695221in}}%
\pgfpathlineto{\pgfqpoint{0.812501in}{0.711872in}}%
\pgfpathlineto{\pgfqpoint{0.798632in}{0.728961in}}%
\pgfpathlineto{\pgfqpoint{0.785300in}{0.746471in}}%
\pgfpathlineto{\pgfqpoint{0.772516in}{0.764386in}}%
\pgfpathlineto{\pgfqpoint{0.760294in}{0.782689in}}%
\pgfpathlineto{\pgfqpoint{0.748645in}{0.801361in}}%
\pgfpathlineto{\pgfqpoint{0.737580in}{0.820385in}}%
\pgfpathlineto{\pgfqpoint{0.727109in}{0.839743in}}%
\pgfpathlineto{\pgfqpoint{0.717244in}{0.859417in}}%
\pgfpathlineto{\pgfqpoint{0.707992in}{0.879386in}}%
\pgfpathlineto{\pgfqpoint{0.699364in}{0.899632in}}%
\pgfpathlineto{\pgfqpoint{0.691367in}{0.920136in}}%
\pgfpathlineto{\pgfqpoint{0.684010in}{0.940878in}}%
\pgfpathlineto{\pgfqpoint{0.677298in}{0.961838in}}%
\pgfpathlineto{\pgfqpoint{0.671240in}{0.982996in}}%
\pgfpathlineto{\pgfqpoint{0.665839in}{1.004332in}}%
\pgfpathlineto{\pgfqpoint{0.661103in}{1.025824in}}%
\pgfpathlineto{\pgfqpoint{0.657034in}{1.047453in}}%
\pgfpathlineto{\pgfqpoint{0.653638in}{1.069198in}}%
\pgfpathlineto{\pgfqpoint{0.650917in}{1.091037in}}%
\pgfpathlineto{\pgfqpoint{0.648874in}{1.112950in}}%
\pgfpathlineto{\pgfqpoint{0.647510in}{1.134917in}}%
\pgfpathlineto{\pgfqpoint{0.646829in}{1.156914in}}%
\pgfpathlineto{\pgfqpoint{0.646829in}{1.178922in}}%
\pgfpathlineto{\pgfqpoint{0.647510in}{1.200920in}}%
\pgfpathlineto{\pgfqpoint{0.648874in}{1.222886in}}%
\pgfpathlineto{\pgfqpoint{0.650917in}{1.244799in}}%
\pgfpathlineto{\pgfqpoint{0.653638in}{1.266639in}}%
\pgfpathlineto{\pgfqpoint{0.657034in}{1.288384in}}%
\pgfpathlineto{\pgfqpoint{0.661103in}{1.310012in}}%
\pgfpathlineto{\pgfqpoint{0.665839in}{1.331505in}}%
\pgfpathlineto{\pgfqpoint{0.671240in}{1.352840in}}%
\pgfpathlineto{\pgfqpoint{0.677298in}{1.373998in}}%
\pgfpathlineto{\pgfqpoint{0.684010in}{1.394958in}}%
\pgfpathlineto{\pgfqpoint{0.691367in}{1.415700in}}%
\pgfpathlineto{\pgfqpoint{0.699364in}{1.436204in}}%
\pgfpathlineto{\pgfqpoint{0.707992in}{1.456451in}}%
\pgfpathlineto{\pgfqpoint{0.717244in}{1.476420in}}%
\pgfpathlineto{\pgfqpoint{0.727109in}{1.496093in}}%
\pgfpathlineto{\pgfqpoint{0.737580in}{1.515451in}}%
\pgfpathlineto{\pgfqpoint{0.748645in}{1.534476in}}%
\pgfpathlineto{\pgfqpoint{0.760294in}{1.553148in}}%
\pgfpathlineto{\pgfqpoint{0.772516in}{1.571451in}}%
\pgfpathlineto{\pgfqpoint{0.785300in}{1.589366in}}%
\pgfpathlineto{\pgfqpoint{0.798632in}{1.606876in}}%
\pgfpathlineto{\pgfqpoint{0.812501in}{1.623965in}}%
\pgfpathlineto{\pgfqpoint{0.826892in}{1.640616in}}%
\pgfpathlineto{\pgfqpoint{0.841792in}{1.656813in}}%
\pgfpathlineto{\pgfqpoint{0.849060in}{1.650728in}}%
\pgfpathlineto{\pgfqpoint{0.917543in}{1.582244in}}%
\pgfpathlineto{\pgfqpoint{0.942654in}{1.607355in}}%
\pgfpathlineto{\pgfqpoint{0.874170in}{1.675838in}}%
\pgfpathlineto{\pgfqpoint{0.874170in}{1.675838in}}%
\pgfusepath{stroke}%
\end{pgfscope}%
\begin{pgfscope}%
\pgfsetrectcap%
\pgfsetmiterjoin%
\pgfsetlinewidth{0.803000pt}%
\definecolor{currentstroke}{rgb}{0.000000,0.000000,0.000000}%
\pgfsetstrokecolor{currentstroke}%
\pgfsetdash{}{0pt}%
\pgfpathmoveto{\pgfqpoint{0.575720in}{0.386658in}}%
\pgfpathlineto{\pgfqpoint{0.575720in}{1.949178in}}%
\pgfusepath{stroke}%
\end{pgfscope}%
\begin{pgfscope}%
\pgfsetrectcap%
\pgfsetmiterjoin%
\pgfsetlinewidth{0.803000pt}%
\definecolor{currentstroke}{rgb}{0.000000,0.000000,0.000000}%
\pgfsetstrokecolor{currentstroke}%
\pgfsetdash{}{0pt}%
\pgfpathmoveto{\pgfqpoint{2.138239in}{0.386658in}}%
\pgfpathlineto{\pgfqpoint{2.138239in}{1.949178in}}%
\pgfusepath{stroke}%
\end{pgfscope}%
\begin{pgfscope}%
\pgfsetrectcap%
\pgfsetmiterjoin%
\pgfsetlinewidth{0.803000pt}%
\definecolor{currentstroke}{rgb}{0.000000,0.000000,0.000000}%
\pgfsetstrokecolor{currentstroke}%
\pgfsetdash{}{0pt}%
\pgfpathmoveto{\pgfqpoint{0.575720in}{0.386658in}}%
\pgfpathlineto{\pgfqpoint{2.138239in}{0.386658in}}%
\pgfusepath{stroke}%
\end{pgfscope}%
\begin{pgfscope}%
\pgfsetrectcap%
\pgfsetmiterjoin%
\pgfsetlinewidth{0.803000pt}%
\definecolor{currentstroke}{rgb}{0.000000,0.000000,0.000000}%
\pgfsetstrokecolor{currentstroke}%
\pgfsetdash{}{0pt}%
\pgfpathmoveto{\pgfqpoint{0.575720in}{1.949178in}}%
\pgfpathlineto{\pgfqpoint{2.138239in}{1.949178in}}%
\pgfusepath{stroke}%
\end{pgfscope}%
\begin{pgfscope}%
\pgfpathrectangle{\pgfqpoint{0.575720in}{0.386658in}}{\pgfqpoint{1.562520in}{1.562520in}}%
\pgfusepath{clip}%
\pgfsetrectcap%
\pgfsetroundjoin%
\pgfsetlinewidth{1.505625pt}%
\definecolor{currentstroke}{rgb}{0.000000,0.000000,0.000000}%
\pgfsetstrokecolor{currentstroke}%
\pgfsetdash{}{0pt}%
\pgfpathmoveto{\pgfqpoint{1.072885in}{1.167918in}}%
\pgfpathlineto{\pgfqpoint{1.641074in}{1.167918in}}%
\pgfusepath{stroke}%
\end{pgfscope}%
\begin{pgfscope}%
\pgfpathrectangle{\pgfqpoint{0.575720in}{0.386658in}}{\pgfqpoint{1.562520in}{1.562520in}}%
\pgfusepath{clip}%
\pgfsetrectcap%
\pgfsetroundjoin%
\pgfsetlinewidth{1.505625pt}%
\definecolor{currentstroke}{rgb}{0.000000,0.000000,0.000000}%
\pgfsetstrokecolor{currentstroke}%
\pgfsetdash{}{0pt}%
\pgfpathmoveto{\pgfqpoint{1.356980in}{0.883824in}}%
\pgfpathlineto{\pgfqpoint{1.356980in}{1.452013in}}%
\pgfusepath{stroke}%
\end{pgfscope}%
\begin{pgfscope}%
\pgfsetbuttcap%
\pgfsetmiterjoin%
\definecolor{currentfill}{rgb}{1.000000,1.000000,1.000000}%
\pgfsetfillcolor{currentfill}%
\pgfsetlinewidth{0.000000pt}%
\definecolor{currentstroke}{rgb}{0.000000,0.000000,0.000000}%
\pgfsetstrokecolor{currentstroke}%
\pgfsetstrokeopacity{0.000000}%
\pgfsetdash{}{0pt}%
\pgfpathmoveto{\pgfqpoint{0.337193in}{2.046836in}}%
\pgfpathlineto{\pgfqpoint{2.376767in}{2.046836in}}%
\pgfpathlineto{\pgfqpoint{2.376767in}{2.148814in}}%
\pgfpathlineto{\pgfqpoint{0.337193in}{2.148814in}}%
\pgfpathlineto{\pgfqpoint{0.337193in}{2.046836in}}%
\pgfpathclose%
\pgfusepath{fill}%
\end{pgfscope}%
\begin{pgfscope}%
\pgfsys@transformshift{0.337500in}{2.046250in}%
\pgftext[left,bottom]{\includegraphics[interpolate=true,width=2.038750in,height=0.102500in]{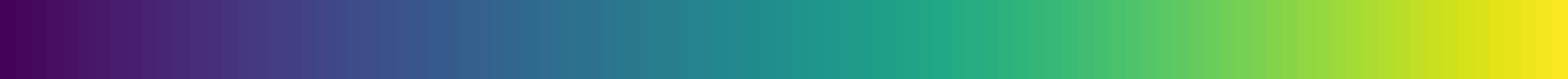}}%
\end{pgfscope}%
\begin{pgfscope}%
\pgfsetbuttcap%
\pgfsetroundjoin%
\definecolor{currentfill}{rgb}{0.000000,0.000000,0.000000}%
\pgfsetfillcolor{currentfill}%
\pgfsetlinewidth{0.803000pt}%
\definecolor{currentstroke}{rgb}{0.000000,0.000000,0.000000}%
\pgfsetstrokecolor{currentstroke}%
\pgfsetdash{}{0pt}%
\pgfsys@defobject{currentmarker}{\pgfqpoint{0.000000in}{0.000000in}}{\pgfqpoint{0.000000in}{0.048611in}}{%
\pgfpathmoveto{\pgfqpoint{0.000000in}{0.000000in}}%
\pgfpathlineto{\pgfqpoint{0.000000in}{0.048611in}}%
\pgfusepath{stroke,fill}%
}%
\begin{pgfscope}%
\pgfsys@transformshift{0.337193in}{2.148814in}%
\pgfsys@useobject{currentmarker}{}%
\end{pgfscope}%
\end{pgfscope}%
\begin{pgfscope}%
\definecolor{textcolor}{rgb}{0.000000,0.000000,0.000000}%
\pgfsetstrokecolor{textcolor}%
\pgfsetfillcolor{textcolor}%
\pgftext[x=0.337193in,y=2.239092in,,bottom]{\color{textcolor}{\rmfamily\fontsize{8.330000}{9.996000}\selectfont\catcode`\^=\active\def^{\ifmmode\sp\else\^{}\fi}\catcode`\%=\active\def%{\%}$\mathdefault{0.000}$}}%
\end{pgfscope}%
\begin{pgfscope}%
\pgfsetbuttcap%
\pgfsetroundjoin%
\definecolor{currentfill}{rgb}{0.000000,0.000000,0.000000}%
\pgfsetfillcolor{currentfill}%
\pgfsetlinewidth{0.803000pt}%
\definecolor{currentstroke}{rgb}{0.000000,0.000000,0.000000}%
\pgfsetstrokecolor{currentstroke}%
\pgfsetdash{}{0pt}%
\pgfsys@defobject{currentmarker}{\pgfqpoint{0.000000in}{0.000000in}}{\pgfqpoint{0.000000in}{0.048611in}}{%
\pgfpathmoveto{\pgfqpoint{0.000000in}{0.000000in}}%
\pgfpathlineto{\pgfqpoint{0.000000in}{0.048611in}}%
\pgfusepath{stroke,fill}%
}%
\begin{pgfscope}%
\pgfsys@transformshift{1.011351in}{2.148814in}%
\pgfsys@useobject{currentmarker}{}%
\end{pgfscope}%
\end{pgfscope}%
\begin{pgfscope}%
\definecolor{textcolor}{rgb}{0.000000,0.000000,0.000000}%
\pgfsetstrokecolor{textcolor}%
\pgfsetfillcolor{textcolor}%
\pgftext[x=1.011351in,y=2.239092in,,bottom]{\color{textcolor}{\rmfamily\fontsize{8.330000}{9.996000}\selectfont\catcode`\^=\active\def^{\ifmmode\sp\else\^{}\fi}\catcode`\%=\active\def%{\%}$\mathdefault{1.309}$}}%
\end{pgfscope}%
\begin{pgfscope}%
\pgfsetbuttcap%
\pgfsetroundjoin%
\definecolor{currentfill}{rgb}{0.000000,0.000000,0.000000}%
\pgfsetfillcolor{currentfill}%
\pgfsetlinewidth{0.803000pt}%
\definecolor{currentstroke}{rgb}{0.000000,0.000000,0.000000}%
\pgfsetstrokecolor{currentstroke}%
\pgfsetdash{}{0pt}%
\pgfsys@defobject{currentmarker}{\pgfqpoint{0.000000in}{0.000000in}}{\pgfqpoint{0.000000in}{0.048611in}}{%
\pgfpathmoveto{\pgfqpoint{0.000000in}{0.000000in}}%
\pgfpathlineto{\pgfqpoint{0.000000in}{0.048611in}}%
\pgfusepath{stroke,fill}%
}%
\begin{pgfscope}%
\pgfsys@transformshift{1.685509in}{2.148814in}%
\pgfsys@useobject{currentmarker}{}%
\end{pgfscope}%
\end{pgfscope}%
\begin{pgfscope}%
\definecolor{textcolor}{rgb}{0.000000,0.000000,0.000000}%
\pgfsetstrokecolor{textcolor}%
\pgfsetfillcolor{textcolor}%
\pgftext[x=1.685509in,y=2.239092in,,bottom]{\color{textcolor}{\rmfamily\fontsize{8.330000}{9.996000}\selectfont\catcode`\^=\active\def^{\ifmmode\sp\else\^{}\fi}\catcode`\%=\active\def%{\%}$\mathdefault{2.618}$}}%
\end{pgfscope}%
\begin{pgfscope}%
\pgfsetbuttcap%
\pgfsetroundjoin%
\definecolor{currentfill}{rgb}{0.000000,0.000000,0.000000}%
\pgfsetfillcolor{currentfill}%
\pgfsetlinewidth{0.803000pt}%
\definecolor{currentstroke}{rgb}{0.000000,0.000000,0.000000}%
\pgfsetstrokecolor{currentstroke}%
\pgfsetdash{}{0pt}%
\pgfsys@defobject{currentmarker}{\pgfqpoint{0.000000in}{0.000000in}}{\pgfqpoint{0.000000in}{0.048611in}}{%
\pgfpathmoveto{\pgfqpoint{0.000000in}{0.000000in}}%
\pgfpathlineto{\pgfqpoint{0.000000in}{0.048611in}}%
\pgfusepath{stroke,fill}%
}%
\begin{pgfscope}%
\pgfsys@transformshift{2.359667in}{2.148814in}%
\pgfsys@useobject{currentmarker}{}%
\end{pgfscope}%
\end{pgfscope}%
\begin{pgfscope}%
\definecolor{textcolor}{rgb}{0.000000,0.000000,0.000000}%
\pgfsetstrokecolor{textcolor}%
\pgfsetfillcolor{textcolor}%
\pgftext[x=2.359667in,y=2.239092in,,bottom]{\color{textcolor}{\rmfamily\fontsize{8.330000}{9.996000}\selectfont\catcode`\^=\active\def^{\ifmmode\sp\else\^{}\fi}\catcode`\%=\active\def%{\%}$\mathdefault{3.927}$}}%
\end{pgfscope}%
\begin{pgfscope}%
\definecolor{textcolor}{rgb}{0.000000,0.000000,0.000000}%
\pgfsetstrokecolor{textcolor}%
\pgfsetfillcolor{textcolor}%
\pgftext[x=1.356980in,y=2.393413in,,base]{\color{textcolor}{\rmfamily\fontsize{10.000000}{12.000000}\selectfont\catcode`\^=\active\def^{\ifmmode\sp\else\^{}\fi}\catcode`\%=\active\def%{\%}$\|\bm{v}\|_2\;[\nicefrac{\mathrm{m}}{\mathrm{s}}]$}}%
\end{pgfscope}%
\begin{pgfscope}%
\definecolor{textcolor}{rgb}{0.000000,0.000000,0.000000}%
\pgfsetstrokecolor{textcolor}%
\pgfsetfillcolor{textcolor}%
\pgftext[x=2.376767in,y=2.379524in,right,bottom]{\color{textcolor}{\rmfamily\fontsize{8.330000}{9.996000}\selectfont\catcode`\^=\active\def^{\ifmmode\sp\else\^{}\fi}\catcode`\%=\active\def%{\%}$\times\mathdefault{10^{\ensuremath{-}2}}\mathdefault{}$}}%
\end{pgfscope}%
\begin{pgfscope}%
\pgfsetrectcap%
\pgfsetmiterjoin%
\pgfsetlinewidth{0.803000pt}%
\definecolor{currentstroke}{rgb}{0.000000,0.000000,0.000000}%
\pgfsetstrokecolor{currentstroke}%
\pgfsetdash{}{0pt}%
\pgfpathmoveto{\pgfqpoint{0.337193in}{2.046836in}}%
\pgfpathlineto{\pgfqpoint{0.337193in}{2.097825in}}%
\pgfpathlineto{\pgfqpoint{0.337193in}{2.148814in}}%
\pgfpathlineto{\pgfqpoint{2.376767in}{2.148814in}}%
\pgfpathlineto{\pgfqpoint{2.376767in}{2.097825in}}%
\pgfpathlineto{\pgfqpoint{2.376767in}{2.046836in}}%
\pgfpathlineto{\pgfqpoint{0.337193in}{2.046836in}}%
\pgfpathclose%
\pgfusepath{stroke}%
\end{pgfscope}%
\end{pgfpicture}%
\makeatother%
\endgroup%

%% file: figures/reference_sol/re_1000/p.pgf
%% Creator: Matplotlib, PGF backend
%%
%% To include the figure in your LaTeX document, write
%%   \input{<filename>.pgf}
%%
%% Make sure the required packages are loaded in your preamble
%%   \usepackage{pgf}
%%
%% Also ensure that all the required font packages are loaded; for instance,
%% the lmodern package is sometimes necessary when using math font.
%%   \usepackage{lmodern}
%%
%% Figures using additional raster images can only be included by \input if
%% they are in the same directory as the main LaTeX file. For loading figures
%% from other directories you can use the `import` package
%% and then include the figures with
%%   \import{<path to file>}{<filename>.pgf}
%%
%% Matplotlib used the following preamble
%%   \def\mathdefault#1{#1}
%%   \everymath=\expandafter{\the\everymath\displaystyle}
%%   \usepackage{amsmath}\usepackage{bm}
%%   \makeatletter\@ifpackageloaded{underscore}{}{\usepackage[strings]{underscore}}\makeatother
%%
\begingroup%
\makeatletter%
\begin{pgfpicture}%
\pgfpathrectangle{\pgfpointorigin}{\pgfqpoint{2.500000in}{2.500000in}}%
\pgfusepath{use as bounding box, clip}%
\begin{pgfscope}%
\pgfsetbuttcap%
\pgfsetmiterjoin%
\definecolor{currentfill}{rgb}{1.000000,1.000000,1.000000}%
\pgfsetfillcolor{currentfill}%
\pgfsetlinewidth{0.000000pt}%
\definecolor{currentstroke}{rgb}{1.000000,1.000000,1.000000}%
\pgfsetstrokecolor{currentstroke}%
\pgfsetdash{}{0pt}%
\pgfpathmoveto{\pgfqpoint{0.000000in}{0.000000in}}%
\pgfpathlineto{\pgfqpoint{2.500000in}{0.000000in}}%
\pgfpathlineto{\pgfqpoint{2.500000in}{2.500000in}}%
\pgfpathlineto{\pgfqpoint{0.000000in}{2.500000in}}%
\pgfpathlineto{\pgfqpoint{0.000000in}{0.000000in}}%
\pgfpathclose%
\pgfusepath{fill}%
\end{pgfscope}%
\begin{pgfscope}%
\pgfsetbuttcap%
\pgfsetmiterjoin%
\definecolor{currentfill}{rgb}{1.000000,1.000000,1.000000}%
\pgfsetfillcolor{currentfill}%
\pgfsetlinewidth{0.000000pt}%
\definecolor{currentstroke}{rgb}{0.000000,0.000000,0.000000}%
\pgfsetstrokecolor{currentstroke}%
\pgfsetstrokeopacity{0.000000}%
\pgfsetdash{}{0pt}%
\pgfpathmoveto{\pgfqpoint{0.595022in}{0.386658in}}%
\pgfpathlineto{\pgfqpoint{2.157542in}{0.386658in}}%
\pgfpathlineto{\pgfqpoint{2.157542in}{1.949179in}}%
\pgfpathlineto{\pgfqpoint{0.595022in}{1.949179in}}%
\pgfpathlineto{\pgfqpoint{0.595022in}{0.386658in}}%
\pgfpathclose%
\pgfusepath{fill}%
\end{pgfscope}%
\begin{pgfscope}%
\pgfsys@transformshift{0.665000in}{0.457500in}%
\pgftext[left,bottom]{\includegraphics[interpolate=true,width=1.422500in,height=1.421250in]{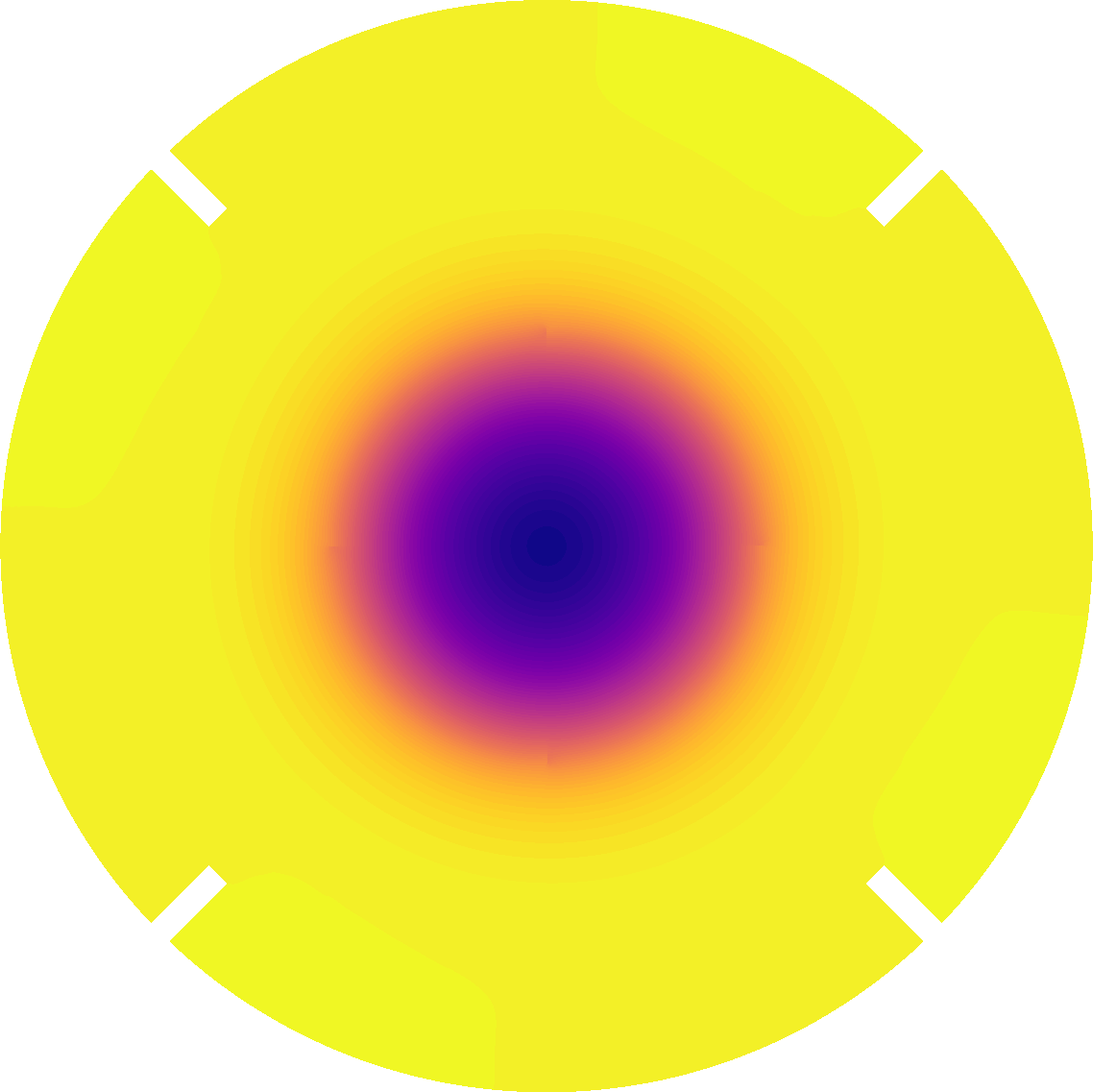}}%
\end{pgfscope}%
\begin{pgfscope}%
\pgfsetbuttcap%
\pgfsetroundjoin%
\definecolor{currentfill}{rgb}{0.000000,0.000000,0.000000}%
\pgfsetfillcolor{currentfill}%
\pgfsetlinewidth{0.803000pt}%
\definecolor{currentstroke}{rgb}{0.000000,0.000000,0.000000}%
\pgfsetstrokecolor{currentstroke}%
\pgfsetdash{}{0pt}%
\pgfsys@defobject{currentmarker}{\pgfqpoint{0.000000in}{-0.048611in}}{\pgfqpoint{0.000000in}{0.000000in}}{%
\pgfpathmoveto{\pgfqpoint{0.000000in}{0.000000in}}%
\pgfpathlineto{\pgfqpoint{0.000000in}{-0.048611in}}%
\pgfusepath{stroke,fill}%
}%
\begin{pgfscope}%
\pgfsys@transformshift{0.666045in}{0.386658in}%
\pgfsys@useobject{currentmarker}{}%
\end{pgfscope}%
\end{pgfscope}%
\begin{pgfscope}%
\definecolor{textcolor}{rgb}{0.000000,0.000000,0.000000}%
\pgfsetstrokecolor{textcolor}%
\pgfsetfillcolor{textcolor}%
\pgftext[x=0.666045in,y=0.296381in,,top]{\color{textcolor}{\rmfamily\fontsize{8.330000}{9.996000}\selectfont\catcode`\^=\active\def^{\ifmmode\sp\else\^{}\fi}\catcode`\%=\active\def%{\%}$\mathdefault{\ensuremath{-}0.1}$}}%
\end{pgfscope}%
\begin{pgfscope}%
\pgfsetbuttcap%
\pgfsetroundjoin%
\definecolor{currentfill}{rgb}{0.000000,0.000000,0.000000}%
\pgfsetfillcolor{currentfill}%
\pgfsetlinewidth{0.803000pt}%
\definecolor{currentstroke}{rgb}{0.000000,0.000000,0.000000}%
\pgfsetstrokecolor{currentstroke}%
\pgfsetdash{}{0pt}%
\pgfsys@defobject{currentmarker}{\pgfqpoint{0.000000in}{-0.048611in}}{\pgfqpoint{0.000000in}{0.000000in}}{%
\pgfpathmoveto{\pgfqpoint{0.000000in}{0.000000in}}%
\pgfpathlineto{\pgfqpoint{0.000000in}{-0.048611in}}%
\pgfusepath{stroke,fill}%
}%
\begin{pgfscope}%
\pgfsys@transformshift{1.376282in}{0.386658in}%
\pgfsys@useobject{currentmarker}{}%
\end{pgfscope}%
\end{pgfscope}%
\begin{pgfscope}%
\definecolor{textcolor}{rgb}{0.000000,0.000000,0.000000}%
\pgfsetstrokecolor{textcolor}%
\pgfsetfillcolor{textcolor}%
\pgftext[x=1.376282in,y=0.296381in,,top]{\color{textcolor}{\rmfamily\fontsize{8.330000}{9.996000}\selectfont\catcode`\^=\active\def^{\ifmmode\sp\else\^{}\fi}\catcode`\%=\active\def%{\%}$\mathdefault{0.0}$}}%
\end{pgfscope}%
\begin{pgfscope}%
\pgfsetbuttcap%
\pgfsetroundjoin%
\definecolor{currentfill}{rgb}{0.000000,0.000000,0.000000}%
\pgfsetfillcolor{currentfill}%
\pgfsetlinewidth{0.803000pt}%
\definecolor{currentstroke}{rgb}{0.000000,0.000000,0.000000}%
\pgfsetstrokecolor{currentstroke}%
\pgfsetdash{}{0pt}%
\pgfsys@defobject{currentmarker}{\pgfqpoint{0.000000in}{-0.048611in}}{\pgfqpoint{0.000000in}{0.000000in}}{%
\pgfpathmoveto{\pgfqpoint{0.000000in}{0.000000in}}%
\pgfpathlineto{\pgfqpoint{0.000000in}{-0.048611in}}%
\pgfusepath{stroke,fill}%
}%
\begin{pgfscope}%
\pgfsys@transformshift{2.086518in}{0.386658in}%
\pgfsys@useobject{currentmarker}{}%
\end{pgfscope}%
\end{pgfscope}%
\begin{pgfscope}%
\definecolor{textcolor}{rgb}{0.000000,0.000000,0.000000}%
\pgfsetstrokecolor{textcolor}%
\pgfsetfillcolor{textcolor}%
\pgftext[x=2.086518in,y=0.296381in,,top]{\color{textcolor}{\rmfamily\fontsize{8.330000}{9.996000}\selectfont\catcode`\^=\active\def^{\ifmmode\sp\else\^{}\fi}\catcode`\%=\active\def%{\%}$\mathdefault{0.1}$}}%
\end{pgfscope}%
\begin{pgfscope}%
\definecolor{textcolor}{rgb}{0.000000,0.000000,0.000000}%
\pgfsetstrokecolor{textcolor}%
\pgfsetfillcolor{textcolor}%
\pgftext[x=1.376282in,y=0.142060in,,top]{\color{textcolor}{\rmfamily\fontsize{10.000000}{12.000000}\selectfont\catcode`\^=\active\def^{\ifmmode\sp\else\^{}\fi}\catcode`\%=\active\def%{\%}$x\;[\text{m}]$}}%
\end{pgfscope}%
\begin{pgfscope}%
\pgfsetbuttcap%
\pgfsetroundjoin%
\definecolor{currentfill}{rgb}{0.000000,0.000000,0.000000}%
\pgfsetfillcolor{currentfill}%
\pgfsetlinewidth{0.803000pt}%
\definecolor{currentstroke}{rgb}{0.000000,0.000000,0.000000}%
\pgfsetstrokecolor{currentstroke}%
\pgfsetdash{}{0pt}%
\pgfsys@defobject{currentmarker}{\pgfqpoint{-0.048611in}{0.000000in}}{\pgfqpoint{-0.000000in}{0.000000in}}{%
\pgfpathmoveto{\pgfqpoint{-0.000000in}{0.000000in}}%
\pgfpathlineto{\pgfqpoint{-0.048611in}{0.000000in}}%
\pgfusepath{stroke,fill}%
}%
\begin{pgfscope}%
\pgfsys@transformshift{0.595022in}{0.457682in}%
\pgfsys@useobject{currentmarker}{}%
\end{pgfscope}%
\end{pgfscope}%
\begin{pgfscope}%
\definecolor{textcolor}{rgb}{0.000000,0.000000,0.000000}%
\pgfsetstrokecolor{textcolor}%
\pgfsetfillcolor{textcolor}%
\pgftext[x=0.262071in, y=0.419102in, left, base]{\color{textcolor}{\rmfamily\fontsize{8.330000}{9.996000}\selectfont\catcode`\^=\active\def^{\ifmmode\sp\else\^{}\fi}\catcode`\%=\active\def%{\%}$\mathdefault{\ensuremath{-}0.1}$}}%
\end{pgfscope}%
\begin{pgfscope}%
\pgfsetbuttcap%
\pgfsetroundjoin%
\definecolor{currentfill}{rgb}{0.000000,0.000000,0.000000}%
\pgfsetfillcolor{currentfill}%
\pgfsetlinewidth{0.803000pt}%
\definecolor{currentstroke}{rgb}{0.000000,0.000000,0.000000}%
\pgfsetstrokecolor{currentstroke}%
\pgfsetdash{}{0pt}%
\pgfsys@defobject{currentmarker}{\pgfqpoint{-0.048611in}{0.000000in}}{\pgfqpoint{-0.000000in}{0.000000in}}{%
\pgfpathmoveto{\pgfqpoint{-0.000000in}{0.000000in}}%
\pgfpathlineto{\pgfqpoint{-0.048611in}{0.000000in}}%
\pgfusepath{stroke,fill}%
}%
\begin{pgfscope}%
\pgfsys@transformshift{0.595022in}{1.167919in}%
\pgfsys@useobject{currentmarker}{}%
\end{pgfscope}%
\end{pgfscope}%
\begin{pgfscope}%
\definecolor{textcolor}{rgb}{0.000000,0.000000,0.000000}%
\pgfsetstrokecolor{textcolor}%
\pgfsetfillcolor{textcolor}%
\pgftext[x=0.353893in, y=1.129338in, left, base]{\color{textcolor}{\rmfamily\fontsize{8.330000}{9.996000}\selectfont\catcode`\^=\active\def^{\ifmmode\sp\else\^{}\fi}\catcode`\%=\active\def%{\%}$\mathdefault{0.0}$}}%
\end{pgfscope}%
\begin{pgfscope}%
\pgfsetbuttcap%
\pgfsetroundjoin%
\definecolor{currentfill}{rgb}{0.000000,0.000000,0.000000}%
\pgfsetfillcolor{currentfill}%
\pgfsetlinewidth{0.803000pt}%
\definecolor{currentstroke}{rgb}{0.000000,0.000000,0.000000}%
\pgfsetstrokecolor{currentstroke}%
\pgfsetdash{}{0pt}%
\pgfsys@defobject{currentmarker}{\pgfqpoint{-0.048611in}{0.000000in}}{\pgfqpoint{-0.000000in}{0.000000in}}{%
\pgfpathmoveto{\pgfqpoint{-0.000000in}{0.000000in}}%
\pgfpathlineto{\pgfqpoint{-0.048611in}{0.000000in}}%
\pgfusepath{stroke,fill}%
}%
\begin{pgfscope}%
\pgfsys@transformshift{0.595022in}{1.878155in}%
\pgfsys@useobject{currentmarker}{}%
\end{pgfscope}%
\end{pgfscope}%
\begin{pgfscope}%
\definecolor{textcolor}{rgb}{0.000000,0.000000,0.000000}%
\pgfsetstrokecolor{textcolor}%
\pgfsetfillcolor{textcolor}%
\pgftext[x=0.353893in, y=1.839575in, left, base]{\color{textcolor}{\rmfamily\fontsize{8.330000}{9.996000}\selectfont\catcode`\^=\active\def^{\ifmmode\sp\else\^{}\fi}\catcode`\%=\active\def%{\%}$\mathdefault{0.1}$}}%
\end{pgfscope}%
\begin{pgfscope}%
\definecolor{textcolor}{rgb}{0.000000,0.000000,0.000000}%
\pgfsetstrokecolor{textcolor}%
\pgfsetfillcolor{textcolor}%
\pgftext[x=0.206515in,y=1.167919in,,bottom,rotate=90.000000]{\color{textcolor}{\rmfamily\fontsize{10.000000}{12.000000}\selectfont\catcode`\^=\active\def^{\ifmmode\sp\else\^{}\fi}\catcode`\%=\active\def%{\%}$y\;[\text{m}]$}}%
\end{pgfscope}%
\begin{pgfscope}%
\pgfpathrectangle{\pgfqpoint{0.595022in}{0.386658in}}{\pgfqpoint{1.562520in}{1.562520in}}%
\pgfusepath{clip}%
\pgfsetbuttcap%
\pgfsetmiterjoin%
\definecolor{currentfill}{rgb}{1.000000,1.000000,1.000000}%
\pgfsetfillcolor{currentfill}%
\pgfsetlinewidth{1.505625pt}%
\definecolor{currentstroke}{rgb}{1.000000,1.000000,1.000000}%
\pgfsetstrokecolor{currentstroke}%
\pgfsetdash{}{0pt}%
\pgfpathmoveto{\pgfqpoint{1.859091in}{1.675839in}}%
\pgfpathlineto{\pgfqpoint{1.852243in}{1.668990in}}%
\pgfpathlineto{\pgfqpoint{1.845395in}{1.662142in}}%
\pgfpathlineto{\pgfqpoint{1.838546in}{1.655294in}}%
\pgfpathlineto{\pgfqpoint{1.831698in}{1.648445in}}%
\pgfpathlineto{\pgfqpoint{1.824850in}{1.641597in}}%
\pgfpathlineto{\pgfqpoint{1.818001in}{1.634749in}}%
\pgfpathlineto{\pgfqpoint{1.811153in}{1.627900in}}%
\pgfpathlineto{\pgfqpoint{1.804304in}{1.621052in}}%
\pgfpathlineto{\pgfqpoint{1.797456in}{1.614204in}}%
\pgfpathlineto{\pgfqpoint{1.790608in}{1.607355in}}%
\pgfpathlineto{\pgfqpoint{1.796885in}{1.601077in}}%
\pgfpathlineto{\pgfqpoint{1.803163in}{1.594800in}}%
\pgfpathlineto{\pgfqpoint{1.809441in}{1.588522in}}%
\pgfpathlineto{\pgfqpoint{1.815718in}{1.582244in}}%
\pgfpathlineto{\pgfqpoint{1.822567in}{1.589093in}}%
\pgfpathlineto{\pgfqpoint{1.829415in}{1.595941in}}%
\pgfpathlineto{\pgfqpoint{1.836263in}{1.602790in}}%
\pgfpathlineto{\pgfqpoint{1.843112in}{1.609638in}}%
\pgfpathlineto{\pgfqpoint{1.849960in}{1.616486in}}%
\pgfpathlineto{\pgfqpoint{1.856809in}{1.623335in}}%
\pgfpathlineto{\pgfqpoint{1.863657in}{1.630183in}}%
\pgfpathlineto{\pgfqpoint{1.870505in}{1.637031in}}%
\pgfpathlineto{\pgfqpoint{1.877354in}{1.643880in}}%
\pgfpathlineto{\pgfqpoint{1.884202in}{1.650728in}}%
\pgfpathlineto{\pgfqpoint{1.859091in}{1.675839in}}%
\pgfpathclose%
\pgfusepath{stroke,fill}%
\end{pgfscope}%
\begin{pgfscope}%
\pgfpathrectangle{\pgfqpoint{0.595022in}{0.386658in}}{\pgfqpoint{1.562520in}{1.562520in}}%
\pgfusepath{clip}%
\pgfsetbuttcap%
\pgfsetmiterjoin%
\definecolor{currentfill}{rgb}{1.000000,1.000000,1.000000}%
\pgfsetfillcolor{currentfill}%
\pgfsetlinewidth{1.505625pt}%
\definecolor{currentstroke}{rgb}{1.000000,1.000000,1.000000}%
\pgfsetstrokecolor{currentstroke}%
\pgfsetdash{}{0pt}%
\pgfpathmoveto{\pgfqpoint{1.884202in}{0.685109in}}%
\pgfpathlineto{\pgfqpoint{1.877354in}{0.691958in}}%
\pgfpathlineto{\pgfqpoint{1.870505in}{0.698806in}}%
\pgfpathlineto{\pgfqpoint{1.863657in}{0.705654in}}%
\pgfpathlineto{\pgfqpoint{1.856809in}{0.712503in}}%
\pgfpathlineto{\pgfqpoint{1.849960in}{0.719351in}}%
\pgfpathlineto{\pgfqpoint{1.843112in}{0.726199in}}%
\pgfpathlineto{\pgfqpoint{1.836263in}{0.733048in}}%
\pgfpathlineto{\pgfqpoint{1.829415in}{0.739896in}}%
\pgfpathlineto{\pgfqpoint{1.822567in}{0.746745in}}%
\pgfpathlineto{\pgfqpoint{1.815718in}{0.753593in}}%
\pgfpathlineto{\pgfqpoint{1.809315in}{0.747190in}}%
\pgfpathlineto{\pgfqpoint{1.802912in}{0.740786in}}%
\pgfpathlineto{\pgfqpoint{1.796509in}{0.734383in}}%
\pgfpathlineto{\pgfqpoint{1.790106in}{0.727980in}}%
\pgfpathlineto{\pgfqpoint{1.790608in}{0.728482in}}%
\pgfpathlineto{\pgfqpoint{1.797456in}{0.721634in}}%
\pgfpathlineto{\pgfqpoint{1.804304in}{0.714786in}}%
\pgfpathlineto{\pgfqpoint{1.811153in}{0.707937in}}%
\pgfpathlineto{\pgfqpoint{1.818001in}{0.701089in}}%
\pgfpathlineto{\pgfqpoint{1.824850in}{0.694240in}}%
\pgfpathlineto{\pgfqpoint{1.831698in}{0.687392in}}%
\pgfpathlineto{\pgfqpoint{1.838546in}{0.680544in}}%
\pgfpathlineto{\pgfqpoint{1.845395in}{0.673695in}}%
\pgfpathlineto{\pgfqpoint{1.852243in}{0.666847in}}%
\pgfpathlineto{\pgfqpoint{1.859091in}{0.659999in}}%
\pgfpathlineto{\pgfqpoint{1.884202in}{0.685109in}}%
\pgfpathclose%
\pgfusepath{stroke,fill}%
\end{pgfscope}%
\begin{pgfscope}%
\pgfpathrectangle{\pgfqpoint{0.595022in}{0.386658in}}{\pgfqpoint{1.562520in}{1.562520in}}%
\pgfusepath{clip}%
\pgfsetbuttcap%
\pgfsetmiterjoin%
\definecolor{currentfill}{rgb}{1.000000,1.000000,1.000000}%
\pgfsetfillcolor{currentfill}%
\pgfsetlinewidth{1.505625pt}%
\definecolor{currentstroke}{rgb}{1.000000,1.000000,1.000000}%
\pgfsetstrokecolor{currentstroke}%
\pgfsetdash{}{0pt}%
\pgfpathmoveto{\pgfqpoint{0.893473in}{0.659999in}}%
\pgfpathlineto{\pgfqpoint{0.900321in}{0.666847in}}%
\pgfpathlineto{\pgfqpoint{0.907169in}{0.673695in}}%
\pgfpathlineto{\pgfqpoint{0.914018in}{0.680544in}}%
\pgfpathlineto{\pgfqpoint{0.920866in}{0.687392in}}%
\pgfpathlineto{\pgfqpoint{0.927714in}{0.694240in}}%
\pgfpathlineto{\pgfqpoint{0.934563in}{0.701089in}}%
\pgfpathlineto{\pgfqpoint{0.941411in}{0.707937in}}%
\pgfpathlineto{\pgfqpoint{0.948259in}{0.714786in}}%
\pgfpathlineto{\pgfqpoint{0.955108in}{0.721634in}}%
\pgfpathlineto{\pgfqpoint{0.961956in}{0.728482in}}%
\pgfpathlineto{\pgfqpoint{0.955553in}{0.734885in}}%
\pgfpathlineto{\pgfqpoint{0.949150in}{0.741289in}}%
\pgfpathlineto{\pgfqpoint{0.942747in}{0.747692in}}%
\pgfpathlineto{\pgfqpoint{0.936343in}{0.754095in}}%
\pgfpathlineto{\pgfqpoint{0.936845in}{0.753593in}}%
\pgfpathlineto{\pgfqpoint{0.929997in}{0.746745in}}%
\pgfpathlineto{\pgfqpoint{0.923149in}{0.739896in}}%
\pgfpathlineto{\pgfqpoint{0.916300in}{0.733048in}}%
\pgfpathlineto{\pgfqpoint{0.909452in}{0.726199in}}%
\pgfpathlineto{\pgfqpoint{0.902604in}{0.719351in}}%
\pgfpathlineto{\pgfqpoint{0.895755in}{0.712503in}}%
\pgfpathlineto{\pgfqpoint{0.888907in}{0.705654in}}%
\pgfpathlineto{\pgfqpoint{0.882059in}{0.698806in}}%
\pgfpathlineto{\pgfqpoint{0.875210in}{0.691958in}}%
\pgfpathlineto{\pgfqpoint{0.868362in}{0.685109in}}%
\pgfpathlineto{\pgfqpoint{0.893473in}{0.659999in}}%
\pgfpathclose%
\pgfusepath{stroke,fill}%
\end{pgfscope}%
\begin{pgfscope}%
\pgfpathrectangle{\pgfqpoint{0.595022in}{0.386658in}}{\pgfqpoint{1.562520in}{1.562520in}}%
\pgfusepath{clip}%
\pgfsetbuttcap%
\pgfsetmiterjoin%
\definecolor{currentfill}{rgb}{1.000000,1.000000,1.000000}%
\pgfsetfillcolor{currentfill}%
\pgfsetlinewidth{1.505625pt}%
\definecolor{currentstroke}{rgb}{1.000000,1.000000,1.000000}%
\pgfsetstrokecolor{currentstroke}%
\pgfsetdash{}{0pt}%
\pgfpathmoveto{\pgfqpoint{0.868362in}{1.650728in}}%
\pgfpathlineto{\pgfqpoint{0.875210in}{1.643880in}}%
\pgfpathlineto{\pgfqpoint{0.882059in}{1.637031in}}%
\pgfpathlineto{\pgfqpoint{0.888907in}{1.630183in}}%
\pgfpathlineto{\pgfqpoint{0.895755in}{1.623335in}}%
\pgfpathlineto{\pgfqpoint{0.902604in}{1.616486in}}%
\pgfpathlineto{\pgfqpoint{0.909452in}{1.609638in}}%
\pgfpathlineto{\pgfqpoint{0.916300in}{1.602790in}}%
\pgfpathlineto{\pgfqpoint{0.923149in}{1.595941in}}%
\pgfpathlineto{\pgfqpoint{0.929997in}{1.589093in}}%
\pgfpathlineto{\pgfqpoint{0.936845in}{1.582244in}}%
\pgfpathlineto{\pgfqpoint{0.943123in}{1.588522in}}%
\pgfpathlineto{\pgfqpoint{0.949401in}{1.594800in}}%
\pgfpathlineto{\pgfqpoint{0.955678in}{1.601077in}}%
\pgfpathlineto{\pgfqpoint{0.961956in}{1.607355in}}%
\pgfpathlineto{\pgfqpoint{0.955108in}{1.614204in}}%
\pgfpathlineto{\pgfqpoint{0.948259in}{1.621052in}}%
\pgfpathlineto{\pgfqpoint{0.941411in}{1.627900in}}%
\pgfpathlineto{\pgfqpoint{0.934563in}{1.634749in}}%
\pgfpathlineto{\pgfqpoint{0.927714in}{1.641597in}}%
\pgfpathlineto{\pgfqpoint{0.920866in}{1.648445in}}%
\pgfpathlineto{\pgfqpoint{0.914018in}{1.655294in}}%
\pgfpathlineto{\pgfqpoint{0.907169in}{1.662142in}}%
\pgfpathlineto{\pgfqpoint{0.900321in}{1.668990in}}%
\pgfpathlineto{\pgfqpoint{0.893473in}{1.675839in}}%
\pgfpathlineto{\pgfqpoint{0.868362in}{1.650728in}}%
\pgfpathclose%
\pgfusepath{stroke,fill}%
\end{pgfscope}%
\begin{pgfscope}%
\pgfpathrectangle{\pgfqpoint{0.595022in}{0.386658in}}{\pgfqpoint{1.562520in}{1.562520in}}%
\pgfusepath{clip}%
\pgfsetrectcap%
\pgfsetroundjoin%
\pgfsetlinewidth{1.505625pt}%
\definecolor{currentstroke}{rgb}{0.000000,0.000000,0.000000}%
\pgfsetstrokecolor{currentstroke}%
\pgfsetdash{}{0pt}%
\pgfpathmoveto{\pgfqpoint{0.868362in}{1.650728in}}%
\pgfpathlineto{\pgfqpoint{0.936845in}{1.582244in}}%
\pgfpathlineto{\pgfqpoint{0.961956in}{1.607355in}}%
\pgfpathlineto{\pgfqpoint{0.893473in}{1.675839in}}%
\pgfpathlineto{\pgfqpoint{0.887387in}{1.683106in}}%
\pgfpathlineto{\pgfqpoint{0.903584in}{1.698006in}}%
\pgfpathlineto{\pgfqpoint{0.920235in}{1.712398in}}%
\pgfpathlineto{\pgfqpoint{0.937324in}{1.726266in}}%
\pgfpathlineto{\pgfqpoint{0.954835in}{1.739599in}}%
\pgfpathlineto{\pgfqpoint{0.972750in}{1.752382in}}%
\pgfpathlineto{\pgfqpoint{0.991052in}{1.764604in}}%
\pgfpathlineto{\pgfqpoint{1.009724in}{1.776254in}}%
\pgfpathlineto{\pgfqpoint{1.028749in}{1.787319in}}%
\pgfpathlineto{\pgfqpoint{1.048107in}{1.797789in}}%
\pgfpathlineto{\pgfqpoint{1.067780in}{1.807655in}}%
\pgfpathlineto{\pgfqpoint{1.087749in}{1.816906in}}%
\pgfpathlineto{\pgfqpoint{1.107996in}{1.825535in}}%
\pgfpathlineto{\pgfqpoint{1.128500in}{1.833531in}}%
\pgfpathlineto{\pgfqpoint{1.149242in}{1.840889in}}%
\pgfpathlineto{\pgfqpoint{1.170202in}{1.847600in}}%
\pgfpathlineto{\pgfqpoint{1.191360in}{1.853659in}}%
\pgfpathlineto{\pgfqpoint{1.212695in}{1.859059in}}%
\pgfpathlineto{\pgfqpoint{1.234188in}{1.863796in}}%
\pgfpathlineto{\pgfqpoint{1.255817in}{1.867864in}}%
\pgfpathlineto{\pgfqpoint{1.277561in}{1.871261in}}%
\pgfpathlineto{\pgfqpoint{1.299401in}{1.873982in}}%
\pgfpathlineto{\pgfqpoint{1.321314in}{1.876025in}}%
\pgfpathlineto{\pgfqpoint{1.343280in}{1.877388in}}%
\pgfpathlineto{\pgfqpoint{1.365278in}{1.878070in}}%
\pgfpathlineto{\pgfqpoint{1.387286in}{1.878070in}}%
\pgfpathlineto{\pgfqpoint{1.409284in}{1.877388in}}%
\pgfpathlineto{\pgfqpoint{1.431250in}{1.876025in}}%
\pgfpathlineto{\pgfqpoint{1.453163in}{1.873982in}}%
\pgfpathlineto{\pgfqpoint{1.475003in}{1.871261in}}%
\pgfpathlineto{\pgfqpoint{1.496747in}{1.867864in}}%
\pgfpathlineto{\pgfqpoint{1.518376in}{1.863796in}}%
\pgfpathlineto{\pgfqpoint{1.539869in}{1.859059in}}%
\pgfpathlineto{\pgfqpoint{1.561204in}{1.853659in}}%
\pgfpathlineto{\pgfqpoint{1.582362in}{1.847600in}}%
\pgfpathlineto{\pgfqpoint{1.603322in}{1.840889in}}%
\pgfpathlineto{\pgfqpoint{1.624064in}{1.833531in}}%
\pgfpathlineto{\pgfqpoint{1.644568in}{1.825535in}}%
\pgfpathlineto{\pgfqpoint{1.664814in}{1.816906in}}%
\pgfpathlineto{\pgfqpoint{1.684784in}{1.807655in}}%
\pgfpathlineto{\pgfqpoint{1.704457in}{1.797789in}}%
\pgfpathlineto{\pgfqpoint{1.723815in}{1.787319in}}%
\pgfpathlineto{\pgfqpoint{1.742839in}{1.776254in}}%
\pgfpathlineto{\pgfqpoint{1.761512in}{1.764604in}}%
\pgfpathlineto{\pgfqpoint{1.779814in}{1.752382in}}%
\pgfpathlineto{\pgfqpoint{1.797729in}{1.739599in}}%
\pgfpathlineto{\pgfqpoint{1.815240in}{1.726266in}}%
\pgfpathlineto{\pgfqpoint{1.832329in}{1.712398in}}%
\pgfpathlineto{\pgfqpoint{1.848979in}{1.698006in}}%
\pgfpathlineto{\pgfqpoint{1.865177in}{1.683106in}}%
\pgfpathlineto{\pgfqpoint{1.859091in}{1.675839in}}%
\pgfpathlineto{\pgfqpoint{1.790608in}{1.607355in}}%
\pgfpathlineto{\pgfqpoint{1.815718in}{1.582244in}}%
\pgfpathlineto{\pgfqpoint{1.884202in}{1.650728in}}%
\pgfpathlineto{\pgfqpoint{1.891469in}{1.656813in}}%
\pgfpathlineto{\pgfqpoint{1.906370in}{1.640616in}}%
\pgfpathlineto{\pgfqpoint{1.920761in}{1.623965in}}%
\pgfpathlineto{\pgfqpoint{1.934630in}{1.606876in}}%
\pgfpathlineto{\pgfqpoint{1.947962in}{1.589366in}}%
\pgfpathlineto{\pgfqpoint{1.960745in}{1.571451in}}%
\pgfpathlineto{\pgfqpoint{1.972968in}{1.553149in}}%
\pgfpathlineto{\pgfqpoint{1.984617in}{1.534476in}}%
\pgfpathlineto{\pgfqpoint{1.995682in}{1.515452in}}%
\pgfpathlineto{\pgfqpoint{2.006153in}{1.496094in}}%
\pgfpathlineto{\pgfqpoint{2.016018in}{1.476421in}}%
\pgfpathlineto{\pgfqpoint{2.025270in}{1.456451in}}%
\pgfpathlineto{\pgfqpoint{2.033898in}{1.436205in}}%
\pgfpathlineto{\pgfqpoint{2.041894in}{1.415701in}}%
\pgfpathlineto{\pgfqpoint{2.049252in}{1.394959in}}%
\pgfpathlineto{\pgfqpoint{2.055963in}{1.373999in}}%
\pgfpathlineto{\pgfqpoint{2.062022in}{1.352841in}}%
\pgfpathlineto{\pgfqpoint{2.067423in}{1.331505in}}%
\pgfpathlineto{\pgfqpoint{2.072159in}{1.310013in}}%
\pgfpathlineto{\pgfqpoint{2.076228in}{1.288384in}}%
\pgfpathlineto{\pgfqpoint{2.079624in}{1.266639in}}%
\pgfpathlineto{\pgfqpoint{2.082345in}{1.244800in}}%
\pgfpathlineto{\pgfqpoint{2.084388in}{1.222887in}}%
\pgfpathlineto{\pgfqpoint{2.085751in}{1.200921in}}%
\pgfpathlineto{\pgfqpoint{2.086433in}{1.178923in}}%
\pgfpathlineto{\pgfqpoint{2.086433in}{1.156915in}}%
\pgfpathlineto{\pgfqpoint{2.085751in}{1.134917in}}%
\pgfpathlineto{\pgfqpoint{2.084388in}{1.112951in}}%
\pgfpathlineto{\pgfqpoint{2.082345in}{1.091038in}}%
\pgfpathlineto{\pgfqpoint{2.079624in}{1.069198in}}%
\pgfpathlineto{\pgfqpoint{2.076228in}{1.047454in}}%
\pgfpathlineto{\pgfqpoint{2.072159in}{1.025825in}}%
\pgfpathlineto{\pgfqpoint{2.067423in}{1.004332in}}%
\pgfpathlineto{\pgfqpoint{2.062022in}{0.982997in}}%
\pgfpathlineto{\pgfqpoint{2.055963in}{0.961839in}}%
\pgfpathlineto{\pgfqpoint{2.049252in}{0.940879in}}%
\pgfpathlineto{\pgfqpoint{2.041894in}{0.920137in}}%
\pgfpathlineto{\pgfqpoint{2.033898in}{0.899633in}}%
\pgfpathlineto{\pgfqpoint{2.025270in}{0.879386in}}%
\pgfpathlineto{\pgfqpoint{2.016018in}{0.859417in}}%
\pgfpathlineto{\pgfqpoint{2.006153in}{0.839744in}}%
\pgfpathlineto{\pgfqpoint{1.995682in}{0.820386in}}%
\pgfpathlineto{\pgfqpoint{1.984617in}{0.801361in}}%
\pgfpathlineto{\pgfqpoint{1.972968in}{0.782689in}}%
\pgfpathlineto{\pgfqpoint{1.960745in}{0.764386in}}%
\pgfpathlineto{\pgfqpoint{1.947962in}{0.746471in}}%
\pgfpathlineto{\pgfqpoint{1.934630in}{0.728961in}}%
\pgfpathlineto{\pgfqpoint{1.920761in}{0.711872in}}%
\pgfpathlineto{\pgfqpoint{1.906370in}{0.695221in}}%
\pgfpathlineto{\pgfqpoint{1.891469in}{0.679024in}}%
\pgfpathlineto{\pgfqpoint{1.884202in}{0.685109in}}%
\pgfpathlineto{\pgfqpoint{1.815718in}{0.753593in}}%
\pgfpathlineto{\pgfqpoint{1.790106in}{0.727980in}}%
\pgfpathlineto{\pgfqpoint{1.790608in}{0.728482in}}%
\pgfpathlineto{\pgfqpoint{1.859091in}{0.659999in}}%
\pgfpathlineto{\pgfqpoint{1.865177in}{0.652731in}}%
\pgfpathlineto{\pgfqpoint{1.848979in}{0.637831in}}%
\pgfpathlineto{\pgfqpoint{1.832329in}{0.623440in}}%
\pgfpathlineto{\pgfqpoint{1.815240in}{0.609571in}}%
\pgfpathlineto{\pgfqpoint{1.797729in}{0.596239in}}%
\pgfpathlineto{\pgfqpoint{1.779814in}{0.583455in}}%
\pgfpathlineto{\pgfqpoint{1.761512in}{0.571233in}}%
\pgfpathlineto{\pgfqpoint{1.742839in}{0.559584in}}%
\pgfpathlineto{\pgfqpoint{1.723815in}{0.548518in}}%
\pgfpathlineto{\pgfqpoint{1.704457in}{0.538048in}}%
\pgfpathlineto{\pgfqpoint{1.684784in}{0.528182in}}%
\pgfpathlineto{\pgfqpoint{1.664814in}{0.518931in}}%
\pgfpathlineto{\pgfqpoint{1.644568in}{0.510303in}}%
\pgfpathlineto{\pgfqpoint{1.624064in}{0.502306in}}%
\pgfpathlineto{\pgfqpoint{1.603322in}{0.494949in}}%
\pgfpathlineto{\pgfqpoint{1.582362in}{0.488237in}}%
\pgfpathlineto{\pgfqpoint{1.561204in}{0.482178in}}%
\pgfpathlineto{\pgfqpoint{1.539869in}{0.476778in}}%
\pgfpathlineto{\pgfqpoint{1.518376in}{0.472041in}}%
\pgfpathlineto{\pgfqpoint{1.496747in}{0.467973in}}%
\pgfpathlineto{\pgfqpoint{1.475003in}{0.464577in}}%
\pgfpathlineto{\pgfqpoint{1.453163in}{0.461855in}}%
\pgfpathlineto{\pgfqpoint{1.431250in}{0.459812in}}%
\pgfpathlineto{\pgfqpoint{1.409284in}{0.458449in}}%
\pgfpathlineto{\pgfqpoint{1.387286in}{0.457767in}}%
\pgfpathlineto{\pgfqpoint{1.365278in}{0.457767in}}%
\pgfpathlineto{\pgfqpoint{1.343280in}{0.458449in}}%
\pgfpathlineto{\pgfqpoint{1.321314in}{0.459812in}}%
\pgfpathlineto{\pgfqpoint{1.299401in}{0.461855in}}%
\pgfpathlineto{\pgfqpoint{1.277561in}{0.464577in}}%
\pgfpathlineto{\pgfqpoint{1.255817in}{0.467973in}}%
\pgfpathlineto{\pgfqpoint{1.234188in}{0.472041in}}%
\pgfpathlineto{\pgfqpoint{1.212695in}{0.476778in}}%
\pgfpathlineto{\pgfqpoint{1.191360in}{0.482178in}}%
\pgfpathlineto{\pgfqpoint{1.170202in}{0.488237in}}%
\pgfpathlineto{\pgfqpoint{1.149242in}{0.494949in}}%
\pgfpathlineto{\pgfqpoint{1.128500in}{0.502306in}}%
\pgfpathlineto{\pgfqpoint{1.107996in}{0.510303in}}%
\pgfpathlineto{\pgfqpoint{1.087749in}{0.518931in}}%
\pgfpathlineto{\pgfqpoint{1.067780in}{0.528182in}}%
\pgfpathlineto{\pgfqpoint{1.048107in}{0.538048in}}%
\pgfpathlineto{\pgfqpoint{1.028749in}{0.548518in}}%
\pgfpathlineto{\pgfqpoint{1.009724in}{0.559584in}}%
\pgfpathlineto{\pgfqpoint{0.991052in}{0.571233in}}%
\pgfpathlineto{\pgfqpoint{0.972750in}{0.583455in}}%
\pgfpathlineto{\pgfqpoint{0.954835in}{0.596239in}}%
\pgfpathlineto{\pgfqpoint{0.937324in}{0.609571in}}%
\pgfpathlineto{\pgfqpoint{0.920235in}{0.623440in}}%
\pgfpathlineto{\pgfqpoint{0.903584in}{0.637831in}}%
\pgfpathlineto{\pgfqpoint{0.887387in}{0.652731in}}%
\pgfpathlineto{\pgfqpoint{0.893473in}{0.659999in}}%
\pgfpathlineto{\pgfqpoint{0.961956in}{0.728482in}}%
\pgfpathlineto{\pgfqpoint{0.936343in}{0.754095in}}%
\pgfpathlineto{\pgfqpoint{0.936845in}{0.753593in}}%
\pgfpathlineto{\pgfqpoint{0.868362in}{0.685109in}}%
\pgfpathlineto{\pgfqpoint{0.861095in}{0.679024in}}%
\pgfpathlineto{\pgfqpoint{0.846194in}{0.695221in}}%
\pgfpathlineto{\pgfqpoint{0.831803in}{0.711872in}}%
\pgfpathlineto{\pgfqpoint{0.817934in}{0.728961in}}%
\pgfpathlineto{\pgfqpoint{0.804602in}{0.746471in}}%
\pgfpathlineto{\pgfqpoint{0.791818in}{0.764386in}}%
\pgfpathlineto{\pgfqpoint{0.779596in}{0.782689in}}%
\pgfpathlineto{\pgfqpoint{0.767947in}{0.801361in}}%
\pgfpathlineto{\pgfqpoint{0.756882in}{0.820386in}}%
\pgfpathlineto{\pgfqpoint{0.746411in}{0.839744in}}%
\pgfpathlineto{\pgfqpoint{0.736546in}{0.859417in}}%
\pgfpathlineto{\pgfqpoint{0.727294in}{0.879386in}}%
\pgfpathlineto{\pgfqpoint{0.718666in}{0.899633in}}%
\pgfpathlineto{\pgfqpoint{0.710669in}{0.920137in}}%
\pgfpathlineto{\pgfqpoint{0.703312in}{0.940879in}}%
\pgfpathlineto{\pgfqpoint{0.696600in}{0.961839in}}%
\pgfpathlineto{\pgfqpoint{0.690542in}{0.982997in}}%
\pgfpathlineto{\pgfqpoint{0.685141in}{1.004332in}}%
\pgfpathlineto{\pgfqpoint{0.680405in}{1.025825in}}%
\pgfpathlineto{\pgfqpoint{0.676336in}{1.047454in}}%
\pgfpathlineto{\pgfqpoint{0.672940in}{1.069198in}}%
\pgfpathlineto{\pgfqpoint{0.670219in}{1.091038in}}%
\pgfpathlineto{\pgfqpoint{0.668176in}{1.112951in}}%
\pgfpathlineto{\pgfqpoint{0.666813in}{1.134917in}}%
\pgfpathlineto{\pgfqpoint{0.666131in}{1.156915in}}%
\pgfpathlineto{\pgfqpoint{0.666131in}{1.178923in}}%
\pgfpathlineto{\pgfqpoint{0.666813in}{1.200921in}}%
\pgfpathlineto{\pgfqpoint{0.668176in}{1.222887in}}%
\pgfpathlineto{\pgfqpoint{0.670219in}{1.244800in}}%
\pgfpathlineto{\pgfqpoint{0.672940in}{1.266639in}}%
\pgfpathlineto{\pgfqpoint{0.676336in}{1.288384in}}%
\pgfpathlineto{\pgfqpoint{0.680405in}{1.310013in}}%
\pgfpathlineto{\pgfqpoint{0.685141in}{1.331505in}}%
\pgfpathlineto{\pgfqpoint{0.690542in}{1.352841in}}%
\pgfpathlineto{\pgfqpoint{0.696600in}{1.373999in}}%
\pgfpathlineto{\pgfqpoint{0.703312in}{1.394959in}}%
\pgfpathlineto{\pgfqpoint{0.710669in}{1.415701in}}%
\pgfpathlineto{\pgfqpoint{0.718666in}{1.436205in}}%
\pgfpathlineto{\pgfqpoint{0.727294in}{1.456451in}}%
\pgfpathlineto{\pgfqpoint{0.736546in}{1.476421in}}%
\pgfpathlineto{\pgfqpoint{0.746411in}{1.496094in}}%
\pgfpathlineto{\pgfqpoint{0.756882in}{1.515452in}}%
\pgfpathlineto{\pgfqpoint{0.767947in}{1.534476in}}%
\pgfpathlineto{\pgfqpoint{0.779596in}{1.553149in}}%
\pgfpathlineto{\pgfqpoint{0.791818in}{1.571451in}}%
\pgfpathlineto{\pgfqpoint{0.804602in}{1.589366in}}%
\pgfpathlineto{\pgfqpoint{0.817934in}{1.606876in}}%
\pgfpathlineto{\pgfqpoint{0.831803in}{1.623965in}}%
\pgfpathlineto{\pgfqpoint{0.846194in}{1.640616in}}%
\pgfpathlineto{\pgfqpoint{0.861095in}{1.656813in}}%
\pgfpathlineto{\pgfqpoint{0.868362in}{1.650728in}}%
\pgfpathlineto{\pgfqpoint{0.936845in}{1.582244in}}%
\pgfpathlineto{\pgfqpoint{0.961956in}{1.607355in}}%
\pgfpathlineto{\pgfqpoint{0.893473in}{1.675839in}}%
\pgfpathlineto{\pgfqpoint{0.893473in}{1.675839in}}%
\pgfusepath{stroke}%
\end{pgfscope}%
\begin{pgfscope}%
\pgfsetrectcap%
\pgfsetmiterjoin%
\pgfsetlinewidth{0.803000pt}%
\definecolor{currentstroke}{rgb}{0.000000,0.000000,0.000000}%
\pgfsetstrokecolor{currentstroke}%
\pgfsetdash{}{0pt}%
\pgfpathmoveto{\pgfqpoint{0.595022in}{0.386658in}}%
\pgfpathlineto{\pgfqpoint{0.595022in}{1.949179in}}%
\pgfusepath{stroke}%
\end{pgfscope}%
\begin{pgfscope}%
\pgfsetrectcap%
\pgfsetmiterjoin%
\pgfsetlinewidth{0.803000pt}%
\definecolor{currentstroke}{rgb}{0.000000,0.000000,0.000000}%
\pgfsetstrokecolor{currentstroke}%
\pgfsetdash{}{0pt}%
\pgfpathmoveto{\pgfqpoint{2.157542in}{0.386658in}}%
\pgfpathlineto{\pgfqpoint{2.157542in}{1.949179in}}%
\pgfusepath{stroke}%
\end{pgfscope}%
\begin{pgfscope}%
\pgfsetrectcap%
\pgfsetmiterjoin%
\pgfsetlinewidth{0.803000pt}%
\definecolor{currentstroke}{rgb}{0.000000,0.000000,0.000000}%
\pgfsetstrokecolor{currentstroke}%
\pgfsetdash{}{0pt}%
\pgfpathmoveto{\pgfqpoint{0.595022in}{0.386658in}}%
\pgfpathlineto{\pgfqpoint{2.157542in}{0.386658in}}%
\pgfusepath{stroke}%
\end{pgfscope}%
\begin{pgfscope}%
\pgfsetrectcap%
\pgfsetmiterjoin%
\pgfsetlinewidth{0.803000pt}%
\definecolor{currentstroke}{rgb}{0.000000,0.000000,0.000000}%
\pgfsetstrokecolor{currentstroke}%
\pgfsetdash{}{0pt}%
\pgfpathmoveto{\pgfqpoint{0.595022in}{1.949179in}}%
\pgfpathlineto{\pgfqpoint{2.157542in}{1.949179in}}%
\pgfusepath{stroke}%
\end{pgfscope}%
\begin{pgfscope}%
\pgfpathrectangle{\pgfqpoint{0.595022in}{0.386658in}}{\pgfqpoint{1.562520in}{1.562520in}}%
\pgfusepath{clip}%
\pgfsetrectcap%
\pgfsetroundjoin%
\pgfsetlinewidth{1.505625pt}%
\definecolor{currentstroke}{rgb}{0.000000,0.000000,0.000000}%
\pgfsetstrokecolor{currentstroke}%
\pgfsetdash{}{0pt}%
\pgfpathmoveto{\pgfqpoint{1.092187in}{1.167919in}}%
\pgfpathlineto{\pgfqpoint{1.660377in}{1.167919in}}%
\pgfusepath{stroke}%
\end{pgfscope}%
\begin{pgfscope}%
\pgfpathrectangle{\pgfqpoint{0.595022in}{0.386658in}}{\pgfqpoint{1.562520in}{1.562520in}}%
\pgfusepath{clip}%
\pgfsetrectcap%
\pgfsetroundjoin%
\pgfsetlinewidth{1.505625pt}%
\definecolor{currentstroke}{rgb}{0.000000,0.000000,0.000000}%
\pgfsetstrokecolor{currentstroke}%
\pgfsetdash{}{0pt}%
\pgfpathmoveto{\pgfqpoint{1.376282in}{0.883824in}}%
\pgfpathlineto{\pgfqpoint{1.376282in}{1.452013in}}%
\pgfusepath{stroke}%
\end{pgfscope}%
\begin{pgfscope}%
\pgfsetbuttcap%
\pgfsetmiterjoin%
\definecolor{currentfill}{rgb}{1.000000,1.000000,1.000000}%
\pgfsetfillcolor{currentfill}%
\pgfsetlinewidth{0.000000pt}%
\definecolor{currentstroke}{rgb}{0.000000,0.000000,0.000000}%
\pgfsetstrokecolor{currentstroke}%
\pgfsetstrokeopacity{0.000000}%
\pgfsetdash{}{0pt}%
\pgfpathmoveto{\pgfqpoint{0.337193in}{2.046836in}}%
\pgfpathlineto{\pgfqpoint{2.415371in}{2.046836in}}%
\pgfpathlineto{\pgfqpoint{2.415371in}{2.150745in}}%
\pgfpathlineto{\pgfqpoint{0.337193in}{2.150745in}}%
\pgfpathlineto{\pgfqpoint{0.337193in}{2.046836in}}%
\pgfpathclose%
\pgfusepath{fill}%
\end{pgfscope}%
\begin{pgfscope}%
\pgfsys@transformshift{0.337500in}{2.046250in}%
\pgftext[left,bottom]{\includegraphics[interpolate=true,width=2.077500in,height=0.105000in]{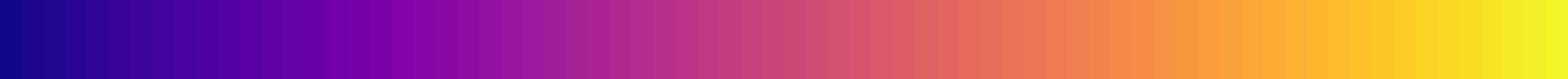}}%
\end{pgfscope}%
\begin{pgfscope}%
\pgfsetbuttcap%
\pgfsetroundjoin%
\definecolor{currentfill}{rgb}{0.000000,0.000000,0.000000}%
\pgfsetfillcolor{currentfill}%
\pgfsetlinewidth{0.803000pt}%
\definecolor{currentstroke}{rgb}{0.000000,0.000000,0.000000}%
\pgfsetstrokecolor{currentstroke}%
\pgfsetdash{}{0pt}%
\pgfsys@defobject{currentmarker}{\pgfqpoint{0.000000in}{0.000000in}}{\pgfqpoint{0.000000in}{0.048611in}}{%
\pgfpathmoveto{\pgfqpoint{0.000000in}{0.000000in}}%
\pgfpathlineto{\pgfqpoint{0.000000in}{0.048611in}}%
\pgfusepath{stroke,fill}%
}%
\begin{pgfscope}%
\pgfsys@transformshift{0.352983in}{2.150745in}%
\pgfsys@useobject{currentmarker}{}%
\end{pgfscope}%
\end{pgfscope}%
\begin{pgfscope}%
\definecolor{textcolor}{rgb}{0.000000,0.000000,0.000000}%
\pgfsetstrokecolor{textcolor}%
\pgfsetfillcolor{textcolor}%
\pgftext[x=0.352983in,y=2.241023in,,bottom]{\color{textcolor}{\rmfamily\fontsize{8.330000}{9.996000}\selectfont\catcode`\^=\active\def^{\ifmmode\sp\else\^{}\fi}\catcode`\%=\active\def%{\%}$\mathdefault{\ensuremath{-}9.97}$}}%
\end{pgfscope}%
\begin{pgfscope}%
\pgfsetbuttcap%
\pgfsetroundjoin%
\definecolor{currentfill}{rgb}{0.000000,0.000000,0.000000}%
\pgfsetfillcolor{currentfill}%
\pgfsetlinewidth{0.803000pt}%
\definecolor{currentstroke}{rgb}{0.000000,0.000000,0.000000}%
\pgfsetstrokecolor{currentstroke}%
\pgfsetdash{}{0pt}%
\pgfsys@defobject{currentmarker}{\pgfqpoint{0.000000in}{0.000000in}}{\pgfqpoint{0.000000in}{0.048611in}}{%
\pgfpathmoveto{\pgfqpoint{0.000000in}{0.000000in}}%
\pgfpathlineto{\pgfqpoint{0.000000in}{0.048611in}}%
\pgfusepath{stroke,fill}%
}%
\begin{pgfscope}%
\pgfsys@transformshift{1.031782in}{2.150745in}%
\pgfsys@useobject{currentmarker}{}%
\end{pgfscope}%
\end{pgfscope}%
\begin{pgfscope}%
\definecolor{textcolor}{rgb}{0.000000,0.000000,0.000000}%
\pgfsetstrokecolor{textcolor}%
\pgfsetfillcolor{textcolor}%
\pgftext[x=1.031782in,y=2.241023in,,bottom]{\color{textcolor}{\rmfamily\fontsize{8.330000}{9.996000}\selectfont\catcode`\^=\active\def^{\ifmmode\sp\else\^{}\fi}\catcode`\%=\active\def%{\%}$\mathdefault{\ensuremath{-}6.44}$}}%
\end{pgfscope}%
\begin{pgfscope}%
\pgfsetbuttcap%
\pgfsetroundjoin%
\definecolor{currentfill}{rgb}{0.000000,0.000000,0.000000}%
\pgfsetfillcolor{currentfill}%
\pgfsetlinewidth{0.803000pt}%
\definecolor{currentstroke}{rgb}{0.000000,0.000000,0.000000}%
\pgfsetstrokecolor{currentstroke}%
\pgfsetdash{}{0pt}%
\pgfsys@defobject{currentmarker}{\pgfqpoint{0.000000in}{0.000000in}}{\pgfqpoint{0.000000in}{0.048611in}}{%
\pgfpathmoveto{\pgfqpoint{0.000000in}{0.000000in}}%
\pgfpathlineto{\pgfqpoint{0.000000in}{0.048611in}}%
\pgfusepath{stroke,fill}%
}%
\begin{pgfscope}%
\pgfsys@transformshift{1.710581in}{2.150745in}%
\pgfsys@useobject{currentmarker}{}%
\end{pgfscope}%
\end{pgfscope}%
\begin{pgfscope}%
\definecolor{textcolor}{rgb}{0.000000,0.000000,0.000000}%
\pgfsetstrokecolor{textcolor}%
\pgfsetfillcolor{textcolor}%
\pgftext[x=1.710581in,y=2.241023in,,bottom]{\color{textcolor}{\rmfamily\fontsize{8.330000}{9.996000}\selectfont\catcode`\^=\active\def^{\ifmmode\sp\else\^{}\fi}\catcode`\%=\active\def%{\%}$\mathdefault{\ensuremath{-}2.91}$}}%
\end{pgfscope}%
\begin{pgfscope}%
\pgfsetbuttcap%
\pgfsetroundjoin%
\definecolor{currentfill}{rgb}{0.000000,0.000000,0.000000}%
\pgfsetfillcolor{currentfill}%
\pgfsetlinewidth{0.803000pt}%
\definecolor{currentstroke}{rgb}{0.000000,0.000000,0.000000}%
\pgfsetstrokecolor{currentstroke}%
\pgfsetdash{}{0pt}%
\pgfsys@defobject{currentmarker}{\pgfqpoint{0.000000in}{0.000000in}}{\pgfqpoint{0.000000in}{0.048611in}}{%
\pgfpathmoveto{\pgfqpoint{0.000000in}{0.000000in}}%
\pgfpathlineto{\pgfqpoint{0.000000in}{0.048611in}}%
\pgfusepath{stroke,fill}%
}%
\begin{pgfscope}%
\pgfsys@transformshift{2.389380in}{2.150745in}%
\pgfsys@useobject{currentmarker}{}%
\end{pgfscope}%
\end{pgfscope}%
\begin{pgfscope}%
\definecolor{textcolor}{rgb}{0.000000,0.000000,0.000000}%
\pgfsetstrokecolor{textcolor}%
\pgfsetfillcolor{textcolor}%
\pgftext[x=2.389380in,y=2.241023in,,bottom]{\color{textcolor}{\rmfamily\fontsize{8.330000}{9.996000}\selectfont\catcode`\^=\active\def^{\ifmmode\sp\else\^{}\fi}\catcode`\%=\active\def%{\%}$\mathdefault{0.61}$}}%
\end{pgfscope}%
\begin{pgfscope}%
\definecolor{textcolor}{rgb}{0.000000,0.000000,0.000000}%
\pgfsetstrokecolor{textcolor}%
\pgfsetfillcolor{textcolor}%
\pgftext[x=1.376282in,y=2.395344in,,base]{\color{textcolor}{\rmfamily\fontsize{10.000000}{12.000000}\selectfont\catcode`\^=\active\def^{\ifmmode\sp\else\^{}\fi}\catcode`\%=\active\def%{\%}$p\;[\mathrm{Pa}]$}}%
\end{pgfscope}%
\begin{pgfscope}%
\definecolor{textcolor}{rgb}{0.000000,0.000000,0.000000}%
\pgfsetstrokecolor{textcolor}%
\pgfsetfillcolor{textcolor}%
\pgftext[x=2.415371in,y=2.381455in,right,bottom]{\color{textcolor}{\rmfamily\fontsize{8.330000}{9.996000}\selectfont\catcode`\^=\active\def^{\ifmmode\sp\else\^{}\fi}\catcode`\%=\active\def%{\%}$\times\mathdefault{10^{\ensuremath{-}1}}\mathdefault{}$}}%
\end{pgfscope}%
\begin{pgfscope}%
\pgfsetrectcap%
\pgfsetmiterjoin%
\pgfsetlinewidth{0.803000pt}%
\definecolor{currentstroke}{rgb}{0.000000,0.000000,0.000000}%
\pgfsetstrokecolor{currentstroke}%
\pgfsetdash{}{0pt}%
\pgfpathmoveto{\pgfqpoint{0.337193in}{2.046836in}}%
\pgfpathlineto{\pgfqpoint{0.337193in}{2.098791in}}%
\pgfpathlineto{\pgfqpoint{0.337193in}{2.150745in}}%
\pgfpathlineto{\pgfqpoint{2.415371in}{2.150745in}}%
\pgfpathlineto{\pgfqpoint{2.415371in}{2.098791in}}%
\pgfpathlineto{\pgfqpoint{2.415371in}{2.046836in}}%
\pgfpathlineto{\pgfqpoint{0.337193in}{2.046836in}}%
\pgfpathclose%
\pgfusepath{stroke}%
\end{pgfscope}%
\end{pgfpicture}%
\makeatother%
\endgroup%

%% file: figures/section1/n_points_per_Re.pgf
%% Creator: Matplotlib, PGF backend
%%
%% To include the figure in your LaTeX document, write
%%   \input{<filename>.pgf}
%%
%% Make sure the required packages are loaded in your preamble
%%   \usepackage{pgf}
%%
%% Also ensure that all the required font packages are loaded; for instance,
%% the lmodern package is sometimes necessary when using math font.
%%   \usepackage{lmodern}
%%
%% Figures using additional raster images can only be included by \input if
%% they are in the same directory as the main LaTeX file. For loading figures
%% from other directories you can use the `import` package
%% and then include the figures with
%%   \import{<path to file>}{<filename>.pgf}
%%
%% Matplotlib used the following preamble
%%   \def\mathdefault#1{#1}
%%   \everymath=\expandafter{\the\everymath\displaystyle}
%%   \usepackage{amsmath}\usepackage{bm}
%%   \makeatletter\@ifpackageloaded{underscore}{}{\usepackage[strings]{underscore}}\makeatother
%%
\begingroup%
\makeatletter%
\begin{pgfpicture}%
\pgfpathrectangle{\pgfpointorigin}{\pgfqpoint{3.000000in}{2.000000in}}%
\pgfusepath{use as bounding box, clip}%
\begin{pgfscope}%
\pgfsetbuttcap%
\pgfsetmiterjoin%
\definecolor{currentfill}{rgb}{1.000000,1.000000,1.000000}%
\pgfsetfillcolor{currentfill}%
\pgfsetlinewidth{0.000000pt}%
\definecolor{currentstroke}{rgb}{1.000000,1.000000,1.000000}%
\pgfsetstrokecolor{currentstroke}%
\pgfsetdash{}{0pt}%
\pgfpathmoveto{\pgfqpoint{0.000000in}{0.000000in}}%
\pgfpathlineto{\pgfqpoint{3.000000in}{0.000000in}}%
\pgfpathlineto{\pgfqpoint{3.000000in}{2.000000in}}%
\pgfpathlineto{\pgfqpoint{0.000000in}{2.000000in}}%
\pgfpathlineto{\pgfqpoint{0.000000in}{0.000000in}}%
\pgfpathclose%
\pgfusepath{fill}%
\end{pgfscope}%
\begin{pgfscope}%
\pgfsetbuttcap%
\pgfsetmiterjoin%
\definecolor{currentfill}{rgb}{1.000000,1.000000,1.000000}%
\pgfsetfillcolor{currentfill}%
\pgfsetlinewidth{0.000000pt}%
\definecolor{currentstroke}{rgb}{0.000000,0.000000,0.000000}%
\pgfsetstrokecolor{currentstroke}%
\pgfsetstrokeopacity{0.000000}%
\pgfsetdash{}{0pt}%
\pgfpathmoveto{\pgfqpoint{0.664468in}{0.521284in}}%
\pgfpathlineto{\pgfqpoint{2.802597in}{0.521284in}}%
\pgfpathlineto{\pgfqpoint{2.802597in}{1.850000in}}%
\pgfpathlineto{\pgfqpoint{0.664468in}{1.850000in}}%
\pgfpathlineto{\pgfqpoint{0.664468in}{0.521284in}}%
\pgfpathclose%
\pgfusepath{fill}%
\end{pgfscope}%
\begin{pgfscope}%
\pgfpathrectangle{\pgfqpoint{0.664468in}{0.521284in}}{\pgfqpoint{2.138130in}{1.328716in}}%
\pgfusepath{clip}%
\pgfsetbuttcap%
\pgfsetroundjoin%
\definecolor{currentfill}{rgb}{0.172549,0.482353,0.713725}%
\pgfsetfillcolor{currentfill}%
\pgfsetlinewidth{0.501875pt}%
\definecolor{currentstroke}{rgb}{0.000000,0.000000,0.000000}%
\pgfsetstrokecolor{currentstroke}%
\pgfsetdash{}{0pt}%
\pgfsys@defobject{currentmarker}{\pgfqpoint{-0.031056in}{-0.031056in}}{\pgfqpoint{0.031056in}{0.031056in}}{%
\pgfpathmoveto{\pgfqpoint{0.000000in}{0.031056in}}%
\pgfpathlineto{\pgfqpoint{-0.031056in}{-0.031056in}}%
\pgfpathlineto{\pgfqpoint{0.031056in}{-0.031056in}}%
\pgfpathlineto{\pgfqpoint{0.000000in}{0.031056in}}%
\pgfpathclose%
\pgfusepath{stroke,fill}%
}%
\begin{pgfscope}%
\pgfsys@transformshift{2.705410in}{0.736487in}%
\pgfsys@useobject{currentmarker}{}%
\end{pgfscope}%
\begin{pgfscope}%
\pgfsys@transformshift{2.305472in}{0.742514in}%
\pgfsys@useobject{currentmarker}{}%
\end{pgfscope}%
\begin{pgfscope}%
\pgfsys@transformshift{2.062179in}{0.734961in}%
\pgfsys@useobject{currentmarker}{}%
\end{pgfscope}%
\begin{pgfscope}%
\pgfsys@transformshift{1.818887in}{0.705190in}%
\pgfsys@useobject{currentmarker}{}%
\end{pgfscope}%
\begin{pgfscope}%
\pgfsys@transformshift{1.453948in}{0.863063in}%
\pgfsys@useobject{currentmarker}{}%
\end{pgfscope}%
\begin{pgfscope}%
\pgfsys@transformshift{1.210655in}{0.955526in}%
\pgfsys@useobject{currentmarker}{}%
\end{pgfscope}%
\begin{pgfscope}%
\pgfsys@transformshift{0.761655in}{1.044119in}%
\pgfsys@useobject{currentmarker}{}%
\end{pgfscope}%
\end{pgfscope}%
\begin{pgfscope}%
\pgfpathrectangle{\pgfqpoint{0.664468in}{0.521284in}}{\pgfqpoint{2.138130in}{1.328716in}}%
\pgfusepath{clip}%
\pgfsetbuttcap%
\pgfsetroundjoin%
\definecolor{currentfill}{rgb}{0.172549,0.482353,0.713725}%
\pgfsetfillcolor{currentfill}%
\pgfsetlinewidth{0.501875pt}%
\definecolor{currentstroke}{rgb}{0.000000,0.000000,0.000000}%
\pgfsetstrokecolor{currentstroke}%
\pgfsetdash{}{0pt}%
\pgfsys@defobject{currentmarker}{\pgfqpoint{-0.043921in}{-0.043921in}}{\pgfqpoint{0.043921in}{0.043921in}}{%
\pgfpathmoveto{\pgfqpoint{-0.000000in}{-0.043921in}}%
\pgfpathlineto{\pgfqpoint{0.043921in}{0.000000in}}%
\pgfpathlineto{\pgfqpoint{0.000000in}{0.043921in}}%
\pgfpathlineto{\pgfqpoint{-0.043921in}{0.000000in}}%
\pgfpathlineto{\pgfqpoint{-0.000000in}{-0.043921in}}%
\pgfpathclose%
\pgfusepath{stroke,fill}%
}%
\begin{pgfscope}%
\pgfsys@transformshift{2.705410in}{0.608787in}%
\pgfsys@useobject{currentmarker}{}%
\end{pgfscope}%
\begin{pgfscope}%
\pgfsys@transformshift{2.305472in}{0.581680in}%
\pgfsys@useobject{currentmarker}{}%
\end{pgfscope}%
\begin{pgfscope}%
\pgfsys@transformshift{2.062179in}{0.601639in}%
\pgfsys@useobject{currentmarker}{}%
\end{pgfscope}%
\begin{pgfscope}%
\pgfsys@transformshift{1.818887in}{0.633596in}%
\pgfsys@useobject{currentmarker}{}%
\end{pgfscope}%
\begin{pgfscope}%
\pgfsys@transformshift{1.453948in}{0.900582in}%
\pgfsys@useobject{currentmarker}{}%
\end{pgfscope}%
\begin{pgfscope}%
\pgfsys@transformshift{1.210655in}{0.850957in}%
\pgfsys@useobject{currentmarker}{}%
\end{pgfscope}%
\begin{pgfscope}%
\pgfsys@transformshift{0.761655in}{1.078041in}%
\pgfsys@useobject{currentmarker}{}%
\end{pgfscope}%
\end{pgfscope}%
\begin{pgfscope}%
\pgfpathrectangle{\pgfqpoint{0.664468in}{0.521284in}}{\pgfqpoint{2.138130in}{1.328716in}}%
\pgfusepath{clip}%
\pgfsetbuttcap%
\pgfsetroundjoin%
\definecolor{currentfill}{rgb}{0.843137,0.098039,0.109804}%
\pgfsetfillcolor{currentfill}%
\pgfsetlinewidth{0.501875pt}%
\definecolor{currentstroke}{rgb}{0.000000,0.000000,0.000000}%
\pgfsetstrokecolor{currentstroke}%
\pgfsetdash{}{0pt}%
\pgfsys@defobject{currentmarker}{\pgfqpoint{-0.031056in}{-0.031056in}}{\pgfqpoint{0.031056in}{0.031056in}}{%
\pgfpathmoveto{\pgfqpoint{0.000000in}{-0.031056in}}%
\pgfpathcurveto{\pgfqpoint{0.008236in}{-0.031056in}}{\pgfqpoint{0.016136in}{-0.027784in}}{\pgfqpoint{0.021960in}{-0.021960in}}%
\pgfpathcurveto{\pgfqpoint{0.027784in}{-0.016136in}}{\pgfqpoint{0.031056in}{-0.008236in}}{\pgfqpoint{0.031056in}{0.000000in}}%
\pgfpathcurveto{\pgfqpoint{0.031056in}{0.008236in}}{\pgfqpoint{0.027784in}{0.016136in}}{\pgfqpoint{0.021960in}{0.021960in}}%
\pgfpathcurveto{\pgfqpoint{0.016136in}{0.027784in}}{\pgfqpoint{0.008236in}{0.031056in}}{\pgfqpoint{0.000000in}{0.031056in}}%
\pgfpathcurveto{\pgfqpoint{-0.008236in}{0.031056in}}{\pgfqpoint{-0.016136in}{0.027784in}}{\pgfqpoint{-0.021960in}{0.021960in}}%
\pgfpathcurveto{\pgfqpoint{-0.027784in}{0.016136in}}{\pgfqpoint{-0.031056in}{0.008236in}}{\pgfqpoint{-0.031056in}{0.000000in}}%
\pgfpathcurveto{\pgfqpoint{-0.031056in}{-0.008236in}}{\pgfqpoint{-0.027784in}{-0.016136in}}{\pgfqpoint{-0.021960in}{-0.021960in}}%
\pgfpathcurveto{\pgfqpoint{-0.016136in}{-0.027784in}}{\pgfqpoint{-0.008236in}{-0.031056in}}{\pgfqpoint{0.000000in}{-0.031056in}}%
\pgfpathlineto{\pgfqpoint{0.000000in}{-0.031056in}}%
\pgfpathclose%
\pgfusepath{stroke,fill}%
}%
\begin{pgfscope}%
\pgfsys@transformshift{2.705410in}{0.821737in}%
\pgfsys@useobject{currentmarker}{}%
\end{pgfscope}%
\begin{pgfscope}%
\pgfsys@transformshift{2.305472in}{0.846808in}%
\pgfsys@useobject{currentmarker}{}%
\end{pgfscope}%
\begin{pgfscope}%
\pgfsys@transformshift{2.062179in}{0.819453in}%
\pgfsys@useobject{currentmarker}{}%
\end{pgfscope}%
\begin{pgfscope}%
\pgfsys@transformshift{1.818887in}{0.997122in}%
\pgfsys@useobject{currentmarker}{}%
\end{pgfscope}%
\begin{pgfscope}%
\pgfsys@transformshift{1.453948in}{1.276371in}%
\pgfsys@useobject{currentmarker}{}%
\end{pgfscope}%
\begin{pgfscope}%
\pgfsys@transformshift{1.210655in}{1.561744in}%
\pgfsys@useobject{currentmarker}{}%
\end{pgfscope}%
\end{pgfscope}%
\begin{pgfscope}%
\pgfpathrectangle{\pgfqpoint{0.664468in}{0.521284in}}{\pgfqpoint{2.138130in}{1.328716in}}%
\pgfusepath{clip}%
\pgfsetbuttcap%
\pgfsetroundjoin%
\definecolor{currentfill}{rgb}{0.843137,0.098039,0.109804}%
\pgfsetfillcolor{currentfill}%
\pgfsetlinewidth{0.501875pt}%
\definecolor{currentstroke}{rgb}{0.000000,0.000000,0.000000}%
\pgfsetstrokecolor{currentstroke}%
\pgfsetdash{}{0pt}%
\pgfsys@defobject{currentmarker}{\pgfqpoint{-0.031056in}{-0.031056in}}{\pgfqpoint{0.031056in}{0.031056in}}{%
\pgfpathmoveto{\pgfqpoint{-0.031056in}{-0.031056in}}%
\pgfpathlineto{\pgfqpoint{0.031056in}{-0.031056in}}%
\pgfpathlineto{\pgfqpoint{0.031056in}{0.031056in}}%
\pgfpathlineto{\pgfqpoint{-0.031056in}{0.031056in}}%
\pgfpathlineto{\pgfqpoint{-0.031056in}{-0.031056in}}%
\pgfpathclose%
\pgfusepath{stroke,fill}%
}%
\begin{pgfscope}%
\pgfsys@transformshift{2.705410in}{0.607607in}%
\pgfsys@useobject{currentmarker}{}%
\end{pgfscope}%
\begin{pgfscope}%
\pgfsys@transformshift{2.305472in}{0.633596in}%
\pgfsys@useobject{currentmarker}{}%
\end{pgfscope}%
\begin{pgfscope}%
\pgfsys@transformshift{2.062179in}{0.657467in}%
\pgfsys@useobject{currentmarker}{}%
\end{pgfscope}%
\begin{pgfscope}%
\pgfsys@transformshift{1.818887in}{1.100857in}%
\pgfsys@useobject{currentmarker}{}%
\end{pgfscope}%
\begin{pgfscope}%
\pgfsys@transformshift{1.453948in}{1.551315in}%
\pgfsys@useobject{currentmarker}{}%
\end{pgfscope}%
\begin{pgfscope}%
\pgfsys@transformshift{1.210655in}{1.789604in}%
\pgfsys@useobject{currentmarker}{}%
\end{pgfscope}%
\end{pgfscope}%
\begin{pgfscope}%
\pgfsetbuttcap%
\pgfsetroundjoin%
\definecolor{currentfill}{rgb}{0.000000,0.000000,0.000000}%
\pgfsetfillcolor{currentfill}%
\pgfsetlinewidth{0.803000pt}%
\definecolor{currentstroke}{rgb}{0.000000,0.000000,0.000000}%
\pgfsetstrokecolor{currentstroke}%
\pgfsetdash{}{0pt}%
\pgfsys@defobject{currentmarker}{\pgfqpoint{0.000000in}{-0.048611in}}{\pgfqpoint{0.000000in}{0.000000in}}{%
\pgfpathmoveto{\pgfqpoint{0.000000in}{0.000000in}}%
\pgfpathlineto{\pgfqpoint{0.000000in}{-0.048611in}}%
\pgfusepath{stroke,fill}%
}%
\begin{pgfscope}%
\pgfsys@transformshift{0.761655in}{0.521284in}%
\pgfsys@useobject{currentmarker}{}%
\end{pgfscope}%
\end{pgfscope}%
\begin{pgfscope}%
\definecolor{textcolor}{rgb}{0.000000,0.000000,0.000000}%
\pgfsetstrokecolor{textcolor}%
\pgfsetfillcolor{textcolor}%
\pgftext[x=0.761655in,y=0.431006in,,top]{\color{textcolor}{\rmfamily\fontsize{8.330000}{9.996000}\selectfont\catcode`\^=\active\def^{\ifmmode\sp\else\^{}\fi}\catcode`\%=\active\def%{\%}0}}%
\end{pgfscope}%
\begin{pgfscope}%
\pgfsetbuttcap%
\pgfsetroundjoin%
\definecolor{currentfill}{rgb}{0.000000,0.000000,0.000000}%
\pgfsetfillcolor{currentfill}%
\pgfsetlinewidth{0.803000pt}%
\definecolor{currentstroke}{rgb}{0.000000,0.000000,0.000000}%
\pgfsetstrokecolor{currentstroke}%
\pgfsetdash{}{0pt}%
\pgfsys@defobject{currentmarker}{\pgfqpoint{0.000000in}{-0.048611in}}{\pgfqpoint{0.000000in}{0.000000in}}{%
\pgfpathmoveto{\pgfqpoint{0.000000in}{0.000000in}}%
\pgfpathlineto{\pgfqpoint{0.000000in}{-0.048611in}}%
\pgfusepath{stroke,fill}%
}%
\begin{pgfscope}%
\pgfsys@transformshift{1.210655in}{0.521284in}%
\pgfsys@useobject{currentmarker}{}%
\end{pgfscope}%
\end{pgfscope}%
\begin{pgfscope}%
\definecolor{textcolor}{rgb}{0.000000,0.000000,0.000000}%
\pgfsetstrokecolor{textcolor}%
\pgfsetfillcolor{textcolor}%
\pgftext[x=1.210655in,y=0.431006in,,top]{\color{textcolor}{\rmfamily\fontsize{8.330000}{9.996000}\selectfont\catcode`\^=\active\def^{\ifmmode\sp\else\^{}\fi}\catcode`\%=\active\def%{\%}2}}%
\end{pgfscope}%
\begin{pgfscope}%
\pgfsetbuttcap%
\pgfsetroundjoin%
\definecolor{currentfill}{rgb}{0.000000,0.000000,0.000000}%
\pgfsetfillcolor{currentfill}%
\pgfsetlinewidth{0.803000pt}%
\definecolor{currentstroke}{rgb}{0.000000,0.000000,0.000000}%
\pgfsetstrokecolor{currentstroke}%
\pgfsetdash{}{0pt}%
\pgfsys@defobject{currentmarker}{\pgfqpoint{0.000000in}{-0.048611in}}{\pgfqpoint{0.000000in}{0.000000in}}{%
\pgfpathmoveto{\pgfqpoint{0.000000in}{0.000000in}}%
\pgfpathlineto{\pgfqpoint{0.000000in}{-0.048611in}}%
\pgfusepath{stroke,fill}%
}%
\begin{pgfscope}%
\pgfsys@transformshift{1.453948in}{0.521284in}%
\pgfsys@useobject{currentmarker}{}%
\end{pgfscope}%
\end{pgfscope}%
\begin{pgfscope}%
\definecolor{textcolor}{rgb}{0.000000,0.000000,0.000000}%
\pgfsetstrokecolor{textcolor}%
\pgfsetfillcolor{textcolor}%
\pgftext[x=1.453948in,y=0.431006in,,top]{\color{textcolor}{\rmfamily\fontsize{8.330000}{9.996000}\selectfont\catcode`\^=\active\def^{\ifmmode\sp\else\^{}\fi}\catcode`\%=\active\def%{\%}8}}%
\end{pgfscope}%
\begin{pgfscope}%
\pgfsetbuttcap%
\pgfsetroundjoin%
\definecolor{currentfill}{rgb}{0.000000,0.000000,0.000000}%
\pgfsetfillcolor{currentfill}%
\pgfsetlinewidth{0.803000pt}%
\definecolor{currentstroke}{rgb}{0.000000,0.000000,0.000000}%
\pgfsetstrokecolor{currentstroke}%
\pgfsetdash{}{0pt}%
\pgfsys@defobject{currentmarker}{\pgfqpoint{0.000000in}{-0.048611in}}{\pgfqpoint{0.000000in}{0.000000in}}{%
\pgfpathmoveto{\pgfqpoint{0.000000in}{0.000000in}}%
\pgfpathlineto{\pgfqpoint{0.000000in}{-0.048611in}}%
\pgfusepath{stroke,fill}%
}%
\begin{pgfscope}%
\pgfsys@transformshift{1.818887in}{0.521284in}%
\pgfsys@useobject{currentmarker}{}%
\end{pgfscope}%
\end{pgfscope}%
\begin{pgfscope}%
\definecolor{textcolor}{rgb}{0.000000,0.000000,0.000000}%
\pgfsetstrokecolor{textcolor}%
\pgfsetfillcolor{textcolor}%
\pgftext[x=1.818887in,y=0.431006in,,top]{\color{textcolor}{\rmfamily\fontsize{8.330000}{9.996000}\selectfont\catcode`\^=\active\def^{\ifmmode\sp\else\^{}\fi}\catcode`\%=\active\def%{\%}64}}%
\end{pgfscope}%
\begin{pgfscope}%
\pgfsetbuttcap%
\pgfsetroundjoin%
\definecolor{currentfill}{rgb}{0.000000,0.000000,0.000000}%
\pgfsetfillcolor{currentfill}%
\pgfsetlinewidth{0.803000pt}%
\definecolor{currentstroke}{rgb}{0.000000,0.000000,0.000000}%
\pgfsetstrokecolor{currentstroke}%
\pgfsetdash{}{0pt}%
\pgfsys@defobject{currentmarker}{\pgfqpoint{0.000000in}{-0.048611in}}{\pgfqpoint{0.000000in}{0.000000in}}{%
\pgfpathmoveto{\pgfqpoint{0.000000in}{0.000000in}}%
\pgfpathlineto{\pgfqpoint{0.000000in}{-0.048611in}}%
\pgfusepath{stroke,fill}%
}%
\begin{pgfscope}%
\pgfsys@transformshift{2.062179in}{0.521284in}%
\pgfsys@useobject{currentmarker}{}%
\end{pgfscope}%
\end{pgfscope}%
\begin{pgfscope}%
\definecolor{textcolor}{rgb}{0.000000,0.000000,0.000000}%
\pgfsetstrokecolor{textcolor}%
\pgfsetfillcolor{textcolor}%
\pgftext[x=2.062179in,y=0.431006in,,top]{\color{textcolor}{\rmfamily\fontsize{8.330000}{9.996000}\selectfont\catcode`\^=\active\def^{\ifmmode\sp\else\^{}\fi}\catcode`\%=\active\def%{\%}256}}%
\end{pgfscope}%
\begin{pgfscope}%
\pgfsetbuttcap%
\pgfsetroundjoin%
\definecolor{currentfill}{rgb}{0.000000,0.000000,0.000000}%
\pgfsetfillcolor{currentfill}%
\pgfsetlinewidth{0.803000pt}%
\definecolor{currentstroke}{rgb}{0.000000,0.000000,0.000000}%
\pgfsetstrokecolor{currentstroke}%
\pgfsetdash{}{0pt}%
\pgfsys@defobject{currentmarker}{\pgfqpoint{0.000000in}{-0.048611in}}{\pgfqpoint{0.000000in}{0.000000in}}{%
\pgfpathmoveto{\pgfqpoint{0.000000in}{0.000000in}}%
\pgfpathlineto{\pgfqpoint{0.000000in}{-0.048611in}}%
\pgfusepath{stroke,fill}%
}%
\begin{pgfscope}%
\pgfsys@transformshift{2.305472in}{0.521284in}%
\pgfsys@useobject{currentmarker}{}%
\end{pgfscope}%
\end{pgfscope}%
\begin{pgfscope}%
\definecolor{textcolor}{rgb}{0.000000,0.000000,0.000000}%
\pgfsetstrokecolor{textcolor}%
\pgfsetfillcolor{textcolor}%
\pgftext[x=2.305472in,y=0.431006in,,top]{\color{textcolor}{\rmfamily\fontsize{8.330000}{9.996000}\selectfont\catcode`\^=\active\def^{\ifmmode\sp\else\^{}\fi}\catcode`\%=\active\def%{\%}1024}}%
\end{pgfscope}%
\begin{pgfscope}%
\pgfsetbuttcap%
\pgfsetroundjoin%
\definecolor{currentfill}{rgb}{0.000000,0.000000,0.000000}%
\pgfsetfillcolor{currentfill}%
\pgfsetlinewidth{0.803000pt}%
\definecolor{currentstroke}{rgb}{0.000000,0.000000,0.000000}%
\pgfsetstrokecolor{currentstroke}%
\pgfsetdash{}{0pt}%
\pgfsys@defobject{currentmarker}{\pgfqpoint{0.000000in}{-0.048611in}}{\pgfqpoint{0.000000in}{0.000000in}}{%
\pgfpathmoveto{\pgfqpoint{0.000000in}{0.000000in}}%
\pgfpathlineto{\pgfqpoint{0.000000in}{-0.048611in}}%
\pgfusepath{stroke,fill}%
}%
\begin{pgfscope}%
\pgfsys@transformshift{2.705410in}{0.521284in}%
\pgfsys@useobject{currentmarker}{}%
\end{pgfscope}%
\end{pgfscope}%
\begin{pgfscope}%
\definecolor{textcolor}{rgb}{0.000000,0.000000,0.000000}%
\pgfsetstrokecolor{textcolor}%
\pgfsetfillcolor{textcolor}%
\pgftext[x=2.705410in,y=0.431006in,,top]{\color{textcolor}{\rmfamily\fontsize{8.330000}{9.996000}\selectfont\catcode`\^=\active\def^{\ifmmode\sp\else\^{}\fi}\catcode`\%=\active\def%{\%}10000}}%
\end{pgfscope}%
\begin{pgfscope}%
\definecolor{textcolor}{rgb}{0.000000,0.000000,0.000000}%
\pgfsetstrokecolor{textcolor}%
\pgfsetfillcolor{textcolor}%
\pgftext[x=1.733532in,y=0.276685in,,top]{\color{textcolor}{\rmfamily\fontsize{10.000000}{12.000000}\selectfont\catcode`\^=\active\def^{\ifmmode\sp\else\^{}\fi}\catcode`\%=\active\def%{\%}Number of Labels per $\mathrm{Re}_\text{train}$}}%
\end{pgfscope}%
\begin{pgfscope}%
\pgfsetbuttcap%
\pgfsetroundjoin%
\definecolor{currentfill}{rgb}{0.000000,0.000000,0.000000}%
\pgfsetfillcolor{currentfill}%
\pgfsetlinewidth{0.803000pt}%
\definecolor{currentstroke}{rgb}{0.000000,0.000000,0.000000}%
\pgfsetstrokecolor{currentstroke}%
\pgfsetdash{}{0pt}%
\pgfsys@defobject{currentmarker}{\pgfqpoint{-0.048611in}{0.000000in}}{\pgfqpoint{-0.000000in}{0.000000in}}{%
\pgfpathmoveto{\pgfqpoint{-0.000000in}{0.000000in}}%
\pgfpathlineto{\pgfqpoint{-0.048611in}{0.000000in}}%
\pgfusepath{stroke,fill}%
}%
\begin{pgfscope}%
\pgfsys@transformshift{0.664468in}{1.016190in}%
\pgfsys@useobject{currentmarker}{}%
\end{pgfscope}%
\end{pgfscope}%
\begin{pgfscope}%
\definecolor{textcolor}{rgb}{0.000000,0.000000,0.000000}%
\pgfsetstrokecolor{textcolor}%
\pgfsetfillcolor{textcolor}%
\pgftext[x=0.398263in, y=0.977062in, left, base]{\color{textcolor}{\rmfamily\fontsize{8.330000}{9.996000}\selectfont\catcode`\^=\active\def^{\ifmmode\sp\else\^{}\fi}\catcode`\%=\active\def%{\%}$\mathdefault{10^{1}}$}}%
\end{pgfscope}%
\begin{pgfscope}%
\pgfsetbuttcap%
\pgfsetroundjoin%
\definecolor{currentfill}{rgb}{0.000000,0.000000,0.000000}%
\pgfsetfillcolor{currentfill}%
\pgfsetlinewidth{0.803000pt}%
\definecolor{currentstroke}{rgb}{0.000000,0.000000,0.000000}%
\pgfsetstrokecolor{currentstroke}%
\pgfsetdash{}{0pt}%
\pgfsys@defobject{currentmarker}{\pgfqpoint{-0.048611in}{0.000000in}}{\pgfqpoint{-0.000000in}{0.000000in}}{%
\pgfpathmoveto{\pgfqpoint{-0.000000in}{0.000000in}}%
\pgfpathlineto{\pgfqpoint{-0.048611in}{0.000000in}}%
\pgfusepath{stroke,fill}%
}%
\begin{pgfscope}%
\pgfsys@transformshift{0.664468in}{1.690923in}%
\pgfsys@useobject{currentmarker}{}%
\end{pgfscope}%
\end{pgfscope}%
\begin{pgfscope}%
\definecolor{textcolor}{rgb}{0.000000,0.000000,0.000000}%
\pgfsetstrokecolor{textcolor}%
\pgfsetfillcolor{textcolor}%
\pgftext[x=0.398263in, y=1.651795in, left, base]{\color{textcolor}{\rmfamily\fontsize{8.330000}{9.996000}\selectfont\catcode`\^=\active\def^{\ifmmode\sp\else\^{}\fi}\catcode`\%=\active\def%{\%}$\mathdefault{10^{2}}$}}%
\end{pgfscope}%
\begin{pgfscope}%
\definecolor{textcolor}{rgb}{0.000000,0.000000,0.000000}%
\pgfsetstrokecolor{textcolor}%
\pgfsetfillcolor{textcolor}%
\pgftext[x=0.342708in,y=1.185642in,,bottom,rotate=90.000000]{\color{textcolor}{\rmfamily\fontsize{10.000000}{12.000000}\selectfont\catcode`\^=\active\def^{\ifmmode\sp\else\^{}\fi}\catcode`\%=\active\def%{\%}$\delta_{\ell^1}^{(q)} \ [\%]$}}%
\end{pgfscope}%
\begin{pgfscope}%
\pgfsetrectcap%
\pgfsetmiterjoin%
\pgfsetlinewidth{0.803000pt}%
\definecolor{currentstroke}{rgb}{0.000000,0.000000,0.000000}%
\pgfsetstrokecolor{currentstroke}%
\pgfsetdash{}{0pt}%
\pgfpathmoveto{\pgfqpoint{0.664468in}{0.521284in}}%
\pgfpathlineto{\pgfqpoint{0.664468in}{1.850000in}}%
\pgfusepath{stroke}%
\end{pgfscope}%
\begin{pgfscope}%
\pgfsetrectcap%
\pgfsetmiterjoin%
\pgfsetlinewidth{0.803000pt}%
\definecolor{currentstroke}{rgb}{0.000000,0.000000,0.000000}%
\pgfsetstrokecolor{currentstroke}%
\pgfsetdash{}{0pt}%
\pgfpathmoveto{\pgfqpoint{2.802597in}{0.521284in}}%
\pgfpathlineto{\pgfqpoint{2.802597in}{1.850000in}}%
\pgfusepath{stroke}%
\end{pgfscope}%
\begin{pgfscope}%
\pgfsetrectcap%
\pgfsetmiterjoin%
\pgfsetlinewidth{0.803000pt}%
\definecolor{currentstroke}{rgb}{0.000000,0.000000,0.000000}%
\pgfsetstrokecolor{currentstroke}%
\pgfsetdash{}{0pt}%
\pgfpathmoveto{\pgfqpoint{0.664468in}{0.521284in}}%
\pgfpathlineto{\pgfqpoint{2.802597in}{0.521284in}}%
\pgfusepath{stroke}%
\end{pgfscope}%
\begin{pgfscope}%
\pgfsetrectcap%
\pgfsetmiterjoin%
\pgfsetlinewidth{0.803000pt}%
\definecolor{currentstroke}{rgb}{0.000000,0.000000,0.000000}%
\pgfsetstrokecolor{currentstroke}%
\pgfsetdash{}{0pt}%
\pgfpathmoveto{\pgfqpoint{0.664468in}{1.850000in}}%
\pgfpathlineto{\pgfqpoint{2.802597in}{1.850000in}}%
\pgfusepath{stroke}%
\end{pgfscope}%
\begin{pgfscope}%
\pgfsetbuttcap%
\pgfsetmiterjoin%
\definecolor{currentfill}{rgb}{1.000000,1.000000,1.000000}%
\pgfsetfillcolor{currentfill}%
\pgfsetfillopacity{0.800000}%
\pgfsetlinewidth{1.003750pt}%
\definecolor{currentstroke}{rgb}{0.800000,0.800000,0.800000}%
\pgfsetstrokecolor{currentstroke}%
\pgfsetstrokeopacity{0.800000}%
\pgfsetdash{}{0pt}%
\pgfpathmoveto{\pgfqpoint{2.018457in}{1.070438in}}%
\pgfpathlineto{\pgfqpoint{2.721611in}{1.070438in}}%
\pgfpathquadraticcurveto{\pgfqpoint{2.744750in}{1.070438in}}{\pgfqpoint{2.744750in}{1.093577in}}%
\pgfpathlineto{\pgfqpoint{2.744750in}{1.769014in}}%
\pgfpathquadraticcurveto{\pgfqpoint{2.744750in}{1.792153in}}{\pgfqpoint{2.721611in}{1.792153in}}%
\pgfpathlineto{\pgfqpoint{2.018457in}{1.792153in}}%
\pgfpathquadraticcurveto{\pgfqpoint{1.995318in}{1.792153in}}{\pgfqpoint{1.995318in}{1.769014in}}%
\pgfpathlineto{\pgfqpoint{1.995318in}{1.093577in}}%
\pgfpathquadraticcurveto{\pgfqpoint{1.995318in}{1.070438in}}{\pgfqpoint{2.018457in}{1.070438in}}%
\pgfpathlineto{\pgfqpoint{2.018457in}{1.070438in}}%
\pgfpathclose%
\pgfusepath{stroke,fill}%
\end{pgfscope}%
\begin{pgfscope}%
\pgfsetbuttcap%
\pgfsetroundjoin%
\definecolor{currentfill}{rgb}{0.172549,0.482353,0.713725}%
\pgfsetfillcolor{currentfill}%
\pgfsetlinewidth{0.501875pt}%
\definecolor{currentstroke}{rgb}{0.000000,0.000000,0.000000}%
\pgfsetstrokecolor{currentstroke}%
\pgfsetdash{}{0pt}%
\pgfsys@defobject{currentmarker}{\pgfqpoint{-0.031056in}{-0.031056in}}{\pgfqpoint{0.031056in}{0.031056in}}{%
\pgfpathmoveto{\pgfqpoint{0.000000in}{0.031056in}}%
\pgfpathlineto{\pgfqpoint{-0.031056in}{-0.031056in}}%
\pgfpathlineto{\pgfqpoint{0.031056in}{-0.031056in}}%
\pgfpathlineto{\pgfqpoint{0.000000in}{0.031056in}}%
\pgfpathclose%
\pgfusepath{stroke,fill}%
}%
\begin{pgfscope}%
\pgfsys@transformshift{2.157290in}{1.694748in}%
\pgfsys@useobject{currentmarker}{}%
\end{pgfscope}%
\end{pgfscope}%
\begin{pgfscope}%
\definecolor{textcolor}{rgb}{0.000000,0.000000,0.000000}%
\pgfsetstrokecolor{textcolor}%
\pgfsetfillcolor{textcolor}%
\pgftext[x=2.365540in,y=1.664378in,left,base]{\color{textcolor}{\rmfamily\fontsize{8.330000}{9.996000}\selectfont\catcode`\^=\active\def^{\ifmmode\sp\else\^{}\fi}\catcode`\%=\active\def%{\%}$\bm{v}^{\mathrm{PINN}}$}}%
\end{pgfscope}%
\begin{pgfscope}%
\pgfsetbuttcap%
\pgfsetroundjoin%
\definecolor{currentfill}{rgb}{0.172549,0.482353,0.713725}%
\pgfsetfillcolor{currentfill}%
\pgfsetlinewidth{0.501875pt}%
\definecolor{currentstroke}{rgb}{0.000000,0.000000,0.000000}%
\pgfsetstrokecolor{currentstroke}%
\pgfsetdash{}{0pt}%
\pgfsys@defobject{currentmarker}{\pgfqpoint{-0.043921in}{-0.043921in}}{\pgfqpoint{0.043921in}{0.043921in}}{%
\pgfpathmoveto{\pgfqpoint{-0.000000in}{-0.043921in}}%
\pgfpathlineto{\pgfqpoint{0.043921in}{0.000000in}}%
\pgfpathlineto{\pgfqpoint{0.000000in}{0.043921in}}%
\pgfpathlineto{\pgfqpoint{-0.043921in}{0.000000in}}%
\pgfpathlineto{\pgfqpoint{-0.000000in}{-0.043921in}}%
\pgfpathclose%
\pgfusepath{stroke,fill}%
}%
\begin{pgfscope}%
\pgfsys@transformshift{2.157290in}{1.512194in}%
\pgfsys@useobject{currentmarker}{}%
\end{pgfscope}%
\end{pgfscope}%
\begin{pgfscope}%
\definecolor{textcolor}{rgb}{0.000000,0.000000,0.000000}%
\pgfsetstrokecolor{textcolor}%
\pgfsetfillcolor{textcolor}%
\pgftext[x=2.365540in,y=1.481824in,left,base]{\color{textcolor}{\rmfamily\fontsize{8.330000}{9.996000}\selectfont\catcode`\^=\active\def^{\ifmmode\sp\else\^{}\fi}\catcode`\%=\active\def%{\%}$p^{\mathrm{PINN}}$}}%
\end{pgfscope}%
\begin{pgfscope}%
\pgfsetbuttcap%
\pgfsetroundjoin%
\definecolor{currentfill}{rgb}{0.843137,0.098039,0.109804}%
\pgfsetfillcolor{currentfill}%
\pgfsetlinewidth{0.501875pt}%
\definecolor{currentstroke}{rgb}{0.000000,0.000000,0.000000}%
\pgfsetstrokecolor{currentstroke}%
\pgfsetdash{}{0pt}%
\pgfsys@defobject{currentmarker}{\pgfqpoint{-0.031056in}{-0.031056in}}{\pgfqpoint{0.031056in}{0.031056in}}{%
\pgfpathmoveto{\pgfqpoint{0.000000in}{-0.031056in}}%
\pgfpathcurveto{\pgfqpoint{0.008236in}{-0.031056in}}{\pgfqpoint{0.016136in}{-0.027784in}}{\pgfqpoint{0.021960in}{-0.021960in}}%
\pgfpathcurveto{\pgfqpoint{0.027784in}{-0.016136in}}{\pgfqpoint{0.031056in}{-0.008236in}}{\pgfqpoint{0.031056in}{0.000000in}}%
\pgfpathcurveto{\pgfqpoint{0.031056in}{0.008236in}}{\pgfqpoint{0.027784in}{0.016136in}}{\pgfqpoint{0.021960in}{0.021960in}}%
\pgfpathcurveto{\pgfqpoint{0.016136in}{0.027784in}}{\pgfqpoint{0.008236in}{0.031056in}}{\pgfqpoint{0.000000in}{0.031056in}}%
\pgfpathcurveto{\pgfqpoint{-0.008236in}{0.031056in}}{\pgfqpoint{-0.016136in}{0.027784in}}{\pgfqpoint{-0.021960in}{0.021960in}}%
\pgfpathcurveto{\pgfqpoint{-0.027784in}{0.016136in}}{\pgfqpoint{-0.031056in}{0.008236in}}{\pgfqpoint{-0.031056in}{0.000000in}}%
\pgfpathcurveto{\pgfqpoint{-0.031056in}{-0.008236in}}{\pgfqpoint{-0.027784in}{-0.016136in}}{\pgfqpoint{-0.021960in}{-0.021960in}}%
\pgfpathcurveto{\pgfqpoint{-0.016136in}{-0.027784in}}{\pgfqpoint{-0.008236in}{-0.031056in}}{\pgfqpoint{0.000000in}{-0.031056in}}%
\pgfpathlineto{\pgfqpoint{0.000000in}{-0.031056in}}%
\pgfpathclose%
\pgfusepath{stroke,fill}%
}%
\begin{pgfscope}%
\pgfsys@transformshift{2.157290in}{1.351245in}%
\pgfsys@useobject{currentmarker}{}%
\end{pgfscope}%
\end{pgfscope}%
\begin{pgfscope}%
\definecolor{textcolor}{rgb}{0.000000,0.000000,0.000000}%
\pgfsetstrokecolor{textcolor}%
\pgfsetfillcolor{textcolor}%
\pgftext[x=2.365540in,y=1.320875in,left,base]{\color{textcolor}{\rmfamily\fontsize{8.330000}{9.996000}\selectfont\catcode`\^=\active\def^{\ifmmode\sp\else\^{}\fi}\catcode`\%=\active\def%{\%}$\bm{v}^{\mathrm{NN}}$}}%
\end{pgfscope}%
\begin{pgfscope}%
\pgfsetbuttcap%
\pgfsetroundjoin%
\definecolor{currentfill}{rgb}{0.843137,0.098039,0.109804}%
\pgfsetfillcolor{currentfill}%
\pgfsetlinewidth{0.501875pt}%
\definecolor{currentstroke}{rgb}{0.000000,0.000000,0.000000}%
\pgfsetstrokecolor{currentstroke}%
\pgfsetdash{}{0pt}%
\pgfsys@defobject{currentmarker}{\pgfqpoint{-0.031056in}{-0.031056in}}{\pgfqpoint{0.031056in}{0.031056in}}{%
\pgfpathmoveto{\pgfqpoint{-0.031056in}{-0.031056in}}%
\pgfpathlineto{\pgfqpoint{0.031056in}{-0.031056in}}%
\pgfpathlineto{\pgfqpoint{0.031056in}{0.031056in}}%
\pgfpathlineto{\pgfqpoint{-0.031056in}{0.031056in}}%
\pgfpathlineto{\pgfqpoint{-0.031056in}{-0.031056in}}%
\pgfpathclose%
\pgfusepath{stroke,fill}%
}%
\begin{pgfscope}%
\pgfsys@transformshift{2.157290in}{1.168690in}%
\pgfsys@useobject{currentmarker}{}%
\end{pgfscope}%
\end{pgfscope}%
\begin{pgfscope}%
\definecolor{textcolor}{rgb}{0.000000,0.000000,0.000000}%
\pgfsetstrokecolor{textcolor}%
\pgfsetfillcolor{textcolor}%
\pgftext[x=2.365540in,y=1.138321in,left,base]{\color{textcolor}{\rmfamily\fontsize{8.330000}{9.996000}\selectfont\catcode`\^=\active\def^{\ifmmode\sp\else\^{}\fi}\catcode`\%=\active\def%{\%}$p^{\mathrm{NN}}$}}%
\end{pgfscope}%
\end{pgfpicture}%
\makeatother%
\endgroup%

%% file: figures/section1/data/10000p/re_1000/err_vmag.pgf
%% Creator: Matplotlib, PGF backend
%%
%% To include the figure in your LaTeX document, write
%%   \input{<filename>.pgf}
%%
%% Make sure the required packages are loaded in your preamble
%%   \usepackage{pgf}
%%
%% Also ensure that all the required font packages are loaded; for instance,
%% the lmodern package is sometimes necessary when using math font.
%%   \usepackage{lmodern}
%%
%% Figures using additional raster images can only be included by \input if
%% they are in the same directory as the main LaTeX file. For loading figures
%% from other directories you can use the `import` package
%% and then include the figures with
%%   \import{<path to file>}{<filename>.pgf}
%%
%% Matplotlib used the following preamble
%%   \def\mathdefault#1{#1}
%%   \everymath=\expandafter{\the\everymath\displaystyle}
%%   \usepackage{amsmath}\usepackage{bm}
%%   \makeatletter\@ifpackageloaded{underscore}{}{\usepackage[strings]{underscore}}\makeatother
%%
\begingroup%
\makeatletter%
\begin{pgfpicture}%
\pgfpathrectangle{\pgfpointorigin}{\pgfqpoint{2.500000in}{2.500000in}}%
\pgfusepath{use as bounding box, clip}%
\begin{pgfscope}%
\pgfsetbuttcap%
\pgfsetmiterjoin%
\definecolor{currentfill}{rgb}{1.000000,1.000000,1.000000}%
\pgfsetfillcolor{currentfill}%
\pgfsetlinewidth{0.000000pt}%
\definecolor{currentstroke}{rgb}{1.000000,1.000000,1.000000}%
\pgfsetstrokecolor{currentstroke}%
\pgfsetdash{}{0pt}%
\pgfpathmoveto{\pgfqpoint{0.000000in}{0.000000in}}%
\pgfpathlineto{\pgfqpoint{2.500000in}{0.000000in}}%
\pgfpathlineto{\pgfqpoint{2.500000in}{2.500000in}}%
\pgfpathlineto{\pgfqpoint{0.000000in}{2.500000in}}%
\pgfpathlineto{\pgfqpoint{0.000000in}{0.000000in}}%
\pgfpathclose%
\pgfusepath{fill}%
\end{pgfscope}%
\begin{pgfscope}%
\pgfsetbuttcap%
\pgfsetmiterjoin%
\definecolor{currentfill}{rgb}{1.000000,1.000000,1.000000}%
\pgfsetfillcolor{currentfill}%
\pgfsetlinewidth{0.000000pt}%
\definecolor{currentstroke}{rgb}{0.000000,0.000000,0.000000}%
\pgfsetstrokecolor{currentstroke}%
\pgfsetstrokeopacity{0.000000}%
\pgfsetdash{}{0pt}%
\pgfpathmoveto{\pgfqpoint{0.584771in}{0.386658in}}%
\pgfpathlineto{\pgfqpoint{2.114698in}{0.386658in}}%
\pgfpathlineto{\pgfqpoint{2.114698in}{1.916585in}}%
\pgfpathlineto{\pgfqpoint{0.584771in}{1.916585in}}%
\pgfpathlineto{\pgfqpoint{0.584771in}{0.386658in}}%
\pgfpathclose%
\pgfusepath{fill}%
\end{pgfscope}%
\begin{pgfscope}%
\pgfsys@transformshift{0.653750in}{0.455000in}%
\pgftext[left,bottom]{\includegraphics[interpolate=true,width=1.391250in,height=1.392500in]{figures/./section1/data/10000p/re_1000//err_vmag-img0.png}}%
\end{pgfscope}%
\begin{pgfscope}%
\pgfsetbuttcap%
\pgfsetroundjoin%
\definecolor{currentfill}{rgb}{0.000000,0.000000,0.000000}%
\pgfsetfillcolor{currentfill}%
\pgfsetlinewidth{0.803000pt}%
\definecolor{currentstroke}{rgb}{0.000000,0.000000,0.000000}%
\pgfsetstrokecolor{currentstroke}%
\pgfsetdash{}{0pt}%
\pgfsys@defobject{currentmarker}{\pgfqpoint{0.000000in}{-0.048611in}}{\pgfqpoint{0.000000in}{0.000000in}}{%
\pgfpathmoveto{\pgfqpoint{0.000000in}{0.000000in}}%
\pgfpathlineto{\pgfqpoint{0.000000in}{-0.048611in}}%
\pgfusepath{stroke,fill}%
}%
\begin{pgfscope}%
\pgfsys@transformshift{0.654313in}{0.386658in}%
\pgfsys@useobject{currentmarker}{}%
\end{pgfscope}%
\end{pgfscope}%
\begin{pgfscope}%
\definecolor{textcolor}{rgb}{0.000000,0.000000,0.000000}%
\pgfsetstrokecolor{textcolor}%
\pgfsetfillcolor{textcolor}%
\pgftext[x=0.654313in,y=0.296381in,,top]{\color{textcolor}{\rmfamily\fontsize{8.330000}{9.996000}\selectfont\catcode`\^=\active\def^{\ifmmode\sp\else\^{}\fi}\catcode`\%=\active\def%{\%}$\mathdefault{\ensuremath{-}0.1}$}}%
\end{pgfscope}%
\begin{pgfscope}%
\pgfsetbuttcap%
\pgfsetroundjoin%
\definecolor{currentfill}{rgb}{0.000000,0.000000,0.000000}%
\pgfsetfillcolor{currentfill}%
\pgfsetlinewidth{0.803000pt}%
\definecolor{currentstroke}{rgb}{0.000000,0.000000,0.000000}%
\pgfsetstrokecolor{currentstroke}%
\pgfsetdash{}{0pt}%
\pgfsys@defobject{currentmarker}{\pgfqpoint{0.000000in}{-0.048611in}}{\pgfqpoint{0.000000in}{0.000000in}}{%
\pgfpathmoveto{\pgfqpoint{0.000000in}{0.000000in}}%
\pgfpathlineto{\pgfqpoint{0.000000in}{-0.048611in}}%
\pgfusepath{stroke,fill}%
}%
\begin{pgfscope}%
\pgfsys@transformshift{1.349734in}{0.386658in}%
\pgfsys@useobject{currentmarker}{}%
\end{pgfscope}%
\end{pgfscope}%
\begin{pgfscope}%
\definecolor{textcolor}{rgb}{0.000000,0.000000,0.000000}%
\pgfsetstrokecolor{textcolor}%
\pgfsetfillcolor{textcolor}%
\pgftext[x=1.349734in,y=0.296381in,,top]{\color{textcolor}{\rmfamily\fontsize{8.330000}{9.996000}\selectfont\catcode`\^=\active\def^{\ifmmode\sp\else\^{}\fi}\catcode`\%=\active\def%{\%}$\mathdefault{0.0}$}}%
\end{pgfscope}%
\begin{pgfscope}%
\pgfsetbuttcap%
\pgfsetroundjoin%
\definecolor{currentfill}{rgb}{0.000000,0.000000,0.000000}%
\pgfsetfillcolor{currentfill}%
\pgfsetlinewidth{0.803000pt}%
\definecolor{currentstroke}{rgb}{0.000000,0.000000,0.000000}%
\pgfsetstrokecolor{currentstroke}%
\pgfsetdash{}{0pt}%
\pgfsys@defobject{currentmarker}{\pgfqpoint{0.000000in}{-0.048611in}}{\pgfqpoint{0.000000in}{0.000000in}}{%
\pgfpathmoveto{\pgfqpoint{0.000000in}{0.000000in}}%
\pgfpathlineto{\pgfqpoint{0.000000in}{-0.048611in}}%
\pgfusepath{stroke,fill}%
}%
\begin{pgfscope}%
\pgfsys@transformshift{2.045155in}{0.386658in}%
\pgfsys@useobject{currentmarker}{}%
\end{pgfscope}%
\end{pgfscope}%
\begin{pgfscope}%
\definecolor{textcolor}{rgb}{0.000000,0.000000,0.000000}%
\pgfsetstrokecolor{textcolor}%
\pgfsetfillcolor{textcolor}%
\pgftext[x=2.045155in,y=0.296381in,,top]{\color{textcolor}{\rmfamily\fontsize{8.330000}{9.996000}\selectfont\catcode`\^=\active\def^{\ifmmode\sp\else\^{}\fi}\catcode`\%=\active\def%{\%}$\mathdefault{0.1}$}}%
\end{pgfscope}%
\begin{pgfscope}%
\definecolor{textcolor}{rgb}{0.000000,0.000000,0.000000}%
\pgfsetstrokecolor{textcolor}%
\pgfsetfillcolor{textcolor}%
\pgftext[x=1.349734in,y=0.142060in,,top]{\color{textcolor}{\rmfamily\fontsize{10.000000}{12.000000}\selectfont\catcode`\^=\active\def^{\ifmmode\sp\else\^{}\fi}\catcode`\%=\active\def%{\%}$x\;[\text{m}]$}}%
\end{pgfscope}%
\begin{pgfscope}%
\pgfsetbuttcap%
\pgfsetroundjoin%
\definecolor{currentfill}{rgb}{0.000000,0.000000,0.000000}%
\pgfsetfillcolor{currentfill}%
\pgfsetlinewidth{0.803000pt}%
\definecolor{currentstroke}{rgb}{0.000000,0.000000,0.000000}%
\pgfsetstrokecolor{currentstroke}%
\pgfsetdash{}{0pt}%
\pgfsys@defobject{currentmarker}{\pgfqpoint{-0.048611in}{0.000000in}}{\pgfqpoint{-0.000000in}{0.000000in}}{%
\pgfpathmoveto{\pgfqpoint{-0.000000in}{0.000000in}}%
\pgfpathlineto{\pgfqpoint{-0.048611in}{0.000000in}}%
\pgfusepath{stroke,fill}%
}%
\begin{pgfscope}%
\pgfsys@transformshift{0.584771in}{0.456201in}%
\pgfsys@useobject{currentmarker}{}%
\end{pgfscope}%
\end{pgfscope}%
\begin{pgfscope}%
\definecolor{textcolor}{rgb}{0.000000,0.000000,0.000000}%
\pgfsetstrokecolor{textcolor}%
\pgfsetfillcolor{textcolor}%
\pgftext[x=0.251820in, y=0.417620in, left, base]{\color{textcolor}{\rmfamily\fontsize{8.330000}{9.996000}\selectfont\catcode`\^=\active\def^{\ifmmode\sp\else\^{}\fi}\catcode`\%=\active\def%{\%}$\mathdefault{\ensuremath{-}0.1}$}}%
\end{pgfscope}%
\begin{pgfscope}%
\pgfsetbuttcap%
\pgfsetroundjoin%
\definecolor{currentfill}{rgb}{0.000000,0.000000,0.000000}%
\pgfsetfillcolor{currentfill}%
\pgfsetlinewidth{0.803000pt}%
\definecolor{currentstroke}{rgb}{0.000000,0.000000,0.000000}%
\pgfsetstrokecolor{currentstroke}%
\pgfsetdash{}{0pt}%
\pgfsys@defobject{currentmarker}{\pgfqpoint{-0.048611in}{0.000000in}}{\pgfqpoint{-0.000000in}{0.000000in}}{%
\pgfpathmoveto{\pgfqpoint{-0.000000in}{0.000000in}}%
\pgfpathlineto{\pgfqpoint{-0.048611in}{0.000000in}}%
\pgfusepath{stroke,fill}%
}%
\begin{pgfscope}%
\pgfsys@transformshift{0.584771in}{1.151622in}%
\pgfsys@useobject{currentmarker}{}%
\end{pgfscope}%
\end{pgfscope}%
\begin{pgfscope}%
\definecolor{textcolor}{rgb}{0.000000,0.000000,0.000000}%
\pgfsetstrokecolor{textcolor}%
\pgfsetfillcolor{textcolor}%
\pgftext[x=0.343642in, y=1.113042in, left, base]{\color{textcolor}{\rmfamily\fontsize{8.330000}{9.996000}\selectfont\catcode`\^=\active\def^{\ifmmode\sp\else\^{}\fi}\catcode`\%=\active\def%{\%}$\mathdefault{0.0}$}}%
\end{pgfscope}%
\begin{pgfscope}%
\pgfsetbuttcap%
\pgfsetroundjoin%
\definecolor{currentfill}{rgb}{0.000000,0.000000,0.000000}%
\pgfsetfillcolor{currentfill}%
\pgfsetlinewidth{0.803000pt}%
\definecolor{currentstroke}{rgb}{0.000000,0.000000,0.000000}%
\pgfsetstrokecolor{currentstroke}%
\pgfsetdash{}{0pt}%
\pgfsys@defobject{currentmarker}{\pgfqpoint{-0.048611in}{0.000000in}}{\pgfqpoint{-0.000000in}{0.000000in}}{%
\pgfpathmoveto{\pgfqpoint{-0.000000in}{0.000000in}}%
\pgfpathlineto{\pgfqpoint{-0.048611in}{0.000000in}}%
\pgfusepath{stroke,fill}%
}%
\begin{pgfscope}%
\pgfsys@transformshift{0.584771in}{1.847043in}%
\pgfsys@useobject{currentmarker}{}%
\end{pgfscope}%
\end{pgfscope}%
\begin{pgfscope}%
\definecolor{textcolor}{rgb}{0.000000,0.000000,0.000000}%
\pgfsetstrokecolor{textcolor}%
\pgfsetfillcolor{textcolor}%
\pgftext[x=0.343642in, y=1.808463in, left, base]{\color{textcolor}{\rmfamily\fontsize{8.330000}{9.996000}\selectfont\catcode`\^=\active\def^{\ifmmode\sp\else\^{}\fi}\catcode`\%=\active\def%{\%}$\mathdefault{0.1}$}}%
\end{pgfscope}%
\begin{pgfscope}%
\definecolor{textcolor}{rgb}{0.000000,0.000000,0.000000}%
\pgfsetstrokecolor{textcolor}%
\pgfsetfillcolor{textcolor}%
\pgftext[x=0.196264in,y=1.151622in,,bottom,rotate=90.000000]{\color{textcolor}{\rmfamily\fontsize{10.000000}{12.000000}\selectfont\catcode`\^=\active\def^{\ifmmode\sp\else\^{}\fi}\catcode`\%=\active\def%{\%}$y\;[\text{m}]$}}%
\end{pgfscope}%
\begin{pgfscope}%
\pgfpathrectangle{\pgfqpoint{0.584771in}{0.386658in}}{\pgfqpoint{1.529927in}{1.529927in}}%
\pgfusepath{clip}%
\pgfsetbuttcap%
\pgfsetmiterjoin%
\definecolor{currentfill}{rgb}{1.000000,1.000000,1.000000}%
\pgfsetfillcolor{currentfill}%
\pgfsetlinewidth{1.505625pt}%
\definecolor{currentstroke}{rgb}{1.000000,1.000000,1.000000}%
\pgfsetstrokecolor{currentstroke}%
\pgfsetdash{}{0pt}%
\pgfpathmoveto{\pgfqpoint{1.822472in}{1.648947in}}%
\pgfpathlineto{\pgfqpoint{1.815767in}{1.642241in}}%
\pgfpathlineto{\pgfqpoint{1.809061in}{1.635536in}}%
\pgfpathlineto{\pgfqpoint{1.802356in}{1.628830in}}%
\pgfpathlineto{\pgfqpoint{1.795650in}{1.622125in}}%
\pgfpathlineto{\pgfqpoint{1.788945in}{1.615419in}}%
\pgfpathlineto{\pgfqpoint{1.782239in}{1.608714in}}%
\pgfpathlineto{\pgfqpoint{1.775534in}{1.602008in}}%
\pgfpathlineto{\pgfqpoint{1.768828in}{1.595303in}}%
\pgfpathlineto{\pgfqpoint{1.762123in}{1.588597in}}%
\pgfpathlineto{\pgfqpoint{1.755417in}{1.581892in}}%
\pgfpathlineto{\pgfqpoint{1.761564in}{1.575745in}}%
\pgfpathlineto{\pgfqpoint{1.767711in}{1.569598in}}%
\pgfpathlineto{\pgfqpoint{1.773857in}{1.563452in}}%
\pgfpathlineto{\pgfqpoint{1.780004in}{1.557305in}}%
\pgfpathlineto{\pgfqpoint{1.786710in}{1.564010in}}%
\pgfpathlineto{\pgfqpoint{1.793415in}{1.570716in}}%
\pgfpathlineto{\pgfqpoint{1.800121in}{1.577421in}}%
\pgfpathlineto{\pgfqpoint{1.806826in}{1.584127in}}%
\pgfpathlineto{\pgfqpoint{1.813532in}{1.590833in}}%
\pgfpathlineto{\pgfqpoint{1.820237in}{1.597538in}}%
\pgfpathlineto{\pgfqpoint{1.826943in}{1.604244in}}%
\pgfpathlineto{\pgfqpoint{1.833648in}{1.610949in}}%
\pgfpathlineto{\pgfqpoint{1.840354in}{1.617655in}}%
\pgfpathlineto{\pgfqpoint{1.847059in}{1.624360in}}%
\pgfpathlineto{\pgfqpoint{1.822472in}{1.648947in}}%
\pgfpathclose%
\pgfusepath{stroke,fill}%
\end{pgfscope}%
\begin{pgfscope}%
\pgfpathrectangle{\pgfqpoint{0.584771in}{0.386658in}}{\pgfqpoint{1.529927in}{1.529927in}}%
\pgfusepath{clip}%
\pgfsetbuttcap%
\pgfsetmiterjoin%
\definecolor{currentfill}{rgb}{1.000000,1.000000,1.000000}%
\pgfsetfillcolor{currentfill}%
\pgfsetlinewidth{1.505625pt}%
\definecolor{currentstroke}{rgb}{1.000000,1.000000,1.000000}%
\pgfsetstrokecolor{currentstroke}%
\pgfsetdash{}{0pt}%
\pgfpathmoveto{\pgfqpoint{1.847059in}{0.678884in}}%
\pgfpathlineto{\pgfqpoint{1.840354in}{0.685589in}}%
\pgfpathlineto{\pgfqpoint{1.833648in}{0.692295in}}%
\pgfpathlineto{\pgfqpoint{1.826943in}{0.699000in}}%
\pgfpathlineto{\pgfqpoint{1.820237in}{0.705706in}}%
\pgfpathlineto{\pgfqpoint{1.813532in}{0.712411in}}%
\pgfpathlineto{\pgfqpoint{1.806826in}{0.719117in}}%
\pgfpathlineto{\pgfqpoint{1.800121in}{0.725822in}}%
\pgfpathlineto{\pgfqpoint{1.793415in}{0.732528in}}%
\pgfpathlineto{\pgfqpoint{1.786710in}{0.739233in}}%
\pgfpathlineto{\pgfqpoint{1.780004in}{0.745939in}}%
\pgfpathlineto{\pgfqpoint{1.773734in}{0.739669in}}%
\pgfpathlineto{\pgfqpoint{1.767465in}{0.733399in}}%
\pgfpathlineto{\pgfqpoint{1.761195in}{0.727130in}}%
\pgfpathlineto{\pgfqpoint{1.754926in}{0.720860in}}%
\pgfpathlineto{\pgfqpoint{1.755417in}{0.721352in}}%
\pgfpathlineto{\pgfqpoint{1.762123in}{0.714646in}}%
\pgfpathlineto{\pgfqpoint{1.768828in}{0.707941in}}%
\pgfpathlineto{\pgfqpoint{1.775534in}{0.701235in}}%
\pgfpathlineto{\pgfqpoint{1.782239in}{0.694530in}}%
\pgfpathlineto{\pgfqpoint{1.788945in}{0.687824in}}%
\pgfpathlineto{\pgfqpoint{1.795650in}{0.681119in}}%
\pgfpathlineto{\pgfqpoint{1.802356in}{0.674413in}}%
\pgfpathlineto{\pgfqpoint{1.809061in}{0.667708in}}%
\pgfpathlineto{\pgfqpoint{1.815767in}{0.661002in}}%
\pgfpathlineto{\pgfqpoint{1.822472in}{0.654297in}}%
\pgfpathlineto{\pgfqpoint{1.847059in}{0.678884in}}%
\pgfpathclose%
\pgfusepath{stroke,fill}%
\end{pgfscope}%
\begin{pgfscope}%
\pgfpathrectangle{\pgfqpoint{0.584771in}{0.386658in}}{\pgfqpoint{1.529927in}{1.529927in}}%
\pgfusepath{clip}%
\pgfsetbuttcap%
\pgfsetmiterjoin%
\definecolor{currentfill}{rgb}{1.000000,1.000000,1.000000}%
\pgfsetfillcolor{currentfill}%
\pgfsetlinewidth{1.505625pt}%
\definecolor{currentstroke}{rgb}{1.000000,1.000000,1.000000}%
\pgfsetstrokecolor{currentstroke}%
\pgfsetdash{}{0pt}%
\pgfpathmoveto{\pgfqpoint{0.876996in}{0.654297in}}%
\pgfpathlineto{\pgfqpoint{0.883702in}{0.661002in}}%
\pgfpathlineto{\pgfqpoint{0.890407in}{0.667708in}}%
\pgfpathlineto{\pgfqpoint{0.897113in}{0.674413in}}%
\pgfpathlineto{\pgfqpoint{0.903818in}{0.681119in}}%
\pgfpathlineto{\pgfqpoint{0.910524in}{0.687824in}}%
\pgfpathlineto{\pgfqpoint{0.917229in}{0.694530in}}%
\pgfpathlineto{\pgfqpoint{0.923935in}{0.701235in}}%
\pgfpathlineto{\pgfqpoint{0.930640in}{0.707941in}}%
\pgfpathlineto{\pgfqpoint{0.937346in}{0.714646in}}%
\pgfpathlineto{\pgfqpoint{0.944051in}{0.721352in}}%
\pgfpathlineto{\pgfqpoint{0.937781in}{0.727622in}}%
\pgfpathlineto{\pgfqpoint{0.931512in}{0.733891in}}%
\pgfpathlineto{\pgfqpoint{0.925242in}{0.740161in}}%
\pgfpathlineto{\pgfqpoint{0.918972in}{0.746431in}}%
\pgfpathlineto{\pgfqpoint{0.919464in}{0.745939in}}%
\pgfpathlineto{\pgfqpoint{0.912759in}{0.739233in}}%
\pgfpathlineto{\pgfqpoint{0.906053in}{0.732528in}}%
\pgfpathlineto{\pgfqpoint{0.899348in}{0.725822in}}%
\pgfpathlineto{\pgfqpoint{0.892642in}{0.719117in}}%
\pgfpathlineto{\pgfqpoint{0.885937in}{0.712411in}}%
\pgfpathlineto{\pgfqpoint{0.879231in}{0.705706in}}%
\pgfpathlineto{\pgfqpoint{0.872526in}{0.699000in}}%
\pgfpathlineto{\pgfqpoint{0.865820in}{0.692295in}}%
\pgfpathlineto{\pgfqpoint{0.859115in}{0.685589in}}%
\pgfpathlineto{\pgfqpoint{0.852409in}{0.678884in}}%
\pgfpathlineto{\pgfqpoint{0.876996in}{0.654297in}}%
\pgfpathclose%
\pgfusepath{stroke,fill}%
\end{pgfscope}%
\begin{pgfscope}%
\pgfpathrectangle{\pgfqpoint{0.584771in}{0.386658in}}{\pgfqpoint{1.529927in}{1.529927in}}%
\pgfusepath{clip}%
\pgfsetbuttcap%
\pgfsetmiterjoin%
\definecolor{currentfill}{rgb}{1.000000,1.000000,1.000000}%
\pgfsetfillcolor{currentfill}%
\pgfsetlinewidth{1.505625pt}%
\definecolor{currentstroke}{rgb}{1.000000,1.000000,1.000000}%
\pgfsetstrokecolor{currentstroke}%
\pgfsetdash{}{0pt}%
\pgfpathmoveto{\pgfqpoint{0.852409in}{1.624360in}}%
\pgfpathlineto{\pgfqpoint{0.859115in}{1.617655in}}%
\pgfpathlineto{\pgfqpoint{0.865820in}{1.610949in}}%
\pgfpathlineto{\pgfqpoint{0.872526in}{1.604244in}}%
\pgfpathlineto{\pgfqpoint{0.879231in}{1.597538in}}%
\pgfpathlineto{\pgfqpoint{0.885937in}{1.590833in}}%
\pgfpathlineto{\pgfqpoint{0.892642in}{1.584127in}}%
\pgfpathlineto{\pgfqpoint{0.899348in}{1.577421in}}%
\pgfpathlineto{\pgfqpoint{0.906053in}{1.570716in}}%
\pgfpathlineto{\pgfqpoint{0.912759in}{1.564010in}}%
\pgfpathlineto{\pgfqpoint{0.919464in}{1.557305in}}%
\pgfpathlineto{\pgfqpoint{0.925611in}{1.563452in}}%
\pgfpathlineto{\pgfqpoint{0.931758in}{1.569598in}}%
\pgfpathlineto{\pgfqpoint{0.937904in}{1.575745in}}%
\pgfpathlineto{\pgfqpoint{0.944051in}{1.581892in}}%
\pgfpathlineto{\pgfqpoint{0.937346in}{1.588597in}}%
\pgfpathlineto{\pgfqpoint{0.930640in}{1.595303in}}%
\pgfpathlineto{\pgfqpoint{0.923935in}{1.602008in}}%
\pgfpathlineto{\pgfqpoint{0.917229in}{1.608714in}}%
\pgfpathlineto{\pgfqpoint{0.910524in}{1.615419in}}%
\pgfpathlineto{\pgfqpoint{0.903818in}{1.622125in}}%
\pgfpathlineto{\pgfqpoint{0.897113in}{1.628830in}}%
\pgfpathlineto{\pgfqpoint{0.890407in}{1.635536in}}%
\pgfpathlineto{\pgfqpoint{0.883702in}{1.642241in}}%
\pgfpathlineto{\pgfqpoint{0.876996in}{1.648947in}}%
\pgfpathlineto{\pgfqpoint{0.852409in}{1.624360in}}%
\pgfpathclose%
\pgfusepath{stroke,fill}%
\end{pgfscope}%
\begin{pgfscope}%
\pgfpathrectangle{\pgfqpoint{0.584771in}{0.386658in}}{\pgfqpoint{1.529927in}{1.529927in}}%
\pgfusepath{clip}%
\pgfsetrectcap%
\pgfsetroundjoin%
\pgfsetlinewidth{1.505625pt}%
\definecolor{currentstroke}{rgb}{0.000000,0.000000,0.000000}%
\pgfsetstrokecolor{currentstroke}%
\pgfsetdash{}{0pt}%
\pgfpathmoveto{\pgfqpoint{0.852409in}{1.624360in}}%
\pgfpathlineto{\pgfqpoint{0.919464in}{1.557305in}}%
\pgfpathlineto{\pgfqpoint{0.944051in}{1.581892in}}%
\pgfpathlineto{\pgfqpoint{0.876996in}{1.648947in}}%
\pgfpathlineto{\pgfqpoint{0.871038in}{1.656063in}}%
\pgfpathlineto{\pgfqpoint{0.886897in}{1.670652in}}%
\pgfpathlineto{\pgfqpoint{0.903201in}{1.684743in}}%
\pgfpathlineto{\pgfqpoint{0.919933in}{1.698323in}}%
\pgfpathlineto{\pgfqpoint{0.937078in}{1.711377in}}%
\pgfpathlineto{\pgfqpoint{0.954619in}{1.723894in}}%
\pgfpathlineto{\pgfqpoint{0.972540in}{1.735861in}}%
\pgfpathlineto{\pgfqpoint{0.990823in}{1.747267in}}%
\pgfpathlineto{\pgfqpoint{1.009450in}{1.758102in}}%
\pgfpathlineto{\pgfqpoint{1.028405in}{1.768354in}}%
\pgfpathlineto{\pgfqpoint{1.047667in}{1.778013in}}%
\pgfpathlineto{\pgfqpoint{1.067220in}{1.787072in}}%
\pgfpathlineto{\pgfqpoint{1.087044in}{1.795520in}}%
\pgfpathlineto{\pgfqpoint{1.107121in}{1.803350in}}%
\pgfpathlineto{\pgfqpoint{1.127430in}{1.810554in}}%
\pgfpathlineto{\pgfqpoint{1.147953in}{1.817125in}}%
\pgfpathlineto{\pgfqpoint{1.168669in}{1.823058in}}%
\pgfpathlineto{\pgfqpoint{1.189560in}{1.828346in}}%
\pgfpathlineto{\pgfqpoint{1.210604in}{1.832983in}}%
\pgfpathlineto{\pgfqpoint{1.231782in}{1.836967in}}%
\pgfpathlineto{\pgfqpoint{1.253073in}{1.840293in}}%
\pgfpathlineto{\pgfqpoint{1.274457in}{1.842957in}}%
\pgfpathlineto{\pgfqpoint{1.295913in}{1.844957in}}%
\pgfpathlineto{\pgfqpoint{1.317421in}{1.846292in}}%
\pgfpathlineto{\pgfqpoint{1.338960in}{1.846960in}}%
\pgfpathlineto{\pgfqpoint{1.360509in}{1.846960in}}%
\pgfpathlineto{\pgfqpoint{1.382048in}{1.846292in}}%
\pgfpathlineto{\pgfqpoint{1.403555in}{1.844957in}}%
\pgfpathlineto{\pgfqpoint{1.425012in}{1.842957in}}%
\pgfpathlineto{\pgfqpoint{1.446395in}{1.840293in}}%
\pgfpathlineto{\pgfqpoint{1.467686in}{1.836967in}}%
\pgfpathlineto{\pgfqpoint{1.488864in}{1.832983in}}%
\pgfpathlineto{\pgfqpoint{1.509908in}{1.828346in}}%
\pgfpathlineto{\pgfqpoint{1.530799in}{1.823058in}}%
\pgfpathlineto{\pgfqpoint{1.551515in}{1.817125in}}%
\pgfpathlineto{\pgfqpoint{1.572038in}{1.810554in}}%
\pgfpathlineto{\pgfqpoint{1.592348in}{1.803350in}}%
\pgfpathlineto{\pgfqpoint{1.612424in}{1.795520in}}%
\pgfpathlineto{\pgfqpoint{1.632248in}{1.787072in}}%
\pgfpathlineto{\pgfqpoint{1.651801in}{1.778013in}}%
\pgfpathlineto{\pgfqpoint{1.671064in}{1.768354in}}%
\pgfpathlineto{\pgfqpoint{1.690018in}{1.758102in}}%
\pgfpathlineto{\pgfqpoint{1.708645in}{1.747267in}}%
\pgfpathlineto{\pgfqpoint{1.726928in}{1.735861in}}%
\pgfpathlineto{\pgfqpoint{1.744849in}{1.723894in}}%
\pgfpathlineto{\pgfqpoint{1.762390in}{1.711377in}}%
\pgfpathlineto{\pgfqpoint{1.779535in}{1.698323in}}%
\pgfpathlineto{\pgfqpoint{1.796268in}{1.684743in}}%
\pgfpathlineto{\pgfqpoint{1.812571in}{1.670652in}}%
\pgfpathlineto{\pgfqpoint{1.828431in}{1.656063in}}%
\pgfpathlineto{\pgfqpoint{1.822472in}{1.648947in}}%
\pgfpathlineto{\pgfqpoint{1.755417in}{1.581892in}}%
\pgfpathlineto{\pgfqpoint{1.780004in}{1.557305in}}%
\pgfpathlineto{\pgfqpoint{1.847059in}{1.624360in}}%
\pgfpathlineto{\pgfqpoint{1.854175in}{1.630318in}}%
\pgfpathlineto{\pgfqpoint{1.868764in}{1.614459in}}%
\pgfpathlineto{\pgfqpoint{1.882856in}{1.598155in}}%
\pgfpathlineto{\pgfqpoint{1.896435in}{1.581423in}}%
\pgfpathlineto{\pgfqpoint{1.909489in}{1.564278in}}%
\pgfpathlineto{\pgfqpoint{1.922006in}{1.546737in}}%
\pgfpathlineto{\pgfqpoint{1.933973in}{1.528816in}}%
\pgfpathlineto{\pgfqpoint{1.945380in}{1.510533in}}%
\pgfpathlineto{\pgfqpoint{1.956214in}{1.491906in}}%
\pgfpathlineto{\pgfqpoint{1.966466in}{1.472951in}}%
\pgfpathlineto{\pgfqpoint{1.976126in}{1.453689in}}%
\pgfpathlineto{\pgfqpoint{1.985184in}{1.434136in}}%
\pgfpathlineto{\pgfqpoint{1.993632in}{1.414312in}}%
\pgfpathlineto{\pgfqpoint{2.001462in}{1.394235in}}%
\pgfpathlineto{\pgfqpoint{2.008666in}{1.373926in}}%
\pgfpathlineto{\pgfqpoint{2.015238in}{1.353403in}}%
\pgfpathlineto{\pgfqpoint{2.021170in}{1.332687in}}%
\pgfpathlineto{\pgfqpoint{2.026458in}{1.311796in}}%
\pgfpathlineto{\pgfqpoint{2.031096in}{1.290752in}}%
\pgfpathlineto{\pgfqpoint{2.035079in}{1.269574in}}%
\pgfpathlineto{\pgfqpoint{2.038405in}{1.248283in}}%
\pgfpathlineto{\pgfqpoint{2.041069in}{1.226899in}}%
\pgfpathlineto{\pgfqpoint{2.043070in}{1.205443in}}%
\pgfpathlineto{\pgfqpoint{2.044404in}{1.183935in}}%
\pgfpathlineto{\pgfqpoint{2.045072in}{1.162396in}}%
\pgfpathlineto{\pgfqpoint{2.045072in}{1.140847in}}%
\pgfpathlineto{\pgfqpoint{2.044404in}{1.119308in}}%
\pgfpathlineto{\pgfqpoint{2.043070in}{1.097801in}}%
\pgfpathlineto{\pgfqpoint{2.041069in}{1.076344in}}%
\pgfpathlineto{\pgfqpoint{2.038405in}{1.054961in}}%
\pgfpathlineto{\pgfqpoint{2.035079in}{1.033670in}}%
\pgfpathlineto{\pgfqpoint{2.031096in}{1.012492in}}%
\pgfpathlineto{\pgfqpoint{2.026458in}{0.991448in}}%
\pgfpathlineto{\pgfqpoint{2.021170in}{0.970557in}}%
\pgfpathlineto{\pgfqpoint{2.015238in}{0.949841in}}%
\pgfpathlineto{\pgfqpoint{2.008666in}{0.929318in}}%
\pgfpathlineto{\pgfqpoint{2.001462in}{0.909009in}}%
\pgfpathlineto{\pgfqpoint{1.993632in}{0.888932in}}%
\pgfpathlineto{\pgfqpoint{1.985184in}{0.869108in}}%
\pgfpathlineto{\pgfqpoint{1.976126in}{0.849555in}}%
\pgfpathlineto{\pgfqpoint{1.966466in}{0.830292in}}%
\pgfpathlineto{\pgfqpoint{1.956214in}{0.811338in}}%
\pgfpathlineto{\pgfqpoint{1.945380in}{0.792711in}}%
\pgfpathlineto{\pgfqpoint{1.933973in}{0.774428in}}%
\pgfpathlineto{\pgfqpoint{1.922006in}{0.756507in}}%
\pgfpathlineto{\pgfqpoint{1.909489in}{0.738966in}}%
\pgfpathlineto{\pgfqpoint{1.896435in}{0.721821in}}%
\pgfpathlineto{\pgfqpoint{1.882856in}{0.705088in}}%
\pgfpathlineto{\pgfqpoint{1.868764in}{0.688785in}}%
\pgfpathlineto{\pgfqpoint{1.854175in}{0.672925in}}%
\pgfpathlineto{\pgfqpoint{1.847059in}{0.678884in}}%
\pgfpathlineto{\pgfqpoint{1.780004in}{0.745939in}}%
\pgfpathlineto{\pgfqpoint{1.754926in}{0.720860in}}%
\pgfpathlineto{\pgfqpoint{1.755417in}{0.721352in}}%
\pgfpathlineto{\pgfqpoint{1.822472in}{0.654297in}}%
\pgfpathlineto{\pgfqpoint{1.828431in}{0.647181in}}%
\pgfpathlineto{\pgfqpoint{1.812571in}{0.632592in}}%
\pgfpathlineto{\pgfqpoint{1.796268in}{0.618500in}}%
\pgfpathlineto{\pgfqpoint{1.779535in}{0.604921in}}%
\pgfpathlineto{\pgfqpoint{1.762390in}{0.591867in}}%
\pgfpathlineto{\pgfqpoint{1.744849in}{0.579350in}}%
\pgfpathlineto{\pgfqpoint{1.726928in}{0.567383in}}%
\pgfpathlineto{\pgfqpoint{1.708645in}{0.555976in}}%
\pgfpathlineto{\pgfqpoint{1.690018in}{0.545142in}}%
\pgfpathlineto{\pgfqpoint{1.671064in}{0.534890in}}%
\pgfpathlineto{\pgfqpoint{1.651801in}{0.525230in}}%
\pgfpathlineto{\pgfqpoint{1.632248in}{0.516172in}}%
\pgfpathlineto{\pgfqpoint{1.612424in}{0.507724in}}%
\pgfpathlineto{\pgfqpoint{1.592348in}{0.499894in}}%
\pgfpathlineto{\pgfqpoint{1.572038in}{0.492690in}}%
\pgfpathlineto{\pgfqpoint{1.551515in}{0.486118in}}%
\pgfpathlineto{\pgfqpoint{1.530799in}{0.480186in}}%
\pgfpathlineto{\pgfqpoint{1.509908in}{0.474898in}}%
\pgfpathlineto{\pgfqpoint{1.488864in}{0.470260in}}%
\pgfpathlineto{\pgfqpoint{1.467686in}{0.466277in}}%
\pgfpathlineto{\pgfqpoint{1.446395in}{0.462951in}}%
\pgfpathlineto{\pgfqpoint{1.425012in}{0.460287in}}%
\pgfpathlineto{\pgfqpoint{1.403555in}{0.458286in}}%
\pgfpathlineto{\pgfqpoint{1.382048in}{0.456952in}}%
\pgfpathlineto{\pgfqpoint{1.360509in}{0.456284in}}%
\pgfpathlineto{\pgfqpoint{1.338960in}{0.456284in}}%
\pgfpathlineto{\pgfqpoint{1.317421in}{0.456952in}}%
\pgfpathlineto{\pgfqpoint{1.295913in}{0.458286in}}%
\pgfpathlineto{\pgfqpoint{1.274457in}{0.460287in}}%
\pgfpathlineto{\pgfqpoint{1.253073in}{0.462951in}}%
\pgfpathlineto{\pgfqpoint{1.231782in}{0.466277in}}%
\pgfpathlineto{\pgfqpoint{1.210604in}{0.470260in}}%
\pgfpathlineto{\pgfqpoint{1.189560in}{0.474898in}}%
\pgfpathlineto{\pgfqpoint{1.168669in}{0.480186in}}%
\pgfpathlineto{\pgfqpoint{1.147953in}{0.486118in}}%
\pgfpathlineto{\pgfqpoint{1.127430in}{0.492690in}}%
\pgfpathlineto{\pgfqpoint{1.107121in}{0.499894in}}%
\pgfpathlineto{\pgfqpoint{1.087044in}{0.507724in}}%
\pgfpathlineto{\pgfqpoint{1.067220in}{0.516172in}}%
\pgfpathlineto{\pgfqpoint{1.047667in}{0.525230in}}%
\pgfpathlineto{\pgfqpoint{1.028405in}{0.534890in}}%
\pgfpathlineto{\pgfqpoint{1.009450in}{0.545142in}}%
\pgfpathlineto{\pgfqpoint{0.990823in}{0.555976in}}%
\pgfpathlineto{\pgfqpoint{0.972540in}{0.567383in}}%
\pgfpathlineto{\pgfqpoint{0.954619in}{0.579350in}}%
\pgfpathlineto{\pgfqpoint{0.937078in}{0.591867in}}%
\pgfpathlineto{\pgfqpoint{0.919933in}{0.604921in}}%
\pgfpathlineto{\pgfqpoint{0.903201in}{0.618500in}}%
\pgfpathlineto{\pgfqpoint{0.886897in}{0.632592in}}%
\pgfpathlineto{\pgfqpoint{0.871038in}{0.647181in}}%
\pgfpathlineto{\pgfqpoint{0.876996in}{0.654297in}}%
\pgfpathlineto{\pgfqpoint{0.944051in}{0.721352in}}%
\pgfpathlineto{\pgfqpoint{0.918972in}{0.746431in}}%
\pgfpathlineto{\pgfqpoint{0.919464in}{0.745939in}}%
\pgfpathlineto{\pgfqpoint{0.852409in}{0.678884in}}%
\pgfpathlineto{\pgfqpoint{0.845293in}{0.672925in}}%
\pgfpathlineto{\pgfqpoint{0.830704in}{0.688785in}}%
\pgfpathlineto{\pgfqpoint{0.816613in}{0.705088in}}%
\pgfpathlineto{\pgfqpoint{0.803033in}{0.721821in}}%
\pgfpathlineto{\pgfqpoint{0.789979in}{0.738966in}}%
\pgfpathlineto{\pgfqpoint{0.777462in}{0.756507in}}%
\pgfpathlineto{\pgfqpoint{0.765495in}{0.774428in}}%
\pgfpathlineto{\pgfqpoint{0.754089in}{0.792711in}}%
\pgfpathlineto{\pgfqpoint{0.743254in}{0.811338in}}%
\pgfpathlineto{\pgfqpoint{0.733002in}{0.830292in}}%
\pgfpathlineto{\pgfqpoint{0.723343in}{0.849555in}}%
\pgfpathlineto{\pgfqpoint{0.714284in}{0.869108in}}%
\pgfpathlineto{\pgfqpoint{0.705836in}{0.888932in}}%
\pgfpathlineto{\pgfqpoint{0.698006in}{0.909009in}}%
\pgfpathlineto{\pgfqpoint{0.690802in}{0.929318in}}%
\pgfpathlineto{\pgfqpoint{0.684231in}{0.949841in}}%
\pgfpathlineto{\pgfqpoint{0.678298in}{0.970557in}}%
\pgfpathlineto{\pgfqpoint{0.673010in}{0.991448in}}%
\pgfpathlineto{\pgfqpoint{0.668373in}{1.012492in}}%
\pgfpathlineto{\pgfqpoint{0.664389in}{1.033670in}}%
\pgfpathlineto{\pgfqpoint{0.661063in}{1.054961in}}%
\pgfpathlineto{\pgfqpoint{0.658399in}{1.076344in}}%
\pgfpathlineto{\pgfqpoint{0.656399in}{1.097801in}}%
\pgfpathlineto{\pgfqpoint{0.655064in}{1.119308in}}%
\pgfpathlineto{\pgfqpoint{0.654396in}{1.140847in}}%
\pgfpathlineto{\pgfqpoint{0.654396in}{1.162396in}}%
\pgfpathlineto{\pgfqpoint{0.655064in}{1.183935in}}%
\pgfpathlineto{\pgfqpoint{0.656399in}{1.205443in}}%
\pgfpathlineto{\pgfqpoint{0.658399in}{1.226899in}}%
\pgfpathlineto{\pgfqpoint{0.661063in}{1.248283in}}%
\pgfpathlineto{\pgfqpoint{0.664389in}{1.269574in}}%
\pgfpathlineto{\pgfqpoint{0.668373in}{1.290752in}}%
\pgfpathlineto{\pgfqpoint{0.673010in}{1.311796in}}%
\pgfpathlineto{\pgfqpoint{0.678298in}{1.332687in}}%
\pgfpathlineto{\pgfqpoint{0.684231in}{1.353403in}}%
\pgfpathlineto{\pgfqpoint{0.690802in}{1.373926in}}%
\pgfpathlineto{\pgfqpoint{0.698006in}{1.394235in}}%
\pgfpathlineto{\pgfqpoint{0.705836in}{1.414312in}}%
\pgfpathlineto{\pgfqpoint{0.714284in}{1.434136in}}%
\pgfpathlineto{\pgfqpoint{0.723343in}{1.453689in}}%
\pgfpathlineto{\pgfqpoint{0.733002in}{1.472951in}}%
\pgfpathlineto{\pgfqpoint{0.743254in}{1.491906in}}%
\pgfpathlineto{\pgfqpoint{0.754089in}{1.510533in}}%
\pgfpathlineto{\pgfqpoint{0.765495in}{1.528816in}}%
\pgfpathlineto{\pgfqpoint{0.777462in}{1.546737in}}%
\pgfpathlineto{\pgfqpoint{0.789979in}{1.564278in}}%
\pgfpathlineto{\pgfqpoint{0.803033in}{1.581423in}}%
\pgfpathlineto{\pgfqpoint{0.816613in}{1.598155in}}%
\pgfpathlineto{\pgfqpoint{0.830704in}{1.614459in}}%
\pgfpathlineto{\pgfqpoint{0.845293in}{1.630318in}}%
\pgfpathlineto{\pgfqpoint{0.852409in}{1.624360in}}%
\pgfpathlineto{\pgfqpoint{0.919464in}{1.557305in}}%
\pgfpathlineto{\pgfqpoint{0.944051in}{1.581892in}}%
\pgfpathlineto{\pgfqpoint{0.876996in}{1.648947in}}%
\pgfpathlineto{\pgfqpoint{0.876996in}{1.648947in}}%
\pgfusepath{stroke}%
\end{pgfscope}%
\begin{pgfscope}%
\pgfsetrectcap%
\pgfsetmiterjoin%
\pgfsetlinewidth{0.803000pt}%
\definecolor{currentstroke}{rgb}{0.000000,0.000000,0.000000}%
\pgfsetstrokecolor{currentstroke}%
\pgfsetdash{}{0pt}%
\pgfpathmoveto{\pgfqpoint{0.584771in}{0.386658in}}%
\pgfpathlineto{\pgfqpoint{0.584771in}{1.916585in}}%
\pgfusepath{stroke}%
\end{pgfscope}%
\begin{pgfscope}%
\pgfsetrectcap%
\pgfsetmiterjoin%
\pgfsetlinewidth{0.803000pt}%
\definecolor{currentstroke}{rgb}{0.000000,0.000000,0.000000}%
\pgfsetstrokecolor{currentstroke}%
\pgfsetdash{}{0pt}%
\pgfpathmoveto{\pgfqpoint{2.114698in}{0.386658in}}%
\pgfpathlineto{\pgfqpoint{2.114698in}{1.916585in}}%
\pgfusepath{stroke}%
\end{pgfscope}%
\begin{pgfscope}%
\pgfsetrectcap%
\pgfsetmiterjoin%
\pgfsetlinewidth{0.803000pt}%
\definecolor{currentstroke}{rgb}{0.000000,0.000000,0.000000}%
\pgfsetstrokecolor{currentstroke}%
\pgfsetdash{}{0pt}%
\pgfpathmoveto{\pgfqpoint{0.584771in}{0.386658in}}%
\pgfpathlineto{\pgfqpoint{2.114698in}{0.386658in}}%
\pgfusepath{stroke}%
\end{pgfscope}%
\begin{pgfscope}%
\pgfsetrectcap%
\pgfsetmiterjoin%
\pgfsetlinewidth{0.803000pt}%
\definecolor{currentstroke}{rgb}{0.000000,0.000000,0.000000}%
\pgfsetstrokecolor{currentstroke}%
\pgfsetdash{}{0pt}%
\pgfpathmoveto{\pgfqpoint{0.584771in}{1.916585in}}%
\pgfpathlineto{\pgfqpoint{2.114698in}{1.916585in}}%
\pgfusepath{stroke}%
\end{pgfscope}%
\begin{pgfscope}%
\pgfpathrectangle{\pgfqpoint{0.584771in}{0.386658in}}{\pgfqpoint{1.529927in}{1.529927in}}%
\pgfusepath{clip}%
\pgfsetrectcap%
\pgfsetroundjoin%
\pgfsetlinewidth{1.505625pt}%
\definecolor{currentstroke}{rgb}{0.000000,0.000000,0.000000}%
\pgfsetstrokecolor{currentstroke}%
\pgfsetdash{}{0pt}%
\pgfpathmoveto{\pgfqpoint{1.071566in}{1.151622in}}%
\pgfpathlineto{\pgfqpoint{1.627903in}{1.151622in}}%
\pgfusepath{stroke}%
\end{pgfscope}%
\begin{pgfscope}%
\pgfpathrectangle{\pgfqpoint{0.584771in}{0.386658in}}{\pgfqpoint{1.529927in}{1.529927in}}%
\pgfusepath{clip}%
\pgfsetrectcap%
\pgfsetroundjoin%
\pgfsetlinewidth{1.505625pt}%
\definecolor{currentstroke}{rgb}{0.000000,0.000000,0.000000}%
\pgfsetstrokecolor{currentstroke}%
\pgfsetdash{}{0pt}%
\pgfpathmoveto{\pgfqpoint{1.349734in}{0.873453in}}%
\pgfpathlineto{\pgfqpoint{1.349734in}{1.429790in}}%
\pgfusepath{stroke}%
\end{pgfscope}%
\begin{pgfscope}%
\pgfsetbuttcap%
\pgfsetmiterjoin%
\definecolor{currentfill}{rgb}{1.000000,1.000000,1.000000}%
\pgfsetfillcolor{currentfill}%
\pgfsetlinewidth{0.000000pt}%
\definecolor{currentstroke}{rgb}{0.000000,0.000000,0.000000}%
\pgfsetstrokecolor{currentstroke}%
\pgfsetstrokeopacity{0.000000}%
\pgfsetdash{}{0pt}%
\pgfpathmoveto{\pgfqpoint{0.337193in}{2.012206in}}%
\pgfpathlineto{\pgfqpoint{2.362276in}{2.012206in}}%
\pgfpathlineto{\pgfqpoint{2.362276in}{2.113460in}}%
\pgfpathlineto{\pgfqpoint{0.337193in}{2.113460in}}%
\pgfpathlineto{\pgfqpoint{0.337193in}{2.012206in}}%
\pgfpathclose%
\pgfusepath{fill}%
\end{pgfscope}%
\begin{pgfscope}%
\pgfsys@transformshift{0.337500in}{2.012500in}%
\pgftext[left,bottom]{\includegraphics[interpolate=true,width=2.025000in,height=0.101250in]{figures/./section1/data/10000p/re_1000//err_vmag-img1.png}}%
\end{pgfscope}%
\begin{pgfscope}%
\pgfsetbuttcap%
\pgfsetroundjoin%
\definecolor{currentfill}{rgb}{0.000000,0.000000,0.000000}%
\pgfsetfillcolor{currentfill}%
\pgfsetlinewidth{0.803000pt}%
\definecolor{currentstroke}{rgb}{0.000000,0.000000,0.000000}%
\pgfsetstrokecolor{currentstroke}%
\pgfsetdash{}{0pt}%
\pgfsys@defobject{currentmarker}{\pgfqpoint{0.000000in}{0.000000in}}{\pgfqpoint{0.000000in}{0.048611in}}{%
\pgfpathmoveto{\pgfqpoint{0.000000in}{0.000000in}}%
\pgfpathlineto{\pgfqpoint{0.000000in}{0.048611in}}%
\pgfusepath{stroke,fill}%
}%
\begin{pgfscope}%
\pgfsys@transformshift{0.338445in}{2.113460in}%
\pgfsys@useobject{currentmarker}{}%
\end{pgfscope}%
\end{pgfscope}%
\begin{pgfscope}%
\definecolor{textcolor}{rgb}{0.000000,0.000000,0.000000}%
\pgfsetstrokecolor{textcolor}%
\pgfsetfillcolor{textcolor}%
\pgftext[x=0.338445in,y=2.203738in,,bottom]{\color{textcolor}{\rmfamily\fontsize{8.330000}{9.996000}\selectfont\catcode`\^=\active\def^{\ifmmode\sp\else\^{}\fi}\catcode`\%=\active\def%{\%}$\mathdefault{\ensuremath{-}1.180}$}}%
\end{pgfscope}%
\begin{pgfscope}%
\pgfsetbuttcap%
\pgfsetroundjoin%
\definecolor{currentfill}{rgb}{0.000000,0.000000,0.000000}%
\pgfsetfillcolor{currentfill}%
\pgfsetlinewidth{0.803000pt}%
\definecolor{currentstroke}{rgb}{0.000000,0.000000,0.000000}%
\pgfsetstrokecolor{currentstroke}%
\pgfsetdash{}{0pt}%
\pgfsys@defobject{currentmarker}{\pgfqpoint{0.000000in}{0.000000in}}{\pgfqpoint{0.000000in}{0.048611in}}{%
\pgfpathmoveto{\pgfqpoint{0.000000in}{0.000000in}}%
\pgfpathlineto{\pgfqpoint{0.000000in}{0.048611in}}%
\pgfusepath{stroke,fill}%
}%
\begin{pgfscope}%
\pgfsys@transformshift{1.012444in}{2.113460in}%
\pgfsys@useobject{currentmarker}{}%
\end{pgfscope}%
\end{pgfscope}%
\begin{pgfscope}%
\definecolor{textcolor}{rgb}{0.000000,0.000000,0.000000}%
\pgfsetstrokecolor{textcolor}%
\pgfsetfillcolor{textcolor}%
\pgftext[x=1.012444in,y=2.203738in,,bottom]{\color{textcolor}{\rmfamily\fontsize{8.330000}{9.996000}\selectfont\catcode`\^=\active\def^{\ifmmode\sp\else\^{}\fi}\catcode`\%=\active\def%{\%}$\mathdefault{\ensuremath{-}0.269}$}}%
\end{pgfscope}%
\begin{pgfscope}%
\pgfsetbuttcap%
\pgfsetroundjoin%
\definecolor{currentfill}{rgb}{0.000000,0.000000,0.000000}%
\pgfsetfillcolor{currentfill}%
\pgfsetlinewidth{0.803000pt}%
\definecolor{currentstroke}{rgb}{0.000000,0.000000,0.000000}%
\pgfsetstrokecolor{currentstroke}%
\pgfsetdash{}{0pt}%
\pgfsys@defobject{currentmarker}{\pgfqpoint{0.000000in}{0.000000in}}{\pgfqpoint{0.000000in}{0.048611in}}{%
\pgfpathmoveto{\pgfqpoint{0.000000in}{0.000000in}}%
\pgfpathlineto{\pgfqpoint{0.000000in}{0.048611in}}%
\pgfusepath{stroke,fill}%
}%
\begin{pgfscope}%
\pgfsys@transformshift{1.686444in}{2.113460in}%
\pgfsys@useobject{currentmarker}{}%
\end{pgfscope}%
\end{pgfscope}%
\begin{pgfscope}%
\definecolor{textcolor}{rgb}{0.000000,0.000000,0.000000}%
\pgfsetstrokecolor{textcolor}%
\pgfsetfillcolor{textcolor}%
\pgftext[x=1.686444in,y=2.203738in,,bottom]{\color{textcolor}{\rmfamily\fontsize{8.330000}{9.996000}\selectfont\catcode`\^=\active\def^{\ifmmode\sp\else\^{}\fi}\catcode`\%=\active\def%{\%}$\mathdefault{0.643}$}}%
\end{pgfscope}%
\begin{pgfscope}%
\pgfsetbuttcap%
\pgfsetroundjoin%
\definecolor{currentfill}{rgb}{0.000000,0.000000,0.000000}%
\pgfsetfillcolor{currentfill}%
\pgfsetlinewidth{0.803000pt}%
\definecolor{currentstroke}{rgb}{0.000000,0.000000,0.000000}%
\pgfsetstrokecolor{currentstroke}%
\pgfsetdash{}{0pt}%
\pgfsys@defobject{currentmarker}{\pgfqpoint{0.000000in}{0.000000in}}{\pgfqpoint{0.000000in}{0.048611in}}{%
\pgfpathmoveto{\pgfqpoint{0.000000in}{0.000000in}}%
\pgfpathlineto{\pgfqpoint{0.000000in}{0.048611in}}%
\pgfusepath{stroke,fill}%
}%
\begin{pgfscope}%
\pgfsys@transformshift{2.360444in}{2.113460in}%
\pgfsys@useobject{currentmarker}{}%
\end{pgfscope}%
\end{pgfscope}%
\begin{pgfscope}%
\definecolor{textcolor}{rgb}{0.000000,0.000000,0.000000}%
\pgfsetstrokecolor{textcolor}%
\pgfsetfillcolor{textcolor}%
\pgftext[x=2.360444in,y=2.203738in,,bottom]{\color{textcolor}{\rmfamily\fontsize{8.330000}{9.996000}\selectfont\catcode`\^=\active\def^{\ifmmode\sp\else\^{}\fi}\catcode`\%=\active\def%{\%}$\mathdefault{1.555}$}}%
\end{pgfscope}%
\begin{pgfscope}%
\definecolor{textcolor}{rgb}{0.000000,0.000000,0.000000}%
\pgfsetstrokecolor{textcolor}%
\pgfsetfillcolor{textcolor}%
\pgftext[x=1.349734in,y=2.358059in,,base]{\color{textcolor}{\rmfamily\fontsize{10.000000}{12.000000}\selectfont\catcode`\^=\active\def^{\ifmmode\sp\else\^{}\fi}\catcode`\%=\active\def%{\%}$f^{(\bm{v})}_\text{err} \ [\%]$}}%
\end{pgfscope}%
\begin{pgfscope}%
\definecolor{textcolor}{rgb}{0.000000,0.000000,0.000000}%
\pgfsetstrokecolor{textcolor}%
\pgfsetfillcolor{textcolor}%
\pgftext[x=2.362276in,y=2.344170in,right,bottom]{\color{textcolor}{\rmfamily\fontsize{8.330000}{9.996000}\selectfont\catcode`\^=\active\def^{\ifmmode\sp\else\^{}\fi}\catcode`\%=\active\def%{\%}$\times\mathdefault{10^{1}}\mathdefault{}$}}%
\end{pgfscope}%
\begin{pgfscope}%
\pgfsetrectcap%
\pgfsetmiterjoin%
\pgfsetlinewidth{0.803000pt}%
\definecolor{currentstroke}{rgb}{0.000000,0.000000,0.000000}%
\pgfsetstrokecolor{currentstroke}%
\pgfsetdash{}{0pt}%
\pgfpathmoveto{\pgfqpoint{0.337193in}{2.012206in}}%
\pgfpathlineto{\pgfqpoint{0.337193in}{2.062833in}}%
\pgfpathlineto{\pgfqpoint{0.337193in}{2.113460in}}%
\pgfpathlineto{\pgfqpoint{2.362276in}{2.113460in}}%
\pgfpathlineto{\pgfqpoint{2.362276in}{2.062833in}}%
\pgfpathlineto{\pgfqpoint{2.362276in}{2.012206in}}%
\pgfpathlineto{\pgfqpoint{0.337193in}{2.012206in}}%
\pgfpathclose%
\pgfusepath{stroke}%
\end{pgfscope}%
\end{pgfpicture}%
\makeatother%
\endgroup%

%% file: figures/section1/errors_all_Re.pgf
%% Creator: Matplotlib, PGF backend
%%
%% To include the figure in your LaTeX document, write
%%   \input{<filename>.pgf}
%%
%% Make sure the required packages are loaded in your preamble
%%   \usepackage{pgf}
%%
%% Also ensure that all the required font packages are loaded; for instance,
%% the lmodern package is sometimes necessary when using math font.
%%   \usepackage{lmodern}
%%
%% Figures using additional raster images can only be included by \input if
%% they are in the same directory as the main LaTeX file. For loading figures
%% from other directories you can use the `import` package
%% and then include the figures with
%%   \import{<path to file>}{<filename>.pgf}
%%
%% Matplotlib used the following preamble
%%   \def\mathdefault#1{#1}
%%   \everymath=\expandafter{\the\everymath\displaystyle}
%%   \usepackage{amsmath}\usepackage{bm}
%%   \makeatletter\@ifpackageloaded{underscore}{}{\usepackage[strings]{underscore}}\makeatother
%%
\begingroup%
\makeatletter%
\begin{pgfpicture}%
\pgfpathrectangle{\pgfpointorigin}{\pgfqpoint{3.000000in}{2.000000in}}%
\pgfusepath{use as bounding box, clip}%
\begin{pgfscope}%
\pgfsetbuttcap%
\pgfsetmiterjoin%
\definecolor{currentfill}{rgb}{1.000000,1.000000,1.000000}%
\pgfsetfillcolor{currentfill}%
\pgfsetlinewidth{0.000000pt}%
\definecolor{currentstroke}{rgb}{1.000000,1.000000,1.000000}%
\pgfsetstrokecolor{currentstroke}%
\pgfsetdash{}{0pt}%
\pgfpathmoveto{\pgfqpoint{0.000000in}{0.000000in}}%
\pgfpathlineto{\pgfqpoint{3.000000in}{0.000000in}}%
\pgfpathlineto{\pgfqpoint{3.000000in}{2.000000in}}%
\pgfpathlineto{\pgfqpoint{0.000000in}{2.000000in}}%
\pgfpathlineto{\pgfqpoint{0.000000in}{0.000000in}}%
\pgfpathclose%
\pgfusepath{fill}%
\end{pgfscope}%
\begin{pgfscope}%
\pgfsetbuttcap%
\pgfsetmiterjoin%
\definecolor{currentfill}{rgb}{1.000000,1.000000,1.000000}%
\pgfsetfillcolor{currentfill}%
\pgfsetlinewidth{0.000000pt}%
\definecolor{currentstroke}{rgb}{0.000000,0.000000,0.000000}%
\pgfsetstrokecolor{currentstroke}%
\pgfsetstrokeopacity{0.000000}%
\pgfsetdash{}{0pt}%
\pgfpathmoveto{\pgfqpoint{0.597016in}{0.498776in}}%
\pgfpathlineto{\pgfqpoint{2.845756in}{0.498776in}}%
\pgfpathlineto{\pgfqpoint{2.845756in}{1.818771in}}%
\pgfpathlineto{\pgfqpoint{0.597016in}{1.818771in}}%
\pgfpathlineto{\pgfqpoint{0.597016in}{0.498776in}}%
\pgfpathclose%
\pgfusepath{fill}%
\end{pgfscope}%
\begin{pgfscope}%
\pgfpathrectangle{\pgfqpoint{0.597016in}{0.498776in}}{\pgfqpoint{2.248740in}{1.319995in}}%
\pgfusepath{clip}%
\pgfsetbuttcap%
\pgfsetroundjoin%
\definecolor{currentfill}{rgb}{0.172549,0.482353,0.713725}%
\pgfsetfillcolor{currentfill}%
\pgfsetlinewidth{0.501875pt}%
\definecolor{currentstroke}{rgb}{0.000000,0.000000,0.000000}%
\pgfsetstrokecolor{currentstroke}%
\pgfsetdash{}{0pt}%
\pgfsys@defobject{currentmarker}{\pgfqpoint{-0.026896in}{-0.026896in}}{\pgfqpoint{0.026896in}{0.026896in}}{%
\pgfpathmoveto{\pgfqpoint{-0.026896in}{-0.026896in}}%
\pgfpathlineto{\pgfqpoint{0.026896in}{-0.026896in}}%
\pgfpathlineto{\pgfqpoint{0.026896in}{0.026896in}}%
\pgfpathlineto{\pgfqpoint{-0.026896in}{0.026896in}}%
\pgfpathlineto{\pgfqpoint{-0.026896in}{-0.026896in}}%
\pgfpathclose%
\pgfusepath{stroke,fill}%
}%
\begin{pgfscope}%
\pgfsys@transformshift{0.699231in}{1.468005in}%
\pgfsys@useobject{currentmarker}{}%
\end{pgfscope}%
\begin{pgfscope}%
\pgfsys@transformshift{0.926377in}{1.301032in}%
\pgfsys@useobject{currentmarker}{}%
\end{pgfscope}%
\begin{pgfscope}%
\pgfsys@transformshift{1.153522in}{0.602596in}%
\pgfsys@useobject{currentmarker}{}%
\end{pgfscope}%
\begin{pgfscope}%
\pgfsys@transformshift{1.380668in}{0.554541in}%
\pgfsys@useobject{currentmarker}{}%
\end{pgfscope}%
\begin{pgfscope}%
\pgfsys@transformshift{1.607813in}{0.617845in}%
\pgfsys@useobject{currentmarker}{}%
\end{pgfscope}%
\begin{pgfscope}%
\pgfsys@transformshift{1.834959in}{0.644003in}%
\pgfsys@useobject{currentmarker}{}%
\end{pgfscope}%
\begin{pgfscope}%
\pgfsys@transformshift{2.062104in}{0.654124in}%
\pgfsys@useobject{currentmarker}{}%
\end{pgfscope}%
\begin{pgfscope}%
\pgfsys@transformshift{2.289250in}{0.664881in}%
\pgfsys@useobject{currentmarker}{}%
\end{pgfscope}%
\begin{pgfscope}%
\pgfsys@transformshift{2.516395in}{0.679716in}%
\pgfsys@useobject{currentmarker}{}%
\end{pgfscope}%
\begin{pgfscope}%
\pgfsys@transformshift{2.743541in}{0.747282in}%
\pgfsys@useobject{currentmarker}{}%
\end{pgfscope}%
\end{pgfscope}%
\begin{pgfscope}%
\pgfpathrectangle{\pgfqpoint{0.597016in}{0.498776in}}{\pgfqpoint{2.248740in}{1.319995in}}%
\pgfusepath{clip}%
\pgfsetbuttcap%
\pgfsetroundjoin%
\definecolor{currentfill}{rgb}{0.843137,0.098039,0.109804}%
\pgfsetfillcolor{currentfill}%
\pgfsetlinewidth{0.501875pt}%
\definecolor{currentstroke}{rgb}{0.000000,0.000000,0.000000}%
\pgfsetstrokecolor{currentstroke}%
\pgfsetdash{}{0pt}%
\pgfsys@defobject{currentmarker}{\pgfqpoint{-0.026896in}{-0.026896in}}{\pgfqpoint{0.026896in}{0.026896in}}{%
\pgfpathmoveto{\pgfqpoint{0.000000in}{-0.026896in}}%
\pgfpathcurveto{\pgfqpoint{0.007133in}{-0.026896in}}{\pgfqpoint{0.013974in}{-0.024062in}}{\pgfqpoint{0.019018in}{-0.019018in}}%
\pgfpathcurveto{\pgfqpoint{0.024062in}{-0.013974in}}{\pgfqpoint{0.026896in}{-0.007133in}}{\pgfqpoint{0.026896in}{0.000000in}}%
\pgfpathcurveto{\pgfqpoint{0.026896in}{0.007133in}}{\pgfqpoint{0.024062in}{0.013974in}}{\pgfqpoint{0.019018in}{0.019018in}}%
\pgfpathcurveto{\pgfqpoint{0.013974in}{0.024062in}}{\pgfqpoint{0.007133in}{0.026896in}}{\pgfqpoint{0.000000in}{0.026896in}}%
\pgfpathcurveto{\pgfqpoint{-0.007133in}{0.026896in}}{\pgfqpoint{-0.013974in}{0.024062in}}{\pgfqpoint{-0.019018in}{0.019018in}}%
\pgfpathcurveto{\pgfqpoint{-0.024062in}{0.013974in}}{\pgfqpoint{-0.026896in}{0.007133in}}{\pgfqpoint{-0.026896in}{0.000000in}}%
\pgfpathcurveto{\pgfqpoint{-0.026896in}{-0.007133in}}{\pgfqpoint{-0.024062in}{-0.013974in}}{\pgfqpoint{-0.019018in}{-0.019018in}}%
\pgfpathcurveto{\pgfqpoint{-0.013974in}{-0.024062in}}{\pgfqpoint{-0.007133in}{-0.026896in}}{\pgfqpoint{0.000000in}{-0.026896in}}%
\pgfpathlineto{\pgfqpoint{0.000000in}{-0.026896in}}%
\pgfpathclose%
\pgfusepath{stroke,fill}%
}%
\begin{pgfscope}%
\pgfsys@transformshift{0.699231in}{0.982369in}%
\pgfsys@useobject{currentmarker}{}%
\end{pgfscope}%
\begin{pgfscope}%
\pgfsys@transformshift{0.926377in}{0.928994in}%
\pgfsys@useobject{currentmarker}{}%
\end{pgfscope}%
\begin{pgfscope}%
\pgfsys@transformshift{1.153522in}{0.625098in}%
\pgfsys@useobject{currentmarker}{}%
\end{pgfscope}%
\begin{pgfscope}%
\pgfsys@transformshift{1.380668in}{0.600982in}%
\pgfsys@useobject{currentmarker}{}%
\end{pgfscope}%
\begin{pgfscope}%
\pgfsys@transformshift{1.607813in}{0.650779in}%
\pgfsys@useobject{currentmarker}{}%
\end{pgfscope}%
\begin{pgfscope}%
\pgfsys@transformshift{1.834959in}{0.681451in}%
\pgfsys@useobject{currentmarker}{}%
\end{pgfscope}%
\begin{pgfscope}%
\pgfsys@transformshift{2.062104in}{0.687075in}%
\pgfsys@useobject{currentmarker}{}%
\end{pgfscope}%
\begin{pgfscope}%
\pgfsys@transformshift{2.289250in}{0.693257in}%
\pgfsys@useobject{currentmarker}{}%
\end{pgfscope}%
\begin{pgfscope}%
\pgfsys@transformshift{2.516395in}{0.700045in}%
\pgfsys@useobject{currentmarker}{}%
\end{pgfscope}%
\begin{pgfscope}%
\pgfsys@transformshift{2.743541in}{0.750646in}%
\pgfsys@useobject{currentmarker}{}%
\end{pgfscope}%
\end{pgfscope}%
\begin{pgfscope}%
\pgfpathrectangle{\pgfqpoint{0.597016in}{0.498776in}}{\pgfqpoint{2.248740in}{1.319995in}}%
\pgfusepath{clip}%
\pgfsetbuttcap%
\pgfsetroundjoin%
\definecolor{currentfill}{rgb}{0.670588,0.850980,0.913725}%
\pgfsetfillcolor{currentfill}%
\pgfsetlinewidth{0.501875pt}%
\definecolor{currentstroke}{rgb}{0.000000,0.000000,0.000000}%
\pgfsetstrokecolor{currentstroke}%
\pgfsetdash{}{0pt}%
\pgfsys@defobject{currentmarker}{\pgfqpoint{-0.038036in}{-0.038036in}}{\pgfqpoint{0.038036in}{0.038036in}}{%
\pgfpathmoveto{\pgfqpoint{-0.000000in}{-0.038036in}}%
\pgfpathlineto{\pgfqpoint{0.038036in}{0.000000in}}%
\pgfpathlineto{\pgfqpoint{0.000000in}{0.038036in}}%
\pgfpathlineto{\pgfqpoint{-0.038036in}{0.000000in}}%
\pgfpathlineto{\pgfqpoint{-0.000000in}{-0.038036in}}%
\pgfpathclose%
\pgfusepath{stroke,fill}%
}%
\begin{pgfscope}%
\pgfsys@transformshift{0.699231in}{1.512670in}%
\pgfsys@useobject{currentmarker}{}%
\end{pgfscope}%
\begin{pgfscope}%
\pgfsys@transformshift{0.926377in}{1.428921in}%
\pgfsys@useobject{currentmarker}{}%
\end{pgfscope}%
\begin{pgfscope}%
\pgfsys@transformshift{1.153522in}{1.099195in}%
\pgfsys@useobject{currentmarker}{}%
\end{pgfscope}%
\begin{pgfscope}%
\pgfsys@transformshift{1.380668in}{0.982757in}%
\pgfsys@useobject{currentmarker}{}%
\end{pgfscope}%
\begin{pgfscope}%
\pgfsys@transformshift{1.607813in}{1.042023in}%
\pgfsys@useobject{currentmarker}{}%
\end{pgfscope}%
\begin{pgfscope}%
\pgfsys@transformshift{1.834959in}{1.040330in}%
\pgfsys@useobject{currentmarker}{}%
\end{pgfscope}%
\begin{pgfscope}%
\pgfsys@transformshift{2.062104in}{1.014541in}%
\pgfsys@useobject{currentmarker}{}%
\end{pgfscope}%
\begin{pgfscope}%
\pgfsys@transformshift{2.289250in}{0.990614in}%
\pgfsys@useobject{currentmarker}{}%
\end{pgfscope}%
\begin{pgfscope}%
\pgfsys@transformshift{2.516395in}{0.963958in}%
\pgfsys@useobject{currentmarker}{}%
\end{pgfscope}%
\begin{pgfscope}%
\pgfsys@transformshift{2.743541in}{0.978733in}%
\pgfsys@useobject{currentmarker}{}%
\end{pgfscope}%
\end{pgfscope}%
\begin{pgfscope}%
\pgfpathrectangle{\pgfqpoint{0.597016in}{0.498776in}}{\pgfqpoint{2.248740in}{1.319995in}}%
\pgfusepath{clip}%
\pgfsetbuttcap%
\pgfsetroundjoin%
\definecolor{currentfill}{rgb}{0.992157,0.682353,0.380392}%
\pgfsetfillcolor{currentfill}%
\pgfsetlinewidth{0.501875pt}%
\definecolor{currentstroke}{rgb}{0.000000,0.000000,0.000000}%
\pgfsetstrokecolor{currentstroke}%
\pgfsetdash{}{0pt}%
\pgfsys@defobject{currentmarker}{\pgfqpoint{-0.026896in}{-0.026896in}}{\pgfqpoint{0.026896in}{0.026896in}}{%
\pgfpathmoveto{\pgfqpoint{0.000000in}{0.026896in}}%
\pgfpathlineto{\pgfqpoint{-0.026896in}{-0.026896in}}%
\pgfpathlineto{\pgfqpoint{0.026896in}{-0.026896in}}%
\pgfpathlineto{\pgfqpoint{0.000000in}{0.026896in}}%
\pgfpathclose%
\pgfusepath{stroke,fill}%
}%
\begin{pgfscope}%
\pgfsys@transformshift{0.699231in}{1.397715in}%
\pgfsys@useobject{currentmarker}{}%
\end{pgfscope}%
\begin{pgfscope}%
\pgfsys@transformshift{0.926377in}{1.337753in}%
\pgfsys@useobject{currentmarker}{}%
\end{pgfscope}%
\begin{pgfscope}%
\pgfsys@transformshift{1.153522in}{1.029030in}%
\pgfsys@useobject{currentmarker}{}%
\end{pgfscope}%
\begin{pgfscope}%
\pgfsys@transformshift{1.380668in}{0.925452in}%
\pgfsys@useobject{currentmarker}{}%
\end{pgfscope}%
\begin{pgfscope}%
\pgfsys@transformshift{1.607813in}{0.982618in}%
\pgfsys@useobject{currentmarker}{}%
\end{pgfscope}%
\begin{pgfscope}%
\pgfsys@transformshift{1.834959in}{0.980304in}%
\pgfsys@useobject{currentmarker}{}%
\end{pgfscope}%
\begin{pgfscope}%
\pgfsys@transformshift{2.062104in}{0.954535in}%
\pgfsys@useobject{currentmarker}{}%
\end{pgfscope}%
\begin{pgfscope}%
\pgfsys@transformshift{2.289250in}{0.929192in}%
\pgfsys@useobject{currentmarker}{}%
\end{pgfscope}%
\begin{pgfscope}%
\pgfsys@transformshift{2.516395in}{0.903750in}%
\pgfsys@useobject{currentmarker}{}%
\end{pgfscope}%
\begin{pgfscope}%
\pgfsys@transformshift{2.743541in}{0.916122in}%
\pgfsys@useobject{currentmarker}{}%
\end{pgfscope}%
\end{pgfscope}%
\begin{pgfscope}%
\pgfsetbuttcap%
\pgfsetroundjoin%
\definecolor{currentfill}{rgb}{0.000000,0.000000,0.000000}%
\pgfsetfillcolor{currentfill}%
\pgfsetlinewidth{0.803000pt}%
\definecolor{currentstroke}{rgb}{0.000000,0.000000,0.000000}%
\pgfsetstrokecolor{currentstroke}%
\pgfsetdash{}{0pt}%
\pgfsys@defobject{currentmarker}{\pgfqpoint{0.000000in}{-0.048611in}}{\pgfqpoint{0.000000in}{0.000000in}}{%
\pgfpathmoveto{\pgfqpoint{0.000000in}{0.000000in}}%
\pgfpathlineto{\pgfqpoint{0.000000in}{-0.048611in}}%
\pgfusepath{stroke,fill}%
}%
\begin{pgfscope}%
\pgfsys@transformshift{0.699231in}{0.498776in}%
\pgfsys@useobject{currentmarker}{}%
\end{pgfscope}%
\end{pgfscope}%
\begin{pgfscope}%
\definecolor{textcolor}{rgb}{0.000000,0.000000,0.000000}%
\pgfsetstrokecolor{textcolor}%
\pgfsetfillcolor{textcolor}%
\pgftext[x=0.699231in,y=0.408498in,,top]{\color{textcolor}{\rmfamily\fontsize{6.500000}{7.800000}\selectfont\catcode`\^=\active\def^{\ifmmode\sp\else\^{}\fi}\catcode`\%=\active\def%{\%}30}}%
\end{pgfscope}%
\begin{pgfscope}%
\pgfsetbuttcap%
\pgfsetroundjoin%
\definecolor{currentfill}{rgb}{0.000000,0.000000,0.000000}%
\pgfsetfillcolor{currentfill}%
\pgfsetlinewidth{0.803000pt}%
\definecolor{currentstroke}{rgb}{0.000000,0.000000,0.000000}%
\pgfsetstrokecolor{currentstroke}%
\pgfsetdash{}{0pt}%
\pgfsys@defobject{currentmarker}{\pgfqpoint{0.000000in}{-0.048611in}}{\pgfqpoint{0.000000in}{0.000000in}}{%
\pgfpathmoveto{\pgfqpoint{0.000000in}{0.000000in}}%
\pgfpathlineto{\pgfqpoint{0.000000in}{-0.048611in}}%
\pgfusepath{stroke,fill}%
}%
\begin{pgfscope}%
\pgfsys@transformshift{0.926377in}{0.498776in}%
\pgfsys@useobject{currentmarker}{}%
\end{pgfscope}%
\end{pgfscope}%
\begin{pgfscope}%
\definecolor{textcolor}{rgb}{0.000000,0.000000,0.000000}%
\pgfsetstrokecolor{textcolor}%
\pgfsetfillcolor{textcolor}%
\pgftext[x=0.926377in,y=0.408498in,,top]{\color{textcolor}{\rmfamily\fontsize{6.500000}{7.800000}\selectfont\catcode`\^=\active\def^{\ifmmode\sp\else\^{}\fi}\catcode`\%=\active\def%{\%}50}}%
\end{pgfscope}%
\begin{pgfscope}%
\pgfsetbuttcap%
\pgfsetroundjoin%
\definecolor{currentfill}{rgb}{0.000000,0.000000,0.000000}%
\pgfsetfillcolor{currentfill}%
\pgfsetlinewidth{0.803000pt}%
\definecolor{currentstroke}{rgb}{0.000000,0.000000,0.000000}%
\pgfsetstrokecolor{currentstroke}%
\pgfsetdash{}{0pt}%
\pgfsys@defobject{currentmarker}{\pgfqpoint{0.000000in}{-0.048611in}}{\pgfqpoint{0.000000in}{0.000000in}}{%
\pgfpathmoveto{\pgfqpoint{0.000000in}{0.000000in}}%
\pgfpathlineto{\pgfqpoint{0.000000in}{-0.048611in}}%
\pgfusepath{stroke,fill}%
}%
\begin{pgfscope}%
\pgfsys@transformshift{1.153522in}{0.498776in}%
\pgfsys@useobject{currentmarker}{}%
\end{pgfscope}%
\end{pgfscope}%
\begin{pgfscope}%
\definecolor{textcolor}{rgb}{0.000000,0.000000,0.000000}%
\pgfsetstrokecolor{textcolor}%
\pgfsetfillcolor{textcolor}%
\pgftext[x=1.153522in,y=0.408498in,,top]{\color{textcolor}{\rmfamily\fontsize{6.500000}{7.800000}\selectfont\catcode`\^=\active\def^{\ifmmode\sp\else\^{}\fi}\catcode`\%=\active\def%{\%}100}}%
\end{pgfscope}%
\begin{pgfscope}%
\pgfsetbuttcap%
\pgfsetroundjoin%
\definecolor{currentfill}{rgb}{0.000000,0.000000,0.000000}%
\pgfsetfillcolor{currentfill}%
\pgfsetlinewidth{0.803000pt}%
\definecolor{currentstroke}{rgb}{0.000000,0.000000,0.000000}%
\pgfsetstrokecolor{currentstroke}%
\pgfsetdash{}{0pt}%
\pgfsys@defobject{currentmarker}{\pgfqpoint{0.000000in}{-0.048611in}}{\pgfqpoint{0.000000in}{0.000000in}}{%
\pgfpathmoveto{\pgfqpoint{0.000000in}{0.000000in}}%
\pgfpathlineto{\pgfqpoint{0.000000in}{-0.048611in}}%
\pgfusepath{stroke,fill}%
}%
\begin{pgfscope}%
\pgfsys@transformshift{1.380668in}{0.498776in}%
\pgfsys@useobject{currentmarker}{}%
\end{pgfscope}%
\end{pgfscope}%
\begin{pgfscope}%
\definecolor{textcolor}{rgb}{0.000000,0.000000,0.000000}%
\pgfsetstrokecolor{textcolor}%
\pgfsetfillcolor{textcolor}%
\pgftext[x=1.380668in,y=0.408498in,,top]{\color{textcolor}{\rmfamily\fontsize{6.500000}{7.800000}\selectfont\catcode`\^=\active\def^{\ifmmode\sp\else\^{}\fi}\catcode`\%=\active\def%{\%}500}}%
\end{pgfscope}%
\begin{pgfscope}%
\pgfsetbuttcap%
\pgfsetroundjoin%
\definecolor{currentfill}{rgb}{0.000000,0.000000,0.000000}%
\pgfsetfillcolor{currentfill}%
\pgfsetlinewidth{0.803000pt}%
\definecolor{currentstroke}{rgb}{0.000000,0.000000,0.000000}%
\pgfsetstrokecolor{currentstroke}%
\pgfsetdash{}{0pt}%
\pgfsys@defobject{currentmarker}{\pgfqpoint{0.000000in}{-0.048611in}}{\pgfqpoint{0.000000in}{0.000000in}}{%
\pgfpathmoveto{\pgfqpoint{0.000000in}{0.000000in}}%
\pgfpathlineto{\pgfqpoint{0.000000in}{-0.048611in}}%
\pgfusepath{stroke,fill}%
}%
\begin{pgfscope}%
\pgfsys@transformshift{1.607813in}{0.498776in}%
\pgfsys@useobject{currentmarker}{}%
\end{pgfscope}%
\end{pgfscope}%
\begin{pgfscope}%
\definecolor{textcolor}{rgb}{0.000000,0.000000,0.000000}%
\pgfsetstrokecolor{textcolor}%
\pgfsetfillcolor{textcolor}%
\pgftext[x=1.607813in,y=0.408498in,,top]{\color{textcolor}{\rmfamily\fontsize{6.500000}{7.800000}\selectfont\catcode`\^=\active\def^{\ifmmode\sp\else\^{}\fi}\catcode`\%=\active\def%{\%}1000}}%
\end{pgfscope}%
\begin{pgfscope}%
\pgfsetbuttcap%
\pgfsetroundjoin%
\definecolor{currentfill}{rgb}{0.000000,0.000000,0.000000}%
\pgfsetfillcolor{currentfill}%
\pgfsetlinewidth{0.803000pt}%
\definecolor{currentstroke}{rgb}{0.000000,0.000000,0.000000}%
\pgfsetstrokecolor{currentstroke}%
\pgfsetdash{}{0pt}%
\pgfsys@defobject{currentmarker}{\pgfqpoint{0.000000in}{-0.048611in}}{\pgfqpoint{0.000000in}{0.000000in}}{%
\pgfpathmoveto{\pgfqpoint{0.000000in}{0.000000in}}%
\pgfpathlineto{\pgfqpoint{0.000000in}{-0.048611in}}%
\pgfusepath{stroke,fill}%
}%
\begin{pgfscope}%
\pgfsys@transformshift{1.834959in}{0.498776in}%
\pgfsys@useobject{currentmarker}{}%
\end{pgfscope}%
\end{pgfscope}%
\begin{pgfscope}%
\definecolor{textcolor}{rgb}{0.000000,0.000000,0.000000}%
\pgfsetstrokecolor{textcolor}%
\pgfsetfillcolor{textcolor}%
\pgftext[x=1.834959in,y=0.408498in,,top]{\color{textcolor}{\rmfamily\fontsize{6.500000}{7.800000}\selectfont\catcode`\^=\active\def^{\ifmmode\sp\else\^{}\fi}\catcode`\%=\active\def%{\%}2000}}%
\end{pgfscope}%
\begin{pgfscope}%
\pgfsetbuttcap%
\pgfsetroundjoin%
\definecolor{currentfill}{rgb}{0.000000,0.000000,0.000000}%
\pgfsetfillcolor{currentfill}%
\pgfsetlinewidth{0.803000pt}%
\definecolor{currentstroke}{rgb}{0.000000,0.000000,0.000000}%
\pgfsetstrokecolor{currentstroke}%
\pgfsetdash{}{0pt}%
\pgfsys@defobject{currentmarker}{\pgfqpoint{0.000000in}{-0.048611in}}{\pgfqpoint{0.000000in}{0.000000in}}{%
\pgfpathmoveto{\pgfqpoint{0.000000in}{0.000000in}}%
\pgfpathlineto{\pgfqpoint{0.000000in}{-0.048611in}}%
\pgfusepath{stroke,fill}%
}%
\begin{pgfscope}%
\pgfsys@transformshift{2.062104in}{0.498776in}%
\pgfsys@useobject{currentmarker}{}%
\end{pgfscope}%
\end{pgfscope}%
\begin{pgfscope}%
\definecolor{textcolor}{rgb}{0.000000,0.000000,0.000000}%
\pgfsetstrokecolor{textcolor}%
\pgfsetfillcolor{textcolor}%
\pgftext[x=2.062104in,y=0.408498in,,top]{\color{textcolor}{\rmfamily\fontsize{6.500000}{7.800000}\selectfont\catcode`\^=\active\def^{\ifmmode\sp\else\^{}\fi}\catcode`\%=\active\def%{\%}3000}}%
\end{pgfscope}%
\begin{pgfscope}%
\pgfsetbuttcap%
\pgfsetroundjoin%
\definecolor{currentfill}{rgb}{0.000000,0.000000,0.000000}%
\pgfsetfillcolor{currentfill}%
\pgfsetlinewidth{0.803000pt}%
\definecolor{currentstroke}{rgb}{0.000000,0.000000,0.000000}%
\pgfsetstrokecolor{currentstroke}%
\pgfsetdash{}{0pt}%
\pgfsys@defobject{currentmarker}{\pgfqpoint{0.000000in}{-0.048611in}}{\pgfqpoint{0.000000in}{0.000000in}}{%
\pgfpathmoveto{\pgfqpoint{0.000000in}{0.000000in}}%
\pgfpathlineto{\pgfqpoint{0.000000in}{-0.048611in}}%
\pgfusepath{stroke,fill}%
}%
\begin{pgfscope}%
\pgfsys@transformshift{2.289250in}{0.498776in}%
\pgfsys@useobject{currentmarker}{}%
\end{pgfscope}%
\end{pgfscope}%
\begin{pgfscope}%
\definecolor{textcolor}{rgb}{0.000000,0.000000,0.000000}%
\pgfsetstrokecolor{textcolor}%
\pgfsetfillcolor{textcolor}%
\pgftext[x=2.289250in,y=0.408498in,,top]{\color{textcolor}{\rmfamily\fontsize{6.500000}{7.800000}\selectfont\catcode`\^=\active\def^{\ifmmode\sp\else\^{}\fi}\catcode`\%=\active\def%{\%}4000}}%
\end{pgfscope}%
\begin{pgfscope}%
\pgfsetbuttcap%
\pgfsetroundjoin%
\definecolor{currentfill}{rgb}{0.000000,0.000000,0.000000}%
\pgfsetfillcolor{currentfill}%
\pgfsetlinewidth{0.803000pt}%
\definecolor{currentstroke}{rgb}{0.000000,0.000000,0.000000}%
\pgfsetstrokecolor{currentstroke}%
\pgfsetdash{}{0pt}%
\pgfsys@defobject{currentmarker}{\pgfqpoint{0.000000in}{-0.048611in}}{\pgfqpoint{0.000000in}{0.000000in}}{%
\pgfpathmoveto{\pgfqpoint{0.000000in}{0.000000in}}%
\pgfpathlineto{\pgfqpoint{0.000000in}{-0.048611in}}%
\pgfusepath{stroke,fill}%
}%
\begin{pgfscope}%
\pgfsys@transformshift{2.516395in}{0.498776in}%
\pgfsys@useobject{currentmarker}{}%
\end{pgfscope}%
\end{pgfscope}%
\begin{pgfscope}%
\definecolor{textcolor}{rgb}{0.000000,0.000000,0.000000}%
\pgfsetstrokecolor{textcolor}%
\pgfsetfillcolor{textcolor}%
\pgftext[x=2.516395in,y=0.408498in,,top]{\color{textcolor}{\rmfamily\fontsize{6.500000}{7.800000}\selectfont\catcode`\^=\active\def^{\ifmmode\sp\else\^{}\fi}\catcode`\%=\active\def%{\%}5000}}%
\end{pgfscope}%
\begin{pgfscope}%
\pgfsetbuttcap%
\pgfsetroundjoin%
\definecolor{currentfill}{rgb}{0.000000,0.000000,0.000000}%
\pgfsetfillcolor{currentfill}%
\pgfsetlinewidth{0.803000pt}%
\definecolor{currentstroke}{rgb}{0.000000,0.000000,0.000000}%
\pgfsetstrokecolor{currentstroke}%
\pgfsetdash{}{0pt}%
\pgfsys@defobject{currentmarker}{\pgfqpoint{0.000000in}{-0.048611in}}{\pgfqpoint{0.000000in}{0.000000in}}{%
\pgfpathmoveto{\pgfqpoint{0.000000in}{0.000000in}}%
\pgfpathlineto{\pgfqpoint{0.000000in}{-0.048611in}}%
\pgfusepath{stroke,fill}%
}%
\begin{pgfscope}%
\pgfsys@transformshift{2.743541in}{0.498776in}%
\pgfsys@useobject{currentmarker}{}%
\end{pgfscope}%
\end{pgfscope}%
\begin{pgfscope}%
\definecolor{textcolor}{rgb}{0.000000,0.000000,0.000000}%
\pgfsetstrokecolor{textcolor}%
\pgfsetfillcolor{textcolor}%
\pgftext[x=2.743541in,y=0.408498in,,top]{\color{textcolor}{\rmfamily\fontsize{6.500000}{7.800000}\selectfont\catcode`\^=\active\def^{\ifmmode\sp\else\^{}\fi}\catcode`\%=\active\def%{\%}6000}}%
\end{pgfscope}%
\begin{pgfscope}%
\definecolor{textcolor}{rgb}{0.000000,0.000000,0.000000}%
\pgfsetstrokecolor{textcolor}%
\pgfsetfillcolor{textcolor}%
\pgftext[x=1.721386in,y=0.278868in,,top]{\color{textcolor}{\rmfamily\fontsize{10.000000}{12.000000}\selectfont\catcode`\^=\active\def^{\ifmmode\sp\else\^{}\fi}\catcode`\%=\active\def%{\%}$\text{Re}$}}%
\end{pgfscope}%
\begin{pgfscope}%
\pgfsetbuttcap%
\pgfsetroundjoin%
\definecolor{currentfill}{rgb}{0.000000,0.000000,0.000000}%
\pgfsetfillcolor{currentfill}%
\pgfsetlinewidth{0.803000pt}%
\definecolor{currentstroke}{rgb}{0.000000,0.000000,0.000000}%
\pgfsetstrokecolor{currentstroke}%
\pgfsetdash{}{0pt}%
\pgfsys@defobject{currentmarker}{\pgfqpoint{-0.048611in}{0.000000in}}{\pgfqpoint{-0.000000in}{0.000000in}}{%
\pgfpathmoveto{\pgfqpoint{-0.000000in}{0.000000in}}%
\pgfpathlineto{\pgfqpoint{-0.048611in}{0.000000in}}%
\pgfusepath{stroke,fill}%
}%
\begin{pgfscope}%
\pgfsys@transformshift{0.597016in}{0.498776in}%
\pgfsys@useobject{currentmarker}{}%
\end{pgfscope}%
\end{pgfscope}%
\begin{pgfscope}%
\definecolor{textcolor}{rgb}{0.000000,0.000000,0.000000}%
\pgfsetstrokecolor{textcolor}%
\pgfsetfillcolor{textcolor}%
\pgftext[x=0.455813in, y=0.469841in, left, base]{\color{textcolor}{\rmfamily\fontsize{6.500000}{7.800000}\selectfont\catcode`\^=\active\def^{\ifmmode\sp\else\^{}\fi}\catcode`\%=\active\def%{\%}$\mathdefault{0}$}}%
\end{pgfscope}%
\begin{pgfscope}%
\pgfsetbuttcap%
\pgfsetroundjoin%
\definecolor{currentfill}{rgb}{0.000000,0.000000,0.000000}%
\pgfsetfillcolor{currentfill}%
\pgfsetlinewidth{0.803000pt}%
\definecolor{currentstroke}{rgb}{0.000000,0.000000,0.000000}%
\pgfsetstrokecolor{currentstroke}%
\pgfsetdash{}{0pt}%
\pgfsys@defobject{currentmarker}{\pgfqpoint{-0.048611in}{0.000000in}}{\pgfqpoint{-0.000000in}{0.000000in}}{%
\pgfpathmoveto{\pgfqpoint{-0.000000in}{0.000000in}}%
\pgfpathlineto{\pgfqpoint{-0.048611in}{0.000000in}}%
\pgfusepath{stroke,fill}%
}%
\begin{pgfscope}%
\pgfsys@transformshift{0.597016in}{0.938774in}%
\pgfsys@useobject{currentmarker}{}%
\end{pgfscope}%
\end{pgfscope}%
\begin{pgfscope}%
\definecolor{textcolor}{rgb}{0.000000,0.000000,0.000000}%
\pgfsetstrokecolor{textcolor}%
\pgfsetfillcolor{textcolor}%
\pgftext[x=0.404888in, y=0.909839in, left, base]{\color{textcolor}{\rmfamily\fontsize{6.500000}{7.800000}\selectfont\catcode`\^=\active\def^{\ifmmode\sp\else\^{}\fi}\catcode`\%=\active\def%{\%}$\mathdefault{10}$}}%
\end{pgfscope}%
\begin{pgfscope}%
\pgfsetbuttcap%
\pgfsetroundjoin%
\definecolor{currentfill}{rgb}{0.000000,0.000000,0.000000}%
\pgfsetfillcolor{currentfill}%
\pgfsetlinewidth{0.803000pt}%
\definecolor{currentstroke}{rgb}{0.000000,0.000000,0.000000}%
\pgfsetstrokecolor{currentstroke}%
\pgfsetdash{}{0pt}%
\pgfsys@defobject{currentmarker}{\pgfqpoint{-0.048611in}{0.000000in}}{\pgfqpoint{-0.000000in}{0.000000in}}{%
\pgfpathmoveto{\pgfqpoint{-0.000000in}{0.000000in}}%
\pgfpathlineto{\pgfqpoint{-0.048611in}{0.000000in}}%
\pgfusepath{stroke,fill}%
}%
\begin{pgfscope}%
\pgfsys@transformshift{0.597016in}{1.378772in}%
\pgfsys@useobject{currentmarker}{}%
\end{pgfscope}%
\end{pgfscope}%
\begin{pgfscope}%
\definecolor{textcolor}{rgb}{0.000000,0.000000,0.000000}%
\pgfsetstrokecolor{textcolor}%
\pgfsetfillcolor{textcolor}%
\pgftext[x=0.404888in, y=1.349837in, left, base]{\color{textcolor}{\rmfamily\fontsize{6.500000}{7.800000}\selectfont\catcode`\^=\active\def^{\ifmmode\sp\else\^{}\fi}\catcode`\%=\active\def%{\%}$\mathdefault{20}$}}%
\end{pgfscope}%
\begin{pgfscope}%
\pgfsetbuttcap%
\pgfsetroundjoin%
\definecolor{currentfill}{rgb}{0.000000,0.000000,0.000000}%
\pgfsetfillcolor{currentfill}%
\pgfsetlinewidth{0.803000pt}%
\definecolor{currentstroke}{rgb}{0.000000,0.000000,0.000000}%
\pgfsetstrokecolor{currentstroke}%
\pgfsetdash{}{0pt}%
\pgfsys@defobject{currentmarker}{\pgfqpoint{-0.048611in}{0.000000in}}{\pgfqpoint{-0.000000in}{0.000000in}}{%
\pgfpathmoveto{\pgfqpoint{-0.000000in}{0.000000in}}%
\pgfpathlineto{\pgfqpoint{-0.048611in}{0.000000in}}%
\pgfusepath{stroke,fill}%
}%
\begin{pgfscope}%
\pgfsys@transformshift{0.597016in}{1.818771in}%
\pgfsys@useobject{currentmarker}{}%
\end{pgfscope}%
\end{pgfscope}%
\begin{pgfscope}%
\definecolor{textcolor}{rgb}{0.000000,0.000000,0.000000}%
\pgfsetstrokecolor{textcolor}%
\pgfsetfillcolor{textcolor}%
\pgftext[x=0.404888in, y=1.789835in, left, base]{\color{textcolor}{\rmfamily\fontsize{6.500000}{7.800000}\selectfont\catcode`\^=\active\def^{\ifmmode\sp\else\^{}\fi}\catcode`\%=\active\def%{\%}$\mathdefault{30}$}}%
\end{pgfscope}%
\begin{pgfscope}%
\definecolor{textcolor}{rgb}{0.000000,0.000000,0.000000}%
\pgfsetstrokecolor{textcolor}%
\pgfsetfillcolor{textcolor}%
\pgftext[x=0.349332in,y=1.158773in,,bottom,rotate=90.000000]{\color{textcolor}{\rmfamily\fontsize{10.000000}{12.000000}\selectfont\catcode`\^=\active\def^{\ifmmode\sp\else\^{}\fi}\catcode`\%=\active\def%{\%}$\delta_{\ell^1}^{(q)} \ [\%]$}}%
\end{pgfscope}%
\begin{pgfscope}%
\pgfsetrectcap%
\pgfsetmiterjoin%
\pgfsetlinewidth{0.803000pt}%
\definecolor{currentstroke}{rgb}{0.000000,0.000000,0.000000}%
\pgfsetstrokecolor{currentstroke}%
\pgfsetdash{}{0pt}%
\pgfpathmoveto{\pgfqpoint{0.597016in}{0.498776in}}%
\pgfpathlineto{\pgfqpoint{0.597016in}{1.818771in}}%
\pgfusepath{stroke}%
\end{pgfscope}%
\begin{pgfscope}%
\pgfsetrectcap%
\pgfsetmiterjoin%
\pgfsetlinewidth{0.803000pt}%
\definecolor{currentstroke}{rgb}{0.000000,0.000000,0.000000}%
\pgfsetstrokecolor{currentstroke}%
\pgfsetdash{}{0pt}%
\pgfpathmoveto{\pgfqpoint{2.845756in}{0.498776in}}%
\pgfpathlineto{\pgfqpoint{2.845756in}{1.818771in}}%
\pgfusepath{stroke}%
\end{pgfscope}%
\begin{pgfscope}%
\pgfsetrectcap%
\pgfsetmiterjoin%
\pgfsetlinewidth{0.803000pt}%
\definecolor{currentstroke}{rgb}{0.000000,0.000000,0.000000}%
\pgfsetstrokecolor{currentstroke}%
\pgfsetdash{}{0pt}%
\pgfpathmoveto{\pgfqpoint{0.597016in}{0.498776in}}%
\pgfpathlineto{\pgfqpoint{2.845756in}{0.498776in}}%
\pgfusepath{stroke}%
\end{pgfscope}%
\begin{pgfscope}%
\pgfsetrectcap%
\pgfsetmiterjoin%
\pgfsetlinewidth{0.803000pt}%
\definecolor{currentstroke}{rgb}{0.000000,0.000000,0.000000}%
\pgfsetstrokecolor{currentstroke}%
\pgfsetdash{}{0pt}%
\pgfpathmoveto{\pgfqpoint{0.597016in}{1.818771in}}%
\pgfpathlineto{\pgfqpoint{2.845756in}{1.818771in}}%
\pgfusepath{stroke}%
\end{pgfscope}%
\begin{pgfscope}%
\pgfsetbuttcap%
\pgfsetmiterjoin%
\definecolor{currentfill}{rgb}{1.000000,1.000000,1.000000}%
\pgfsetfillcolor{currentfill}%
\pgfsetfillopacity{0.800000}%
\pgfsetlinewidth{1.003750pt}%
\definecolor{currentstroke}{rgb}{0.800000,0.800000,0.800000}%
\pgfsetstrokecolor{currentstroke}%
\pgfsetstrokeopacity{0.800000}%
\pgfsetdash{}{0pt}%
\pgfpathmoveto{\pgfqpoint{1.173350in}{1.348761in}}%
\pgfpathlineto{\pgfqpoint{2.764770in}{1.348761in}}%
\pgfpathquadraticcurveto{\pgfqpoint{2.787909in}{1.348761in}}{\pgfqpoint{2.787909in}{1.371900in}}%
\pgfpathlineto{\pgfqpoint{2.787909in}{1.737785in}}%
\pgfpathquadraticcurveto{\pgfqpoint{2.787909in}{1.760923in}}{\pgfqpoint{2.764770in}{1.760923in}}%
\pgfpathlineto{\pgfqpoint{1.173350in}{1.760923in}}%
\pgfpathquadraticcurveto{\pgfqpoint{1.150212in}{1.760923in}}{\pgfqpoint{1.150212in}{1.737785in}}%
\pgfpathlineto{\pgfqpoint{1.150212in}{1.371900in}}%
\pgfpathquadraticcurveto{\pgfqpoint{1.150212in}{1.348761in}}{\pgfqpoint{1.173350in}{1.348761in}}%
\pgfpathlineto{\pgfqpoint{1.173350in}{1.348761in}}%
\pgfpathclose%
\pgfusepath{stroke,fill}%
\end{pgfscope}%
\begin{pgfscope}%
\pgfsetbuttcap%
\pgfsetroundjoin%
\definecolor{currentfill}{rgb}{0.172549,0.482353,0.713725}%
\pgfsetfillcolor{currentfill}%
\pgfsetlinewidth{0.501875pt}%
\definecolor{currentstroke}{rgb}{0.000000,0.000000,0.000000}%
\pgfsetstrokecolor{currentstroke}%
\pgfsetdash{}{0pt}%
\pgfsys@defobject{currentmarker}{\pgfqpoint{-0.026896in}{-0.026896in}}{\pgfqpoint{0.026896in}{0.026896in}}{%
\pgfpathmoveto{\pgfqpoint{-0.026896in}{-0.026896in}}%
\pgfpathlineto{\pgfqpoint{0.026896in}{-0.026896in}}%
\pgfpathlineto{\pgfqpoint{0.026896in}{0.026896in}}%
\pgfpathlineto{\pgfqpoint{-0.026896in}{0.026896in}}%
\pgfpathlineto{\pgfqpoint{-0.026896in}{-0.026896in}}%
\pgfpathclose%
\pgfusepath{stroke,fill}%
}%
\begin{pgfscope}%
\pgfsys@transformshift{1.312184in}{1.641914in}%
\pgfsys@useobject{currentmarker}{}%
\end{pgfscope}%
\end{pgfscope}%
\begin{pgfscope}%
\definecolor{textcolor}{rgb}{0.000000,0.000000,0.000000}%
\pgfsetstrokecolor{textcolor}%
\pgfsetfillcolor{textcolor}%
\pgftext[x=1.520434in,y=1.611544in,left,base]{\color{textcolor}{\rmfamily\fontsize{8.330000}{9.996000}\selectfont\catcode`\^=\active\def^{\ifmmode\sp\else\^{}\fi}\catcode`\%=\active\def%{\%}$p^\text{PINN}_{64}$}}%
\end{pgfscope}%
\begin{pgfscope}%
\pgfsetbuttcap%
\pgfsetroundjoin%
\definecolor{currentfill}{rgb}{0.843137,0.098039,0.109804}%
\pgfsetfillcolor{currentfill}%
\pgfsetlinewidth{0.501875pt}%
\definecolor{currentstroke}{rgb}{0.000000,0.000000,0.000000}%
\pgfsetstrokecolor{currentstroke}%
\pgfsetdash{}{0pt}%
\pgfsys@defobject{currentmarker}{\pgfqpoint{-0.026896in}{-0.026896in}}{\pgfqpoint{0.026896in}{0.026896in}}{%
\pgfpathmoveto{\pgfqpoint{0.000000in}{-0.026896in}}%
\pgfpathcurveto{\pgfqpoint{0.007133in}{-0.026896in}}{\pgfqpoint{0.013974in}{-0.024062in}}{\pgfqpoint{0.019018in}{-0.019018in}}%
\pgfpathcurveto{\pgfqpoint{0.024062in}{-0.013974in}}{\pgfqpoint{0.026896in}{-0.007133in}}{\pgfqpoint{0.026896in}{0.000000in}}%
\pgfpathcurveto{\pgfqpoint{0.026896in}{0.007133in}}{\pgfqpoint{0.024062in}{0.013974in}}{\pgfqpoint{0.019018in}{0.019018in}}%
\pgfpathcurveto{\pgfqpoint{0.013974in}{0.024062in}}{\pgfqpoint{0.007133in}{0.026896in}}{\pgfqpoint{0.000000in}{0.026896in}}%
\pgfpathcurveto{\pgfqpoint{-0.007133in}{0.026896in}}{\pgfqpoint{-0.013974in}{0.024062in}}{\pgfqpoint{-0.019018in}{0.019018in}}%
\pgfpathcurveto{\pgfqpoint{-0.024062in}{0.013974in}}{\pgfqpoint{-0.026896in}{0.007133in}}{\pgfqpoint{-0.026896in}{0.000000in}}%
\pgfpathcurveto{\pgfqpoint{-0.026896in}{-0.007133in}}{\pgfqpoint{-0.024062in}{-0.013974in}}{\pgfqpoint{-0.019018in}{-0.019018in}}%
\pgfpathcurveto{\pgfqpoint{-0.013974in}{-0.024062in}}{\pgfqpoint{-0.007133in}{-0.026896in}}{\pgfqpoint{0.000000in}{-0.026896in}}%
\pgfpathlineto{\pgfqpoint{0.000000in}{-0.026896in}}%
\pgfpathclose%
\pgfusepath{stroke,fill}%
}%
\begin{pgfscope}%
\pgfsys@transformshift{1.312184in}{1.453187in}%
\pgfsys@useobject{currentmarker}{}%
\end{pgfscope}%
\end{pgfscope}%
\begin{pgfscope}%
\definecolor{textcolor}{rgb}{0.000000,0.000000,0.000000}%
\pgfsetstrokecolor{textcolor}%
\pgfsetfillcolor{textcolor}%
\pgftext[x=1.520434in,y=1.422817in,left,base]{\color{textcolor}{\rmfamily\fontsize{8.330000}{9.996000}\selectfont\catcode`\^=\active\def^{\ifmmode\sp\else\^{}\fi}\catcode`\%=\active\def%{\%}$\bm{v}^\text{PINN}_{64}$}}%
\end{pgfscope}%
\begin{pgfscope}%
\pgfsetbuttcap%
\pgfsetroundjoin%
\definecolor{currentfill}{rgb}{0.670588,0.850980,0.913725}%
\pgfsetfillcolor{currentfill}%
\pgfsetlinewidth{0.501875pt}%
\definecolor{currentstroke}{rgb}{0.000000,0.000000,0.000000}%
\pgfsetstrokecolor{currentstroke}%
\pgfsetdash{}{0pt}%
\pgfsys@defobject{currentmarker}{\pgfqpoint{-0.038036in}{-0.038036in}}{\pgfqpoint{0.038036in}{0.038036in}}{%
\pgfpathmoveto{\pgfqpoint{-0.000000in}{-0.038036in}}%
\pgfpathlineto{\pgfqpoint{0.038036in}{0.000000in}}%
\pgfpathlineto{\pgfqpoint{0.000000in}{0.038036in}}%
\pgfpathlineto{\pgfqpoint{-0.038036in}{0.000000in}}%
\pgfpathlineto{\pgfqpoint{-0.000000in}{-0.038036in}}%
\pgfpathclose%
\pgfusepath{stroke,fill}%
}%
\begin{pgfscope}%
\pgfsys@transformshift{2.200449in}{1.641914in}%
\pgfsys@useobject{currentmarker}{}%
\end{pgfscope}%
\end{pgfscope}%
\begin{pgfscope}%
\definecolor{textcolor}{rgb}{0.000000,0.000000,0.000000}%
\pgfsetstrokecolor{textcolor}%
\pgfsetfillcolor{textcolor}%
\pgftext[x=2.408699in,y=1.611544in,left,base]{\color{textcolor}{\rmfamily\fontsize{8.330000}{9.996000}\selectfont\catcode`\^=\active\def^{\ifmmode\sp\else\^{}\fi}\catcode`\%=\active\def%{\%}$p^\text{PINN}_{0}$}}%
\end{pgfscope}%
\begin{pgfscope}%
\pgfsetbuttcap%
\pgfsetroundjoin%
\definecolor{currentfill}{rgb}{0.992157,0.682353,0.380392}%
\pgfsetfillcolor{currentfill}%
\pgfsetlinewidth{0.501875pt}%
\definecolor{currentstroke}{rgb}{0.000000,0.000000,0.000000}%
\pgfsetstrokecolor{currentstroke}%
\pgfsetdash{}{0pt}%
\pgfsys@defobject{currentmarker}{\pgfqpoint{-0.026896in}{-0.026896in}}{\pgfqpoint{0.026896in}{0.026896in}}{%
\pgfpathmoveto{\pgfqpoint{0.000000in}{0.026896in}}%
\pgfpathlineto{\pgfqpoint{-0.026896in}{-0.026896in}}%
\pgfpathlineto{\pgfqpoint{0.026896in}{-0.026896in}}%
\pgfpathlineto{\pgfqpoint{0.000000in}{0.026896in}}%
\pgfpathclose%
\pgfusepath{stroke,fill}%
}%
\begin{pgfscope}%
\pgfsys@transformshift{2.200449in}{1.453187in}%
\pgfsys@useobject{currentmarker}{}%
\end{pgfscope}%
\end{pgfscope}%
\begin{pgfscope}%
\definecolor{textcolor}{rgb}{0.000000,0.000000,0.000000}%
\pgfsetstrokecolor{textcolor}%
\pgfsetfillcolor{textcolor}%
\pgftext[x=2.408699in,y=1.422817in,left,base]{\color{textcolor}{\rmfamily\fontsize{8.330000}{9.996000}\selectfont\catcode`\^=\active\def^{\ifmmode\sp\else\^{}\fi}\catcode`\%=\active\def%{\%}$\bm{v}^\text{PINN}_{0}$}}%
\end{pgfscope}%
\end{pgfpicture}%
\makeatother%
\endgroup%

%% file: figures/section1/n_points_per_Re_velonly.pgf
%% Creator: Matplotlib, PGF backend
%%
%% To include the figure in your LaTeX document, write
%%   \input{<filename>.pgf}
%%
%% Make sure the required packages are loaded in your preamble
%%   \usepackage{pgf}
%%
%% Also ensure that all the required font packages are loaded; for instance,
%% the lmodern package is sometimes necessary when using math font.
%%   \usepackage{lmodern}
%%
%% Figures using additional raster images can only be included by \input if
%% they are in the same directory as the main LaTeX file. For loading figures
%% from other directories you can use the `import` package
%% and then include the figures with
%%   \import{<path to file>}{<filename>.pgf}
%%
%% Matplotlib used the following preamble
%%   \def\mathdefault#1{#1}
%%   \everymath=\expandafter{\the\everymath\displaystyle}
%%   \usepackage{amsmath}\usepackage{bm}
%%   \makeatletter\@ifpackageloaded{underscore}{}{\usepackage[strings]{underscore}}\makeatother
%%
\begingroup%
\makeatletter%
\begin{pgfpicture}%
\pgfpathrectangle{\pgfpointorigin}{\pgfqpoint{3.000000in}{2.000000in}}%
\pgfusepath{use as bounding box, clip}%
\begin{pgfscope}%
\pgfsetbuttcap%
\pgfsetmiterjoin%
\definecolor{currentfill}{rgb}{1.000000,1.000000,1.000000}%
\pgfsetfillcolor{currentfill}%
\pgfsetlinewidth{0.000000pt}%
\definecolor{currentstroke}{rgb}{1.000000,1.000000,1.000000}%
\pgfsetstrokecolor{currentstroke}%
\pgfsetdash{}{0pt}%
\pgfpathmoveto{\pgfqpoint{0.000000in}{0.000000in}}%
\pgfpathlineto{\pgfqpoint{3.000000in}{0.000000in}}%
\pgfpathlineto{\pgfqpoint{3.000000in}{2.000000in}}%
\pgfpathlineto{\pgfqpoint{0.000000in}{2.000000in}}%
\pgfpathlineto{\pgfqpoint{0.000000in}{0.000000in}}%
\pgfpathclose%
\pgfusepath{fill}%
\end{pgfscope}%
\begin{pgfscope}%
\pgfsetbuttcap%
\pgfsetmiterjoin%
\definecolor{currentfill}{rgb}{1.000000,1.000000,1.000000}%
\pgfsetfillcolor{currentfill}%
\pgfsetlinewidth{0.000000pt}%
\definecolor{currentstroke}{rgb}{0.000000,0.000000,0.000000}%
\pgfsetstrokecolor{currentstroke}%
\pgfsetstrokeopacity{0.000000}%
\pgfsetdash{}{0pt}%
\pgfpathmoveto{\pgfqpoint{0.548324in}{0.521284in}}%
\pgfpathlineto{\pgfqpoint{2.802597in}{0.521284in}}%
\pgfpathlineto{\pgfqpoint{2.802597in}{1.850000in}}%
\pgfpathlineto{\pgfqpoint{0.548324in}{1.850000in}}%
\pgfpathlineto{\pgfqpoint{0.548324in}{0.521284in}}%
\pgfpathclose%
\pgfusepath{fill}%
\end{pgfscope}%
\begin{pgfscope}%
\pgfpathrectangle{\pgfqpoint{0.548324in}{0.521284in}}{\pgfqpoint{2.254273in}{1.328716in}}%
\pgfusepath{clip}%
\pgfsetbuttcap%
\pgfsetroundjoin%
\definecolor{currentfill}{rgb}{0.172549,0.482353,0.713725}%
\pgfsetfillcolor{currentfill}%
\pgfsetlinewidth{0.501875pt}%
\definecolor{currentstroke}{rgb}{0.000000,0.000000,0.000000}%
\pgfsetstrokecolor{currentstroke}%
\pgfsetdash{}{0pt}%
\pgfsys@defobject{currentmarker}{\pgfqpoint{-0.031056in}{-0.031056in}}{\pgfqpoint{0.031056in}{0.031056in}}{%
\pgfpathmoveto{\pgfqpoint{0.000000in}{0.031056in}}%
\pgfpathlineto{\pgfqpoint{-0.031056in}{-0.031056in}}%
\pgfpathlineto{\pgfqpoint{0.031056in}{-0.031056in}}%
\pgfpathlineto{\pgfqpoint{0.000000in}{0.031056in}}%
\pgfpathclose%
\pgfusepath{stroke,fill}%
}%
\begin{pgfscope}%
\pgfsys@transformshift{2.700130in}{0.907366in}%
\pgfsys@useobject{currentmarker}{}%
\end{pgfscope}%
\begin{pgfscope}%
\pgfsys@transformshift{2.151807in}{0.923857in}%
\pgfsys@useobject{currentmarker}{}%
\end{pgfscope}%
\begin{pgfscope}%
\pgfsys@transformshift{1.818248in}{0.903243in}%
\pgfsys@useobject{currentmarker}{}%
\end{pgfscope}%
\begin{pgfscope}%
\pgfsys@transformshift{1.484689in}{0.826975in}%
\pgfsys@useobject{currentmarker}{}%
\end{pgfscope}%
\begin{pgfscope}%
\pgfsys@transformshift{0.984350in}{1.336117in}%
\pgfsys@useobject{currentmarker}{}%
\end{pgfscope}%
\begin{pgfscope}%
\pgfsys@transformshift{0.650791in}{1.789604in}%
\pgfsys@useobject{currentmarker}{}%
\end{pgfscope}%
\end{pgfscope}%
\begin{pgfscope}%
\pgfpathrectangle{\pgfqpoint{0.548324in}{0.521284in}}{\pgfqpoint{2.254273in}{1.328716in}}%
\pgfusepath{clip}%
\pgfsetbuttcap%
\pgfsetroundjoin%
\definecolor{currentfill}{rgb}{0.172549,0.482353,0.713725}%
\pgfsetfillcolor{currentfill}%
\pgfsetlinewidth{0.501875pt}%
\definecolor{currentstroke}{rgb}{0.000000,0.000000,0.000000}%
\pgfsetstrokecolor{currentstroke}%
\pgfsetdash{}{0pt}%
\pgfsys@defobject{currentmarker}{\pgfqpoint{-0.043921in}{-0.043921in}}{\pgfqpoint{0.043921in}{0.043921in}}{%
\pgfpathmoveto{\pgfqpoint{-0.000000in}{-0.043921in}}%
\pgfpathlineto{\pgfqpoint{0.043921in}{0.000000in}}%
\pgfpathlineto{\pgfqpoint{0.000000in}{0.043921in}}%
\pgfpathlineto{\pgfqpoint{-0.043921in}{0.000000in}}%
\pgfpathlineto{\pgfqpoint{-0.000000in}{-0.043921in}}%
\pgfpathclose%
\pgfusepath{stroke,fill}%
}%
\begin{pgfscope}%
\pgfsys@transformshift{2.700130in}{0.627029in}%
\pgfsys@useobject{currentmarker}{}%
\end{pgfscope}%
\begin{pgfscope}%
\pgfsys@transformshift{2.151807in}{0.581680in}%
\pgfsys@useobject{currentmarker}{}%
\end{pgfscope}%
\begin{pgfscope}%
\pgfsys@transformshift{1.818248in}{0.614661in}%
\pgfsys@useobject{currentmarker}{}%
\end{pgfscope}%
\begin{pgfscope}%
\pgfsys@transformshift{1.484689in}{0.672378in}%
\pgfsys@useobject{currentmarker}{}%
\end{pgfscope}%
\begin{pgfscope}%
\pgfsys@transformshift{0.984350in}{1.503083in}%
\pgfsys@useobject{currentmarker}{}%
\end{pgfscope}%
\begin{pgfscope}%
\pgfsys@transformshift{0.650791in}{1.286646in}%
\pgfsys@useobject{currentmarker}{}%
\end{pgfscope}%
\end{pgfscope}%
\begin{pgfscope}%
\pgfpathrectangle{\pgfqpoint{0.548324in}{0.521284in}}{\pgfqpoint{2.254273in}{1.328716in}}%
\pgfusepath{clip}%
\pgfsetbuttcap%
\pgfsetroundjoin%
\definecolor{currentfill}{rgb}{0.992157,0.682353,0.380392}%
\pgfsetfillcolor{currentfill}%
\pgfsetlinewidth{0.501875pt}%
\definecolor{currentstroke}{rgb}{0.000000,0.000000,0.000000}%
\pgfsetstrokecolor{currentstroke}%
\pgfsetdash{}{0pt}%
\pgfsys@defobject{currentmarker}{\pgfqpoint{-0.031056in}{-0.031056in}}{\pgfqpoint{0.031056in}{0.031056in}}{%
\pgfpathmoveto{\pgfqpoint{0.000000in}{-0.031056in}}%
\pgfpathcurveto{\pgfqpoint{0.008236in}{-0.031056in}}{\pgfqpoint{0.016136in}{-0.027784in}}{\pgfqpoint{0.021960in}{-0.021960in}}%
\pgfpathcurveto{\pgfqpoint{0.027784in}{-0.016136in}}{\pgfqpoint{0.031056in}{-0.008236in}}{\pgfqpoint{0.031056in}{0.000000in}}%
\pgfpathcurveto{\pgfqpoint{0.031056in}{0.008236in}}{\pgfqpoint{0.027784in}{0.016136in}}{\pgfqpoint{0.021960in}{0.021960in}}%
\pgfpathcurveto{\pgfqpoint{0.016136in}{0.027784in}}{\pgfqpoint{0.008236in}{0.031056in}}{\pgfqpoint{0.000000in}{0.031056in}}%
\pgfpathcurveto{\pgfqpoint{-0.008236in}{0.031056in}}{\pgfqpoint{-0.016136in}{0.027784in}}{\pgfqpoint{-0.021960in}{0.021960in}}%
\pgfpathcurveto{\pgfqpoint{-0.027784in}{0.016136in}}{\pgfqpoint{-0.031056in}{0.008236in}}{\pgfqpoint{-0.031056in}{0.000000in}}%
\pgfpathcurveto{\pgfqpoint{-0.031056in}{-0.008236in}}{\pgfqpoint{-0.027784in}{-0.016136in}}{\pgfqpoint{-0.021960in}{-0.021960in}}%
\pgfpathcurveto{\pgfqpoint{-0.016136in}{-0.027784in}}{\pgfqpoint{-0.008236in}{-0.031056in}}{\pgfqpoint{0.000000in}{-0.031056in}}%
\pgfpathlineto{\pgfqpoint{0.000000in}{-0.031056in}}%
\pgfpathclose%
\pgfusepath{stroke,fill}%
}%
\begin{pgfscope}%
\pgfsys@transformshift{2.700130in}{0.851711in}%
\pgfsys@useobject{currentmarker}{}%
\end{pgfscope}%
\begin{pgfscope}%
\pgfsys@transformshift{2.151807in}{0.874385in}%
\pgfsys@useobject{currentmarker}{}%
\end{pgfscope}%
\begin{pgfscope}%
\pgfsys@transformshift{1.818248in}{0.833159in}%
\pgfsys@useobject{currentmarker}{}%
\end{pgfscope}%
\begin{pgfscope}%
\pgfsys@transformshift{1.484689in}{0.872324in}%
\pgfsys@useobject{currentmarker}{}%
\end{pgfscope}%
\begin{pgfscope}%
\pgfsys@transformshift{0.984350in}{1.222745in}%
\pgfsys@useobject{currentmarker}{}%
\end{pgfscope}%
\begin{pgfscope}%
\pgfsys@transformshift{0.650791in}{1.323749in}%
\pgfsys@useobject{currentmarker}{}%
\end{pgfscope}%
\end{pgfscope}%
\begin{pgfscope}%
\pgfpathrectangle{\pgfqpoint{0.548324in}{0.521284in}}{\pgfqpoint{2.254273in}{1.328716in}}%
\pgfusepath{clip}%
\pgfsetbuttcap%
\pgfsetroundjoin%
\definecolor{currentfill}{rgb}{0.992157,0.682353,0.380392}%
\pgfsetfillcolor{currentfill}%
\pgfsetlinewidth{0.501875pt}%
\definecolor{currentstroke}{rgb}{0.000000,0.000000,0.000000}%
\pgfsetstrokecolor{currentstroke}%
\pgfsetdash{}{0pt}%
\pgfsys@defobject{currentmarker}{\pgfqpoint{-0.031056in}{-0.031056in}}{\pgfqpoint{0.031056in}{0.031056in}}{%
\pgfpathmoveto{\pgfqpoint{-0.031056in}{-0.031056in}}%
\pgfpathlineto{\pgfqpoint{0.031056in}{-0.031056in}}%
\pgfpathlineto{\pgfqpoint{0.031056in}{0.031056in}}%
\pgfpathlineto{\pgfqpoint{-0.031056in}{0.031056in}}%
\pgfpathlineto{\pgfqpoint{-0.031056in}{-0.031056in}}%
\pgfpathclose%
\pgfusepath{stroke,fill}%
}%
\begin{pgfscope}%
\pgfsys@transformshift{2.700130in}{0.831098in}%
\pgfsys@useobject{currentmarker}{}%
\end{pgfscope}%
\begin{pgfscope}%
\pgfsys@transformshift{2.151807in}{0.779565in}%
\pgfsys@useobject{currentmarker}{}%
\end{pgfscope}%
\begin{pgfscope}%
\pgfsys@transformshift{1.818248in}{0.758952in}%
\pgfsys@useobject{currentmarker}{}%
\end{pgfscope}%
\begin{pgfscope}%
\pgfsys@transformshift{1.484689in}{0.839343in}%
\pgfsys@useobject{currentmarker}{}%
\end{pgfscope}%
\begin{pgfscope}%
\pgfsys@transformshift{0.984350in}{1.181519in}%
\pgfsys@useobject{currentmarker}{}%
\end{pgfscope}%
\begin{pgfscope}%
\pgfsys@transformshift{0.650791in}{1.230991in}%
\pgfsys@useobject{currentmarker}{}%
\end{pgfscope}%
\end{pgfscope}%
\begin{pgfscope}%
\pgfsetbuttcap%
\pgfsetroundjoin%
\definecolor{currentfill}{rgb}{0.000000,0.000000,0.000000}%
\pgfsetfillcolor{currentfill}%
\pgfsetlinewidth{0.803000pt}%
\definecolor{currentstroke}{rgb}{0.000000,0.000000,0.000000}%
\pgfsetstrokecolor{currentstroke}%
\pgfsetdash{}{0pt}%
\pgfsys@defobject{currentmarker}{\pgfqpoint{0.000000in}{-0.048611in}}{\pgfqpoint{0.000000in}{0.000000in}}{%
\pgfpathmoveto{\pgfqpoint{0.000000in}{0.000000in}}%
\pgfpathlineto{\pgfqpoint{0.000000in}{-0.048611in}}%
\pgfusepath{stroke,fill}%
}%
\begin{pgfscope}%
\pgfsys@transformshift{0.650791in}{0.521284in}%
\pgfsys@useobject{currentmarker}{}%
\end{pgfscope}%
\end{pgfscope}%
\begin{pgfscope}%
\definecolor{textcolor}{rgb}{0.000000,0.000000,0.000000}%
\pgfsetstrokecolor{textcolor}%
\pgfsetfillcolor{textcolor}%
\pgftext[x=0.650791in,y=0.431006in,,top]{\color{textcolor}{\rmfamily\fontsize{8.330000}{9.996000}\selectfont\catcode`\^=\active\def^{\ifmmode\sp\else\^{}\fi}\catcode`\%=\active\def%{\%}2}}%
\end{pgfscope}%
\begin{pgfscope}%
\pgfsetbuttcap%
\pgfsetroundjoin%
\definecolor{currentfill}{rgb}{0.000000,0.000000,0.000000}%
\pgfsetfillcolor{currentfill}%
\pgfsetlinewidth{0.803000pt}%
\definecolor{currentstroke}{rgb}{0.000000,0.000000,0.000000}%
\pgfsetstrokecolor{currentstroke}%
\pgfsetdash{}{0pt}%
\pgfsys@defobject{currentmarker}{\pgfqpoint{0.000000in}{-0.048611in}}{\pgfqpoint{0.000000in}{0.000000in}}{%
\pgfpathmoveto{\pgfqpoint{0.000000in}{0.000000in}}%
\pgfpathlineto{\pgfqpoint{0.000000in}{-0.048611in}}%
\pgfusepath{stroke,fill}%
}%
\begin{pgfscope}%
\pgfsys@transformshift{0.984350in}{0.521284in}%
\pgfsys@useobject{currentmarker}{}%
\end{pgfscope}%
\end{pgfscope}%
\begin{pgfscope}%
\definecolor{textcolor}{rgb}{0.000000,0.000000,0.000000}%
\pgfsetstrokecolor{textcolor}%
\pgfsetfillcolor{textcolor}%
\pgftext[x=0.984350in,y=0.431006in,,top]{\color{textcolor}{\rmfamily\fontsize{8.330000}{9.996000}\selectfont\catcode`\^=\active\def^{\ifmmode\sp\else\^{}\fi}\catcode`\%=\active\def%{\%}8}}%
\end{pgfscope}%
\begin{pgfscope}%
\pgfsetbuttcap%
\pgfsetroundjoin%
\definecolor{currentfill}{rgb}{0.000000,0.000000,0.000000}%
\pgfsetfillcolor{currentfill}%
\pgfsetlinewidth{0.803000pt}%
\definecolor{currentstroke}{rgb}{0.000000,0.000000,0.000000}%
\pgfsetstrokecolor{currentstroke}%
\pgfsetdash{}{0pt}%
\pgfsys@defobject{currentmarker}{\pgfqpoint{0.000000in}{-0.048611in}}{\pgfqpoint{0.000000in}{0.000000in}}{%
\pgfpathmoveto{\pgfqpoint{0.000000in}{0.000000in}}%
\pgfpathlineto{\pgfqpoint{0.000000in}{-0.048611in}}%
\pgfusepath{stroke,fill}%
}%
\begin{pgfscope}%
\pgfsys@transformshift{1.484689in}{0.521284in}%
\pgfsys@useobject{currentmarker}{}%
\end{pgfscope}%
\end{pgfscope}%
\begin{pgfscope}%
\definecolor{textcolor}{rgb}{0.000000,0.000000,0.000000}%
\pgfsetstrokecolor{textcolor}%
\pgfsetfillcolor{textcolor}%
\pgftext[x=1.484689in,y=0.431006in,,top]{\color{textcolor}{\rmfamily\fontsize{8.330000}{9.996000}\selectfont\catcode`\^=\active\def^{\ifmmode\sp\else\^{}\fi}\catcode`\%=\active\def%{\%}64}}%
\end{pgfscope}%
\begin{pgfscope}%
\pgfsetbuttcap%
\pgfsetroundjoin%
\definecolor{currentfill}{rgb}{0.000000,0.000000,0.000000}%
\pgfsetfillcolor{currentfill}%
\pgfsetlinewidth{0.803000pt}%
\definecolor{currentstroke}{rgb}{0.000000,0.000000,0.000000}%
\pgfsetstrokecolor{currentstroke}%
\pgfsetdash{}{0pt}%
\pgfsys@defobject{currentmarker}{\pgfqpoint{0.000000in}{-0.048611in}}{\pgfqpoint{0.000000in}{0.000000in}}{%
\pgfpathmoveto{\pgfqpoint{0.000000in}{0.000000in}}%
\pgfpathlineto{\pgfqpoint{0.000000in}{-0.048611in}}%
\pgfusepath{stroke,fill}%
}%
\begin{pgfscope}%
\pgfsys@transformshift{1.818248in}{0.521284in}%
\pgfsys@useobject{currentmarker}{}%
\end{pgfscope}%
\end{pgfscope}%
\begin{pgfscope}%
\definecolor{textcolor}{rgb}{0.000000,0.000000,0.000000}%
\pgfsetstrokecolor{textcolor}%
\pgfsetfillcolor{textcolor}%
\pgftext[x=1.818248in,y=0.431006in,,top]{\color{textcolor}{\rmfamily\fontsize{8.330000}{9.996000}\selectfont\catcode`\^=\active\def^{\ifmmode\sp\else\^{}\fi}\catcode`\%=\active\def%{\%}256}}%
\end{pgfscope}%
\begin{pgfscope}%
\pgfsetbuttcap%
\pgfsetroundjoin%
\definecolor{currentfill}{rgb}{0.000000,0.000000,0.000000}%
\pgfsetfillcolor{currentfill}%
\pgfsetlinewidth{0.803000pt}%
\definecolor{currentstroke}{rgb}{0.000000,0.000000,0.000000}%
\pgfsetstrokecolor{currentstroke}%
\pgfsetdash{}{0pt}%
\pgfsys@defobject{currentmarker}{\pgfqpoint{0.000000in}{-0.048611in}}{\pgfqpoint{0.000000in}{0.000000in}}{%
\pgfpathmoveto{\pgfqpoint{0.000000in}{0.000000in}}%
\pgfpathlineto{\pgfqpoint{0.000000in}{-0.048611in}}%
\pgfusepath{stroke,fill}%
}%
\begin{pgfscope}%
\pgfsys@transformshift{2.151807in}{0.521284in}%
\pgfsys@useobject{currentmarker}{}%
\end{pgfscope}%
\end{pgfscope}%
\begin{pgfscope}%
\definecolor{textcolor}{rgb}{0.000000,0.000000,0.000000}%
\pgfsetstrokecolor{textcolor}%
\pgfsetfillcolor{textcolor}%
\pgftext[x=2.151807in,y=0.431006in,,top]{\color{textcolor}{\rmfamily\fontsize{8.330000}{9.996000}\selectfont\catcode`\^=\active\def^{\ifmmode\sp\else\^{}\fi}\catcode`\%=\active\def%{\%}1024}}%
\end{pgfscope}%
\begin{pgfscope}%
\pgfsetbuttcap%
\pgfsetroundjoin%
\definecolor{currentfill}{rgb}{0.000000,0.000000,0.000000}%
\pgfsetfillcolor{currentfill}%
\pgfsetlinewidth{0.803000pt}%
\definecolor{currentstroke}{rgb}{0.000000,0.000000,0.000000}%
\pgfsetstrokecolor{currentstroke}%
\pgfsetdash{}{0pt}%
\pgfsys@defobject{currentmarker}{\pgfqpoint{0.000000in}{-0.048611in}}{\pgfqpoint{0.000000in}{0.000000in}}{%
\pgfpathmoveto{\pgfqpoint{0.000000in}{0.000000in}}%
\pgfpathlineto{\pgfqpoint{0.000000in}{-0.048611in}}%
\pgfusepath{stroke,fill}%
}%
\begin{pgfscope}%
\pgfsys@transformshift{2.700130in}{0.521284in}%
\pgfsys@useobject{currentmarker}{}%
\end{pgfscope}%
\end{pgfscope}%
\begin{pgfscope}%
\definecolor{textcolor}{rgb}{0.000000,0.000000,0.000000}%
\pgfsetstrokecolor{textcolor}%
\pgfsetfillcolor{textcolor}%
\pgftext[x=2.700130in,y=0.431006in,,top]{\color{textcolor}{\rmfamily\fontsize{8.330000}{9.996000}\selectfont\catcode`\^=\active\def^{\ifmmode\sp\else\^{}\fi}\catcode`\%=\active\def%{\%}10000}}%
\end{pgfscope}%
\begin{pgfscope}%
\definecolor{textcolor}{rgb}{0.000000,0.000000,0.000000}%
\pgfsetstrokecolor{textcolor}%
\pgfsetfillcolor{textcolor}%
\pgftext[x=1.675460in,y=0.276685in,,top]{\color{textcolor}{\rmfamily\fontsize{10.000000}{12.000000}\selectfont\catcode`\^=\active\def^{\ifmmode\sp\else\^{}\fi}\catcode`\%=\active\def%{\%}Number of Labels per $\mathrm{Re}_\text{train}$}}%
\end{pgfscope}%
\begin{pgfscope}%
\pgfsetbuttcap%
\pgfsetroundjoin%
\definecolor{currentfill}{rgb}{0.000000,0.000000,0.000000}%
\pgfsetfillcolor{currentfill}%
\pgfsetlinewidth{0.803000pt}%
\definecolor{currentstroke}{rgb}{0.000000,0.000000,0.000000}%
\pgfsetstrokecolor{currentstroke}%
\pgfsetdash{}{0pt}%
\pgfsys@defobject{currentmarker}{\pgfqpoint{-0.048611in}{0.000000in}}{\pgfqpoint{-0.000000in}{0.000000in}}{%
\pgfpathmoveto{\pgfqpoint{-0.000000in}{0.000000in}}%
\pgfpathlineto{\pgfqpoint{-0.048611in}{0.000000in}}%
\pgfusepath{stroke,fill}%
}%
\begin{pgfscope}%
\pgfsys@transformshift{0.548324in}{0.526025in}%
\pgfsys@useobject{currentmarker}{}%
\end{pgfscope}%
\end{pgfscope}%
\begin{pgfscope}%
\definecolor{textcolor}{rgb}{0.000000,0.000000,0.000000}%
\pgfsetstrokecolor{textcolor}%
\pgfsetfillcolor{textcolor}%
\pgftext[x=0.399017in, y=0.487445in, left, base]{\color{textcolor}{\rmfamily\fontsize{8.330000}{9.996000}\selectfont\catcode`\^=\active\def^{\ifmmode\sp\else\^{}\fi}\catcode`\%=\active\def%{\%}$\mathdefault{2}$}}%
\end{pgfscope}%
\begin{pgfscope}%
\pgfsetbuttcap%
\pgfsetroundjoin%
\definecolor{currentfill}{rgb}{0.000000,0.000000,0.000000}%
\pgfsetfillcolor{currentfill}%
\pgfsetlinewidth{0.803000pt}%
\definecolor{currentstroke}{rgb}{0.000000,0.000000,0.000000}%
\pgfsetstrokecolor{currentstroke}%
\pgfsetdash{}{0pt}%
\pgfsys@defobject{currentmarker}{\pgfqpoint{-0.048611in}{0.000000in}}{\pgfqpoint{-0.000000in}{0.000000in}}{%
\pgfpathmoveto{\pgfqpoint{-0.000000in}{0.000000in}}%
\pgfpathlineto{\pgfqpoint{-0.048611in}{0.000000in}}%
\pgfusepath{stroke,fill}%
}%
\begin{pgfscope}%
\pgfsys@transformshift{0.548324in}{0.938286in}%
\pgfsys@useobject{currentmarker}{}%
\end{pgfscope}%
\end{pgfscope}%
\begin{pgfscope}%
\definecolor{textcolor}{rgb}{0.000000,0.000000,0.000000}%
\pgfsetstrokecolor{textcolor}%
\pgfsetfillcolor{textcolor}%
\pgftext[x=0.399017in, y=0.899705in, left, base]{\color{textcolor}{\rmfamily\fontsize{8.330000}{9.996000}\selectfont\catcode`\^=\active\def^{\ifmmode\sp\else\^{}\fi}\catcode`\%=\active\def%{\%}$\mathdefault{4}$}}%
\end{pgfscope}%
\begin{pgfscope}%
\pgfsetbuttcap%
\pgfsetroundjoin%
\definecolor{currentfill}{rgb}{0.000000,0.000000,0.000000}%
\pgfsetfillcolor{currentfill}%
\pgfsetlinewidth{0.803000pt}%
\definecolor{currentstroke}{rgb}{0.000000,0.000000,0.000000}%
\pgfsetstrokecolor{currentstroke}%
\pgfsetdash{}{0pt}%
\pgfsys@defobject{currentmarker}{\pgfqpoint{-0.048611in}{0.000000in}}{\pgfqpoint{-0.000000in}{0.000000in}}{%
\pgfpathmoveto{\pgfqpoint{-0.000000in}{0.000000in}}%
\pgfpathlineto{\pgfqpoint{-0.048611in}{0.000000in}}%
\pgfusepath{stroke,fill}%
}%
\begin{pgfscope}%
\pgfsys@transformshift{0.548324in}{1.350546in}%
\pgfsys@useobject{currentmarker}{}%
\end{pgfscope}%
\end{pgfscope}%
\begin{pgfscope}%
\definecolor{textcolor}{rgb}{0.000000,0.000000,0.000000}%
\pgfsetstrokecolor{textcolor}%
\pgfsetfillcolor{textcolor}%
\pgftext[x=0.399017in, y=1.311966in, left, base]{\color{textcolor}{\rmfamily\fontsize{8.330000}{9.996000}\selectfont\catcode`\^=\active\def^{\ifmmode\sp\else\^{}\fi}\catcode`\%=\active\def%{\%}$\mathdefault{6}$}}%
\end{pgfscope}%
\begin{pgfscope}%
\pgfsetbuttcap%
\pgfsetroundjoin%
\definecolor{currentfill}{rgb}{0.000000,0.000000,0.000000}%
\pgfsetfillcolor{currentfill}%
\pgfsetlinewidth{0.803000pt}%
\definecolor{currentstroke}{rgb}{0.000000,0.000000,0.000000}%
\pgfsetstrokecolor{currentstroke}%
\pgfsetdash{}{0pt}%
\pgfsys@defobject{currentmarker}{\pgfqpoint{-0.048611in}{0.000000in}}{\pgfqpoint{-0.000000in}{0.000000in}}{%
\pgfpathmoveto{\pgfqpoint{-0.000000in}{0.000000in}}%
\pgfpathlineto{\pgfqpoint{-0.048611in}{0.000000in}}%
\pgfusepath{stroke,fill}%
}%
\begin{pgfscope}%
\pgfsys@transformshift{0.548324in}{1.762807in}%
\pgfsys@useobject{currentmarker}{}%
\end{pgfscope}%
\end{pgfscope}%
\begin{pgfscope}%
\definecolor{textcolor}{rgb}{0.000000,0.000000,0.000000}%
\pgfsetstrokecolor{textcolor}%
\pgfsetfillcolor{textcolor}%
\pgftext[x=0.399017in, y=1.724227in, left, base]{\color{textcolor}{\rmfamily\fontsize{8.330000}{9.996000}\selectfont\catcode`\^=\active\def^{\ifmmode\sp\else\^{}\fi}\catcode`\%=\active\def%{\%}$\mathdefault{8}$}}%
\end{pgfscope}%
\begin{pgfscope}%
\definecolor{textcolor}{rgb}{0.000000,0.000000,0.000000}%
\pgfsetstrokecolor{textcolor}%
\pgfsetfillcolor{textcolor}%
\pgftext[x=0.343462in,y=1.185642in,,bottom,rotate=90.000000]{\color{textcolor}{\rmfamily\fontsize{10.000000}{12.000000}\selectfont\catcode`\^=\active\def^{\ifmmode\sp\else\^{}\fi}\catcode`\%=\active\def%{\%}$\delta_{\ell^1}^{(q)} \ [\%]$}}%
\end{pgfscope}%
\begin{pgfscope}%
\pgfsetrectcap%
\pgfsetmiterjoin%
\pgfsetlinewidth{0.803000pt}%
\definecolor{currentstroke}{rgb}{0.000000,0.000000,0.000000}%
\pgfsetstrokecolor{currentstroke}%
\pgfsetdash{}{0pt}%
\pgfpathmoveto{\pgfqpoint{0.548324in}{0.521284in}}%
\pgfpathlineto{\pgfqpoint{0.548324in}{1.850000in}}%
\pgfusepath{stroke}%
\end{pgfscope}%
\begin{pgfscope}%
\pgfsetrectcap%
\pgfsetmiterjoin%
\pgfsetlinewidth{0.803000pt}%
\definecolor{currentstroke}{rgb}{0.000000,0.000000,0.000000}%
\pgfsetstrokecolor{currentstroke}%
\pgfsetdash{}{0pt}%
\pgfpathmoveto{\pgfqpoint{2.802597in}{0.521284in}}%
\pgfpathlineto{\pgfqpoint{2.802597in}{1.850000in}}%
\pgfusepath{stroke}%
\end{pgfscope}%
\begin{pgfscope}%
\pgfsetrectcap%
\pgfsetmiterjoin%
\pgfsetlinewidth{0.803000pt}%
\definecolor{currentstroke}{rgb}{0.000000,0.000000,0.000000}%
\pgfsetstrokecolor{currentstroke}%
\pgfsetdash{}{0pt}%
\pgfpathmoveto{\pgfqpoint{0.548324in}{0.521284in}}%
\pgfpathlineto{\pgfqpoint{2.802597in}{0.521284in}}%
\pgfusepath{stroke}%
\end{pgfscope}%
\begin{pgfscope}%
\pgfsetrectcap%
\pgfsetmiterjoin%
\pgfsetlinewidth{0.803000pt}%
\definecolor{currentstroke}{rgb}{0.000000,0.000000,0.000000}%
\pgfsetstrokecolor{currentstroke}%
\pgfsetdash{}{0pt}%
\pgfpathmoveto{\pgfqpoint{0.548324in}{1.850000in}}%
\pgfpathlineto{\pgfqpoint{2.802597in}{1.850000in}}%
\pgfusepath{stroke}%
\end{pgfscope}%
\begin{pgfscope}%
\pgfsetbuttcap%
\pgfsetmiterjoin%
\definecolor{currentfill}{rgb}{1.000000,1.000000,1.000000}%
\pgfsetfillcolor{currentfill}%
\pgfsetfillopacity{0.800000}%
\pgfsetlinewidth{1.003750pt}%
\definecolor{currentstroke}{rgb}{0.800000,0.800000,0.800000}%
\pgfsetstrokecolor{currentstroke}%
\pgfsetstrokeopacity{0.800000}%
\pgfsetdash{}{0pt}%
\pgfpathmoveto{\pgfqpoint{1.893842in}{1.070438in}}%
\pgfpathlineto{\pgfqpoint{2.721611in}{1.070438in}}%
\pgfpathquadraticcurveto{\pgfqpoint{2.744750in}{1.070438in}}{\pgfqpoint{2.744750in}{1.093577in}}%
\pgfpathlineto{\pgfqpoint{2.744750in}{1.769014in}}%
\pgfpathquadraticcurveto{\pgfqpoint{2.744750in}{1.792153in}}{\pgfqpoint{2.721611in}{1.792153in}}%
\pgfpathlineto{\pgfqpoint{1.893842in}{1.792153in}}%
\pgfpathquadraticcurveto{\pgfqpoint{1.870703in}{1.792153in}}{\pgfqpoint{1.870703in}{1.769014in}}%
\pgfpathlineto{\pgfqpoint{1.870703in}{1.093577in}}%
\pgfpathquadraticcurveto{\pgfqpoint{1.870703in}{1.070438in}}{\pgfqpoint{1.893842in}{1.070438in}}%
\pgfpathlineto{\pgfqpoint{1.893842in}{1.070438in}}%
\pgfpathclose%
\pgfusepath{stroke,fill}%
\end{pgfscope}%
\begin{pgfscope}%
\pgfsetbuttcap%
\pgfsetroundjoin%
\definecolor{currentfill}{rgb}{0.172549,0.482353,0.713725}%
\pgfsetfillcolor{currentfill}%
\pgfsetlinewidth{0.501875pt}%
\definecolor{currentstroke}{rgb}{0.000000,0.000000,0.000000}%
\pgfsetstrokecolor{currentstroke}%
\pgfsetdash{}{0pt}%
\pgfsys@defobject{currentmarker}{\pgfqpoint{-0.031056in}{-0.031056in}}{\pgfqpoint{0.031056in}{0.031056in}}{%
\pgfpathmoveto{\pgfqpoint{0.000000in}{0.031056in}}%
\pgfpathlineto{\pgfqpoint{-0.031056in}{-0.031056in}}%
\pgfpathlineto{\pgfqpoint{0.031056in}{-0.031056in}}%
\pgfpathlineto{\pgfqpoint{0.000000in}{0.031056in}}%
\pgfpathclose%
\pgfusepath{stroke,fill}%
}%
\begin{pgfscope}%
\pgfsys@transformshift{2.032675in}{1.694748in}%
\pgfsys@useobject{currentmarker}{}%
\end{pgfscope}%
\end{pgfscope}%
\begin{pgfscope}%
\definecolor{textcolor}{rgb}{0.000000,0.000000,0.000000}%
\pgfsetstrokecolor{textcolor}%
\pgfsetfillcolor{textcolor}%
\pgftext[x=2.240925in,y=1.664378in,left,base]{\color{textcolor}{\rmfamily\fontsize{8.330000}{9.996000}\selectfont\catcode`\^=\active\def^{\ifmmode\sp\else\^{}\fi}\catcode`\%=\active\def%{\%}$\bm{v}^{\mathrm{PINN}}$}}%
\end{pgfscope}%
\begin{pgfscope}%
\pgfsetbuttcap%
\pgfsetroundjoin%
\definecolor{currentfill}{rgb}{0.172549,0.482353,0.713725}%
\pgfsetfillcolor{currentfill}%
\pgfsetlinewidth{0.501875pt}%
\definecolor{currentstroke}{rgb}{0.000000,0.000000,0.000000}%
\pgfsetstrokecolor{currentstroke}%
\pgfsetdash{}{0pt}%
\pgfsys@defobject{currentmarker}{\pgfqpoint{-0.043921in}{-0.043921in}}{\pgfqpoint{0.043921in}{0.043921in}}{%
\pgfpathmoveto{\pgfqpoint{-0.000000in}{-0.043921in}}%
\pgfpathlineto{\pgfqpoint{0.043921in}{0.000000in}}%
\pgfpathlineto{\pgfqpoint{0.000000in}{0.043921in}}%
\pgfpathlineto{\pgfqpoint{-0.043921in}{0.000000in}}%
\pgfpathlineto{\pgfqpoint{-0.000000in}{-0.043921in}}%
\pgfpathclose%
\pgfusepath{stroke,fill}%
}%
\begin{pgfscope}%
\pgfsys@transformshift{2.032675in}{1.512194in}%
\pgfsys@useobject{currentmarker}{}%
\end{pgfscope}%
\end{pgfscope}%
\begin{pgfscope}%
\definecolor{textcolor}{rgb}{0.000000,0.000000,0.000000}%
\pgfsetstrokecolor{textcolor}%
\pgfsetfillcolor{textcolor}%
\pgftext[x=2.240925in,y=1.481824in,left,base]{\color{textcolor}{\rmfamily\fontsize{8.330000}{9.996000}\selectfont\catcode`\^=\active\def^{\ifmmode\sp\else\^{}\fi}\catcode`\%=\active\def%{\%}$p^{\mathrm{PINN}}$}}%
\end{pgfscope}%
\begin{pgfscope}%
\pgfsetbuttcap%
\pgfsetroundjoin%
\definecolor{currentfill}{rgb}{0.992157,0.682353,0.380392}%
\pgfsetfillcolor{currentfill}%
\pgfsetlinewidth{0.501875pt}%
\definecolor{currentstroke}{rgb}{0.000000,0.000000,0.000000}%
\pgfsetstrokecolor{currentstroke}%
\pgfsetdash{}{0pt}%
\pgfsys@defobject{currentmarker}{\pgfqpoint{-0.031056in}{-0.031056in}}{\pgfqpoint{0.031056in}{0.031056in}}{%
\pgfpathmoveto{\pgfqpoint{0.000000in}{-0.031056in}}%
\pgfpathcurveto{\pgfqpoint{0.008236in}{-0.031056in}}{\pgfqpoint{0.016136in}{-0.027784in}}{\pgfqpoint{0.021960in}{-0.021960in}}%
\pgfpathcurveto{\pgfqpoint{0.027784in}{-0.016136in}}{\pgfqpoint{0.031056in}{-0.008236in}}{\pgfqpoint{0.031056in}{0.000000in}}%
\pgfpathcurveto{\pgfqpoint{0.031056in}{0.008236in}}{\pgfqpoint{0.027784in}{0.016136in}}{\pgfqpoint{0.021960in}{0.021960in}}%
\pgfpathcurveto{\pgfqpoint{0.016136in}{0.027784in}}{\pgfqpoint{0.008236in}{0.031056in}}{\pgfqpoint{0.000000in}{0.031056in}}%
\pgfpathcurveto{\pgfqpoint{-0.008236in}{0.031056in}}{\pgfqpoint{-0.016136in}{0.027784in}}{\pgfqpoint{-0.021960in}{0.021960in}}%
\pgfpathcurveto{\pgfqpoint{-0.027784in}{0.016136in}}{\pgfqpoint{-0.031056in}{0.008236in}}{\pgfqpoint{-0.031056in}{0.000000in}}%
\pgfpathcurveto{\pgfqpoint{-0.031056in}{-0.008236in}}{\pgfqpoint{-0.027784in}{-0.016136in}}{\pgfqpoint{-0.021960in}{-0.021960in}}%
\pgfpathcurveto{\pgfqpoint{-0.016136in}{-0.027784in}}{\pgfqpoint{-0.008236in}{-0.031056in}}{\pgfqpoint{0.000000in}{-0.031056in}}%
\pgfpathlineto{\pgfqpoint{0.000000in}{-0.031056in}}%
\pgfpathclose%
\pgfusepath{stroke,fill}%
}%
\begin{pgfscope}%
\pgfsys@transformshift{2.032675in}{1.351245in}%
\pgfsys@useobject{currentmarker}{}%
\end{pgfscope}%
\end{pgfscope}%
\begin{pgfscope}%
\definecolor{textcolor}{rgb}{0.000000,0.000000,0.000000}%
\pgfsetstrokecolor{textcolor}%
\pgfsetfillcolor{textcolor}%
\pgftext[x=2.240925in,y=1.320875in,left,base]{\color{textcolor}{\rmfamily\fontsize{8.330000}{9.996000}\selectfont\catcode`\^=\active\def^{\ifmmode\sp\else\^{}\fi}\catcode`\%=\active\def%{\%}$\bm{v}^{\mathrm{PINN}_\mathrm{vel}}$}}%
\end{pgfscope}%
\begin{pgfscope}%
\pgfsetbuttcap%
\pgfsetroundjoin%
\definecolor{currentfill}{rgb}{0.992157,0.682353,0.380392}%
\pgfsetfillcolor{currentfill}%
\pgfsetlinewidth{0.501875pt}%
\definecolor{currentstroke}{rgb}{0.000000,0.000000,0.000000}%
\pgfsetstrokecolor{currentstroke}%
\pgfsetdash{}{0pt}%
\pgfsys@defobject{currentmarker}{\pgfqpoint{-0.031056in}{-0.031056in}}{\pgfqpoint{0.031056in}{0.031056in}}{%
\pgfpathmoveto{\pgfqpoint{-0.031056in}{-0.031056in}}%
\pgfpathlineto{\pgfqpoint{0.031056in}{-0.031056in}}%
\pgfpathlineto{\pgfqpoint{0.031056in}{0.031056in}}%
\pgfpathlineto{\pgfqpoint{-0.031056in}{0.031056in}}%
\pgfpathlineto{\pgfqpoint{-0.031056in}{-0.031056in}}%
\pgfpathclose%
\pgfusepath{stroke,fill}%
}%
\begin{pgfscope}%
\pgfsys@transformshift{2.032675in}{1.168690in}%
\pgfsys@useobject{currentmarker}{}%
\end{pgfscope}%
\end{pgfscope}%
\begin{pgfscope}%
\definecolor{textcolor}{rgb}{0.000000,0.000000,0.000000}%
\pgfsetstrokecolor{textcolor}%
\pgfsetfillcolor{textcolor}%
\pgftext[x=2.240925in,y=1.138321in,left,base]{\color{textcolor}{\rmfamily\fontsize{8.330000}{9.996000}\selectfont\catcode`\^=\active\def^{\ifmmode\sp\else\^{}\fi}\catcode`\%=\active\def%{\%}$p^{\mathrm{PINN}_\mathrm{vel}}$}}%
\end{pgfscope}%
\end{pgfpicture}%
\makeatother%
\endgroup%

%% file: figures/section2/dataBC/10000p/re_1000/err_vmag.pgf
%% Creator: Matplotlib, PGF backend
%%
%% To include the figure in your LaTeX document, write
%%   \input{<filename>.pgf}
%%
%% Make sure the required packages are loaded in your preamble
%%   \usepackage{pgf}
%%
%% Also ensure that all the required font packages are loaded; for instance,
%% the lmodern package is sometimes necessary when using math font.
%%   \usepackage{lmodern}
%%
%% Figures using additional raster images can only be included by \input if
%% they are in the same directory as the main LaTeX file. For loading figures
%% from other directories you can use the `import` package
%% and then include the figures with
%%   \import{<path to file>}{<filename>.pgf}
%%
%% Matplotlib used the following preamble
%%   \def\mathdefault#1{#1}
%%   \everymath=\expandafter{\the\everymath\displaystyle}
%%   \usepackage{amsmath}\usepackage{bm}
%%   \makeatletter\@ifpackageloaded{underscore}{}{\usepackage[strings]{underscore}}\makeatother
%%
\begingroup%
\makeatletter%
\begin{pgfpicture}%
\pgfpathrectangle{\pgfpointorigin}{\pgfqpoint{2.500000in}{2.500000in}}%
\pgfusepath{use as bounding box, clip}%
\begin{pgfscope}%
\pgfsetbuttcap%
\pgfsetmiterjoin%
\definecolor{currentfill}{rgb}{1.000000,1.000000,1.000000}%
\pgfsetfillcolor{currentfill}%
\pgfsetlinewidth{0.000000pt}%
\definecolor{currentstroke}{rgb}{1.000000,1.000000,1.000000}%
\pgfsetstrokecolor{currentstroke}%
\pgfsetdash{}{0pt}%
\pgfpathmoveto{\pgfqpoint{0.000000in}{0.000000in}}%
\pgfpathlineto{\pgfqpoint{2.500000in}{0.000000in}}%
\pgfpathlineto{\pgfqpoint{2.500000in}{2.500000in}}%
\pgfpathlineto{\pgfqpoint{0.000000in}{2.500000in}}%
\pgfpathlineto{\pgfqpoint{0.000000in}{0.000000in}}%
\pgfpathclose%
\pgfusepath{fill}%
\end{pgfscope}%
\begin{pgfscope}%
\pgfsetbuttcap%
\pgfsetmiterjoin%
\definecolor{currentfill}{rgb}{1.000000,1.000000,1.000000}%
\pgfsetfillcolor{currentfill}%
\pgfsetlinewidth{0.000000pt}%
\definecolor{currentstroke}{rgb}{0.000000,0.000000,0.000000}%
\pgfsetstrokecolor{currentstroke}%
\pgfsetstrokeopacity{0.000000}%
\pgfsetdash{}{0pt}%
\pgfpathmoveto{\pgfqpoint{0.584719in}{0.386658in}}%
\pgfpathlineto{\pgfqpoint{2.114646in}{0.386658in}}%
\pgfpathlineto{\pgfqpoint{2.114646in}{1.916585in}}%
\pgfpathlineto{\pgfqpoint{0.584719in}{1.916585in}}%
\pgfpathlineto{\pgfqpoint{0.584719in}{0.386658in}}%
\pgfpathclose%
\pgfusepath{fill}%
\end{pgfscope}%
\begin{pgfscope}%
\pgfsys@transformshift{0.653750in}{0.455000in}%
\pgftext[left,bottom]{\includegraphics[interpolate=true,width=1.391250in,height=1.392500in]{figures/./section2/dataBC/10000p/re_1000//err_vmag-img0.png}}%
\end{pgfscope}%
\begin{pgfscope}%
\pgfsetbuttcap%
\pgfsetroundjoin%
\definecolor{currentfill}{rgb}{0.000000,0.000000,0.000000}%
\pgfsetfillcolor{currentfill}%
\pgfsetlinewidth{0.803000pt}%
\definecolor{currentstroke}{rgb}{0.000000,0.000000,0.000000}%
\pgfsetstrokecolor{currentstroke}%
\pgfsetdash{}{0pt}%
\pgfsys@defobject{currentmarker}{\pgfqpoint{0.000000in}{-0.048611in}}{\pgfqpoint{0.000000in}{0.000000in}}{%
\pgfpathmoveto{\pgfqpoint{0.000000in}{0.000000in}}%
\pgfpathlineto{\pgfqpoint{0.000000in}{-0.048611in}}%
\pgfusepath{stroke,fill}%
}%
\begin{pgfscope}%
\pgfsys@transformshift{0.654261in}{0.386658in}%
\pgfsys@useobject{currentmarker}{}%
\end{pgfscope}%
\end{pgfscope}%
\begin{pgfscope}%
\definecolor{textcolor}{rgb}{0.000000,0.000000,0.000000}%
\pgfsetstrokecolor{textcolor}%
\pgfsetfillcolor{textcolor}%
\pgftext[x=0.654261in,y=0.296381in,,top]{\color{textcolor}{\rmfamily\fontsize{8.330000}{9.996000}\selectfont\catcode`\^=\active\def^{\ifmmode\sp\else\^{}\fi}\catcode`\%=\active\def%{\%}$\mathdefault{\ensuremath{-}0.1}$}}%
\end{pgfscope}%
\begin{pgfscope}%
\pgfsetbuttcap%
\pgfsetroundjoin%
\definecolor{currentfill}{rgb}{0.000000,0.000000,0.000000}%
\pgfsetfillcolor{currentfill}%
\pgfsetlinewidth{0.803000pt}%
\definecolor{currentstroke}{rgb}{0.000000,0.000000,0.000000}%
\pgfsetstrokecolor{currentstroke}%
\pgfsetdash{}{0pt}%
\pgfsys@defobject{currentmarker}{\pgfqpoint{0.000000in}{-0.048611in}}{\pgfqpoint{0.000000in}{0.000000in}}{%
\pgfpathmoveto{\pgfqpoint{0.000000in}{0.000000in}}%
\pgfpathlineto{\pgfqpoint{0.000000in}{-0.048611in}}%
\pgfusepath{stroke,fill}%
}%
\begin{pgfscope}%
\pgfsys@transformshift{1.349682in}{0.386658in}%
\pgfsys@useobject{currentmarker}{}%
\end{pgfscope}%
\end{pgfscope}%
\begin{pgfscope}%
\definecolor{textcolor}{rgb}{0.000000,0.000000,0.000000}%
\pgfsetstrokecolor{textcolor}%
\pgfsetfillcolor{textcolor}%
\pgftext[x=1.349682in,y=0.296381in,,top]{\color{textcolor}{\rmfamily\fontsize{8.330000}{9.996000}\selectfont\catcode`\^=\active\def^{\ifmmode\sp\else\^{}\fi}\catcode`\%=\active\def%{\%}$\mathdefault{0.0}$}}%
\end{pgfscope}%
\begin{pgfscope}%
\pgfsetbuttcap%
\pgfsetroundjoin%
\definecolor{currentfill}{rgb}{0.000000,0.000000,0.000000}%
\pgfsetfillcolor{currentfill}%
\pgfsetlinewidth{0.803000pt}%
\definecolor{currentstroke}{rgb}{0.000000,0.000000,0.000000}%
\pgfsetstrokecolor{currentstroke}%
\pgfsetdash{}{0pt}%
\pgfsys@defobject{currentmarker}{\pgfqpoint{0.000000in}{-0.048611in}}{\pgfqpoint{0.000000in}{0.000000in}}{%
\pgfpathmoveto{\pgfqpoint{0.000000in}{0.000000in}}%
\pgfpathlineto{\pgfqpoint{0.000000in}{-0.048611in}}%
\pgfusepath{stroke,fill}%
}%
\begin{pgfscope}%
\pgfsys@transformshift{2.045104in}{0.386658in}%
\pgfsys@useobject{currentmarker}{}%
\end{pgfscope}%
\end{pgfscope}%
\begin{pgfscope}%
\definecolor{textcolor}{rgb}{0.000000,0.000000,0.000000}%
\pgfsetstrokecolor{textcolor}%
\pgfsetfillcolor{textcolor}%
\pgftext[x=2.045104in,y=0.296381in,,top]{\color{textcolor}{\rmfamily\fontsize{8.330000}{9.996000}\selectfont\catcode`\^=\active\def^{\ifmmode\sp\else\^{}\fi}\catcode`\%=\active\def%{\%}$\mathdefault{0.1}$}}%
\end{pgfscope}%
\begin{pgfscope}%
\definecolor{textcolor}{rgb}{0.000000,0.000000,0.000000}%
\pgfsetstrokecolor{textcolor}%
\pgfsetfillcolor{textcolor}%
\pgftext[x=1.349682in,y=0.142060in,,top]{\color{textcolor}{\rmfamily\fontsize{10.000000}{12.000000}\selectfont\catcode`\^=\active\def^{\ifmmode\sp\else\^{}\fi}\catcode`\%=\active\def%{\%}$x\;[\text{m}]$}}%
\end{pgfscope}%
\begin{pgfscope}%
\pgfsetbuttcap%
\pgfsetroundjoin%
\definecolor{currentfill}{rgb}{0.000000,0.000000,0.000000}%
\pgfsetfillcolor{currentfill}%
\pgfsetlinewidth{0.803000pt}%
\definecolor{currentstroke}{rgb}{0.000000,0.000000,0.000000}%
\pgfsetstrokecolor{currentstroke}%
\pgfsetdash{}{0pt}%
\pgfsys@defobject{currentmarker}{\pgfqpoint{-0.048611in}{0.000000in}}{\pgfqpoint{-0.000000in}{0.000000in}}{%
\pgfpathmoveto{\pgfqpoint{-0.000000in}{0.000000in}}%
\pgfpathlineto{\pgfqpoint{-0.048611in}{0.000000in}}%
\pgfusepath{stroke,fill}%
}%
\begin{pgfscope}%
\pgfsys@transformshift{0.584719in}{0.456201in}%
\pgfsys@useobject{currentmarker}{}%
\end{pgfscope}%
\end{pgfscope}%
\begin{pgfscope}%
\definecolor{textcolor}{rgb}{0.000000,0.000000,0.000000}%
\pgfsetstrokecolor{textcolor}%
\pgfsetfillcolor{textcolor}%
\pgftext[x=0.251768in, y=0.417620in, left, base]{\color{textcolor}{\rmfamily\fontsize{8.330000}{9.996000}\selectfont\catcode`\^=\active\def^{\ifmmode\sp\else\^{}\fi}\catcode`\%=\active\def%{\%}$\mathdefault{\ensuremath{-}0.1}$}}%
\end{pgfscope}%
\begin{pgfscope}%
\pgfsetbuttcap%
\pgfsetroundjoin%
\definecolor{currentfill}{rgb}{0.000000,0.000000,0.000000}%
\pgfsetfillcolor{currentfill}%
\pgfsetlinewidth{0.803000pt}%
\definecolor{currentstroke}{rgb}{0.000000,0.000000,0.000000}%
\pgfsetstrokecolor{currentstroke}%
\pgfsetdash{}{0pt}%
\pgfsys@defobject{currentmarker}{\pgfqpoint{-0.048611in}{0.000000in}}{\pgfqpoint{-0.000000in}{0.000000in}}{%
\pgfpathmoveto{\pgfqpoint{-0.000000in}{0.000000in}}%
\pgfpathlineto{\pgfqpoint{-0.048611in}{0.000000in}}%
\pgfusepath{stroke,fill}%
}%
\begin{pgfscope}%
\pgfsys@transformshift{0.584719in}{1.151622in}%
\pgfsys@useobject{currentmarker}{}%
\end{pgfscope}%
\end{pgfscope}%
\begin{pgfscope}%
\definecolor{textcolor}{rgb}{0.000000,0.000000,0.000000}%
\pgfsetstrokecolor{textcolor}%
\pgfsetfillcolor{textcolor}%
\pgftext[x=0.343590in, y=1.113042in, left, base]{\color{textcolor}{\rmfamily\fontsize{8.330000}{9.996000}\selectfont\catcode`\^=\active\def^{\ifmmode\sp\else\^{}\fi}\catcode`\%=\active\def%{\%}$\mathdefault{0.0}$}}%
\end{pgfscope}%
\begin{pgfscope}%
\pgfsetbuttcap%
\pgfsetroundjoin%
\definecolor{currentfill}{rgb}{0.000000,0.000000,0.000000}%
\pgfsetfillcolor{currentfill}%
\pgfsetlinewidth{0.803000pt}%
\definecolor{currentstroke}{rgb}{0.000000,0.000000,0.000000}%
\pgfsetstrokecolor{currentstroke}%
\pgfsetdash{}{0pt}%
\pgfsys@defobject{currentmarker}{\pgfqpoint{-0.048611in}{0.000000in}}{\pgfqpoint{-0.000000in}{0.000000in}}{%
\pgfpathmoveto{\pgfqpoint{-0.000000in}{0.000000in}}%
\pgfpathlineto{\pgfqpoint{-0.048611in}{0.000000in}}%
\pgfusepath{stroke,fill}%
}%
\begin{pgfscope}%
\pgfsys@transformshift{0.584719in}{1.847043in}%
\pgfsys@useobject{currentmarker}{}%
\end{pgfscope}%
\end{pgfscope}%
\begin{pgfscope}%
\definecolor{textcolor}{rgb}{0.000000,0.000000,0.000000}%
\pgfsetstrokecolor{textcolor}%
\pgfsetfillcolor{textcolor}%
\pgftext[x=0.343590in, y=1.808463in, left, base]{\color{textcolor}{\rmfamily\fontsize{8.330000}{9.996000}\selectfont\catcode`\^=\active\def^{\ifmmode\sp\else\^{}\fi}\catcode`\%=\active\def%{\%}$\mathdefault{0.1}$}}%
\end{pgfscope}%
\begin{pgfscope}%
\definecolor{textcolor}{rgb}{0.000000,0.000000,0.000000}%
\pgfsetstrokecolor{textcolor}%
\pgfsetfillcolor{textcolor}%
\pgftext[x=0.196212in,y=1.151622in,,bottom,rotate=90.000000]{\color{textcolor}{\rmfamily\fontsize{10.000000}{12.000000}\selectfont\catcode`\^=\active\def^{\ifmmode\sp\else\^{}\fi}\catcode`\%=\active\def%{\%}$y\;[\text{m}]$}}%
\end{pgfscope}%
\begin{pgfscope}%
\pgfpathrectangle{\pgfqpoint{0.584719in}{0.386658in}}{\pgfqpoint{1.529927in}{1.529927in}}%
\pgfusepath{clip}%
\pgfsetbuttcap%
\pgfsetmiterjoin%
\definecolor{currentfill}{rgb}{1.000000,1.000000,1.000000}%
\pgfsetfillcolor{currentfill}%
\pgfsetlinewidth{1.505625pt}%
\definecolor{currentstroke}{rgb}{1.000000,1.000000,1.000000}%
\pgfsetstrokecolor{currentstroke}%
\pgfsetdash{}{0pt}%
\pgfpathmoveto{\pgfqpoint{1.822421in}{1.648947in}}%
\pgfpathlineto{\pgfqpoint{1.815715in}{1.642241in}}%
\pgfpathlineto{\pgfqpoint{1.809010in}{1.635536in}}%
\pgfpathlineto{\pgfqpoint{1.802304in}{1.628830in}}%
\pgfpathlineto{\pgfqpoint{1.795599in}{1.622125in}}%
\pgfpathlineto{\pgfqpoint{1.788893in}{1.615419in}}%
\pgfpathlineto{\pgfqpoint{1.782187in}{1.608714in}}%
\pgfpathlineto{\pgfqpoint{1.775482in}{1.602008in}}%
\pgfpathlineto{\pgfqpoint{1.768776in}{1.595303in}}%
\pgfpathlineto{\pgfqpoint{1.762071in}{1.588597in}}%
\pgfpathlineto{\pgfqpoint{1.755365in}{1.581892in}}%
\pgfpathlineto{\pgfqpoint{1.761512in}{1.575745in}}%
\pgfpathlineto{\pgfqpoint{1.767659in}{1.569598in}}%
\pgfpathlineto{\pgfqpoint{1.773806in}{1.563452in}}%
\pgfpathlineto{\pgfqpoint{1.779952in}{1.557305in}}%
\pgfpathlineto{\pgfqpoint{1.786658in}{1.564010in}}%
\pgfpathlineto{\pgfqpoint{1.793363in}{1.570716in}}%
\pgfpathlineto{\pgfqpoint{1.800069in}{1.577421in}}%
\pgfpathlineto{\pgfqpoint{1.806774in}{1.584127in}}%
\pgfpathlineto{\pgfqpoint{1.813480in}{1.590833in}}%
\pgfpathlineto{\pgfqpoint{1.820185in}{1.597538in}}%
\pgfpathlineto{\pgfqpoint{1.826891in}{1.604244in}}%
\pgfpathlineto{\pgfqpoint{1.833596in}{1.610949in}}%
\pgfpathlineto{\pgfqpoint{1.840302in}{1.617655in}}%
\pgfpathlineto{\pgfqpoint{1.847007in}{1.624360in}}%
\pgfpathlineto{\pgfqpoint{1.822421in}{1.648947in}}%
\pgfpathclose%
\pgfusepath{stroke,fill}%
\end{pgfscope}%
\begin{pgfscope}%
\pgfpathrectangle{\pgfqpoint{0.584719in}{0.386658in}}{\pgfqpoint{1.529927in}{1.529927in}}%
\pgfusepath{clip}%
\pgfsetbuttcap%
\pgfsetmiterjoin%
\definecolor{currentfill}{rgb}{1.000000,1.000000,1.000000}%
\pgfsetfillcolor{currentfill}%
\pgfsetlinewidth{1.505625pt}%
\definecolor{currentstroke}{rgb}{1.000000,1.000000,1.000000}%
\pgfsetstrokecolor{currentstroke}%
\pgfsetdash{}{0pt}%
\pgfpathmoveto{\pgfqpoint{1.847007in}{0.678884in}}%
\pgfpathlineto{\pgfqpoint{1.840302in}{0.685589in}}%
\pgfpathlineto{\pgfqpoint{1.833596in}{0.692295in}}%
\pgfpathlineto{\pgfqpoint{1.826891in}{0.699000in}}%
\pgfpathlineto{\pgfqpoint{1.820185in}{0.705706in}}%
\pgfpathlineto{\pgfqpoint{1.813480in}{0.712411in}}%
\pgfpathlineto{\pgfqpoint{1.806774in}{0.719117in}}%
\pgfpathlineto{\pgfqpoint{1.800069in}{0.725822in}}%
\pgfpathlineto{\pgfqpoint{1.793363in}{0.732528in}}%
\pgfpathlineto{\pgfqpoint{1.786658in}{0.739233in}}%
\pgfpathlineto{\pgfqpoint{1.779952in}{0.745939in}}%
\pgfpathlineto{\pgfqpoint{1.773683in}{0.739669in}}%
\pgfpathlineto{\pgfqpoint{1.767413in}{0.733399in}}%
\pgfpathlineto{\pgfqpoint{1.761143in}{0.727130in}}%
\pgfpathlineto{\pgfqpoint{1.754874in}{0.720860in}}%
\pgfpathlineto{\pgfqpoint{1.755365in}{0.721352in}}%
\pgfpathlineto{\pgfqpoint{1.762071in}{0.714646in}}%
\pgfpathlineto{\pgfqpoint{1.768776in}{0.707941in}}%
\pgfpathlineto{\pgfqpoint{1.775482in}{0.701235in}}%
\pgfpathlineto{\pgfqpoint{1.782187in}{0.694530in}}%
\pgfpathlineto{\pgfqpoint{1.788893in}{0.687824in}}%
\pgfpathlineto{\pgfqpoint{1.795599in}{0.681119in}}%
\pgfpathlineto{\pgfqpoint{1.802304in}{0.674413in}}%
\pgfpathlineto{\pgfqpoint{1.809010in}{0.667708in}}%
\pgfpathlineto{\pgfqpoint{1.815715in}{0.661002in}}%
\pgfpathlineto{\pgfqpoint{1.822421in}{0.654297in}}%
\pgfpathlineto{\pgfqpoint{1.847007in}{0.678884in}}%
\pgfpathclose%
\pgfusepath{stroke,fill}%
\end{pgfscope}%
\begin{pgfscope}%
\pgfpathrectangle{\pgfqpoint{0.584719in}{0.386658in}}{\pgfqpoint{1.529927in}{1.529927in}}%
\pgfusepath{clip}%
\pgfsetbuttcap%
\pgfsetmiterjoin%
\definecolor{currentfill}{rgb}{1.000000,1.000000,1.000000}%
\pgfsetfillcolor{currentfill}%
\pgfsetlinewidth{1.505625pt}%
\definecolor{currentstroke}{rgb}{1.000000,1.000000,1.000000}%
\pgfsetstrokecolor{currentstroke}%
\pgfsetdash{}{0pt}%
\pgfpathmoveto{\pgfqpoint{0.876944in}{0.654297in}}%
\pgfpathlineto{\pgfqpoint{0.883650in}{0.661002in}}%
\pgfpathlineto{\pgfqpoint{0.890355in}{0.667708in}}%
\pgfpathlineto{\pgfqpoint{0.897061in}{0.674413in}}%
\pgfpathlineto{\pgfqpoint{0.903766in}{0.681119in}}%
\pgfpathlineto{\pgfqpoint{0.910472in}{0.687824in}}%
\pgfpathlineto{\pgfqpoint{0.917177in}{0.694530in}}%
\pgfpathlineto{\pgfqpoint{0.923883in}{0.701235in}}%
\pgfpathlineto{\pgfqpoint{0.930588in}{0.707941in}}%
\pgfpathlineto{\pgfqpoint{0.937294in}{0.714646in}}%
\pgfpathlineto{\pgfqpoint{0.943999in}{0.721352in}}%
\pgfpathlineto{\pgfqpoint{0.937730in}{0.727622in}}%
\pgfpathlineto{\pgfqpoint{0.931460in}{0.733891in}}%
\pgfpathlineto{\pgfqpoint{0.925190in}{0.740161in}}%
\pgfpathlineto{\pgfqpoint{0.918921in}{0.746431in}}%
\pgfpathlineto{\pgfqpoint{0.919412in}{0.745939in}}%
\pgfpathlineto{\pgfqpoint{0.912707in}{0.739233in}}%
\pgfpathlineto{\pgfqpoint{0.906001in}{0.732528in}}%
\pgfpathlineto{\pgfqpoint{0.899296in}{0.725822in}}%
\pgfpathlineto{\pgfqpoint{0.892590in}{0.719117in}}%
\pgfpathlineto{\pgfqpoint{0.885885in}{0.712411in}}%
\pgfpathlineto{\pgfqpoint{0.879179in}{0.705706in}}%
\pgfpathlineto{\pgfqpoint{0.872474in}{0.699000in}}%
\pgfpathlineto{\pgfqpoint{0.865768in}{0.692295in}}%
\pgfpathlineto{\pgfqpoint{0.859063in}{0.685589in}}%
\pgfpathlineto{\pgfqpoint{0.852357in}{0.678884in}}%
\pgfpathlineto{\pgfqpoint{0.876944in}{0.654297in}}%
\pgfpathclose%
\pgfusepath{stroke,fill}%
\end{pgfscope}%
\begin{pgfscope}%
\pgfpathrectangle{\pgfqpoint{0.584719in}{0.386658in}}{\pgfqpoint{1.529927in}{1.529927in}}%
\pgfusepath{clip}%
\pgfsetbuttcap%
\pgfsetmiterjoin%
\definecolor{currentfill}{rgb}{1.000000,1.000000,1.000000}%
\pgfsetfillcolor{currentfill}%
\pgfsetlinewidth{1.505625pt}%
\definecolor{currentstroke}{rgb}{1.000000,1.000000,1.000000}%
\pgfsetstrokecolor{currentstroke}%
\pgfsetdash{}{0pt}%
\pgfpathmoveto{\pgfqpoint{0.852357in}{1.624360in}}%
\pgfpathlineto{\pgfqpoint{0.859063in}{1.617655in}}%
\pgfpathlineto{\pgfqpoint{0.865768in}{1.610949in}}%
\pgfpathlineto{\pgfqpoint{0.872474in}{1.604244in}}%
\pgfpathlineto{\pgfqpoint{0.879179in}{1.597538in}}%
\pgfpathlineto{\pgfqpoint{0.885885in}{1.590833in}}%
\pgfpathlineto{\pgfqpoint{0.892590in}{1.584127in}}%
\pgfpathlineto{\pgfqpoint{0.899296in}{1.577421in}}%
\pgfpathlineto{\pgfqpoint{0.906001in}{1.570716in}}%
\pgfpathlineto{\pgfqpoint{0.912707in}{1.564010in}}%
\pgfpathlineto{\pgfqpoint{0.919412in}{1.557305in}}%
\pgfpathlineto{\pgfqpoint{0.925559in}{1.563452in}}%
\pgfpathlineto{\pgfqpoint{0.931706in}{1.569598in}}%
\pgfpathlineto{\pgfqpoint{0.937853in}{1.575745in}}%
\pgfpathlineto{\pgfqpoint{0.943999in}{1.581892in}}%
\pgfpathlineto{\pgfqpoint{0.937294in}{1.588597in}}%
\pgfpathlineto{\pgfqpoint{0.930588in}{1.595303in}}%
\pgfpathlineto{\pgfqpoint{0.923883in}{1.602008in}}%
\pgfpathlineto{\pgfqpoint{0.917177in}{1.608714in}}%
\pgfpathlineto{\pgfqpoint{0.910472in}{1.615419in}}%
\pgfpathlineto{\pgfqpoint{0.903766in}{1.622125in}}%
\pgfpathlineto{\pgfqpoint{0.897061in}{1.628830in}}%
\pgfpathlineto{\pgfqpoint{0.890355in}{1.635536in}}%
\pgfpathlineto{\pgfqpoint{0.883650in}{1.642241in}}%
\pgfpathlineto{\pgfqpoint{0.876944in}{1.648947in}}%
\pgfpathlineto{\pgfqpoint{0.852357in}{1.624360in}}%
\pgfpathclose%
\pgfusepath{stroke,fill}%
\end{pgfscope}%
\begin{pgfscope}%
\pgfpathrectangle{\pgfqpoint{0.584719in}{0.386658in}}{\pgfqpoint{1.529927in}{1.529927in}}%
\pgfusepath{clip}%
\pgfsetrectcap%
\pgfsetroundjoin%
\pgfsetlinewidth{1.505625pt}%
\definecolor{currentstroke}{rgb}{0.000000,0.000000,0.000000}%
\pgfsetstrokecolor{currentstroke}%
\pgfsetdash{}{0pt}%
\pgfpathmoveto{\pgfqpoint{0.852357in}{1.624360in}}%
\pgfpathlineto{\pgfqpoint{0.919412in}{1.557305in}}%
\pgfpathlineto{\pgfqpoint{0.943999in}{1.581892in}}%
\pgfpathlineto{\pgfqpoint{0.876944in}{1.648947in}}%
\pgfpathlineto{\pgfqpoint{0.870986in}{1.656063in}}%
\pgfpathlineto{\pgfqpoint{0.886845in}{1.670652in}}%
\pgfpathlineto{\pgfqpoint{0.903149in}{1.684743in}}%
\pgfpathlineto{\pgfqpoint{0.919881in}{1.698323in}}%
\pgfpathlineto{\pgfqpoint{0.937026in}{1.711377in}}%
\pgfpathlineto{\pgfqpoint{0.954568in}{1.723894in}}%
\pgfpathlineto{\pgfqpoint{0.972488in}{1.735861in}}%
\pgfpathlineto{\pgfqpoint{0.990771in}{1.747267in}}%
\pgfpathlineto{\pgfqpoint{1.009399in}{1.758102in}}%
\pgfpathlineto{\pgfqpoint{1.028353in}{1.768354in}}%
\pgfpathlineto{\pgfqpoint{1.047616in}{1.778013in}}%
\pgfpathlineto{\pgfqpoint{1.067169in}{1.787072in}}%
\pgfpathlineto{\pgfqpoint{1.086993in}{1.795520in}}%
\pgfpathlineto{\pgfqpoint{1.107069in}{1.803350in}}%
\pgfpathlineto{\pgfqpoint{1.127378in}{1.810554in}}%
\pgfpathlineto{\pgfqpoint{1.147901in}{1.817125in}}%
\pgfpathlineto{\pgfqpoint{1.168618in}{1.823058in}}%
\pgfpathlineto{\pgfqpoint{1.189508in}{1.828346in}}%
\pgfpathlineto{\pgfqpoint{1.210552in}{1.832983in}}%
\pgfpathlineto{\pgfqpoint{1.231730in}{1.836967in}}%
\pgfpathlineto{\pgfqpoint{1.253021in}{1.840293in}}%
\pgfpathlineto{\pgfqpoint{1.274405in}{1.842957in}}%
\pgfpathlineto{\pgfqpoint{1.295861in}{1.844957in}}%
\pgfpathlineto{\pgfqpoint{1.317369in}{1.846292in}}%
\pgfpathlineto{\pgfqpoint{1.338908in}{1.846960in}}%
\pgfpathlineto{\pgfqpoint{1.360457in}{1.846960in}}%
\pgfpathlineto{\pgfqpoint{1.381996in}{1.846292in}}%
\pgfpathlineto{\pgfqpoint{1.403504in}{1.844957in}}%
\pgfpathlineto{\pgfqpoint{1.424960in}{1.842957in}}%
\pgfpathlineto{\pgfqpoint{1.446344in}{1.840293in}}%
\pgfpathlineto{\pgfqpoint{1.467635in}{1.836967in}}%
\pgfpathlineto{\pgfqpoint{1.488812in}{1.832983in}}%
\pgfpathlineto{\pgfqpoint{1.509857in}{1.828346in}}%
\pgfpathlineto{\pgfqpoint{1.530747in}{1.823058in}}%
\pgfpathlineto{\pgfqpoint{1.551464in}{1.817125in}}%
\pgfpathlineto{\pgfqpoint{1.571986in}{1.810554in}}%
\pgfpathlineto{\pgfqpoint{1.592296in}{1.803350in}}%
\pgfpathlineto{\pgfqpoint{1.612372in}{1.795520in}}%
\pgfpathlineto{\pgfqpoint{1.632196in}{1.787072in}}%
\pgfpathlineto{\pgfqpoint{1.651749in}{1.778013in}}%
\pgfpathlineto{\pgfqpoint{1.671012in}{1.768354in}}%
\pgfpathlineto{\pgfqpoint{1.689966in}{1.758102in}}%
\pgfpathlineto{\pgfqpoint{1.708594in}{1.747267in}}%
\pgfpathlineto{\pgfqpoint{1.726877in}{1.735861in}}%
\pgfpathlineto{\pgfqpoint{1.744797in}{1.723894in}}%
\pgfpathlineto{\pgfqpoint{1.762339in}{1.711377in}}%
\pgfpathlineto{\pgfqpoint{1.779484in}{1.698323in}}%
\pgfpathlineto{\pgfqpoint{1.796216in}{1.684743in}}%
\pgfpathlineto{\pgfqpoint{1.812520in}{1.670652in}}%
\pgfpathlineto{\pgfqpoint{1.828379in}{1.656063in}}%
\pgfpathlineto{\pgfqpoint{1.822421in}{1.648947in}}%
\pgfpathlineto{\pgfqpoint{1.755365in}{1.581892in}}%
\pgfpathlineto{\pgfqpoint{1.779952in}{1.557305in}}%
\pgfpathlineto{\pgfqpoint{1.847007in}{1.624360in}}%
\pgfpathlineto{\pgfqpoint{1.854123in}{1.630318in}}%
\pgfpathlineto{\pgfqpoint{1.868713in}{1.614459in}}%
\pgfpathlineto{\pgfqpoint{1.882804in}{1.598155in}}%
\pgfpathlineto{\pgfqpoint{1.896383in}{1.581423in}}%
\pgfpathlineto{\pgfqpoint{1.909437in}{1.564278in}}%
\pgfpathlineto{\pgfqpoint{1.921954in}{1.546737in}}%
\pgfpathlineto{\pgfqpoint{1.933921in}{1.528816in}}%
\pgfpathlineto{\pgfqpoint{1.945328in}{1.510533in}}%
\pgfpathlineto{\pgfqpoint{1.956162in}{1.491906in}}%
\pgfpathlineto{\pgfqpoint{1.966414in}{1.472951in}}%
\pgfpathlineto{\pgfqpoint{1.976074in}{1.453689in}}%
\pgfpathlineto{\pgfqpoint{1.985132in}{1.434136in}}%
\pgfpathlineto{\pgfqpoint{1.993581in}{1.414312in}}%
\pgfpathlineto{\pgfqpoint{2.001410in}{1.394235in}}%
\pgfpathlineto{\pgfqpoint{2.008615in}{1.373926in}}%
\pgfpathlineto{\pgfqpoint{2.015186in}{1.353403in}}%
\pgfpathlineto{\pgfqpoint{2.021118in}{1.332687in}}%
\pgfpathlineto{\pgfqpoint{2.026406in}{1.311796in}}%
\pgfpathlineto{\pgfqpoint{2.031044in}{1.290752in}}%
\pgfpathlineto{\pgfqpoint{2.035028in}{1.269574in}}%
\pgfpathlineto{\pgfqpoint{2.038353in}{1.248283in}}%
\pgfpathlineto{\pgfqpoint{2.041017in}{1.226899in}}%
\pgfpathlineto{\pgfqpoint{2.043018in}{1.205443in}}%
\pgfpathlineto{\pgfqpoint{2.044352in}{1.183935in}}%
\pgfpathlineto{\pgfqpoint{2.045020in}{1.162396in}}%
\pgfpathlineto{\pgfqpoint{2.045020in}{1.140847in}}%
\pgfpathlineto{\pgfqpoint{2.044352in}{1.119308in}}%
\pgfpathlineto{\pgfqpoint{2.043018in}{1.097801in}}%
\pgfpathlineto{\pgfqpoint{2.041017in}{1.076344in}}%
\pgfpathlineto{\pgfqpoint{2.038353in}{1.054961in}}%
\pgfpathlineto{\pgfqpoint{2.035028in}{1.033670in}}%
\pgfpathlineto{\pgfqpoint{2.031044in}{1.012492in}}%
\pgfpathlineto{\pgfqpoint{2.026406in}{0.991448in}}%
\pgfpathlineto{\pgfqpoint{2.021118in}{0.970557in}}%
\pgfpathlineto{\pgfqpoint{2.015186in}{0.949841in}}%
\pgfpathlineto{\pgfqpoint{2.008615in}{0.929318in}}%
\pgfpathlineto{\pgfqpoint{2.001410in}{0.909009in}}%
\pgfpathlineto{\pgfqpoint{1.993581in}{0.888932in}}%
\pgfpathlineto{\pgfqpoint{1.985132in}{0.869108in}}%
\pgfpathlineto{\pgfqpoint{1.976074in}{0.849555in}}%
\pgfpathlineto{\pgfqpoint{1.966414in}{0.830292in}}%
\pgfpathlineto{\pgfqpoint{1.956162in}{0.811338in}}%
\pgfpathlineto{\pgfqpoint{1.945328in}{0.792711in}}%
\pgfpathlineto{\pgfqpoint{1.933921in}{0.774428in}}%
\pgfpathlineto{\pgfqpoint{1.921954in}{0.756507in}}%
\pgfpathlineto{\pgfqpoint{1.909437in}{0.738966in}}%
\pgfpathlineto{\pgfqpoint{1.896383in}{0.721821in}}%
\pgfpathlineto{\pgfqpoint{1.882804in}{0.705088in}}%
\pgfpathlineto{\pgfqpoint{1.868713in}{0.688785in}}%
\pgfpathlineto{\pgfqpoint{1.854123in}{0.672925in}}%
\pgfpathlineto{\pgfqpoint{1.847007in}{0.678884in}}%
\pgfpathlineto{\pgfqpoint{1.779952in}{0.745939in}}%
\pgfpathlineto{\pgfqpoint{1.754874in}{0.720860in}}%
\pgfpathlineto{\pgfqpoint{1.755365in}{0.721352in}}%
\pgfpathlineto{\pgfqpoint{1.822421in}{0.654297in}}%
\pgfpathlineto{\pgfqpoint{1.828379in}{0.647181in}}%
\pgfpathlineto{\pgfqpoint{1.812520in}{0.632592in}}%
\pgfpathlineto{\pgfqpoint{1.796216in}{0.618500in}}%
\pgfpathlineto{\pgfqpoint{1.779484in}{0.604921in}}%
\pgfpathlineto{\pgfqpoint{1.762339in}{0.591867in}}%
\pgfpathlineto{\pgfqpoint{1.744797in}{0.579350in}}%
\pgfpathlineto{\pgfqpoint{1.726877in}{0.567383in}}%
\pgfpathlineto{\pgfqpoint{1.708594in}{0.555976in}}%
\pgfpathlineto{\pgfqpoint{1.689966in}{0.545142in}}%
\pgfpathlineto{\pgfqpoint{1.671012in}{0.534890in}}%
\pgfpathlineto{\pgfqpoint{1.651749in}{0.525230in}}%
\pgfpathlineto{\pgfqpoint{1.632196in}{0.516172in}}%
\pgfpathlineto{\pgfqpoint{1.612372in}{0.507724in}}%
\pgfpathlineto{\pgfqpoint{1.592296in}{0.499894in}}%
\pgfpathlineto{\pgfqpoint{1.571986in}{0.492690in}}%
\pgfpathlineto{\pgfqpoint{1.551464in}{0.486118in}}%
\pgfpathlineto{\pgfqpoint{1.530747in}{0.480186in}}%
\pgfpathlineto{\pgfqpoint{1.509857in}{0.474898in}}%
\pgfpathlineto{\pgfqpoint{1.488812in}{0.470260in}}%
\pgfpathlineto{\pgfqpoint{1.467635in}{0.466277in}}%
\pgfpathlineto{\pgfqpoint{1.446344in}{0.462951in}}%
\pgfpathlineto{\pgfqpoint{1.424960in}{0.460287in}}%
\pgfpathlineto{\pgfqpoint{1.403504in}{0.458286in}}%
\pgfpathlineto{\pgfqpoint{1.381996in}{0.456952in}}%
\pgfpathlineto{\pgfqpoint{1.360457in}{0.456284in}}%
\pgfpathlineto{\pgfqpoint{1.338908in}{0.456284in}}%
\pgfpathlineto{\pgfqpoint{1.317369in}{0.456952in}}%
\pgfpathlineto{\pgfqpoint{1.295861in}{0.458286in}}%
\pgfpathlineto{\pgfqpoint{1.274405in}{0.460287in}}%
\pgfpathlineto{\pgfqpoint{1.253021in}{0.462951in}}%
\pgfpathlineto{\pgfqpoint{1.231730in}{0.466277in}}%
\pgfpathlineto{\pgfqpoint{1.210552in}{0.470260in}}%
\pgfpathlineto{\pgfqpoint{1.189508in}{0.474898in}}%
\pgfpathlineto{\pgfqpoint{1.168618in}{0.480186in}}%
\pgfpathlineto{\pgfqpoint{1.147901in}{0.486118in}}%
\pgfpathlineto{\pgfqpoint{1.127378in}{0.492690in}}%
\pgfpathlineto{\pgfqpoint{1.107069in}{0.499894in}}%
\pgfpathlineto{\pgfqpoint{1.086993in}{0.507724in}}%
\pgfpathlineto{\pgfqpoint{1.067169in}{0.516172in}}%
\pgfpathlineto{\pgfqpoint{1.047616in}{0.525230in}}%
\pgfpathlineto{\pgfqpoint{1.028353in}{0.534890in}}%
\pgfpathlineto{\pgfqpoint{1.009399in}{0.545142in}}%
\pgfpathlineto{\pgfqpoint{0.990771in}{0.555976in}}%
\pgfpathlineto{\pgfqpoint{0.972488in}{0.567383in}}%
\pgfpathlineto{\pgfqpoint{0.954568in}{0.579350in}}%
\pgfpathlineto{\pgfqpoint{0.937026in}{0.591867in}}%
\pgfpathlineto{\pgfqpoint{0.919881in}{0.604921in}}%
\pgfpathlineto{\pgfqpoint{0.903149in}{0.618500in}}%
\pgfpathlineto{\pgfqpoint{0.886845in}{0.632592in}}%
\pgfpathlineto{\pgfqpoint{0.870986in}{0.647181in}}%
\pgfpathlineto{\pgfqpoint{0.876944in}{0.654297in}}%
\pgfpathlineto{\pgfqpoint{0.943999in}{0.721352in}}%
\pgfpathlineto{\pgfqpoint{0.918921in}{0.746431in}}%
\pgfpathlineto{\pgfqpoint{0.919412in}{0.745939in}}%
\pgfpathlineto{\pgfqpoint{0.852357in}{0.678884in}}%
\pgfpathlineto{\pgfqpoint{0.845242in}{0.672925in}}%
\pgfpathlineto{\pgfqpoint{0.830652in}{0.688785in}}%
\pgfpathlineto{\pgfqpoint{0.816561in}{0.705088in}}%
\pgfpathlineto{\pgfqpoint{0.802982in}{0.721821in}}%
\pgfpathlineto{\pgfqpoint{0.789927in}{0.738966in}}%
\pgfpathlineto{\pgfqpoint{0.777411in}{0.756507in}}%
\pgfpathlineto{\pgfqpoint{0.765443in}{0.774428in}}%
\pgfpathlineto{\pgfqpoint{0.754037in}{0.792711in}}%
\pgfpathlineto{\pgfqpoint{0.743203in}{0.811338in}}%
\pgfpathlineto{\pgfqpoint{0.732951in}{0.830292in}}%
\pgfpathlineto{\pgfqpoint{0.723291in}{0.849555in}}%
\pgfpathlineto{\pgfqpoint{0.714232in}{0.869108in}}%
\pgfpathlineto{\pgfqpoint{0.705784in}{0.888932in}}%
\pgfpathlineto{\pgfqpoint{0.697954in}{0.909009in}}%
\pgfpathlineto{\pgfqpoint{0.690750in}{0.929318in}}%
\pgfpathlineto{\pgfqpoint{0.684179in}{0.949841in}}%
\pgfpathlineto{\pgfqpoint{0.678246in}{0.970557in}}%
\pgfpathlineto{\pgfqpoint{0.672959in}{0.991448in}}%
\pgfpathlineto{\pgfqpoint{0.668321in}{1.012492in}}%
\pgfpathlineto{\pgfqpoint{0.664337in}{1.033670in}}%
\pgfpathlineto{\pgfqpoint{0.661012in}{1.054961in}}%
\pgfpathlineto{\pgfqpoint{0.658347in}{1.076344in}}%
\pgfpathlineto{\pgfqpoint{0.656347in}{1.097801in}}%
\pgfpathlineto{\pgfqpoint{0.655012in}{1.119308in}}%
\pgfpathlineto{\pgfqpoint{0.654345in}{1.140847in}}%
\pgfpathlineto{\pgfqpoint{0.654345in}{1.162396in}}%
\pgfpathlineto{\pgfqpoint{0.655012in}{1.183935in}}%
\pgfpathlineto{\pgfqpoint{0.656347in}{1.205443in}}%
\pgfpathlineto{\pgfqpoint{0.658347in}{1.226899in}}%
\pgfpathlineto{\pgfqpoint{0.661012in}{1.248283in}}%
\pgfpathlineto{\pgfqpoint{0.664337in}{1.269574in}}%
\pgfpathlineto{\pgfqpoint{0.668321in}{1.290752in}}%
\pgfpathlineto{\pgfqpoint{0.672959in}{1.311796in}}%
\pgfpathlineto{\pgfqpoint{0.678246in}{1.332687in}}%
\pgfpathlineto{\pgfqpoint{0.684179in}{1.353403in}}%
\pgfpathlineto{\pgfqpoint{0.690750in}{1.373926in}}%
\pgfpathlineto{\pgfqpoint{0.697954in}{1.394235in}}%
\pgfpathlineto{\pgfqpoint{0.705784in}{1.414312in}}%
\pgfpathlineto{\pgfqpoint{0.714232in}{1.434136in}}%
\pgfpathlineto{\pgfqpoint{0.723291in}{1.453689in}}%
\pgfpathlineto{\pgfqpoint{0.732951in}{1.472951in}}%
\pgfpathlineto{\pgfqpoint{0.743203in}{1.491906in}}%
\pgfpathlineto{\pgfqpoint{0.754037in}{1.510533in}}%
\pgfpathlineto{\pgfqpoint{0.765443in}{1.528816in}}%
\pgfpathlineto{\pgfqpoint{0.777411in}{1.546737in}}%
\pgfpathlineto{\pgfqpoint{0.789927in}{1.564278in}}%
\pgfpathlineto{\pgfqpoint{0.802982in}{1.581423in}}%
\pgfpathlineto{\pgfqpoint{0.816561in}{1.598155in}}%
\pgfpathlineto{\pgfqpoint{0.830652in}{1.614459in}}%
\pgfpathlineto{\pgfqpoint{0.845242in}{1.630318in}}%
\pgfpathlineto{\pgfqpoint{0.852357in}{1.624360in}}%
\pgfpathlineto{\pgfqpoint{0.919412in}{1.557305in}}%
\pgfpathlineto{\pgfqpoint{0.943999in}{1.581892in}}%
\pgfpathlineto{\pgfqpoint{0.876944in}{1.648947in}}%
\pgfpathlineto{\pgfqpoint{0.876944in}{1.648947in}}%
\pgfusepath{stroke}%
\end{pgfscope}%
\begin{pgfscope}%
\pgfsetrectcap%
\pgfsetmiterjoin%
\pgfsetlinewidth{0.803000pt}%
\definecolor{currentstroke}{rgb}{0.000000,0.000000,0.000000}%
\pgfsetstrokecolor{currentstroke}%
\pgfsetdash{}{0pt}%
\pgfpathmoveto{\pgfqpoint{0.584719in}{0.386658in}}%
\pgfpathlineto{\pgfqpoint{0.584719in}{1.916585in}}%
\pgfusepath{stroke}%
\end{pgfscope}%
\begin{pgfscope}%
\pgfsetrectcap%
\pgfsetmiterjoin%
\pgfsetlinewidth{0.803000pt}%
\definecolor{currentstroke}{rgb}{0.000000,0.000000,0.000000}%
\pgfsetstrokecolor{currentstroke}%
\pgfsetdash{}{0pt}%
\pgfpathmoveto{\pgfqpoint{2.114646in}{0.386658in}}%
\pgfpathlineto{\pgfqpoint{2.114646in}{1.916585in}}%
\pgfusepath{stroke}%
\end{pgfscope}%
\begin{pgfscope}%
\pgfsetrectcap%
\pgfsetmiterjoin%
\pgfsetlinewidth{0.803000pt}%
\definecolor{currentstroke}{rgb}{0.000000,0.000000,0.000000}%
\pgfsetstrokecolor{currentstroke}%
\pgfsetdash{}{0pt}%
\pgfpathmoveto{\pgfqpoint{0.584719in}{0.386658in}}%
\pgfpathlineto{\pgfqpoint{2.114646in}{0.386658in}}%
\pgfusepath{stroke}%
\end{pgfscope}%
\begin{pgfscope}%
\pgfsetrectcap%
\pgfsetmiterjoin%
\pgfsetlinewidth{0.803000pt}%
\definecolor{currentstroke}{rgb}{0.000000,0.000000,0.000000}%
\pgfsetstrokecolor{currentstroke}%
\pgfsetdash{}{0pt}%
\pgfpathmoveto{\pgfqpoint{0.584719in}{1.916585in}}%
\pgfpathlineto{\pgfqpoint{2.114646in}{1.916585in}}%
\pgfusepath{stroke}%
\end{pgfscope}%
\begin{pgfscope}%
\pgfpathrectangle{\pgfqpoint{0.584719in}{0.386658in}}{\pgfqpoint{1.529927in}{1.529927in}}%
\pgfusepath{clip}%
\pgfsetrectcap%
\pgfsetroundjoin%
\pgfsetlinewidth{1.505625pt}%
\definecolor{currentstroke}{rgb}{0.000000,0.000000,0.000000}%
\pgfsetstrokecolor{currentstroke}%
\pgfsetdash{}{0pt}%
\pgfpathmoveto{\pgfqpoint{1.071514in}{1.151622in}}%
\pgfpathlineto{\pgfqpoint{1.627851in}{1.151622in}}%
\pgfusepath{stroke}%
\end{pgfscope}%
\begin{pgfscope}%
\pgfpathrectangle{\pgfqpoint{0.584719in}{0.386658in}}{\pgfqpoint{1.529927in}{1.529927in}}%
\pgfusepath{clip}%
\pgfsetrectcap%
\pgfsetroundjoin%
\pgfsetlinewidth{1.505625pt}%
\definecolor{currentstroke}{rgb}{0.000000,0.000000,0.000000}%
\pgfsetstrokecolor{currentstroke}%
\pgfsetdash{}{0pt}%
\pgfpathmoveto{\pgfqpoint{1.349682in}{0.873453in}}%
\pgfpathlineto{\pgfqpoint{1.349682in}{1.429790in}}%
\pgfusepath{stroke}%
\end{pgfscope}%
\begin{pgfscope}%
\pgfsetbuttcap%
\pgfsetmiterjoin%
\definecolor{currentfill}{rgb}{1.000000,1.000000,1.000000}%
\pgfsetfillcolor{currentfill}%
\pgfsetlinewidth{0.000000pt}%
\definecolor{currentstroke}{rgb}{0.000000,0.000000,0.000000}%
\pgfsetstrokecolor{currentstroke}%
\pgfsetstrokeopacity{0.000000}%
\pgfsetdash{}{0pt}%
\pgfpathmoveto{\pgfqpoint{0.337193in}{2.012206in}}%
\pgfpathlineto{\pgfqpoint{2.362172in}{2.012206in}}%
\pgfpathlineto{\pgfqpoint{2.362172in}{2.113455in}}%
\pgfpathlineto{\pgfqpoint{0.337193in}{2.113455in}}%
\pgfpathlineto{\pgfqpoint{0.337193in}{2.012206in}}%
\pgfpathclose%
\pgfusepath{fill}%
\end{pgfscope}%
\begin{pgfscope}%
\pgfsys@transformshift{0.337500in}{2.012500in}%
\pgftext[left,bottom]{\includegraphics[interpolate=true,width=2.025000in,height=0.101250in]{figures/./section2/dataBC/10000p/re_1000//err_vmag-img1.png}}%
\end{pgfscope}%
\begin{pgfscope}%
\pgfsetbuttcap%
\pgfsetroundjoin%
\definecolor{currentfill}{rgb}{0.000000,0.000000,0.000000}%
\pgfsetfillcolor{currentfill}%
\pgfsetlinewidth{0.803000pt}%
\definecolor{currentstroke}{rgb}{0.000000,0.000000,0.000000}%
\pgfsetstrokecolor{currentstroke}%
\pgfsetdash{}{0pt}%
\pgfsys@defobject{currentmarker}{\pgfqpoint{0.000000in}{0.000000in}}{\pgfqpoint{0.000000in}{0.048611in}}{%
\pgfpathmoveto{\pgfqpoint{0.000000in}{0.000000in}}%
\pgfpathlineto{\pgfqpoint{0.000000in}{0.048611in}}%
\pgfusepath{stroke,fill}%
}%
\begin{pgfscope}%
\pgfsys@transformshift{0.337694in}{2.113455in}%
\pgfsys@useobject{currentmarker}{}%
\end{pgfscope}%
\end{pgfscope}%
\begin{pgfscope}%
\definecolor{textcolor}{rgb}{0.000000,0.000000,0.000000}%
\pgfsetstrokecolor{textcolor}%
\pgfsetfillcolor{textcolor}%
\pgftext[x=0.337694in,y=2.203732in,,bottom]{\color{textcolor}{\rmfamily\fontsize{8.330000}{9.996000}\selectfont\catcode`\^=\active\def^{\ifmmode\sp\else\^{}\fi}\catcode`\%=\active\def%{\%}$\mathdefault{\ensuremath{-}1.103}$}}%
\end{pgfscope}%
\begin{pgfscope}%
\pgfsetbuttcap%
\pgfsetroundjoin%
\definecolor{currentfill}{rgb}{0.000000,0.000000,0.000000}%
\pgfsetfillcolor{currentfill}%
\pgfsetlinewidth{0.803000pt}%
\definecolor{currentstroke}{rgb}{0.000000,0.000000,0.000000}%
\pgfsetstrokecolor{currentstroke}%
\pgfsetdash{}{0pt}%
\pgfsys@defobject{currentmarker}{\pgfqpoint{0.000000in}{0.000000in}}{\pgfqpoint{0.000000in}{0.048611in}}{%
\pgfpathmoveto{\pgfqpoint{0.000000in}{0.000000in}}%
\pgfpathlineto{\pgfqpoint{0.000000in}{0.048611in}}%
\pgfusepath{stroke,fill}%
}%
\begin{pgfscope}%
\pgfsys@transformshift{1.011945in}{2.113455in}%
\pgfsys@useobject{currentmarker}{}%
\end{pgfscope}%
\end{pgfscope}%
\begin{pgfscope}%
\definecolor{textcolor}{rgb}{0.000000,0.000000,0.000000}%
\pgfsetstrokecolor{textcolor}%
\pgfsetfillcolor{textcolor}%
\pgftext[x=1.011945in,y=2.203732in,,bottom]{\color{textcolor}{\rmfamily\fontsize{8.330000}{9.996000}\selectfont\catcode`\^=\active\def^{\ifmmode\sp\else\^{}\fi}\catcode`\%=\active\def%{\%}$\mathdefault{\ensuremath{-}0.095}$}}%
\end{pgfscope}%
\begin{pgfscope}%
\pgfsetbuttcap%
\pgfsetroundjoin%
\definecolor{currentfill}{rgb}{0.000000,0.000000,0.000000}%
\pgfsetfillcolor{currentfill}%
\pgfsetlinewidth{0.803000pt}%
\definecolor{currentstroke}{rgb}{0.000000,0.000000,0.000000}%
\pgfsetstrokecolor{currentstroke}%
\pgfsetdash{}{0pt}%
\pgfsys@defobject{currentmarker}{\pgfqpoint{0.000000in}{0.000000in}}{\pgfqpoint{0.000000in}{0.048611in}}{%
\pgfpathmoveto{\pgfqpoint{0.000000in}{0.000000in}}%
\pgfpathlineto{\pgfqpoint{0.000000in}{0.048611in}}%
\pgfusepath{stroke,fill}%
}%
\begin{pgfscope}%
\pgfsys@transformshift{1.686197in}{2.113455in}%
\pgfsys@useobject{currentmarker}{}%
\end{pgfscope}%
\end{pgfscope}%
\begin{pgfscope}%
\definecolor{textcolor}{rgb}{0.000000,0.000000,0.000000}%
\pgfsetstrokecolor{textcolor}%
\pgfsetfillcolor{textcolor}%
\pgftext[x=1.686197in,y=2.203732in,,bottom]{\color{textcolor}{\rmfamily\fontsize{8.330000}{9.996000}\selectfont\catcode`\^=\active\def^{\ifmmode\sp\else\^{}\fi}\catcode`\%=\active\def%{\%}$\mathdefault{0.913}$}}%
\end{pgfscope}%
\begin{pgfscope}%
\pgfsetbuttcap%
\pgfsetroundjoin%
\definecolor{currentfill}{rgb}{0.000000,0.000000,0.000000}%
\pgfsetfillcolor{currentfill}%
\pgfsetlinewidth{0.803000pt}%
\definecolor{currentstroke}{rgb}{0.000000,0.000000,0.000000}%
\pgfsetstrokecolor{currentstroke}%
\pgfsetdash{}{0pt}%
\pgfsys@defobject{currentmarker}{\pgfqpoint{0.000000in}{0.000000in}}{\pgfqpoint{0.000000in}{0.048611in}}{%
\pgfpathmoveto{\pgfqpoint{0.000000in}{0.000000in}}%
\pgfpathlineto{\pgfqpoint{0.000000in}{0.048611in}}%
\pgfusepath{stroke,fill}%
}%
\begin{pgfscope}%
\pgfsys@transformshift{2.360448in}{2.113455in}%
\pgfsys@useobject{currentmarker}{}%
\end{pgfscope}%
\end{pgfscope}%
\begin{pgfscope}%
\definecolor{textcolor}{rgb}{0.000000,0.000000,0.000000}%
\pgfsetstrokecolor{textcolor}%
\pgfsetfillcolor{textcolor}%
\pgftext[x=2.360448in,y=2.203732in,,bottom]{\color{textcolor}{\rmfamily\fontsize{8.330000}{9.996000}\selectfont\catcode`\^=\active\def^{\ifmmode\sp\else\^{}\fi}\catcode`\%=\active\def%{\%}$\mathdefault{1.921}$}}%
\end{pgfscope}%
\begin{pgfscope}%
\definecolor{textcolor}{rgb}{0.000000,0.000000,0.000000}%
\pgfsetstrokecolor{textcolor}%
\pgfsetfillcolor{textcolor}%
\pgftext[x=1.349682in,y=2.358054in,,base]{\color{textcolor}{\rmfamily\fontsize{10.000000}{12.000000}\selectfont\catcode`\^=\active\def^{\ifmmode\sp\else\^{}\fi}\catcode`\%=\active\def%{\%}$f^{(\bm{v})}_\text{err} \ [\%]$}}%
\end{pgfscope}%
\begin{pgfscope}%
\definecolor{textcolor}{rgb}{0.000000,0.000000,0.000000}%
\pgfsetstrokecolor{textcolor}%
\pgfsetfillcolor{textcolor}%
\pgftext[x=2.362172in,y=2.344165in,right,bottom]{\color{textcolor}{\rmfamily\fontsize{8.330000}{9.996000}\selectfont\catcode`\^=\active\def^{\ifmmode\sp\else\^{}\fi}\catcode`\%=\active\def%{\%}$\times\mathdefault{10^{1}}\mathdefault{}$}}%
\end{pgfscope}%
\begin{pgfscope}%
\pgfsetrectcap%
\pgfsetmiterjoin%
\pgfsetlinewidth{0.803000pt}%
\definecolor{currentstroke}{rgb}{0.000000,0.000000,0.000000}%
\pgfsetstrokecolor{currentstroke}%
\pgfsetdash{}{0pt}%
\pgfpathmoveto{\pgfqpoint{0.337193in}{2.012206in}}%
\pgfpathlineto{\pgfqpoint{0.337193in}{2.062830in}}%
\pgfpathlineto{\pgfqpoint{0.337193in}{2.113455in}}%
\pgfpathlineto{\pgfqpoint{2.362172in}{2.113455in}}%
\pgfpathlineto{\pgfqpoint{2.362172in}{2.062830in}}%
\pgfpathlineto{\pgfqpoint{2.362172in}{2.012206in}}%
\pgfpathlineto{\pgfqpoint{0.337193in}{2.012206in}}%
\pgfpathclose%
\pgfusepath{stroke}%
\end{pgfscope}%
\end{pgfpicture}%
\makeatother%
\endgroup%

%% file: figures/section2/errors_all_Re_64.pgf
%% Creator: Matplotlib, PGF backend
%%
%% To include the figure in your LaTeX document, write
%%   \input{<filename>.pgf}
%%
%% Make sure the required packages are loaded in your preamble
%%   \usepackage{pgf}
%%
%% Also ensure that all the required font packages are loaded; for instance,
%% the lmodern package is sometimes necessary when using math font.
%%   \usepackage{lmodern}
%%
%% Figures using additional raster images can only be included by \input if
%% they are in the same directory as the main LaTeX file. For loading figures
%% from other directories you can use the `import` package
%% and then include the figures with
%%   \import{<path to file>}{<filename>.pgf}
%%
%% Matplotlib used the following preamble
%%   \def\mathdefault#1{#1}
%%   \everymath=\expandafter{\the\everymath\displaystyle}
%%   \usepackage{amsmath}\usepackage{bm}
%%   \makeatletter\@ifpackageloaded{underscore}{}{\usepackage[strings]{underscore}}\makeatother
%%
\begingroup%
\makeatletter%
\begin{pgfpicture}%
\pgfpathrectangle{\pgfpointorigin}{\pgfqpoint{3.000000in}{2.000000in}}%
\pgfusepath{use as bounding box, clip}%
\begin{pgfscope}%
\pgfsetbuttcap%
\pgfsetmiterjoin%
\definecolor{currentfill}{rgb}{1.000000,1.000000,1.000000}%
\pgfsetfillcolor{currentfill}%
\pgfsetlinewidth{0.000000pt}%
\definecolor{currentstroke}{rgb}{1.000000,1.000000,1.000000}%
\pgfsetstrokecolor{currentstroke}%
\pgfsetdash{}{0pt}%
\pgfpathmoveto{\pgfqpoint{0.000000in}{0.000000in}}%
\pgfpathlineto{\pgfqpoint{3.000000in}{0.000000in}}%
\pgfpathlineto{\pgfqpoint{3.000000in}{2.000000in}}%
\pgfpathlineto{\pgfqpoint{0.000000in}{2.000000in}}%
\pgfpathlineto{\pgfqpoint{0.000000in}{0.000000in}}%
\pgfpathclose%
\pgfusepath{fill}%
\end{pgfscope}%
\begin{pgfscope}%
\pgfsetbuttcap%
\pgfsetmiterjoin%
\definecolor{currentfill}{rgb}{1.000000,1.000000,1.000000}%
\pgfsetfillcolor{currentfill}%
\pgfsetlinewidth{0.000000pt}%
\definecolor{currentstroke}{rgb}{0.000000,0.000000,0.000000}%
\pgfsetstrokecolor{currentstroke}%
\pgfsetstrokeopacity{0.000000}%
\pgfsetdash{}{0pt}%
\pgfpathmoveto{\pgfqpoint{0.597016in}{0.498776in}}%
\pgfpathlineto{\pgfqpoint{2.845756in}{0.498776in}}%
\pgfpathlineto{\pgfqpoint{2.845756in}{1.818771in}}%
\pgfpathlineto{\pgfqpoint{0.597016in}{1.818771in}}%
\pgfpathlineto{\pgfqpoint{0.597016in}{0.498776in}}%
\pgfpathclose%
\pgfusepath{fill}%
\end{pgfscope}%
\begin{pgfscope}%
\pgfpathrectangle{\pgfqpoint{0.597016in}{0.498776in}}{\pgfqpoint{2.248740in}{1.319995in}}%
\pgfusepath{clip}%
\pgfsetbuttcap%
\pgfsetroundjoin%
\definecolor{currentfill}{rgb}{0.172549,0.482353,0.713725}%
\pgfsetfillcolor{currentfill}%
\pgfsetlinewidth{0.501875pt}%
\definecolor{currentstroke}{rgb}{0.000000,0.000000,0.000000}%
\pgfsetstrokecolor{currentstroke}%
\pgfsetdash{}{0pt}%
\pgfsys@defobject{currentmarker}{\pgfqpoint{-0.026896in}{-0.026896in}}{\pgfqpoint{0.026896in}{0.026896in}}{%
\pgfpathmoveto{\pgfqpoint{-0.026896in}{-0.026896in}}%
\pgfpathlineto{\pgfqpoint{0.026896in}{-0.026896in}}%
\pgfpathlineto{\pgfqpoint{0.026896in}{0.026896in}}%
\pgfpathlineto{\pgfqpoint{-0.026896in}{0.026896in}}%
\pgfpathlineto{\pgfqpoint{-0.026896in}{-0.026896in}}%
\pgfpathclose%
\pgfusepath{stroke,fill}%
}%
\begin{pgfscope}%
\pgfsys@transformshift{0.699231in}{1.468005in}%
\pgfsys@useobject{currentmarker}{}%
\end{pgfscope}%
\begin{pgfscope}%
\pgfsys@transformshift{0.926377in}{1.301032in}%
\pgfsys@useobject{currentmarker}{}%
\end{pgfscope}%
\begin{pgfscope}%
\pgfsys@transformshift{1.153522in}{0.602596in}%
\pgfsys@useobject{currentmarker}{}%
\end{pgfscope}%
\begin{pgfscope}%
\pgfsys@transformshift{1.380668in}{0.554541in}%
\pgfsys@useobject{currentmarker}{}%
\end{pgfscope}%
\begin{pgfscope}%
\pgfsys@transformshift{1.607813in}{0.617845in}%
\pgfsys@useobject{currentmarker}{}%
\end{pgfscope}%
\begin{pgfscope}%
\pgfsys@transformshift{1.834959in}{0.644003in}%
\pgfsys@useobject{currentmarker}{}%
\end{pgfscope}%
\begin{pgfscope}%
\pgfsys@transformshift{2.062104in}{0.654124in}%
\pgfsys@useobject{currentmarker}{}%
\end{pgfscope}%
\begin{pgfscope}%
\pgfsys@transformshift{2.289250in}{0.664881in}%
\pgfsys@useobject{currentmarker}{}%
\end{pgfscope}%
\begin{pgfscope}%
\pgfsys@transformshift{2.516395in}{0.679716in}%
\pgfsys@useobject{currentmarker}{}%
\end{pgfscope}%
\begin{pgfscope}%
\pgfsys@transformshift{2.743541in}{0.747282in}%
\pgfsys@useobject{currentmarker}{}%
\end{pgfscope}%
\end{pgfscope}%
\begin{pgfscope}%
\pgfpathrectangle{\pgfqpoint{0.597016in}{0.498776in}}{\pgfqpoint{2.248740in}{1.319995in}}%
\pgfusepath{clip}%
\pgfsetbuttcap%
\pgfsetroundjoin%
\definecolor{currentfill}{rgb}{0.843137,0.098039,0.109804}%
\pgfsetfillcolor{currentfill}%
\pgfsetlinewidth{0.501875pt}%
\definecolor{currentstroke}{rgb}{0.000000,0.000000,0.000000}%
\pgfsetstrokecolor{currentstroke}%
\pgfsetdash{}{0pt}%
\pgfsys@defobject{currentmarker}{\pgfqpoint{-0.026896in}{-0.026896in}}{\pgfqpoint{0.026896in}{0.026896in}}{%
\pgfpathmoveto{\pgfqpoint{0.000000in}{-0.026896in}}%
\pgfpathcurveto{\pgfqpoint{0.007133in}{-0.026896in}}{\pgfqpoint{0.013974in}{-0.024062in}}{\pgfqpoint{0.019018in}{-0.019018in}}%
\pgfpathcurveto{\pgfqpoint{0.024062in}{-0.013974in}}{\pgfqpoint{0.026896in}{-0.007133in}}{\pgfqpoint{0.026896in}{0.000000in}}%
\pgfpathcurveto{\pgfqpoint{0.026896in}{0.007133in}}{\pgfqpoint{0.024062in}{0.013974in}}{\pgfqpoint{0.019018in}{0.019018in}}%
\pgfpathcurveto{\pgfqpoint{0.013974in}{0.024062in}}{\pgfqpoint{0.007133in}{0.026896in}}{\pgfqpoint{0.000000in}{0.026896in}}%
\pgfpathcurveto{\pgfqpoint{-0.007133in}{0.026896in}}{\pgfqpoint{-0.013974in}{0.024062in}}{\pgfqpoint{-0.019018in}{0.019018in}}%
\pgfpathcurveto{\pgfqpoint{-0.024062in}{0.013974in}}{\pgfqpoint{-0.026896in}{0.007133in}}{\pgfqpoint{-0.026896in}{0.000000in}}%
\pgfpathcurveto{\pgfqpoint{-0.026896in}{-0.007133in}}{\pgfqpoint{-0.024062in}{-0.013974in}}{\pgfqpoint{-0.019018in}{-0.019018in}}%
\pgfpathcurveto{\pgfqpoint{-0.013974in}{-0.024062in}}{\pgfqpoint{-0.007133in}{-0.026896in}}{\pgfqpoint{0.000000in}{-0.026896in}}%
\pgfpathlineto{\pgfqpoint{0.000000in}{-0.026896in}}%
\pgfpathclose%
\pgfusepath{stroke,fill}%
}%
\begin{pgfscope}%
\pgfsys@transformshift{0.699231in}{0.982369in}%
\pgfsys@useobject{currentmarker}{}%
\end{pgfscope}%
\begin{pgfscope}%
\pgfsys@transformshift{0.926377in}{0.928994in}%
\pgfsys@useobject{currentmarker}{}%
\end{pgfscope}%
\begin{pgfscope}%
\pgfsys@transformshift{1.153522in}{0.625098in}%
\pgfsys@useobject{currentmarker}{}%
\end{pgfscope}%
\begin{pgfscope}%
\pgfsys@transformshift{1.380668in}{0.600982in}%
\pgfsys@useobject{currentmarker}{}%
\end{pgfscope}%
\begin{pgfscope}%
\pgfsys@transformshift{1.607813in}{0.650779in}%
\pgfsys@useobject{currentmarker}{}%
\end{pgfscope}%
\begin{pgfscope}%
\pgfsys@transformshift{1.834959in}{0.681451in}%
\pgfsys@useobject{currentmarker}{}%
\end{pgfscope}%
\begin{pgfscope}%
\pgfsys@transformshift{2.062104in}{0.687075in}%
\pgfsys@useobject{currentmarker}{}%
\end{pgfscope}%
\begin{pgfscope}%
\pgfsys@transformshift{2.289250in}{0.693257in}%
\pgfsys@useobject{currentmarker}{}%
\end{pgfscope}%
\begin{pgfscope}%
\pgfsys@transformshift{2.516395in}{0.700045in}%
\pgfsys@useobject{currentmarker}{}%
\end{pgfscope}%
\begin{pgfscope}%
\pgfsys@transformshift{2.743541in}{0.750646in}%
\pgfsys@useobject{currentmarker}{}%
\end{pgfscope}%
\end{pgfscope}%
\begin{pgfscope}%
\pgfpathrectangle{\pgfqpoint{0.597016in}{0.498776in}}{\pgfqpoint{2.248740in}{1.319995in}}%
\pgfusepath{clip}%
\pgfsetbuttcap%
\pgfsetroundjoin%
\definecolor{currentfill}{rgb}{0.670588,0.850980,0.913725}%
\pgfsetfillcolor{currentfill}%
\pgfsetlinewidth{0.501875pt}%
\definecolor{currentstroke}{rgb}{0.000000,0.000000,0.000000}%
\pgfsetstrokecolor{currentstroke}%
\pgfsetdash{}{0pt}%
\pgfsys@defobject{currentmarker}{\pgfqpoint{-0.038036in}{-0.038036in}}{\pgfqpoint{0.038036in}{0.038036in}}{%
\pgfpathmoveto{\pgfqpoint{-0.000000in}{-0.038036in}}%
\pgfpathlineto{\pgfqpoint{0.038036in}{0.000000in}}%
\pgfpathlineto{\pgfqpoint{0.000000in}{0.038036in}}%
\pgfpathlineto{\pgfqpoint{-0.038036in}{0.000000in}}%
\pgfpathlineto{\pgfqpoint{-0.000000in}{-0.038036in}}%
\pgfpathclose%
\pgfusepath{stroke,fill}%
}%
\begin{pgfscope}%
\pgfsys@transformshift{0.699231in}{1.327841in}%
\pgfsys@useobject{currentmarker}{}%
\end{pgfscope}%
\begin{pgfscope}%
\pgfsys@transformshift{0.926377in}{1.167478in}%
\pgfsys@useobject{currentmarker}{}%
\end{pgfscope}%
\begin{pgfscope}%
\pgfsys@transformshift{1.153522in}{0.814202in}%
\pgfsys@useobject{currentmarker}{}%
\end{pgfscope}%
\begin{pgfscope}%
\pgfsys@transformshift{1.380668in}{0.725710in}%
\pgfsys@useobject{currentmarker}{}%
\end{pgfscope}%
\begin{pgfscope}%
\pgfsys@transformshift{1.607813in}{0.800583in}%
\pgfsys@useobject{currentmarker}{}%
\end{pgfscope}%
\begin{pgfscope}%
\pgfsys@transformshift{1.834959in}{1.061357in}%
\pgfsys@useobject{currentmarker}{}%
\end{pgfscope}%
\begin{pgfscope}%
\pgfsys@transformshift{2.062104in}{1.142475in}%
\pgfsys@useobject{currentmarker}{}%
\end{pgfscope}%
\begin{pgfscope}%
\pgfsys@transformshift{2.289250in}{1.023114in}%
\pgfsys@useobject{currentmarker}{}%
\end{pgfscope}%
\begin{pgfscope}%
\pgfsys@transformshift{2.516395in}{1.127230in}%
\pgfsys@useobject{currentmarker}{}%
\end{pgfscope}%
\begin{pgfscope}%
\pgfsys@transformshift{2.743541in}{1.238132in}%
\pgfsys@useobject{currentmarker}{}%
\end{pgfscope}%
\end{pgfscope}%
\begin{pgfscope}%
\pgfpathrectangle{\pgfqpoint{0.597016in}{0.498776in}}{\pgfqpoint{2.248740in}{1.319995in}}%
\pgfusepath{clip}%
\pgfsetbuttcap%
\pgfsetroundjoin%
\definecolor{currentfill}{rgb}{0.992157,0.682353,0.380392}%
\pgfsetfillcolor{currentfill}%
\pgfsetlinewidth{0.501875pt}%
\definecolor{currentstroke}{rgb}{0.000000,0.000000,0.000000}%
\pgfsetstrokecolor{currentstroke}%
\pgfsetdash{}{0pt}%
\pgfsys@defobject{currentmarker}{\pgfqpoint{-0.026896in}{-0.026896in}}{\pgfqpoint{0.026896in}{0.026896in}}{%
\pgfpathmoveto{\pgfqpoint{0.000000in}{0.026896in}}%
\pgfpathlineto{\pgfqpoint{-0.026896in}{-0.026896in}}%
\pgfpathlineto{\pgfqpoint{0.026896in}{-0.026896in}}%
\pgfpathlineto{\pgfqpoint{0.000000in}{0.026896in}}%
\pgfpathclose%
\pgfusepath{stroke,fill}%
}%
\begin{pgfscope}%
\pgfsys@transformshift{0.699231in}{1.100070in}%
\pgfsys@useobject{currentmarker}{}%
\end{pgfscope}%
\begin{pgfscope}%
\pgfsys@transformshift{0.926377in}{1.041174in}%
\pgfsys@useobject{currentmarker}{}%
\end{pgfscope}%
\begin{pgfscope}%
\pgfsys@transformshift{1.153522in}{0.732059in}%
\pgfsys@useobject{currentmarker}{}%
\end{pgfscope}%
\begin{pgfscope}%
\pgfsys@transformshift{1.380668in}{0.655824in}%
\pgfsys@useobject{currentmarker}{}%
\end{pgfscope}%
\begin{pgfscope}%
\pgfsys@transformshift{1.607813in}{0.701280in}%
\pgfsys@useobject{currentmarker}{}%
\end{pgfscope}%
\begin{pgfscope}%
\pgfsys@transformshift{1.834959in}{0.693829in}%
\pgfsys@useobject{currentmarker}{}%
\end{pgfscope}%
\begin{pgfscope}%
\pgfsys@transformshift{2.062104in}{0.686746in}%
\pgfsys@useobject{currentmarker}{}%
\end{pgfscope}%
\begin{pgfscope}%
\pgfsys@transformshift{2.289250in}{0.695282in}%
\pgfsys@useobject{currentmarker}{}%
\end{pgfscope}%
\begin{pgfscope}%
\pgfsys@transformshift{2.516395in}{0.727081in}%
\pgfsys@useobject{currentmarker}{}%
\end{pgfscope}%
\begin{pgfscope}%
\pgfsys@transformshift{2.743541in}{0.862121in}%
\pgfsys@useobject{currentmarker}{}%
\end{pgfscope}%
\end{pgfscope}%
\begin{pgfscope}%
\pgfsetbuttcap%
\pgfsetroundjoin%
\definecolor{currentfill}{rgb}{0.000000,0.000000,0.000000}%
\pgfsetfillcolor{currentfill}%
\pgfsetlinewidth{0.803000pt}%
\definecolor{currentstroke}{rgb}{0.000000,0.000000,0.000000}%
\pgfsetstrokecolor{currentstroke}%
\pgfsetdash{}{0pt}%
\pgfsys@defobject{currentmarker}{\pgfqpoint{0.000000in}{-0.048611in}}{\pgfqpoint{0.000000in}{0.000000in}}{%
\pgfpathmoveto{\pgfqpoint{0.000000in}{0.000000in}}%
\pgfpathlineto{\pgfqpoint{0.000000in}{-0.048611in}}%
\pgfusepath{stroke,fill}%
}%
\begin{pgfscope}%
\pgfsys@transformshift{0.699231in}{0.498776in}%
\pgfsys@useobject{currentmarker}{}%
\end{pgfscope}%
\end{pgfscope}%
\begin{pgfscope}%
\definecolor{textcolor}{rgb}{0.000000,0.000000,0.000000}%
\pgfsetstrokecolor{textcolor}%
\pgfsetfillcolor{textcolor}%
\pgftext[x=0.699231in,y=0.408498in,,top]{\color{textcolor}{\rmfamily\fontsize{6.500000}{7.800000}\selectfont\catcode`\^=\active\def^{\ifmmode\sp\else\^{}\fi}\catcode`\%=\active\def%{\%}30}}%
\end{pgfscope}%
\begin{pgfscope}%
\pgfsetbuttcap%
\pgfsetroundjoin%
\definecolor{currentfill}{rgb}{0.000000,0.000000,0.000000}%
\pgfsetfillcolor{currentfill}%
\pgfsetlinewidth{0.803000pt}%
\definecolor{currentstroke}{rgb}{0.000000,0.000000,0.000000}%
\pgfsetstrokecolor{currentstroke}%
\pgfsetdash{}{0pt}%
\pgfsys@defobject{currentmarker}{\pgfqpoint{0.000000in}{-0.048611in}}{\pgfqpoint{0.000000in}{0.000000in}}{%
\pgfpathmoveto{\pgfqpoint{0.000000in}{0.000000in}}%
\pgfpathlineto{\pgfqpoint{0.000000in}{-0.048611in}}%
\pgfusepath{stroke,fill}%
}%
\begin{pgfscope}%
\pgfsys@transformshift{0.926377in}{0.498776in}%
\pgfsys@useobject{currentmarker}{}%
\end{pgfscope}%
\end{pgfscope}%
\begin{pgfscope}%
\definecolor{textcolor}{rgb}{0.000000,0.000000,0.000000}%
\pgfsetstrokecolor{textcolor}%
\pgfsetfillcolor{textcolor}%
\pgftext[x=0.926377in,y=0.408498in,,top]{\color{textcolor}{\rmfamily\fontsize{6.500000}{7.800000}\selectfont\catcode`\^=\active\def^{\ifmmode\sp\else\^{}\fi}\catcode`\%=\active\def%{\%}50}}%
\end{pgfscope}%
\begin{pgfscope}%
\pgfsetbuttcap%
\pgfsetroundjoin%
\definecolor{currentfill}{rgb}{0.000000,0.000000,0.000000}%
\pgfsetfillcolor{currentfill}%
\pgfsetlinewidth{0.803000pt}%
\definecolor{currentstroke}{rgb}{0.000000,0.000000,0.000000}%
\pgfsetstrokecolor{currentstroke}%
\pgfsetdash{}{0pt}%
\pgfsys@defobject{currentmarker}{\pgfqpoint{0.000000in}{-0.048611in}}{\pgfqpoint{0.000000in}{0.000000in}}{%
\pgfpathmoveto{\pgfqpoint{0.000000in}{0.000000in}}%
\pgfpathlineto{\pgfqpoint{0.000000in}{-0.048611in}}%
\pgfusepath{stroke,fill}%
}%
\begin{pgfscope}%
\pgfsys@transformshift{1.153522in}{0.498776in}%
\pgfsys@useobject{currentmarker}{}%
\end{pgfscope}%
\end{pgfscope}%
\begin{pgfscope}%
\definecolor{textcolor}{rgb}{0.000000,0.000000,0.000000}%
\pgfsetstrokecolor{textcolor}%
\pgfsetfillcolor{textcolor}%
\pgftext[x=1.153522in,y=0.408498in,,top]{\color{textcolor}{\rmfamily\fontsize{6.500000}{7.800000}\selectfont\catcode`\^=\active\def^{\ifmmode\sp\else\^{}\fi}\catcode`\%=\active\def%{\%}100}}%
\end{pgfscope}%
\begin{pgfscope}%
\pgfsetbuttcap%
\pgfsetroundjoin%
\definecolor{currentfill}{rgb}{0.000000,0.000000,0.000000}%
\pgfsetfillcolor{currentfill}%
\pgfsetlinewidth{0.803000pt}%
\definecolor{currentstroke}{rgb}{0.000000,0.000000,0.000000}%
\pgfsetstrokecolor{currentstroke}%
\pgfsetdash{}{0pt}%
\pgfsys@defobject{currentmarker}{\pgfqpoint{0.000000in}{-0.048611in}}{\pgfqpoint{0.000000in}{0.000000in}}{%
\pgfpathmoveto{\pgfqpoint{0.000000in}{0.000000in}}%
\pgfpathlineto{\pgfqpoint{0.000000in}{-0.048611in}}%
\pgfusepath{stroke,fill}%
}%
\begin{pgfscope}%
\pgfsys@transformshift{1.380668in}{0.498776in}%
\pgfsys@useobject{currentmarker}{}%
\end{pgfscope}%
\end{pgfscope}%
\begin{pgfscope}%
\definecolor{textcolor}{rgb}{0.000000,0.000000,0.000000}%
\pgfsetstrokecolor{textcolor}%
\pgfsetfillcolor{textcolor}%
\pgftext[x=1.380668in,y=0.408498in,,top]{\color{textcolor}{\rmfamily\fontsize{6.500000}{7.800000}\selectfont\catcode`\^=\active\def^{\ifmmode\sp\else\^{}\fi}\catcode`\%=\active\def%{\%}500}}%
\end{pgfscope}%
\begin{pgfscope}%
\pgfsetbuttcap%
\pgfsetroundjoin%
\definecolor{currentfill}{rgb}{0.000000,0.000000,0.000000}%
\pgfsetfillcolor{currentfill}%
\pgfsetlinewidth{0.803000pt}%
\definecolor{currentstroke}{rgb}{0.000000,0.000000,0.000000}%
\pgfsetstrokecolor{currentstroke}%
\pgfsetdash{}{0pt}%
\pgfsys@defobject{currentmarker}{\pgfqpoint{0.000000in}{-0.048611in}}{\pgfqpoint{0.000000in}{0.000000in}}{%
\pgfpathmoveto{\pgfqpoint{0.000000in}{0.000000in}}%
\pgfpathlineto{\pgfqpoint{0.000000in}{-0.048611in}}%
\pgfusepath{stroke,fill}%
}%
\begin{pgfscope}%
\pgfsys@transformshift{1.607813in}{0.498776in}%
\pgfsys@useobject{currentmarker}{}%
\end{pgfscope}%
\end{pgfscope}%
\begin{pgfscope}%
\definecolor{textcolor}{rgb}{0.000000,0.000000,0.000000}%
\pgfsetstrokecolor{textcolor}%
\pgfsetfillcolor{textcolor}%
\pgftext[x=1.607813in,y=0.408498in,,top]{\color{textcolor}{\rmfamily\fontsize{6.500000}{7.800000}\selectfont\catcode`\^=\active\def^{\ifmmode\sp\else\^{}\fi}\catcode`\%=\active\def%{\%}1000}}%
\end{pgfscope}%
\begin{pgfscope}%
\pgfsetbuttcap%
\pgfsetroundjoin%
\definecolor{currentfill}{rgb}{0.000000,0.000000,0.000000}%
\pgfsetfillcolor{currentfill}%
\pgfsetlinewidth{0.803000pt}%
\definecolor{currentstroke}{rgb}{0.000000,0.000000,0.000000}%
\pgfsetstrokecolor{currentstroke}%
\pgfsetdash{}{0pt}%
\pgfsys@defobject{currentmarker}{\pgfqpoint{0.000000in}{-0.048611in}}{\pgfqpoint{0.000000in}{0.000000in}}{%
\pgfpathmoveto{\pgfqpoint{0.000000in}{0.000000in}}%
\pgfpathlineto{\pgfqpoint{0.000000in}{-0.048611in}}%
\pgfusepath{stroke,fill}%
}%
\begin{pgfscope}%
\pgfsys@transformshift{1.834959in}{0.498776in}%
\pgfsys@useobject{currentmarker}{}%
\end{pgfscope}%
\end{pgfscope}%
\begin{pgfscope}%
\definecolor{textcolor}{rgb}{0.000000,0.000000,0.000000}%
\pgfsetstrokecolor{textcolor}%
\pgfsetfillcolor{textcolor}%
\pgftext[x=1.834959in,y=0.408498in,,top]{\color{textcolor}{\rmfamily\fontsize{6.500000}{7.800000}\selectfont\catcode`\^=\active\def^{\ifmmode\sp\else\^{}\fi}\catcode`\%=\active\def%{\%}2000}}%
\end{pgfscope}%
\begin{pgfscope}%
\pgfsetbuttcap%
\pgfsetroundjoin%
\definecolor{currentfill}{rgb}{0.000000,0.000000,0.000000}%
\pgfsetfillcolor{currentfill}%
\pgfsetlinewidth{0.803000pt}%
\definecolor{currentstroke}{rgb}{0.000000,0.000000,0.000000}%
\pgfsetstrokecolor{currentstroke}%
\pgfsetdash{}{0pt}%
\pgfsys@defobject{currentmarker}{\pgfqpoint{0.000000in}{-0.048611in}}{\pgfqpoint{0.000000in}{0.000000in}}{%
\pgfpathmoveto{\pgfqpoint{0.000000in}{0.000000in}}%
\pgfpathlineto{\pgfqpoint{0.000000in}{-0.048611in}}%
\pgfusepath{stroke,fill}%
}%
\begin{pgfscope}%
\pgfsys@transformshift{2.062104in}{0.498776in}%
\pgfsys@useobject{currentmarker}{}%
\end{pgfscope}%
\end{pgfscope}%
\begin{pgfscope}%
\definecolor{textcolor}{rgb}{0.000000,0.000000,0.000000}%
\pgfsetstrokecolor{textcolor}%
\pgfsetfillcolor{textcolor}%
\pgftext[x=2.062104in,y=0.408498in,,top]{\color{textcolor}{\rmfamily\fontsize{6.500000}{7.800000}\selectfont\catcode`\^=\active\def^{\ifmmode\sp\else\^{}\fi}\catcode`\%=\active\def%{\%}3000}}%
\end{pgfscope}%
\begin{pgfscope}%
\pgfsetbuttcap%
\pgfsetroundjoin%
\definecolor{currentfill}{rgb}{0.000000,0.000000,0.000000}%
\pgfsetfillcolor{currentfill}%
\pgfsetlinewidth{0.803000pt}%
\definecolor{currentstroke}{rgb}{0.000000,0.000000,0.000000}%
\pgfsetstrokecolor{currentstroke}%
\pgfsetdash{}{0pt}%
\pgfsys@defobject{currentmarker}{\pgfqpoint{0.000000in}{-0.048611in}}{\pgfqpoint{0.000000in}{0.000000in}}{%
\pgfpathmoveto{\pgfqpoint{0.000000in}{0.000000in}}%
\pgfpathlineto{\pgfqpoint{0.000000in}{-0.048611in}}%
\pgfusepath{stroke,fill}%
}%
\begin{pgfscope}%
\pgfsys@transformshift{2.289250in}{0.498776in}%
\pgfsys@useobject{currentmarker}{}%
\end{pgfscope}%
\end{pgfscope}%
\begin{pgfscope}%
\definecolor{textcolor}{rgb}{0.000000,0.000000,0.000000}%
\pgfsetstrokecolor{textcolor}%
\pgfsetfillcolor{textcolor}%
\pgftext[x=2.289250in,y=0.408498in,,top]{\color{textcolor}{\rmfamily\fontsize{6.500000}{7.800000}\selectfont\catcode`\^=\active\def^{\ifmmode\sp\else\^{}\fi}\catcode`\%=\active\def%{\%}4000}}%
\end{pgfscope}%
\begin{pgfscope}%
\pgfsetbuttcap%
\pgfsetroundjoin%
\definecolor{currentfill}{rgb}{0.000000,0.000000,0.000000}%
\pgfsetfillcolor{currentfill}%
\pgfsetlinewidth{0.803000pt}%
\definecolor{currentstroke}{rgb}{0.000000,0.000000,0.000000}%
\pgfsetstrokecolor{currentstroke}%
\pgfsetdash{}{0pt}%
\pgfsys@defobject{currentmarker}{\pgfqpoint{0.000000in}{-0.048611in}}{\pgfqpoint{0.000000in}{0.000000in}}{%
\pgfpathmoveto{\pgfqpoint{0.000000in}{0.000000in}}%
\pgfpathlineto{\pgfqpoint{0.000000in}{-0.048611in}}%
\pgfusepath{stroke,fill}%
}%
\begin{pgfscope}%
\pgfsys@transformshift{2.516395in}{0.498776in}%
\pgfsys@useobject{currentmarker}{}%
\end{pgfscope}%
\end{pgfscope}%
\begin{pgfscope}%
\definecolor{textcolor}{rgb}{0.000000,0.000000,0.000000}%
\pgfsetstrokecolor{textcolor}%
\pgfsetfillcolor{textcolor}%
\pgftext[x=2.516395in,y=0.408498in,,top]{\color{textcolor}{\rmfamily\fontsize{6.500000}{7.800000}\selectfont\catcode`\^=\active\def^{\ifmmode\sp\else\^{}\fi}\catcode`\%=\active\def%{\%}5000}}%
\end{pgfscope}%
\begin{pgfscope}%
\pgfsetbuttcap%
\pgfsetroundjoin%
\definecolor{currentfill}{rgb}{0.000000,0.000000,0.000000}%
\pgfsetfillcolor{currentfill}%
\pgfsetlinewidth{0.803000pt}%
\definecolor{currentstroke}{rgb}{0.000000,0.000000,0.000000}%
\pgfsetstrokecolor{currentstroke}%
\pgfsetdash{}{0pt}%
\pgfsys@defobject{currentmarker}{\pgfqpoint{0.000000in}{-0.048611in}}{\pgfqpoint{0.000000in}{0.000000in}}{%
\pgfpathmoveto{\pgfqpoint{0.000000in}{0.000000in}}%
\pgfpathlineto{\pgfqpoint{0.000000in}{-0.048611in}}%
\pgfusepath{stroke,fill}%
}%
\begin{pgfscope}%
\pgfsys@transformshift{2.743541in}{0.498776in}%
\pgfsys@useobject{currentmarker}{}%
\end{pgfscope}%
\end{pgfscope}%
\begin{pgfscope}%
\definecolor{textcolor}{rgb}{0.000000,0.000000,0.000000}%
\pgfsetstrokecolor{textcolor}%
\pgfsetfillcolor{textcolor}%
\pgftext[x=2.743541in,y=0.408498in,,top]{\color{textcolor}{\rmfamily\fontsize{6.500000}{7.800000}\selectfont\catcode`\^=\active\def^{\ifmmode\sp\else\^{}\fi}\catcode`\%=\active\def%{\%}6000}}%
\end{pgfscope}%
\begin{pgfscope}%
\definecolor{textcolor}{rgb}{0.000000,0.000000,0.000000}%
\pgfsetstrokecolor{textcolor}%
\pgfsetfillcolor{textcolor}%
\pgftext[x=1.721386in,y=0.278868in,,top]{\color{textcolor}{\rmfamily\fontsize{10.000000}{12.000000}\selectfont\catcode`\^=\active\def^{\ifmmode\sp\else\^{}\fi}\catcode`\%=\active\def%{\%}$\text{Re}$}}%
\end{pgfscope}%
\begin{pgfscope}%
\pgfsetbuttcap%
\pgfsetroundjoin%
\definecolor{currentfill}{rgb}{0.000000,0.000000,0.000000}%
\pgfsetfillcolor{currentfill}%
\pgfsetlinewidth{0.803000pt}%
\definecolor{currentstroke}{rgb}{0.000000,0.000000,0.000000}%
\pgfsetstrokecolor{currentstroke}%
\pgfsetdash{}{0pt}%
\pgfsys@defobject{currentmarker}{\pgfqpoint{-0.048611in}{0.000000in}}{\pgfqpoint{-0.000000in}{0.000000in}}{%
\pgfpathmoveto{\pgfqpoint{-0.000000in}{0.000000in}}%
\pgfpathlineto{\pgfqpoint{-0.048611in}{0.000000in}}%
\pgfusepath{stroke,fill}%
}%
\begin{pgfscope}%
\pgfsys@transformshift{0.597016in}{0.498776in}%
\pgfsys@useobject{currentmarker}{}%
\end{pgfscope}%
\end{pgfscope}%
\begin{pgfscope}%
\definecolor{textcolor}{rgb}{0.000000,0.000000,0.000000}%
\pgfsetstrokecolor{textcolor}%
\pgfsetfillcolor{textcolor}%
\pgftext[x=0.455813in, y=0.469841in, left, base]{\color{textcolor}{\rmfamily\fontsize{6.500000}{7.800000}\selectfont\catcode`\^=\active\def^{\ifmmode\sp\else\^{}\fi}\catcode`\%=\active\def%{\%}$\mathdefault{0}$}}%
\end{pgfscope}%
\begin{pgfscope}%
\pgfsetbuttcap%
\pgfsetroundjoin%
\definecolor{currentfill}{rgb}{0.000000,0.000000,0.000000}%
\pgfsetfillcolor{currentfill}%
\pgfsetlinewidth{0.803000pt}%
\definecolor{currentstroke}{rgb}{0.000000,0.000000,0.000000}%
\pgfsetstrokecolor{currentstroke}%
\pgfsetdash{}{0pt}%
\pgfsys@defobject{currentmarker}{\pgfqpoint{-0.048611in}{0.000000in}}{\pgfqpoint{-0.000000in}{0.000000in}}{%
\pgfpathmoveto{\pgfqpoint{-0.000000in}{0.000000in}}%
\pgfpathlineto{\pgfqpoint{-0.048611in}{0.000000in}}%
\pgfusepath{stroke,fill}%
}%
\begin{pgfscope}%
\pgfsys@transformshift{0.597016in}{0.938774in}%
\pgfsys@useobject{currentmarker}{}%
\end{pgfscope}%
\end{pgfscope}%
\begin{pgfscope}%
\definecolor{textcolor}{rgb}{0.000000,0.000000,0.000000}%
\pgfsetstrokecolor{textcolor}%
\pgfsetfillcolor{textcolor}%
\pgftext[x=0.404888in, y=0.909839in, left, base]{\color{textcolor}{\rmfamily\fontsize{6.500000}{7.800000}\selectfont\catcode`\^=\active\def^{\ifmmode\sp\else\^{}\fi}\catcode`\%=\active\def%{\%}$\mathdefault{10}$}}%
\end{pgfscope}%
\begin{pgfscope}%
\pgfsetbuttcap%
\pgfsetroundjoin%
\definecolor{currentfill}{rgb}{0.000000,0.000000,0.000000}%
\pgfsetfillcolor{currentfill}%
\pgfsetlinewidth{0.803000pt}%
\definecolor{currentstroke}{rgb}{0.000000,0.000000,0.000000}%
\pgfsetstrokecolor{currentstroke}%
\pgfsetdash{}{0pt}%
\pgfsys@defobject{currentmarker}{\pgfqpoint{-0.048611in}{0.000000in}}{\pgfqpoint{-0.000000in}{0.000000in}}{%
\pgfpathmoveto{\pgfqpoint{-0.000000in}{0.000000in}}%
\pgfpathlineto{\pgfqpoint{-0.048611in}{0.000000in}}%
\pgfusepath{stroke,fill}%
}%
\begin{pgfscope}%
\pgfsys@transformshift{0.597016in}{1.378772in}%
\pgfsys@useobject{currentmarker}{}%
\end{pgfscope}%
\end{pgfscope}%
\begin{pgfscope}%
\definecolor{textcolor}{rgb}{0.000000,0.000000,0.000000}%
\pgfsetstrokecolor{textcolor}%
\pgfsetfillcolor{textcolor}%
\pgftext[x=0.404888in, y=1.349837in, left, base]{\color{textcolor}{\rmfamily\fontsize{6.500000}{7.800000}\selectfont\catcode`\^=\active\def^{\ifmmode\sp\else\^{}\fi}\catcode`\%=\active\def%{\%}$\mathdefault{20}$}}%
\end{pgfscope}%
\begin{pgfscope}%
\pgfsetbuttcap%
\pgfsetroundjoin%
\definecolor{currentfill}{rgb}{0.000000,0.000000,0.000000}%
\pgfsetfillcolor{currentfill}%
\pgfsetlinewidth{0.803000pt}%
\definecolor{currentstroke}{rgb}{0.000000,0.000000,0.000000}%
\pgfsetstrokecolor{currentstroke}%
\pgfsetdash{}{0pt}%
\pgfsys@defobject{currentmarker}{\pgfqpoint{-0.048611in}{0.000000in}}{\pgfqpoint{-0.000000in}{0.000000in}}{%
\pgfpathmoveto{\pgfqpoint{-0.000000in}{0.000000in}}%
\pgfpathlineto{\pgfqpoint{-0.048611in}{0.000000in}}%
\pgfusepath{stroke,fill}%
}%
\begin{pgfscope}%
\pgfsys@transformshift{0.597016in}{1.818771in}%
\pgfsys@useobject{currentmarker}{}%
\end{pgfscope}%
\end{pgfscope}%
\begin{pgfscope}%
\definecolor{textcolor}{rgb}{0.000000,0.000000,0.000000}%
\pgfsetstrokecolor{textcolor}%
\pgfsetfillcolor{textcolor}%
\pgftext[x=0.404888in, y=1.789835in, left, base]{\color{textcolor}{\rmfamily\fontsize{6.500000}{7.800000}\selectfont\catcode`\^=\active\def^{\ifmmode\sp\else\^{}\fi}\catcode`\%=\active\def%{\%}$\mathdefault{30}$}}%
\end{pgfscope}%
\begin{pgfscope}%
\definecolor{textcolor}{rgb}{0.000000,0.000000,0.000000}%
\pgfsetstrokecolor{textcolor}%
\pgfsetfillcolor{textcolor}%
\pgftext[x=0.349332in,y=1.158773in,,bottom,rotate=90.000000]{\color{textcolor}{\rmfamily\fontsize{10.000000}{12.000000}\selectfont\catcode`\^=\active\def^{\ifmmode\sp\else\^{}\fi}\catcode`\%=\active\def%{\%}$\delta_{\ell^1}^{(q)} \ [\%]$}}%
\end{pgfscope}%
\begin{pgfscope}%
\pgfsetrectcap%
\pgfsetmiterjoin%
\pgfsetlinewidth{0.803000pt}%
\definecolor{currentstroke}{rgb}{0.000000,0.000000,0.000000}%
\pgfsetstrokecolor{currentstroke}%
\pgfsetdash{}{0pt}%
\pgfpathmoveto{\pgfqpoint{0.597016in}{0.498776in}}%
\pgfpathlineto{\pgfqpoint{0.597016in}{1.818771in}}%
\pgfusepath{stroke}%
\end{pgfscope}%
\begin{pgfscope}%
\pgfsetrectcap%
\pgfsetmiterjoin%
\pgfsetlinewidth{0.803000pt}%
\definecolor{currentstroke}{rgb}{0.000000,0.000000,0.000000}%
\pgfsetstrokecolor{currentstroke}%
\pgfsetdash{}{0pt}%
\pgfpathmoveto{\pgfqpoint{2.845756in}{0.498776in}}%
\pgfpathlineto{\pgfqpoint{2.845756in}{1.818771in}}%
\pgfusepath{stroke}%
\end{pgfscope}%
\begin{pgfscope}%
\pgfsetrectcap%
\pgfsetmiterjoin%
\pgfsetlinewidth{0.803000pt}%
\definecolor{currentstroke}{rgb}{0.000000,0.000000,0.000000}%
\pgfsetstrokecolor{currentstroke}%
\pgfsetdash{}{0pt}%
\pgfpathmoveto{\pgfqpoint{0.597016in}{0.498776in}}%
\pgfpathlineto{\pgfqpoint{2.845756in}{0.498776in}}%
\pgfusepath{stroke}%
\end{pgfscope}%
\begin{pgfscope}%
\pgfsetrectcap%
\pgfsetmiterjoin%
\pgfsetlinewidth{0.803000pt}%
\definecolor{currentstroke}{rgb}{0.000000,0.000000,0.000000}%
\pgfsetstrokecolor{currentstroke}%
\pgfsetdash{}{0pt}%
\pgfpathmoveto{\pgfqpoint{0.597016in}{1.818771in}}%
\pgfpathlineto{\pgfqpoint{2.845756in}{1.818771in}}%
\pgfusepath{stroke}%
\end{pgfscope}%
\begin{pgfscope}%
\pgfsetbuttcap%
\pgfsetmiterjoin%
\definecolor{currentfill}{rgb}{1.000000,1.000000,1.000000}%
\pgfsetfillcolor{currentfill}%
\pgfsetfillopacity{0.800000}%
\pgfsetlinewidth{1.003750pt}%
\definecolor{currentstroke}{rgb}{0.800000,0.800000,0.800000}%
\pgfsetstrokecolor{currentstroke}%
\pgfsetstrokeopacity{0.800000}%
\pgfsetdash{}{0pt}%
\pgfpathmoveto{\pgfqpoint{1.468512in}{1.348761in}}%
\pgfpathlineto{\pgfqpoint{2.764770in}{1.348761in}}%
\pgfpathquadraticcurveto{\pgfqpoint{2.787909in}{1.348761in}}{\pgfqpoint{2.787909in}{1.371900in}}%
\pgfpathlineto{\pgfqpoint{2.787909in}{1.737785in}}%
\pgfpathquadraticcurveto{\pgfqpoint{2.787909in}{1.760923in}}{\pgfqpoint{2.764770in}{1.760923in}}%
\pgfpathlineto{\pgfqpoint{1.468512in}{1.760923in}}%
\pgfpathquadraticcurveto{\pgfqpoint{1.445374in}{1.760923in}}{\pgfqpoint{1.445374in}{1.737785in}}%
\pgfpathlineto{\pgfqpoint{1.445374in}{1.371900in}}%
\pgfpathquadraticcurveto{\pgfqpoint{1.445374in}{1.348761in}}{\pgfqpoint{1.468512in}{1.348761in}}%
\pgfpathlineto{\pgfqpoint{1.468512in}{1.348761in}}%
\pgfpathclose%
\pgfusepath{stroke,fill}%
\end{pgfscope}%
\begin{pgfscope}%
\pgfsetbuttcap%
\pgfsetroundjoin%
\definecolor{currentfill}{rgb}{0.172549,0.482353,0.713725}%
\pgfsetfillcolor{currentfill}%
\pgfsetlinewidth{0.501875pt}%
\definecolor{currentstroke}{rgb}{0.000000,0.000000,0.000000}%
\pgfsetstrokecolor{currentstroke}%
\pgfsetdash{}{0pt}%
\pgfsys@defobject{currentmarker}{\pgfqpoint{-0.026896in}{-0.026896in}}{\pgfqpoint{0.026896in}{0.026896in}}{%
\pgfpathmoveto{\pgfqpoint{-0.026896in}{-0.026896in}}%
\pgfpathlineto{\pgfqpoint{0.026896in}{-0.026896in}}%
\pgfpathlineto{\pgfqpoint{0.026896in}{0.026896in}}%
\pgfpathlineto{\pgfqpoint{-0.026896in}{0.026896in}}%
\pgfpathlineto{\pgfqpoint{-0.026896in}{-0.026896in}}%
\pgfpathclose%
\pgfusepath{stroke,fill}%
}%
\begin{pgfscope}%
\pgfsys@transformshift{1.607346in}{1.641914in}%
\pgfsys@useobject{currentmarker}{}%
\end{pgfscope}%
\end{pgfscope}%
\begin{pgfscope}%
\definecolor{textcolor}{rgb}{0.000000,0.000000,0.000000}%
\pgfsetstrokecolor{textcolor}%
\pgfsetfillcolor{textcolor}%
\pgftext[x=1.815596in,y=1.611544in,left,base]{\color{textcolor}{\rmfamily\fontsize{8.330000}{9.996000}\selectfont\catcode`\^=\active\def^{\ifmmode\sp\else\^{}\fi}\catcode`\%=\active\def%{\%}$p^\text{II}_{64}$}}%
\end{pgfscope}%
\begin{pgfscope}%
\pgfsetbuttcap%
\pgfsetroundjoin%
\definecolor{currentfill}{rgb}{0.843137,0.098039,0.109804}%
\pgfsetfillcolor{currentfill}%
\pgfsetlinewidth{0.501875pt}%
\definecolor{currentstroke}{rgb}{0.000000,0.000000,0.000000}%
\pgfsetstrokecolor{currentstroke}%
\pgfsetdash{}{0pt}%
\pgfsys@defobject{currentmarker}{\pgfqpoint{-0.026896in}{-0.026896in}}{\pgfqpoint{0.026896in}{0.026896in}}{%
\pgfpathmoveto{\pgfqpoint{0.000000in}{-0.026896in}}%
\pgfpathcurveto{\pgfqpoint{0.007133in}{-0.026896in}}{\pgfqpoint{0.013974in}{-0.024062in}}{\pgfqpoint{0.019018in}{-0.019018in}}%
\pgfpathcurveto{\pgfqpoint{0.024062in}{-0.013974in}}{\pgfqpoint{0.026896in}{-0.007133in}}{\pgfqpoint{0.026896in}{0.000000in}}%
\pgfpathcurveto{\pgfqpoint{0.026896in}{0.007133in}}{\pgfqpoint{0.024062in}{0.013974in}}{\pgfqpoint{0.019018in}{0.019018in}}%
\pgfpathcurveto{\pgfqpoint{0.013974in}{0.024062in}}{\pgfqpoint{0.007133in}{0.026896in}}{\pgfqpoint{0.000000in}{0.026896in}}%
\pgfpathcurveto{\pgfqpoint{-0.007133in}{0.026896in}}{\pgfqpoint{-0.013974in}{0.024062in}}{\pgfqpoint{-0.019018in}{0.019018in}}%
\pgfpathcurveto{\pgfqpoint{-0.024062in}{0.013974in}}{\pgfqpoint{-0.026896in}{0.007133in}}{\pgfqpoint{-0.026896in}{0.000000in}}%
\pgfpathcurveto{\pgfqpoint{-0.026896in}{-0.007133in}}{\pgfqpoint{-0.024062in}{-0.013974in}}{\pgfqpoint{-0.019018in}{-0.019018in}}%
\pgfpathcurveto{\pgfqpoint{-0.013974in}{-0.024062in}}{\pgfqpoint{-0.007133in}{-0.026896in}}{\pgfqpoint{0.000000in}{-0.026896in}}%
\pgfpathlineto{\pgfqpoint{0.000000in}{-0.026896in}}%
\pgfpathclose%
\pgfusepath{stroke,fill}%
}%
\begin{pgfscope}%
\pgfsys@transformshift{1.607346in}{1.453187in}%
\pgfsys@useobject{currentmarker}{}%
\end{pgfscope}%
\end{pgfscope}%
\begin{pgfscope}%
\definecolor{textcolor}{rgb}{0.000000,0.000000,0.000000}%
\pgfsetstrokecolor{textcolor}%
\pgfsetfillcolor{textcolor}%
\pgftext[x=1.815596in,y=1.422817in,left,base]{\color{textcolor}{\rmfamily\fontsize{8.330000}{9.996000}\selectfont\catcode`\^=\active\def^{\ifmmode\sp\else\^{}\fi}\catcode`\%=\active\def%{\%}$\bm{v}^\text{II}_{64}$}}%
\end{pgfscope}%
\begin{pgfscope}%
\pgfsetbuttcap%
\pgfsetroundjoin%
\definecolor{currentfill}{rgb}{0.670588,0.850980,0.913725}%
\pgfsetfillcolor{currentfill}%
\pgfsetlinewidth{0.501875pt}%
\definecolor{currentstroke}{rgb}{0.000000,0.000000,0.000000}%
\pgfsetstrokecolor{currentstroke}%
\pgfsetdash{}{0pt}%
\pgfsys@defobject{currentmarker}{\pgfqpoint{-0.038036in}{-0.038036in}}{\pgfqpoint{0.038036in}{0.038036in}}{%
\pgfpathmoveto{\pgfqpoint{-0.000000in}{-0.038036in}}%
\pgfpathlineto{\pgfqpoint{0.038036in}{0.000000in}}%
\pgfpathlineto{\pgfqpoint{0.000000in}{0.038036in}}%
\pgfpathlineto{\pgfqpoint{-0.038036in}{0.000000in}}%
\pgfpathlineto{\pgfqpoint{-0.000000in}{-0.038036in}}%
\pgfpathclose%
\pgfusepath{stroke,fill}%
}%
\begin{pgfscope}%
\pgfsys@transformshift{2.338874in}{1.641914in}%
\pgfsys@useobject{currentmarker}{}%
\end{pgfscope}%
\end{pgfscope}%
\begin{pgfscope}%
\definecolor{textcolor}{rgb}{0.000000,0.000000,0.000000}%
\pgfsetstrokecolor{textcolor}%
\pgfsetfillcolor{textcolor}%
\pgftext[x=2.547124in,y=1.611544in,left,base]{\color{textcolor}{\rmfamily\fontsize{8.330000}{9.996000}\selectfont\catcode`\^=\active\def^{\ifmmode\sp\else\^{}\fi}\catcode`\%=\active\def%{\%}$p^\text{III}_{64}$}}%
\end{pgfscope}%
\begin{pgfscope}%
\pgfsetbuttcap%
\pgfsetroundjoin%
\definecolor{currentfill}{rgb}{0.992157,0.682353,0.380392}%
\pgfsetfillcolor{currentfill}%
\pgfsetlinewidth{0.501875pt}%
\definecolor{currentstroke}{rgb}{0.000000,0.000000,0.000000}%
\pgfsetstrokecolor{currentstroke}%
\pgfsetdash{}{0pt}%
\pgfsys@defobject{currentmarker}{\pgfqpoint{-0.026896in}{-0.026896in}}{\pgfqpoint{0.026896in}{0.026896in}}{%
\pgfpathmoveto{\pgfqpoint{0.000000in}{0.026896in}}%
\pgfpathlineto{\pgfqpoint{-0.026896in}{-0.026896in}}%
\pgfpathlineto{\pgfqpoint{0.026896in}{-0.026896in}}%
\pgfpathlineto{\pgfqpoint{0.000000in}{0.026896in}}%
\pgfpathclose%
\pgfusepath{stroke,fill}%
}%
\begin{pgfscope}%
\pgfsys@transformshift{2.338874in}{1.453187in}%
\pgfsys@useobject{currentmarker}{}%
\end{pgfscope}%
\end{pgfscope}%
\begin{pgfscope}%
\definecolor{textcolor}{rgb}{0.000000,0.000000,0.000000}%
\pgfsetstrokecolor{textcolor}%
\pgfsetfillcolor{textcolor}%
\pgftext[x=2.547124in,y=1.422817in,left,base]{\color{textcolor}{\rmfamily\fontsize{8.330000}{9.996000}\selectfont\catcode`\^=\active\def^{\ifmmode\sp\else\^{}\fi}\catcode`\%=\active\def%{\%}$\bm{v}^\text{III}_{64}$}}%
\end{pgfscope}%
\end{pgfpicture}%
\makeatother%
\endgroup%

%% file: figures/section2/errors_all_Re_10k.pgf
%% Creator: Matplotlib, PGF backend
%%
%% To include the figure in your LaTeX document, write
%%   \input{<filename>.pgf}
%%
%% Make sure the required packages are loaded in your preamble
%%   \usepackage{pgf}
%%
%% Also ensure that all the required font packages are loaded; for instance,
%% the lmodern package is sometimes necessary when using math font.
%%   \usepackage{lmodern}
%%
%% Figures using additional raster images can only be included by \input if
%% they are in the same directory as the main LaTeX file. For loading figures
%% from other directories you can use the `import` package
%% and then include the figures with
%%   \import{<path to file>}{<filename>.pgf}
%%
%% Matplotlib used the following preamble
%%   \def\mathdefault#1{#1}
%%   \everymath=\expandafter{\the\everymath\displaystyle}
%%   \usepackage{amsmath}\usepackage{bm}
%%   \makeatletter\@ifpackageloaded{underscore}{}{\usepackage[strings]{underscore}}\makeatother
%%
\begingroup%
\makeatletter%
\begin{pgfpicture}%
\pgfpathrectangle{\pgfpointorigin}{\pgfqpoint{3.000000in}{2.000000in}}%
\pgfusepath{use as bounding box, clip}%
\begin{pgfscope}%
\pgfsetbuttcap%
\pgfsetmiterjoin%
\definecolor{currentfill}{rgb}{1.000000,1.000000,1.000000}%
\pgfsetfillcolor{currentfill}%
\pgfsetlinewidth{0.000000pt}%
\definecolor{currentstroke}{rgb}{1.000000,1.000000,1.000000}%
\pgfsetstrokecolor{currentstroke}%
\pgfsetdash{}{0pt}%
\pgfpathmoveto{\pgfqpoint{0.000000in}{0.000000in}}%
\pgfpathlineto{\pgfqpoint{3.000000in}{0.000000in}}%
\pgfpathlineto{\pgfqpoint{3.000000in}{2.000000in}}%
\pgfpathlineto{\pgfqpoint{0.000000in}{2.000000in}}%
\pgfpathlineto{\pgfqpoint{0.000000in}{0.000000in}}%
\pgfpathclose%
\pgfusepath{fill}%
\end{pgfscope}%
\begin{pgfscope}%
\pgfsetbuttcap%
\pgfsetmiterjoin%
\definecolor{currentfill}{rgb}{1.000000,1.000000,1.000000}%
\pgfsetfillcolor{currentfill}%
\pgfsetlinewidth{0.000000pt}%
\definecolor{currentstroke}{rgb}{0.000000,0.000000,0.000000}%
\pgfsetstrokecolor{currentstroke}%
\pgfsetstrokeopacity{0.000000}%
\pgfsetdash{}{0pt}%
\pgfpathmoveto{\pgfqpoint{0.597016in}{0.498776in}}%
\pgfpathlineto{\pgfqpoint{2.845756in}{0.498776in}}%
\pgfpathlineto{\pgfqpoint{2.845756in}{1.818771in}}%
\pgfpathlineto{\pgfqpoint{0.597016in}{1.818771in}}%
\pgfpathlineto{\pgfqpoint{0.597016in}{0.498776in}}%
\pgfpathclose%
\pgfusepath{fill}%
\end{pgfscope}%
\begin{pgfscope}%
\pgfpathrectangle{\pgfqpoint{0.597016in}{0.498776in}}{\pgfqpoint{2.248740in}{1.319995in}}%
\pgfusepath{clip}%
\pgfsetbuttcap%
\pgfsetroundjoin%
\definecolor{currentfill}{rgb}{0.172549,0.482353,0.713725}%
\pgfsetfillcolor{currentfill}%
\pgfsetlinewidth{0.501875pt}%
\definecolor{currentstroke}{rgb}{0.000000,0.000000,0.000000}%
\pgfsetstrokecolor{currentstroke}%
\pgfsetdash{}{0pt}%
\pgfsys@defobject{currentmarker}{\pgfqpoint{-0.026896in}{-0.026896in}}{\pgfqpoint{0.026896in}{0.026896in}}{%
\pgfpathmoveto{\pgfqpoint{-0.026896in}{-0.026896in}}%
\pgfpathlineto{\pgfqpoint{0.026896in}{-0.026896in}}%
\pgfpathlineto{\pgfqpoint{0.026896in}{0.026896in}}%
\pgfpathlineto{\pgfqpoint{-0.026896in}{0.026896in}}%
\pgfpathlineto{\pgfqpoint{-0.026896in}{-0.026896in}}%
\pgfpathclose%
\pgfusepath{stroke,fill}%
}%
\begin{pgfscope}%
\pgfsys@transformshift{0.699231in}{1.484879in}%
\pgfsys@useobject{currentmarker}{}%
\end{pgfscope}%
\begin{pgfscope}%
\pgfsys@transformshift{0.926377in}{1.318327in}%
\pgfsys@useobject{currentmarker}{}%
\end{pgfscope}%
\begin{pgfscope}%
\pgfsys@transformshift{1.153522in}{0.609418in}%
\pgfsys@useobject{currentmarker}{}%
\end{pgfscope}%
\begin{pgfscope}%
\pgfsys@transformshift{1.380668in}{0.538627in}%
\pgfsys@useobject{currentmarker}{}%
\end{pgfscope}%
\begin{pgfscope}%
\pgfsys@transformshift{1.607813in}{0.608277in}%
\pgfsys@useobject{currentmarker}{}%
\end{pgfscope}%
\begin{pgfscope}%
\pgfsys@transformshift{1.834959in}{0.638649in}%
\pgfsys@useobject{currentmarker}{}%
\end{pgfscope}%
\begin{pgfscope}%
\pgfsys@transformshift{2.062104in}{0.651877in}%
\pgfsys@useobject{currentmarker}{}%
\end{pgfscope}%
\begin{pgfscope}%
\pgfsys@transformshift{2.289250in}{0.667148in}%
\pgfsys@useobject{currentmarker}{}%
\end{pgfscope}%
\begin{pgfscope}%
\pgfsys@transformshift{2.516395in}{0.676101in}%
\pgfsys@useobject{currentmarker}{}%
\end{pgfscope}%
\begin{pgfscope}%
\pgfsys@transformshift{2.743541in}{0.717253in}%
\pgfsys@useobject{currentmarker}{}%
\end{pgfscope}%
\end{pgfscope}%
\begin{pgfscope}%
\pgfpathrectangle{\pgfqpoint{0.597016in}{0.498776in}}{\pgfqpoint{2.248740in}{1.319995in}}%
\pgfusepath{clip}%
\pgfsetbuttcap%
\pgfsetroundjoin%
\definecolor{currentfill}{rgb}{0.843137,0.098039,0.109804}%
\pgfsetfillcolor{currentfill}%
\pgfsetlinewidth{0.501875pt}%
\definecolor{currentstroke}{rgb}{0.000000,0.000000,0.000000}%
\pgfsetstrokecolor{currentstroke}%
\pgfsetdash{}{0pt}%
\pgfsys@defobject{currentmarker}{\pgfqpoint{-0.026896in}{-0.026896in}}{\pgfqpoint{0.026896in}{0.026896in}}{%
\pgfpathmoveto{\pgfqpoint{0.000000in}{-0.026896in}}%
\pgfpathcurveto{\pgfqpoint{0.007133in}{-0.026896in}}{\pgfqpoint{0.013974in}{-0.024062in}}{\pgfqpoint{0.019018in}{-0.019018in}}%
\pgfpathcurveto{\pgfqpoint{0.024062in}{-0.013974in}}{\pgfqpoint{0.026896in}{-0.007133in}}{\pgfqpoint{0.026896in}{0.000000in}}%
\pgfpathcurveto{\pgfqpoint{0.026896in}{0.007133in}}{\pgfqpoint{0.024062in}{0.013974in}}{\pgfqpoint{0.019018in}{0.019018in}}%
\pgfpathcurveto{\pgfqpoint{0.013974in}{0.024062in}}{\pgfqpoint{0.007133in}{0.026896in}}{\pgfqpoint{0.000000in}{0.026896in}}%
\pgfpathcurveto{\pgfqpoint{-0.007133in}{0.026896in}}{\pgfqpoint{-0.013974in}{0.024062in}}{\pgfqpoint{-0.019018in}{0.019018in}}%
\pgfpathcurveto{\pgfqpoint{-0.024062in}{0.013974in}}{\pgfqpoint{-0.026896in}{0.007133in}}{\pgfqpoint{-0.026896in}{0.000000in}}%
\pgfpathcurveto{\pgfqpoint{-0.026896in}{-0.007133in}}{\pgfqpoint{-0.024062in}{-0.013974in}}{\pgfqpoint{-0.019018in}{-0.019018in}}%
\pgfpathcurveto{\pgfqpoint{-0.013974in}{-0.024062in}}{\pgfqpoint{-0.007133in}{-0.026896in}}{\pgfqpoint{0.000000in}{-0.026896in}}%
\pgfpathlineto{\pgfqpoint{0.000000in}{-0.026896in}}%
\pgfpathclose%
\pgfusepath{stroke,fill}%
}%
\begin{pgfscope}%
\pgfsys@transformshift{0.699231in}{1.023179in}%
\pgfsys@useobject{currentmarker}{}%
\end{pgfscope}%
\begin{pgfscope}%
\pgfsys@transformshift{0.926377in}{0.963481in}%
\pgfsys@useobject{currentmarker}{}%
\end{pgfscope}%
\begin{pgfscope}%
\pgfsys@transformshift{1.153522in}{0.654674in}%
\pgfsys@useobject{currentmarker}{}%
\end{pgfscope}%
\begin{pgfscope}%
\pgfsys@transformshift{1.380668in}{0.590219in}%
\pgfsys@useobject{currentmarker}{}%
\end{pgfscope}%
\begin{pgfscope}%
\pgfsys@transformshift{1.607813in}{0.668052in}%
\pgfsys@useobject{currentmarker}{}%
\end{pgfscope}%
\begin{pgfscope}%
\pgfsys@transformshift{1.834959in}{0.697927in}%
\pgfsys@useobject{currentmarker}{}%
\end{pgfscope}%
\begin{pgfscope}%
\pgfsys@transformshift{2.062104in}{0.700376in}%
\pgfsys@useobject{currentmarker}{}%
\end{pgfscope}%
\begin{pgfscope}%
\pgfsys@transformshift{2.289250in}{0.700398in}%
\pgfsys@useobject{currentmarker}{}%
\end{pgfscope}%
\begin{pgfscope}%
\pgfsys@transformshift{2.516395in}{0.697868in}%
\pgfsys@useobject{currentmarker}{}%
\end{pgfscope}%
\begin{pgfscope}%
\pgfsys@transformshift{2.743541in}{0.735512in}%
\pgfsys@useobject{currentmarker}{}%
\end{pgfscope}%
\end{pgfscope}%
\begin{pgfscope}%
\pgfpathrectangle{\pgfqpoint{0.597016in}{0.498776in}}{\pgfqpoint{2.248740in}{1.319995in}}%
\pgfusepath{clip}%
\pgfsetbuttcap%
\pgfsetroundjoin%
\definecolor{currentfill}{rgb}{0.670588,0.850980,0.913725}%
\pgfsetfillcolor{currentfill}%
\pgfsetlinewidth{0.501875pt}%
\definecolor{currentstroke}{rgb}{0.000000,0.000000,0.000000}%
\pgfsetstrokecolor{currentstroke}%
\pgfsetdash{}{0pt}%
\pgfsys@defobject{currentmarker}{\pgfqpoint{-0.038036in}{-0.038036in}}{\pgfqpoint{0.038036in}{0.038036in}}{%
\pgfpathmoveto{\pgfqpoint{-0.000000in}{-0.038036in}}%
\pgfpathlineto{\pgfqpoint{0.038036in}{0.000000in}}%
\pgfpathlineto{\pgfqpoint{0.000000in}{0.038036in}}%
\pgfpathlineto{\pgfqpoint{-0.038036in}{0.000000in}}%
\pgfpathlineto{\pgfqpoint{-0.000000in}{-0.038036in}}%
\pgfpathclose%
\pgfusepath{stroke,fill}%
}%
\begin{pgfscope}%
\pgfsys@transformshift{0.699231in}{1.475366in}%
\pgfsys@useobject{currentmarker}{}%
\end{pgfscope}%
\begin{pgfscope}%
\pgfsys@transformshift{0.926377in}{1.309158in}%
\pgfsys@useobject{currentmarker}{}%
\end{pgfscope}%
\begin{pgfscope}%
\pgfsys@transformshift{1.153522in}{0.609205in}%
\pgfsys@useobject{currentmarker}{}%
\end{pgfscope}%
\begin{pgfscope}%
\pgfsys@transformshift{1.380668in}{0.545558in}%
\pgfsys@useobject{currentmarker}{}%
\end{pgfscope}%
\begin{pgfscope}%
\pgfsys@transformshift{1.607813in}{0.609103in}%
\pgfsys@useobject{currentmarker}{}%
\end{pgfscope}%
\begin{pgfscope}%
\pgfsys@transformshift{1.834959in}{0.635408in}%
\pgfsys@useobject{currentmarker}{}%
\end{pgfscope}%
\begin{pgfscope}%
\pgfsys@transformshift{2.062104in}{0.649222in}%
\pgfsys@useobject{currentmarker}{}%
\end{pgfscope}%
\begin{pgfscope}%
\pgfsys@transformshift{2.289250in}{0.661990in}%
\pgfsys@useobject{currentmarker}{}%
\end{pgfscope}%
\begin{pgfscope}%
\pgfsys@transformshift{2.516395in}{0.666938in}%
\pgfsys@useobject{currentmarker}{}%
\end{pgfscope}%
\begin{pgfscope}%
\pgfsys@transformshift{2.743541in}{0.725384in}%
\pgfsys@useobject{currentmarker}{}%
\end{pgfscope}%
\end{pgfscope}%
\begin{pgfscope}%
\pgfpathrectangle{\pgfqpoint{0.597016in}{0.498776in}}{\pgfqpoint{2.248740in}{1.319995in}}%
\pgfusepath{clip}%
\pgfsetbuttcap%
\pgfsetroundjoin%
\definecolor{currentfill}{rgb}{0.992157,0.682353,0.380392}%
\pgfsetfillcolor{currentfill}%
\pgfsetlinewidth{0.501875pt}%
\definecolor{currentstroke}{rgb}{0.000000,0.000000,0.000000}%
\pgfsetstrokecolor{currentstroke}%
\pgfsetdash{}{0pt}%
\pgfsys@defobject{currentmarker}{\pgfqpoint{-0.026896in}{-0.026896in}}{\pgfqpoint{0.026896in}{0.026896in}}{%
\pgfpathmoveto{\pgfqpoint{0.000000in}{0.026896in}}%
\pgfpathlineto{\pgfqpoint{-0.026896in}{-0.026896in}}%
\pgfpathlineto{\pgfqpoint{0.026896in}{-0.026896in}}%
\pgfpathlineto{\pgfqpoint{0.000000in}{0.026896in}}%
\pgfpathclose%
\pgfusepath{stroke,fill}%
}%
\begin{pgfscope}%
\pgfsys@transformshift{0.699231in}{1.048511in}%
\pgfsys@useobject{currentmarker}{}%
\end{pgfscope}%
\begin{pgfscope}%
\pgfsys@transformshift{0.926377in}{0.988231in}%
\pgfsys@useobject{currentmarker}{}%
\end{pgfscope}%
\begin{pgfscope}%
\pgfsys@transformshift{1.153522in}{0.682248in}%
\pgfsys@useobject{currentmarker}{}%
\end{pgfscope}%
\begin{pgfscope}%
\pgfsys@transformshift{1.380668in}{0.617952in}%
\pgfsys@useobject{currentmarker}{}%
\end{pgfscope}%
\begin{pgfscope}%
\pgfsys@transformshift{1.607813in}{0.685162in}%
\pgfsys@useobject{currentmarker}{}%
\end{pgfscope}%
\begin{pgfscope}%
\pgfsys@transformshift{1.834959in}{0.708261in}%
\pgfsys@useobject{currentmarker}{}%
\end{pgfscope}%
\begin{pgfscope}%
\pgfsys@transformshift{2.062104in}{0.707482in}%
\pgfsys@useobject{currentmarker}{}%
\end{pgfscope}%
\begin{pgfscope}%
\pgfsys@transformshift{2.289250in}{0.707145in}%
\pgfsys@useobject{currentmarker}{}%
\end{pgfscope}%
\begin{pgfscope}%
\pgfsys@transformshift{2.516395in}{0.706724in}%
\pgfsys@useobject{currentmarker}{}%
\end{pgfscope}%
\begin{pgfscope}%
\pgfsys@transformshift{2.743541in}{0.748796in}%
\pgfsys@useobject{currentmarker}{}%
\end{pgfscope}%
\end{pgfscope}%
\begin{pgfscope}%
\pgfsetbuttcap%
\pgfsetroundjoin%
\definecolor{currentfill}{rgb}{0.000000,0.000000,0.000000}%
\pgfsetfillcolor{currentfill}%
\pgfsetlinewidth{0.803000pt}%
\definecolor{currentstroke}{rgb}{0.000000,0.000000,0.000000}%
\pgfsetstrokecolor{currentstroke}%
\pgfsetdash{}{0pt}%
\pgfsys@defobject{currentmarker}{\pgfqpoint{0.000000in}{-0.048611in}}{\pgfqpoint{0.000000in}{0.000000in}}{%
\pgfpathmoveto{\pgfqpoint{0.000000in}{0.000000in}}%
\pgfpathlineto{\pgfqpoint{0.000000in}{-0.048611in}}%
\pgfusepath{stroke,fill}%
}%
\begin{pgfscope}%
\pgfsys@transformshift{0.699231in}{0.498776in}%
\pgfsys@useobject{currentmarker}{}%
\end{pgfscope}%
\end{pgfscope}%
\begin{pgfscope}%
\definecolor{textcolor}{rgb}{0.000000,0.000000,0.000000}%
\pgfsetstrokecolor{textcolor}%
\pgfsetfillcolor{textcolor}%
\pgftext[x=0.699231in,y=0.408498in,,top]{\color{textcolor}{\rmfamily\fontsize{6.500000}{7.800000}\selectfont\catcode`\^=\active\def^{\ifmmode\sp\else\^{}\fi}\catcode`\%=\active\def%{\%}30}}%
\end{pgfscope}%
\begin{pgfscope}%
\pgfsetbuttcap%
\pgfsetroundjoin%
\definecolor{currentfill}{rgb}{0.000000,0.000000,0.000000}%
\pgfsetfillcolor{currentfill}%
\pgfsetlinewidth{0.803000pt}%
\definecolor{currentstroke}{rgb}{0.000000,0.000000,0.000000}%
\pgfsetstrokecolor{currentstroke}%
\pgfsetdash{}{0pt}%
\pgfsys@defobject{currentmarker}{\pgfqpoint{0.000000in}{-0.048611in}}{\pgfqpoint{0.000000in}{0.000000in}}{%
\pgfpathmoveto{\pgfqpoint{0.000000in}{0.000000in}}%
\pgfpathlineto{\pgfqpoint{0.000000in}{-0.048611in}}%
\pgfusepath{stroke,fill}%
}%
\begin{pgfscope}%
\pgfsys@transformshift{0.926377in}{0.498776in}%
\pgfsys@useobject{currentmarker}{}%
\end{pgfscope}%
\end{pgfscope}%
\begin{pgfscope}%
\definecolor{textcolor}{rgb}{0.000000,0.000000,0.000000}%
\pgfsetstrokecolor{textcolor}%
\pgfsetfillcolor{textcolor}%
\pgftext[x=0.926377in,y=0.408498in,,top]{\color{textcolor}{\rmfamily\fontsize{6.500000}{7.800000}\selectfont\catcode`\^=\active\def^{\ifmmode\sp\else\^{}\fi}\catcode`\%=\active\def%{\%}50}}%
\end{pgfscope}%
\begin{pgfscope}%
\pgfsetbuttcap%
\pgfsetroundjoin%
\definecolor{currentfill}{rgb}{0.000000,0.000000,0.000000}%
\pgfsetfillcolor{currentfill}%
\pgfsetlinewidth{0.803000pt}%
\definecolor{currentstroke}{rgb}{0.000000,0.000000,0.000000}%
\pgfsetstrokecolor{currentstroke}%
\pgfsetdash{}{0pt}%
\pgfsys@defobject{currentmarker}{\pgfqpoint{0.000000in}{-0.048611in}}{\pgfqpoint{0.000000in}{0.000000in}}{%
\pgfpathmoveto{\pgfqpoint{0.000000in}{0.000000in}}%
\pgfpathlineto{\pgfqpoint{0.000000in}{-0.048611in}}%
\pgfusepath{stroke,fill}%
}%
\begin{pgfscope}%
\pgfsys@transformshift{1.153522in}{0.498776in}%
\pgfsys@useobject{currentmarker}{}%
\end{pgfscope}%
\end{pgfscope}%
\begin{pgfscope}%
\definecolor{textcolor}{rgb}{0.000000,0.000000,0.000000}%
\pgfsetstrokecolor{textcolor}%
\pgfsetfillcolor{textcolor}%
\pgftext[x=1.153522in,y=0.408498in,,top]{\color{textcolor}{\rmfamily\fontsize{6.500000}{7.800000}\selectfont\catcode`\^=\active\def^{\ifmmode\sp\else\^{}\fi}\catcode`\%=\active\def%{\%}100}}%
\end{pgfscope}%
\begin{pgfscope}%
\pgfsetbuttcap%
\pgfsetroundjoin%
\definecolor{currentfill}{rgb}{0.000000,0.000000,0.000000}%
\pgfsetfillcolor{currentfill}%
\pgfsetlinewidth{0.803000pt}%
\definecolor{currentstroke}{rgb}{0.000000,0.000000,0.000000}%
\pgfsetstrokecolor{currentstroke}%
\pgfsetdash{}{0pt}%
\pgfsys@defobject{currentmarker}{\pgfqpoint{0.000000in}{-0.048611in}}{\pgfqpoint{0.000000in}{0.000000in}}{%
\pgfpathmoveto{\pgfqpoint{0.000000in}{0.000000in}}%
\pgfpathlineto{\pgfqpoint{0.000000in}{-0.048611in}}%
\pgfusepath{stroke,fill}%
}%
\begin{pgfscope}%
\pgfsys@transformshift{1.380668in}{0.498776in}%
\pgfsys@useobject{currentmarker}{}%
\end{pgfscope}%
\end{pgfscope}%
\begin{pgfscope}%
\definecolor{textcolor}{rgb}{0.000000,0.000000,0.000000}%
\pgfsetstrokecolor{textcolor}%
\pgfsetfillcolor{textcolor}%
\pgftext[x=1.380668in,y=0.408498in,,top]{\color{textcolor}{\rmfamily\fontsize{6.500000}{7.800000}\selectfont\catcode`\^=\active\def^{\ifmmode\sp\else\^{}\fi}\catcode`\%=\active\def%{\%}500}}%
\end{pgfscope}%
\begin{pgfscope}%
\pgfsetbuttcap%
\pgfsetroundjoin%
\definecolor{currentfill}{rgb}{0.000000,0.000000,0.000000}%
\pgfsetfillcolor{currentfill}%
\pgfsetlinewidth{0.803000pt}%
\definecolor{currentstroke}{rgb}{0.000000,0.000000,0.000000}%
\pgfsetstrokecolor{currentstroke}%
\pgfsetdash{}{0pt}%
\pgfsys@defobject{currentmarker}{\pgfqpoint{0.000000in}{-0.048611in}}{\pgfqpoint{0.000000in}{0.000000in}}{%
\pgfpathmoveto{\pgfqpoint{0.000000in}{0.000000in}}%
\pgfpathlineto{\pgfqpoint{0.000000in}{-0.048611in}}%
\pgfusepath{stroke,fill}%
}%
\begin{pgfscope}%
\pgfsys@transformshift{1.607813in}{0.498776in}%
\pgfsys@useobject{currentmarker}{}%
\end{pgfscope}%
\end{pgfscope}%
\begin{pgfscope}%
\definecolor{textcolor}{rgb}{0.000000,0.000000,0.000000}%
\pgfsetstrokecolor{textcolor}%
\pgfsetfillcolor{textcolor}%
\pgftext[x=1.607813in,y=0.408498in,,top]{\color{textcolor}{\rmfamily\fontsize{6.500000}{7.800000}\selectfont\catcode`\^=\active\def^{\ifmmode\sp\else\^{}\fi}\catcode`\%=\active\def%{\%}1000}}%
\end{pgfscope}%
\begin{pgfscope}%
\pgfsetbuttcap%
\pgfsetroundjoin%
\definecolor{currentfill}{rgb}{0.000000,0.000000,0.000000}%
\pgfsetfillcolor{currentfill}%
\pgfsetlinewidth{0.803000pt}%
\definecolor{currentstroke}{rgb}{0.000000,0.000000,0.000000}%
\pgfsetstrokecolor{currentstroke}%
\pgfsetdash{}{0pt}%
\pgfsys@defobject{currentmarker}{\pgfqpoint{0.000000in}{-0.048611in}}{\pgfqpoint{0.000000in}{0.000000in}}{%
\pgfpathmoveto{\pgfqpoint{0.000000in}{0.000000in}}%
\pgfpathlineto{\pgfqpoint{0.000000in}{-0.048611in}}%
\pgfusepath{stroke,fill}%
}%
\begin{pgfscope}%
\pgfsys@transformshift{1.834959in}{0.498776in}%
\pgfsys@useobject{currentmarker}{}%
\end{pgfscope}%
\end{pgfscope}%
\begin{pgfscope}%
\definecolor{textcolor}{rgb}{0.000000,0.000000,0.000000}%
\pgfsetstrokecolor{textcolor}%
\pgfsetfillcolor{textcolor}%
\pgftext[x=1.834959in,y=0.408498in,,top]{\color{textcolor}{\rmfamily\fontsize{6.500000}{7.800000}\selectfont\catcode`\^=\active\def^{\ifmmode\sp\else\^{}\fi}\catcode`\%=\active\def%{\%}2000}}%
\end{pgfscope}%
\begin{pgfscope}%
\pgfsetbuttcap%
\pgfsetroundjoin%
\definecolor{currentfill}{rgb}{0.000000,0.000000,0.000000}%
\pgfsetfillcolor{currentfill}%
\pgfsetlinewidth{0.803000pt}%
\definecolor{currentstroke}{rgb}{0.000000,0.000000,0.000000}%
\pgfsetstrokecolor{currentstroke}%
\pgfsetdash{}{0pt}%
\pgfsys@defobject{currentmarker}{\pgfqpoint{0.000000in}{-0.048611in}}{\pgfqpoint{0.000000in}{0.000000in}}{%
\pgfpathmoveto{\pgfqpoint{0.000000in}{0.000000in}}%
\pgfpathlineto{\pgfqpoint{0.000000in}{-0.048611in}}%
\pgfusepath{stroke,fill}%
}%
\begin{pgfscope}%
\pgfsys@transformshift{2.062104in}{0.498776in}%
\pgfsys@useobject{currentmarker}{}%
\end{pgfscope}%
\end{pgfscope}%
\begin{pgfscope}%
\definecolor{textcolor}{rgb}{0.000000,0.000000,0.000000}%
\pgfsetstrokecolor{textcolor}%
\pgfsetfillcolor{textcolor}%
\pgftext[x=2.062104in,y=0.408498in,,top]{\color{textcolor}{\rmfamily\fontsize{6.500000}{7.800000}\selectfont\catcode`\^=\active\def^{\ifmmode\sp\else\^{}\fi}\catcode`\%=\active\def%{\%}3000}}%
\end{pgfscope}%
\begin{pgfscope}%
\pgfsetbuttcap%
\pgfsetroundjoin%
\definecolor{currentfill}{rgb}{0.000000,0.000000,0.000000}%
\pgfsetfillcolor{currentfill}%
\pgfsetlinewidth{0.803000pt}%
\definecolor{currentstroke}{rgb}{0.000000,0.000000,0.000000}%
\pgfsetstrokecolor{currentstroke}%
\pgfsetdash{}{0pt}%
\pgfsys@defobject{currentmarker}{\pgfqpoint{0.000000in}{-0.048611in}}{\pgfqpoint{0.000000in}{0.000000in}}{%
\pgfpathmoveto{\pgfqpoint{0.000000in}{0.000000in}}%
\pgfpathlineto{\pgfqpoint{0.000000in}{-0.048611in}}%
\pgfusepath{stroke,fill}%
}%
\begin{pgfscope}%
\pgfsys@transformshift{2.289250in}{0.498776in}%
\pgfsys@useobject{currentmarker}{}%
\end{pgfscope}%
\end{pgfscope}%
\begin{pgfscope}%
\definecolor{textcolor}{rgb}{0.000000,0.000000,0.000000}%
\pgfsetstrokecolor{textcolor}%
\pgfsetfillcolor{textcolor}%
\pgftext[x=2.289250in,y=0.408498in,,top]{\color{textcolor}{\rmfamily\fontsize{6.500000}{7.800000}\selectfont\catcode`\^=\active\def^{\ifmmode\sp\else\^{}\fi}\catcode`\%=\active\def%{\%}4000}}%
\end{pgfscope}%
\begin{pgfscope}%
\pgfsetbuttcap%
\pgfsetroundjoin%
\definecolor{currentfill}{rgb}{0.000000,0.000000,0.000000}%
\pgfsetfillcolor{currentfill}%
\pgfsetlinewidth{0.803000pt}%
\definecolor{currentstroke}{rgb}{0.000000,0.000000,0.000000}%
\pgfsetstrokecolor{currentstroke}%
\pgfsetdash{}{0pt}%
\pgfsys@defobject{currentmarker}{\pgfqpoint{0.000000in}{-0.048611in}}{\pgfqpoint{0.000000in}{0.000000in}}{%
\pgfpathmoveto{\pgfqpoint{0.000000in}{0.000000in}}%
\pgfpathlineto{\pgfqpoint{0.000000in}{-0.048611in}}%
\pgfusepath{stroke,fill}%
}%
\begin{pgfscope}%
\pgfsys@transformshift{2.516395in}{0.498776in}%
\pgfsys@useobject{currentmarker}{}%
\end{pgfscope}%
\end{pgfscope}%
\begin{pgfscope}%
\definecolor{textcolor}{rgb}{0.000000,0.000000,0.000000}%
\pgfsetstrokecolor{textcolor}%
\pgfsetfillcolor{textcolor}%
\pgftext[x=2.516395in,y=0.408498in,,top]{\color{textcolor}{\rmfamily\fontsize{6.500000}{7.800000}\selectfont\catcode`\^=\active\def^{\ifmmode\sp\else\^{}\fi}\catcode`\%=\active\def%{\%}5000}}%
\end{pgfscope}%
\begin{pgfscope}%
\pgfsetbuttcap%
\pgfsetroundjoin%
\definecolor{currentfill}{rgb}{0.000000,0.000000,0.000000}%
\pgfsetfillcolor{currentfill}%
\pgfsetlinewidth{0.803000pt}%
\definecolor{currentstroke}{rgb}{0.000000,0.000000,0.000000}%
\pgfsetstrokecolor{currentstroke}%
\pgfsetdash{}{0pt}%
\pgfsys@defobject{currentmarker}{\pgfqpoint{0.000000in}{-0.048611in}}{\pgfqpoint{0.000000in}{0.000000in}}{%
\pgfpathmoveto{\pgfqpoint{0.000000in}{0.000000in}}%
\pgfpathlineto{\pgfqpoint{0.000000in}{-0.048611in}}%
\pgfusepath{stroke,fill}%
}%
\begin{pgfscope}%
\pgfsys@transformshift{2.743541in}{0.498776in}%
\pgfsys@useobject{currentmarker}{}%
\end{pgfscope}%
\end{pgfscope}%
\begin{pgfscope}%
\definecolor{textcolor}{rgb}{0.000000,0.000000,0.000000}%
\pgfsetstrokecolor{textcolor}%
\pgfsetfillcolor{textcolor}%
\pgftext[x=2.743541in,y=0.408498in,,top]{\color{textcolor}{\rmfamily\fontsize{6.500000}{7.800000}\selectfont\catcode`\^=\active\def^{\ifmmode\sp\else\^{}\fi}\catcode`\%=\active\def%{\%}6000}}%
\end{pgfscope}%
\begin{pgfscope}%
\definecolor{textcolor}{rgb}{0.000000,0.000000,0.000000}%
\pgfsetstrokecolor{textcolor}%
\pgfsetfillcolor{textcolor}%
\pgftext[x=1.721386in,y=0.278868in,,top]{\color{textcolor}{\rmfamily\fontsize{10.000000}{12.000000}\selectfont\catcode`\^=\active\def^{\ifmmode\sp\else\^{}\fi}\catcode`\%=\active\def%{\%}$\text{Re}$}}%
\end{pgfscope}%
\begin{pgfscope}%
\pgfsetbuttcap%
\pgfsetroundjoin%
\definecolor{currentfill}{rgb}{0.000000,0.000000,0.000000}%
\pgfsetfillcolor{currentfill}%
\pgfsetlinewidth{0.803000pt}%
\definecolor{currentstroke}{rgb}{0.000000,0.000000,0.000000}%
\pgfsetstrokecolor{currentstroke}%
\pgfsetdash{}{0pt}%
\pgfsys@defobject{currentmarker}{\pgfqpoint{-0.048611in}{0.000000in}}{\pgfqpoint{-0.000000in}{0.000000in}}{%
\pgfpathmoveto{\pgfqpoint{-0.000000in}{0.000000in}}%
\pgfpathlineto{\pgfqpoint{-0.048611in}{0.000000in}}%
\pgfusepath{stroke,fill}%
}%
\begin{pgfscope}%
\pgfsys@transformshift{0.597016in}{0.498776in}%
\pgfsys@useobject{currentmarker}{}%
\end{pgfscope}%
\end{pgfscope}%
\begin{pgfscope}%
\definecolor{textcolor}{rgb}{0.000000,0.000000,0.000000}%
\pgfsetstrokecolor{textcolor}%
\pgfsetfillcolor{textcolor}%
\pgftext[x=0.455813in, y=0.469841in, left, base]{\color{textcolor}{\rmfamily\fontsize{6.500000}{7.800000}\selectfont\catcode`\^=\active\def^{\ifmmode\sp\else\^{}\fi}\catcode`\%=\active\def%{\%}$\mathdefault{0}$}}%
\end{pgfscope}%
\begin{pgfscope}%
\pgfsetbuttcap%
\pgfsetroundjoin%
\definecolor{currentfill}{rgb}{0.000000,0.000000,0.000000}%
\pgfsetfillcolor{currentfill}%
\pgfsetlinewidth{0.803000pt}%
\definecolor{currentstroke}{rgb}{0.000000,0.000000,0.000000}%
\pgfsetstrokecolor{currentstroke}%
\pgfsetdash{}{0pt}%
\pgfsys@defobject{currentmarker}{\pgfqpoint{-0.048611in}{0.000000in}}{\pgfqpoint{-0.000000in}{0.000000in}}{%
\pgfpathmoveto{\pgfqpoint{-0.000000in}{0.000000in}}%
\pgfpathlineto{\pgfqpoint{-0.048611in}{0.000000in}}%
\pgfusepath{stroke,fill}%
}%
\begin{pgfscope}%
\pgfsys@transformshift{0.597016in}{0.938774in}%
\pgfsys@useobject{currentmarker}{}%
\end{pgfscope}%
\end{pgfscope}%
\begin{pgfscope}%
\definecolor{textcolor}{rgb}{0.000000,0.000000,0.000000}%
\pgfsetstrokecolor{textcolor}%
\pgfsetfillcolor{textcolor}%
\pgftext[x=0.404888in, y=0.909839in, left, base]{\color{textcolor}{\rmfamily\fontsize{6.500000}{7.800000}\selectfont\catcode`\^=\active\def^{\ifmmode\sp\else\^{}\fi}\catcode`\%=\active\def%{\%}$\mathdefault{10}$}}%
\end{pgfscope}%
\begin{pgfscope}%
\pgfsetbuttcap%
\pgfsetroundjoin%
\definecolor{currentfill}{rgb}{0.000000,0.000000,0.000000}%
\pgfsetfillcolor{currentfill}%
\pgfsetlinewidth{0.803000pt}%
\definecolor{currentstroke}{rgb}{0.000000,0.000000,0.000000}%
\pgfsetstrokecolor{currentstroke}%
\pgfsetdash{}{0pt}%
\pgfsys@defobject{currentmarker}{\pgfqpoint{-0.048611in}{0.000000in}}{\pgfqpoint{-0.000000in}{0.000000in}}{%
\pgfpathmoveto{\pgfqpoint{-0.000000in}{0.000000in}}%
\pgfpathlineto{\pgfqpoint{-0.048611in}{0.000000in}}%
\pgfusepath{stroke,fill}%
}%
\begin{pgfscope}%
\pgfsys@transformshift{0.597016in}{1.378772in}%
\pgfsys@useobject{currentmarker}{}%
\end{pgfscope}%
\end{pgfscope}%
\begin{pgfscope}%
\definecolor{textcolor}{rgb}{0.000000,0.000000,0.000000}%
\pgfsetstrokecolor{textcolor}%
\pgfsetfillcolor{textcolor}%
\pgftext[x=0.404888in, y=1.349837in, left, base]{\color{textcolor}{\rmfamily\fontsize{6.500000}{7.800000}\selectfont\catcode`\^=\active\def^{\ifmmode\sp\else\^{}\fi}\catcode`\%=\active\def%{\%}$\mathdefault{20}$}}%
\end{pgfscope}%
\begin{pgfscope}%
\pgfsetbuttcap%
\pgfsetroundjoin%
\definecolor{currentfill}{rgb}{0.000000,0.000000,0.000000}%
\pgfsetfillcolor{currentfill}%
\pgfsetlinewidth{0.803000pt}%
\definecolor{currentstroke}{rgb}{0.000000,0.000000,0.000000}%
\pgfsetstrokecolor{currentstroke}%
\pgfsetdash{}{0pt}%
\pgfsys@defobject{currentmarker}{\pgfqpoint{-0.048611in}{0.000000in}}{\pgfqpoint{-0.000000in}{0.000000in}}{%
\pgfpathmoveto{\pgfqpoint{-0.000000in}{0.000000in}}%
\pgfpathlineto{\pgfqpoint{-0.048611in}{0.000000in}}%
\pgfusepath{stroke,fill}%
}%
\begin{pgfscope}%
\pgfsys@transformshift{0.597016in}{1.818771in}%
\pgfsys@useobject{currentmarker}{}%
\end{pgfscope}%
\end{pgfscope}%
\begin{pgfscope}%
\definecolor{textcolor}{rgb}{0.000000,0.000000,0.000000}%
\pgfsetstrokecolor{textcolor}%
\pgfsetfillcolor{textcolor}%
\pgftext[x=0.404888in, y=1.789835in, left, base]{\color{textcolor}{\rmfamily\fontsize{6.500000}{7.800000}\selectfont\catcode`\^=\active\def^{\ifmmode\sp\else\^{}\fi}\catcode`\%=\active\def%{\%}$\mathdefault{30}$}}%
\end{pgfscope}%
\begin{pgfscope}%
\definecolor{textcolor}{rgb}{0.000000,0.000000,0.000000}%
\pgfsetstrokecolor{textcolor}%
\pgfsetfillcolor{textcolor}%
\pgftext[x=0.349332in,y=1.158773in,,bottom,rotate=90.000000]{\color{textcolor}{\rmfamily\fontsize{10.000000}{12.000000}\selectfont\catcode`\^=\active\def^{\ifmmode\sp\else\^{}\fi}\catcode`\%=\active\def%{\%}$\delta_{\ell^1}^{(q)} \ [\%]$}}%
\end{pgfscope}%
\begin{pgfscope}%
\pgfsetrectcap%
\pgfsetmiterjoin%
\pgfsetlinewidth{0.803000pt}%
\definecolor{currentstroke}{rgb}{0.000000,0.000000,0.000000}%
\pgfsetstrokecolor{currentstroke}%
\pgfsetdash{}{0pt}%
\pgfpathmoveto{\pgfqpoint{0.597016in}{0.498776in}}%
\pgfpathlineto{\pgfqpoint{0.597016in}{1.818771in}}%
\pgfusepath{stroke}%
\end{pgfscope}%
\begin{pgfscope}%
\pgfsetrectcap%
\pgfsetmiterjoin%
\pgfsetlinewidth{0.803000pt}%
\definecolor{currentstroke}{rgb}{0.000000,0.000000,0.000000}%
\pgfsetstrokecolor{currentstroke}%
\pgfsetdash{}{0pt}%
\pgfpathmoveto{\pgfqpoint{2.845756in}{0.498776in}}%
\pgfpathlineto{\pgfqpoint{2.845756in}{1.818771in}}%
\pgfusepath{stroke}%
\end{pgfscope}%
\begin{pgfscope}%
\pgfsetrectcap%
\pgfsetmiterjoin%
\pgfsetlinewidth{0.803000pt}%
\definecolor{currentstroke}{rgb}{0.000000,0.000000,0.000000}%
\pgfsetstrokecolor{currentstroke}%
\pgfsetdash{}{0pt}%
\pgfpathmoveto{\pgfqpoint{0.597016in}{0.498776in}}%
\pgfpathlineto{\pgfqpoint{2.845756in}{0.498776in}}%
\pgfusepath{stroke}%
\end{pgfscope}%
\begin{pgfscope}%
\pgfsetrectcap%
\pgfsetmiterjoin%
\pgfsetlinewidth{0.803000pt}%
\definecolor{currentstroke}{rgb}{0.000000,0.000000,0.000000}%
\pgfsetstrokecolor{currentstroke}%
\pgfsetdash{}{0pt}%
\pgfpathmoveto{\pgfqpoint{0.597016in}{1.818771in}}%
\pgfpathlineto{\pgfqpoint{2.845756in}{1.818771in}}%
\pgfusepath{stroke}%
\end{pgfscope}%
\begin{pgfscope}%
\pgfsetbuttcap%
\pgfsetmiterjoin%
\definecolor{currentfill}{rgb}{1.000000,1.000000,1.000000}%
\pgfsetfillcolor{currentfill}%
\pgfsetfillopacity{0.800000}%
\pgfsetlinewidth{1.003750pt}%
\definecolor{currentstroke}{rgb}{0.800000,0.800000,0.800000}%
\pgfsetstrokecolor{currentstroke}%
\pgfsetstrokeopacity{0.800000}%
\pgfsetdash{}{0pt}%
\pgfpathmoveto{\pgfqpoint{1.181273in}{1.348761in}}%
\pgfpathlineto{\pgfqpoint{2.764770in}{1.348761in}}%
\pgfpathquadraticcurveto{\pgfqpoint{2.787909in}{1.348761in}}{\pgfqpoint{2.787909in}{1.371900in}}%
\pgfpathlineto{\pgfqpoint{2.787909in}{1.737785in}}%
\pgfpathquadraticcurveto{\pgfqpoint{2.787909in}{1.760923in}}{\pgfqpoint{2.764770in}{1.760923in}}%
\pgfpathlineto{\pgfqpoint{1.181273in}{1.760923in}}%
\pgfpathquadraticcurveto{\pgfqpoint{1.158134in}{1.760923in}}{\pgfqpoint{1.158134in}{1.737785in}}%
\pgfpathlineto{\pgfqpoint{1.158134in}{1.371900in}}%
\pgfpathquadraticcurveto{\pgfqpoint{1.158134in}{1.348761in}}{\pgfqpoint{1.181273in}{1.348761in}}%
\pgfpathlineto{\pgfqpoint{1.181273in}{1.348761in}}%
\pgfpathclose%
\pgfusepath{stroke,fill}%
\end{pgfscope}%
\begin{pgfscope}%
\pgfsetbuttcap%
\pgfsetroundjoin%
\definecolor{currentfill}{rgb}{0.172549,0.482353,0.713725}%
\pgfsetfillcolor{currentfill}%
\pgfsetlinewidth{0.501875pt}%
\definecolor{currentstroke}{rgb}{0.000000,0.000000,0.000000}%
\pgfsetstrokecolor{currentstroke}%
\pgfsetdash{}{0pt}%
\pgfsys@defobject{currentmarker}{\pgfqpoint{-0.026896in}{-0.026896in}}{\pgfqpoint{0.026896in}{0.026896in}}{%
\pgfpathmoveto{\pgfqpoint{-0.026896in}{-0.026896in}}%
\pgfpathlineto{\pgfqpoint{0.026896in}{-0.026896in}}%
\pgfpathlineto{\pgfqpoint{0.026896in}{0.026896in}}%
\pgfpathlineto{\pgfqpoint{-0.026896in}{0.026896in}}%
\pgfpathlineto{\pgfqpoint{-0.026896in}{-0.026896in}}%
\pgfpathclose%
\pgfusepath{stroke,fill}%
}%
\begin{pgfscope}%
\pgfsys@transformshift{1.320107in}{1.641914in}%
\pgfsys@useobject{currentmarker}{}%
\end{pgfscope}%
\end{pgfscope}%
\begin{pgfscope}%
\definecolor{textcolor}{rgb}{0.000000,0.000000,0.000000}%
\pgfsetstrokecolor{textcolor}%
\pgfsetfillcolor{textcolor}%
\pgftext[x=1.528357in,y=1.611544in,left,base]{\color{textcolor}{\rmfamily\fontsize{8.330000}{9.996000}\selectfont\catcode`\^=\active\def^{\ifmmode\sp\else\^{}\fi}\catcode`\%=\active\def%{\%}$p^\text{II}_{10000}$}}%
\end{pgfscope}%
\begin{pgfscope}%
\pgfsetbuttcap%
\pgfsetroundjoin%
\definecolor{currentfill}{rgb}{0.843137,0.098039,0.109804}%
\pgfsetfillcolor{currentfill}%
\pgfsetlinewidth{0.501875pt}%
\definecolor{currentstroke}{rgb}{0.000000,0.000000,0.000000}%
\pgfsetstrokecolor{currentstroke}%
\pgfsetdash{}{0pt}%
\pgfsys@defobject{currentmarker}{\pgfqpoint{-0.026896in}{-0.026896in}}{\pgfqpoint{0.026896in}{0.026896in}}{%
\pgfpathmoveto{\pgfqpoint{0.000000in}{-0.026896in}}%
\pgfpathcurveto{\pgfqpoint{0.007133in}{-0.026896in}}{\pgfqpoint{0.013974in}{-0.024062in}}{\pgfqpoint{0.019018in}{-0.019018in}}%
\pgfpathcurveto{\pgfqpoint{0.024062in}{-0.013974in}}{\pgfqpoint{0.026896in}{-0.007133in}}{\pgfqpoint{0.026896in}{0.000000in}}%
\pgfpathcurveto{\pgfqpoint{0.026896in}{0.007133in}}{\pgfqpoint{0.024062in}{0.013974in}}{\pgfqpoint{0.019018in}{0.019018in}}%
\pgfpathcurveto{\pgfqpoint{0.013974in}{0.024062in}}{\pgfqpoint{0.007133in}{0.026896in}}{\pgfqpoint{0.000000in}{0.026896in}}%
\pgfpathcurveto{\pgfqpoint{-0.007133in}{0.026896in}}{\pgfqpoint{-0.013974in}{0.024062in}}{\pgfqpoint{-0.019018in}{0.019018in}}%
\pgfpathcurveto{\pgfqpoint{-0.024062in}{0.013974in}}{\pgfqpoint{-0.026896in}{0.007133in}}{\pgfqpoint{-0.026896in}{0.000000in}}%
\pgfpathcurveto{\pgfqpoint{-0.026896in}{-0.007133in}}{\pgfqpoint{-0.024062in}{-0.013974in}}{\pgfqpoint{-0.019018in}{-0.019018in}}%
\pgfpathcurveto{\pgfqpoint{-0.013974in}{-0.024062in}}{\pgfqpoint{-0.007133in}{-0.026896in}}{\pgfqpoint{0.000000in}{-0.026896in}}%
\pgfpathlineto{\pgfqpoint{0.000000in}{-0.026896in}}%
\pgfpathclose%
\pgfusepath{stroke,fill}%
}%
\begin{pgfscope}%
\pgfsys@transformshift{1.320107in}{1.453187in}%
\pgfsys@useobject{currentmarker}{}%
\end{pgfscope}%
\end{pgfscope}%
\begin{pgfscope}%
\definecolor{textcolor}{rgb}{0.000000,0.000000,0.000000}%
\pgfsetstrokecolor{textcolor}%
\pgfsetfillcolor{textcolor}%
\pgftext[x=1.528357in,y=1.422817in,left,base]{\color{textcolor}{\rmfamily\fontsize{8.330000}{9.996000}\selectfont\catcode`\^=\active\def^{\ifmmode\sp\else\^{}\fi}\catcode`\%=\active\def%{\%}$\bm{v}^\text{II}_{10000}$}}%
\end{pgfscope}%
\begin{pgfscope}%
\pgfsetbuttcap%
\pgfsetroundjoin%
\definecolor{currentfill}{rgb}{0.670588,0.850980,0.913725}%
\pgfsetfillcolor{currentfill}%
\pgfsetlinewidth{0.501875pt}%
\definecolor{currentstroke}{rgb}{0.000000,0.000000,0.000000}%
\pgfsetstrokecolor{currentstroke}%
\pgfsetdash{}{0pt}%
\pgfsys@defobject{currentmarker}{\pgfqpoint{-0.038036in}{-0.038036in}}{\pgfqpoint{0.038036in}{0.038036in}}{%
\pgfpathmoveto{\pgfqpoint{-0.000000in}{-0.038036in}}%
\pgfpathlineto{\pgfqpoint{0.038036in}{0.000000in}}%
\pgfpathlineto{\pgfqpoint{0.000000in}{0.038036in}}%
\pgfpathlineto{\pgfqpoint{-0.038036in}{0.000000in}}%
\pgfpathlineto{\pgfqpoint{-0.000000in}{-0.038036in}}%
\pgfpathclose%
\pgfusepath{stroke,fill}%
}%
\begin{pgfscope}%
\pgfsys@transformshift{2.204410in}{1.641914in}%
\pgfsys@useobject{currentmarker}{}%
\end{pgfscope}%
\end{pgfscope}%
\begin{pgfscope}%
\definecolor{textcolor}{rgb}{0.000000,0.000000,0.000000}%
\pgfsetstrokecolor{textcolor}%
\pgfsetfillcolor{textcolor}%
\pgftext[x=2.412660in,y=1.611544in,left,base]{\color{textcolor}{\rmfamily\fontsize{8.330000}{9.996000}\selectfont\catcode`\^=\active\def^{\ifmmode\sp\else\^{}\fi}\catcode`\%=\active\def%{\%}$p^\text{III}_{10000}$}}%
\end{pgfscope}%
\begin{pgfscope}%
\pgfsetbuttcap%
\pgfsetroundjoin%
\definecolor{currentfill}{rgb}{0.992157,0.682353,0.380392}%
\pgfsetfillcolor{currentfill}%
\pgfsetlinewidth{0.501875pt}%
\definecolor{currentstroke}{rgb}{0.000000,0.000000,0.000000}%
\pgfsetstrokecolor{currentstroke}%
\pgfsetdash{}{0pt}%
\pgfsys@defobject{currentmarker}{\pgfqpoint{-0.026896in}{-0.026896in}}{\pgfqpoint{0.026896in}{0.026896in}}{%
\pgfpathmoveto{\pgfqpoint{0.000000in}{0.026896in}}%
\pgfpathlineto{\pgfqpoint{-0.026896in}{-0.026896in}}%
\pgfpathlineto{\pgfqpoint{0.026896in}{-0.026896in}}%
\pgfpathlineto{\pgfqpoint{0.000000in}{0.026896in}}%
\pgfpathclose%
\pgfusepath{stroke,fill}%
}%
\begin{pgfscope}%
\pgfsys@transformshift{2.204410in}{1.453187in}%
\pgfsys@useobject{currentmarker}{}%
\end{pgfscope}%
\end{pgfscope}%
\begin{pgfscope}%
\definecolor{textcolor}{rgb}{0.000000,0.000000,0.000000}%
\pgfsetstrokecolor{textcolor}%
\pgfsetfillcolor{textcolor}%
\pgftext[x=2.412660in,y=1.422817in,left,base]{\color{textcolor}{\rmfamily\fontsize{8.330000}{9.996000}\selectfont\catcode`\^=\active\def^{\ifmmode\sp\else\^{}\fi}\catcode`\%=\active\def%{\%}$\bm{v}^\text{III}_{10000}$}}%
\end{pgfscope}%
\end{pgfpicture}%
\makeatother%
\endgroup%

%% file: figures/section3/n_Re_velocity.pgf
%% Creator: Matplotlib, PGF backend
%%
%% To include the figure in your LaTeX document, write
%%   \input{<filename>.pgf}
%%
%% Make sure the required packages are loaded in your preamble
%%   \usepackage{pgf}
%%
%% Also ensure that all the required font packages are loaded; for instance,
%% the lmodern package is sometimes necessary when using math font.
%%   \usepackage{lmodern}
%%
%% Figures using additional raster images can only be included by \input if
%% they are in the same directory as the main LaTeX file. For loading figures
%% from other directories you can use the `import` package
%% and then include the figures with
%%   \import{<path to file>}{<filename>.pgf}
%%
%% Matplotlib used the following preamble
%%   \def\mathdefault#1{#1}
%%   \everymath=\expandafter{\the\everymath\displaystyle}
%%   \usepackage{amsmath}\usepackage{bm}
%%   \makeatletter\@ifpackageloaded{underscore}{}{\usepackage[strings]{underscore}}\makeatother
%%
\begingroup%
\makeatletter%
\begin{pgfpicture}%
\pgfpathrectangle{\pgfpointorigin}{\pgfqpoint{3.000000in}{2.000000in}}%
\pgfusepath{use as bounding box, clip}%
\begin{pgfscope}%
\pgfsetbuttcap%
\pgfsetmiterjoin%
\definecolor{currentfill}{rgb}{1.000000,1.000000,1.000000}%
\pgfsetfillcolor{currentfill}%
\pgfsetlinewidth{0.000000pt}%
\definecolor{currentstroke}{rgb}{1.000000,1.000000,1.000000}%
\pgfsetstrokecolor{currentstroke}%
\pgfsetdash{}{0pt}%
\pgfpathmoveto{\pgfqpoint{0.000000in}{0.000000in}}%
\pgfpathlineto{\pgfqpoint{3.000000in}{0.000000in}}%
\pgfpathlineto{\pgfqpoint{3.000000in}{2.000000in}}%
\pgfpathlineto{\pgfqpoint{0.000000in}{2.000000in}}%
\pgfpathlineto{\pgfqpoint{0.000000in}{0.000000in}}%
\pgfpathclose%
\pgfusepath{fill}%
\end{pgfscope}%
\begin{pgfscope}%
\pgfsetbuttcap%
\pgfsetmiterjoin%
\definecolor{currentfill}{rgb}{1.000000,1.000000,1.000000}%
\pgfsetfillcolor{currentfill}%
\pgfsetlinewidth{0.000000pt}%
\definecolor{currentstroke}{rgb}{0.000000,0.000000,0.000000}%
\pgfsetstrokecolor{currentstroke}%
\pgfsetstrokeopacity{0.000000}%
\pgfsetdash{}{0pt}%
\pgfpathmoveto{\pgfqpoint{0.664468in}{0.521284in}}%
\pgfpathlineto{\pgfqpoint{2.850000in}{0.521284in}}%
\pgfpathlineto{\pgfqpoint{2.850000in}{1.850000in}}%
\pgfpathlineto{\pgfqpoint{0.664468in}{1.850000in}}%
\pgfpathlineto{\pgfqpoint{0.664468in}{0.521284in}}%
\pgfpathclose%
\pgfusepath{fill}%
\end{pgfscope}%
\begin{pgfscope}%
\pgfpathrectangle{\pgfqpoint{0.664468in}{0.521284in}}{\pgfqpoint{2.185532in}{1.328716in}}%
\pgfusepath{clip}%
\pgfsetbuttcap%
\pgfsetroundjoin%
\definecolor{currentfill}{rgb}{0.843137,0.098039,0.109804}%
\pgfsetfillcolor{currentfill}%
\pgfsetlinewidth{0.501875pt}%
\definecolor{currentstroke}{rgb}{0.000000,0.000000,0.000000}%
\pgfsetstrokecolor{currentstroke}%
\pgfsetdash{}{0pt}%
\pgfsys@defobject{currentmarker}{\pgfqpoint{-0.026896in}{-0.026896in}}{\pgfqpoint{0.026896in}{0.026896in}}{%
\pgfpathmoveto{\pgfqpoint{0.000000in}{0.026896in}}%
\pgfpathlineto{\pgfqpoint{-0.026896in}{-0.026896in}}%
\pgfpathlineto{\pgfqpoint{0.026896in}{-0.026896in}}%
\pgfpathlineto{\pgfqpoint{0.000000in}{0.026896in}}%
\pgfpathclose%
\pgfusepath{stroke,fill}%
}%
\begin{pgfscope}%
\pgfsys@transformshift{0.763810in}{0.883816in}%
\pgfsys@useobject{currentmarker}{}%
\end{pgfscope}%
\begin{pgfscope}%
\pgfsys@transformshift{1.757234in}{0.847563in}%
\pgfsys@useobject{currentmarker}{}%
\end{pgfscope}%
\begin{pgfscope}%
\pgfsys@transformshift{2.750658in}{0.786674in}%
\pgfsys@useobject{currentmarker}{}%
\end{pgfscope}%
\end{pgfscope}%
\begin{pgfscope}%
\pgfpathrectangle{\pgfqpoint{0.664468in}{0.521284in}}{\pgfqpoint{2.185532in}{1.328716in}}%
\pgfusepath{clip}%
\pgfsetbuttcap%
\pgfsetroundjoin%
\definecolor{currentfill}{rgb}{0.172549,0.482353,0.713725}%
\pgfsetfillcolor{currentfill}%
\pgfsetlinewidth{0.501875pt}%
\definecolor{currentstroke}{rgb}{0.000000,0.000000,0.000000}%
\pgfsetstrokecolor{currentstroke}%
\pgfsetdash{}{0pt}%
\pgfsys@defobject{currentmarker}{\pgfqpoint{-0.026896in}{-0.026896in}}{\pgfqpoint{0.026896in}{0.026896in}}{%
\pgfpathmoveto{\pgfqpoint{-0.000000in}{-0.026896in}}%
\pgfpathlineto{\pgfqpoint{0.026896in}{0.026896in}}%
\pgfpathlineto{\pgfqpoint{-0.026896in}{0.026896in}}%
\pgfpathlineto{\pgfqpoint{-0.000000in}{-0.026896in}}%
\pgfpathclose%
\pgfusepath{stroke,fill}%
}%
\begin{pgfscope}%
\pgfsys@transformshift{0.763810in}{1.029647in}%
\pgfsys@useobject{currentmarker}{}%
\end{pgfscope}%
\begin{pgfscope}%
\pgfsys@transformshift{1.757234in}{0.995994in}%
\pgfsys@useobject{currentmarker}{}%
\end{pgfscope}%
\begin{pgfscope}%
\pgfsys@transformshift{2.750658in}{0.789131in}%
\pgfsys@useobject{currentmarker}{}%
\end{pgfscope}%
\end{pgfscope}%
\begin{pgfscope}%
\pgfpathrectangle{\pgfqpoint{0.664468in}{0.521284in}}{\pgfqpoint{2.185532in}{1.328716in}}%
\pgfusepath{clip}%
\pgfsetbuttcap%
\pgfsetroundjoin%
\definecolor{currentfill}{rgb}{0.992157,0.682353,0.380392}%
\pgfsetfillcolor{currentfill}%
\pgfsetlinewidth{0.501875pt}%
\definecolor{currentstroke}{rgb}{0.000000,0.000000,0.000000}%
\pgfsetstrokecolor{currentstroke}%
\pgfsetdash{}{0pt}%
\pgfsys@defobject{currentmarker}{\pgfqpoint{-0.026896in}{-0.026896in}}{\pgfqpoint{0.026896in}{0.026896in}}{%
\pgfpathmoveto{\pgfqpoint{0.000000in}{-0.026896in}}%
\pgfpathcurveto{\pgfqpoint{0.007133in}{-0.026896in}}{\pgfqpoint{0.013974in}{-0.024062in}}{\pgfqpoint{0.019018in}{-0.019018in}}%
\pgfpathcurveto{\pgfqpoint{0.024062in}{-0.013974in}}{\pgfqpoint{0.026896in}{-0.007133in}}{\pgfqpoint{0.026896in}{0.000000in}}%
\pgfpathcurveto{\pgfqpoint{0.026896in}{0.007133in}}{\pgfqpoint{0.024062in}{0.013974in}}{\pgfqpoint{0.019018in}{0.019018in}}%
\pgfpathcurveto{\pgfqpoint{0.013974in}{0.024062in}}{\pgfqpoint{0.007133in}{0.026896in}}{\pgfqpoint{0.000000in}{0.026896in}}%
\pgfpathcurveto{\pgfqpoint{-0.007133in}{0.026896in}}{\pgfqpoint{-0.013974in}{0.024062in}}{\pgfqpoint{-0.019018in}{0.019018in}}%
\pgfpathcurveto{\pgfqpoint{-0.024062in}{0.013974in}}{\pgfqpoint{-0.026896in}{0.007133in}}{\pgfqpoint{-0.026896in}{0.000000in}}%
\pgfpathcurveto{\pgfqpoint{-0.026896in}{-0.007133in}}{\pgfqpoint{-0.024062in}{-0.013974in}}{\pgfqpoint{-0.019018in}{-0.019018in}}%
\pgfpathcurveto{\pgfqpoint{-0.013974in}{-0.024062in}}{\pgfqpoint{-0.007133in}{-0.026896in}}{\pgfqpoint{0.000000in}{-0.026896in}}%
\pgfpathlineto{\pgfqpoint{0.000000in}{-0.026896in}}%
\pgfpathclose%
\pgfusepath{stroke,fill}%
}%
\begin{pgfscope}%
\pgfsys@transformshift{0.763810in}{1.254729in}%
\pgfsys@useobject{currentmarker}{}%
\end{pgfscope}%
\begin{pgfscope}%
\pgfsys@transformshift{1.757234in}{1.123105in}%
\pgfsys@useobject{currentmarker}{}%
\end{pgfscope}%
\begin{pgfscope}%
\pgfsys@transformshift{2.750658in}{0.871710in}%
\pgfsys@useobject{currentmarker}{}%
\end{pgfscope}%
\end{pgfscope}%
\begin{pgfscope}%
\pgfsetbuttcap%
\pgfsetroundjoin%
\definecolor{currentfill}{rgb}{0.000000,0.000000,0.000000}%
\pgfsetfillcolor{currentfill}%
\pgfsetlinewidth{0.803000pt}%
\definecolor{currentstroke}{rgb}{0.000000,0.000000,0.000000}%
\pgfsetstrokecolor{currentstroke}%
\pgfsetdash{}{0pt}%
\pgfsys@defobject{currentmarker}{\pgfqpoint{0.000000in}{-0.048611in}}{\pgfqpoint{0.000000in}{0.000000in}}{%
\pgfpathmoveto{\pgfqpoint{0.000000in}{0.000000in}}%
\pgfpathlineto{\pgfqpoint{0.000000in}{-0.048611in}}%
\pgfusepath{stroke,fill}%
}%
\begin{pgfscope}%
\pgfsys@transformshift{0.763810in}{0.521284in}%
\pgfsys@useobject{currentmarker}{}%
\end{pgfscope}%
\end{pgfscope}%
\begin{pgfscope}%
\definecolor{textcolor}{rgb}{0.000000,0.000000,0.000000}%
\pgfsetstrokecolor{textcolor}%
\pgfsetfillcolor{textcolor}%
\pgftext[x=0.763810in,y=0.431006in,,top]{\color{textcolor}{\rmfamily\fontsize{8.330000}{9.996000}\selectfont\catcode`\^=\active\def^{\ifmmode\sp\else\^{}\fi}\catcode`\%=\active\def%{\%}2}}%
\end{pgfscope}%
\begin{pgfscope}%
\pgfsetbuttcap%
\pgfsetroundjoin%
\definecolor{currentfill}{rgb}{0.000000,0.000000,0.000000}%
\pgfsetfillcolor{currentfill}%
\pgfsetlinewidth{0.803000pt}%
\definecolor{currentstroke}{rgb}{0.000000,0.000000,0.000000}%
\pgfsetstrokecolor{currentstroke}%
\pgfsetdash{}{0pt}%
\pgfsys@defobject{currentmarker}{\pgfqpoint{0.000000in}{-0.048611in}}{\pgfqpoint{0.000000in}{0.000000in}}{%
\pgfpathmoveto{\pgfqpoint{0.000000in}{0.000000in}}%
\pgfpathlineto{\pgfqpoint{0.000000in}{-0.048611in}}%
\pgfusepath{stroke,fill}%
}%
\begin{pgfscope}%
\pgfsys@transformshift{1.757234in}{0.521284in}%
\pgfsys@useobject{currentmarker}{}%
\end{pgfscope}%
\end{pgfscope}%
\begin{pgfscope}%
\definecolor{textcolor}{rgb}{0.000000,0.000000,0.000000}%
\pgfsetstrokecolor{textcolor}%
\pgfsetfillcolor{textcolor}%
\pgftext[x=1.757234in,y=0.431006in,,top]{\color{textcolor}{\rmfamily\fontsize{8.330000}{9.996000}\selectfont\catcode`\^=\active\def^{\ifmmode\sp\else\^{}\fi}\catcode`\%=\active\def%{\%}3}}%
\end{pgfscope}%
\begin{pgfscope}%
\pgfsetbuttcap%
\pgfsetroundjoin%
\definecolor{currentfill}{rgb}{0.000000,0.000000,0.000000}%
\pgfsetfillcolor{currentfill}%
\pgfsetlinewidth{0.803000pt}%
\definecolor{currentstroke}{rgb}{0.000000,0.000000,0.000000}%
\pgfsetstrokecolor{currentstroke}%
\pgfsetdash{}{0pt}%
\pgfsys@defobject{currentmarker}{\pgfqpoint{0.000000in}{-0.048611in}}{\pgfqpoint{0.000000in}{0.000000in}}{%
\pgfpathmoveto{\pgfqpoint{0.000000in}{0.000000in}}%
\pgfpathlineto{\pgfqpoint{0.000000in}{-0.048611in}}%
\pgfusepath{stroke,fill}%
}%
\begin{pgfscope}%
\pgfsys@transformshift{2.750658in}{0.521284in}%
\pgfsys@useobject{currentmarker}{}%
\end{pgfscope}%
\end{pgfscope}%
\begin{pgfscope}%
\definecolor{textcolor}{rgb}{0.000000,0.000000,0.000000}%
\pgfsetstrokecolor{textcolor}%
\pgfsetfillcolor{textcolor}%
\pgftext[x=2.750658in,y=0.431006in,,top]{\color{textcolor}{\rmfamily\fontsize{8.330000}{9.996000}\selectfont\catcode`\^=\active\def^{\ifmmode\sp\else\^{}\fi}\catcode`\%=\active\def%{\%}4}}%
\end{pgfscope}%
\begin{pgfscope}%
\definecolor{textcolor}{rgb}{0.000000,0.000000,0.000000}%
\pgfsetstrokecolor{textcolor}%
\pgfsetfillcolor{textcolor}%
\pgftext[x=1.757234in,y=0.276685in,,top]{\color{textcolor}{\rmfamily\fontsize{10.000000}{12.000000}\selectfont\catcode`\^=\active\def^{\ifmmode\sp\else\^{}\fi}\catcode`\%=\active\def%{\%}Number of sampled values of $\text{Re}$}}%
\end{pgfscope}%
\begin{pgfscope}%
\pgfsetbuttcap%
\pgfsetroundjoin%
\definecolor{currentfill}{rgb}{0.000000,0.000000,0.000000}%
\pgfsetfillcolor{currentfill}%
\pgfsetlinewidth{0.803000pt}%
\definecolor{currentstroke}{rgb}{0.000000,0.000000,0.000000}%
\pgfsetstrokecolor{currentstroke}%
\pgfsetdash{}{0pt}%
\pgfsys@defobject{currentmarker}{\pgfqpoint{-0.048611in}{0.000000in}}{\pgfqpoint{-0.000000in}{0.000000in}}{%
\pgfpathmoveto{\pgfqpoint{-0.000000in}{0.000000in}}%
\pgfpathlineto{\pgfqpoint{-0.048611in}{0.000000in}}%
\pgfusepath{stroke,fill}%
}%
\begin{pgfscope}%
\pgfsys@transformshift{0.664468in}{1.013589in}%
\pgfsys@useobject{currentmarker}{}%
\end{pgfscope}%
\end{pgfscope}%
\begin{pgfscope}%
\definecolor{textcolor}{rgb}{0.000000,0.000000,0.000000}%
\pgfsetstrokecolor{textcolor}%
\pgfsetfillcolor{textcolor}%
\pgftext[x=0.398263in, y=0.974461in, left, base]{\color{textcolor}{\rmfamily\fontsize{8.330000}{9.996000}\selectfont\catcode`\^=\active\def^{\ifmmode\sp\else\^{}\fi}\catcode`\%=\active\def%{\%}$\mathdefault{10^{1}}$}}%
\end{pgfscope}%
\begin{pgfscope}%
\pgfsetbuttcap%
\pgfsetroundjoin%
\definecolor{currentfill}{rgb}{0.000000,0.000000,0.000000}%
\pgfsetfillcolor{currentfill}%
\pgfsetlinewidth{0.803000pt}%
\definecolor{currentstroke}{rgb}{0.000000,0.000000,0.000000}%
\pgfsetstrokecolor{currentstroke}%
\pgfsetdash{}{0pt}%
\pgfsys@defobject{currentmarker}{\pgfqpoint{-0.048611in}{0.000000in}}{\pgfqpoint{-0.000000in}{0.000000in}}{%
\pgfpathmoveto{\pgfqpoint{-0.000000in}{0.000000in}}%
\pgfpathlineto{\pgfqpoint{-0.048611in}{0.000000in}}%
\pgfusepath{stroke,fill}%
}%
\begin{pgfscope}%
\pgfsys@transformshift{0.664468in}{1.505894in}%
\pgfsys@useobject{currentmarker}{}%
\end{pgfscope}%
\end{pgfscope}%
\begin{pgfscope}%
\definecolor{textcolor}{rgb}{0.000000,0.000000,0.000000}%
\pgfsetstrokecolor{textcolor}%
\pgfsetfillcolor{textcolor}%
\pgftext[x=0.398263in, y=1.466766in, left, base]{\color{textcolor}{\rmfamily\fontsize{8.330000}{9.996000}\selectfont\catcode`\^=\active\def^{\ifmmode\sp\else\^{}\fi}\catcode`\%=\active\def%{\%}$\mathdefault{10^{2}}$}}%
\end{pgfscope}%
\begin{pgfscope}%
\definecolor{textcolor}{rgb}{0.000000,0.000000,0.000000}%
\pgfsetstrokecolor{textcolor}%
\pgfsetfillcolor{textcolor}%
\pgftext[x=0.342708in,y=1.185642in,,bottom,rotate=90.000000]{\color{textcolor}{\rmfamily\fontsize{10.000000}{12.000000}\selectfont\catcode`\^=\active\def^{\ifmmode\sp\else\^{}\fi}\catcode`\%=\active\def%{\%}$\delta_{\ell^1}^{(\bm{v})} \ [\%]$}}%
\end{pgfscope}%
\begin{pgfscope}%
\pgfsetrectcap%
\pgfsetmiterjoin%
\pgfsetlinewidth{0.803000pt}%
\definecolor{currentstroke}{rgb}{0.000000,0.000000,0.000000}%
\pgfsetstrokecolor{currentstroke}%
\pgfsetdash{}{0pt}%
\pgfpathmoveto{\pgfqpoint{0.664468in}{0.521284in}}%
\pgfpathlineto{\pgfqpoint{0.664468in}{1.850000in}}%
\pgfusepath{stroke}%
\end{pgfscope}%
\begin{pgfscope}%
\pgfsetrectcap%
\pgfsetmiterjoin%
\pgfsetlinewidth{0.803000pt}%
\definecolor{currentstroke}{rgb}{0.000000,0.000000,0.000000}%
\pgfsetstrokecolor{currentstroke}%
\pgfsetdash{}{0pt}%
\pgfpathmoveto{\pgfqpoint{2.850000in}{0.521284in}}%
\pgfpathlineto{\pgfqpoint{2.850000in}{1.850000in}}%
\pgfusepath{stroke}%
\end{pgfscope}%
\begin{pgfscope}%
\pgfsetrectcap%
\pgfsetmiterjoin%
\pgfsetlinewidth{0.803000pt}%
\definecolor{currentstroke}{rgb}{0.000000,0.000000,0.000000}%
\pgfsetstrokecolor{currentstroke}%
\pgfsetdash{}{0pt}%
\pgfpathmoveto{\pgfqpoint{0.664468in}{0.521284in}}%
\pgfpathlineto{\pgfqpoint{2.850000in}{0.521284in}}%
\pgfusepath{stroke}%
\end{pgfscope}%
\begin{pgfscope}%
\pgfsetrectcap%
\pgfsetmiterjoin%
\pgfsetlinewidth{0.803000pt}%
\definecolor{currentstroke}{rgb}{0.000000,0.000000,0.000000}%
\pgfsetstrokecolor{currentstroke}%
\pgfsetdash{}{0pt}%
\pgfpathmoveto{\pgfqpoint{0.664468in}{1.850000in}}%
\pgfpathlineto{\pgfqpoint{2.850000in}{1.850000in}}%
\pgfusepath{stroke}%
\end{pgfscope}%
\begin{pgfscope}%
\pgfsetbuttcap%
\pgfsetmiterjoin%
\definecolor{currentfill}{rgb}{1.000000,1.000000,1.000000}%
\pgfsetfillcolor{currentfill}%
\pgfsetfillopacity{0.800000}%
\pgfsetlinewidth{1.003750pt}%
\definecolor{currentstroke}{rgb}{0.800000,0.800000,0.800000}%
\pgfsetstrokecolor{currentstroke}%
\pgfsetstrokeopacity{0.800000}%
\pgfsetdash{}{0pt}%
\pgfpathmoveto{\pgfqpoint{1.279984in}{1.435546in}}%
\pgfpathlineto{\pgfqpoint{2.769014in}{1.435546in}}%
\pgfpathquadraticcurveto{\pgfqpoint{2.792153in}{1.435546in}}{\pgfqpoint{2.792153in}{1.458685in}}%
\pgfpathlineto{\pgfqpoint{2.792153in}{1.769014in}}%
\pgfpathquadraticcurveto{\pgfqpoint{2.792153in}{1.792153in}}{\pgfqpoint{2.769014in}{1.792153in}}%
\pgfpathlineto{\pgfqpoint{1.279984in}{1.792153in}}%
\pgfpathquadraticcurveto{\pgfqpoint{1.256846in}{1.792153in}}{\pgfqpoint{1.256846in}{1.769014in}}%
\pgfpathlineto{\pgfqpoint{1.256846in}{1.458685in}}%
\pgfpathquadraticcurveto{\pgfqpoint{1.256846in}{1.435546in}}{\pgfqpoint{1.279984in}{1.435546in}}%
\pgfpathlineto{\pgfqpoint{1.279984in}{1.435546in}}%
\pgfpathclose%
\pgfusepath{stroke,fill}%
\end{pgfscope}%
\begin{pgfscope}%
\pgfsetbuttcap%
\pgfsetroundjoin%
\definecolor{currentfill}{rgb}{0.843137,0.098039,0.109804}%
\pgfsetfillcolor{currentfill}%
\pgfsetlinewidth{0.501875pt}%
\definecolor{currentstroke}{rgb}{0.000000,0.000000,0.000000}%
\pgfsetstrokecolor{currentstroke}%
\pgfsetdash{}{0pt}%
\pgfsys@defobject{currentmarker}{\pgfqpoint{-0.026896in}{-0.026896in}}{\pgfqpoint{0.026896in}{0.026896in}}{%
\pgfpathmoveto{\pgfqpoint{0.000000in}{0.026896in}}%
\pgfpathlineto{\pgfqpoint{-0.026896in}{-0.026896in}}%
\pgfpathlineto{\pgfqpoint{0.026896in}{-0.026896in}}%
\pgfpathlineto{\pgfqpoint{0.000000in}{0.026896in}}%
\pgfpathclose%
\pgfusepath{stroke,fill}%
}%
\begin{pgfscope}%
\pgfsys@transformshift{1.418818in}{1.694748in}%
\pgfsys@useobject{currentmarker}{}%
\end{pgfscope}%
\end{pgfscope}%
\begin{pgfscope}%
\definecolor{textcolor}{rgb}{0.000000,0.000000,0.000000}%
\pgfsetstrokecolor{textcolor}%
\pgfsetfillcolor{textcolor}%
\pgftext[x=1.627068in,y=1.664378in,left,base]{\color{textcolor}{\rmfamily\fontsize{8.330000}{9.996000}\selectfont\catcode`\^=\active\def^{\ifmmode\sp\else\^{}\fi}\catcode`\%=\active\def%{\%}$\bm{v}^\text{PINN}$}}%
\end{pgfscope}%
\begin{pgfscope}%
\pgfsetbuttcap%
\pgfsetroundjoin%
\definecolor{currentfill}{rgb}{0.172549,0.482353,0.713725}%
\pgfsetfillcolor{currentfill}%
\pgfsetlinewidth{0.501875pt}%
\definecolor{currentstroke}{rgb}{0.000000,0.000000,0.000000}%
\pgfsetstrokecolor{currentstroke}%
\pgfsetdash{}{0pt}%
\pgfsys@defobject{currentmarker}{\pgfqpoint{-0.026896in}{-0.026896in}}{\pgfqpoint{0.026896in}{0.026896in}}{%
\pgfpathmoveto{\pgfqpoint{-0.000000in}{-0.026896in}}%
\pgfpathlineto{\pgfqpoint{0.026896in}{0.026896in}}%
\pgfpathlineto{\pgfqpoint{-0.026896in}{0.026896in}}%
\pgfpathlineto{\pgfqpoint{-0.000000in}{-0.026896in}}%
\pgfpathclose%
\pgfusepath{stroke,fill}%
}%
\begin{pgfscope}%
\pgfsys@transformshift{1.418818in}{1.533799in}%
\pgfsys@useobject{currentmarker}{}%
\end{pgfscope}%
\end{pgfscope}%
\begin{pgfscope}%
\definecolor{textcolor}{rgb}{0.000000,0.000000,0.000000}%
\pgfsetstrokecolor{textcolor}%
\pgfsetfillcolor{textcolor}%
\pgftext[x=1.627068in,y=1.503429in,left,base]{\color{textcolor}{\rmfamily\fontsize{8.330000}{9.996000}\selectfont\catcode`\^=\active\def^{\ifmmode\sp\else\^{}\fi}\catcode`\%=\active\def%{\%}$\bm{v}^\text{BINN}$}}%
\end{pgfscope}%
\begin{pgfscope}%
\pgfsetbuttcap%
\pgfsetroundjoin%
\definecolor{currentfill}{rgb}{0.992157,0.682353,0.380392}%
\pgfsetfillcolor{currentfill}%
\pgfsetlinewidth{0.501875pt}%
\definecolor{currentstroke}{rgb}{0.000000,0.000000,0.000000}%
\pgfsetstrokecolor{currentstroke}%
\pgfsetdash{}{0pt}%
\pgfsys@defobject{currentmarker}{\pgfqpoint{-0.026896in}{-0.026896in}}{\pgfqpoint{0.026896in}{0.026896in}}{%
\pgfpathmoveto{\pgfqpoint{0.000000in}{-0.026896in}}%
\pgfpathcurveto{\pgfqpoint{0.007133in}{-0.026896in}}{\pgfqpoint{0.013974in}{-0.024062in}}{\pgfqpoint{0.019018in}{-0.019018in}}%
\pgfpathcurveto{\pgfqpoint{0.024062in}{-0.013974in}}{\pgfqpoint{0.026896in}{-0.007133in}}{\pgfqpoint{0.026896in}{0.000000in}}%
\pgfpathcurveto{\pgfqpoint{0.026896in}{0.007133in}}{\pgfqpoint{0.024062in}{0.013974in}}{\pgfqpoint{0.019018in}{0.019018in}}%
\pgfpathcurveto{\pgfqpoint{0.013974in}{0.024062in}}{\pgfqpoint{0.007133in}{0.026896in}}{\pgfqpoint{0.000000in}{0.026896in}}%
\pgfpathcurveto{\pgfqpoint{-0.007133in}{0.026896in}}{\pgfqpoint{-0.013974in}{0.024062in}}{\pgfqpoint{-0.019018in}{0.019018in}}%
\pgfpathcurveto{\pgfqpoint{-0.024062in}{0.013974in}}{\pgfqpoint{-0.026896in}{0.007133in}}{\pgfqpoint{-0.026896in}{0.000000in}}%
\pgfpathcurveto{\pgfqpoint{-0.026896in}{-0.007133in}}{\pgfqpoint{-0.024062in}{-0.013974in}}{\pgfqpoint{-0.019018in}{-0.019018in}}%
\pgfpathcurveto{\pgfqpoint{-0.013974in}{-0.024062in}}{\pgfqpoint{-0.007133in}{-0.026896in}}{\pgfqpoint{0.000000in}{-0.026896in}}%
\pgfpathlineto{\pgfqpoint{0.000000in}{-0.026896in}}%
\pgfpathclose%
\pgfusepath{stroke,fill}%
}%
\begin{pgfscope}%
\pgfsys@transformshift{2.309784in}{1.694748in}%
\pgfsys@useobject{currentmarker}{}%
\end{pgfscope}%
\end{pgfscope}%
\begin{pgfscope}%
\definecolor{textcolor}{rgb}{0.000000,0.000000,0.000000}%
\pgfsetstrokecolor{textcolor}%
\pgfsetfillcolor{textcolor}%
\pgftext[x=2.518034in,y=1.664378in,left,base]{\color{textcolor}{\rmfamily\fontsize{8.330000}{9.996000}\selectfont\catcode`\^=\active\def^{\ifmmode\sp\else\^{}\fi}\catcode`\%=\active\def%{\%}$\bm{v}^{\text{NN}}$}}%
\end{pgfscope}%
\end{pgfpicture}%
\makeatother%
\endgroup%

%% file: figures/section3/n_Re_pressure.pgf
%% Creator: Matplotlib, PGF backend
%%
%% To include the figure in your LaTeX document, write
%%   \input{<filename>.pgf}
%%
%% Make sure the required packages are loaded in your preamble
%%   \usepackage{pgf}
%%
%% Also ensure that all the required font packages are loaded; for instance,
%% the lmodern package is sometimes necessary when using math font.
%%   \usepackage{lmodern}
%%
%% Figures using additional raster images can only be included by \input if
%% they are in the same directory as the main LaTeX file. For loading figures
%% from other directories you can use the `import` package
%% and then include the figures with
%%   \import{<path to file>}{<filename>.pgf}
%%
%% Matplotlib used the following preamble
%%   \def\mathdefault#1{#1}
%%   \everymath=\expandafter{\the\everymath\displaystyle}
%%   \usepackage{amsmath}\usepackage{bm}
%%   \makeatletter\@ifpackageloaded{underscore}{}{\usepackage[strings]{underscore}}\makeatother
%%
\begingroup%
\makeatletter%
\begin{pgfpicture}%
\pgfpathrectangle{\pgfpointorigin}{\pgfqpoint{3.000000in}{2.000000in}}%
\pgfusepath{use as bounding box, clip}%
\begin{pgfscope}%
\pgfsetbuttcap%
\pgfsetmiterjoin%
\definecolor{currentfill}{rgb}{1.000000,1.000000,1.000000}%
\pgfsetfillcolor{currentfill}%
\pgfsetlinewidth{0.000000pt}%
\definecolor{currentstroke}{rgb}{1.000000,1.000000,1.000000}%
\pgfsetstrokecolor{currentstroke}%
\pgfsetdash{}{0pt}%
\pgfpathmoveto{\pgfqpoint{0.000000in}{0.000000in}}%
\pgfpathlineto{\pgfqpoint{3.000000in}{0.000000in}}%
\pgfpathlineto{\pgfqpoint{3.000000in}{2.000000in}}%
\pgfpathlineto{\pgfqpoint{0.000000in}{2.000000in}}%
\pgfpathlineto{\pgfqpoint{0.000000in}{0.000000in}}%
\pgfpathclose%
\pgfusepath{fill}%
\end{pgfscope}%
\begin{pgfscope}%
\pgfsetbuttcap%
\pgfsetmiterjoin%
\definecolor{currentfill}{rgb}{1.000000,1.000000,1.000000}%
\pgfsetfillcolor{currentfill}%
\pgfsetlinewidth{0.000000pt}%
\definecolor{currentstroke}{rgb}{0.000000,0.000000,0.000000}%
\pgfsetstrokecolor{currentstroke}%
\pgfsetstrokeopacity{0.000000}%
\pgfsetdash{}{0pt}%
\pgfpathmoveto{\pgfqpoint{0.664468in}{0.521284in}}%
\pgfpathlineto{\pgfqpoint{2.850000in}{0.521284in}}%
\pgfpathlineto{\pgfqpoint{2.850000in}{1.850000in}}%
\pgfpathlineto{\pgfqpoint{0.664468in}{1.850000in}}%
\pgfpathlineto{\pgfqpoint{0.664468in}{0.521284in}}%
\pgfpathclose%
\pgfusepath{fill}%
\end{pgfscope}%
\begin{pgfscope}%
\pgfpathrectangle{\pgfqpoint{0.664468in}{0.521284in}}{\pgfqpoint{2.185532in}{1.328716in}}%
\pgfusepath{clip}%
\pgfsetbuttcap%
\pgfsetroundjoin%
\definecolor{currentfill}{rgb}{0.843137,0.098039,0.109804}%
\pgfsetfillcolor{currentfill}%
\pgfsetlinewidth{0.501875pt}%
\definecolor{currentstroke}{rgb}{0.000000,0.000000,0.000000}%
\pgfsetstrokecolor{currentstroke}%
\pgfsetdash{}{0pt}%
\pgfsys@defobject{currentmarker}{\pgfqpoint{-0.038036in}{-0.038036in}}{\pgfqpoint{0.038036in}{0.038036in}}{%
\pgfpathmoveto{\pgfqpoint{-0.000000in}{-0.038036in}}%
\pgfpathlineto{\pgfqpoint{0.038036in}{0.000000in}}%
\pgfpathlineto{\pgfqpoint{0.000000in}{0.038036in}}%
\pgfpathlineto{\pgfqpoint{-0.038036in}{0.000000in}}%
\pgfpathlineto{\pgfqpoint{-0.000000in}{-0.038036in}}%
\pgfpathclose%
\pgfusepath{stroke,fill}%
}%
\begin{pgfscope}%
\pgfsys@transformshift{0.763810in}{0.837085in}%
\pgfsys@useobject{currentmarker}{}%
\end{pgfscope}%
\begin{pgfscope}%
\pgfsys@transformshift{1.757234in}{0.807276in}%
\pgfsys@useobject{currentmarker}{}%
\end{pgfscope}%
\begin{pgfscope}%
\pgfsys@transformshift{2.750658in}{0.734437in}%
\pgfsys@useobject{currentmarker}{}%
\end{pgfscope}%
\end{pgfscope}%
\begin{pgfscope}%
\pgfpathrectangle{\pgfqpoint{0.664468in}{0.521284in}}{\pgfqpoint{2.185532in}{1.328716in}}%
\pgfusepath{clip}%
\pgfsetbuttcap%
\pgfsetroundjoin%
\definecolor{currentfill}{rgb}{0.172549,0.482353,0.713725}%
\pgfsetfillcolor{currentfill}%
\pgfsetlinewidth{0.501875pt}%
\definecolor{currentstroke}{rgb}{0.000000,0.000000,0.000000}%
\pgfsetstrokecolor{currentstroke}%
\pgfsetdash{}{0pt}%
\pgfsys@defobject{currentmarker}{\pgfqpoint{-0.025579in}{-0.021759in}}{\pgfqpoint{0.025579in}{0.026896in}}{%
\pgfpathmoveto{\pgfqpoint{0.000000in}{0.026896in}}%
\pgfpathlineto{\pgfqpoint{-0.025579in}{0.008311in}}%
\pgfpathlineto{\pgfqpoint{-0.015809in}{-0.021759in}}%
\pgfpathlineto{\pgfqpoint{0.015809in}{-0.021759in}}%
\pgfpathlineto{\pgfqpoint{0.025579in}{0.008311in}}%
\pgfpathlineto{\pgfqpoint{0.000000in}{0.026896in}}%
\pgfpathclose%
\pgfusepath{stroke,fill}%
}%
\begin{pgfscope}%
\pgfsys@transformshift{0.763810in}{1.415102in}%
\pgfsys@useobject{currentmarker}{}%
\end{pgfscope}%
\begin{pgfscope}%
\pgfsys@transformshift{1.757234in}{1.022795in}%
\pgfsys@useobject{currentmarker}{}%
\end{pgfscope}%
\begin{pgfscope}%
\pgfsys@transformshift{2.750658in}{0.706675in}%
\pgfsys@useobject{currentmarker}{}%
\end{pgfscope}%
\end{pgfscope}%
\begin{pgfscope}%
\pgfpathrectangle{\pgfqpoint{0.664468in}{0.521284in}}{\pgfqpoint{2.185532in}{1.328716in}}%
\pgfusepath{clip}%
\pgfsetbuttcap%
\pgfsetroundjoin%
\definecolor{currentfill}{rgb}{0.992157,0.682353,0.380392}%
\pgfsetfillcolor{currentfill}%
\pgfsetlinewidth{0.501875pt}%
\definecolor{currentstroke}{rgb}{0.000000,0.000000,0.000000}%
\pgfsetstrokecolor{currentstroke}%
\pgfsetdash{}{0pt}%
\pgfsys@defobject{currentmarker}{\pgfqpoint{-0.026896in}{-0.026896in}}{\pgfqpoint{0.026896in}{0.026896in}}{%
\pgfpathmoveto{\pgfqpoint{-0.026896in}{-0.026896in}}%
\pgfpathlineto{\pgfqpoint{0.026896in}{-0.026896in}}%
\pgfpathlineto{\pgfqpoint{0.026896in}{0.026896in}}%
\pgfpathlineto{\pgfqpoint{-0.026896in}{0.026896in}}%
\pgfpathlineto{\pgfqpoint{-0.026896in}{-0.026896in}}%
\pgfpathclose%
\pgfusepath{stroke,fill}%
}%
\begin{pgfscope}%
\pgfsys@transformshift{0.763810in}{1.389059in}%
\pgfsys@useobject{currentmarker}{}%
\end{pgfscope}%
\begin{pgfscope}%
\pgfsys@transformshift{1.757234in}{1.049158in}%
\pgfsys@useobject{currentmarker}{}%
\end{pgfscope}%
\begin{pgfscope}%
\pgfsys@transformshift{2.750658in}{0.723926in}%
\pgfsys@useobject{currentmarker}{}%
\end{pgfscope}%
\end{pgfscope}%
\begin{pgfscope}%
\pgfsetbuttcap%
\pgfsetroundjoin%
\definecolor{currentfill}{rgb}{0.000000,0.000000,0.000000}%
\pgfsetfillcolor{currentfill}%
\pgfsetlinewidth{0.803000pt}%
\definecolor{currentstroke}{rgb}{0.000000,0.000000,0.000000}%
\pgfsetstrokecolor{currentstroke}%
\pgfsetdash{}{0pt}%
\pgfsys@defobject{currentmarker}{\pgfqpoint{0.000000in}{-0.048611in}}{\pgfqpoint{0.000000in}{0.000000in}}{%
\pgfpathmoveto{\pgfqpoint{0.000000in}{0.000000in}}%
\pgfpathlineto{\pgfqpoint{0.000000in}{-0.048611in}}%
\pgfusepath{stroke,fill}%
}%
\begin{pgfscope}%
\pgfsys@transformshift{0.763810in}{0.521284in}%
\pgfsys@useobject{currentmarker}{}%
\end{pgfscope}%
\end{pgfscope}%
\begin{pgfscope}%
\definecolor{textcolor}{rgb}{0.000000,0.000000,0.000000}%
\pgfsetstrokecolor{textcolor}%
\pgfsetfillcolor{textcolor}%
\pgftext[x=0.763810in,y=0.431006in,,top]{\color{textcolor}{\rmfamily\fontsize{8.330000}{9.996000}\selectfont\catcode`\^=\active\def^{\ifmmode\sp\else\^{}\fi}\catcode`\%=\active\def%{\%}2}}%
\end{pgfscope}%
\begin{pgfscope}%
\pgfsetbuttcap%
\pgfsetroundjoin%
\definecolor{currentfill}{rgb}{0.000000,0.000000,0.000000}%
\pgfsetfillcolor{currentfill}%
\pgfsetlinewidth{0.803000pt}%
\definecolor{currentstroke}{rgb}{0.000000,0.000000,0.000000}%
\pgfsetstrokecolor{currentstroke}%
\pgfsetdash{}{0pt}%
\pgfsys@defobject{currentmarker}{\pgfqpoint{0.000000in}{-0.048611in}}{\pgfqpoint{0.000000in}{0.000000in}}{%
\pgfpathmoveto{\pgfqpoint{0.000000in}{0.000000in}}%
\pgfpathlineto{\pgfqpoint{0.000000in}{-0.048611in}}%
\pgfusepath{stroke,fill}%
}%
\begin{pgfscope}%
\pgfsys@transformshift{1.757234in}{0.521284in}%
\pgfsys@useobject{currentmarker}{}%
\end{pgfscope}%
\end{pgfscope}%
\begin{pgfscope}%
\definecolor{textcolor}{rgb}{0.000000,0.000000,0.000000}%
\pgfsetstrokecolor{textcolor}%
\pgfsetfillcolor{textcolor}%
\pgftext[x=1.757234in,y=0.431006in,,top]{\color{textcolor}{\rmfamily\fontsize{8.330000}{9.996000}\selectfont\catcode`\^=\active\def^{\ifmmode\sp\else\^{}\fi}\catcode`\%=\active\def%{\%}3}}%
\end{pgfscope}%
\begin{pgfscope}%
\pgfsetbuttcap%
\pgfsetroundjoin%
\definecolor{currentfill}{rgb}{0.000000,0.000000,0.000000}%
\pgfsetfillcolor{currentfill}%
\pgfsetlinewidth{0.803000pt}%
\definecolor{currentstroke}{rgb}{0.000000,0.000000,0.000000}%
\pgfsetstrokecolor{currentstroke}%
\pgfsetdash{}{0pt}%
\pgfsys@defobject{currentmarker}{\pgfqpoint{0.000000in}{-0.048611in}}{\pgfqpoint{0.000000in}{0.000000in}}{%
\pgfpathmoveto{\pgfqpoint{0.000000in}{0.000000in}}%
\pgfpathlineto{\pgfqpoint{0.000000in}{-0.048611in}}%
\pgfusepath{stroke,fill}%
}%
\begin{pgfscope}%
\pgfsys@transformshift{2.750658in}{0.521284in}%
\pgfsys@useobject{currentmarker}{}%
\end{pgfscope}%
\end{pgfscope}%
\begin{pgfscope}%
\definecolor{textcolor}{rgb}{0.000000,0.000000,0.000000}%
\pgfsetstrokecolor{textcolor}%
\pgfsetfillcolor{textcolor}%
\pgftext[x=2.750658in,y=0.431006in,,top]{\color{textcolor}{\rmfamily\fontsize{8.330000}{9.996000}\selectfont\catcode`\^=\active\def^{\ifmmode\sp\else\^{}\fi}\catcode`\%=\active\def%{\%}4}}%
\end{pgfscope}%
\begin{pgfscope}%
\definecolor{textcolor}{rgb}{0.000000,0.000000,0.000000}%
\pgfsetstrokecolor{textcolor}%
\pgfsetfillcolor{textcolor}%
\pgftext[x=1.757234in,y=0.276685in,,top]{\color{textcolor}{\rmfamily\fontsize{10.000000}{12.000000}\selectfont\catcode`\^=\active\def^{\ifmmode\sp\else\^{}\fi}\catcode`\%=\active\def%{\%}Number of sampled values of $\text{Re}$}}%
\end{pgfscope}%
\begin{pgfscope}%
\pgfsetbuttcap%
\pgfsetroundjoin%
\definecolor{currentfill}{rgb}{0.000000,0.000000,0.000000}%
\pgfsetfillcolor{currentfill}%
\pgfsetlinewidth{0.803000pt}%
\definecolor{currentstroke}{rgb}{0.000000,0.000000,0.000000}%
\pgfsetstrokecolor{currentstroke}%
\pgfsetdash{}{0pt}%
\pgfsys@defobject{currentmarker}{\pgfqpoint{-0.048611in}{0.000000in}}{\pgfqpoint{-0.000000in}{0.000000in}}{%
\pgfpathmoveto{\pgfqpoint{-0.000000in}{0.000000in}}%
\pgfpathlineto{\pgfqpoint{-0.048611in}{0.000000in}}%
\pgfusepath{stroke,fill}%
}%
\begin{pgfscope}%
\pgfsys@transformshift{0.664468in}{1.013589in}%
\pgfsys@useobject{currentmarker}{}%
\end{pgfscope}%
\end{pgfscope}%
\begin{pgfscope}%
\definecolor{textcolor}{rgb}{0.000000,0.000000,0.000000}%
\pgfsetstrokecolor{textcolor}%
\pgfsetfillcolor{textcolor}%
\pgftext[x=0.398263in, y=0.974461in, left, base]{\color{textcolor}{\rmfamily\fontsize{8.330000}{9.996000}\selectfont\catcode`\^=\active\def^{\ifmmode\sp\else\^{}\fi}\catcode`\%=\active\def%{\%}$\mathdefault{10^{1}}$}}%
\end{pgfscope}%
\begin{pgfscope}%
\pgfsetbuttcap%
\pgfsetroundjoin%
\definecolor{currentfill}{rgb}{0.000000,0.000000,0.000000}%
\pgfsetfillcolor{currentfill}%
\pgfsetlinewidth{0.803000pt}%
\definecolor{currentstroke}{rgb}{0.000000,0.000000,0.000000}%
\pgfsetstrokecolor{currentstroke}%
\pgfsetdash{}{0pt}%
\pgfsys@defobject{currentmarker}{\pgfqpoint{-0.048611in}{0.000000in}}{\pgfqpoint{-0.000000in}{0.000000in}}{%
\pgfpathmoveto{\pgfqpoint{-0.000000in}{0.000000in}}%
\pgfpathlineto{\pgfqpoint{-0.048611in}{0.000000in}}%
\pgfusepath{stroke,fill}%
}%
\begin{pgfscope}%
\pgfsys@transformshift{0.664468in}{1.505894in}%
\pgfsys@useobject{currentmarker}{}%
\end{pgfscope}%
\end{pgfscope}%
\begin{pgfscope}%
\definecolor{textcolor}{rgb}{0.000000,0.000000,0.000000}%
\pgfsetstrokecolor{textcolor}%
\pgfsetfillcolor{textcolor}%
\pgftext[x=0.398263in, y=1.466766in, left, base]{\color{textcolor}{\rmfamily\fontsize{8.330000}{9.996000}\selectfont\catcode`\^=\active\def^{\ifmmode\sp\else\^{}\fi}\catcode`\%=\active\def%{\%}$\mathdefault{10^{2}}$}}%
\end{pgfscope}%
\begin{pgfscope}%
\definecolor{textcolor}{rgb}{0.000000,0.000000,0.000000}%
\pgfsetstrokecolor{textcolor}%
\pgfsetfillcolor{textcolor}%
\pgftext[x=0.342708in,y=1.185642in,,bottom,rotate=90.000000]{\color{textcolor}{\rmfamily\fontsize{10.000000}{12.000000}\selectfont\catcode`\^=\active\def^{\ifmmode\sp\else\^{}\fi}\catcode`\%=\active\def%{\%}$\delta_{\ell^1}^{(p)} \ [\%]$}}%
\end{pgfscope}%
\begin{pgfscope}%
\pgfsetrectcap%
\pgfsetmiterjoin%
\pgfsetlinewidth{0.803000pt}%
\definecolor{currentstroke}{rgb}{0.000000,0.000000,0.000000}%
\pgfsetstrokecolor{currentstroke}%
\pgfsetdash{}{0pt}%
\pgfpathmoveto{\pgfqpoint{0.664468in}{0.521284in}}%
\pgfpathlineto{\pgfqpoint{0.664468in}{1.850000in}}%
\pgfusepath{stroke}%
\end{pgfscope}%
\begin{pgfscope}%
\pgfsetrectcap%
\pgfsetmiterjoin%
\pgfsetlinewidth{0.803000pt}%
\definecolor{currentstroke}{rgb}{0.000000,0.000000,0.000000}%
\pgfsetstrokecolor{currentstroke}%
\pgfsetdash{}{0pt}%
\pgfpathmoveto{\pgfqpoint{2.850000in}{0.521284in}}%
\pgfpathlineto{\pgfqpoint{2.850000in}{1.850000in}}%
\pgfusepath{stroke}%
\end{pgfscope}%
\begin{pgfscope}%
\pgfsetrectcap%
\pgfsetmiterjoin%
\pgfsetlinewidth{0.803000pt}%
\definecolor{currentstroke}{rgb}{0.000000,0.000000,0.000000}%
\pgfsetstrokecolor{currentstroke}%
\pgfsetdash{}{0pt}%
\pgfpathmoveto{\pgfqpoint{0.664468in}{0.521284in}}%
\pgfpathlineto{\pgfqpoint{2.850000in}{0.521284in}}%
\pgfusepath{stroke}%
\end{pgfscope}%
\begin{pgfscope}%
\pgfsetrectcap%
\pgfsetmiterjoin%
\pgfsetlinewidth{0.803000pt}%
\definecolor{currentstroke}{rgb}{0.000000,0.000000,0.000000}%
\pgfsetstrokecolor{currentstroke}%
\pgfsetdash{}{0pt}%
\pgfpathmoveto{\pgfqpoint{0.664468in}{1.850000in}}%
\pgfpathlineto{\pgfqpoint{2.850000in}{1.850000in}}%
\pgfusepath{stroke}%
\end{pgfscope}%
\begin{pgfscope}%
\pgfsetbuttcap%
\pgfsetmiterjoin%
\definecolor{currentfill}{rgb}{1.000000,1.000000,1.000000}%
\pgfsetfillcolor{currentfill}%
\pgfsetfillopacity{0.800000}%
\pgfsetlinewidth{1.003750pt}%
\definecolor{currentstroke}{rgb}{0.800000,0.800000,0.800000}%
\pgfsetstrokecolor{currentstroke}%
\pgfsetstrokeopacity{0.800000}%
\pgfsetdash{}{0pt}%
\pgfpathmoveto{\pgfqpoint{1.304161in}{1.392336in}}%
\pgfpathlineto{\pgfqpoint{2.769014in}{1.392336in}}%
\pgfpathquadraticcurveto{\pgfqpoint{2.792153in}{1.392336in}}{\pgfqpoint{2.792153in}{1.415475in}}%
\pgfpathlineto{\pgfqpoint{2.792153in}{1.769014in}}%
\pgfpathquadraticcurveto{\pgfqpoint{2.792153in}{1.792153in}}{\pgfqpoint{2.769014in}{1.792153in}}%
\pgfpathlineto{\pgfqpoint{1.304161in}{1.792153in}}%
\pgfpathquadraticcurveto{\pgfqpoint{1.281022in}{1.792153in}}{\pgfqpoint{1.281022in}{1.769014in}}%
\pgfpathlineto{\pgfqpoint{1.281022in}{1.415475in}}%
\pgfpathquadraticcurveto{\pgfqpoint{1.281022in}{1.392336in}}{\pgfqpoint{1.304161in}{1.392336in}}%
\pgfpathlineto{\pgfqpoint{1.304161in}{1.392336in}}%
\pgfpathclose%
\pgfusepath{stroke,fill}%
\end{pgfscope}%
\begin{pgfscope}%
\pgfsetbuttcap%
\pgfsetroundjoin%
\definecolor{currentfill}{rgb}{0.843137,0.098039,0.109804}%
\pgfsetfillcolor{currentfill}%
\pgfsetlinewidth{0.501875pt}%
\definecolor{currentstroke}{rgb}{0.000000,0.000000,0.000000}%
\pgfsetstrokecolor{currentstroke}%
\pgfsetdash{}{0pt}%
\pgfsys@defobject{currentmarker}{\pgfqpoint{-0.038036in}{-0.038036in}}{\pgfqpoint{0.038036in}{0.038036in}}{%
\pgfpathmoveto{\pgfqpoint{-0.000000in}{-0.038036in}}%
\pgfpathlineto{\pgfqpoint{0.038036in}{0.000000in}}%
\pgfpathlineto{\pgfqpoint{0.000000in}{0.038036in}}%
\pgfpathlineto{\pgfqpoint{-0.038036in}{0.000000in}}%
\pgfpathlineto{\pgfqpoint{-0.000000in}{-0.038036in}}%
\pgfpathclose%
\pgfusepath{stroke,fill}%
}%
\begin{pgfscope}%
\pgfsys@transformshift{1.442994in}{1.673143in}%
\pgfsys@useobject{currentmarker}{}%
\end{pgfscope}%
\end{pgfscope}%
\begin{pgfscope}%
\definecolor{textcolor}{rgb}{0.000000,0.000000,0.000000}%
\pgfsetstrokecolor{textcolor}%
\pgfsetfillcolor{textcolor}%
\pgftext[x=1.651244in,y=1.642773in,left,base]{\color{textcolor}{\rmfamily\fontsize{8.330000}{9.996000}\selectfont\catcode`\^=\active\def^{\ifmmode\sp\else\^{}\fi}\catcode`\%=\active\def%{\%}$p^\text{PINN}$}}%
\end{pgfscope}%
\begin{pgfscope}%
\pgfsetbuttcap%
\pgfsetroundjoin%
\definecolor{currentfill}{rgb}{0.172549,0.482353,0.713725}%
\pgfsetfillcolor{currentfill}%
\pgfsetlinewidth{0.501875pt}%
\definecolor{currentstroke}{rgb}{0.000000,0.000000,0.000000}%
\pgfsetstrokecolor{currentstroke}%
\pgfsetdash{}{0pt}%
\pgfsys@defobject{currentmarker}{\pgfqpoint{-0.025579in}{-0.021759in}}{\pgfqpoint{0.025579in}{0.026896in}}{%
\pgfpathmoveto{\pgfqpoint{0.000000in}{0.026896in}}%
\pgfpathlineto{\pgfqpoint{-0.025579in}{0.008311in}}%
\pgfpathlineto{\pgfqpoint{-0.015809in}{-0.021759in}}%
\pgfpathlineto{\pgfqpoint{0.015809in}{-0.021759in}}%
\pgfpathlineto{\pgfqpoint{0.025579in}{0.008311in}}%
\pgfpathlineto{\pgfqpoint{0.000000in}{0.026896in}}%
\pgfpathclose%
\pgfusepath{stroke,fill}%
}%
\begin{pgfscope}%
\pgfsys@transformshift{1.442994in}{1.490589in}%
\pgfsys@useobject{currentmarker}{}%
\end{pgfscope}%
\end{pgfscope}%
\begin{pgfscope}%
\definecolor{textcolor}{rgb}{0.000000,0.000000,0.000000}%
\pgfsetstrokecolor{textcolor}%
\pgfsetfillcolor{textcolor}%
\pgftext[x=1.651244in,y=1.460219in,left,base]{\color{textcolor}{\rmfamily\fontsize{8.330000}{9.996000}\selectfont\catcode`\^=\active\def^{\ifmmode\sp\else\^{}\fi}\catcode`\%=\active\def%{\%}$p^\text{BINN}$}}%
\end{pgfscope}%
\begin{pgfscope}%
\pgfsetbuttcap%
\pgfsetroundjoin%
\definecolor{currentfill}{rgb}{0.992157,0.682353,0.380392}%
\pgfsetfillcolor{currentfill}%
\pgfsetlinewidth{0.501875pt}%
\definecolor{currentstroke}{rgb}{0.000000,0.000000,0.000000}%
\pgfsetstrokecolor{currentstroke}%
\pgfsetdash{}{0pt}%
\pgfsys@defobject{currentmarker}{\pgfqpoint{-0.026896in}{-0.026896in}}{\pgfqpoint{0.026896in}{0.026896in}}{%
\pgfpathmoveto{\pgfqpoint{-0.026896in}{-0.026896in}}%
\pgfpathlineto{\pgfqpoint{0.026896in}{-0.026896in}}%
\pgfpathlineto{\pgfqpoint{0.026896in}{0.026896in}}%
\pgfpathlineto{\pgfqpoint{-0.026896in}{0.026896in}}%
\pgfpathlineto{\pgfqpoint{-0.026896in}{-0.026896in}}%
\pgfpathclose%
\pgfusepath{stroke,fill}%
}%
\begin{pgfscope}%
\pgfsys@transformshift{2.321872in}{1.673143in}%
\pgfsys@useobject{currentmarker}{}%
\end{pgfscope}%
\end{pgfscope}%
\begin{pgfscope}%
\definecolor{textcolor}{rgb}{0.000000,0.000000,0.000000}%
\pgfsetstrokecolor{textcolor}%
\pgfsetfillcolor{textcolor}%
\pgftext[x=2.530122in,y=1.642773in,left,base]{\color{textcolor}{\rmfamily\fontsize{8.330000}{9.996000}\selectfont\catcode`\^=\active\def^{\ifmmode\sp\else\^{}\fi}\catcode`\%=\active\def%{\%}$p^{\text{NN}}$}}%
\end{pgfscope}%
\end{pgfpicture}%
\makeatother%
\endgroup%

%% file: figures/section3/dataBC/3Re/13500p/re_1000/vmag.pgf
%% Creator: Matplotlib, PGF backend
%%
%% To include the figure in your LaTeX document, write
%%   \input{<filename>.pgf}
%%
%% Make sure the required packages are loaded in your preamble
%%   \usepackage{pgf}
%%
%% Also ensure that all the required font packages are loaded; for instance,
%% the lmodern package is sometimes necessary when using math font.
%%   \usepackage{lmodern}
%%
%% Figures using additional raster images can only be included by \input if
%% they are in the same directory as the main LaTeX file. For loading figures
%% from other directories you can use the `import` package
%% and then include the figures with
%%   \import{<path to file>}{<filename>.pgf}
%%
%% Matplotlib used the following preamble
%%   \def\mathdefault#1{#1}
%%   \everymath=\expandafter{\the\everymath\displaystyle}
%%   \usepackage{amsmath}\usepackage{bm}
%%   \makeatletter\@ifpackageloaded{underscore}{}{\usepackage[strings]{underscore}}\makeatother
%%
\begingroup%
\makeatletter%
\begin{pgfpicture}%
\pgfpathrectangle{\pgfpointorigin}{\pgfqpoint{2.500000in}{2.500000in}}%
\pgfusepath{use as bounding box, clip}%
\begin{pgfscope}%
\pgfsetbuttcap%
\pgfsetmiterjoin%
\definecolor{currentfill}{rgb}{1.000000,1.000000,1.000000}%
\pgfsetfillcolor{currentfill}%
\pgfsetlinewidth{0.000000pt}%
\definecolor{currentstroke}{rgb}{1.000000,1.000000,1.000000}%
\pgfsetstrokecolor{currentstroke}%
\pgfsetdash{}{0pt}%
\pgfpathmoveto{\pgfqpoint{0.000000in}{0.000000in}}%
\pgfpathlineto{\pgfqpoint{2.500000in}{0.000000in}}%
\pgfpathlineto{\pgfqpoint{2.500000in}{2.500000in}}%
\pgfpathlineto{\pgfqpoint{0.000000in}{2.500000in}}%
\pgfpathlineto{\pgfqpoint{0.000000in}{0.000000in}}%
\pgfpathclose%
\pgfusepath{fill}%
\end{pgfscope}%
\begin{pgfscope}%
\pgfsetbuttcap%
\pgfsetmiterjoin%
\definecolor{currentfill}{rgb}{1.000000,1.000000,1.000000}%
\pgfsetfillcolor{currentfill}%
\pgfsetlinewidth{0.000000pt}%
\definecolor{currentstroke}{rgb}{0.000000,0.000000,0.000000}%
\pgfsetstrokecolor{currentstroke}%
\pgfsetstrokeopacity{0.000000}%
\pgfsetdash{}{0pt}%
\pgfpathmoveto{\pgfqpoint{0.568385in}{0.386658in}}%
\pgfpathlineto{\pgfqpoint{2.130905in}{0.386658in}}%
\pgfpathlineto{\pgfqpoint{2.130905in}{1.949178in}}%
\pgfpathlineto{\pgfqpoint{0.568385in}{1.949178in}}%
\pgfpathlineto{\pgfqpoint{0.568385in}{0.386658in}}%
\pgfpathclose%
\pgfusepath{fill}%
\end{pgfscope}%
\begin{pgfscope}%
\pgfsys@transformshift{0.638750in}{0.457500in}%
\pgftext[left,bottom]{\includegraphics[interpolate=true,width=1.421250in,height=1.421250in]{figures/./section3/dataBC/3Re/13500p/re_1000//vmag-img0.png}}%
\end{pgfscope}%
\begin{pgfscope}%
\pgfsetbuttcap%
\pgfsetroundjoin%
\definecolor{currentfill}{rgb}{0.000000,0.000000,0.000000}%
\pgfsetfillcolor{currentfill}%
\pgfsetlinewidth{0.803000pt}%
\definecolor{currentstroke}{rgb}{0.000000,0.000000,0.000000}%
\pgfsetstrokecolor{currentstroke}%
\pgfsetdash{}{0pt}%
\pgfsys@defobject{currentmarker}{\pgfqpoint{0.000000in}{-0.048611in}}{\pgfqpoint{0.000000in}{0.000000in}}{%
\pgfpathmoveto{\pgfqpoint{0.000000in}{0.000000in}}%
\pgfpathlineto{\pgfqpoint{0.000000in}{-0.048611in}}%
\pgfusepath{stroke,fill}%
}%
\begin{pgfscope}%
\pgfsys@transformshift{0.639408in}{0.386658in}%
\pgfsys@useobject{currentmarker}{}%
\end{pgfscope}%
\end{pgfscope}%
\begin{pgfscope}%
\definecolor{textcolor}{rgb}{0.000000,0.000000,0.000000}%
\pgfsetstrokecolor{textcolor}%
\pgfsetfillcolor{textcolor}%
\pgftext[x=0.639408in,y=0.296381in,,top]{\color{textcolor}{\rmfamily\fontsize{8.330000}{9.996000}\selectfont\catcode`\^=\active\def^{\ifmmode\sp\else\^{}\fi}\catcode`\%=\active\def%{\%}$\mathdefault{\ensuremath{-}0.1}$}}%
\end{pgfscope}%
\begin{pgfscope}%
\pgfsetbuttcap%
\pgfsetroundjoin%
\definecolor{currentfill}{rgb}{0.000000,0.000000,0.000000}%
\pgfsetfillcolor{currentfill}%
\pgfsetlinewidth{0.803000pt}%
\definecolor{currentstroke}{rgb}{0.000000,0.000000,0.000000}%
\pgfsetstrokecolor{currentstroke}%
\pgfsetdash{}{0pt}%
\pgfsys@defobject{currentmarker}{\pgfqpoint{0.000000in}{-0.048611in}}{\pgfqpoint{0.000000in}{0.000000in}}{%
\pgfpathmoveto{\pgfqpoint{0.000000in}{0.000000in}}%
\pgfpathlineto{\pgfqpoint{0.000000in}{-0.048611in}}%
\pgfusepath{stroke,fill}%
}%
\begin{pgfscope}%
\pgfsys@transformshift{1.349645in}{0.386658in}%
\pgfsys@useobject{currentmarker}{}%
\end{pgfscope}%
\end{pgfscope}%
\begin{pgfscope}%
\definecolor{textcolor}{rgb}{0.000000,0.000000,0.000000}%
\pgfsetstrokecolor{textcolor}%
\pgfsetfillcolor{textcolor}%
\pgftext[x=1.349645in,y=0.296381in,,top]{\color{textcolor}{\rmfamily\fontsize{8.330000}{9.996000}\selectfont\catcode`\^=\active\def^{\ifmmode\sp\else\^{}\fi}\catcode`\%=\active\def%{\%}$\mathdefault{0.0}$}}%
\end{pgfscope}%
\begin{pgfscope}%
\pgfsetbuttcap%
\pgfsetroundjoin%
\definecolor{currentfill}{rgb}{0.000000,0.000000,0.000000}%
\pgfsetfillcolor{currentfill}%
\pgfsetlinewidth{0.803000pt}%
\definecolor{currentstroke}{rgb}{0.000000,0.000000,0.000000}%
\pgfsetstrokecolor{currentstroke}%
\pgfsetdash{}{0pt}%
\pgfsys@defobject{currentmarker}{\pgfqpoint{0.000000in}{-0.048611in}}{\pgfqpoint{0.000000in}{0.000000in}}{%
\pgfpathmoveto{\pgfqpoint{0.000000in}{0.000000in}}%
\pgfpathlineto{\pgfqpoint{0.000000in}{-0.048611in}}%
\pgfusepath{stroke,fill}%
}%
\begin{pgfscope}%
\pgfsys@transformshift{2.059881in}{0.386658in}%
\pgfsys@useobject{currentmarker}{}%
\end{pgfscope}%
\end{pgfscope}%
\begin{pgfscope}%
\definecolor{textcolor}{rgb}{0.000000,0.000000,0.000000}%
\pgfsetstrokecolor{textcolor}%
\pgfsetfillcolor{textcolor}%
\pgftext[x=2.059881in,y=0.296381in,,top]{\color{textcolor}{\rmfamily\fontsize{8.330000}{9.996000}\selectfont\catcode`\^=\active\def^{\ifmmode\sp\else\^{}\fi}\catcode`\%=\active\def%{\%}$\mathdefault{0.1}$}}%
\end{pgfscope}%
\begin{pgfscope}%
\definecolor{textcolor}{rgb}{0.000000,0.000000,0.000000}%
\pgfsetstrokecolor{textcolor}%
\pgfsetfillcolor{textcolor}%
\pgftext[x=1.349645in,y=0.142060in,,top]{\color{textcolor}{\rmfamily\fontsize{10.000000}{12.000000}\selectfont\catcode`\^=\active\def^{\ifmmode\sp\else\^{}\fi}\catcode`\%=\active\def%{\%}$x\;[\text{m}]$}}%
\end{pgfscope}%
\begin{pgfscope}%
\pgfsetbuttcap%
\pgfsetroundjoin%
\definecolor{currentfill}{rgb}{0.000000,0.000000,0.000000}%
\pgfsetfillcolor{currentfill}%
\pgfsetlinewidth{0.803000pt}%
\definecolor{currentstroke}{rgb}{0.000000,0.000000,0.000000}%
\pgfsetstrokecolor{currentstroke}%
\pgfsetdash{}{0pt}%
\pgfsys@defobject{currentmarker}{\pgfqpoint{-0.048611in}{0.000000in}}{\pgfqpoint{-0.000000in}{0.000000in}}{%
\pgfpathmoveto{\pgfqpoint{-0.000000in}{0.000000in}}%
\pgfpathlineto{\pgfqpoint{-0.048611in}{0.000000in}}%
\pgfusepath{stroke,fill}%
}%
\begin{pgfscope}%
\pgfsys@transformshift{0.568385in}{0.457682in}%
\pgfsys@useobject{currentmarker}{}%
\end{pgfscope}%
\end{pgfscope}%
\begin{pgfscope}%
\definecolor{textcolor}{rgb}{0.000000,0.000000,0.000000}%
\pgfsetstrokecolor{textcolor}%
\pgfsetfillcolor{textcolor}%
\pgftext[x=0.235434in, y=0.419102in, left, base]{\color{textcolor}{\rmfamily\fontsize{8.330000}{9.996000}\selectfont\catcode`\^=\active\def^{\ifmmode\sp\else\^{}\fi}\catcode`\%=\active\def%{\%}$\mathdefault{\ensuremath{-}0.1}$}}%
\end{pgfscope}%
\begin{pgfscope}%
\pgfsetbuttcap%
\pgfsetroundjoin%
\definecolor{currentfill}{rgb}{0.000000,0.000000,0.000000}%
\pgfsetfillcolor{currentfill}%
\pgfsetlinewidth{0.803000pt}%
\definecolor{currentstroke}{rgb}{0.000000,0.000000,0.000000}%
\pgfsetstrokecolor{currentstroke}%
\pgfsetdash{}{0pt}%
\pgfsys@defobject{currentmarker}{\pgfqpoint{-0.048611in}{0.000000in}}{\pgfqpoint{-0.000000in}{0.000000in}}{%
\pgfpathmoveto{\pgfqpoint{-0.000000in}{0.000000in}}%
\pgfpathlineto{\pgfqpoint{-0.048611in}{0.000000in}}%
\pgfusepath{stroke,fill}%
}%
\begin{pgfscope}%
\pgfsys@transformshift{0.568385in}{1.167918in}%
\pgfsys@useobject{currentmarker}{}%
\end{pgfscope}%
\end{pgfscope}%
\begin{pgfscope}%
\definecolor{textcolor}{rgb}{0.000000,0.000000,0.000000}%
\pgfsetstrokecolor{textcolor}%
\pgfsetfillcolor{textcolor}%
\pgftext[x=0.327256in, y=1.129338in, left, base]{\color{textcolor}{\rmfamily\fontsize{8.330000}{9.996000}\selectfont\catcode`\^=\active\def^{\ifmmode\sp\else\^{}\fi}\catcode`\%=\active\def%{\%}$\mathdefault{0.0}$}}%
\end{pgfscope}%
\begin{pgfscope}%
\pgfsetbuttcap%
\pgfsetroundjoin%
\definecolor{currentfill}{rgb}{0.000000,0.000000,0.000000}%
\pgfsetfillcolor{currentfill}%
\pgfsetlinewidth{0.803000pt}%
\definecolor{currentstroke}{rgb}{0.000000,0.000000,0.000000}%
\pgfsetstrokecolor{currentstroke}%
\pgfsetdash{}{0pt}%
\pgfsys@defobject{currentmarker}{\pgfqpoint{-0.048611in}{0.000000in}}{\pgfqpoint{-0.000000in}{0.000000in}}{%
\pgfpathmoveto{\pgfqpoint{-0.000000in}{0.000000in}}%
\pgfpathlineto{\pgfqpoint{-0.048611in}{0.000000in}}%
\pgfusepath{stroke,fill}%
}%
\begin{pgfscope}%
\pgfsys@transformshift{0.568385in}{1.878155in}%
\pgfsys@useobject{currentmarker}{}%
\end{pgfscope}%
\end{pgfscope}%
\begin{pgfscope}%
\definecolor{textcolor}{rgb}{0.000000,0.000000,0.000000}%
\pgfsetstrokecolor{textcolor}%
\pgfsetfillcolor{textcolor}%
\pgftext[x=0.327256in, y=1.839574in, left, base]{\color{textcolor}{\rmfamily\fontsize{8.330000}{9.996000}\selectfont\catcode`\^=\active\def^{\ifmmode\sp\else\^{}\fi}\catcode`\%=\active\def%{\%}$\mathdefault{0.1}$}}%
\end{pgfscope}%
\begin{pgfscope}%
\definecolor{textcolor}{rgb}{0.000000,0.000000,0.000000}%
\pgfsetstrokecolor{textcolor}%
\pgfsetfillcolor{textcolor}%
\pgftext[x=0.179878in,y=1.167918in,,bottom,rotate=90.000000]{\color{textcolor}{\rmfamily\fontsize{10.000000}{12.000000}\selectfont\catcode`\^=\active\def^{\ifmmode\sp\else\^{}\fi}\catcode`\%=\active\def%{\%}$y\;[\text{m}]$}}%
\end{pgfscope}%
\begin{pgfscope}%
\pgfpathrectangle{\pgfqpoint{0.568385in}{0.386658in}}{\pgfqpoint{1.562520in}{1.562520in}}%
\pgfusepath{clip}%
\pgfsetbuttcap%
\pgfsetmiterjoin%
\definecolor{currentfill}{rgb}{1.000000,1.000000,1.000000}%
\pgfsetfillcolor{currentfill}%
\pgfsetlinewidth{1.505625pt}%
\definecolor{currentstroke}{rgb}{1.000000,1.000000,1.000000}%
\pgfsetstrokecolor{currentstroke}%
\pgfsetdash{}{0pt}%
\pgfpathmoveto{\pgfqpoint{1.832454in}{1.675838in}}%
\pgfpathlineto{\pgfqpoint{1.825606in}{1.668990in}}%
\pgfpathlineto{\pgfqpoint{1.818757in}{1.662141in}}%
\pgfpathlineto{\pgfqpoint{1.811909in}{1.655293in}}%
\pgfpathlineto{\pgfqpoint{1.805060in}{1.648445in}}%
\pgfpathlineto{\pgfqpoint{1.798212in}{1.641596in}}%
\pgfpathlineto{\pgfqpoint{1.791364in}{1.634748in}}%
\pgfpathlineto{\pgfqpoint{1.784515in}{1.627900in}}%
\pgfpathlineto{\pgfqpoint{1.777667in}{1.621051in}}%
\pgfpathlineto{\pgfqpoint{1.770819in}{1.614203in}}%
\pgfpathlineto{\pgfqpoint{1.763970in}{1.607355in}}%
\pgfpathlineto{\pgfqpoint{1.770248in}{1.601077in}}%
\pgfpathlineto{\pgfqpoint{1.776526in}{1.594799in}}%
\pgfpathlineto{\pgfqpoint{1.782803in}{1.588522in}}%
\pgfpathlineto{\pgfqpoint{1.789081in}{1.582244in}}%
\pgfpathlineto{\pgfqpoint{1.795929in}{1.589092in}}%
\pgfpathlineto{\pgfqpoint{1.802778in}{1.595941in}}%
\pgfpathlineto{\pgfqpoint{1.809626in}{1.602789in}}%
\pgfpathlineto{\pgfqpoint{1.816474in}{1.609637in}}%
\pgfpathlineto{\pgfqpoint{1.823323in}{1.616486in}}%
\pgfpathlineto{\pgfqpoint{1.830171in}{1.623334in}}%
\pgfpathlineto{\pgfqpoint{1.837019in}{1.630182in}}%
\pgfpathlineto{\pgfqpoint{1.843868in}{1.637031in}}%
\pgfpathlineto{\pgfqpoint{1.850716in}{1.643879in}}%
\pgfpathlineto{\pgfqpoint{1.857565in}{1.650728in}}%
\pgfpathlineto{\pgfqpoint{1.832454in}{1.675838in}}%
\pgfpathclose%
\pgfusepath{stroke,fill}%
\end{pgfscope}%
\begin{pgfscope}%
\pgfpathrectangle{\pgfqpoint{0.568385in}{0.386658in}}{\pgfqpoint{1.562520in}{1.562520in}}%
\pgfusepath{clip}%
\pgfsetbuttcap%
\pgfsetmiterjoin%
\definecolor{currentfill}{rgb}{1.000000,1.000000,1.000000}%
\pgfsetfillcolor{currentfill}%
\pgfsetlinewidth{1.505625pt}%
\definecolor{currentstroke}{rgb}{1.000000,1.000000,1.000000}%
\pgfsetstrokecolor{currentstroke}%
\pgfsetdash{}{0pt}%
\pgfpathmoveto{\pgfqpoint{1.857565in}{0.685109in}}%
\pgfpathlineto{\pgfqpoint{1.850716in}{0.691958in}}%
\pgfpathlineto{\pgfqpoint{1.843868in}{0.698806in}}%
\pgfpathlineto{\pgfqpoint{1.837019in}{0.705654in}}%
\pgfpathlineto{\pgfqpoint{1.830171in}{0.712503in}}%
\pgfpathlineto{\pgfqpoint{1.823323in}{0.719351in}}%
\pgfpathlineto{\pgfqpoint{1.816474in}{0.726199in}}%
\pgfpathlineto{\pgfqpoint{1.809626in}{0.733048in}}%
\pgfpathlineto{\pgfqpoint{1.802778in}{0.739896in}}%
\pgfpathlineto{\pgfqpoint{1.795929in}{0.746744in}}%
\pgfpathlineto{\pgfqpoint{1.789081in}{0.753593in}}%
\pgfpathlineto{\pgfqpoint{1.782678in}{0.747190in}}%
\pgfpathlineto{\pgfqpoint{1.776275in}{0.740786in}}%
\pgfpathlineto{\pgfqpoint{1.769871in}{0.734383in}}%
\pgfpathlineto{\pgfqpoint{1.763468in}{0.727980in}}%
\pgfpathlineto{\pgfqpoint{1.763970in}{0.728482in}}%
\pgfpathlineto{\pgfqpoint{1.770819in}{0.721634in}}%
\pgfpathlineto{\pgfqpoint{1.777667in}{0.714785in}}%
\pgfpathlineto{\pgfqpoint{1.784515in}{0.707937in}}%
\pgfpathlineto{\pgfqpoint{1.791364in}{0.701089in}}%
\pgfpathlineto{\pgfqpoint{1.798212in}{0.694240in}}%
\pgfpathlineto{\pgfqpoint{1.805060in}{0.687392in}}%
\pgfpathlineto{\pgfqpoint{1.811909in}{0.680544in}}%
\pgfpathlineto{\pgfqpoint{1.818757in}{0.673695in}}%
\pgfpathlineto{\pgfqpoint{1.825606in}{0.666847in}}%
\pgfpathlineto{\pgfqpoint{1.832454in}{0.659999in}}%
\pgfpathlineto{\pgfqpoint{1.857565in}{0.685109in}}%
\pgfpathclose%
\pgfusepath{stroke,fill}%
\end{pgfscope}%
\begin{pgfscope}%
\pgfpathrectangle{\pgfqpoint{0.568385in}{0.386658in}}{\pgfqpoint{1.562520in}{1.562520in}}%
\pgfusepath{clip}%
\pgfsetbuttcap%
\pgfsetmiterjoin%
\definecolor{currentfill}{rgb}{1.000000,1.000000,1.000000}%
\pgfsetfillcolor{currentfill}%
\pgfsetlinewidth{1.505625pt}%
\definecolor{currentstroke}{rgb}{1.000000,1.000000,1.000000}%
\pgfsetstrokecolor{currentstroke}%
\pgfsetdash{}{0pt}%
\pgfpathmoveto{\pgfqpoint{0.866836in}{0.659999in}}%
\pgfpathlineto{\pgfqpoint{0.873684in}{0.666847in}}%
\pgfpathlineto{\pgfqpoint{0.880532in}{0.673695in}}%
\pgfpathlineto{\pgfqpoint{0.887381in}{0.680544in}}%
\pgfpathlineto{\pgfqpoint{0.894229in}{0.687392in}}%
\pgfpathlineto{\pgfqpoint{0.901077in}{0.694240in}}%
\pgfpathlineto{\pgfqpoint{0.907926in}{0.701089in}}%
\pgfpathlineto{\pgfqpoint{0.914774in}{0.707937in}}%
\pgfpathlineto{\pgfqpoint{0.921622in}{0.714785in}}%
\pgfpathlineto{\pgfqpoint{0.928471in}{0.721634in}}%
\pgfpathlineto{\pgfqpoint{0.935319in}{0.728482in}}%
\pgfpathlineto{\pgfqpoint{0.928916in}{0.734885in}}%
\pgfpathlineto{\pgfqpoint{0.922513in}{0.741289in}}%
\pgfpathlineto{\pgfqpoint{0.916109in}{0.747692in}}%
\pgfpathlineto{\pgfqpoint{0.909706in}{0.754095in}}%
\pgfpathlineto{\pgfqpoint{0.910208in}{0.753593in}}%
\pgfpathlineto{\pgfqpoint{0.903360in}{0.746744in}}%
\pgfpathlineto{\pgfqpoint{0.896512in}{0.739896in}}%
\pgfpathlineto{\pgfqpoint{0.889663in}{0.733048in}}%
\pgfpathlineto{\pgfqpoint{0.882815in}{0.726199in}}%
\pgfpathlineto{\pgfqpoint{0.875967in}{0.719351in}}%
\pgfpathlineto{\pgfqpoint{0.869118in}{0.712503in}}%
\pgfpathlineto{\pgfqpoint{0.862270in}{0.705654in}}%
\pgfpathlineto{\pgfqpoint{0.855422in}{0.698806in}}%
\pgfpathlineto{\pgfqpoint{0.848573in}{0.691958in}}%
\pgfpathlineto{\pgfqpoint{0.841725in}{0.685109in}}%
\pgfpathlineto{\pgfqpoint{0.866836in}{0.659999in}}%
\pgfpathclose%
\pgfusepath{stroke,fill}%
\end{pgfscope}%
\begin{pgfscope}%
\pgfpathrectangle{\pgfqpoint{0.568385in}{0.386658in}}{\pgfqpoint{1.562520in}{1.562520in}}%
\pgfusepath{clip}%
\pgfsetbuttcap%
\pgfsetmiterjoin%
\definecolor{currentfill}{rgb}{1.000000,1.000000,1.000000}%
\pgfsetfillcolor{currentfill}%
\pgfsetlinewidth{1.505625pt}%
\definecolor{currentstroke}{rgb}{1.000000,1.000000,1.000000}%
\pgfsetstrokecolor{currentstroke}%
\pgfsetdash{}{0pt}%
\pgfpathmoveto{\pgfqpoint{0.841725in}{1.650728in}}%
\pgfpathlineto{\pgfqpoint{0.848573in}{1.643879in}}%
\pgfpathlineto{\pgfqpoint{0.855422in}{1.637031in}}%
\pgfpathlineto{\pgfqpoint{0.862270in}{1.630182in}}%
\pgfpathlineto{\pgfqpoint{0.869118in}{1.623334in}}%
\pgfpathlineto{\pgfqpoint{0.875967in}{1.616486in}}%
\pgfpathlineto{\pgfqpoint{0.882815in}{1.609637in}}%
\pgfpathlineto{\pgfqpoint{0.889663in}{1.602789in}}%
\pgfpathlineto{\pgfqpoint{0.896512in}{1.595941in}}%
\pgfpathlineto{\pgfqpoint{0.903360in}{1.589092in}}%
\pgfpathlineto{\pgfqpoint{0.910208in}{1.582244in}}%
\pgfpathlineto{\pgfqpoint{0.916486in}{1.588522in}}%
\pgfpathlineto{\pgfqpoint{0.922764in}{1.594799in}}%
\pgfpathlineto{\pgfqpoint{0.929041in}{1.601077in}}%
\pgfpathlineto{\pgfqpoint{0.935319in}{1.607355in}}%
\pgfpathlineto{\pgfqpoint{0.928471in}{1.614203in}}%
\pgfpathlineto{\pgfqpoint{0.921622in}{1.621051in}}%
\pgfpathlineto{\pgfqpoint{0.914774in}{1.627900in}}%
\pgfpathlineto{\pgfqpoint{0.907926in}{1.634748in}}%
\pgfpathlineto{\pgfqpoint{0.901077in}{1.641596in}}%
\pgfpathlineto{\pgfqpoint{0.894229in}{1.648445in}}%
\pgfpathlineto{\pgfqpoint{0.887381in}{1.655293in}}%
\pgfpathlineto{\pgfqpoint{0.880532in}{1.662141in}}%
\pgfpathlineto{\pgfqpoint{0.873684in}{1.668990in}}%
\pgfpathlineto{\pgfqpoint{0.866836in}{1.675838in}}%
\pgfpathlineto{\pgfqpoint{0.841725in}{1.650728in}}%
\pgfpathclose%
\pgfusepath{stroke,fill}%
\end{pgfscope}%
\begin{pgfscope}%
\pgfpathrectangle{\pgfqpoint{0.568385in}{0.386658in}}{\pgfqpoint{1.562520in}{1.562520in}}%
\pgfusepath{clip}%
\pgfsetrectcap%
\pgfsetroundjoin%
\pgfsetlinewidth{1.505625pt}%
\definecolor{currentstroke}{rgb}{0.000000,0.000000,0.000000}%
\pgfsetstrokecolor{currentstroke}%
\pgfsetdash{}{0pt}%
\pgfpathmoveto{\pgfqpoint{0.841725in}{1.650728in}}%
\pgfpathlineto{\pgfqpoint{0.910208in}{1.582244in}}%
\pgfpathlineto{\pgfqpoint{0.935319in}{1.607355in}}%
\pgfpathlineto{\pgfqpoint{0.866836in}{1.675838in}}%
\pgfpathlineto{\pgfqpoint{0.860750in}{1.683106in}}%
\pgfpathlineto{\pgfqpoint{0.876947in}{1.698006in}}%
\pgfpathlineto{\pgfqpoint{0.893598in}{1.712397in}}%
\pgfpathlineto{\pgfqpoint{0.910687in}{1.726266in}}%
\pgfpathlineto{\pgfqpoint{0.928197in}{1.739598in}}%
\pgfpathlineto{\pgfqpoint{0.946112in}{1.752382in}}%
\pgfpathlineto{\pgfqpoint{0.964415in}{1.764604in}}%
\pgfpathlineto{\pgfqpoint{0.983087in}{1.776253in}}%
\pgfpathlineto{\pgfqpoint{1.002112in}{1.787318in}}%
\pgfpathlineto{\pgfqpoint{1.021470in}{1.797789in}}%
\pgfpathlineto{\pgfqpoint{1.041143in}{1.807654in}}%
\pgfpathlineto{\pgfqpoint{1.061112in}{1.816906in}}%
\pgfpathlineto{\pgfqpoint{1.081359in}{1.825534in}}%
\pgfpathlineto{\pgfqpoint{1.101863in}{1.833531in}}%
\pgfpathlineto{\pgfqpoint{1.122605in}{1.840888in}}%
\pgfpathlineto{\pgfqpoint{1.143565in}{1.847600in}}%
\pgfpathlineto{\pgfqpoint{1.164723in}{1.853658in}}%
\pgfpathlineto{\pgfqpoint{1.186058in}{1.859059in}}%
\pgfpathlineto{\pgfqpoint{1.207551in}{1.863795in}}%
\pgfpathlineto{\pgfqpoint{1.229180in}{1.867864in}}%
\pgfpathlineto{\pgfqpoint{1.250924in}{1.871260in}}%
\pgfpathlineto{\pgfqpoint{1.272764in}{1.873981in}}%
\pgfpathlineto{\pgfqpoint{1.294677in}{1.876024in}}%
\pgfpathlineto{\pgfqpoint{1.316643in}{1.877387in}}%
\pgfpathlineto{\pgfqpoint{1.338641in}{1.878069in}}%
\pgfpathlineto{\pgfqpoint{1.360649in}{1.878069in}}%
\pgfpathlineto{\pgfqpoint{1.382647in}{1.877387in}}%
\pgfpathlineto{\pgfqpoint{1.404613in}{1.876024in}}%
\pgfpathlineto{\pgfqpoint{1.426526in}{1.873981in}}%
\pgfpathlineto{\pgfqpoint{1.448365in}{1.871260in}}%
\pgfpathlineto{\pgfqpoint{1.470110in}{1.867864in}}%
\pgfpathlineto{\pgfqpoint{1.491739in}{1.863795in}}%
\pgfpathlineto{\pgfqpoint{1.513231in}{1.859059in}}%
\pgfpathlineto{\pgfqpoint{1.534567in}{1.853658in}}%
\pgfpathlineto{\pgfqpoint{1.555725in}{1.847600in}}%
\pgfpathlineto{\pgfqpoint{1.576685in}{1.840888in}}%
\pgfpathlineto{\pgfqpoint{1.597427in}{1.833531in}}%
\pgfpathlineto{\pgfqpoint{1.617931in}{1.825534in}}%
\pgfpathlineto{\pgfqpoint{1.638177in}{1.816906in}}%
\pgfpathlineto{\pgfqpoint{1.658147in}{1.807654in}}%
\pgfpathlineto{\pgfqpoint{1.677820in}{1.797789in}}%
\pgfpathlineto{\pgfqpoint{1.697178in}{1.787318in}}%
\pgfpathlineto{\pgfqpoint{1.716202in}{1.776253in}}%
\pgfpathlineto{\pgfqpoint{1.734874in}{1.764604in}}%
\pgfpathlineto{\pgfqpoint{1.753177in}{1.752382in}}%
\pgfpathlineto{\pgfqpoint{1.771092in}{1.739598in}}%
\pgfpathlineto{\pgfqpoint{1.788602in}{1.726266in}}%
\pgfpathlineto{\pgfqpoint{1.805691in}{1.712397in}}%
\pgfpathlineto{\pgfqpoint{1.822342in}{1.698006in}}%
\pgfpathlineto{\pgfqpoint{1.838539in}{1.683106in}}%
\pgfpathlineto{\pgfqpoint{1.832454in}{1.675838in}}%
\pgfpathlineto{\pgfqpoint{1.763970in}{1.607355in}}%
\pgfpathlineto{\pgfqpoint{1.789081in}{1.582244in}}%
\pgfpathlineto{\pgfqpoint{1.857565in}{1.650728in}}%
\pgfpathlineto{\pgfqpoint{1.864832in}{1.656813in}}%
\pgfpathlineto{\pgfqpoint{1.879732in}{1.640616in}}%
\pgfpathlineto{\pgfqpoint{1.894124in}{1.623965in}}%
\pgfpathlineto{\pgfqpoint{1.907992in}{1.606876in}}%
\pgfpathlineto{\pgfqpoint{1.921324in}{1.589366in}}%
\pgfpathlineto{\pgfqpoint{1.934108in}{1.571451in}}%
\pgfpathlineto{\pgfqpoint{1.946330in}{1.553148in}}%
\pgfpathlineto{\pgfqpoint{1.957980in}{1.534476in}}%
\pgfpathlineto{\pgfqpoint{1.969045in}{1.515451in}}%
\pgfpathlineto{\pgfqpoint{1.979515in}{1.496093in}}%
\pgfpathlineto{\pgfqpoint{1.989381in}{1.476420in}}%
\pgfpathlineto{\pgfqpoint{1.998632in}{1.456451in}}%
\pgfpathlineto{\pgfqpoint{2.007260in}{1.436204in}}%
\pgfpathlineto{\pgfqpoint{2.015257in}{1.415700in}}%
\pgfpathlineto{\pgfqpoint{2.022615in}{1.394958in}}%
\pgfpathlineto{\pgfqpoint{2.029326in}{1.373998in}}%
\pgfpathlineto{\pgfqpoint{2.035385in}{1.352840in}}%
\pgfpathlineto{\pgfqpoint{2.040785in}{1.331505in}}%
\pgfpathlineto{\pgfqpoint{2.045522in}{1.310012in}}%
\pgfpathlineto{\pgfqpoint{2.049590in}{1.288384in}}%
\pgfpathlineto{\pgfqpoint{2.052987in}{1.266639in}}%
\pgfpathlineto{\pgfqpoint{2.055708in}{1.244799in}}%
\pgfpathlineto{\pgfqpoint{2.057751in}{1.222886in}}%
\pgfpathlineto{\pgfqpoint{2.059114in}{1.200920in}}%
\pgfpathlineto{\pgfqpoint{2.059796in}{1.178922in}}%
\pgfpathlineto{\pgfqpoint{2.059796in}{1.156914in}}%
\pgfpathlineto{\pgfqpoint{2.059114in}{1.134917in}}%
\pgfpathlineto{\pgfqpoint{2.057751in}{1.112950in}}%
\pgfpathlineto{\pgfqpoint{2.055708in}{1.091037in}}%
\pgfpathlineto{\pgfqpoint{2.052987in}{1.069198in}}%
\pgfpathlineto{\pgfqpoint{2.049590in}{1.047453in}}%
\pgfpathlineto{\pgfqpoint{2.045522in}{1.025824in}}%
\pgfpathlineto{\pgfqpoint{2.040785in}{1.004332in}}%
\pgfpathlineto{\pgfqpoint{2.035385in}{0.982996in}}%
\pgfpathlineto{\pgfqpoint{2.029326in}{0.961838in}}%
\pgfpathlineto{\pgfqpoint{2.022615in}{0.940878in}}%
\pgfpathlineto{\pgfqpoint{2.015257in}{0.920136in}}%
\pgfpathlineto{\pgfqpoint{2.007260in}{0.899632in}}%
\pgfpathlineto{\pgfqpoint{1.998632in}{0.879386in}}%
\pgfpathlineto{\pgfqpoint{1.989381in}{0.859417in}}%
\pgfpathlineto{\pgfqpoint{1.979515in}{0.839743in}}%
\pgfpathlineto{\pgfqpoint{1.969045in}{0.820385in}}%
\pgfpathlineto{\pgfqpoint{1.957980in}{0.801361in}}%
\pgfpathlineto{\pgfqpoint{1.946330in}{0.782689in}}%
\pgfpathlineto{\pgfqpoint{1.934108in}{0.764386in}}%
\pgfpathlineto{\pgfqpoint{1.921324in}{0.746471in}}%
\pgfpathlineto{\pgfqpoint{1.907992in}{0.728961in}}%
\pgfpathlineto{\pgfqpoint{1.894124in}{0.711872in}}%
\pgfpathlineto{\pgfqpoint{1.879732in}{0.695221in}}%
\pgfpathlineto{\pgfqpoint{1.864832in}{0.679024in}}%
\pgfpathlineto{\pgfqpoint{1.857565in}{0.685109in}}%
\pgfpathlineto{\pgfqpoint{1.789081in}{0.753593in}}%
\pgfpathlineto{\pgfqpoint{1.763468in}{0.727980in}}%
\pgfpathlineto{\pgfqpoint{1.763970in}{0.728482in}}%
\pgfpathlineto{\pgfqpoint{1.832454in}{0.659999in}}%
\pgfpathlineto{\pgfqpoint{1.838539in}{0.652731in}}%
\pgfpathlineto{\pgfqpoint{1.822342in}{0.637831in}}%
\pgfpathlineto{\pgfqpoint{1.805691in}{0.623440in}}%
\pgfpathlineto{\pgfqpoint{1.788602in}{0.609571in}}%
\pgfpathlineto{\pgfqpoint{1.771092in}{0.596239in}}%
\pgfpathlineto{\pgfqpoint{1.753177in}{0.583455in}}%
\pgfpathlineto{\pgfqpoint{1.734874in}{0.571233in}}%
\pgfpathlineto{\pgfqpoint{1.716202in}{0.559584in}}%
\pgfpathlineto{\pgfqpoint{1.697178in}{0.548518in}}%
\pgfpathlineto{\pgfqpoint{1.677820in}{0.538048in}}%
\pgfpathlineto{\pgfqpoint{1.658147in}{0.528182in}}%
\pgfpathlineto{\pgfqpoint{1.638177in}{0.518931in}}%
\pgfpathlineto{\pgfqpoint{1.617931in}{0.510303in}}%
\pgfpathlineto{\pgfqpoint{1.597427in}{0.502306in}}%
\pgfpathlineto{\pgfqpoint{1.576685in}{0.494949in}}%
\pgfpathlineto{\pgfqpoint{1.555725in}{0.488237in}}%
\pgfpathlineto{\pgfqpoint{1.534567in}{0.482178in}}%
\pgfpathlineto{\pgfqpoint{1.513231in}{0.476778in}}%
\pgfpathlineto{\pgfqpoint{1.491739in}{0.472041in}}%
\pgfpathlineto{\pgfqpoint{1.470110in}{0.467973in}}%
\pgfpathlineto{\pgfqpoint{1.448365in}{0.464576in}}%
\pgfpathlineto{\pgfqpoint{1.426526in}{0.461855in}}%
\pgfpathlineto{\pgfqpoint{1.404613in}{0.459812in}}%
\pgfpathlineto{\pgfqpoint{1.382647in}{0.458449in}}%
\pgfpathlineto{\pgfqpoint{1.360649in}{0.457767in}}%
\pgfpathlineto{\pgfqpoint{1.338641in}{0.457767in}}%
\pgfpathlineto{\pgfqpoint{1.316643in}{0.458449in}}%
\pgfpathlineto{\pgfqpoint{1.294677in}{0.459812in}}%
\pgfpathlineto{\pgfqpoint{1.272764in}{0.461855in}}%
\pgfpathlineto{\pgfqpoint{1.250924in}{0.464576in}}%
\pgfpathlineto{\pgfqpoint{1.229180in}{0.467973in}}%
\pgfpathlineto{\pgfqpoint{1.207551in}{0.472041in}}%
\pgfpathlineto{\pgfqpoint{1.186058in}{0.476778in}}%
\pgfpathlineto{\pgfqpoint{1.164723in}{0.482178in}}%
\pgfpathlineto{\pgfqpoint{1.143565in}{0.488237in}}%
\pgfpathlineto{\pgfqpoint{1.122605in}{0.494949in}}%
\pgfpathlineto{\pgfqpoint{1.101863in}{0.502306in}}%
\pgfpathlineto{\pgfqpoint{1.081359in}{0.510303in}}%
\pgfpathlineto{\pgfqpoint{1.061112in}{0.518931in}}%
\pgfpathlineto{\pgfqpoint{1.041143in}{0.528182in}}%
\pgfpathlineto{\pgfqpoint{1.021470in}{0.538048in}}%
\pgfpathlineto{\pgfqpoint{1.002112in}{0.548518in}}%
\pgfpathlineto{\pgfqpoint{0.983087in}{0.559584in}}%
\pgfpathlineto{\pgfqpoint{0.964415in}{0.571233in}}%
\pgfpathlineto{\pgfqpoint{0.946112in}{0.583455in}}%
\pgfpathlineto{\pgfqpoint{0.928197in}{0.596239in}}%
\pgfpathlineto{\pgfqpoint{0.910687in}{0.609571in}}%
\pgfpathlineto{\pgfqpoint{0.893598in}{0.623440in}}%
\pgfpathlineto{\pgfqpoint{0.876947in}{0.637831in}}%
\pgfpathlineto{\pgfqpoint{0.860750in}{0.652731in}}%
\pgfpathlineto{\pgfqpoint{0.866836in}{0.659999in}}%
\pgfpathlineto{\pgfqpoint{0.935319in}{0.728482in}}%
\pgfpathlineto{\pgfqpoint{0.909706in}{0.754095in}}%
\pgfpathlineto{\pgfqpoint{0.910208in}{0.753593in}}%
\pgfpathlineto{\pgfqpoint{0.841725in}{0.685109in}}%
\pgfpathlineto{\pgfqpoint{0.834458in}{0.679024in}}%
\pgfpathlineto{\pgfqpoint{0.819557in}{0.695221in}}%
\pgfpathlineto{\pgfqpoint{0.805166in}{0.711872in}}%
\pgfpathlineto{\pgfqpoint{0.791297in}{0.728961in}}%
\pgfpathlineto{\pgfqpoint{0.777965in}{0.746471in}}%
\pgfpathlineto{\pgfqpoint{0.765181in}{0.764386in}}%
\pgfpathlineto{\pgfqpoint{0.752959in}{0.782689in}}%
\pgfpathlineto{\pgfqpoint{0.741310in}{0.801361in}}%
\pgfpathlineto{\pgfqpoint{0.730245in}{0.820385in}}%
\pgfpathlineto{\pgfqpoint{0.719774in}{0.839743in}}%
\pgfpathlineto{\pgfqpoint{0.709909in}{0.859417in}}%
\pgfpathlineto{\pgfqpoint{0.700657in}{0.879386in}}%
\pgfpathlineto{\pgfqpoint{0.692029in}{0.899632in}}%
\pgfpathlineto{\pgfqpoint{0.684032in}{0.920136in}}%
\pgfpathlineto{\pgfqpoint{0.676675in}{0.940878in}}%
\pgfpathlineto{\pgfqpoint{0.669963in}{0.961838in}}%
\pgfpathlineto{\pgfqpoint{0.663905in}{0.982996in}}%
\pgfpathlineto{\pgfqpoint{0.658504in}{1.004332in}}%
\pgfpathlineto{\pgfqpoint{0.653768in}{1.025824in}}%
\pgfpathlineto{\pgfqpoint{0.649699in}{1.047453in}}%
\pgfpathlineto{\pgfqpoint{0.646303in}{1.069198in}}%
\pgfpathlineto{\pgfqpoint{0.643582in}{1.091037in}}%
\pgfpathlineto{\pgfqpoint{0.641539in}{1.112950in}}%
\pgfpathlineto{\pgfqpoint{0.640176in}{1.134917in}}%
\pgfpathlineto{\pgfqpoint{0.639494in}{1.156914in}}%
\pgfpathlineto{\pgfqpoint{0.639494in}{1.178922in}}%
\pgfpathlineto{\pgfqpoint{0.640176in}{1.200920in}}%
\pgfpathlineto{\pgfqpoint{0.641539in}{1.222886in}}%
\pgfpathlineto{\pgfqpoint{0.643582in}{1.244799in}}%
\pgfpathlineto{\pgfqpoint{0.646303in}{1.266639in}}%
\pgfpathlineto{\pgfqpoint{0.649699in}{1.288384in}}%
\pgfpathlineto{\pgfqpoint{0.653768in}{1.310012in}}%
\pgfpathlineto{\pgfqpoint{0.658504in}{1.331505in}}%
\pgfpathlineto{\pgfqpoint{0.663905in}{1.352840in}}%
\pgfpathlineto{\pgfqpoint{0.669963in}{1.373998in}}%
\pgfpathlineto{\pgfqpoint{0.676675in}{1.394958in}}%
\pgfpathlineto{\pgfqpoint{0.684032in}{1.415700in}}%
\pgfpathlineto{\pgfqpoint{0.692029in}{1.436204in}}%
\pgfpathlineto{\pgfqpoint{0.700657in}{1.456451in}}%
\pgfpathlineto{\pgfqpoint{0.709909in}{1.476420in}}%
\pgfpathlineto{\pgfqpoint{0.719774in}{1.496093in}}%
\pgfpathlineto{\pgfqpoint{0.730245in}{1.515451in}}%
\pgfpathlineto{\pgfqpoint{0.741310in}{1.534476in}}%
\pgfpathlineto{\pgfqpoint{0.752959in}{1.553148in}}%
\pgfpathlineto{\pgfqpoint{0.765181in}{1.571451in}}%
\pgfpathlineto{\pgfqpoint{0.777965in}{1.589366in}}%
\pgfpathlineto{\pgfqpoint{0.791297in}{1.606876in}}%
\pgfpathlineto{\pgfqpoint{0.805166in}{1.623965in}}%
\pgfpathlineto{\pgfqpoint{0.819557in}{1.640616in}}%
\pgfpathlineto{\pgfqpoint{0.834458in}{1.656813in}}%
\pgfpathlineto{\pgfqpoint{0.841725in}{1.650728in}}%
\pgfpathlineto{\pgfqpoint{0.910208in}{1.582244in}}%
\pgfpathlineto{\pgfqpoint{0.935319in}{1.607355in}}%
\pgfpathlineto{\pgfqpoint{0.866836in}{1.675838in}}%
\pgfpathlineto{\pgfqpoint{0.866836in}{1.675838in}}%
\pgfusepath{stroke}%
\end{pgfscope}%
\begin{pgfscope}%
\pgfsetrectcap%
\pgfsetmiterjoin%
\pgfsetlinewidth{0.803000pt}%
\definecolor{currentstroke}{rgb}{0.000000,0.000000,0.000000}%
\pgfsetstrokecolor{currentstroke}%
\pgfsetdash{}{0pt}%
\pgfpathmoveto{\pgfqpoint{0.568385in}{0.386658in}}%
\pgfpathlineto{\pgfqpoint{0.568385in}{1.949178in}}%
\pgfusepath{stroke}%
\end{pgfscope}%
\begin{pgfscope}%
\pgfsetrectcap%
\pgfsetmiterjoin%
\pgfsetlinewidth{0.803000pt}%
\definecolor{currentstroke}{rgb}{0.000000,0.000000,0.000000}%
\pgfsetstrokecolor{currentstroke}%
\pgfsetdash{}{0pt}%
\pgfpathmoveto{\pgfqpoint{2.130905in}{0.386658in}}%
\pgfpathlineto{\pgfqpoint{2.130905in}{1.949178in}}%
\pgfusepath{stroke}%
\end{pgfscope}%
\begin{pgfscope}%
\pgfsetrectcap%
\pgfsetmiterjoin%
\pgfsetlinewidth{0.803000pt}%
\definecolor{currentstroke}{rgb}{0.000000,0.000000,0.000000}%
\pgfsetstrokecolor{currentstroke}%
\pgfsetdash{}{0pt}%
\pgfpathmoveto{\pgfqpoint{0.568385in}{0.386658in}}%
\pgfpathlineto{\pgfqpoint{2.130905in}{0.386658in}}%
\pgfusepath{stroke}%
\end{pgfscope}%
\begin{pgfscope}%
\pgfsetrectcap%
\pgfsetmiterjoin%
\pgfsetlinewidth{0.803000pt}%
\definecolor{currentstroke}{rgb}{0.000000,0.000000,0.000000}%
\pgfsetstrokecolor{currentstroke}%
\pgfsetdash{}{0pt}%
\pgfpathmoveto{\pgfqpoint{0.568385in}{1.949178in}}%
\pgfpathlineto{\pgfqpoint{2.130905in}{1.949178in}}%
\pgfusepath{stroke}%
\end{pgfscope}%
\begin{pgfscope}%
\pgfpathrectangle{\pgfqpoint{0.568385in}{0.386658in}}{\pgfqpoint{1.562520in}{1.562520in}}%
\pgfusepath{clip}%
\pgfsetrectcap%
\pgfsetroundjoin%
\pgfsetlinewidth{1.505625pt}%
\definecolor{currentstroke}{rgb}{0.000000,0.000000,0.000000}%
\pgfsetstrokecolor{currentstroke}%
\pgfsetdash{}{0pt}%
\pgfpathmoveto{\pgfqpoint{1.065550in}{1.167918in}}%
\pgfpathlineto{\pgfqpoint{1.633739in}{1.167918in}}%
\pgfusepath{stroke}%
\end{pgfscope}%
\begin{pgfscope}%
\pgfpathrectangle{\pgfqpoint{0.568385in}{0.386658in}}{\pgfqpoint{1.562520in}{1.562520in}}%
\pgfusepath{clip}%
\pgfsetrectcap%
\pgfsetroundjoin%
\pgfsetlinewidth{1.505625pt}%
\definecolor{currentstroke}{rgb}{0.000000,0.000000,0.000000}%
\pgfsetstrokecolor{currentstroke}%
\pgfsetdash{}{0pt}%
\pgfpathmoveto{\pgfqpoint{1.349645in}{0.883824in}}%
\pgfpathlineto{\pgfqpoint{1.349645in}{1.452013in}}%
\pgfusepath{stroke}%
\end{pgfscope}%
\begin{pgfscope}%
\pgfsetbuttcap%
\pgfsetmiterjoin%
\definecolor{currentfill}{rgb}{1.000000,1.000000,1.000000}%
\pgfsetfillcolor{currentfill}%
\pgfsetlinewidth{0.000000pt}%
\definecolor{currentstroke}{rgb}{0.000000,0.000000,0.000000}%
\pgfsetstrokecolor{currentstroke}%
\pgfsetstrokeopacity{0.000000}%
\pgfsetdash{}{0pt}%
\pgfpathmoveto{\pgfqpoint{0.337193in}{2.046836in}}%
\pgfpathlineto{\pgfqpoint{2.362097in}{2.046836in}}%
\pgfpathlineto{\pgfqpoint{2.362097in}{2.148081in}}%
\pgfpathlineto{\pgfqpoint{0.337193in}{2.148081in}}%
\pgfpathlineto{\pgfqpoint{0.337193in}{2.046836in}}%
\pgfpathclose%
\pgfusepath{fill}%
\end{pgfscope}%
\begin{pgfscope}%
\pgfsys@transformshift{0.337500in}{2.046250in}%
\pgftext[left,bottom]{\includegraphics[interpolate=true,width=2.025000in,height=0.101250in]{figures/./section3/dataBC/3Re/13500p/re_1000//vmag-img1.png}}%
\end{pgfscope}%
\begin{pgfscope}%
\pgfsetbuttcap%
\pgfsetroundjoin%
\definecolor{currentfill}{rgb}{0.000000,0.000000,0.000000}%
\pgfsetfillcolor{currentfill}%
\pgfsetlinewidth{0.803000pt}%
\definecolor{currentstroke}{rgb}{0.000000,0.000000,0.000000}%
\pgfsetstrokecolor{currentstroke}%
\pgfsetdash{}{0pt}%
\pgfsys@defobject{currentmarker}{\pgfqpoint{0.000000in}{0.000000in}}{\pgfqpoint{0.000000in}{0.048611in}}{%
\pgfpathmoveto{\pgfqpoint{0.000000in}{0.000000in}}%
\pgfpathlineto{\pgfqpoint{0.000000in}{0.048611in}}%
\pgfusepath{stroke,fill}%
}%
\begin{pgfscope}%
\pgfsys@transformshift{0.337809in}{2.148081in}%
\pgfsys@useobject{currentmarker}{}%
\end{pgfscope}%
\end{pgfscope}%
\begin{pgfscope}%
\definecolor{textcolor}{rgb}{0.000000,0.000000,0.000000}%
\pgfsetstrokecolor{textcolor}%
\pgfsetfillcolor{textcolor}%
\pgftext[x=0.337809in,y=2.238359in,,bottom]{\color{textcolor}{\rmfamily\fontsize{8.330000}{9.996000}\selectfont\catcode`\^=\active\def^{\ifmmode\sp\else\^{}\fi}\catcode`\%=\active\def%{\%}$\mathdefault{0.001}$}}%
\end{pgfscope}%
\begin{pgfscope}%
\pgfsetbuttcap%
\pgfsetroundjoin%
\definecolor{currentfill}{rgb}{0.000000,0.000000,0.000000}%
\pgfsetfillcolor{currentfill}%
\pgfsetlinewidth{0.803000pt}%
\definecolor{currentstroke}{rgb}{0.000000,0.000000,0.000000}%
\pgfsetstrokecolor{currentstroke}%
\pgfsetdash{}{0pt}%
\pgfsys@defobject{currentmarker}{\pgfqpoint{0.000000in}{0.000000in}}{\pgfqpoint{0.000000in}{0.048611in}}{%
\pgfpathmoveto{\pgfqpoint{0.000000in}{0.000000in}}%
\pgfpathlineto{\pgfqpoint{0.000000in}{0.048611in}}%
\pgfusepath{stroke,fill}%
}%
\begin{pgfscope}%
\pgfsys@transformshift{1.012024in}{2.148081in}%
\pgfsys@useobject{currentmarker}{}%
\end{pgfscope}%
\end{pgfscope}%
\begin{pgfscope}%
\definecolor{textcolor}{rgb}{0.000000,0.000000,0.000000}%
\pgfsetstrokecolor{textcolor}%
\pgfsetfillcolor{textcolor}%
\pgftext[x=1.012024in,y=2.238359in,,bottom]{\color{textcolor}{\rmfamily\fontsize{8.330000}{9.996000}\selectfont\catcode`\^=\active\def^{\ifmmode\sp\else\^{}\fi}\catcode`\%=\active\def%{\%}$\mathdefault{1.170}$}}%
\end{pgfscope}%
\begin{pgfscope}%
\pgfsetbuttcap%
\pgfsetroundjoin%
\definecolor{currentfill}{rgb}{0.000000,0.000000,0.000000}%
\pgfsetfillcolor{currentfill}%
\pgfsetlinewidth{0.803000pt}%
\definecolor{currentstroke}{rgb}{0.000000,0.000000,0.000000}%
\pgfsetstrokecolor{currentstroke}%
\pgfsetdash{}{0pt}%
\pgfsys@defobject{currentmarker}{\pgfqpoint{0.000000in}{0.000000in}}{\pgfqpoint{0.000000in}{0.048611in}}{%
\pgfpathmoveto{\pgfqpoint{0.000000in}{0.000000in}}%
\pgfpathlineto{\pgfqpoint{0.000000in}{0.048611in}}%
\pgfusepath{stroke,fill}%
}%
\begin{pgfscope}%
\pgfsys@transformshift{1.686238in}{2.148081in}%
\pgfsys@useobject{currentmarker}{}%
\end{pgfscope}%
\end{pgfscope}%
\begin{pgfscope}%
\definecolor{textcolor}{rgb}{0.000000,0.000000,0.000000}%
\pgfsetstrokecolor{textcolor}%
\pgfsetfillcolor{textcolor}%
\pgftext[x=1.686238in,y=2.238359in,,bottom]{\color{textcolor}{\rmfamily\fontsize{8.330000}{9.996000}\selectfont\catcode`\^=\active\def^{\ifmmode\sp\else\^{}\fi}\catcode`\%=\active\def%{\%}$\mathdefault{2.340}$}}%
\end{pgfscope}%
\begin{pgfscope}%
\pgfsetbuttcap%
\pgfsetroundjoin%
\definecolor{currentfill}{rgb}{0.000000,0.000000,0.000000}%
\pgfsetfillcolor{currentfill}%
\pgfsetlinewidth{0.803000pt}%
\definecolor{currentstroke}{rgb}{0.000000,0.000000,0.000000}%
\pgfsetstrokecolor{currentstroke}%
\pgfsetdash{}{0pt}%
\pgfsys@defobject{currentmarker}{\pgfqpoint{0.000000in}{0.000000in}}{\pgfqpoint{0.000000in}{0.048611in}}{%
\pgfpathmoveto{\pgfqpoint{0.000000in}{0.000000in}}%
\pgfpathlineto{\pgfqpoint{0.000000in}{0.048611in}}%
\pgfusepath{stroke,fill}%
}%
\begin{pgfscope}%
\pgfsys@transformshift{2.360452in}{2.148081in}%
\pgfsys@useobject{currentmarker}{}%
\end{pgfscope}%
\end{pgfscope}%
\begin{pgfscope}%
\definecolor{textcolor}{rgb}{0.000000,0.000000,0.000000}%
\pgfsetstrokecolor{textcolor}%
\pgfsetfillcolor{textcolor}%
\pgftext[x=2.360452in,y=2.238359in,,bottom]{\color{textcolor}{\rmfamily\fontsize{8.330000}{9.996000}\selectfont\catcode`\^=\active\def^{\ifmmode\sp\else\^{}\fi}\catcode`\%=\active\def%{\%}$\mathdefault{3.509}$}}%
\end{pgfscope}%
\begin{pgfscope}%
\definecolor{textcolor}{rgb}{0.000000,0.000000,0.000000}%
\pgfsetstrokecolor{textcolor}%
\pgfsetfillcolor{textcolor}%
\pgftext[x=1.349645in,y=2.392680in,,base]{\color{textcolor}{\rmfamily\fontsize{10.000000}{12.000000}\selectfont\catcode`\^=\active\def^{\ifmmode\sp\else\^{}\fi}\catcode`\%=\active\def%{\%}$\|\bm{v}\|_2\;[\nicefrac{\mathrm{m}}{\mathrm{s}}]$}}%
\end{pgfscope}%
\begin{pgfscope}%
\definecolor{textcolor}{rgb}{0.000000,0.000000,0.000000}%
\pgfsetstrokecolor{textcolor}%
\pgfsetfillcolor{textcolor}%
\pgftext[x=2.362097in,y=2.378791in,right,bottom]{\color{textcolor}{\rmfamily\fontsize{8.330000}{9.996000}\selectfont\catcode`\^=\active\def^{\ifmmode\sp\else\^{}\fi}\catcode`\%=\active\def%{\%}$\times\mathdefault{10^{\ensuremath{-}2}}\mathdefault{}$}}%
\end{pgfscope}%
\begin{pgfscope}%
\pgfsetrectcap%
\pgfsetmiterjoin%
\pgfsetlinewidth{0.803000pt}%
\definecolor{currentstroke}{rgb}{0.000000,0.000000,0.000000}%
\pgfsetstrokecolor{currentstroke}%
\pgfsetdash{}{0pt}%
\pgfpathmoveto{\pgfqpoint{0.337193in}{2.046836in}}%
\pgfpathlineto{\pgfqpoint{0.337193in}{2.097458in}}%
\pgfpathlineto{\pgfqpoint{0.337193in}{2.148081in}}%
\pgfpathlineto{\pgfqpoint{2.362097in}{2.148081in}}%
\pgfpathlineto{\pgfqpoint{2.362097in}{2.097458in}}%
\pgfpathlineto{\pgfqpoint{2.362097in}{2.046836in}}%
\pgfpathlineto{\pgfqpoint{0.337193in}{2.046836in}}%
\pgfpathclose%
\pgfusepath{stroke}%
\end{pgfscope}%
\end{pgfpicture}%
\makeatother%
\endgroup%

%% file: figures/section3/dataBC/3Re/13500p/re_1000/err_vmag.pgf
%% Creator: Matplotlib, PGF backend
%%
%% To include the figure in your LaTeX document, write
%%   \input{<filename>.pgf}
%%
%% Make sure the required packages are loaded in your preamble
%%   \usepackage{pgf}
%%
%% Also ensure that all the required font packages are loaded; for instance,
%% the lmodern package is sometimes necessary when using math font.
%%   \usepackage{lmodern}
%%
%% Figures using additional raster images can only be included by \input if
%% they are in the same directory as the main LaTeX file. For loading figures
%% from other directories you can use the `import` package
%% and then include the figures with
%%   \import{<path to file>}{<filename>.pgf}
%%
%% Matplotlib used the following preamble
%%   \def\mathdefault#1{#1}
%%   \everymath=\expandafter{\the\everymath\displaystyle}
%%   \usepackage{amsmath}\usepackage{bm}
%%   \makeatletter\@ifpackageloaded{underscore}{}{\usepackage[strings]{underscore}}\makeatother
%%
\begingroup%
\makeatletter%
\begin{pgfpicture}%
\pgfpathrectangle{\pgfpointorigin}{\pgfqpoint{2.500000in}{2.500000in}}%
\pgfusepath{use as bounding box, clip}%
\begin{pgfscope}%
\pgfsetbuttcap%
\pgfsetmiterjoin%
\definecolor{currentfill}{rgb}{1.000000,1.000000,1.000000}%
\pgfsetfillcolor{currentfill}%
\pgfsetlinewidth{0.000000pt}%
\definecolor{currentstroke}{rgb}{1.000000,1.000000,1.000000}%
\pgfsetstrokecolor{currentstroke}%
\pgfsetdash{}{0pt}%
\pgfpathmoveto{\pgfqpoint{0.000000in}{0.000000in}}%
\pgfpathlineto{\pgfqpoint{2.500000in}{0.000000in}}%
\pgfpathlineto{\pgfqpoint{2.500000in}{2.500000in}}%
\pgfpathlineto{\pgfqpoint{0.000000in}{2.500000in}}%
\pgfpathlineto{\pgfqpoint{0.000000in}{0.000000in}}%
\pgfpathclose%
\pgfusepath{fill}%
\end{pgfscope}%
\begin{pgfscope}%
\pgfsetbuttcap%
\pgfsetmiterjoin%
\definecolor{currentfill}{rgb}{1.000000,1.000000,1.000000}%
\pgfsetfillcolor{currentfill}%
\pgfsetlinewidth{0.000000pt}%
\definecolor{currentstroke}{rgb}{0.000000,0.000000,0.000000}%
\pgfsetstrokecolor{currentstroke}%
\pgfsetstrokeopacity{0.000000}%
\pgfsetdash{}{0pt}%
\pgfpathmoveto{\pgfqpoint{0.584626in}{0.386658in}}%
\pgfpathlineto{\pgfqpoint{2.114552in}{0.386658in}}%
\pgfpathlineto{\pgfqpoint{2.114552in}{1.916585in}}%
\pgfpathlineto{\pgfqpoint{0.584626in}{1.916585in}}%
\pgfpathlineto{\pgfqpoint{0.584626in}{0.386658in}}%
\pgfpathclose%
\pgfusepath{fill}%
\end{pgfscope}%
\begin{pgfscope}%
\pgfsys@transformshift{0.653750in}{0.455000in}%
\pgftext[left,bottom]{\includegraphics[interpolate=true,width=1.391250in,height=1.392500in]{figures/./section3/dataBC/3Re/13500p/re_1000//err_vmag-img0.png}}%
\end{pgfscope}%
\begin{pgfscope}%
\pgfsetbuttcap%
\pgfsetroundjoin%
\definecolor{currentfill}{rgb}{0.000000,0.000000,0.000000}%
\pgfsetfillcolor{currentfill}%
\pgfsetlinewidth{0.803000pt}%
\definecolor{currentstroke}{rgb}{0.000000,0.000000,0.000000}%
\pgfsetstrokecolor{currentstroke}%
\pgfsetdash{}{0pt}%
\pgfsys@defobject{currentmarker}{\pgfqpoint{0.000000in}{-0.048611in}}{\pgfqpoint{0.000000in}{0.000000in}}{%
\pgfpathmoveto{\pgfqpoint{0.000000in}{0.000000in}}%
\pgfpathlineto{\pgfqpoint{0.000000in}{-0.048611in}}%
\pgfusepath{stroke,fill}%
}%
\begin{pgfscope}%
\pgfsys@transformshift{0.654168in}{0.386658in}%
\pgfsys@useobject{currentmarker}{}%
\end{pgfscope}%
\end{pgfscope}%
\begin{pgfscope}%
\definecolor{textcolor}{rgb}{0.000000,0.000000,0.000000}%
\pgfsetstrokecolor{textcolor}%
\pgfsetfillcolor{textcolor}%
\pgftext[x=0.654168in,y=0.296381in,,top]{\color{textcolor}{\rmfamily\fontsize{8.330000}{9.996000}\selectfont\catcode`\^=\active\def^{\ifmmode\sp\else\^{}\fi}\catcode`\%=\active\def%{\%}$\mathdefault{\ensuremath{-}0.1}$}}%
\end{pgfscope}%
\begin{pgfscope}%
\pgfsetbuttcap%
\pgfsetroundjoin%
\definecolor{currentfill}{rgb}{0.000000,0.000000,0.000000}%
\pgfsetfillcolor{currentfill}%
\pgfsetlinewidth{0.803000pt}%
\definecolor{currentstroke}{rgb}{0.000000,0.000000,0.000000}%
\pgfsetstrokecolor{currentstroke}%
\pgfsetdash{}{0pt}%
\pgfsys@defobject{currentmarker}{\pgfqpoint{0.000000in}{-0.048611in}}{\pgfqpoint{0.000000in}{0.000000in}}{%
\pgfpathmoveto{\pgfqpoint{0.000000in}{0.000000in}}%
\pgfpathlineto{\pgfqpoint{0.000000in}{-0.048611in}}%
\pgfusepath{stroke,fill}%
}%
\begin{pgfscope}%
\pgfsys@transformshift{1.349589in}{0.386658in}%
\pgfsys@useobject{currentmarker}{}%
\end{pgfscope}%
\end{pgfscope}%
\begin{pgfscope}%
\definecolor{textcolor}{rgb}{0.000000,0.000000,0.000000}%
\pgfsetstrokecolor{textcolor}%
\pgfsetfillcolor{textcolor}%
\pgftext[x=1.349589in,y=0.296381in,,top]{\color{textcolor}{\rmfamily\fontsize{8.330000}{9.996000}\selectfont\catcode`\^=\active\def^{\ifmmode\sp\else\^{}\fi}\catcode`\%=\active\def%{\%}$\mathdefault{0.0}$}}%
\end{pgfscope}%
\begin{pgfscope}%
\pgfsetbuttcap%
\pgfsetroundjoin%
\definecolor{currentfill}{rgb}{0.000000,0.000000,0.000000}%
\pgfsetfillcolor{currentfill}%
\pgfsetlinewidth{0.803000pt}%
\definecolor{currentstroke}{rgb}{0.000000,0.000000,0.000000}%
\pgfsetstrokecolor{currentstroke}%
\pgfsetdash{}{0pt}%
\pgfsys@defobject{currentmarker}{\pgfqpoint{0.000000in}{-0.048611in}}{\pgfqpoint{0.000000in}{0.000000in}}{%
\pgfpathmoveto{\pgfqpoint{0.000000in}{0.000000in}}%
\pgfpathlineto{\pgfqpoint{0.000000in}{-0.048611in}}%
\pgfusepath{stroke,fill}%
}%
\begin{pgfscope}%
\pgfsys@transformshift{2.045010in}{0.386658in}%
\pgfsys@useobject{currentmarker}{}%
\end{pgfscope}%
\end{pgfscope}%
\begin{pgfscope}%
\definecolor{textcolor}{rgb}{0.000000,0.000000,0.000000}%
\pgfsetstrokecolor{textcolor}%
\pgfsetfillcolor{textcolor}%
\pgftext[x=2.045010in,y=0.296381in,,top]{\color{textcolor}{\rmfamily\fontsize{8.330000}{9.996000}\selectfont\catcode`\^=\active\def^{\ifmmode\sp\else\^{}\fi}\catcode`\%=\active\def%{\%}$\mathdefault{0.1}$}}%
\end{pgfscope}%
\begin{pgfscope}%
\definecolor{textcolor}{rgb}{0.000000,0.000000,0.000000}%
\pgfsetstrokecolor{textcolor}%
\pgfsetfillcolor{textcolor}%
\pgftext[x=1.349589in,y=0.142060in,,top]{\color{textcolor}{\rmfamily\fontsize{10.000000}{12.000000}\selectfont\catcode`\^=\active\def^{\ifmmode\sp\else\^{}\fi}\catcode`\%=\active\def%{\%}$x\;[\text{m}]$}}%
\end{pgfscope}%
\begin{pgfscope}%
\pgfsetbuttcap%
\pgfsetroundjoin%
\definecolor{currentfill}{rgb}{0.000000,0.000000,0.000000}%
\pgfsetfillcolor{currentfill}%
\pgfsetlinewidth{0.803000pt}%
\definecolor{currentstroke}{rgb}{0.000000,0.000000,0.000000}%
\pgfsetstrokecolor{currentstroke}%
\pgfsetdash{}{0pt}%
\pgfsys@defobject{currentmarker}{\pgfqpoint{-0.048611in}{0.000000in}}{\pgfqpoint{-0.000000in}{0.000000in}}{%
\pgfpathmoveto{\pgfqpoint{-0.000000in}{0.000000in}}%
\pgfpathlineto{\pgfqpoint{-0.048611in}{0.000000in}}%
\pgfusepath{stroke,fill}%
}%
\begin{pgfscope}%
\pgfsys@transformshift{0.584626in}{0.456201in}%
\pgfsys@useobject{currentmarker}{}%
\end{pgfscope}%
\end{pgfscope}%
\begin{pgfscope}%
\definecolor{textcolor}{rgb}{0.000000,0.000000,0.000000}%
\pgfsetstrokecolor{textcolor}%
\pgfsetfillcolor{textcolor}%
\pgftext[x=0.251675in, y=0.417620in, left, base]{\color{textcolor}{\rmfamily\fontsize{8.330000}{9.996000}\selectfont\catcode`\^=\active\def^{\ifmmode\sp\else\^{}\fi}\catcode`\%=\active\def%{\%}$\mathdefault{\ensuremath{-}0.1}$}}%
\end{pgfscope}%
\begin{pgfscope}%
\pgfsetbuttcap%
\pgfsetroundjoin%
\definecolor{currentfill}{rgb}{0.000000,0.000000,0.000000}%
\pgfsetfillcolor{currentfill}%
\pgfsetlinewidth{0.803000pt}%
\definecolor{currentstroke}{rgb}{0.000000,0.000000,0.000000}%
\pgfsetstrokecolor{currentstroke}%
\pgfsetdash{}{0pt}%
\pgfsys@defobject{currentmarker}{\pgfqpoint{-0.048611in}{0.000000in}}{\pgfqpoint{-0.000000in}{0.000000in}}{%
\pgfpathmoveto{\pgfqpoint{-0.000000in}{0.000000in}}%
\pgfpathlineto{\pgfqpoint{-0.048611in}{0.000000in}}%
\pgfusepath{stroke,fill}%
}%
\begin{pgfscope}%
\pgfsys@transformshift{0.584626in}{1.151622in}%
\pgfsys@useobject{currentmarker}{}%
\end{pgfscope}%
\end{pgfscope}%
\begin{pgfscope}%
\definecolor{textcolor}{rgb}{0.000000,0.000000,0.000000}%
\pgfsetstrokecolor{textcolor}%
\pgfsetfillcolor{textcolor}%
\pgftext[x=0.343497in, y=1.113042in, left, base]{\color{textcolor}{\rmfamily\fontsize{8.330000}{9.996000}\selectfont\catcode`\^=\active\def^{\ifmmode\sp\else\^{}\fi}\catcode`\%=\active\def%{\%}$\mathdefault{0.0}$}}%
\end{pgfscope}%
\begin{pgfscope}%
\pgfsetbuttcap%
\pgfsetroundjoin%
\definecolor{currentfill}{rgb}{0.000000,0.000000,0.000000}%
\pgfsetfillcolor{currentfill}%
\pgfsetlinewidth{0.803000pt}%
\definecolor{currentstroke}{rgb}{0.000000,0.000000,0.000000}%
\pgfsetstrokecolor{currentstroke}%
\pgfsetdash{}{0pt}%
\pgfsys@defobject{currentmarker}{\pgfqpoint{-0.048611in}{0.000000in}}{\pgfqpoint{-0.000000in}{0.000000in}}{%
\pgfpathmoveto{\pgfqpoint{-0.000000in}{0.000000in}}%
\pgfpathlineto{\pgfqpoint{-0.048611in}{0.000000in}}%
\pgfusepath{stroke,fill}%
}%
\begin{pgfscope}%
\pgfsys@transformshift{0.584626in}{1.847043in}%
\pgfsys@useobject{currentmarker}{}%
\end{pgfscope}%
\end{pgfscope}%
\begin{pgfscope}%
\definecolor{textcolor}{rgb}{0.000000,0.000000,0.000000}%
\pgfsetstrokecolor{textcolor}%
\pgfsetfillcolor{textcolor}%
\pgftext[x=0.343497in, y=1.808463in, left, base]{\color{textcolor}{\rmfamily\fontsize{8.330000}{9.996000}\selectfont\catcode`\^=\active\def^{\ifmmode\sp\else\^{}\fi}\catcode`\%=\active\def%{\%}$\mathdefault{0.1}$}}%
\end{pgfscope}%
\begin{pgfscope}%
\definecolor{textcolor}{rgb}{0.000000,0.000000,0.000000}%
\pgfsetstrokecolor{textcolor}%
\pgfsetfillcolor{textcolor}%
\pgftext[x=0.196119in,y=1.151622in,,bottom,rotate=90.000000]{\color{textcolor}{\rmfamily\fontsize{10.000000}{12.000000}\selectfont\catcode`\^=\active\def^{\ifmmode\sp\else\^{}\fi}\catcode`\%=\active\def%{\%}$y\;[\text{m}]$}}%
\end{pgfscope}%
\begin{pgfscope}%
\pgfpathrectangle{\pgfqpoint{0.584626in}{0.386658in}}{\pgfqpoint{1.529927in}{1.529927in}}%
\pgfusepath{clip}%
\pgfsetbuttcap%
\pgfsetmiterjoin%
\definecolor{currentfill}{rgb}{1.000000,1.000000,1.000000}%
\pgfsetfillcolor{currentfill}%
\pgfsetlinewidth{1.505625pt}%
\definecolor{currentstroke}{rgb}{1.000000,1.000000,1.000000}%
\pgfsetstrokecolor{currentstroke}%
\pgfsetdash{}{0pt}%
\pgfpathmoveto{\pgfqpoint{1.822327in}{1.648947in}}%
\pgfpathlineto{\pgfqpoint{1.815622in}{1.642241in}}%
\pgfpathlineto{\pgfqpoint{1.808916in}{1.635536in}}%
\pgfpathlineto{\pgfqpoint{1.802211in}{1.628830in}}%
\pgfpathlineto{\pgfqpoint{1.795505in}{1.622125in}}%
\pgfpathlineto{\pgfqpoint{1.788800in}{1.615419in}}%
\pgfpathlineto{\pgfqpoint{1.782094in}{1.608714in}}%
\pgfpathlineto{\pgfqpoint{1.775389in}{1.602008in}}%
\pgfpathlineto{\pgfqpoint{1.768683in}{1.595303in}}%
\pgfpathlineto{\pgfqpoint{1.761978in}{1.588597in}}%
\pgfpathlineto{\pgfqpoint{1.755272in}{1.581892in}}%
\pgfpathlineto{\pgfqpoint{1.761419in}{1.575745in}}%
\pgfpathlineto{\pgfqpoint{1.767566in}{1.569598in}}%
\pgfpathlineto{\pgfqpoint{1.773712in}{1.563452in}}%
\pgfpathlineto{\pgfqpoint{1.779859in}{1.557305in}}%
\pgfpathlineto{\pgfqpoint{1.786565in}{1.564010in}}%
\pgfpathlineto{\pgfqpoint{1.793270in}{1.570716in}}%
\pgfpathlineto{\pgfqpoint{1.799976in}{1.577421in}}%
\pgfpathlineto{\pgfqpoint{1.806681in}{1.584127in}}%
\pgfpathlineto{\pgfqpoint{1.813387in}{1.590833in}}%
\pgfpathlineto{\pgfqpoint{1.820092in}{1.597538in}}%
\pgfpathlineto{\pgfqpoint{1.826798in}{1.604244in}}%
\pgfpathlineto{\pgfqpoint{1.833503in}{1.610949in}}%
\pgfpathlineto{\pgfqpoint{1.840209in}{1.617655in}}%
\pgfpathlineto{\pgfqpoint{1.846914in}{1.624360in}}%
\pgfpathlineto{\pgfqpoint{1.822327in}{1.648947in}}%
\pgfpathclose%
\pgfusepath{stroke,fill}%
\end{pgfscope}%
\begin{pgfscope}%
\pgfpathrectangle{\pgfqpoint{0.584626in}{0.386658in}}{\pgfqpoint{1.529927in}{1.529927in}}%
\pgfusepath{clip}%
\pgfsetbuttcap%
\pgfsetmiterjoin%
\definecolor{currentfill}{rgb}{1.000000,1.000000,1.000000}%
\pgfsetfillcolor{currentfill}%
\pgfsetlinewidth{1.505625pt}%
\definecolor{currentstroke}{rgb}{1.000000,1.000000,1.000000}%
\pgfsetstrokecolor{currentstroke}%
\pgfsetdash{}{0pt}%
\pgfpathmoveto{\pgfqpoint{1.846914in}{0.678884in}}%
\pgfpathlineto{\pgfqpoint{1.840209in}{0.685589in}}%
\pgfpathlineto{\pgfqpoint{1.833503in}{0.692295in}}%
\pgfpathlineto{\pgfqpoint{1.826798in}{0.699000in}}%
\pgfpathlineto{\pgfqpoint{1.820092in}{0.705706in}}%
\pgfpathlineto{\pgfqpoint{1.813387in}{0.712411in}}%
\pgfpathlineto{\pgfqpoint{1.806681in}{0.719117in}}%
\pgfpathlineto{\pgfqpoint{1.799976in}{0.725822in}}%
\pgfpathlineto{\pgfqpoint{1.793270in}{0.732528in}}%
\pgfpathlineto{\pgfqpoint{1.786565in}{0.739233in}}%
\pgfpathlineto{\pgfqpoint{1.779859in}{0.745939in}}%
\pgfpathlineto{\pgfqpoint{1.773589in}{0.739669in}}%
\pgfpathlineto{\pgfqpoint{1.767320in}{0.733399in}}%
\pgfpathlineto{\pgfqpoint{1.761050in}{0.727130in}}%
\pgfpathlineto{\pgfqpoint{1.754780in}{0.720860in}}%
\pgfpathlineto{\pgfqpoint{1.755272in}{0.721352in}}%
\pgfpathlineto{\pgfqpoint{1.761978in}{0.714646in}}%
\pgfpathlineto{\pgfqpoint{1.768683in}{0.707941in}}%
\pgfpathlineto{\pgfqpoint{1.775389in}{0.701235in}}%
\pgfpathlineto{\pgfqpoint{1.782094in}{0.694530in}}%
\pgfpathlineto{\pgfqpoint{1.788800in}{0.687824in}}%
\pgfpathlineto{\pgfqpoint{1.795505in}{0.681119in}}%
\pgfpathlineto{\pgfqpoint{1.802211in}{0.674413in}}%
\pgfpathlineto{\pgfqpoint{1.808916in}{0.667708in}}%
\pgfpathlineto{\pgfqpoint{1.815622in}{0.661002in}}%
\pgfpathlineto{\pgfqpoint{1.822327in}{0.654297in}}%
\pgfpathlineto{\pgfqpoint{1.846914in}{0.678884in}}%
\pgfpathclose%
\pgfusepath{stroke,fill}%
\end{pgfscope}%
\begin{pgfscope}%
\pgfpathrectangle{\pgfqpoint{0.584626in}{0.386658in}}{\pgfqpoint{1.529927in}{1.529927in}}%
\pgfusepath{clip}%
\pgfsetbuttcap%
\pgfsetmiterjoin%
\definecolor{currentfill}{rgb}{1.000000,1.000000,1.000000}%
\pgfsetfillcolor{currentfill}%
\pgfsetlinewidth{1.505625pt}%
\definecolor{currentstroke}{rgb}{1.000000,1.000000,1.000000}%
\pgfsetstrokecolor{currentstroke}%
\pgfsetdash{}{0pt}%
\pgfpathmoveto{\pgfqpoint{0.876851in}{0.654297in}}%
\pgfpathlineto{\pgfqpoint{0.883556in}{0.661002in}}%
\pgfpathlineto{\pgfqpoint{0.890262in}{0.667708in}}%
\pgfpathlineto{\pgfqpoint{0.896967in}{0.674413in}}%
\pgfpathlineto{\pgfqpoint{0.903673in}{0.681119in}}%
\pgfpathlineto{\pgfqpoint{0.910378in}{0.687824in}}%
\pgfpathlineto{\pgfqpoint{0.917084in}{0.694530in}}%
\pgfpathlineto{\pgfqpoint{0.923789in}{0.701235in}}%
\pgfpathlineto{\pgfqpoint{0.930495in}{0.707941in}}%
\pgfpathlineto{\pgfqpoint{0.937200in}{0.714646in}}%
\pgfpathlineto{\pgfqpoint{0.943906in}{0.721352in}}%
\pgfpathlineto{\pgfqpoint{0.937636in}{0.727622in}}%
\pgfpathlineto{\pgfqpoint{0.931367in}{0.733891in}}%
\pgfpathlineto{\pgfqpoint{0.925097in}{0.740161in}}%
\pgfpathlineto{\pgfqpoint{0.918827in}{0.746431in}}%
\pgfpathlineto{\pgfqpoint{0.919319in}{0.745939in}}%
\pgfpathlineto{\pgfqpoint{0.912614in}{0.739233in}}%
\pgfpathlineto{\pgfqpoint{0.905908in}{0.732528in}}%
\pgfpathlineto{\pgfqpoint{0.899203in}{0.725822in}}%
\pgfpathlineto{\pgfqpoint{0.892497in}{0.719117in}}%
\pgfpathlineto{\pgfqpoint{0.885792in}{0.712411in}}%
\pgfpathlineto{\pgfqpoint{0.879086in}{0.705706in}}%
\pgfpathlineto{\pgfqpoint{0.872381in}{0.699000in}}%
\pgfpathlineto{\pgfqpoint{0.865675in}{0.692295in}}%
\pgfpathlineto{\pgfqpoint{0.858970in}{0.685589in}}%
\pgfpathlineto{\pgfqpoint{0.852264in}{0.678884in}}%
\pgfpathlineto{\pgfqpoint{0.876851in}{0.654297in}}%
\pgfpathclose%
\pgfusepath{stroke,fill}%
\end{pgfscope}%
\begin{pgfscope}%
\pgfpathrectangle{\pgfqpoint{0.584626in}{0.386658in}}{\pgfqpoint{1.529927in}{1.529927in}}%
\pgfusepath{clip}%
\pgfsetbuttcap%
\pgfsetmiterjoin%
\definecolor{currentfill}{rgb}{1.000000,1.000000,1.000000}%
\pgfsetfillcolor{currentfill}%
\pgfsetlinewidth{1.505625pt}%
\definecolor{currentstroke}{rgb}{1.000000,1.000000,1.000000}%
\pgfsetstrokecolor{currentstroke}%
\pgfsetdash{}{0pt}%
\pgfpathmoveto{\pgfqpoint{0.852264in}{1.624360in}}%
\pgfpathlineto{\pgfqpoint{0.858970in}{1.617655in}}%
\pgfpathlineto{\pgfqpoint{0.865675in}{1.610949in}}%
\pgfpathlineto{\pgfqpoint{0.872381in}{1.604244in}}%
\pgfpathlineto{\pgfqpoint{0.879086in}{1.597538in}}%
\pgfpathlineto{\pgfqpoint{0.885792in}{1.590833in}}%
\pgfpathlineto{\pgfqpoint{0.892497in}{1.584127in}}%
\pgfpathlineto{\pgfqpoint{0.899203in}{1.577421in}}%
\pgfpathlineto{\pgfqpoint{0.905908in}{1.570716in}}%
\pgfpathlineto{\pgfqpoint{0.912614in}{1.564010in}}%
\pgfpathlineto{\pgfqpoint{0.919319in}{1.557305in}}%
\pgfpathlineto{\pgfqpoint{0.925466in}{1.563452in}}%
\pgfpathlineto{\pgfqpoint{0.931613in}{1.569598in}}%
\pgfpathlineto{\pgfqpoint{0.937759in}{1.575745in}}%
\pgfpathlineto{\pgfqpoint{0.943906in}{1.581892in}}%
\pgfpathlineto{\pgfqpoint{0.937200in}{1.588597in}}%
\pgfpathlineto{\pgfqpoint{0.930495in}{1.595303in}}%
\pgfpathlineto{\pgfqpoint{0.923789in}{1.602008in}}%
\pgfpathlineto{\pgfqpoint{0.917084in}{1.608714in}}%
\pgfpathlineto{\pgfqpoint{0.910378in}{1.615419in}}%
\pgfpathlineto{\pgfqpoint{0.903673in}{1.622125in}}%
\pgfpathlineto{\pgfqpoint{0.896967in}{1.628830in}}%
\pgfpathlineto{\pgfqpoint{0.890262in}{1.635536in}}%
\pgfpathlineto{\pgfqpoint{0.883556in}{1.642241in}}%
\pgfpathlineto{\pgfqpoint{0.876851in}{1.648947in}}%
\pgfpathlineto{\pgfqpoint{0.852264in}{1.624360in}}%
\pgfpathclose%
\pgfusepath{stroke,fill}%
\end{pgfscope}%
\begin{pgfscope}%
\pgfpathrectangle{\pgfqpoint{0.584626in}{0.386658in}}{\pgfqpoint{1.529927in}{1.529927in}}%
\pgfusepath{clip}%
\pgfsetrectcap%
\pgfsetroundjoin%
\pgfsetlinewidth{1.505625pt}%
\definecolor{currentstroke}{rgb}{0.000000,0.000000,0.000000}%
\pgfsetstrokecolor{currentstroke}%
\pgfsetdash{}{0pt}%
\pgfpathmoveto{\pgfqpoint{0.852264in}{1.624360in}}%
\pgfpathlineto{\pgfqpoint{0.919319in}{1.557305in}}%
\pgfpathlineto{\pgfqpoint{0.943906in}{1.581892in}}%
\pgfpathlineto{\pgfqpoint{0.876851in}{1.648947in}}%
\pgfpathlineto{\pgfqpoint{0.870893in}{1.656063in}}%
\pgfpathlineto{\pgfqpoint{0.886752in}{1.670652in}}%
\pgfpathlineto{\pgfqpoint{0.903055in}{1.684743in}}%
\pgfpathlineto{\pgfqpoint{0.919788in}{1.698323in}}%
\pgfpathlineto{\pgfqpoint{0.936933in}{1.711377in}}%
\pgfpathlineto{\pgfqpoint{0.954474in}{1.723894in}}%
\pgfpathlineto{\pgfqpoint{0.972395in}{1.735861in}}%
\pgfpathlineto{\pgfqpoint{0.990678in}{1.747267in}}%
\pgfpathlineto{\pgfqpoint{1.009305in}{1.758102in}}%
\pgfpathlineto{\pgfqpoint{1.028260in}{1.768354in}}%
\pgfpathlineto{\pgfqpoint{1.047522in}{1.778013in}}%
\pgfpathlineto{\pgfqpoint{1.067075in}{1.787072in}}%
\pgfpathlineto{\pgfqpoint{1.086899in}{1.795520in}}%
\pgfpathlineto{\pgfqpoint{1.106976in}{1.803350in}}%
\pgfpathlineto{\pgfqpoint{1.127285in}{1.810554in}}%
\pgfpathlineto{\pgfqpoint{1.147808in}{1.817125in}}%
\pgfpathlineto{\pgfqpoint{1.168524in}{1.823058in}}%
\pgfpathlineto{\pgfqpoint{1.189415in}{1.828346in}}%
\pgfpathlineto{\pgfqpoint{1.210459in}{1.832983in}}%
\pgfpathlineto{\pgfqpoint{1.231637in}{1.836967in}}%
\pgfpathlineto{\pgfqpoint{1.252928in}{1.840293in}}%
\pgfpathlineto{\pgfqpoint{1.274312in}{1.842957in}}%
\pgfpathlineto{\pgfqpoint{1.295768in}{1.844957in}}%
\pgfpathlineto{\pgfqpoint{1.317276in}{1.846292in}}%
\pgfpathlineto{\pgfqpoint{1.338814in}{1.846960in}}%
\pgfpathlineto{\pgfqpoint{1.360364in}{1.846960in}}%
\pgfpathlineto{\pgfqpoint{1.381903in}{1.846292in}}%
\pgfpathlineto{\pgfqpoint{1.403410in}{1.844957in}}%
\pgfpathlineto{\pgfqpoint{1.424867in}{1.842957in}}%
\pgfpathlineto{\pgfqpoint{1.446250in}{1.840293in}}%
\pgfpathlineto{\pgfqpoint{1.467541in}{1.836967in}}%
\pgfpathlineto{\pgfqpoint{1.488719in}{1.832983in}}%
\pgfpathlineto{\pgfqpoint{1.509763in}{1.828346in}}%
\pgfpathlineto{\pgfqpoint{1.530654in}{1.823058in}}%
\pgfpathlineto{\pgfqpoint{1.551370in}{1.817125in}}%
\pgfpathlineto{\pgfqpoint{1.571893in}{1.810554in}}%
\pgfpathlineto{\pgfqpoint{1.592202in}{1.803350in}}%
\pgfpathlineto{\pgfqpoint{1.612279in}{1.795520in}}%
\pgfpathlineto{\pgfqpoint{1.632103in}{1.787072in}}%
\pgfpathlineto{\pgfqpoint{1.651656in}{1.778013in}}%
\pgfpathlineto{\pgfqpoint{1.670919in}{1.768354in}}%
\pgfpathlineto{\pgfqpoint{1.689873in}{1.758102in}}%
\pgfpathlineto{\pgfqpoint{1.708500in}{1.747267in}}%
\pgfpathlineto{\pgfqpoint{1.726783in}{1.735861in}}%
\pgfpathlineto{\pgfqpoint{1.744704in}{1.723894in}}%
\pgfpathlineto{\pgfqpoint{1.762245in}{1.711377in}}%
\pgfpathlineto{\pgfqpoint{1.779390in}{1.698323in}}%
\pgfpathlineto{\pgfqpoint{1.796123in}{1.684743in}}%
\pgfpathlineto{\pgfqpoint{1.812426in}{1.670652in}}%
\pgfpathlineto{\pgfqpoint{1.828286in}{1.656063in}}%
\pgfpathlineto{\pgfqpoint{1.822327in}{1.648947in}}%
\pgfpathlineto{\pgfqpoint{1.755272in}{1.581892in}}%
\pgfpathlineto{\pgfqpoint{1.779859in}{1.557305in}}%
\pgfpathlineto{\pgfqpoint{1.846914in}{1.624360in}}%
\pgfpathlineto{\pgfqpoint{1.854030in}{1.630318in}}%
\pgfpathlineto{\pgfqpoint{1.868619in}{1.614459in}}%
\pgfpathlineto{\pgfqpoint{1.882711in}{1.598155in}}%
\pgfpathlineto{\pgfqpoint{1.896290in}{1.581423in}}%
\pgfpathlineto{\pgfqpoint{1.909344in}{1.564278in}}%
\pgfpathlineto{\pgfqpoint{1.921861in}{1.546737in}}%
\pgfpathlineto{\pgfqpoint{1.933828in}{1.528816in}}%
\pgfpathlineto{\pgfqpoint{1.945234in}{1.510533in}}%
\pgfpathlineto{\pgfqpoint{1.956069in}{1.491906in}}%
\pgfpathlineto{\pgfqpoint{1.966321in}{1.472951in}}%
\pgfpathlineto{\pgfqpoint{1.975981in}{1.453689in}}%
\pgfpathlineto{\pgfqpoint{1.985039in}{1.434136in}}%
\pgfpathlineto{\pgfqpoint{1.993487in}{1.414312in}}%
\pgfpathlineto{\pgfqpoint{2.001317in}{1.394235in}}%
\pgfpathlineto{\pgfqpoint{2.008521in}{1.373926in}}%
\pgfpathlineto{\pgfqpoint{2.015093in}{1.353403in}}%
\pgfpathlineto{\pgfqpoint{2.021025in}{1.332687in}}%
\pgfpathlineto{\pgfqpoint{2.026313in}{1.311796in}}%
\pgfpathlineto{\pgfqpoint{2.030951in}{1.290752in}}%
\pgfpathlineto{\pgfqpoint{2.034934in}{1.269574in}}%
\pgfpathlineto{\pgfqpoint{2.038260in}{1.248283in}}%
\pgfpathlineto{\pgfqpoint{2.040924in}{1.226899in}}%
\pgfpathlineto{\pgfqpoint{2.042925in}{1.205443in}}%
\pgfpathlineto{\pgfqpoint{2.044259in}{1.183935in}}%
\pgfpathlineto{\pgfqpoint{2.044927in}{1.162396in}}%
\pgfpathlineto{\pgfqpoint{2.044927in}{1.140847in}}%
\pgfpathlineto{\pgfqpoint{2.044259in}{1.119308in}}%
\pgfpathlineto{\pgfqpoint{2.042925in}{1.097801in}}%
\pgfpathlineto{\pgfqpoint{2.040924in}{1.076344in}}%
\pgfpathlineto{\pgfqpoint{2.038260in}{1.054961in}}%
\pgfpathlineto{\pgfqpoint{2.034934in}{1.033670in}}%
\pgfpathlineto{\pgfqpoint{2.030951in}{1.012492in}}%
\pgfpathlineto{\pgfqpoint{2.026313in}{0.991448in}}%
\pgfpathlineto{\pgfqpoint{2.021025in}{0.970557in}}%
\pgfpathlineto{\pgfqpoint{2.015093in}{0.949841in}}%
\pgfpathlineto{\pgfqpoint{2.008521in}{0.929318in}}%
\pgfpathlineto{\pgfqpoint{2.001317in}{0.909009in}}%
\pgfpathlineto{\pgfqpoint{1.993487in}{0.888932in}}%
\pgfpathlineto{\pgfqpoint{1.985039in}{0.869108in}}%
\pgfpathlineto{\pgfqpoint{1.975981in}{0.849555in}}%
\pgfpathlineto{\pgfqpoint{1.966321in}{0.830292in}}%
\pgfpathlineto{\pgfqpoint{1.956069in}{0.811338in}}%
\pgfpathlineto{\pgfqpoint{1.945234in}{0.792711in}}%
\pgfpathlineto{\pgfqpoint{1.933828in}{0.774428in}}%
\pgfpathlineto{\pgfqpoint{1.921861in}{0.756507in}}%
\pgfpathlineto{\pgfqpoint{1.909344in}{0.738966in}}%
\pgfpathlineto{\pgfqpoint{1.896290in}{0.721821in}}%
\pgfpathlineto{\pgfqpoint{1.882711in}{0.705088in}}%
\pgfpathlineto{\pgfqpoint{1.868619in}{0.688785in}}%
\pgfpathlineto{\pgfqpoint{1.854030in}{0.672925in}}%
\pgfpathlineto{\pgfqpoint{1.846914in}{0.678884in}}%
\pgfpathlineto{\pgfqpoint{1.779859in}{0.745939in}}%
\pgfpathlineto{\pgfqpoint{1.754780in}{0.720860in}}%
\pgfpathlineto{\pgfqpoint{1.755272in}{0.721352in}}%
\pgfpathlineto{\pgfqpoint{1.822327in}{0.654297in}}%
\pgfpathlineto{\pgfqpoint{1.828286in}{0.647181in}}%
\pgfpathlineto{\pgfqpoint{1.812426in}{0.632592in}}%
\pgfpathlineto{\pgfqpoint{1.796123in}{0.618500in}}%
\pgfpathlineto{\pgfqpoint{1.779390in}{0.604921in}}%
\pgfpathlineto{\pgfqpoint{1.762245in}{0.591867in}}%
\pgfpathlineto{\pgfqpoint{1.744704in}{0.579350in}}%
\pgfpathlineto{\pgfqpoint{1.726783in}{0.567383in}}%
\pgfpathlineto{\pgfqpoint{1.708500in}{0.555976in}}%
\pgfpathlineto{\pgfqpoint{1.689873in}{0.545142in}}%
\pgfpathlineto{\pgfqpoint{1.670919in}{0.534890in}}%
\pgfpathlineto{\pgfqpoint{1.651656in}{0.525230in}}%
\pgfpathlineto{\pgfqpoint{1.632103in}{0.516172in}}%
\pgfpathlineto{\pgfqpoint{1.612279in}{0.507724in}}%
\pgfpathlineto{\pgfqpoint{1.592202in}{0.499894in}}%
\pgfpathlineto{\pgfqpoint{1.571893in}{0.492690in}}%
\pgfpathlineto{\pgfqpoint{1.551370in}{0.486118in}}%
\pgfpathlineto{\pgfqpoint{1.530654in}{0.480186in}}%
\pgfpathlineto{\pgfqpoint{1.509763in}{0.474898in}}%
\pgfpathlineto{\pgfqpoint{1.488719in}{0.470260in}}%
\pgfpathlineto{\pgfqpoint{1.467541in}{0.466277in}}%
\pgfpathlineto{\pgfqpoint{1.446250in}{0.462951in}}%
\pgfpathlineto{\pgfqpoint{1.424867in}{0.460287in}}%
\pgfpathlineto{\pgfqpoint{1.403410in}{0.458286in}}%
\pgfpathlineto{\pgfqpoint{1.381903in}{0.456952in}}%
\pgfpathlineto{\pgfqpoint{1.360364in}{0.456284in}}%
\pgfpathlineto{\pgfqpoint{1.338814in}{0.456284in}}%
\pgfpathlineto{\pgfqpoint{1.317276in}{0.456952in}}%
\pgfpathlineto{\pgfqpoint{1.295768in}{0.458286in}}%
\pgfpathlineto{\pgfqpoint{1.274312in}{0.460287in}}%
\pgfpathlineto{\pgfqpoint{1.252928in}{0.462951in}}%
\pgfpathlineto{\pgfqpoint{1.231637in}{0.466277in}}%
\pgfpathlineto{\pgfqpoint{1.210459in}{0.470260in}}%
\pgfpathlineto{\pgfqpoint{1.189415in}{0.474898in}}%
\pgfpathlineto{\pgfqpoint{1.168524in}{0.480186in}}%
\pgfpathlineto{\pgfqpoint{1.147808in}{0.486118in}}%
\pgfpathlineto{\pgfqpoint{1.127285in}{0.492690in}}%
\pgfpathlineto{\pgfqpoint{1.106976in}{0.499894in}}%
\pgfpathlineto{\pgfqpoint{1.086899in}{0.507724in}}%
\pgfpathlineto{\pgfqpoint{1.067075in}{0.516172in}}%
\pgfpathlineto{\pgfqpoint{1.047522in}{0.525230in}}%
\pgfpathlineto{\pgfqpoint{1.028260in}{0.534890in}}%
\pgfpathlineto{\pgfqpoint{1.009305in}{0.545142in}}%
\pgfpathlineto{\pgfqpoint{0.990678in}{0.555976in}}%
\pgfpathlineto{\pgfqpoint{0.972395in}{0.567383in}}%
\pgfpathlineto{\pgfqpoint{0.954474in}{0.579350in}}%
\pgfpathlineto{\pgfqpoint{0.936933in}{0.591867in}}%
\pgfpathlineto{\pgfqpoint{0.919788in}{0.604921in}}%
\pgfpathlineto{\pgfqpoint{0.903055in}{0.618500in}}%
\pgfpathlineto{\pgfqpoint{0.886752in}{0.632592in}}%
\pgfpathlineto{\pgfqpoint{0.870893in}{0.647181in}}%
\pgfpathlineto{\pgfqpoint{0.876851in}{0.654297in}}%
\pgfpathlineto{\pgfqpoint{0.943906in}{0.721352in}}%
\pgfpathlineto{\pgfqpoint{0.918827in}{0.746431in}}%
\pgfpathlineto{\pgfqpoint{0.919319in}{0.745939in}}%
\pgfpathlineto{\pgfqpoint{0.852264in}{0.678884in}}%
\pgfpathlineto{\pgfqpoint{0.845148in}{0.672925in}}%
\pgfpathlineto{\pgfqpoint{0.830559in}{0.688785in}}%
\pgfpathlineto{\pgfqpoint{0.816468in}{0.705088in}}%
\pgfpathlineto{\pgfqpoint{0.802888in}{0.721821in}}%
\pgfpathlineto{\pgfqpoint{0.789834in}{0.738966in}}%
\pgfpathlineto{\pgfqpoint{0.777317in}{0.756507in}}%
\pgfpathlineto{\pgfqpoint{0.765350in}{0.774428in}}%
\pgfpathlineto{\pgfqpoint{0.753944in}{0.792711in}}%
\pgfpathlineto{\pgfqpoint{0.743109in}{0.811338in}}%
\pgfpathlineto{\pgfqpoint{0.732857in}{0.830292in}}%
\pgfpathlineto{\pgfqpoint{0.723197in}{0.849555in}}%
\pgfpathlineto{\pgfqpoint{0.714139in}{0.869108in}}%
\pgfpathlineto{\pgfqpoint{0.705691in}{0.888932in}}%
\pgfpathlineto{\pgfqpoint{0.697861in}{0.909009in}}%
\pgfpathlineto{\pgfqpoint{0.690657in}{0.929318in}}%
\pgfpathlineto{\pgfqpoint{0.684085in}{0.949841in}}%
\pgfpathlineto{\pgfqpoint{0.678153in}{0.970557in}}%
\pgfpathlineto{\pgfqpoint{0.672865in}{0.991448in}}%
\pgfpathlineto{\pgfqpoint{0.668228in}{1.012492in}}%
\pgfpathlineto{\pgfqpoint{0.664244in}{1.033670in}}%
\pgfpathlineto{\pgfqpoint{0.660918in}{1.054961in}}%
\pgfpathlineto{\pgfqpoint{0.658254in}{1.076344in}}%
\pgfpathlineto{\pgfqpoint{0.656254in}{1.097801in}}%
\pgfpathlineto{\pgfqpoint{0.654919in}{1.119308in}}%
\pgfpathlineto{\pgfqpoint{0.654251in}{1.140847in}}%
\pgfpathlineto{\pgfqpoint{0.654251in}{1.162396in}}%
\pgfpathlineto{\pgfqpoint{0.654919in}{1.183935in}}%
\pgfpathlineto{\pgfqpoint{0.656254in}{1.205443in}}%
\pgfpathlineto{\pgfqpoint{0.658254in}{1.226899in}}%
\pgfpathlineto{\pgfqpoint{0.660918in}{1.248283in}}%
\pgfpathlineto{\pgfqpoint{0.664244in}{1.269574in}}%
\pgfpathlineto{\pgfqpoint{0.668228in}{1.290752in}}%
\pgfpathlineto{\pgfqpoint{0.672865in}{1.311796in}}%
\pgfpathlineto{\pgfqpoint{0.678153in}{1.332687in}}%
\pgfpathlineto{\pgfqpoint{0.684085in}{1.353403in}}%
\pgfpathlineto{\pgfqpoint{0.690657in}{1.373926in}}%
\pgfpathlineto{\pgfqpoint{0.697861in}{1.394235in}}%
\pgfpathlineto{\pgfqpoint{0.705691in}{1.414312in}}%
\pgfpathlineto{\pgfqpoint{0.714139in}{1.434136in}}%
\pgfpathlineto{\pgfqpoint{0.723197in}{1.453689in}}%
\pgfpathlineto{\pgfqpoint{0.732857in}{1.472951in}}%
\pgfpathlineto{\pgfqpoint{0.743109in}{1.491906in}}%
\pgfpathlineto{\pgfqpoint{0.753944in}{1.510533in}}%
\pgfpathlineto{\pgfqpoint{0.765350in}{1.528816in}}%
\pgfpathlineto{\pgfqpoint{0.777317in}{1.546737in}}%
\pgfpathlineto{\pgfqpoint{0.789834in}{1.564278in}}%
\pgfpathlineto{\pgfqpoint{0.802888in}{1.581423in}}%
\pgfpathlineto{\pgfqpoint{0.816468in}{1.598155in}}%
\pgfpathlineto{\pgfqpoint{0.830559in}{1.614459in}}%
\pgfpathlineto{\pgfqpoint{0.845148in}{1.630318in}}%
\pgfpathlineto{\pgfqpoint{0.852264in}{1.624360in}}%
\pgfpathlineto{\pgfqpoint{0.919319in}{1.557305in}}%
\pgfpathlineto{\pgfqpoint{0.943906in}{1.581892in}}%
\pgfpathlineto{\pgfqpoint{0.876851in}{1.648947in}}%
\pgfpathlineto{\pgfqpoint{0.876851in}{1.648947in}}%
\pgfusepath{stroke}%
\end{pgfscope}%
\begin{pgfscope}%
\pgfsetrectcap%
\pgfsetmiterjoin%
\pgfsetlinewidth{0.803000pt}%
\definecolor{currentstroke}{rgb}{0.000000,0.000000,0.000000}%
\pgfsetstrokecolor{currentstroke}%
\pgfsetdash{}{0pt}%
\pgfpathmoveto{\pgfqpoint{0.584626in}{0.386658in}}%
\pgfpathlineto{\pgfqpoint{0.584626in}{1.916585in}}%
\pgfusepath{stroke}%
\end{pgfscope}%
\begin{pgfscope}%
\pgfsetrectcap%
\pgfsetmiterjoin%
\pgfsetlinewidth{0.803000pt}%
\definecolor{currentstroke}{rgb}{0.000000,0.000000,0.000000}%
\pgfsetstrokecolor{currentstroke}%
\pgfsetdash{}{0pt}%
\pgfpathmoveto{\pgfqpoint{2.114552in}{0.386658in}}%
\pgfpathlineto{\pgfqpoint{2.114552in}{1.916585in}}%
\pgfusepath{stroke}%
\end{pgfscope}%
\begin{pgfscope}%
\pgfsetrectcap%
\pgfsetmiterjoin%
\pgfsetlinewidth{0.803000pt}%
\definecolor{currentstroke}{rgb}{0.000000,0.000000,0.000000}%
\pgfsetstrokecolor{currentstroke}%
\pgfsetdash{}{0pt}%
\pgfpathmoveto{\pgfqpoint{0.584626in}{0.386658in}}%
\pgfpathlineto{\pgfqpoint{2.114552in}{0.386658in}}%
\pgfusepath{stroke}%
\end{pgfscope}%
\begin{pgfscope}%
\pgfsetrectcap%
\pgfsetmiterjoin%
\pgfsetlinewidth{0.803000pt}%
\definecolor{currentstroke}{rgb}{0.000000,0.000000,0.000000}%
\pgfsetstrokecolor{currentstroke}%
\pgfsetdash{}{0pt}%
\pgfpathmoveto{\pgfqpoint{0.584626in}{1.916585in}}%
\pgfpathlineto{\pgfqpoint{2.114552in}{1.916585in}}%
\pgfusepath{stroke}%
\end{pgfscope}%
\begin{pgfscope}%
\pgfpathrectangle{\pgfqpoint{0.584626in}{0.386658in}}{\pgfqpoint{1.529927in}{1.529927in}}%
\pgfusepath{clip}%
\pgfsetrectcap%
\pgfsetroundjoin%
\pgfsetlinewidth{1.505625pt}%
\definecolor{currentstroke}{rgb}{0.000000,0.000000,0.000000}%
\pgfsetstrokecolor{currentstroke}%
\pgfsetdash{}{0pt}%
\pgfpathmoveto{\pgfqpoint{1.071421in}{1.151622in}}%
\pgfpathlineto{\pgfqpoint{1.627758in}{1.151622in}}%
\pgfusepath{stroke}%
\end{pgfscope}%
\begin{pgfscope}%
\pgfpathrectangle{\pgfqpoint{0.584626in}{0.386658in}}{\pgfqpoint{1.529927in}{1.529927in}}%
\pgfusepath{clip}%
\pgfsetrectcap%
\pgfsetroundjoin%
\pgfsetlinewidth{1.505625pt}%
\definecolor{currentstroke}{rgb}{0.000000,0.000000,0.000000}%
\pgfsetstrokecolor{currentstroke}%
\pgfsetdash{}{0pt}%
\pgfpathmoveto{\pgfqpoint{1.349589in}{0.873453in}}%
\pgfpathlineto{\pgfqpoint{1.349589in}{1.429790in}}%
\pgfusepath{stroke}%
\end{pgfscope}%
\begin{pgfscope}%
\pgfsetbuttcap%
\pgfsetmiterjoin%
\definecolor{currentfill}{rgb}{1.000000,1.000000,1.000000}%
\pgfsetfillcolor{currentfill}%
\pgfsetlinewidth{0.000000pt}%
\definecolor{currentstroke}{rgb}{0.000000,0.000000,0.000000}%
\pgfsetstrokecolor{currentstroke}%
\pgfsetstrokeopacity{0.000000}%
\pgfsetdash{}{0pt}%
\pgfpathmoveto{\pgfqpoint{0.337193in}{2.012206in}}%
\pgfpathlineto{\pgfqpoint{2.361986in}{2.012206in}}%
\pgfpathlineto{\pgfqpoint{2.361986in}{2.113445in}}%
\pgfpathlineto{\pgfqpoint{0.337193in}{2.113445in}}%
\pgfpathlineto{\pgfqpoint{0.337193in}{2.012206in}}%
\pgfpathclose%
\pgfusepath{fill}%
\end{pgfscope}%
\begin{pgfscope}%
\pgfsys@transformshift{0.337500in}{2.012500in}%
\pgftext[left,bottom]{\includegraphics[interpolate=true,width=2.025000in,height=0.101250in]{figures/./section3/dataBC/3Re/13500p/re_1000//err_vmag-img1.png}}%
\end{pgfscope}%
\begin{pgfscope}%
\pgfsetbuttcap%
\pgfsetroundjoin%
\definecolor{currentfill}{rgb}{0.000000,0.000000,0.000000}%
\pgfsetfillcolor{currentfill}%
\pgfsetlinewidth{0.803000pt}%
\definecolor{currentstroke}{rgb}{0.000000,0.000000,0.000000}%
\pgfsetstrokecolor{currentstroke}%
\pgfsetdash{}{0pt}%
\pgfsys@defobject{currentmarker}{\pgfqpoint{0.000000in}{0.000000in}}{\pgfqpoint{0.000000in}{0.048611in}}{%
\pgfpathmoveto{\pgfqpoint{0.000000in}{0.000000in}}%
\pgfpathlineto{\pgfqpoint{0.000000in}{0.048611in}}%
\pgfusepath{stroke,fill}%
}%
\begin{pgfscope}%
\pgfsys@transformshift{0.337595in}{2.113445in}%
\pgfsys@useobject{currentmarker}{}%
\end{pgfscope}%
\end{pgfscope}%
\begin{pgfscope}%
\definecolor{textcolor}{rgb}{0.000000,0.000000,0.000000}%
\pgfsetstrokecolor{textcolor}%
\pgfsetfillcolor{textcolor}%
\pgftext[x=0.337595in,y=2.203723in,,bottom]{\color{textcolor}{\rmfamily\fontsize{8.330000}{9.996000}\selectfont\catcode`\^=\active\def^{\ifmmode\sp\else\^{}\fi}\catcode`\%=\active\def%{\%}$\mathdefault{\ensuremath{-}5.188}$}}%
\end{pgfscope}%
\begin{pgfscope}%
\pgfsetbuttcap%
\pgfsetroundjoin%
\definecolor{currentfill}{rgb}{0.000000,0.000000,0.000000}%
\pgfsetfillcolor{currentfill}%
\pgfsetlinewidth{0.803000pt}%
\definecolor{currentstroke}{rgb}{0.000000,0.000000,0.000000}%
\pgfsetstrokecolor{currentstroke}%
\pgfsetdash{}{0pt}%
\pgfsys@defobject{currentmarker}{\pgfqpoint{0.000000in}{0.000000in}}{\pgfqpoint{0.000000in}{0.048611in}}{%
\pgfpathmoveto{\pgfqpoint{0.000000in}{0.000000in}}%
\pgfpathlineto{\pgfqpoint{0.000000in}{0.048611in}}%
\pgfusepath{stroke,fill}%
}%
\begin{pgfscope}%
\pgfsys@transformshift{1.011883in}{2.113445in}%
\pgfsys@useobject{currentmarker}{}%
\end{pgfscope}%
\end{pgfscope}%
\begin{pgfscope}%
\definecolor{textcolor}{rgb}{0.000000,0.000000,0.000000}%
\pgfsetstrokecolor{textcolor}%
\pgfsetfillcolor{textcolor}%
\pgftext[x=1.011883in,y=2.203723in,,bottom]{\color{textcolor}{\rmfamily\fontsize{8.330000}{9.996000}\selectfont\catcode`\^=\active\def^{\ifmmode\sp\else\^{}\fi}\catcode`\%=\active\def%{\%}$\mathdefault{\ensuremath{-}2.134}$}}%
\end{pgfscope}%
\begin{pgfscope}%
\pgfsetbuttcap%
\pgfsetroundjoin%
\definecolor{currentfill}{rgb}{0.000000,0.000000,0.000000}%
\pgfsetfillcolor{currentfill}%
\pgfsetlinewidth{0.803000pt}%
\definecolor{currentstroke}{rgb}{0.000000,0.000000,0.000000}%
\pgfsetstrokecolor{currentstroke}%
\pgfsetdash{}{0pt}%
\pgfsys@defobject{currentmarker}{\pgfqpoint{0.000000in}{0.000000in}}{\pgfqpoint{0.000000in}{0.048611in}}{%
\pgfpathmoveto{\pgfqpoint{0.000000in}{0.000000in}}%
\pgfpathlineto{\pgfqpoint{0.000000in}{0.048611in}}%
\pgfusepath{stroke,fill}%
}%
\begin{pgfscope}%
\pgfsys@transformshift{1.686170in}{2.113445in}%
\pgfsys@useobject{currentmarker}{}%
\end{pgfscope}%
\end{pgfscope}%
\begin{pgfscope}%
\definecolor{textcolor}{rgb}{0.000000,0.000000,0.000000}%
\pgfsetstrokecolor{textcolor}%
\pgfsetfillcolor{textcolor}%
\pgftext[x=1.686170in,y=2.203723in,,bottom]{\color{textcolor}{\rmfamily\fontsize{8.330000}{9.996000}\selectfont\catcode`\^=\active\def^{\ifmmode\sp\else\^{}\fi}\catcode`\%=\active\def%{\%}$\mathdefault{0.919}$}}%
\end{pgfscope}%
\begin{pgfscope}%
\pgfsetbuttcap%
\pgfsetroundjoin%
\definecolor{currentfill}{rgb}{0.000000,0.000000,0.000000}%
\pgfsetfillcolor{currentfill}%
\pgfsetlinewidth{0.803000pt}%
\definecolor{currentstroke}{rgb}{0.000000,0.000000,0.000000}%
\pgfsetstrokecolor{currentstroke}%
\pgfsetdash{}{0pt}%
\pgfsys@defobject{currentmarker}{\pgfqpoint{0.000000in}{0.000000in}}{\pgfqpoint{0.000000in}{0.048611in}}{%
\pgfpathmoveto{\pgfqpoint{0.000000in}{0.000000in}}%
\pgfpathlineto{\pgfqpoint{0.000000in}{0.048611in}}%
\pgfusepath{stroke,fill}%
}%
\begin{pgfscope}%
\pgfsys@transformshift{2.360457in}{2.113445in}%
\pgfsys@useobject{currentmarker}{}%
\end{pgfscope}%
\end{pgfscope}%
\begin{pgfscope}%
\definecolor{textcolor}{rgb}{0.000000,0.000000,0.000000}%
\pgfsetstrokecolor{textcolor}%
\pgfsetfillcolor{textcolor}%
\pgftext[x=2.360457in,y=2.203723in,,bottom]{\color{textcolor}{\rmfamily\fontsize{8.330000}{9.996000}\selectfont\catcode`\^=\active\def^{\ifmmode\sp\else\^{}\fi}\catcode`\%=\active\def%{\%}$\mathdefault{3.973}$}}%
\end{pgfscope}%
\begin{pgfscope}%
\definecolor{textcolor}{rgb}{0.000000,0.000000,0.000000}%
\pgfsetstrokecolor{textcolor}%
\pgfsetfillcolor{textcolor}%
\pgftext[x=1.349589in,y=2.358044in,,base]{\color{textcolor}{\rmfamily\fontsize{10.000000}{12.000000}\selectfont\catcode`\^=\active\def^{\ifmmode\sp\else\^{}\fi}\catcode`\%=\active\def%{\%}$f^{(\bm{v})}_\text{err} \ [\%]$}}%
\end{pgfscope}%
\begin{pgfscope}%
\definecolor{textcolor}{rgb}{0.000000,0.000000,0.000000}%
\pgfsetstrokecolor{textcolor}%
\pgfsetfillcolor{textcolor}%
\pgftext[x=2.361986in,y=2.344155in,right,bottom]{\color{textcolor}{\rmfamily\fontsize{8.330000}{9.996000}\selectfont\catcode`\^=\active\def^{\ifmmode\sp\else\^{}\fi}\catcode`\%=\active\def%{\%}$\times\mathdefault{10^{1}}\mathdefault{}$}}%
\end{pgfscope}%
\begin{pgfscope}%
\pgfsetrectcap%
\pgfsetmiterjoin%
\pgfsetlinewidth{0.803000pt}%
\definecolor{currentstroke}{rgb}{0.000000,0.000000,0.000000}%
\pgfsetstrokecolor{currentstroke}%
\pgfsetdash{}{0pt}%
\pgfpathmoveto{\pgfqpoint{0.337193in}{2.012206in}}%
\pgfpathlineto{\pgfqpoint{0.337193in}{2.062826in}}%
\pgfpathlineto{\pgfqpoint{0.337193in}{2.113445in}}%
\pgfpathlineto{\pgfqpoint{2.361986in}{2.113445in}}%
\pgfpathlineto{\pgfqpoint{2.361986in}{2.062826in}}%
\pgfpathlineto{\pgfqpoint{2.361986in}{2.012206in}}%
\pgfpathlineto{\pgfqpoint{0.337193in}{2.012206in}}%
\pgfpathclose%
\pgfusepath{stroke}%
\end{pgfscope}%
\end{pgfpicture}%
\makeatother%
\endgroup%

%% file: figures/section4/3Re/avg_errors_Re.pgf
%% Creator: Matplotlib, PGF backend
%%
%% To include the figure in your LaTeX document, write
%%   \input{<filename>.pgf}
%%
%% Make sure the required packages are loaded in your preamble
%%   \usepackage{pgf}
%%
%% Also ensure that all the required font packages are loaded; for instance,
%% the lmodern package is sometimes necessary when using math font.
%%   \usepackage{lmodern}
%%
%% Figures using additional raster images can only be included by \input if
%% they are in the same directory as the main LaTeX file. For loading figures
%% from other directories you can use the `import` package
%% and then include the figures with
%%   \import{<path to file>}{<filename>.pgf}
%%
%% Matplotlib used the following preamble
%%   \def\mathdefault#1{#1}
%%   \everymath=\expandafter{\the\everymath\displaystyle}
%%   \usepackage{amsmath}\usepackage{bm}
%%   \makeatletter\@ifpackageloaded{underscore}{}{\usepackage[strings]{underscore}}\makeatother
%%
\begingroup%
\makeatletter%
\begin{pgfpicture}%
\pgfpathrectangle{\pgfpointorigin}{\pgfqpoint{3.000000in}{2.000000in}}%
\pgfusepath{use as bounding box, clip}%
\begin{pgfscope}%
\pgfsetbuttcap%
\pgfsetmiterjoin%
\definecolor{currentfill}{rgb}{1.000000,1.000000,1.000000}%
\pgfsetfillcolor{currentfill}%
\pgfsetlinewidth{0.000000pt}%
\definecolor{currentstroke}{rgb}{1.000000,1.000000,1.000000}%
\pgfsetstrokecolor{currentstroke}%
\pgfsetdash{}{0pt}%
\pgfpathmoveto{\pgfqpoint{0.000000in}{0.000000in}}%
\pgfpathlineto{\pgfqpoint{3.000000in}{0.000000in}}%
\pgfpathlineto{\pgfqpoint{3.000000in}{2.000000in}}%
\pgfpathlineto{\pgfqpoint{0.000000in}{2.000000in}}%
\pgfpathlineto{\pgfqpoint{0.000000in}{0.000000in}}%
\pgfpathclose%
\pgfusepath{fill}%
\end{pgfscope}%
\begin{pgfscope}%
\pgfsetbuttcap%
\pgfsetmiterjoin%
\definecolor{currentfill}{rgb}{1.000000,1.000000,1.000000}%
\pgfsetfillcolor{currentfill}%
\pgfsetlinewidth{0.000000pt}%
\definecolor{currentstroke}{rgb}{0.000000,0.000000,0.000000}%
\pgfsetstrokecolor{currentstroke}%
\pgfsetstrokeopacity{0.000000}%
\pgfsetdash{}{0pt}%
\pgfpathmoveto{\pgfqpoint{0.597016in}{0.498776in}}%
\pgfpathlineto{\pgfqpoint{2.845756in}{0.498776in}}%
\pgfpathlineto{\pgfqpoint{2.845756in}{1.850000in}}%
\pgfpathlineto{\pgfqpoint{0.597016in}{1.850000in}}%
\pgfpathlineto{\pgfqpoint{0.597016in}{0.498776in}}%
\pgfpathclose%
\pgfusepath{fill}%
\end{pgfscope}%
\begin{pgfscope}%
\pgfsetbuttcap%
\pgfsetroundjoin%
\definecolor{currentfill}{rgb}{0.000000,0.000000,0.000000}%
\pgfsetfillcolor{currentfill}%
\pgfsetlinewidth{0.803000pt}%
\definecolor{currentstroke}{rgb}{0.000000,0.000000,0.000000}%
\pgfsetstrokecolor{currentstroke}%
\pgfsetdash{}{0pt}%
\pgfsys@defobject{currentmarker}{\pgfqpoint{0.000000in}{-0.048611in}}{\pgfqpoint{0.000000in}{0.000000in}}{%
\pgfpathmoveto{\pgfqpoint{0.000000in}{0.000000in}}%
\pgfpathlineto{\pgfqpoint{0.000000in}{-0.048611in}}%
\pgfusepath{stroke,fill}%
}%
\begin{pgfscope}%
\pgfsys@transformshift{0.699231in}{0.498776in}%
\pgfsys@useobject{currentmarker}{}%
\end{pgfscope}%
\end{pgfscope}%
\begin{pgfscope}%
\definecolor{textcolor}{rgb}{0.000000,0.000000,0.000000}%
\pgfsetstrokecolor{textcolor}%
\pgfsetfillcolor{textcolor}%
\pgftext[x=0.699231in,y=0.408498in,,top]{\color{textcolor}{\rmfamily\fontsize{6.500000}{7.800000}\selectfont\catcode`\^=\active\def^{\ifmmode\sp\else\^{}\fi}\catcode`\%=\active\def%{\%}30}}%
\end{pgfscope}%
\begin{pgfscope}%
\pgfsetbuttcap%
\pgfsetroundjoin%
\definecolor{currentfill}{rgb}{0.000000,0.000000,0.000000}%
\pgfsetfillcolor{currentfill}%
\pgfsetlinewidth{0.803000pt}%
\definecolor{currentstroke}{rgb}{0.000000,0.000000,0.000000}%
\pgfsetstrokecolor{currentstroke}%
\pgfsetdash{}{0pt}%
\pgfsys@defobject{currentmarker}{\pgfqpoint{0.000000in}{-0.048611in}}{\pgfqpoint{0.000000in}{0.000000in}}{%
\pgfpathmoveto{\pgfqpoint{0.000000in}{0.000000in}}%
\pgfpathlineto{\pgfqpoint{0.000000in}{-0.048611in}}%
\pgfusepath{stroke,fill}%
}%
\begin{pgfscope}%
\pgfsys@transformshift{0.926377in}{0.498776in}%
\pgfsys@useobject{currentmarker}{}%
\end{pgfscope}%
\end{pgfscope}%
\begin{pgfscope}%
\definecolor{textcolor}{rgb}{0.000000,0.000000,0.000000}%
\pgfsetstrokecolor{textcolor}%
\pgfsetfillcolor{textcolor}%
\pgftext[x=0.926377in,y=0.408498in,,top]{\color{textcolor}{\rmfamily\fontsize{6.500000}{7.800000}\selectfont\catcode`\^=\active\def^{\ifmmode\sp\else\^{}\fi}\catcode`\%=\active\def%{\%}50}}%
\end{pgfscope}%
\begin{pgfscope}%
\pgfsetbuttcap%
\pgfsetroundjoin%
\definecolor{currentfill}{rgb}{0.000000,0.000000,0.000000}%
\pgfsetfillcolor{currentfill}%
\pgfsetlinewidth{0.803000pt}%
\definecolor{currentstroke}{rgb}{0.000000,0.000000,0.000000}%
\pgfsetstrokecolor{currentstroke}%
\pgfsetdash{}{0pt}%
\pgfsys@defobject{currentmarker}{\pgfqpoint{0.000000in}{-0.048611in}}{\pgfqpoint{0.000000in}{0.000000in}}{%
\pgfpathmoveto{\pgfqpoint{0.000000in}{0.000000in}}%
\pgfpathlineto{\pgfqpoint{0.000000in}{-0.048611in}}%
\pgfusepath{stroke,fill}%
}%
\begin{pgfscope}%
\pgfsys@transformshift{1.153522in}{0.498776in}%
\pgfsys@useobject{currentmarker}{}%
\end{pgfscope}%
\end{pgfscope}%
\begin{pgfscope}%
\definecolor{textcolor}{rgb}{0.000000,0.000000,0.000000}%
\pgfsetstrokecolor{textcolor}%
\pgfsetfillcolor{textcolor}%
\pgftext[x=1.153522in,y=0.408498in,,top]{\color{textcolor}{\rmfamily\fontsize{6.500000}{7.800000}\selectfont\catcode`\^=\active\def^{\ifmmode\sp\else\^{}\fi}\catcode`\%=\active\def%{\%}100}}%
\end{pgfscope}%
\begin{pgfscope}%
\pgfsetbuttcap%
\pgfsetroundjoin%
\definecolor{currentfill}{rgb}{0.000000,0.000000,0.000000}%
\pgfsetfillcolor{currentfill}%
\pgfsetlinewidth{0.803000pt}%
\definecolor{currentstroke}{rgb}{0.000000,0.000000,0.000000}%
\pgfsetstrokecolor{currentstroke}%
\pgfsetdash{}{0pt}%
\pgfsys@defobject{currentmarker}{\pgfqpoint{0.000000in}{-0.048611in}}{\pgfqpoint{0.000000in}{0.000000in}}{%
\pgfpathmoveto{\pgfqpoint{0.000000in}{0.000000in}}%
\pgfpathlineto{\pgfqpoint{0.000000in}{-0.048611in}}%
\pgfusepath{stroke,fill}%
}%
\begin{pgfscope}%
\pgfsys@transformshift{1.380668in}{0.498776in}%
\pgfsys@useobject{currentmarker}{}%
\end{pgfscope}%
\end{pgfscope}%
\begin{pgfscope}%
\definecolor{textcolor}{rgb}{0.000000,0.000000,0.000000}%
\pgfsetstrokecolor{textcolor}%
\pgfsetfillcolor{textcolor}%
\pgftext[x=1.380668in,y=0.408498in,,top]{\color{textcolor}{\rmfamily\fontsize{6.500000}{7.800000}\selectfont\catcode`\^=\active\def^{\ifmmode\sp\else\^{}\fi}\catcode`\%=\active\def%{\%}500}}%
\end{pgfscope}%
\begin{pgfscope}%
\pgfsetbuttcap%
\pgfsetroundjoin%
\definecolor{currentfill}{rgb}{0.000000,0.000000,0.000000}%
\pgfsetfillcolor{currentfill}%
\pgfsetlinewidth{0.803000pt}%
\definecolor{currentstroke}{rgb}{0.000000,0.000000,0.000000}%
\pgfsetstrokecolor{currentstroke}%
\pgfsetdash{}{0pt}%
\pgfsys@defobject{currentmarker}{\pgfqpoint{0.000000in}{-0.048611in}}{\pgfqpoint{0.000000in}{0.000000in}}{%
\pgfpathmoveto{\pgfqpoint{0.000000in}{0.000000in}}%
\pgfpathlineto{\pgfqpoint{0.000000in}{-0.048611in}}%
\pgfusepath{stroke,fill}%
}%
\begin{pgfscope}%
\pgfsys@transformshift{1.607813in}{0.498776in}%
\pgfsys@useobject{currentmarker}{}%
\end{pgfscope}%
\end{pgfscope}%
\begin{pgfscope}%
\definecolor{textcolor}{rgb}{0.000000,0.000000,0.000000}%
\pgfsetstrokecolor{textcolor}%
\pgfsetfillcolor{textcolor}%
\pgftext[x=1.607813in,y=0.408498in,,top]{\color{textcolor}{\rmfamily\fontsize{6.500000}{7.800000}\selectfont\catcode`\^=\active\def^{\ifmmode\sp\else\^{}\fi}\catcode`\%=\active\def%{\%}1000}}%
\end{pgfscope}%
\begin{pgfscope}%
\pgfsetbuttcap%
\pgfsetroundjoin%
\definecolor{currentfill}{rgb}{0.000000,0.000000,0.000000}%
\pgfsetfillcolor{currentfill}%
\pgfsetlinewidth{0.803000pt}%
\definecolor{currentstroke}{rgb}{0.000000,0.000000,0.000000}%
\pgfsetstrokecolor{currentstroke}%
\pgfsetdash{}{0pt}%
\pgfsys@defobject{currentmarker}{\pgfqpoint{0.000000in}{-0.048611in}}{\pgfqpoint{0.000000in}{0.000000in}}{%
\pgfpathmoveto{\pgfqpoint{0.000000in}{0.000000in}}%
\pgfpathlineto{\pgfqpoint{0.000000in}{-0.048611in}}%
\pgfusepath{stroke,fill}%
}%
\begin{pgfscope}%
\pgfsys@transformshift{1.834959in}{0.498776in}%
\pgfsys@useobject{currentmarker}{}%
\end{pgfscope}%
\end{pgfscope}%
\begin{pgfscope}%
\definecolor{textcolor}{rgb}{0.000000,0.000000,0.000000}%
\pgfsetstrokecolor{textcolor}%
\pgfsetfillcolor{textcolor}%
\pgftext[x=1.834959in,y=0.408498in,,top]{\color{textcolor}{\rmfamily\fontsize{6.500000}{7.800000}\selectfont\catcode`\^=\active\def^{\ifmmode\sp\else\^{}\fi}\catcode`\%=\active\def%{\%}2000}}%
\end{pgfscope}%
\begin{pgfscope}%
\pgfsetbuttcap%
\pgfsetroundjoin%
\definecolor{currentfill}{rgb}{0.000000,0.000000,0.000000}%
\pgfsetfillcolor{currentfill}%
\pgfsetlinewidth{0.803000pt}%
\definecolor{currentstroke}{rgb}{0.000000,0.000000,0.000000}%
\pgfsetstrokecolor{currentstroke}%
\pgfsetdash{}{0pt}%
\pgfsys@defobject{currentmarker}{\pgfqpoint{0.000000in}{-0.048611in}}{\pgfqpoint{0.000000in}{0.000000in}}{%
\pgfpathmoveto{\pgfqpoint{0.000000in}{0.000000in}}%
\pgfpathlineto{\pgfqpoint{0.000000in}{-0.048611in}}%
\pgfusepath{stroke,fill}%
}%
\begin{pgfscope}%
\pgfsys@transformshift{2.062104in}{0.498776in}%
\pgfsys@useobject{currentmarker}{}%
\end{pgfscope}%
\end{pgfscope}%
\begin{pgfscope}%
\definecolor{textcolor}{rgb}{0.000000,0.000000,0.000000}%
\pgfsetstrokecolor{textcolor}%
\pgfsetfillcolor{textcolor}%
\pgftext[x=2.062104in,y=0.408498in,,top]{\color{textcolor}{\rmfamily\fontsize{6.500000}{7.800000}\selectfont\catcode`\^=\active\def^{\ifmmode\sp\else\^{}\fi}\catcode`\%=\active\def%{\%}3000}}%
\end{pgfscope}%
\begin{pgfscope}%
\pgfsetbuttcap%
\pgfsetroundjoin%
\definecolor{currentfill}{rgb}{0.000000,0.000000,0.000000}%
\pgfsetfillcolor{currentfill}%
\pgfsetlinewidth{0.803000pt}%
\definecolor{currentstroke}{rgb}{0.000000,0.000000,0.000000}%
\pgfsetstrokecolor{currentstroke}%
\pgfsetdash{}{0pt}%
\pgfsys@defobject{currentmarker}{\pgfqpoint{0.000000in}{-0.048611in}}{\pgfqpoint{0.000000in}{0.000000in}}{%
\pgfpathmoveto{\pgfqpoint{0.000000in}{0.000000in}}%
\pgfpathlineto{\pgfqpoint{0.000000in}{-0.048611in}}%
\pgfusepath{stroke,fill}%
}%
\begin{pgfscope}%
\pgfsys@transformshift{2.289250in}{0.498776in}%
\pgfsys@useobject{currentmarker}{}%
\end{pgfscope}%
\end{pgfscope}%
\begin{pgfscope}%
\definecolor{textcolor}{rgb}{0.000000,0.000000,0.000000}%
\pgfsetstrokecolor{textcolor}%
\pgfsetfillcolor{textcolor}%
\pgftext[x=2.289250in,y=0.408498in,,top]{\color{textcolor}{\rmfamily\fontsize{6.500000}{7.800000}\selectfont\catcode`\^=\active\def^{\ifmmode\sp\else\^{}\fi}\catcode`\%=\active\def%{\%}4000}}%
\end{pgfscope}%
\begin{pgfscope}%
\pgfsetbuttcap%
\pgfsetroundjoin%
\definecolor{currentfill}{rgb}{0.000000,0.000000,0.000000}%
\pgfsetfillcolor{currentfill}%
\pgfsetlinewidth{0.803000pt}%
\definecolor{currentstroke}{rgb}{0.000000,0.000000,0.000000}%
\pgfsetstrokecolor{currentstroke}%
\pgfsetdash{}{0pt}%
\pgfsys@defobject{currentmarker}{\pgfqpoint{0.000000in}{-0.048611in}}{\pgfqpoint{0.000000in}{0.000000in}}{%
\pgfpathmoveto{\pgfqpoint{0.000000in}{0.000000in}}%
\pgfpathlineto{\pgfqpoint{0.000000in}{-0.048611in}}%
\pgfusepath{stroke,fill}%
}%
\begin{pgfscope}%
\pgfsys@transformshift{2.516395in}{0.498776in}%
\pgfsys@useobject{currentmarker}{}%
\end{pgfscope}%
\end{pgfscope}%
\begin{pgfscope}%
\definecolor{textcolor}{rgb}{0.000000,0.000000,0.000000}%
\pgfsetstrokecolor{textcolor}%
\pgfsetfillcolor{textcolor}%
\pgftext[x=2.516395in,y=0.408498in,,top]{\color{textcolor}{\rmfamily\fontsize{6.500000}{7.800000}\selectfont\catcode`\^=\active\def^{\ifmmode\sp\else\^{}\fi}\catcode`\%=\active\def%{\%}5000}}%
\end{pgfscope}%
\begin{pgfscope}%
\pgfsetbuttcap%
\pgfsetroundjoin%
\definecolor{currentfill}{rgb}{0.000000,0.000000,0.000000}%
\pgfsetfillcolor{currentfill}%
\pgfsetlinewidth{0.803000pt}%
\definecolor{currentstroke}{rgb}{0.000000,0.000000,0.000000}%
\pgfsetstrokecolor{currentstroke}%
\pgfsetdash{}{0pt}%
\pgfsys@defobject{currentmarker}{\pgfqpoint{0.000000in}{-0.048611in}}{\pgfqpoint{0.000000in}{0.000000in}}{%
\pgfpathmoveto{\pgfqpoint{0.000000in}{0.000000in}}%
\pgfpathlineto{\pgfqpoint{0.000000in}{-0.048611in}}%
\pgfusepath{stroke,fill}%
}%
\begin{pgfscope}%
\pgfsys@transformshift{2.743541in}{0.498776in}%
\pgfsys@useobject{currentmarker}{}%
\end{pgfscope}%
\end{pgfscope}%
\begin{pgfscope}%
\definecolor{textcolor}{rgb}{0.000000,0.000000,0.000000}%
\pgfsetstrokecolor{textcolor}%
\pgfsetfillcolor{textcolor}%
\pgftext[x=2.743541in,y=0.408498in,,top]{\color{textcolor}{\rmfamily\fontsize{6.500000}{7.800000}\selectfont\catcode`\^=\active\def^{\ifmmode\sp\else\^{}\fi}\catcode`\%=\active\def%{\%}6000}}%
\end{pgfscope}%
\begin{pgfscope}%
\definecolor{textcolor}{rgb}{0.000000,0.000000,0.000000}%
\pgfsetstrokecolor{textcolor}%
\pgfsetfillcolor{textcolor}%
\pgftext[x=1.721386in,y=0.278868in,,top]{\color{textcolor}{\rmfamily\fontsize{10.000000}{12.000000}\selectfont\catcode`\^=\active\def^{\ifmmode\sp\else\^{}\fi}\catcode`\%=\active\def%{\%}$\text{Re}$}}%
\end{pgfscope}%
\begin{pgfscope}%
\pgfsetbuttcap%
\pgfsetroundjoin%
\definecolor{currentfill}{rgb}{0.000000,0.000000,0.000000}%
\pgfsetfillcolor{currentfill}%
\pgfsetlinewidth{0.803000pt}%
\definecolor{currentstroke}{rgb}{0.000000,0.000000,0.000000}%
\pgfsetstrokecolor{currentstroke}%
\pgfsetdash{}{0pt}%
\pgfsys@defobject{currentmarker}{\pgfqpoint{-0.048611in}{0.000000in}}{\pgfqpoint{-0.000000in}{0.000000in}}{%
\pgfpathmoveto{\pgfqpoint{-0.000000in}{0.000000in}}%
\pgfpathlineto{\pgfqpoint{-0.048611in}{0.000000in}}%
\pgfusepath{stroke,fill}%
}%
\begin{pgfscope}%
\pgfsys@transformshift{0.597016in}{0.773990in}%
\pgfsys@useobject{currentmarker}{}%
\end{pgfscope}%
\end{pgfscope}%
\begin{pgfscope}%
\definecolor{textcolor}{rgb}{0.000000,0.000000,0.000000}%
\pgfsetstrokecolor{textcolor}%
\pgfsetfillcolor{textcolor}%
\pgftext[x=0.455813in, y=0.745055in, left, base]{\color{textcolor}{\rmfamily\fontsize{6.500000}{7.800000}\selectfont\catcode`\^=\active\def^{\ifmmode\sp\else\^{}\fi}\catcode`\%=\active\def%{\%}$\mathdefault{5}$}}%
\end{pgfscope}%
\begin{pgfscope}%
\pgfsetbuttcap%
\pgfsetroundjoin%
\definecolor{currentfill}{rgb}{0.000000,0.000000,0.000000}%
\pgfsetfillcolor{currentfill}%
\pgfsetlinewidth{0.803000pt}%
\definecolor{currentstroke}{rgb}{0.000000,0.000000,0.000000}%
\pgfsetstrokecolor{currentstroke}%
\pgfsetdash{}{0pt}%
\pgfsys@defobject{currentmarker}{\pgfqpoint{-0.048611in}{0.000000in}}{\pgfqpoint{-0.000000in}{0.000000in}}{%
\pgfpathmoveto{\pgfqpoint{-0.000000in}{0.000000in}}%
\pgfpathlineto{\pgfqpoint{-0.048611in}{0.000000in}}%
\pgfusepath{stroke,fill}%
}%
\begin{pgfscope}%
\pgfsys@transformshift{0.597016in}{1.073581in}%
\pgfsys@useobject{currentmarker}{}%
\end{pgfscope}%
\end{pgfscope}%
\begin{pgfscope}%
\definecolor{textcolor}{rgb}{0.000000,0.000000,0.000000}%
\pgfsetstrokecolor{textcolor}%
\pgfsetfillcolor{textcolor}%
\pgftext[x=0.404888in, y=1.044646in, left, base]{\color{textcolor}{\rmfamily\fontsize{6.500000}{7.800000}\selectfont\catcode`\^=\active\def^{\ifmmode\sp\else\^{}\fi}\catcode`\%=\active\def%{\%}$\mathdefault{10}$}}%
\end{pgfscope}%
\begin{pgfscope}%
\pgfsetbuttcap%
\pgfsetroundjoin%
\definecolor{currentfill}{rgb}{0.000000,0.000000,0.000000}%
\pgfsetfillcolor{currentfill}%
\pgfsetlinewidth{0.803000pt}%
\definecolor{currentstroke}{rgb}{0.000000,0.000000,0.000000}%
\pgfsetstrokecolor{currentstroke}%
\pgfsetdash{}{0pt}%
\pgfsys@defobject{currentmarker}{\pgfqpoint{-0.048611in}{0.000000in}}{\pgfqpoint{-0.000000in}{0.000000in}}{%
\pgfpathmoveto{\pgfqpoint{-0.000000in}{0.000000in}}%
\pgfpathlineto{\pgfqpoint{-0.048611in}{0.000000in}}%
\pgfusepath{stroke,fill}%
}%
\begin{pgfscope}%
\pgfsys@transformshift{0.597016in}{1.373173in}%
\pgfsys@useobject{currentmarker}{}%
\end{pgfscope}%
\end{pgfscope}%
\begin{pgfscope}%
\definecolor{textcolor}{rgb}{0.000000,0.000000,0.000000}%
\pgfsetstrokecolor{textcolor}%
\pgfsetfillcolor{textcolor}%
\pgftext[x=0.404888in, y=1.344238in, left, base]{\color{textcolor}{\rmfamily\fontsize{6.500000}{7.800000}\selectfont\catcode`\^=\active\def^{\ifmmode\sp\else\^{}\fi}\catcode`\%=\active\def%{\%}$\mathdefault{15}$}}%
\end{pgfscope}%
\begin{pgfscope}%
\pgfsetbuttcap%
\pgfsetroundjoin%
\definecolor{currentfill}{rgb}{0.000000,0.000000,0.000000}%
\pgfsetfillcolor{currentfill}%
\pgfsetlinewidth{0.803000pt}%
\definecolor{currentstroke}{rgb}{0.000000,0.000000,0.000000}%
\pgfsetstrokecolor{currentstroke}%
\pgfsetdash{}{0pt}%
\pgfsys@defobject{currentmarker}{\pgfqpoint{-0.048611in}{0.000000in}}{\pgfqpoint{-0.000000in}{0.000000in}}{%
\pgfpathmoveto{\pgfqpoint{-0.000000in}{0.000000in}}%
\pgfpathlineto{\pgfqpoint{-0.048611in}{0.000000in}}%
\pgfusepath{stroke,fill}%
}%
\begin{pgfscope}%
\pgfsys@transformshift{0.597016in}{1.672765in}%
\pgfsys@useobject{currentmarker}{}%
\end{pgfscope}%
\end{pgfscope}%
\begin{pgfscope}%
\definecolor{textcolor}{rgb}{0.000000,0.000000,0.000000}%
\pgfsetstrokecolor{textcolor}%
\pgfsetfillcolor{textcolor}%
\pgftext[x=0.404888in, y=1.643829in, left, base]{\color{textcolor}{\rmfamily\fontsize{6.500000}{7.800000}\selectfont\catcode`\^=\active\def^{\ifmmode\sp\else\^{}\fi}\catcode`\%=\active\def%{\%}$\mathdefault{20}$}}%
\end{pgfscope}%
\begin{pgfscope}%
\definecolor{textcolor}{rgb}{0.000000,0.000000,0.000000}%
\pgfsetstrokecolor{textcolor}%
\pgfsetfillcolor{textcolor}%
\pgftext[x=0.349332in,y=1.174388in,,bottom,rotate=90.000000]{\color{textcolor}{\rmfamily\fontsize{10.000000}{12.000000}\selectfont\catcode`\^=\active\def^{\ifmmode\sp\else\^{}\fi}\catcode`\%=\active\def%{\%}$\delta_{\ell^1}^{(q)} \ [\%]$}}%
\end{pgfscope}%
\begin{pgfscope}%
\pgfpathrectangle{\pgfqpoint{0.597016in}{0.498776in}}{\pgfqpoint{2.248740in}{1.351224in}}%
\pgfusepath{clip}%
\pgfsetbuttcap%
\pgfsetroundjoin%
\pgfsetlinewidth{0.501875pt}%
\definecolor{currentstroke}{rgb}{0.843137,0.098039,0.109804}%
\pgfsetstrokecolor{currentstroke}%
\pgfsetdash{}{0pt}%
\pgfpathmoveto{\pgfqpoint{0.699231in}{1.161503in}}%
\pgfpathlineto{\pgfqpoint{0.699231in}{1.193162in}}%
\pgfusepath{stroke}%
\end{pgfscope}%
\begin{pgfscope}%
\pgfpathrectangle{\pgfqpoint{0.597016in}{0.498776in}}{\pgfqpoint{2.248740in}{1.351224in}}%
\pgfusepath{clip}%
\pgfsetbuttcap%
\pgfsetroundjoin%
\pgfsetlinewidth{0.501875pt}%
\definecolor{currentstroke}{rgb}{0.843137,0.098039,0.109804}%
\pgfsetstrokecolor{currentstroke}%
\pgfsetdash{}{0pt}%
\pgfpathmoveto{\pgfqpoint{0.926377in}{1.088343in}}%
\pgfpathlineto{\pgfqpoint{0.926377in}{1.119200in}}%
\pgfusepath{stroke}%
\end{pgfscope}%
\begin{pgfscope}%
\pgfpathrectangle{\pgfqpoint{0.597016in}{0.498776in}}{\pgfqpoint{2.248740in}{1.351224in}}%
\pgfusepath{clip}%
\pgfsetbuttcap%
\pgfsetroundjoin%
\pgfsetlinewidth{0.501875pt}%
\definecolor{currentstroke}{rgb}{0.843137,0.098039,0.109804}%
\pgfsetstrokecolor{currentstroke}%
\pgfsetdash{}{0pt}%
\pgfpathmoveto{\pgfqpoint{1.153522in}{0.676030in}}%
\pgfpathlineto{\pgfqpoint{1.153522in}{0.710414in}}%
\pgfusepath{stroke}%
\end{pgfscope}%
\begin{pgfscope}%
\pgfpathrectangle{\pgfqpoint{0.597016in}{0.498776in}}{\pgfqpoint{2.248740in}{1.351224in}}%
\pgfusepath{clip}%
\pgfsetbuttcap%
\pgfsetroundjoin%
\pgfsetlinewidth{0.501875pt}%
\definecolor{currentstroke}{rgb}{0.843137,0.098039,0.109804}%
\pgfsetstrokecolor{currentstroke}%
\pgfsetdash{}{0pt}%
\pgfpathmoveto{\pgfqpoint{1.380668in}{0.629173in}}%
\pgfpathlineto{\pgfqpoint{1.380668in}{0.652092in}}%
\pgfusepath{stroke}%
\end{pgfscope}%
\begin{pgfscope}%
\pgfpathrectangle{\pgfqpoint{0.597016in}{0.498776in}}{\pgfqpoint{2.248740in}{1.351224in}}%
\pgfusepath{clip}%
\pgfsetbuttcap%
\pgfsetroundjoin%
\pgfsetlinewidth{0.501875pt}%
\definecolor{currentstroke}{rgb}{0.843137,0.098039,0.109804}%
\pgfsetstrokecolor{currentstroke}%
\pgfsetdash{}{0pt}%
\pgfpathmoveto{\pgfqpoint{1.607813in}{0.710107in}}%
\pgfpathlineto{\pgfqpoint{1.607813in}{0.741092in}}%
\pgfusepath{stroke}%
\end{pgfscope}%
\begin{pgfscope}%
\pgfpathrectangle{\pgfqpoint{0.597016in}{0.498776in}}{\pgfqpoint{2.248740in}{1.351224in}}%
\pgfusepath{clip}%
\pgfsetbuttcap%
\pgfsetroundjoin%
\pgfsetlinewidth{0.501875pt}%
\definecolor{currentstroke}{rgb}{0.843137,0.098039,0.109804}%
\pgfsetstrokecolor{currentstroke}%
\pgfsetdash{}{0pt}%
\pgfpathmoveto{\pgfqpoint{1.834959in}{0.743290in}}%
\pgfpathlineto{\pgfqpoint{1.834959in}{0.771400in}}%
\pgfusepath{stroke}%
\end{pgfscope}%
\begin{pgfscope}%
\pgfpathrectangle{\pgfqpoint{0.597016in}{0.498776in}}{\pgfqpoint{2.248740in}{1.351224in}}%
\pgfusepath{clip}%
\pgfsetbuttcap%
\pgfsetroundjoin%
\pgfsetlinewidth{0.501875pt}%
\definecolor{currentstroke}{rgb}{0.843137,0.098039,0.109804}%
\pgfsetstrokecolor{currentstroke}%
\pgfsetdash{}{0pt}%
\pgfpathmoveto{\pgfqpoint{2.062104in}{0.742218in}}%
\pgfpathlineto{\pgfqpoint{2.062104in}{0.766963in}}%
\pgfusepath{stroke}%
\end{pgfscope}%
\begin{pgfscope}%
\pgfpathrectangle{\pgfqpoint{0.597016in}{0.498776in}}{\pgfqpoint{2.248740in}{1.351224in}}%
\pgfusepath{clip}%
\pgfsetbuttcap%
\pgfsetroundjoin%
\pgfsetlinewidth{0.501875pt}%
\definecolor{currentstroke}{rgb}{0.843137,0.098039,0.109804}%
\pgfsetstrokecolor{currentstroke}%
\pgfsetdash{}{0pt}%
\pgfpathmoveto{\pgfqpoint{2.289250in}{0.739566in}}%
\pgfpathlineto{\pgfqpoint{2.289250in}{0.762604in}}%
\pgfusepath{stroke}%
\end{pgfscope}%
\begin{pgfscope}%
\pgfpathrectangle{\pgfqpoint{0.597016in}{0.498776in}}{\pgfqpoint{2.248740in}{1.351224in}}%
\pgfusepath{clip}%
\pgfsetbuttcap%
\pgfsetroundjoin%
\pgfsetlinewidth{0.501875pt}%
\definecolor{currentstroke}{rgb}{0.843137,0.098039,0.109804}%
\pgfsetstrokecolor{currentstroke}%
\pgfsetdash{}{0pt}%
\pgfpathmoveto{\pgfqpoint{2.516395in}{0.734252in}}%
\pgfpathlineto{\pgfqpoint{2.516395in}{0.760174in}}%
\pgfusepath{stroke}%
\end{pgfscope}%
\begin{pgfscope}%
\pgfpathrectangle{\pgfqpoint{0.597016in}{0.498776in}}{\pgfqpoint{2.248740in}{1.351224in}}%
\pgfusepath{clip}%
\pgfsetbuttcap%
\pgfsetroundjoin%
\pgfsetlinewidth{0.501875pt}%
\definecolor{currentstroke}{rgb}{0.843137,0.098039,0.109804}%
\pgfsetstrokecolor{currentstroke}%
\pgfsetdash{}{0pt}%
\pgfpathmoveto{\pgfqpoint{2.743541in}{0.784047in}}%
\pgfpathlineto{\pgfqpoint{2.743541in}{0.819034in}}%
\pgfusepath{stroke}%
\end{pgfscope}%
\begin{pgfscope}%
\pgfpathrectangle{\pgfqpoint{0.597016in}{0.498776in}}{\pgfqpoint{2.248740in}{1.351224in}}%
\pgfusepath{clip}%
\pgfsetbuttcap%
\pgfsetroundjoin%
\definecolor{currentfill}{rgb}{0.843137,0.098039,0.109804}%
\pgfsetfillcolor{currentfill}%
\pgfsetlinewidth{1.003750pt}%
\definecolor{currentstroke}{rgb}{0.843137,0.098039,0.109804}%
\pgfsetstrokecolor{currentstroke}%
\pgfsetdash{}{0pt}%
\pgfsys@defobject{currentmarker}{\pgfqpoint{-0.069444in}{-0.000000in}}{\pgfqpoint{0.069444in}{0.000000in}}{%
\pgfpathmoveto{\pgfqpoint{0.069444in}{-0.000000in}}%
\pgfpathlineto{\pgfqpoint{-0.069444in}{0.000000in}}%
\pgfusepath{stroke,fill}%
}%
\begin{pgfscope}%
\pgfsys@transformshift{0.699231in}{1.161503in}%
\pgfsys@useobject{currentmarker}{}%
\end{pgfscope}%
\begin{pgfscope}%
\pgfsys@transformshift{0.926377in}{1.088343in}%
\pgfsys@useobject{currentmarker}{}%
\end{pgfscope}%
\begin{pgfscope}%
\pgfsys@transformshift{1.153522in}{0.676030in}%
\pgfsys@useobject{currentmarker}{}%
\end{pgfscope}%
\begin{pgfscope}%
\pgfsys@transformshift{1.380668in}{0.629173in}%
\pgfsys@useobject{currentmarker}{}%
\end{pgfscope}%
\begin{pgfscope}%
\pgfsys@transformshift{1.607813in}{0.710107in}%
\pgfsys@useobject{currentmarker}{}%
\end{pgfscope}%
\begin{pgfscope}%
\pgfsys@transformshift{1.834959in}{0.743290in}%
\pgfsys@useobject{currentmarker}{}%
\end{pgfscope}%
\begin{pgfscope}%
\pgfsys@transformshift{2.062104in}{0.742218in}%
\pgfsys@useobject{currentmarker}{}%
\end{pgfscope}%
\begin{pgfscope}%
\pgfsys@transformshift{2.289250in}{0.739566in}%
\pgfsys@useobject{currentmarker}{}%
\end{pgfscope}%
\begin{pgfscope}%
\pgfsys@transformshift{2.516395in}{0.734252in}%
\pgfsys@useobject{currentmarker}{}%
\end{pgfscope}%
\begin{pgfscope}%
\pgfsys@transformshift{2.743541in}{0.784047in}%
\pgfsys@useobject{currentmarker}{}%
\end{pgfscope}%
\end{pgfscope}%
\begin{pgfscope}%
\pgfpathrectangle{\pgfqpoint{0.597016in}{0.498776in}}{\pgfqpoint{2.248740in}{1.351224in}}%
\pgfusepath{clip}%
\pgfsetbuttcap%
\pgfsetroundjoin%
\definecolor{currentfill}{rgb}{0.843137,0.098039,0.109804}%
\pgfsetfillcolor{currentfill}%
\pgfsetlinewidth{1.003750pt}%
\definecolor{currentstroke}{rgb}{0.843137,0.098039,0.109804}%
\pgfsetstrokecolor{currentstroke}%
\pgfsetdash{}{0pt}%
\pgfsys@defobject{currentmarker}{\pgfqpoint{-0.069444in}{-0.000000in}}{\pgfqpoint{0.069444in}{0.000000in}}{%
\pgfpathmoveto{\pgfqpoint{0.069444in}{-0.000000in}}%
\pgfpathlineto{\pgfqpoint{-0.069444in}{0.000000in}}%
\pgfusepath{stroke,fill}%
}%
\begin{pgfscope}%
\pgfsys@transformshift{0.699231in}{1.193162in}%
\pgfsys@useobject{currentmarker}{}%
\end{pgfscope}%
\begin{pgfscope}%
\pgfsys@transformshift{0.926377in}{1.119200in}%
\pgfsys@useobject{currentmarker}{}%
\end{pgfscope}%
\begin{pgfscope}%
\pgfsys@transformshift{1.153522in}{0.710414in}%
\pgfsys@useobject{currentmarker}{}%
\end{pgfscope}%
\begin{pgfscope}%
\pgfsys@transformshift{1.380668in}{0.652092in}%
\pgfsys@useobject{currentmarker}{}%
\end{pgfscope}%
\begin{pgfscope}%
\pgfsys@transformshift{1.607813in}{0.741092in}%
\pgfsys@useobject{currentmarker}{}%
\end{pgfscope}%
\begin{pgfscope}%
\pgfsys@transformshift{1.834959in}{0.771400in}%
\pgfsys@useobject{currentmarker}{}%
\end{pgfscope}%
\begin{pgfscope}%
\pgfsys@transformshift{2.062104in}{0.766963in}%
\pgfsys@useobject{currentmarker}{}%
\end{pgfscope}%
\begin{pgfscope}%
\pgfsys@transformshift{2.289250in}{0.762604in}%
\pgfsys@useobject{currentmarker}{}%
\end{pgfscope}%
\begin{pgfscope}%
\pgfsys@transformshift{2.516395in}{0.760174in}%
\pgfsys@useobject{currentmarker}{}%
\end{pgfscope}%
\begin{pgfscope}%
\pgfsys@transformshift{2.743541in}{0.819034in}%
\pgfsys@useobject{currentmarker}{}%
\end{pgfscope}%
\end{pgfscope}%
\begin{pgfscope}%
\pgfpathrectangle{\pgfqpoint{0.597016in}{0.498776in}}{\pgfqpoint{2.248740in}{1.351224in}}%
\pgfusepath{clip}%
\pgfsetbuttcap%
\pgfsetroundjoin%
\pgfsetlinewidth{0.501875pt}%
\definecolor{currentstroke}{rgb}{0.172549,0.482353,0.713725}%
\pgfsetstrokecolor{currentstroke}%
\pgfsetdash{}{0pt}%
\pgfpathmoveto{\pgfqpoint{0.699231in}{1.766154in}}%
\pgfpathlineto{\pgfqpoint{0.699231in}{1.788581in}}%
\pgfusepath{stroke}%
\end{pgfscope}%
\begin{pgfscope}%
\pgfpathrectangle{\pgfqpoint{0.597016in}{0.498776in}}{\pgfqpoint{2.248740in}{1.351224in}}%
\pgfusepath{clip}%
\pgfsetbuttcap%
\pgfsetroundjoin%
\pgfsetlinewidth{0.501875pt}%
\definecolor{currentstroke}{rgb}{0.172549,0.482353,0.713725}%
\pgfsetstrokecolor{currentstroke}%
\pgfsetdash{}{0pt}%
\pgfpathmoveto{\pgfqpoint{0.926377in}{1.540677in}}%
\pgfpathlineto{\pgfqpoint{0.926377in}{1.562932in}}%
\pgfusepath{stroke}%
\end{pgfscope}%
\begin{pgfscope}%
\pgfpathrectangle{\pgfqpoint{0.597016in}{0.498776in}}{\pgfqpoint{2.248740in}{1.351224in}}%
\pgfusepath{clip}%
\pgfsetbuttcap%
\pgfsetroundjoin%
\pgfsetlinewidth{0.501875pt}%
\definecolor{currentstroke}{rgb}{0.172549,0.482353,0.713725}%
\pgfsetstrokecolor{currentstroke}%
\pgfsetdash{}{0pt}%
\pgfpathmoveto{\pgfqpoint{1.153522in}{0.632641in}}%
\pgfpathlineto{\pgfqpoint{1.153522in}{0.654374in}}%
\pgfusepath{stroke}%
\end{pgfscope}%
\begin{pgfscope}%
\pgfpathrectangle{\pgfqpoint{0.597016in}{0.498776in}}{\pgfqpoint{2.248740in}{1.351224in}}%
\pgfusepath{clip}%
\pgfsetbuttcap%
\pgfsetroundjoin%
\pgfsetlinewidth{0.501875pt}%
\definecolor{currentstroke}{rgb}{0.172549,0.482353,0.713725}%
\pgfsetstrokecolor{currentstroke}%
\pgfsetdash{}{0pt}%
\pgfpathmoveto{\pgfqpoint{1.380668in}{0.560195in}}%
\pgfpathlineto{\pgfqpoint{1.380668in}{0.589405in}}%
\pgfusepath{stroke}%
\end{pgfscope}%
\begin{pgfscope}%
\pgfpathrectangle{\pgfqpoint{0.597016in}{0.498776in}}{\pgfqpoint{2.248740in}{1.351224in}}%
\pgfusepath{clip}%
\pgfsetbuttcap%
\pgfsetroundjoin%
\pgfsetlinewidth{0.501875pt}%
\definecolor{currentstroke}{rgb}{0.172549,0.482353,0.713725}%
\pgfsetstrokecolor{currentstroke}%
\pgfsetdash{}{0pt}%
\pgfpathmoveto{\pgfqpoint{1.607813in}{0.646148in}}%
\pgfpathlineto{\pgfqpoint{1.607813in}{0.671678in}}%
\pgfusepath{stroke}%
\end{pgfscope}%
\begin{pgfscope}%
\pgfpathrectangle{\pgfqpoint{0.597016in}{0.498776in}}{\pgfqpoint{2.248740in}{1.351224in}}%
\pgfusepath{clip}%
\pgfsetbuttcap%
\pgfsetroundjoin%
\pgfsetlinewidth{0.501875pt}%
\definecolor{currentstroke}{rgb}{0.172549,0.482353,0.713725}%
\pgfsetstrokecolor{currentstroke}%
\pgfsetdash{}{0pt}%
\pgfpathmoveto{\pgfqpoint{1.834959in}{0.676871in}}%
\pgfpathlineto{\pgfqpoint{1.834959in}{0.694929in}}%
\pgfusepath{stroke}%
\end{pgfscope}%
\begin{pgfscope}%
\pgfpathrectangle{\pgfqpoint{0.597016in}{0.498776in}}{\pgfqpoint{2.248740in}{1.351224in}}%
\pgfusepath{clip}%
\pgfsetbuttcap%
\pgfsetroundjoin%
\pgfsetlinewidth{0.501875pt}%
\definecolor{currentstroke}{rgb}{0.172549,0.482353,0.713725}%
\pgfsetstrokecolor{currentstroke}%
\pgfsetdash{}{0pt}%
\pgfpathmoveto{\pgfqpoint{2.062104in}{0.684903in}}%
\pgfpathlineto{\pgfqpoint{2.062104in}{0.698154in}}%
\pgfusepath{stroke}%
\end{pgfscope}%
\begin{pgfscope}%
\pgfpathrectangle{\pgfqpoint{0.597016in}{0.498776in}}{\pgfqpoint{2.248740in}{1.351224in}}%
\pgfusepath{clip}%
\pgfsetbuttcap%
\pgfsetroundjoin%
\pgfsetlinewidth{0.501875pt}%
\definecolor{currentstroke}{rgb}{0.172549,0.482353,0.713725}%
\pgfsetstrokecolor{currentstroke}%
\pgfsetdash{}{0pt}%
\pgfpathmoveto{\pgfqpoint{2.289250in}{0.695676in}}%
\pgfpathlineto{\pgfqpoint{2.289250in}{0.705597in}}%
\pgfusepath{stroke}%
\end{pgfscope}%
\begin{pgfscope}%
\pgfpathrectangle{\pgfqpoint{0.597016in}{0.498776in}}{\pgfqpoint{2.248740in}{1.351224in}}%
\pgfusepath{clip}%
\pgfsetbuttcap%
\pgfsetroundjoin%
\pgfsetlinewidth{0.501875pt}%
\definecolor{currentstroke}{rgb}{0.172549,0.482353,0.713725}%
\pgfsetstrokecolor{currentstroke}%
\pgfsetdash{}{0pt}%
\pgfpathmoveto{\pgfqpoint{2.516395in}{0.708417in}}%
\pgfpathlineto{\pgfqpoint{2.516395in}{0.724227in}}%
\pgfusepath{stroke}%
\end{pgfscope}%
\begin{pgfscope}%
\pgfpathrectangle{\pgfqpoint{0.597016in}{0.498776in}}{\pgfqpoint{2.248740in}{1.351224in}}%
\pgfusepath{clip}%
\pgfsetbuttcap%
\pgfsetroundjoin%
\pgfsetlinewidth{0.501875pt}%
\definecolor{currentstroke}{rgb}{0.172549,0.482353,0.713725}%
\pgfsetstrokecolor{currentstroke}%
\pgfsetdash{}{0pt}%
\pgfpathmoveto{\pgfqpoint{2.743541in}{0.782503in}}%
\pgfpathlineto{\pgfqpoint{2.743541in}{0.810438in}}%
\pgfusepath{stroke}%
\end{pgfscope}%
\begin{pgfscope}%
\pgfpathrectangle{\pgfqpoint{0.597016in}{0.498776in}}{\pgfqpoint{2.248740in}{1.351224in}}%
\pgfusepath{clip}%
\pgfsetbuttcap%
\pgfsetroundjoin%
\definecolor{currentfill}{rgb}{0.172549,0.482353,0.713725}%
\pgfsetfillcolor{currentfill}%
\pgfsetlinewidth{1.003750pt}%
\definecolor{currentstroke}{rgb}{0.172549,0.482353,0.713725}%
\pgfsetstrokecolor{currentstroke}%
\pgfsetdash{}{0pt}%
\pgfsys@defobject{currentmarker}{\pgfqpoint{-0.069444in}{-0.000000in}}{\pgfqpoint{0.069444in}{0.000000in}}{%
\pgfpathmoveto{\pgfqpoint{0.069444in}{-0.000000in}}%
\pgfpathlineto{\pgfqpoint{-0.069444in}{0.000000in}}%
\pgfusepath{stroke,fill}%
}%
\begin{pgfscope}%
\pgfsys@transformshift{0.699231in}{1.766154in}%
\pgfsys@useobject{currentmarker}{}%
\end{pgfscope}%
\begin{pgfscope}%
\pgfsys@transformshift{0.926377in}{1.540677in}%
\pgfsys@useobject{currentmarker}{}%
\end{pgfscope}%
\begin{pgfscope}%
\pgfsys@transformshift{1.153522in}{0.632641in}%
\pgfsys@useobject{currentmarker}{}%
\end{pgfscope}%
\begin{pgfscope}%
\pgfsys@transformshift{1.380668in}{0.560195in}%
\pgfsys@useobject{currentmarker}{}%
\end{pgfscope}%
\begin{pgfscope}%
\pgfsys@transformshift{1.607813in}{0.646148in}%
\pgfsys@useobject{currentmarker}{}%
\end{pgfscope}%
\begin{pgfscope}%
\pgfsys@transformshift{1.834959in}{0.676871in}%
\pgfsys@useobject{currentmarker}{}%
\end{pgfscope}%
\begin{pgfscope}%
\pgfsys@transformshift{2.062104in}{0.684903in}%
\pgfsys@useobject{currentmarker}{}%
\end{pgfscope}%
\begin{pgfscope}%
\pgfsys@transformshift{2.289250in}{0.695676in}%
\pgfsys@useobject{currentmarker}{}%
\end{pgfscope}%
\begin{pgfscope}%
\pgfsys@transformshift{2.516395in}{0.708417in}%
\pgfsys@useobject{currentmarker}{}%
\end{pgfscope}%
\begin{pgfscope}%
\pgfsys@transformshift{2.743541in}{0.782503in}%
\pgfsys@useobject{currentmarker}{}%
\end{pgfscope}%
\end{pgfscope}%
\begin{pgfscope}%
\pgfpathrectangle{\pgfqpoint{0.597016in}{0.498776in}}{\pgfqpoint{2.248740in}{1.351224in}}%
\pgfusepath{clip}%
\pgfsetbuttcap%
\pgfsetroundjoin%
\definecolor{currentfill}{rgb}{0.172549,0.482353,0.713725}%
\pgfsetfillcolor{currentfill}%
\pgfsetlinewidth{1.003750pt}%
\definecolor{currentstroke}{rgb}{0.172549,0.482353,0.713725}%
\pgfsetstrokecolor{currentstroke}%
\pgfsetdash{}{0pt}%
\pgfsys@defobject{currentmarker}{\pgfqpoint{-0.069444in}{-0.000000in}}{\pgfqpoint{0.069444in}{0.000000in}}{%
\pgfpathmoveto{\pgfqpoint{0.069444in}{-0.000000in}}%
\pgfpathlineto{\pgfqpoint{-0.069444in}{0.000000in}}%
\pgfusepath{stroke,fill}%
}%
\begin{pgfscope}%
\pgfsys@transformshift{0.699231in}{1.788581in}%
\pgfsys@useobject{currentmarker}{}%
\end{pgfscope}%
\begin{pgfscope}%
\pgfsys@transformshift{0.926377in}{1.562932in}%
\pgfsys@useobject{currentmarker}{}%
\end{pgfscope}%
\begin{pgfscope}%
\pgfsys@transformshift{1.153522in}{0.654374in}%
\pgfsys@useobject{currentmarker}{}%
\end{pgfscope}%
\begin{pgfscope}%
\pgfsys@transformshift{1.380668in}{0.589405in}%
\pgfsys@useobject{currentmarker}{}%
\end{pgfscope}%
\begin{pgfscope}%
\pgfsys@transformshift{1.607813in}{0.671678in}%
\pgfsys@useobject{currentmarker}{}%
\end{pgfscope}%
\begin{pgfscope}%
\pgfsys@transformshift{1.834959in}{0.694929in}%
\pgfsys@useobject{currentmarker}{}%
\end{pgfscope}%
\begin{pgfscope}%
\pgfsys@transformshift{2.062104in}{0.698154in}%
\pgfsys@useobject{currentmarker}{}%
\end{pgfscope}%
\begin{pgfscope}%
\pgfsys@transformshift{2.289250in}{0.705597in}%
\pgfsys@useobject{currentmarker}{}%
\end{pgfscope}%
\begin{pgfscope}%
\pgfsys@transformshift{2.516395in}{0.724227in}%
\pgfsys@useobject{currentmarker}{}%
\end{pgfscope}%
\begin{pgfscope}%
\pgfsys@transformshift{2.743541in}{0.810438in}%
\pgfsys@useobject{currentmarker}{}%
\end{pgfscope}%
\end{pgfscope}%
\begin{pgfscope}%
\pgfpathrectangle{\pgfqpoint{0.597016in}{0.498776in}}{\pgfqpoint{2.248740in}{1.351224in}}%
\pgfusepath{clip}%
\pgfsetbuttcap%
\pgfsetroundjoin%
\definecolor{currentfill}{rgb}{0.843137,0.098039,0.109804}%
\pgfsetfillcolor{currentfill}%
\pgfsetlinewidth{1.003750pt}%
\definecolor{currentstroke}{rgb}{0.000000,0.000000,0.000000}%
\pgfsetstrokecolor{currentstroke}%
\pgfsetdash{}{0pt}%
\pgfsys@defobject{currentmarker}{\pgfqpoint{-0.027778in}{-0.027778in}}{\pgfqpoint{0.027778in}{0.027778in}}{%
\pgfpathmoveto{\pgfqpoint{0.000000in}{-0.027778in}}%
\pgfpathcurveto{\pgfqpoint{0.007367in}{-0.027778in}}{\pgfqpoint{0.014433in}{-0.024851in}}{\pgfqpoint{0.019642in}{-0.019642in}}%
\pgfpathcurveto{\pgfqpoint{0.024851in}{-0.014433in}}{\pgfqpoint{0.027778in}{-0.007367in}}{\pgfqpoint{0.027778in}{0.000000in}}%
\pgfpathcurveto{\pgfqpoint{0.027778in}{0.007367in}}{\pgfqpoint{0.024851in}{0.014433in}}{\pgfqpoint{0.019642in}{0.019642in}}%
\pgfpathcurveto{\pgfqpoint{0.014433in}{0.024851in}}{\pgfqpoint{0.007367in}{0.027778in}}{\pgfqpoint{0.000000in}{0.027778in}}%
\pgfpathcurveto{\pgfqpoint{-0.007367in}{0.027778in}}{\pgfqpoint{-0.014433in}{0.024851in}}{\pgfqpoint{-0.019642in}{0.019642in}}%
\pgfpathcurveto{\pgfqpoint{-0.024851in}{0.014433in}}{\pgfqpoint{-0.027778in}{0.007367in}}{\pgfqpoint{-0.027778in}{0.000000in}}%
\pgfpathcurveto{\pgfqpoint{-0.027778in}{-0.007367in}}{\pgfqpoint{-0.024851in}{-0.014433in}}{\pgfqpoint{-0.019642in}{-0.019642in}}%
\pgfpathcurveto{\pgfqpoint{-0.014433in}{-0.024851in}}{\pgfqpoint{-0.007367in}{-0.027778in}}{\pgfqpoint{0.000000in}{-0.027778in}}%
\pgfpathlineto{\pgfqpoint{0.000000in}{-0.027778in}}%
\pgfpathclose%
\pgfusepath{stroke,fill}%
}%
\begin{pgfscope}%
\pgfsys@transformshift{0.699231in}{1.177332in}%
\pgfsys@useobject{currentmarker}{}%
\end{pgfscope}%
\begin{pgfscope}%
\pgfsys@transformshift{0.926377in}{1.103772in}%
\pgfsys@useobject{currentmarker}{}%
\end{pgfscope}%
\begin{pgfscope}%
\pgfsys@transformshift{1.153522in}{0.693222in}%
\pgfsys@useobject{currentmarker}{}%
\end{pgfscope}%
\begin{pgfscope}%
\pgfsys@transformshift{1.380668in}{0.640632in}%
\pgfsys@useobject{currentmarker}{}%
\end{pgfscope}%
\begin{pgfscope}%
\pgfsys@transformshift{1.607813in}{0.725600in}%
\pgfsys@useobject{currentmarker}{}%
\end{pgfscope}%
\begin{pgfscope}%
\pgfsys@transformshift{1.834959in}{0.757345in}%
\pgfsys@useobject{currentmarker}{}%
\end{pgfscope}%
\begin{pgfscope}%
\pgfsys@transformshift{2.062104in}{0.754591in}%
\pgfsys@useobject{currentmarker}{}%
\end{pgfscope}%
\begin{pgfscope}%
\pgfsys@transformshift{2.289250in}{0.751085in}%
\pgfsys@useobject{currentmarker}{}%
\end{pgfscope}%
\begin{pgfscope}%
\pgfsys@transformshift{2.516395in}{0.747213in}%
\pgfsys@useobject{currentmarker}{}%
\end{pgfscope}%
\begin{pgfscope}%
\pgfsys@transformshift{2.743541in}{0.801540in}%
\pgfsys@useobject{currentmarker}{}%
\end{pgfscope}%
\end{pgfscope}%
\begin{pgfscope}%
\pgfpathrectangle{\pgfqpoint{0.597016in}{0.498776in}}{\pgfqpoint{2.248740in}{1.351224in}}%
\pgfusepath{clip}%
\pgfsetbuttcap%
\pgfsetmiterjoin%
\definecolor{currentfill}{rgb}{0.172549,0.482353,0.713725}%
\pgfsetfillcolor{currentfill}%
\pgfsetlinewidth{1.003750pt}%
\definecolor{currentstroke}{rgb}{0.000000,0.000000,0.000000}%
\pgfsetstrokecolor{currentstroke}%
\pgfsetdash{}{0pt}%
\pgfsys@defobject{currentmarker}{\pgfqpoint{-0.029463in}{-0.029463in}}{\pgfqpoint{0.029463in}{0.029463in}}{%
\pgfpathmoveto{\pgfqpoint{-0.000000in}{-0.029463in}}%
\pgfpathlineto{\pgfqpoint{0.029463in}{0.000000in}}%
\pgfpathlineto{\pgfqpoint{0.000000in}{0.029463in}}%
\pgfpathlineto{\pgfqpoint{-0.029463in}{0.000000in}}%
\pgfpathlineto{\pgfqpoint{-0.000000in}{-0.029463in}}%
\pgfpathclose%
\pgfusepath{stroke,fill}%
}%
\begin{pgfscope}%
\pgfsys@transformshift{0.699231in}{1.777368in}%
\pgfsys@useobject{currentmarker}{}%
\end{pgfscope}%
\begin{pgfscope}%
\pgfsys@transformshift{0.926377in}{1.551804in}%
\pgfsys@useobject{currentmarker}{}%
\end{pgfscope}%
\begin{pgfscope}%
\pgfsys@transformshift{1.153522in}{0.643508in}%
\pgfsys@useobject{currentmarker}{}%
\end{pgfscope}%
\begin{pgfscope}%
\pgfsys@transformshift{1.380668in}{0.574800in}%
\pgfsys@useobject{currentmarker}{}%
\end{pgfscope}%
\begin{pgfscope}%
\pgfsys@transformshift{1.607813in}{0.658913in}%
\pgfsys@useobject{currentmarker}{}%
\end{pgfscope}%
\begin{pgfscope}%
\pgfsys@transformshift{1.834959in}{0.685900in}%
\pgfsys@useobject{currentmarker}{}%
\end{pgfscope}%
\begin{pgfscope}%
\pgfsys@transformshift{2.062104in}{0.691529in}%
\pgfsys@useobject{currentmarker}{}%
\end{pgfscope}%
\begin{pgfscope}%
\pgfsys@transformshift{2.289250in}{0.700637in}%
\pgfsys@useobject{currentmarker}{}%
\end{pgfscope}%
\begin{pgfscope}%
\pgfsys@transformshift{2.516395in}{0.716322in}%
\pgfsys@useobject{currentmarker}{}%
\end{pgfscope}%
\begin{pgfscope}%
\pgfsys@transformshift{2.743541in}{0.796471in}%
\pgfsys@useobject{currentmarker}{}%
\end{pgfscope}%
\end{pgfscope}%
\begin{pgfscope}%
\pgfsetrectcap%
\pgfsetmiterjoin%
\pgfsetlinewidth{0.803000pt}%
\definecolor{currentstroke}{rgb}{0.000000,0.000000,0.000000}%
\pgfsetstrokecolor{currentstroke}%
\pgfsetdash{}{0pt}%
\pgfpathmoveto{\pgfqpoint{0.597016in}{0.498776in}}%
\pgfpathlineto{\pgfqpoint{0.597016in}{1.850000in}}%
\pgfusepath{stroke}%
\end{pgfscope}%
\begin{pgfscope}%
\pgfsetrectcap%
\pgfsetmiterjoin%
\pgfsetlinewidth{0.803000pt}%
\definecolor{currentstroke}{rgb}{0.000000,0.000000,0.000000}%
\pgfsetstrokecolor{currentstroke}%
\pgfsetdash{}{0pt}%
\pgfpathmoveto{\pgfqpoint{2.845756in}{0.498776in}}%
\pgfpathlineto{\pgfqpoint{2.845756in}{1.850000in}}%
\pgfusepath{stroke}%
\end{pgfscope}%
\begin{pgfscope}%
\pgfsetrectcap%
\pgfsetmiterjoin%
\pgfsetlinewidth{0.803000pt}%
\definecolor{currentstroke}{rgb}{0.000000,0.000000,0.000000}%
\pgfsetstrokecolor{currentstroke}%
\pgfsetdash{}{0pt}%
\pgfpathmoveto{\pgfqpoint{0.597016in}{0.498776in}}%
\pgfpathlineto{\pgfqpoint{2.845756in}{0.498776in}}%
\pgfusepath{stroke}%
\end{pgfscope}%
\begin{pgfscope}%
\pgfsetrectcap%
\pgfsetmiterjoin%
\pgfsetlinewidth{0.803000pt}%
\definecolor{currentstroke}{rgb}{0.000000,0.000000,0.000000}%
\pgfsetstrokecolor{currentstroke}%
\pgfsetdash{}{0pt}%
\pgfpathmoveto{\pgfqpoint{0.597016in}{1.850000in}}%
\pgfpathlineto{\pgfqpoint{2.845756in}{1.850000in}}%
\pgfusepath{stroke}%
\end{pgfscope}%
\begin{pgfscope}%
\pgfsetbuttcap%
\pgfsetmiterjoin%
\definecolor{currentfill}{rgb}{1.000000,1.000000,1.000000}%
\pgfsetfillcolor{currentfill}%
\pgfsetfillopacity{0.800000}%
\pgfsetlinewidth{1.003750pt}%
\definecolor{currentstroke}{rgb}{0.800000,0.800000,0.800000}%
\pgfsetstrokecolor{currentstroke}%
\pgfsetstrokeopacity{0.800000}%
\pgfsetdash{}{0pt}%
\pgfpathmoveto{\pgfqpoint{2.061616in}{1.373972in}}%
\pgfpathlineto{\pgfqpoint{2.764770in}{1.373972in}}%
\pgfpathquadraticcurveto{\pgfqpoint{2.787909in}{1.373972in}}{\pgfqpoint{2.787909in}{1.397111in}}%
\pgfpathlineto{\pgfqpoint{2.787909in}{1.769014in}}%
\pgfpathquadraticcurveto{\pgfqpoint{2.787909in}{1.792153in}}{\pgfqpoint{2.764770in}{1.792153in}}%
\pgfpathlineto{\pgfqpoint{2.061616in}{1.792153in}}%
\pgfpathquadraticcurveto{\pgfqpoint{2.038477in}{1.792153in}}{\pgfqpoint{2.038477in}{1.769014in}}%
\pgfpathlineto{\pgfqpoint{2.038477in}{1.397111in}}%
\pgfpathquadraticcurveto{\pgfqpoint{2.038477in}{1.373972in}}{\pgfqpoint{2.061616in}{1.373972in}}%
\pgfpathlineto{\pgfqpoint{2.061616in}{1.373972in}}%
\pgfpathclose%
\pgfusepath{stroke,fill}%
\end{pgfscope}%
\begin{pgfscope}%
\pgfsetbuttcap%
\pgfsetroundjoin%
\pgfsetlinewidth{0.501875pt}%
\definecolor{currentstroke}{rgb}{0.843137,0.098039,0.109804}%
\pgfsetstrokecolor{currentstroke}%
\pgfsetdash{}{0pt}%
\pgfpathmoveto{\pgfqpoint{2.200449in}{1.625419in}}%
\pgfpathlineto{\pgfqpoint{2.200449in}{1.741113in}}%
\pgfusepath{stroke}%
\end{pgfscope}%
\begin{pgfscope}%
\pgfsetbuttcap%
\pgfsetroundjoin%
\definecolor{currentfill}{rgb}{0.843137,0.098039,0.109804}%
\pgfsetfillcolor{currentfill}%
\pgfsetlinewidth{1.003750pt}%
\definecolor{currentstroke}{rgb}{0.843137,0.098039,0.109804}%
\pgfsetstrokecolor{currentstroke}%
\pgfsetdash{}{0pt}%
\pgfsys@defobject{currentmarker}{\pgfqpoint{-0.069444in}{-0.000000in}}{\pgfqpoint{0.069444in}{0.000000in}}{%
\pgfpathmoveto{\pgfqpoint{0.069444in}{-0.000000in}}%
\pgfpathlineto{\pgfqpoint{-0.069444in}{0.000000in}}%
\pgfusepath{stroke,fill}%
}%
\begin{pgfscope}%
\pgfsys@transformshift{2.200449in}{1.625419in}%
\pgfsys@useobject{currentmarker}{}%
\end{pgfscope}%
\end{pgfscope}%
\begin{pgfscope}%
\pgfsetbuttcap%
\pgfsetroundjoin%
\definecolor{currentfill}{rgb}{0.843137,0.098039,0.109804}%
\pgfsetfillcolor{currentfill}%
\pgfsetlinewidth{1.003750pt}%
\definecolor{currentstroke}{rgb}{0.843137,0.098039,0.109804}%
\pgfsetstrokecolor{currentstroke}%
\pgfsetdash{}{0pt}%
\pgfsys@defobject{currentmarker}{\pgfqpoint{-0.069444in}{-0.000000in}}{\pgfqpoint{0.069444in}{0.000000in}}{%
\pgfpathmoveto{\pgfqpoint{0.069444in}{-0.000000in}}%
\pgfpathlineto{\pgfqpoint{-0.069444in}{0.000000in}}%
\pgfusepath{stroke,fill}%
}%
\begin{pgfscope}%
\pgfsys@transformshift{2.200449in}{1.741113in}%
\pgfsys@useobject{currentmarker}{}%
\end{pgfscope}%
\end{pgfscope}%
\begin{pgfscope}%
\pgfsetbuttcap%
\pgfsetroundjoin%
\definecolor{currentfill}{rgb}{0.843137,0.098039,0.109804}%
\pgfsetfillcolor{currentfill}%
\pgfsetlinewidth{1.003750pt}%
\definecolor{currentstroke}{rgb}{0.000000,0.000000,0.000000}%
\pgfsetstrokecolor{currentstroke}%
\pgfsetdash{}{0pt}%
\pgfsys@defobject{currentmarker}{\pgfqpoint{-0.027778in}{-0.027778in}}{\pgfqpoint{0.027778in}{0.027778in}}{%
\pgfpathmoveto{\pgfqpoint{0.000000in}{-0.027778in}}%
\pgfpathcurveto{\pgfqpoint{0.007367in}{-0.027778in}}{\pgfqpoint{0.014433in}{-0.024851in}}{\pgfqpoint{0.019642in}{-0.019642in}}%
\pgfpathcurveto{\pgfqpoint{0.024851in}{-0.014433in}}{\pgfqpoint{0.027778in}{-0.007367in}}{\pgfqpoint{0.027778in}{0.000000in}}%
\pgfpathcurveto{\pgfqpoint{0.027778in}{0.007367in}}{\pgfqpoint{0.024851in}{0.014433in}}{\pgfqpoint{0.019642in}{0.019642in}}%
\pgfpathcurveto{\pgfqpoint{0.014433in}{0.024851in}}{\pgfqpoint{0.007367in}{0.027778in}}{\pgfqpoint{0.000000in}{0.027778in}}%
\pgfpathcurveto{\pgfqpoint{-0.007367in}{0.027778in}}{\pgfqpoint{-0.014433in}{0.024851in}}{\pgfqpoint{-0.019642in}{0.019642in}}%
\pgfpathcurveto{\pgfqpoint{-0.024851in}{0.014433in}}{\pgfqpoint{-0.027778in}{0.007367in}}{\pgfqpoint{-0.027778in}{0.000000in}}%
\pgfpathcurveto{\pgfqpoint{-0.027778in}{-0.007367in}}{\pgfqpoint{-0.024851in}{-0.014433in}}{\pgfqpoint{-0.019642in}{-0.019642in}}%
\pgfpathcurveto{\pgfqpoint{-0.014433in}{-0.024851in}}{\pgfqpoint{-0.007367in}{-0.027778in}}{\pgfqpoint{0.000000in}{-0.027778in}}%
\pgfpathlineto{\pgfqpoint{0.000000in}{-0.027778in}}%
\pgfpathclose%
\pgfusepath{stroke,fill}%
}%
\begin{pgfscope}%
\pgfsys@transformshift{2.200449in}{1.683266in}%
\pgfsys@useobject{currentmarker}{}%
\end{pgfscope}%
\end{pgfscope}%
\begin{pgfscope}%
\definecolor{textcolor}{rgb}{0.000000,0.000000,0.000000}%
\pgfsetstrokecolor{textcolor}%
\pgfsetfillcolor{textcolor}%
\pgftext[x=2.408699in,y=1.642773in,left,base]{\color{textcolor}{\rmfamily\fontsize{8.330000}{9.996000}\selectfont\catcode`\^=\active\def^{\ifmmode\sp\else\^{}\fi}\catcode`\%=\active\def%{\%}$\bm{v}^\text{PINN}_{3\text{Re}}$}}%
\end{pgfscope}%
\begin{pgfscope}%
\pgfsetbuttcap%
\pgfsetroundjoin%
\pgfsetlinewidth{0.501875pt}%
\definecolor{currentstroke}{rgb}{0.172549,0.482353,0.713725}%
\pgfsetstrokecolor{currentstroke}%
\pgfsetdash{}{0pt}%
\pgfpathmoveto{\pgfqpoint{2.200449in}{1.433683in}}%
\pgfpathlineto{\pgfqpoint{2.200449in}{1.549377in}}%
\pgfusepath{stroke}%
\end{pgfscope}%
\begin{pgfscope}%
\pgfsetbuttcap%
\pgfsetroundjoin%
\definecolor{currentfill}{rgb}{0.172549,0.482353,0.713725}%
\pgfsetfillcolor{currentfill}%
\pgfsetlinewidth{1.003750pt}%
\definecolor{currentstroke}{rgb}{0.172549,0.482353,0.713725}%
\pgfsetstrokecolor{currentstroke}%
\pgfsetdash{}{0pt}%
\pgfsys@defobject{currentmarker}{\pgfqpoint{-0.069444in}{-0.000000in}}{\pgfqpoint{0.069444in}{0.000000in}}{%
\pgfpathmoveto{\pgfqpoint{0.069444in}{-0.000000in}}%
\pgfpathlineto{\pgfqpoint{-0.069444in}{0.000000in}}%
\pgfusepath{stroke,fill}%
}%
\begin{pgfscope}%
\pgfsys@transformshift{2.200449in}{1.433683in}%
\pgfsys@useobject{currentmarker}{}%
\end{pgfscope}%
\end{pgfscope}%
\begin{pgfscope}%
\pgfsetbuttcap%
\pgfsetroundjoin%
\definecolor{currentfill}{rgb}{0.172549,0.482353,0.713725}%
\pgfsetfillcolor{currentfill}%
\pgfsetlinewidth{1.003750pt}%
\definecolor{currentstroke}{rgb}{0.172549,0.482353,0.713725}%
\pgfsetstrokecolor{currentstroke}%
\pgfsetdash{}{0pt}%
\pgfsys@defobject{currentmarker}{\pgfqpoint{-0.069444in}{-0.000000in}}{\pgfqpoint{0.069444in}{0.000000in}}{%
\pgfpathmoveto{\pgfqpoint{0.069444in}{-0.000000in}}%
\pgfpathlineto{\pgfqpoint{-0.069444in}{0.000000in}}%
\pgfusepath{stroke,fill}%
}%
\begin{pgfscope}%
\pgfsys@transformshift{2.200449in}{1.549377in}%
\pgfsys@useobject{currentmarker}{}%
\end{pgfscope}%
\end{pgfscope}%
\begin{pgfscope}%
\pgfsetbuttcap%
\pgfsetmiterjoin%
\definecolor{currentfill}{rgb}{0.172549,0.482353,0.713725}%
\pgfsetfillcolor{currentfill}%
\pgfsetlinewidth{1.003750pt}%
\definecolor{currentstroke}{rgb}{0.000000,0.000000,0.000000}%
\pgfsetstrokecolor{currentstroke}%
\pgfsetdash{}{0pt}%
\pgfsys@defobject{currentmarker}{\pgfqpoint{-0.029463in}{-0.029463in}}{\pgfqpoint{0.029463in}{0.029463in}}{%
\pgfpathmoveto{\pgfqpoint{-0.000000in}{-0.029463in}}%
\pgfpathlineto{\pgfqpoint{0.029463in}{0.000000in}}%
\pgfpathlineto{\pgfqpoint{0.000000in}{0.029463in}}%
\pgfpathlineto{\pgfqpoint{-0.029463in}{0.000000in}}%
\pgfpathlineto{\pgfqpoint{-0.000000in}{-0.029463in}}%
\pgfpathclose%
\pgfusepath{stroke,fill}%
}%
\begin{pgfscope}%
\pgfsys@transformshift{2.200449in}{1.491530in}%
\pgfsys@useobject{currentmarker}{}%
\end{pgfscope}%
\end{pgfscope}%
\begin{pgfscope}%
\definecolor{textcolor}{rgb}{0.000000,0.000000,0.000000}%
\pgfsetstrokecolor{textcolor}%
\pgfsetfillcolor{textcolor}%
\pgftext[x=2.408699in,y=1.451037in,left,base]{\color{textcolor}{\rmfamily\fontsize{8.330000}{9.996000}\selectfont\catcode`\^=\active\def^{\ifmmode\sp\else\^{}\fi}\catcode`\%=\active\def%{\%}$p^\text{PINN}_{3\text{Re}}$}}%
\end{pgfscope}%
\end{pgfpicture}%
\makeatother%
\endgroup%

%% file: figures/section4/2Re/avg_errors_Re.pgf
%% Creator: Matplotlib, PGF backend
%%
%% To include the figure in your LaTeX document, write
%%   \input{<filename>.pgf}
%%
%% Make sure the required packages are loaded in your preamble
%%   \usepackage{pgf}
%%
%% Also ensure that all the required font packages are loaded; for instance,
%% the lmodern package is sometimes necessary when using math font.
%%   \usepackage{lmodern}
%%
%% Figures using additional raster images can only be included by \input if
%% they are in the same directory as the main LaTeX file. For loading figures
%% from other directories you can use the `import` package
%% and then include the figures with
%%   \import{<path to file>}{<filename>.pgf}
%%
%% Matplotlib used the following preamble
%%   \def\mathdefault#1{#1}
%%   \everymath=\expandafter{\the\everymath\displaystyle}
%%   \usepackage{amsmath}\usepackage{bm}
%%   \makeatletter\@ifpackageloaded{underscore}{}{\usepackage[strings]{underscore}}\makeatother
%%
\begingroup%
\makeatletter%
\begin{pgfpicture}%
\pgfpathrectangle{\pgfpointorigin}{\pgfqpoint{3.000000in}{2.000000in}}%
\pgfusepath{use as bounding box, clip}%
\begin{pgfscope}%
\pgfsetbuttcap%
\pgfsetmiterjoin%
\definecolor{currentfill}{rgb}{1.000000,1.000000,1.000000}%
\pgfsetfillcolor{currentfill}%
\pgfsetlinewidth{0.000000pt}%
\definecolor{currentstroke}{rgb}{1.000000,1.000000,1.000000}%
\pgfsetstrokecolor{currentstroke}%
\pgfsetdash{}{0pt}%
\pgfpathmoveto{\pgfqpoint{0.000000in}{0.000000in}}%
\pgfpathlineto{\pgfqpoint{3.000000in}{0.000000in}}%
\pgfpathlineto{\pgfqpoint{3.000000in}{2.000000in}}%
\pgfpathlineto{\pgfqpoint{0.000000in}{2.000000in}}%
\pgfpathlineto{\pgfqpoint{0.000000in}{0.000000in}}%
\pgfpathclose%
\pgfusepath{fill}%
\end{pgfscope}%
\begin{pgfscope}%
\pgfsetbuttcap%
\pgfsetmiterjoin%
\definecolor{currentfill}{rgb}{1.000000,1.000000,1.000000}%
\pgfsetfillcolor{currentfill}%
\pgfsetlinewidth{0.000000pt}%
\definecolor{currentstroke}{rgb}{0.000000,0.000000,0.000000}%
\pgfsetstrokecolor{currentstroke}%
\pgfsetstrokeopacity{0.000000}%
\pgfsetdash{}{0pt}%
\pgfpathmoveto{\pgfqpoint{0.597016in}{0.498776in}}%
\pgfpathlineto{\pgfqpoint{2.845756in}{0.498776in}}%
\pgfpathlineto{\pgfqpoint{2.845756in}{1.850000in}}%
\pgfpathlineto{\pgfqpoint{0.597016in}{1.850000in}}%
\pgfpathlineto{\pgfqpoint{0.597016in}{0.498776in}}%
\pgfpathclose%
\pgfusepath{fill}%
\end{pgfscope}%
\begin{pgfscope}%
\pgfsetbuttcap%
\pgfsetroundjoin%
\definecolor{currentfill}{rgb}{0.000000,0.000000,0.000000}%
\pgfsetfillcolor{currentfill}%
\pgfsetlinewidth{0.803000pt}%
\definecolor{currentstroke}{rgb}{0.000000,0.000000,0.000000}%
\pgfsetstrokecolor{currentstroke}%
\pgfsetdash{}{0pt}%
\pgfsys@defobject{currentmarker}{\pgfqpoint{0.000000in}{-0.048611in}}{\pgfqpoint{0.000000in}{0.000000in}}{%
\pgfpathmoveto{\pgfqpoint{0.000000in}{0.000000in}}%
\pgfpathlineto{\pgfqpoint{0.000000in}{-0.048611in}}%
\pgfusepath{stroke,fill}%
}%
\begin{pgfscope}%
\pgfsys@transformshift{0.699231in}{0.498776in}%
\pgfsys@useobject{currentmarker}{}%
\end{pgfscope}%
\end{pgfscope}%
\begin{pgfscope}%
\definecolor{textcolor}{rgb}{0.000000,0.000000,0.000000}%
\pgfsetstrokecolor{textcolor}%
\pgfsetfillcolor{textcolor}%
\pgftext[x=0.699231in,y=0.408498in,,top]{\color{textcolor}{\rmfamily\fontsize{6.500000}{7.800000}\selectfont\catcode`\^=\active\def^{\ifmmode\sp\else\^{}\fi}\catcode`\%=\active\def%{\%}30}}%
\end{pgfscope}%
\begin{pgfscope}%
\pgfsetbuttcap%
\pgfsetroundjoin%
\definecolor{currentfill}{rgb}{0.000000,0.000000,0.000000}%
\pgfsetfillcolor{currentfill}%
\pgfsetlinewidth{0.803000pt}%
\definecolor{currentstroke}{rgb}{0.000000,0.000000,0.000000}%
\pgfsetstrokecolor{currentstroke}%
\pgfsetdash{}{0pt}%
\pgfsys@defobject{currentmarker}{\pgfqpoint{0.000000in}{-0.048611in}}{\pgfqpoint{0.000000in}{0.000000in}}{%
\pgfpathmoveto{\pgfqpoint{0.000000in}{0.000000in}}%
\pgfpathlineto{\pgfqpoint{0.000000in}{-0.048611in}}%
\pgfusepath{stroke,fill}%
}%
\begin{pgfscope}%
\pgfsys@transformshift{0.926377in}{0.498776in}%
\pgfsys@useobject{currentmarker}{}%
\end{pgfscope}%
\end{pgfscope}%
\begin{pgfscope}%
\definecolor{textcolor}{rgb}{0.000000,0.000000,0.000000}%
\pgfsetstrokecolor{textcolor}%
\pgfsetfillcolor{textcolor}%
\pgftext[x=0.926377in,y=0.408498in,,top]{\color{textcolor}{\rmfamily\fontsize{6.500000}{7.800000}\selectfont\catcode`\^=\active\def^{\ifmmode\sp\else\^{}\fi}\catcode`\%=\active\def%{\%}50}}%
\end{pgfscope}%
\begin{pgfscope}%
\pgfsetbuttcap%
\pgfsetroundjoin%
\definecolor{currentfill}{rgb}{0.000000,0.000000,0.000000}%
\pgfsetfillcolor{currentfill}%
\pgfsetlinewidth{0.803000pt}%
\definecolor{currentstroke}{rgb}{0.000000,0.000000,0.000000}%
\pgfsetstrokecolor{currentstroke}%
\pgfsetdash{}{0pt}%
\pgfsys@defobject{currentmarker}{\pgfqpoint{0.000000in}{-0.048611in}}{\pgfqpoint{0.000000in}{0.000000in}}{%
\pgfpathmoveto{\pgfqpoint{0.000000in}{0.000000in}}%
\pgfpathlineto{\pgfqpoint{0.000000in}{-0.048611in}}%
\pgfusepath{stroke,fill}%
}%
\begin{pgfscope}%
\pgfsys@transformshift{1.153522in}{0.498776in}%
\pgfsys@useobject{currentmarker}{}%
\end{pgfscope}%
\end{pgfscope}%
\begin{pgfscope}%
\definecolor{textcolor}{rgb}{0.000000,0.000000,0.000000}%
\pgfsetstrokecolor{textcolor}%
\pgfsetfillcolor{textcolor}%
\pgftext[x=1.153522in,y=0.408498in,,top]{\color{textcolor}{\rmfamily\fontsize{6.500000}{7.800000}\selectfont\catcode`\^=\active\def^{\ifmmode\sp\else\^{}\fi}\catcode`\%=\active\def%{\%}100}}%
\end{pgfscope}%
\begin{pgfscope}%
\pgfsetbuttcap%
\pgfsetroundjoin%
\definecolor{currentfill}{rgb}{0.000000,0.000000,0.000000}%
\pgfsetfillcolor{currentfill}%
\pgfsetlinewidth{0.803000pt}%
\definecolor{currentstroke}{rgb}{0.000000,0.000000,0.000000}%
\pgfsetstrokecolor{currentstroke}%
\pgfsetdash{}{0pt}%
\pgfsys@defobject{currentmarker}{\pgfqpoint{0.000000in}{-0.048611in}}{\pgfqpoint{0.000000in}{0.000000in}}{%
\pgfpathmoveto{\pgfqpoint{0.000000in}{0.000000in}}%
\pgfpathlineto{\pgfqpoint{0.000000in}{-0.048611in}}%
\pgfusepath{stroke,fill}%
}%
\begin{pgfscope}%
\pgfsys@transformshift{1.380668in}{0.498776in}%
\pgfsys@useobject{currentmarker}{}%
\end{pgfscope}%
\end{pgfscope}%
\begin{pgfscope}%
\definecolor{textcolor}{rgb}{0.000000,0.000000,0.000000}%
\pgfsetstrokecolor{textcolor}%
\pgfsetfillcolor{textcolor}%
\pgftext[x=1.380668in,y=0.408498in,,top]{\color{textcolor}{\rmfamily\fontsize{6.500000}{7.800000}\selectfont\catcode`\^=\active\def^{\ifmmode\sp\else\^{}\fi}\catcode`\%=\active\def%{\%}500}}%
\end{pgfscope}%
\begin{pgfscope}%
\pgfsetbuttcap%
\pgfsetroundjoin%
\definecolor{currentfill}{rgb}{0.000000,0.000000,0.000000}%
\pgfsetfillcolor{currentfill}%
\pgfsetlinewidth{0.803000pt}%
\definecolor{currentstroke}{rgb}{0.000000,0.000000,0.000000}%
\pgfsetstrokecolor{currentstroke}%
\pgfsetdash{}{0pt}%
\pgfsys@defobject{currentmarker}{\pgfqpoint{0.000000in}{-0.048611in}}{\pgfqpoint{0.000000in}{0.000000in}}{%
\pgfpathmoveto{\pgfqpoint{0.000000in}{0.000000in}}%
\pgfpathlineto{\pgfqpoint{0.000000in}{-0.048611in}}%
\pgfusepath{stroke,fill}%
}%
\begin{pgfscope}%
\pgfsys@transformshift{1.607813in}{0.498776in}%
\pgfsys@useobject{currentmarker}{}%
\end{pgfscope}%
\end{pgfscope}%
\begin{pgfscope}%
\definecolor{textcolor}{rgb}{0.000000,0.000000,0.000000}%
\pgfsetstrokecolor{textcolor}%
\pgfsetfillcolor{textcolor}%
\pgftext[x=1.607813in,y=0.408498in,,top]{\color{textcolor}{\rmfamily\fontsize{6.500000}{7.800000}\selectfont\catcode`\^=\active\def^{\ifmmode\sp\else\^{}\fi}\catcode`\%=\active\def%{\%}1000}}%
\end{pgfscope}%
\begin{pgfscope}%
\pgfsetbuttcap%
\pgfsetroundjoin%
\definecolor{currentfill}{rgb}{0.000000,0.000000,0.000000}%
\pgfsetfillcolor{currentfill}%
\pgfsetlinewidth{0.803000pt}%
\definecolor{currentstroke}{rgb}{0.000000,0.000000,0.000000}%
\pgfsetstrokecolor{currentstroke}%
\pgfsetdash{}{0pt}%
\pgfsys@defobject{currentmarker}{\pgfqpoint{0.000000in}{-0.048611in}}{\pgfqpoint{0.000000in}{0.000000in}}{%
\pgfpathmoveto{\pgfqpoint{0.000000in}{0.000000in}}%
\pgfpathlineto{\pgfqpoint{0.000000in}{-0.048611in}}%
\pgfusepath{stroke,fill}%
}%
\begin{pgfscope}%
\pgfsys@transformshift{1.834959in}{0.498776in}%
\pgfsys@useobject{currentmarker}{}%
\end{pgfscope}%
\end{pgfscope}%
\begin{pgfscope}%
\definecolor{textcolor}{rgb}{0.000000,0.000000,0.000000}%
\pgfsetstrokecolor{textcolor}%
\pgfsetfillcolor{textcolor}%
\pgftext[x=1.834959in,y=0.408498in,,top]{\color{textcolor}{\rmfamily\fontsize{6.500000}{7.800000}\selectfont\catcode`\^=\active\def^{\ifmmode\sp\else\^{}\fi}\catcode`\%=\active\def%{\%}2000}}%
\end{pgfscope}%
\begin{pgfscope}%
\pgfsetbuttcap%
\pgfsetroundjoin%
\definecolor{currentfill}{rgb}{0.000000,0.000000,0.000000}%
\pgfsetfillcolor{currentfill}%
\pgfsetlinewidth{0.803000pt}%
\definecolor{currentstroke}{rgb}{0.000000,0.000000,0.000000}%
\pgfsetstrokecolor{currentstroke}%
\pgfsetdash{}{0pt}%
\pgfsys@defobject{currentmarker}{\pgfqpoint{0.000000in}{-0.048611in}}{\pgfqpoint{0.000000in}{0.000000in}}{%
\pgfpathmoveto{\pgfqpoint{0.000000in}{0.000000in}}%
\pgfpathlineto{\pgfqpoint{0.000000in}{-0.048611in}}%
\pgfusepath{stroke,fill}%
}%
\begin{pgfscope}%
\pgfsys@transformshift{2.062104in}{0.498776in}%
\pgfsys@useobject{currentmarker}{}%
\end{pgfscope}%
\end{pgfscope}%
\begin{pgfscope}%
\definecolor{textcolor}{rgb}{0.000000,0.000000,0.000000}%
\pgfsetstrokecolor{textcolor}%
\pgfsetfillcolor{textcolor}%
\pgftext[x=2.062104in,y=0.408498in,,top]{\color{textcolor}{\rmfamily\fontsize{6.500000}{7.800000}\selectfont\catcode`\^=\active\def^{\ifmmode\sp\else\^{}\fi}\catcode`\%=\active\def%{\%}3000}}%
\end{pgfscope}%
\begin{pgfscope}%
\pgfsetbuttcap%
\pgfsetroundjoin%
\definecolor{currentfill}{rgb}{0.000000,0.000000,0.000000}%
\pgfsetfillcolor{currentfill}%
\pgfsetlinewidth{0.803000pt}%
\definecolor{currentstroke}{rgb}{0.000000,0.000000,0.000000}%
\pgfsetstrokecolor{currentstroke}%
\pgfsetdash{}{0pt}%
\pgfsys@defobject{currentmarker}{\pgfqpoint{0.000000in}{-0.048611in}}{\pgfqpoint{0.000000in}{0.000000in}}{%
\pgfpathmoveto{\pgfqpoint{0.000000in}{0.000000in}}%
\pgfpathlineto{\pgfqpoint{0.000000in}{-0.048611in}}%
\pgfusepath{stroke,fill}%
}%
\begin{pgfscope}%
\pgfsys@transformshift{2.289250in}{0.498776in}%
\pgfsys@useobject{currentmarker}{}%
\end{pgfscope}%
\end{pgfscope}%
\begin{pgfscope}%
\definecolor{textcolor}{rgb}{0.000000,0.000000,0.000000}%
\pgfsetstrokecolor{textcolor}%
\pgfsetfillcolor{textcolor}%
\pgftext[x=2.289250in,y=0.408498in,,top]{\color{textcolor}{\rmfamily\fontsize{6.500000}{7.800000}\selectfont\catcode`\^=\active\def^{\ifmmode\sp\else\^{}\fi}\catcode`\%=\active\def%{\%}4000}}%
\end{pgfscope}%
\begin{pgfscope}%
\pgfsetbuttcap%
\pgfsetroundjoin%
\definecolor{currentfill}{rgb}{0.000000,0.000000,0.000000}%
\pgfsetfillcolor{currentfill}%
\pgfsetlinewidth{0.803000pt}%
\definecolor{currentstroke}{rgb}{0.000000,0.000000,0.000000}%
\pgfsetstrokecolor{currentstroke}%
\pgfsetdash{}{0pt}%
\pgfsys@defobject{currentmarker}{\pgfqpoint{0.000000in}{-0.048611in}}{\pgfqpoint{0.000000in}{0.000000in}}{%
\pgfpathmoveto{\pgfqpoint{0.000000in}{0.000000in}}%
\pgfpathlineto{\pgfqpoint{0.000000in}{-0.048611in}}%
\pgfusepath{stroke,fill}%
}%
\begin{pgfscope}%
\pgfsys@transformshift{2.516395in}{0.498776in}%
\pgfsys@useobject{currentmarker}{}%
\end{pgfscope}%
\end{pgfscope}%
\begin{pgfscope}%
\definecolor{textcolor}{rgb}{0.000000,0.000000,0.000000}%
\pgfsetstrokecolor{textcolor}%
\pgfsetfillcolor{textcolor}%
\pgftext[x=2.516395in,y=0.408498in,,top]{\color{textcolor}{\rmfamily\fontsize{6.500000}{7.800000}\selectfont\catcode`\^=\active\def^{\ifmmode\sp\else\^{}\fi}\catcode`\%=\active\def%{\%}5000}}%
\end{pgfscope}%
\begin{pgfscope}%
\pgfsetbuttcap%
\pgfsetroundjoin%
\definecolor{currentfill}{rgb}{0.000000,0.000000,0.000000}%
\pgfsetfillcolor{currentfill}%
\pgfsetlinewidth{0.803000pt}%
\definecolor{currentstroke}{rgb}{0.000000,0.000000,0.000000}%
\pgfsetstrokecolor{currentstroke}%
\pgfsetdash{}{0pt}%
\pgfsys@defobject{currentmarker}{\pgfqpoint{0.000000in}{-0.048611in}}{\pgfqpoint{0.000000in}{0.000000in}}{%
\pgfpathmoveto{\pgfqpoint{0.000000in}{0.000000in}}%
\pgfpathlineto{\pgfqpoint{0.000000in}{-0.048611in}}%
\pgfusepath{stroke,fill}%
}%
\begin{pgfscope}%
\pgfsys@transformshift{2.743541in}{0.498776in}%
\pgfsys@useobject{currentmarker}{}%
\end{pgfscope}%
\end{pgfscope}%
\begin{pgfscope}%
\definecolor{textcolor}{rgb}{0.000000,0.000000,0.000000}%
\pgfsetstrokecolor{textcolor}%
\pgfsetfillcolor{textcolor}%
\pgftext[x=2.743541in,y=0.408498in,,top]{\color{textcolor}{\rmfamily\fontsize{6.500000}{7.800000}\selectfont\catcode`\^=\active\def^{\ifmmode\sp\else\^{}\fi}\catcode`\%=\active\def%{\%}6000}}%
\end{pgfscope}%
\begin{pgfscope}%
\definecolor{textcolor}{rgb}{0.000000,0.000000,0.000000}%
\pgfsetstrokecolor{textcolor}%
\pgfsetfillcolor{textcolor}%
\pgftext[x=1.721386in,y=0.278868in,,top]{\color{textcolor}{\rmfamily\fontsize{10.000000}{12.000000}\selectfont\catcode`\^=\active\def^{\ifmmode\sp\else\^{}\fi}\catcode`\%=\active\def%{\%}$\text{Re}$}}%
\end{pgfscope}%
\begin{pgfscope}%
\pgfsetbuttcap%
\pgfsetroundjoin%
\definecolor{currentfill}{rgb}{0.000000,0.000000,0.000000}%
\pgfsetfillcolor{currentfill}%
\pgfsetlinewidth{0.803000pt}%
\definecolor{currentstroke}{rgb}{0.000000,0.000000,0.000000}%
\pgfsetstrokecolor{currentstroke}%
\pgfsetdash{}{0pt}%
\pgfsys@defobject{currentmarker}{\pgfqpoint{-0.048611in}{0.000000in}}{\pgfqpoint{-0.000000in}{0.000000in}}{%
\pgfpathmoveto{\pgfqpoint{-0.000000in}{0.000000in}}%
\pgfpathlineto{\pgfqpoint{-0.048611in}{0.000000in}}%
\pgfusepath{stroke,fill}%
}%
\begin{pgfscope}%
\pgfsys@transformshift{0.597016in}{0.697601in}%
\pgfsys@useobject{currentmarker}{}%
\end{pgfscope}%
\end{pgfscope}%
\begin{pgfscope}%
\definecolor{textcolor}{rgb}{0.000000,0.000000,0.000000}%
\pgfsetstrokecolor{textcolor}%
\pgfsetfillcolor{textcolor}%
\pgftext[x=0.455813in, y=0.668666in, left, base]{\color{textcolor}{\rmfamily\fontsize{6.500000}{7.800000}\selectfont\catcode`\^=\active\def^{\ifmmode\sp\else\^{}\fi}\catcode`\%=\active\def%{\%}$\mathdefault{5}$}}%
\end{pgfscope}%
\begin{pgfscope}%
\pgfsetbuttcap%
\pgfsetroundjoin%
\definecolor{currentfill}{rgb}{0.000000,0.000000,0.000000}%
\pgfsetfillcolor{currentfill}%
\pgfsetlinewidth{0.803000pt}%
\definecolor{currentstroke}{rgb}{0.000000,0.000000,0.000000}%
\pgfsetstrokecolor{currentstroke}%
\pgfsetdash{}{0pt}%
\pgfsys@defobject{currentmarker}{\pgfqpoint{-0.048611in}{0.000000in}}{\pgfqpoint{-0.000000in}{0.000000in}}{%
\pgfpathmoveto{\pgfqpoint{-0.000000in}{0.000000in}}%
\pgfpathlineto{\pgfqpoint{-0.048611in}{0.000000in}}%
\pgfusepath{stroke,fill}%
}%
\begin{pgfscope}%
\pgfsys@transformshift{0.597016in}{1.043143in}%
\pgfsys@useobject{currentmarker}{}%
\end{pgfscope}%
\end{pgfscope}%
\begin{pgfscope}%
\definecolor{textcolor}{rgb}{0.000000,0.000000,0.000000}%
\pgfsetstrokecolor{textcolor}%
\pgfsetfillcolor{textcolor}%
\pgftext[x=0.404888in, y=1.014207in, left, base]{\color{textcolor}{\rmfamily\fontsize{6.500000}{7.800000}\selectfont\catcode`\^=\active\def^{\ifmmode\sp\else\^{}\fi}\catcode`\%=\active\def%{\%}$\mathdefault{10}$}}%
\end{pgfscope}%
\begin{pgfscope}%
\pgfsetbuttcap%
\pgfsetroundjoin%
\definecolor{currentfill}{rgb}{0.000000,0.000000,0.000000}%
\pgfsetfillcolor{currentfill}%
\pgfsetlinewidth{0.803000pt}%
\definecolor{currentstroke}{rgb}{0.000000,0.000000,0.000000}%
\pgfsetstrokecolor{currentstroke}%
\pgfsetdash{}{0pt}%
\pgfsys@defobject{currentmarker}{\pgfqpoint{-0.048611in}{0.000000in}}{\pgfqpoint{-0.000000in}{0.000000in}}{%
\pgfpathmoveto{\pgfqpoint{-0.000000in}{0.000000in}}%
\pgfpathlineto{\pgfqpoint{-0.048611in}{0.000000in}}%
\pgfusepath{stroke,fill}%
}%
\begin{pgfscope}%
\pgfsys@transformshift{0.597016in}{1.388684in}%
\pgfsys@useobject{currentmarker}{}%
\end{pgfscope}%
\end{pgfscope}%
\begin{pgfscope}%
\definecolor{textcolor}{rgb}{0.000000,0.000000,0.000000}%
\pgfsetstrokecolor{textcolor}%
\pgfsetfillcolor{textcolor}%
\pgftext[x=0.404888in, y=1.359749in, left, base]{\color{textcolor}{\rmfamily\fontsize{6.500000}{7.800000}\selectfont\catcode`\^=\active\def^{\ifmmode\sp\else\^{}\fi}\catcode`\%=\active\def%{\%}$\mathdefault{15}$}}%
\end{pgfscope}%
\begin{pgfscope}%
\pgfsetbuttcap%
\pgfsetroundjoin%
\definecolor{currentfill}{rgb}{0.000000,0.000000,0.000000}%
\pgfsetfillcolor{currentfill}%
\pgfsetlinewidth{0.803000pt}%
\definecolor{currentstroke}{rgb}{0.000000,0.000000,0.000000}%
\pgfsetstrokecolor{currentstroke}%
\pgfsetdash{}{0pt}%
\pgfsys@defobject{currentmarker}{\pgfqpoint{-0.048611in}{0.000000in}}{\pgfqpoint{-0.000000in}{0.000000in}}{%
\pgfpathmoveto{\pgfqpoint{-0.000000in}{0.000000in}}%
\pgfpathlineto{\pgfqpoint{-0.048611in}{0.000000in}}%
\pgfusepath{stroke,fill}%
}%
\begin{pgfscope}%
\pgfsys@transformshift{0.597016in}{1.734225in}%
\pgfsys@useobject{currentmarker}{}%
\end{pgfscope}%
\end{pgfscope}%
\begin{pgfscope}%
\definecolor{textcolor}{rgb}{0.000000,0.000000,0.000000}%
\pgfsetstrokecolor{textcolor}%
\pgfsetfillcolor{textcolor}%
\pgftext[x=0.404888in, y=1.705290in, left, base]{\color{textcolor}{\rmfamily\fontsize{6.500000}{7.800000}\selectfont\catcode`\^=\active\def^{\ifmmode\sp\else\^{}\fi}\catcode`\%=\active\def%{\%}$\mathdefault{20}$}}%
\end{pgfscope}%
\begin{pgfscope}%
\definecolor{textcolor}{rgb}{0.000000,0.000000,0.000000}%
\pgfsetstrokecolor{textcolor}%
\pgfsetfillcolor{textcolor}%
\pgftext[x=0.349332in,y=1.174388in,,bottom,rotate=90.000000]{\color{textcolor}{\rmfamily\fontsize{10.000000}{12.000000}\selectfont\catcode`\^=\active\def^{\ifmmode\sp\else\^{}\fi}\catcode`\%=\active\def%{\%}$\delta_{\ell^1}^{(q)} \ [\%]$}}%
\end{pgfscope}%
\begin{pgfscope}%
\pgfpathrectangle{\pgfqpoint{0.597016in}{0.498776in}}{\pgfqpoint{2.248740in}{1.351224in}}%
\pgfusepath{clip}%
\pgfsetbuttcap%
\pgfsetroundjoin%
\pgfsetlinewidth{0.501875pt}%
\definecolor{currentstroke}{rgb}{0.843137,0.098039,0.109804}%
\pgfsetstrokecolor{currentstroke}%
\pgfsetdash{}{0pt}%
\pgfpathmoveto{\pgfqpoint{0.699231in}{1.200725in}}%
\pgfpathlineto{\pgfqpoint{0.699231in}{1.299700in}}%
\pgfusepath{stroke}%
\end{pgfscope}%
\begin{pgfscope}%
\pgfpathrectangle{\pgfqpoint{0.597016in}{0.498776in}}{\pgfqpoint{2.248740in}{1.351224in}}%
\pgfusepath{clip}%
\pgfsetbuttcap%
\pgfsetroundjoin%
\pgfsetlinewidth{0.501875pt}%
\definecolor{currentstroke}{rgb}{0.843137,0.098039,0.109804}%
\pgfsetstrokecolor{currentstroke}%
\pgfsetdash{}{0pt}%
\pgfpathmoveto{\pgfqpoint{0.926377in}{1.112692in}}%
\pgfpathlineto{\pgfqpoint{0.926377in}{1.209115in}}%
\pgfusepath{stroke}%
\end{pgfscope}%
\begin{pgfscope}%
\pgfpathrectangle{\pgfqpoint{0.597016in}{0.498776in}}{\pgfqpoint{2.248740in}{1.351224in}}%
\pgfusepath{clip}%
\pgfsetbuttcap%
\pgfsetroundjoin%
\pgfsetlinewidth{0.501875pt}%
\definecolor{currentstroke}{rgb}{0.843137,0.098039,0.109804}%
\pgfsetstrokecolor{currentstroke}%
\pgfsetdash{}{0pt}%
\pgfpathmoveto{\pgfqpoint{1.153522in}{0.636076in}}%
\pgfpathlineto{\pgfqpoint{1.153522in}{0.735260in}}%
\pgfusepath{stroke}%
\end{pgfscope}%
\begin{pgfscope}%
\pgfpathrectangle{\pgfqpoint{0.597016in}{0.498776in}}{\pgfqpoint{2.248740in}{1.351224in}}%
\pgfusepath{clip}%
\pgfsetbuttcap%
\pgfsetroundjoin%
\pgfsetlinewidth{0.501875pt}%
\definecolor{currentstroke}{rgb}{0.843137,0.098039,0.109804}%
\pgfsetstrokecolor{currentstroke}%
\pgfsetdash{}{0pt}%
\pgfpathmoveto{\pgfqpoint{1.380668in}{0.560195in}}%
\pgfpathlineto{\pgfqpoint{1.380668in}{0.636870in}}%
\pgfusepath{stroke}%
\end{pgfscope}%
\begin{pgfscope}%
\pgfpathrectangle{\pgfqpoint{0.597016in}{0.498776in}}{\pgfqpoint{2.248740in}{1.351224in}}%
\pgfusepath{clip}%
\pgfsetbuttcap%
\pgfsetroundjoin%
\pgfsetlinewidth{0.501875pt}%
\definecolor{currentstroke}{rgb}{0.843137,0.098039,0.109804}%
\pgfsetstrokecolor{currentstroke}%
\pgfsetdash{}{0pt}%
\pgfpathmoveto{\pgfqpoint{1.607813in}{0.669540in}}%
\pgfpathlineto{\pgfqpoint{1.607813in}{0.765300in}}%
\pgfusepath{stroke}%
\end{pgfscope}%
\begin{pgfscope}%
\pgfpathrectangle{\pgfqpoint{0.597016in}{0.498776in}}{\pgfqpoint{2.248740in}{1.351224in}}%
\pgfusepath{clip}%
\pgfsetbuttcap%
\pgfsetroundjoin%
\pgfsetlinewidth{0.501875pt}%
\definecolor{currentstroke}{rgb}{0.843137,0.098039,0.109804}%
\pgfsetstrokecolor{currentstroke}%
\pgfsetdash{}{0pt}%
\pgfpathmoveto{\pgfqpoint{1.834959in}{0.699091in}}%
\pgfpathlineto{\pgfqpoint{1.834959in}{0.785086in}}%
\pgfusepath{stroke}%
\end{pgfscope}%
\begin{pgfscope}%
\pgfpathrectangle{\pgfqpoint{0.597016in}{0.498776in}}{\pgfqpoint{2.248740in}{1.351224in}}%
\pgfusepath{clip}%
\pgfsetbuttcap%
\pgfsetroundjoin%
\pgfsetlinewidth{0.501875pt}%
\definecolor{currentstroke}{rgb}{0.843137,0.098039,0.109804}%
\pgfsetstrokecolor{currentstroke}%
\pgfsetdash{}{0pt}%
\pgfpathmoveto{\pgfqpoint{2.062104in}{0.687995in}}%
\pgfpathlineto{\pgfqpoint{2.062104in}{0.758946in}}%
\pgfusepath{stroke}%
\end{pgfscope}%
\begin{pgfscope}%
\pgfpathrectangle{\pgfqpoint{0.597016in}{0.498776in}}{\pgfqpoint{2.248740in}{1.351224in}}%
\pgfusepath{clip}%
\pgfsetbuttcap%
\pgfsetroundjoin%
\pgfsetlinewidth{0.501875pt}%
\definecolor{currentstroke}{rgb}{0.843137,0.098039,0.109804}%
\pgfsetstrokecolor{currentstroke}%
\pgfsetdash{}{0pt}%
\pgfpathmoveto{\pgfqpoint{2.289250in}{0.678943in}}%
\pgfpathlineto{\pgfqpoint{2.289250in}{0.731134in}}%
\pgfusepath{stroke}%
\end{pgfscope}%
\begin{pgfscope}%
\pgfpathrectangle{\pgfqpoint{0.597016in}{0.498776in}}{\pgfqpoint{2.248740in}{1.351224in}}%
\pgfusepath{clip}%
\pgfsetbuttcap%
\pgfsetroundjoin%
\pgfsetlinewidth{0.501875pt}%
\definecolor{currentstroke}{rgb}{0.843137,0.098039,0.109804}%
\pgfsetstrokecolor{currentstroke}%
\pgfsetdash{}{0pt}%
\pgfpathmoveto{\pgfqpoint{2.516395in}{0.671662in}}%
\pgfpathlineto{\pgfqpoint{2.516395in}{0.706354in}}%
\pgfusepath{stroke}%
\end{pgfscope}%
\begin{pgfscope}%
\pgfpathrectangle{\pgfqpoint{0.597016in}{0.498776in}}{\pgfqpoint{2.248740in}{1.351224in}}%
\pgfusepath{clip}%
\pgfsetbuttcap%
\pgfsetroundjoin%
\pgfsetlinewidth{0.501875pt}%
\definecolor{currentstroke}{rgb}{0.843137,0.098039,0.109804}%
\pgfsetstrokecolor{currentstroke}%
\pgfsetdash{}{0pt}%
\pgfpathmoveto{\pgfqpoint{2.743541in}{0.729419in}}%
\pgfpathlineto{\pgfqpoint{2.743541in}{0.757447in}}%
\pgfusepath{stroke}%
\end{pgfscope}%
\begin{pgfscope}%
\pgfpathrectangle{\pgfqpoint{0.597016in}{0.498776in}}{\pgfqpoint{2.248740in}{1.351224in}}%
\pgfusepath{clip}%
\pgfsetbuttcap%
\pgfsetroundjoin%
\definecolor{currentfill}{rgb}{0.843137,0.098039,0.109804}%
\pgfsetfillcolor{currentfill}%
\pgfsetlinewidth{1.003750pt}%
\definecolor{currentstroke}{rgb}{0.843137,0.098039,0.109804}%
\pgfsetstrokecolor{currentstroke}%
\pgfsetdash{}{0pt}%
\pgfsys@defobject{currentmarker}{\pgfqpoint{-0.069444in}{-0.000000in}}{\pgfqpoint{0.069444in}{0.000000in}}{%
\pgfpathmoveto{\pgfqpoint{0.069444in}{-0.000000in}}%
\pgfpathlineto{\pgfqpoint{-0.069444in}{0.000000in}}%
\pgfusepath{stroke,fill}%
}%
\begin{pgfscope}%
\pgfsys@transformshift{0.699231in}{1.200725in}%
\pgfsys@useobject{currentmarker}{}%
\end{pgfscope}%
\begin{pgfscope}%
\pgfsys@transformshift{0.926377in}{1.112692in}%
\pgfsys@useobject{currentmarker}{}%
\end{pgfscope}%
\begin{pgfscope}%
\pgfsys@transformshift{1.153522in}{0.636076in}%
\pgfsys@useobject{currentmarker}{}%
\end{pgfscope}%
\begin{pgfscope}%
\pgfsys@transformshift{1.380668in}{0.560195in}%
\pgfsys@useobject{currentmarker}{}%
\end{pgfscope}%
\begin{pgfscope}%
\pgfsys@transformshift{1.607813in}{0.669540in}%
\pgfsys@useobject{currentmarker}{}%
\end{pgfscope}%
\begin{pgfscope}%
\pgfsys@transformshift{1.834959in}{0.699091in}%
\pgfsys@useobject{currentmarker}{}%
\end{pgfscope}%
\begin{pgfscope}%
\pgfsys@transformshift{2.062104in}{0.687995in}%
\pgfsys@useobject{currentmarker}{}%
\end{pgfscope}%
\begin{pgfscope}%
\pgfsys@transformshift{2.289250in}{0.678943in}%
\pgfsys@useobject{currentmarker}{}%
\end{pgfscope}%
\begin{pgfscope}%
\pgfsys@transformshift{2.516395in}{0.671662in}%
\pgfsys@useobject{currentmarker}{}%
\end{pgfscope}%
\begin{pgfscope}%
\pgfsys@transformshift{2.743541in}{0.729419in}%
\pgfsys@useobject{currentmarker}{}%
\end{pgfscope}%
\end{pgfscope}%
\begin{pgfscope}%
\pgfpathrectangle{\pgfqpoint{0.597016in}{0.498776in}}{\pgfqpoint{2.248740in}{1.351224in}}%
\pgfusepath{clip}%
\pgfsetbuttcap%
\pgfsetroundjoin%
\definecolor{currentfill}{rgb}{0.843137,0.098039,0.109804}%
\pgfsetfillcolor{currentfill}%
\pgfsetlinewidth{1.003750pt}%
\definecolor{currentstroke}{rgb}{0.843137,0.098039,0.109804}%
\pgfsetstrokecolor{currentstroke}%
\pgfsetdash{}{0pt}%
\pgfsys@defobject{currentmarker}{\pgfqpoint{-0.069444in}{-0.000000in}}{\pgfqpoint{0.069444in}{0.000000in}}{%
\pgfpathmoveto{\pgfqpoint{0.069444in}{-0.000000in}}%
\pgfpathlineto{\pgfqpoint{-0.069444in}{0.000000in}}%
\pgfusepath{stroke,fill}%
}%
\begin{pgfscope}%
\pgfsys@transformshift{0.699231in}{1.299700in}%
\pgfsys@useobject{currentmarker}{}%
\end{pgfscope}%
\begin{pgfscope}%
\pgfsys@transformshift{0.926377in}{1.209115in}%
\pgfsys@useobject{currentmarker}{}%
\end{pgfscope}%
\begin{pgfscope}%
\pgfsys@transformshift{1.153522in}{0.735260in}%
\pgfsys@useobject{currentmarker}{}%
\end{pgfscope}%
\begin{pgfscope}%
\pgfsys@transformshift{1.380668in}{0.636870in}%
\pgfsys@useobject{currentmarker}{}%
\end{pgfscope}%
\begin{pgfscope}%
\pgfsys@transformshift{1.607813in}{0.765300in}%
\pgfsys@useobject{currentmarker}{}%
\end{pgfscope}%
\begin{pgfscope}%
\pgfsys@transformshift{1.834959in}{0.785086in}%
\pgfsys@useobject{currentmarker}{}%
\end{pgfscope}%
\begin{pgfscope}%
\pgfsys@transformshift{2.062104in}{0.758946in}%
\pgfsys@useobject{currentmarker}{}%
\end{pgfscope}%
\begin{pgfscope}%
\pgfsys@transformshift{2.289250in}{0.731134in}%
\pgfsys@useobject{currentmarker}{}%
\end{pgfscope}%
\begin{pgfscope}%
\pgfsys@transformshift{2.516395in}{0.706354in}%
\pgfsys@useobject{currentmarker}{}%
\end{pgfscope}%
\begin{pgfscope}%
\pgfsys@transformshift{2.743541in}{0.757447in}%
\pgfsys@useobject{currentmarker}{}%
\end{pgfscope}%
\end{pgfscope}%
\begin{pgfscope}%
\pgfpathrectangle{\pgfqpoint{0.597016in}{0.498776in}}{\pgfqpoint{2.248740in}{1.351224in}}%
\pgfusepath{clip}%
\pgfsetbuttcap%
\pgfsetroundjoin%
\pgfsetlinewidth{0.501875pt}%
\definecolor{currentstroke}{rgb}{0.172549,0.482353,0.713725}%
\pgfsetstrokecolor{currentstroke}%
\pgfsetdash{}{0pt}%
\pgfpathmoveto{\pgfqpoint{0.699231in}{1.718687in}}%
\pgfpathlineto{\pgfqpoint{0.699231in}{1.788581in}}%
\pgfusepath{stroke}%
\end{pgfscope}%
\begin{pgfscope}%
\pgfpathrectangle{\pgfqpoint{0.597016in}{0.498776in}}{\pgfqpoint{2.248740in}{1.351224in}}%
\pgfusepath{clip}%
\pgfsetbuttcap%
\pgfsetroundjoin%
\pgfsetlinewidth{0.501875pt}%
\definecolor{currentstroke}{rgb}{0.172549,0.482353,0.713725}%
\pgfsetstrokecolor{currentstroke}%
\pgfsetdash{}{0pt}%
\pgfpathmoveto{\pgfqpoint{0.926377in}{1.479968in}}%
\pgfpathlineto{\pgfqpoint{0.926377in}{1.542696in}}%
\pgfusepath{stroke}%
\end{pgfscope}%
\begin{pgfscope}%
\pgfpathrectangle{\pgfqpoint{0.597016in}{0.498776in}}{\pgfqpoint{2.248740in}{1.351224in}}%
\pgfusepath{clip}%
\pgfsetbuttcap%
\pgfsetroundjoin%
\pgfsetlinewidth{0.501875pt}%
\definecolor{currentstroke}{rgb}{0.172549,0.482353,0.713725}%
\pgfsetstrokecolor{currentstroke}%
\pgfsetdash{}{0pt}%
\pgfpathmoveto{\pgfqpoint{1.153522in}{0.652833in}}%
\pgfpathlineto{\pgfqpoint{1.153522in}{0.786710in}}%
\pgfusepath{stroke}%
\end{pgfscope}%
\begin{pgfscope}%
\pgfpathrectangle{\pgfqpoint{0.597016in}{0.498776in}}{\pgfqpoint{2.248740in}{1.351224in}}%
\pgfusepath{clip}%
\pgfsetbuttcap%
\pgfsetroundjoin%
\pgfsetlinewidth{0.501875pt}%
\definecolor{currentstroke}{rgb}{0.172549,0.482353,0.713725}%
\pgfsetstrokecolor{currentstroke}%
\pgfsetdash{}{0pt}%
\pgfpathmoveto{\pgfqpoint{1.380668in}{0.570032in}}%
\pgfpathlineto{\pgfqpoint{1.380668in}{0.689852in}}%
\pgfusepath{stroke}%
\end{pgfscope}%
\begin{pgfscope}%
\pgfpathrectangle{\pgfqpoint{0.597016in}{0.498776in}}{\pgfqpoint{2.248740in}{1.351224in}}%
\pgfusepath{clip}%
\pgfsetbuttcap%
\pgfsetroundjoin%
\pgfsetlinewidth{0.501875pt}%
\definecolor{currentstroke}{rgb}{0.172549,0.482353,0.713725}%
\pgfsetstrokecolor{currentstroke}%
\pgfsetdash{}{0pt}%
\pgfpathmoveto{\pgfqpoint{1.607813in}{0.654901in}}%
\pgfpathlineto{\pgfqpoint{1.607813in}{0.757784in}}%
\pgfusepath{stroke}%
\end{pgfscope}%
\begin{pgfscope}%
\pgfpathrectangle{\pgfqpoint{0.597016in}{0.498776in}}{\pgfqpoint{2.248740in}{1.351224in}}%
\pgfusepath{clip}%
\pgfsetbuttcap%
\pgfsetroundjoin%
\pgfsetlinewidth{0.501875pt}%
\definecolor{currentstroke}{rgb}{0.172549,0.482353,0.713725}%
\pgfsetstrokecolor{currentstroke}%
\pgfsetdash{}{0pt}%
\pgfpathmoveto{\pgfqpoint{1.834959in}{0.650734in}}%
\pgfpathlineto{\pgfqpoint{1.834959in}{0.717624in}}%
\pgfusepath{stroke}%
\end{pgfscope}%
\begin{pgfscope}%
\pgfpathrectangle{\pgfqpoint{0.597016in}{0.498776in}}{\pgfqpoint{2.248740in}{1.351224in}}%
\pgfusepath{clip}%
\pgfsetbuttcap%
\pgfsetroundjoin%
\pgfsetlinewidth{0.501875pt}%
\definecolor{currentstroke}{rgb}{0.172549,0.482353,0.713725}%
\pgfsetstrokecolor{currentstroke}%
\pgfsetdash{}{0pt}%
\pgfpathmoveto{\pgfqpoint{2.062104in}{0.628443in}}%
\pgfpathlineto{\pgfqpoint{2.062104in}{0.667368in}}%
\pgfusepath{stroke}%
\end{pgfscope}%
\begin{pgfscope}%
\pgfpathrectangle{\pgfqpoint{0.597016in}{0.498776in}}{\pgfqpoint{2.248740in}{1.351224in}}%
\pgfusepath{clip}%
\pgfsetbuttcap%
\pgfsetroundjoin%
\pgfsetlinewidth{0.501875pt}%
\definecolor{currentstroke}{rgb}{0.172549,0.482353,0.713725}%
\pgfsetstrokecolor{currentstroke}%
\pgfsetdash{}{0pt}%
\pgfpathmoveto{\pgfqpoint{2.289250in}{0.622846in}}%
\pgfpathlineto{\pgfqpoint{2.289250in}{0.646345in}}%
\pgfusepath{stroke}%
\end{pgfscope}%
\begin{pgfscope}%
\pgfpathrectangle{\pgfqpoint{0.597016in}{0.498776in}}{\pgfqpoint{2.248740in}{1.351224in}}%
\pgfusepath{clip}%
\pgfsetbuttcap%
\pgfsetroundjoin%
\pgfsetlinewidth{0.501875pt}%
\definecolor{currentstroke}{rgb}{0.172549,0.482353,0.713725}%
\pgfsetstrokecolor{currentstroke}%
\pgfsetdash{}{0pt}%
\pgfpathmoveto{\pgfqpoint{2.516395in}{0.621367in}}%
\pgfpathlineto{\pgfqpoint{2.516395in}{0.649912in}}%
\pgfusepath{stroke}%
\end{pgfscope}%
\begin{pgfscope}%
\pgfpathrectangle{\pgfqpoint{0.597016in}{0.498776in}}{\pgfqpoint{2.248740in}{1.351224in}}%
\pgfusepath{clip}%
\pgfsetbuttcap%
\pgfsetroundjoin%
\pgfsetlinewidth{0.501875pt}%
\definecolor{currentstroke}{rgb}{0.172549,0.482353,0.713725}%
\pgfsetstrokecolor{currentstroke}%
\pgfsetdash{}{0pt}%
\pgfpathmoveto{\pgfqpoint{2.743541in}{0.693418in}}%
\pgfpathlineto{\pgfqpoint{2.743541in}{0.741296in}}%
\pgfusepath{stroke}%
\end{pgfscope}%
\begin{pgfscope}%
\pgfpathrectangle{\pgfqpoint{0.597016in}{0.498776in}}{\pgfqpoint{2.248740in}{1.351224in}}%
\pgfusepath{clip}%
\pgfsetbuttcap%
\pgfsetroundjoin%
\definecolor{currentfill}{rgb}{0.172549,0.482353,0.713725}%
\pgfsetfillcolor{currentfill}%
\pgfsetlinewidth{1.003750pt}%
\definecolor{currentstroke}{rgb}{0.172549,0.482353,0.713725}%
\pgfsetstrokecolor{currentstroke}%
\pgfsetdash{}{0pt}%
\pgfsys@defobject{currentmarker}{\pgfqpoint{-0.069444in}{-0.000000in}}{\pgfqpoint{0.069444in}{0.000000in}}{%
\pgfpathmoveto{\pgfqpoint{0.069444in}{-0.000000in}}%
\pgfpathlineto{\pgfqpoint{-0.069444in}{0.000000in}}%
\pgfusepath{stroke,fill}%
}%
\begin{pgfscope}%
\pgfsys@transformshift{0.699231in}{1.718687in}%
\pgfsys@useobject{currentmarker}{}%
\end{pgfscope}%
\begin{pgfscope}%
\pgfsys@transformshift{0.926377in}{1.479968in}%
\pgfsys@useobject{currentmarker}{}%
\end{pgfscope}%
\begin{pgfscope}%
\pgfsys@transformshift{1.153522in}{0.652833in}%
\pgfsys@useobject{currentmarker}{}%
\end{pgfscope}%
\begin{pgfscope}%
\pgfsys@transformshift{1.380668in}{0.570032in}%
\pgfsys@useobject{currentmarker}{}%
\end{pgfscope}%
\begin{pgfscope}%
\pgfsys@transformshift{1.607813in}{0.654901in}%
\pgfsys@useobject{currentmarker}{}%
\end{pgfscope}%
\begin{pgfscope}%
\pgfsys@transformshift{1.834959in}{0.650734in}%
\pgfsys@useobject{currentmarker}{}%
\end{pgfscope}%
\begin{pgfscope}%
\pgfsys@transformshift{2.062104in}{0.628443in}%
\pgfsys@useobject{currentmarker}{}%
\end{pgfscope}%
\begin{pgfscope}%
\pgfsys@transformshift{2.289250in}{0.622846in}%
\pgfsys@useobject{currentmarker}{}%
\end{pgfscope}%
\begin{pgfscope}%
\pgfsys@transformshift{2.516395in}{0.621367in}%
\pgfsys@useobject{currentmarker}{}%
\end{pgfscope}%
\begin{pgfscope}%
\pgfsys@transformshift{2.743541in}{0.693418in}%
\pgfsys@useobject{currentmarker}{}%
\end{pgfscope}%
\end{pgfscope}%
\begin{pgfscope}%
\pgfpathrectangle{\pgfqpoint{0.597016in}{0.498776in}}{\pgfqpoint{2.248740in}{1.351224in}}%
\pgfusepath{clip}%
\pgfsetbuttcap%
\pgfsetroundjoin%
\definecolor{currentfill}{rgb}{0.172549,0.482353,0.713725}%
\pgfsetfillcolor{currentfill}%
\pgfsetlinewidth{1.003750pt}%
\definecolor{currentstroke}{rgb}{0.172549,0.482353,0.713725}%
\pgfsetstrokecolor{currentstroke}%
\pgfsetdash{}{0pt}%
\pgfsys@defobject{currentmarker}{\pgfqpoint{-0.069444in}{-0.000000in}}{\pgfqpoint{0.069444in}{0.000000in}}{%
\pgfpathmoveto{\pgfqpoint{0.069444in}{-0.000000in}}%
\pgfpathlineto{\pgfqpoint{-0.069444in}{0.000000in}}%
\pgfusepath{stroke,fill}%
}%
\begin{pgfscope}%
\pgfsys@transformshift{0.699231in}{1.788581in}%
\pgfsys@useobject{currentmarker}{}%
\end{pgfscope}%
\begin{pgfscope}%
\pgfsys@transformshift{0.926377in}{1.542696in}%
\pgfsys@useobject{currentmarker}{}%
\end{pgfscope}%
\begin{pgfscope}%
\pgfsys@transformshift{1.153522in}{0.786710in}%
\pgfsys@useobject{currentmarker}{}%
\end{pgfscope}%
\begin{pgfscope}%
\pgfsys@transformshift{1.380668in}{0.689852in}%
\pgfsys@useobject{currentmarker}{}%
\end{pgfscope}%
\begin{pgfscope}%
\pgfsys@transformshift{1.607813in}{0.757784in}%
\pgfsys@useobject{currentmarker}{}%
\end{pgfscope}%
\begin{pgfscope}%
\pgfsys@transformshift{1.834959in}{0.717624in}%
\pgfsys@useobject{currentmarker}{}%
\end{pgfscope}%
\begin{pgfscope}%
\pgfsys@transformshift{2.062104in}{0.667368in}%
\pgfsys@useobject{currentmarker}{}%
\end{pgfscope}%
\begin{pgfscope}%
\pgfsys@transformshift{2.289250in}{0.646345in}%
\pgfsys@useobject{currentmarker}{}%
\end{pgfscope}%
\begin{pgfscope}%
\pgfsys@transformshift{2.516395in}{0.649912in}%
\pgfsys@useobject{currentmarker}{}%
\end{pgfscope}%
\begin{pgfscope}%
\pgfsys@transformshift{2.743541in}{0.741296in}%
\pgfsys@useobject{currentmarker}{}%
\end{pgfscope}%
\end{pgfscope}%
\begin{pgfscope}%
\pgfpathrectangle{\pgfqpoint{0.597016in}{0.498776in}}{\pgfqpoint{2.248740in}{1.351224in}}%
\pgfusepath{clip}%
\pgfsetbuttcap%
\pgfsetroundjoin%
\definecolor{currentfill}{rgb}{0.843137,0.098039,0.109804}%
\pgfsetfillcolor{currentfill}%
\pgfsetlinewidth{1.003750pt}%
\definecolor{currentstroke}{rgb}{0.000000,0.000000,0.000000}%
\pgfsetstrokecolor{currentstroke}%
\pgfsetdash{}{0pt}%
\pgfsys@defobject{currentmarker}{\pgfqpoint{-0.027778in}{-0.027778in}}{\pgfqpoint{0.027778in}{0.027778in}}{%
\pgfpathmoveto{\pgfqpoint{0.000000in}{-0.027778in}}%
\pgfpathcurveto{\pgfqpoint{0.007367in}{-0.027778in}}{\pgfqpoint{0.014433in}{-0.024851in}}{\pgfqpoint{0.019642in}{-0.019642in}}%
\pgfpathcurveto{\pgfqpoint{0.024851in}{-0.014433in}}{\pgfqpoint{0.027778in}{-0.007367in}}{\pgfqpoint{0.027778in}{0.000000in}}%
\pgfpathcurveto{\pgfqpoint{0.027778in}{0.007367in}}{\pgfqpoint{0.024851in}{0.014433in}}{\pgfqpoint{0.019642in}{0.019642in}}%
\pgfpathcurveto{\pgfqpoint{0.014433in}{0.024851in}}{\pgfqpoint{0.007367in}{0.027778in}}{\pgfqpoint{0.000000in}{0.027778in}}%
\pgfpathcurveto{\pgfqpoint{-0.007367in}{0.027778in}}{\pgfqpoint{-0.014433in}{0.024851in}}{\pgfqpoint{-0.019642in}{0.019642in}}%
\pgfpathcurveto{\pgfqpoint{-0.024851in}{0.014433in}}{\pgfqpoint{-0.027778in}{0.007367in}}{\pgfqpoint{-0.027778in}{0.000000in}}%
\pgfpathcurveto{\pgfqpoint{-0.027778in}{-0.007367in}}{\pgfqpoint{-0.024851in}{-0.014433in}}{\pgfqpoint{-0.019642in}{-0.019642in}}%
\pgfpathcurveto{\pgfqpoint{-0.014433in}{-0.024851in}}{\pgfqpoint{-0.007367in}{-0.027778in}}{\pgfqpoint{0.000000in}{-0.027778in}}%
\pgfpathlineto{\pgfqpoint{0.000000in}{-0.027778in}}%
\pgfpathclose%
\pgfusepath{stroke,fill}%
}%
\begin{pgfscope}%
\pgfsys@transformshift{0.699231in}{1.250213in}%
\pgfsys@useobject{currentmarker}{}%
\end{pgfscope}%
\begin{pgfscope}%
\pgfsys@transformshift{0.926377in}{1.160904in}%
\pgfsys@useobject{currentmarker}{}%
\end{pgfscope}%
\begin{pgfscope}%
\pgfsys@transformshift{1.153522in}{0.685668in}%
\pgfsys@useobject{currentmarker}{}%
\end{pgfscope}%
\begin{pgfscope}%
\pgfsys@transformshift{1.380668in}{0.598533in}%
\pgfsys@useobject{currentmarker}{}%
\end{pgfscope}%
\begin{pgfscope}%
\pgfsys@transformshift{1.607813in}{0.717420in}%
\pgfsys@useobject{currentmarker}{}%
\end{pgfscope}%
\begin{pgfscope}%
\pgfsys@transformshift{1.834959in}{0.742089in}%
\pgfsys@useobject{currentmarker}{}%
\end{pgfscope}%
\begin{pgfscope}%
\pgfsys@transformshift{2.062104in}{0.723470in}%
\pgfsys@useobject{currentmarker}{}%
\end{pgfscope}%
\begin{pgfscope}%
\pgfsys@transformshift{2.289250in}{0.705039in}%
\pgfsys@useobject{currentmarker}{}%
\end{pgfscope}%
\begin{pgfscope}%
\pgfsys@transformshift{2.516395in}{0.689008in}%
\pgfsys@useobject{currentmarker}{}%
\end{pgfscope}%
\begin{pgfscope}%
\pgfsys@transformshift{2.743541in}{0.743433in}%
\pgfsys@useobject{currentmarker}{}%
\end{pgfscope}%
\end{pgfscope}%
\begin{pgfscope}%
\pgfpathrectangle{\pgfqpoint{0.597016in}{0.498776in}}{\pgfqpoint{2.248740in}{1.351224in}}%
\pgfusepath{clip}%
\pgfsetbuttcap%
\pgfsetmiterjoin%
\definecolor{currentfill}{rgb}{0.172549,0.482353,0.713725}%
\pgfsetfillcolor{currentfill}%
\pgfsetlinewidth{1.003750pt}%
\definecolor{currentstroke}{rgb}{0.000000,0.000000,0.000000}%
\pgfsetstrokecolor{currentstroke}%
\pgfsetdash{}{0pt}%
\pgfsys@defobject{currentmarker}{\pgfqpoint{-0.029463in}{-0.029463in}}{\pgfqpoint{0.029463in}{0.029463in}}{%
\pgfpathmoveto{\pgfqpoint{-0.000000in}{-0.029463in}}%
\pgfpathlineto{\pgfqpoint{0.029463in}{0.000000in}}%
\pgfpathlineto{\pgfqpoint{0.000000in}{0.029463in}}%
\pgfpathlineto{\pgfqpoint{-0.029463in}{0.000000in}}%
\pgfpathlineto{\pgfqpoint{-0.000000in}{-0.029463in}}%
\pgfpathclose%
\pgfusepath{stroke,fill}%
}%
\begin{pgfscope}%
\pgfsys@transformshift{0.699231in}{1.753634in}%
\pgfsys@useobject{currentmarker}{}%
\end{pgfscope}%
\begin{pgfscope}%
\pgfsys@transformshift{0.926377in}{1.511332in}%
\pgfsys@useobject{currentmarker}{}%
\end{pgfscope}%
\begin{pgfscope}%
\pgfsys@transformshift{1.153522in}{0.719771in}%
\pgfsys@useobject{currentmarker}{}%
\end{pgfscope}%
\begin{pgfscope}%
\pgfsys@transformshift{1.380668in}{0.629942in}%
\pgfsys@useobject{currentmarker}{}%
\end{pgfscope}%
\begin{pgfscope}%
\pgfsys@transformshift{1.607813in}{0.706343in}%
\pgfsys@useobject{currentmarker}{}%
\end{pgfscope}%
\begin{pgfscope}%
\pgfsys@transformshift{1.834959in}{0.684179in}%
\pgfsys@useobject{currentmarker}{}%
\end{pgfscope}%
\begin{pgfscope}%
\pgfsys@transformshift{2.062104in}{0.647906in}%
\pgfsys@useobject{currentmarker}{}%
\end{pgfscope}%
\begin{pgfscope}%
\pgfsys@transformshift{2.289250in}{0.634595in}%
\pgfsys@useobject{currentmarker}{}%
\end{pgfscope}%
\begin{pgfscope}%
\pgfsys@transformshift{2.516395in}{0.635639in}%
\pgfsys@useobject{currentmarker}{}%
\end{pgfscope}%
\begin{pgfscope}%
\pgfsys@transformshift{2.743541in}{0.717357in}%
\pgfsys@useobject{currentmarker}{}%
\end{pgfscope}%
\end{pgfscope}%
\begin{pgfscope}%
\pgfsetrectcap%
\pgfsetmiterjoin%
\pgfsetlinewidth{0.803000pt}%
\definecolor{currentstroke}{rgb}{0.000000,0.000000,0.000000}%
\pgfsetstrokecolor{currentstroke}%
\pgfsetdash{}{0pt}%
\pgfpathmoveto{\pgfqpoint{0.597016in}{0.498776in}}%
\pgfpathlineto{\pgfqpoint{0.597016in}{1.850000in}}%
\pgfusepath{stroke}%
\end{pgfscope}%
\begin{pgfscope}%
\pgfsetrectcap%
\pgfsetmiterjoin%
\pgfsetlinewidth{0.803000pt}%
\definecolor{currentstroke}{rgb}{0.000000,0.000000,0.000000}%
\pgfsetstrokecolor{currentstroke}%
\pgfsetdash{}{0pt}%
\pgfpathmoveto{\pgfqpoint{2.845756in}{0.498776in}}%
\pgfpathlineto{\pgfqpoint{2.845756in}{1.850000in}}%
\pgfusepath{stroke}%
\end{pgfscope}%
\begin{pgfscope}%
\pgfsetrectcap%
\pgfsetmiterjoin%
\pgfsetlinewidth{0.803000pt}%
\definecolor{currentstroke}{rgb}{0.000000,0.000000,0.000000}%
\pgfsetstrokecolor{currentstroke}%
\pgfsetdash{}{0pt}%
\pgfpathmoveto{\pgfqpoint{0.597016in}{0.498776in}}%
\pgfpathlineto{\pgfqpoint{2.845756in}{0.498776in}}%
\pgfusepath{stroke}%
\end{pgfscope}%
\begin{pgfscope}%
\pgfsetrectcap%
\pgfsetmiterjoin%
\pgfsetlinewidth{0.803000pt}%
\definecolor{currentstroke}{rgb}{0.000000,0.000000,0.000000}%
\pgfsetstrokecolor{currentstroke}%
\pgfsetdash{}{0pt}%
\pgfpathmoveto{\pgfqpoint{0.597016in}{1.850000in}}%
\pgfpathlineto{\pgfqpoint{2.845756in}{1.850000in}}%
\pgfusepath{stroke}%
\end{pgfscope}%
\begin{pgfscope}%
\pgfsetbuttcap%
\pgfsetmiterjoin%
\definecolor{currentfill}{rgb}{1.000000,1.000000,1.000000}%
\pgfsetfillcolor{currentfill}%
\pgfsetfillopacity{0.800000}%
\pgfsetlinewidth{1.003750pt}%
\definecolor{currentstroke}{rgb}{0.800000,0.800000,0.800000}%
\pgfsetstrokecolor{currentstroke}%
\pgfsetstrokeopacity{0.800000}%
\pgfsetdash{}{0pt}%
\pgfpathmoveto{\pgfqpoint{2.061616in}{1.373972in}}%
\pgfpathlineto{\pgfqpoint{2.764770in}{1.373972in}}%
\pgfpathquadraticcurveto{\pgfqpoint{2.787909in}{1.373972in}}{\pgfqpoint{2.787909in}{1.397111in}}%
\pgfpathlineto{\pgfqpoint{2.787909in}{1.769014in}}%
\pgfpathquadraticcurveto{\pgfqpoint{2.787909in}{1.792153in}}{\pgfqpoint{2.764770in}{1.792153in}}%
\pgfpathlineto{\pgfqpoint{2.061616in}{1.792153in}}%
\pgfpathquadraticcurveto{\pgfqpoint{2.038477in}{1.792153in}}{\pgfqpoint{2.038477in}{1.769014in}}%
\pgfpathlineto{\pgfqpoint{2.038477in}{1.397111in}}%
\pgfpathquadraticcurveto{\pgfqpoint{2.038477in}{1.373972in}}{\pgfqpoint{2.061616in}{1.373972in}}%
\pgfpathlineto{\pgfqpoint{2.061616in}{1.373972in}}%
\pgfpathclose%
\pgfusepath{stroke,fill}%
\end{pgfscope}%
\begin{pgfscope}%
\pgfsetbuttcap%
\pgfsetroundjoin%
\pgfsetlinewidth{0.501875pt}%
\definecolor{currentstroke}{rgb}{0.843137,0.098039,0.109804}%
\pgfsetstrokecolor{currentstroke}%
\pgfsetdash{}{0pt}%
\pgfpathmoveto{\pgfqpoint{2.200449in}{1.625419in}}%
\pgfpathlineto{\pgfqpoint{2.200449in}{1.741113in}}%
\pgfusepath{stroke}%
\end{pgfscope}%
\begin{pgfscope}%
\pgfsetbuttcap%
\pgfsetroundjoin%
\definecolor{currentfill}{rgb}{0.843137,0.098039,0.109804}%
\pgfsetfillcolor{currentfill}%
\pgfsetlinewidth{1.003750pt}%
\definecolor{currentstroke}{rgb}{0.843137,0.098039,0.109804}%
\pgfsetstrokecolor{currentstroke}%
\pgfsetdash{}{0pt}%
\pgfsys@defobject{currentmarker}{\pgfqpoint{-0.069444in}{-0.000000in}}{\pgfqpoint{0.069444in}{0.000000in}}{%
\pgfpathmoveto{\pgfqpoint{0.069444in}{-0.000000in}}%
\pgfpathlineto{\pgfqpoint{-0.069444in}{0.000000in}}%
\pgfusepath{stroke,fill}%
}%
\begin{pgfscope}%
\pgfsys@transformshift{2.200449in}{1.625419in}%
\pgfsys@useobject{currentmarker}{}%
\end{pgfscope}%
\end{pgfscope}%
\begin{pgfscope}%
\pgfsetbuttcap%
\pgfsetroundjoin%
\definecolor{currentfill}{rgb}{0.843137,0.098039,0.109804}%
\pgfsetfillcolor{currentfill}%
\pgfsetlinewidth{1.003750pt}%
\definecolor{currentstroke}{rgb}{0.843137,0.098039,0.109804}%
\pgfsetstrokecolor{currentstroke}%
\pgfsetdash{}{0pt}%
\pgfsys@defobject{currentmarker}{\pgfqpoint{-0.069444in}{-0.000000in}}{\pgfqpoint{0.069444in}{0.000000in}}{%
\pgfpathmoveto{\pgfqpoint{0.069444in}{-0.000000in}}%
\pgfpathlineto{\pgfqpoint{-0.069444in}{0.000000in}}%
\pgfusepath{stroke,fill}%
}%
\begin{pgfscope}%
\pgfsys@transformshift{2.200449in}{1.741113in}%
\pgfsys@useobject{currentmarker}{}%
\end{pgfscope}%
\end{pgfscope}%
\begin{pgfscope}%
\pgfsetbuttcap%
\pgfsetroundjoin%
\definecolor{currentfill}{rgb}{0.843137,0.098039,0.109804}%
\pgfsetfillcolor{currentfill}%
\pgfsetlinewidth{1.003750pt}%
\definecolor{currentstroke}{rgb}{0.000000,0.000000,0.000000}%
\pgfsetstrokecolor{currentstroke}%
\pgfsetdash{}{0pt}%
\pgfsys@defobject{currentmarker}{\pgfqpoint{-0.027778in}{-0.027778in}}{\pgfqpoint{0.027778in}{0.027778in}}{%
\pgfpathmoveto{\pgfqpoint{0.000000in}{-0.027778in}}%
\pgfpathcurveto{\pgfqpoint{0.007367in}{-0.027778in}}{\pgfqpoint{0.014433in}{-0.024851in}}{\pgfqpoint{0.019642in}{-0.019642in}}%
\pgfpathcurveto{\pgfqpoint{0.024851in}{-0.014433in}}{\pgfqpoint{0.027778in}{-0.007367in}}{\pgfqpoint{0.027778in}{0.000000in}}%
\pgfpathcurveto{\pgfqpoint{0.027778in}{0.007367in}}{\pgfqpoint{0.024851in}{0.014433in}}{\pgfqpoint{0.019642in}{0.019642in}}%
\pgfpathcurveto{\pgfqpoint{0.014433in}{0.024851in}}{\pgfqpoint{0.007367in}{0.027778in}}{\pgfqpoint{0.000000in}{0.027778in}}%
\pgfpathcurveto{\pgfqpoint{-0.007367in}{0.027778in}}{\pgfqpoint{-0.014433in}{0.024851in}}{\pgfqpoint{-0.019642in}{0.019642in}}%
\pgfpathcurveto{\pgfqpoint{-0.024851in}{0.014433in}}{\pgfqpoint{-0.027778in}{0.007367in}}{\pgfqpoint{-0.027778in}{0.000000in}}%
\pgfpathcurveto{\pgfqpoint{-0.027778in}{-0.007367in}}{\pgfqpoint{-0.024851in}{-0.014433in}}{\pgfqpoint{-0.019642in}{-0.019642in}}%
\pgfpathcurveto{\pgfqpoint{-0.014433in}{-0.024851in}}{\pgfqpoint{-0.007367in}{-0.027778in}}{\pgfqpoint{0.000000in}{-0.027778in}}%
\pgfpathlineto{\pgfqpoint{0.000000in}{-0.027778in}}%
\pgfpathclose%
\pgfusepath{stroke,fill}%
}%
\begin{pgfscope}%
\pgfsys@transformshift{2.200449in}{1.683266in}%
\pgfsys@useobject{currentmarker}{}%
\end{pgfscope}%
\end{pgfscope}%
\begin{pgfscope}%
\definecolor{textcolor}{rgb}{0.000000,0.000000,0.000000}%
\pgfsetstrokecolor{textcolor}%
\pgfsetfillcolor{textcolor}%
\pgftext[x=2.408699in,y=1.642773in,left,base]{\color{textcolor}{\rmfamily\fontsize{8.330000}{9.996000}\selectfont\catcode`\^=\active\def^{\ifmmode\sp\else\^{}\fi}\catcode`\%=\active\def%{\%}$\bm{v}^\text{PINN}_{2\text{Re}}$}}%
\end{pgfscope}%
\begin{pgfscope}%
\pgfsetbuttcap%
\pgfsetroundjoin%
\pgfsetlinewidth{0.501875pt}%
\definecolor{currentstroke}{rgb}{0.172549,0.482353,0.713725}%
\pgfsetstrokecolor{currentstroke}%
\pgfsetdash{}{0pt}%
\pgfpathmoveto{\pgfqpoint{2.200449in}{1.433683in}}%
\pgfpathlineto{\pgfqpoint{2.200449in}{1.549377in}}%
\pgfusepath{stroke}%
\end{pgfscope}%
\begin{pgfscope}%
\pgfsetbuttcap%
\pgfsetroundjoin%
\definecolor{currentfill}{rgb}{0.172549,0.482353,0.713725}%
\pgfsetfillcolor{currentfill}%
\pgfsetlinewidth{1.003750pt}%
\definecolor{currentstroke}{rgb}{0.172549,0.482353,0.713725}%
\pgfsetstrokecolor{currentstroke}%
\pgfsetdash{}{0pt}%
\pgfsys@defobject{currentmarker}{\pgfqpoint{-0.069444in}{-0.000000in}}{\pgfqpoint{0.069444in}{0.000000in}}{%
\pgfpathmoveto{\pgfqpoint{0.069444in}{-0.000000in}}%
\pgfpathlineto{\pgfqpoint{-0.069444in}{0.000000in}}%
\pgfusepath{stroke,fill}%
}%
\begin{pgfscope}%
\pgfsys@transformshift{2.200449in}{1.433683in}%
\pgfsys@useobject{currentmarker}{}%
\end{pgfscope}%
\end{pgfscope}%
\begin{pgfscope}%
\pgfsetbuttcap%
\pgfsetroundjoin%
\definecolor{currentfill}{rgb}{0.172549,0.482353,0.713725}%
\pgfsetfillcolor{currentfill}%
\pgfsetlinewidth{1.003750pt}%
\definecolor{currentstroke}{rgb}{0.172549,0.482353,0.713725}%
\pgfsetstrokecolor{currentstroke}%
\pgfsetdash{}{0pt}%
\pgfsys@defobject{currentmarker}{\pgfqpoint{-0.069444in}{-0.000000in}}{\pgfqpoint{0.069444in}{0.000000in}}{%
\pgfpathmoveto{\pgfqpoint{0.069444in}{-0.000000in}}%
\pgfpathlineto{\pgfqpoint{-0.069444in}{0.000000in}}%
\pgfusepath{stroke,fill}%
}%
\begin{pgfscope}%
\pgfsys@transformshift{2.200449in}{1.549377in}%
\pgfsys@useobject{currentmarker}{}%
\end{pgfscope}%
\end{pgfscope}%
\begin{pgfscope}%
\pgfsetbuttcap%
\pgfsetmiterjoin%
\definecolor{currentfill}{rgb}{0.172549,0.482353,0.713725}%
\pgfsetfillcolor{currentfill}%
\pgfsetlinewidth{1.003750pt}%
\definecolor{currentstroke}{rgb}{0.000000,0.000000,0.000000}%
\pgfsetstrokecolor{currentstroke}%
\pgfsetdash{}{0pt}%
\pgfsys@defobject{currentmarker}{\pgfqpoint{-0.029463in}{-0.029463in}}{\pgfqpoint{0.029463in}{0.029463in}}{%
\pgfpathmoveto{\pgfqpoint{-0.000000in}{-0.029463in}}%
\pgfpathlineto{\pgfqpoint{0.029463in}{0.000000in}}%
\pgfpathlineto{\pgfqpoint{0.000000in}{0.029463in}}%
\pgfpathlineto{\pgfqpoint{-0.029463in}{0.000000in}}%
\pgfpathlineto{\pgfqpoint{-0.000000in}{-0.029463in}}%
\pgfpathclose%
\pgfusepath{stroke,fill}%
}%
\begin{pgfscope}%
\pgfsys@transformshift{2.200449in}{1.491530in}%
\pgfsys@useobject{currentmarker}{}%
\end{pgfscope}%
\end{pgfscope}%
\begin{pgfscope}%
\definecolor{textcolor}{rgb}{0.000000,0.000000,0.000000}%
\pgfsetstrokecolor{textcolor}%
\pgfsetfillcolor{textcolor}%
\pgftext[x=2.408699in,y=1.451037in,left,base]{\color{textcolor}{\rmfamily\fontsize{8.330000}{9.996000}\selectfont\catcode`\^=\active\def^{\ifmmode\sp\else\^{}\fi}\catcode`\%=\active\def%{\%}$p^\text{PINN}_{2\text{Re}}$}}%
\end{pgfscope}%
\end{pgfpicture}%
\makeatother%
\endgroup%

%% file: figures/section5/n_points_per_Re_targeted.pgf
%% Creator: Matplotlib, PGF backend
%%
%% To include the figure in your LaTeX document, write
%%   \input{<filename>.pgf}
%%
%% Make sure the required packages are loaded in your preamble
%%   \usepackage{pgf}
%%
%% Also ensure that all the required font packages are loaded; for instance,
%% the lmodern package is sometimes necessary when using math font.
%%   \usepackage{lmodern}
%%
%% Figures using additional raster images can only be included by \input if
%% they are in the same directory as the main LaTeX file. For loading figures
%% from other directories you can use the `import` package
%% and then include the figures with
%%   \import{<path to file>}{<filename>.pgf}
%%
%% Matplotlib used the following preamble
%%   \def\mathdefault#1{#1}
%%   \everymath=\expandafter{\the\everymath\displaystyle}
%%   \usepackage{amsmath}\usepackage{bm}
%%   \makeatletter\@ifpackageloaded{underscore}{}{\usepackage[strings]{underscore}}\makeatother
%%
\begingroup%
\makeatletter%
\begin{pgfpicture}%
\pgfpathrectangle{\pgfpointorigin}{\pgfqpoint{3.000000in}{2.000000in}}%
\pgfusepath{use as bounding box, clip}%
\begin{pgfscope}%
\pgfsetbuttcap%
\pgfsetmiterjoin%
\definecolor{currentfill}{rgb}{1.000000,1.000000,1.000000}%
\pgfsetfillcolor{currentfill}%
\pgfsetlinewidth{0.000000pt}%
\definecolor{currentstroke}{rgb}{1.000000,1.000000,1.000000}%
\pgfsetstrokecolor{currentstroke}%
\pgfsetdash{}{0pt}%
\pgfpathmoveto{\pgfqpoint{0.000000in}{0.000000in}}%
\pgfpathlineto{\pgfqpoint{3.000000in}{0.000000in}}%
\pgfpathlineto{\pgfqpoint{3.000000in}{2.000000in}}%
\pgfpathlineto{\pgfqpoint{0.000000in}{2.000000in}}%
\pgfpathlineto{\pgfqpoint{0.000000in}{0.000000in}}%
\pgfpathclose%
\pgfusepath{fill}%
\end{pgfscope}%
\begin{pgfscope}%
\pgfsetbuttcap%
\pgfsetmiterjoin%
\definecolor{currentfill}{rgb}{1.000000,1.000000,1.000000}%
\pgfsetfillcolor{currentfill}%
\pgfsetlinewidth{0.000000pt}%
\definecolor{currentstroke}{rgb}{0.000000,0.000000,0.000000}%
\pgfsetstrokecolor{currentstroke}%
\pgfsetstrokeopacity{0.000000}%
\pgfsetdash{}{0pt}%
\pgfpathmoveto{\pgfqpoint{0.548324in}{0.521284in}}%
\pgfpathlineto{\pgfqpoint{2.850000in}{0.521284in}}%
\pgfpathlineto{\pgfqpoint{2.850000in}{1.850000in}}%
\pgfpathlineto{\pgfqpoint{0.548324in}{1.850000in}}%
\pgfpathlineto{\pgfqpoint{0.548324in}{0.521284in}}%
\pgfpathclose%
\pgfusepath{fill}%
\end{pgfscope}%
\begin{pgfscope}%
\pgfpathrectangle{\pgfqpoint{0.548324in}{0.521284in}}{\pgfqpoint{2.301676in}{1.328716in}}%
\pgfusepath{clip}%
\pgfsetbuttcap%
\pgfsetroundjoin%
\definecolor{currentfill}{rgb}{0.172549,0.482353,0.713725}%
\pgfsetfillcolor{currentfill}%
\pgfsetlinewidth{0.501875pt}%
\definecolor{currentstroke}{rgb}{0.000000,0.000000,0.000000}%
\pgfsetstrokecolor{currentstroke}%
\pgfsetdash{}{0pt}%
\pgfsys@defobject{currentmarker}{\pgfqpoint{-0.026896in}{-0.026896in}}{\pgfqpoint{0.026896in}{0.026896in}}{%
\pgfpathmoveto{\pgfqpoint{0.000000in}{0.026896in}}%
\pgfpathlineto{\pgfqpoint{-0.026896in}{-0.026896in}}%
\pgfpathlineto{\pgfqpoint{0.026896in}{-0.026896in}}%
\pgfpathlineto{\pgfqpoint{0.000000in}{0.026896in}}%
\pgfpathclose%
\pgfusepath{stroke,fill}%
}%
\begin{pgfscope}%
\pgfsys@transformshift{0.652945in}{1.789604in}%
\pgfsys@useobject{currentmarker}{}%
\end{pgfscope}%
\begin{pgfscope}%
\pgfsys@transformshift{1.699162in}{1.624546in}%
\pgfsys@useobject{currentmarker}{}%
\end{pgfscope}%
\begin{pgfscope}%
\pgfsys@transformshift{2.745378in}{0.664209in}%
\pgfsys@useobject{currentmarker}{}%
\end{pgfscope}%
\end{pgfscope}%
\begin{pgfscope}%
\pgfpathrectangle{\pgfqpoint{0.548324in}{0.521284in}}{\pgfqpoint{2.301676in}{1.328716in}}%
\pgfusepath{clip}%
\pgfsetbuttcap%
\pgfsetroundjoin%
\definecolor{currentfill}{rgb}{0.172549,0.482353,0.713725}%
\pgfsetfillcolor{currentfill}%
\pgfsetlinewidth{0.501875pt}%
\definecolor{currentstroke}{rgb}{0.000000,0.000000,0.000000}%
\pgfsetstrokecolor{currentstroke}%
\pgfsetdash{}{0pt}%
\pgfsys@defobject{currentmarker}{\pgfqpoint{-0.038036in}{-0.038036in}}{\pgfqpoint{0.038036in}{0.038036in}}{%
\pgfpathmoveto{\pgfqpoint{-0.000000in}{-0.038036in}}%
\pgfpathlineto{\pgfqpoint{0.038036in}{0.000000in}}%
\pgfpathlineto{\pgfqpoint{0.000000in}{0.038036in}}%
\pgfpathlineto{\pgfqpoint{-0.038036in}{0.000000in}}%
\pgfpathlineto{\pgfqpoint{-0.000000in}{-0.038036in}}%
\pgfpathclose%
\pgfusepath{stroke,fill}%
}%
\begin{pgfscope}%
\pgfsys@transformshift{0.652945in}{1.534514in}%
\pgfsys@useobject{currentmarker}{}%
\end{pgfscope}%
\begin{pgfscope}%
\pgfsys@transformshift{1.699162in}{1.343197in}%
\pgfsys@useobject{currentmarker}{}%
\end{pgfscope}%
\begin{pgfscope}%
\pgfsys@transformshift{2.745378in}{0.581680in}%
\pgfsys@useobject{currentmarker}{}%
\end{pgfscope}%
\end{pgfscope}%
\begin{pgfscope}%
\pgfsetbuttcap%
\pgfsetroundjoin%
\definecolor{currentfill}{rgb}{0.000000,0.000000,0.000000}%
\pgfsetfillcolor{currentfill}%
\pgfsetlinewidth{0.803000pt}%
\definecolor{currentstroke}{rgb}{0.000000,0.000000,0.000000}%
\pgfsetstrokecolor{currentstroke}%
\pgfsetdash{}{0pt}%
\pgfsys@defobject{currentmarker}{\pgfqpoint{0.000000in}{-0.048611in}}{\pgfqpoint{0.000000in}{0.000000in}}{%
\pgfpathmoveto{\pgfqpoint{0.000000in}{0.000000in}}%
\pgfpathlineto{\pgfqpoint{0.000000in}{-0.048611in}}%
\pgfusepath{stroke,fill}%
}%
\begin{pgfscope}%
\pgfsys@transformshift{0.652945in}{0.521284in}%
\pgfsys@useobject{currentmarker}{}%
\end{pgfscope}%
\end{pgfscope}%
\begin{pgfscope}%
\definecolor{textcolor}{rgb}{0.000000,0.000000,0.000000}%
\pgfsetstrokecolor{textcolor}%
\pgfsetfillcolor{textcolor}%
\pgftext[x=0.652945in,y=0.431006in,,top]{\color{textcolor}{\rmfamily\fontsize{8.330000}{9.996000}\selectfont\catcode`\^=\active\def^{\ifmmode\sp\else\^{}\fi}\catcode`\%=\active\def%{\%}1}}%
\end{pgfscope}%
\begin{pgfscope}%
\pgfsetbuttcap%
\pgfsetroundjoin%
\definecolor{currentfill}{rgb}{0.000000,0.000000,0.000000}%
\pgfsetfillcolor{currentfill}%
\pgfsetlinewidth{0.803000pt}%
\definecolor{currentstroke}{rgb}{0.000000,0.000000,0.000000}%
\pgfsetstrokecolor{currentstroke}%
\pgfsetdash{}{0pt}%
\pgfsys@defobject{currentmarker}{\pgfqpoint{0.000000in}{-0.048611in}}{\pgfqpoint{0.000000in}{0.000000in}}{%
\pgfpathmoveto{\pgfqpoint{0.000000in}{0.000000in}}%
\pgfpathlineto{\pgfqpoint{0.000000in}{-0.048611in}}%
\pgfusepath{stroke,fill}%
}%
\begin{pgfscope}%
\pgfsys@transformshift{1.699162in}{0.521284in}%
\pgfsys@useobject{currentmarker}{}%
\end{pgfscope}%
\end{pgfscope}%
\begin{pgfscope}%
\definecolor{textcolor}{rgb}{0.000000,0.000000,0.000000}%
\pgfsetstrokecolor{textcolor}%
\pgfsetfillcolor{textcolor}%
\pgftext[x=1.699162in,y=0.431006in,,top]{\color{textcolor}{\rmfamily\fontsize{8.330000}{9.996000}\selectfont\catcode`\^=\active\def^{\ifmmode\sp\else\^{}\fi}\catcode`\%=\active\def%{\%}2}}%
\end{pgfscope}%
\begin{pgfscope}%
\pgfsetbuttcap%
\pgfsetroundjoin%
\definecolor{currentfill}{rgb}{0.000000,0.000000,0.000000}%
\pgfsetfillcolor{currentfill}%
\pgfsetlinewidth{0.803000pt}%
\definecolor{currentstroke}{rgb}{0.000000,0.000000,0.000000}%
\pgfsetstrokecolor{currentstroke}%
\pgfsetdash{}{0pt}%
\pgfsys@defobject{currentmarker}{\pgfqpoint{0.000000in}{-0.048611in}}{\pgfqpoint{0.000000in}{0.000000in}}{%
\pgfpathmoveto{\pgfqpoint{0.000000in}{0.000000in}}%
\pgfpathlineto{\pgfqpoint{0.000000in}{-0.048611in}}%
\pgfusepath{stroke,fill}%
}%
\begin{pgfscope}%
\pgfsys@transformshift{2.745378in}{0.521284in}%
\pgfsys@useobject{currentmarker}{}%
\end{pgfscope}%
\end{pgfscope}%
\begin{pgfscope}%
\definecolor{textcolor}{rgb}{0.000000,0.000000,0.000000}%
\pgfsetstrokecolor{textcolor}%
\pgfsetfillcolor{textcolor}%
\pgftext[x=2.745378in,y=0.431006in,,top]{\color{textcolor}{\rmfamily\fontsize{8.330000}{9.996000}\selectfont\catcode`\^=\active\def^{\ifmmode\sp\else\^{}\fi}\catcode`\%=\active\def%{\%}3}}%
\end{pgfscope}%
\begin{pgfscope}%
\definecolor{textcolor}{rgb}{0.000000,0.000000,0.000000}%
\pgfsetstrokecolor{textcolor}%
\pgfsetfillcolor{textcolor}%
\pgftext[x=1.699162in,y=0.276685in,,top]{\color{textcolor}{\rmfamily\fontsize{10.000000}{12.000000}\selectfont\catcode`\^=\active\def^{\ifmmode\sp\else\^{}\fi}\catcode`\%=\active\def%{\%}Number of labels per $\text{Re}$}}%
\end{pgfscope}%
\begin{pgfscope}%
\pgfsetbuttcap%
\pgfsetroundjoin%
\definecolor{currentfill}{rgb}{0.000000,0.000000,0.000000}%
\pgfsetfillcolor{currentfill}%
\pgfsetlinewidth{0.803000pt}%
\definecolor{currentstroke}{rgb}{0.000000,0.000000,0.000000}%
\pgfsetstrokecolor{currentstroke}%
\pgfsetdash{}{0pt}%
\pgfsys@defobject{currentmarker}{\pgfqpoint{-0.048611in}{0.000000in}}{\pgfqpoint{-0.000000in}{0.000000in}}{%
\pgfpathmoveto{\pgfqpoint{-0.000000in}{0.000000in}}%
\pgfpathlineto{\pgfqpoint{-0.048611in}{0.000000in}}%
\pgfusepath{stroke,fill}%
}%
\begin{pgfscope}%
\pgfsys@transformshift{0.548324in}{0.607939in}%
\pgfsys@useobject{currentmarker}{}%
\end{pgfscope}%
\end{pgfscope}%
\begin{pgfscope}%
\definecolor{textcolor}{rgb}{0.000000,0.000000,0.000000}%
\pgfsetstrokecolor{textcolor}%
\pgfsetfillcolor{textcolor}%
\pgftext[x=0.399017in, y=0.569359in, left, base]{\color{textcolor}{\rmfamily\fontsize{8.330000}{9.996000}\selectfont\catcode`\^=\active\def^{\ifmmode\sp\else\^{}\fi}\catcode`\%=\active\def%{\%}$\mathdefault{3}$}}%
\end{pgfscope}%
\begin{pgfscope}%
\pgfsetbuttcap%
\pgfsetroundjoin%
\definecolor{currentfill}{rgb}{0.000000,0.000000,0.000000}%
\pgfsetfillcolor{currentfill}%
\pgfsetlinewidth{0.803000pt}%
\definecolor{currentstroke}{rgb}{0.000000,0.000000,0.000000}%
\pgfsetstrokecolor{currentstroke}%
\pgfsetdash{}{0pt}%
\pgfsys@defobject{currentmarker}{\pgfqpoint{-0.048611in}{0.000000in}}{\pgfqpoint{-0.000000in}{0.000000in}}{%
\pgfpathmoveto{\pgfqpoint{-0.000000in}{0.000000in}}%
\pgfpathlineto{\pgfqpoint{-0.048611in}{0.000000in}}%
\pgfusepath{stroke,fill}%
}%
\begin{pgfscope}%
\pgfsys@transformshift{0.548324in}{0.983071in}%
\pgfsys@useobject{currentmarker}{}%
\end{pgfscope}%
\end{pgfscope}%
\begin{pgfscope}%
\definecolor{textcolor}{rgb}{0.000000,0.000000,0.000000}%
\pgfsetstrokecolor{textcolor}%
\pgfsetfillcolor{textcolor}%
\pgftext[x=0.399017in, y=0.944491in, left, base]{\color{textcolor}{\rmfamily\fontsize{8.330000}{9.996000}\selectfont\catcode`\^=\active\def^{\ifmmode\sp\else\^{}\fi}\catcode`\%=\active\def%{\%}$\mathdefault{4}$}}%
\end{pgfscope}%
\begin{pgfscope}%
\pgfsetbuttcap%
\pgfsetroundjoin%
\definecolor{currentfill}{rgb}{0.000000,0.000000,0.000000}%
\pgfsetfillcolor{currentfill}%
\pgfsetlinewidth{0.803000pt}%
\definecolor{currentstroke}{rgb}{0.000000,0.000000,0.000000}%
\pgfsetstrokecolor{currentstroke}%
\pgfsetdash{}{0pt}%
\pgfsys@defobject{currentmarker}{\pgfqpoint{-0.048611in}{0.000000in}}{\pgfqpoint{-0.000000in}{0.000000in}}{%
\pgfpathmoveto{\pgfqpoint{-0.000000in}{0.000000in}}%
\pgfpathlineto{\pgfqpoint{-0.048611in}{0.000000in}}%
\pgfusepath{stroke,fill}%
}%
\begin{pgfscope}%
\pgfsys@transformshift{0.548324in}{1.358203in}%
\pgfsys@useobject{currentmarker}{}%
\end{pgfscope}%
\end{pgfscope}%
\begin{pgfscope}%
\definecolor{textcolor}{rgb}{0.000000,0.000000,0.000000}%
\pgfsetstrokecolor{textcolor}%
\pgfsetfillcolor{textcolor}%
\pgftext[x=0.399017in, y=1.319622in, left, base]{\color{textcolor}{\rmfamily\fontsize{8.330000}{9.996000}\selectfont\catcode`\^=\active\def^{\ifmmode\sp\else\^{}\fi}\catcode`\%=\active\def%{\%}$\mathdefault{5}$}}%
\end{pgfscope}%
\begin{pgfscope}%
\pgfsetbuttcap%
\pgfsetroundjoin%
\definecolor{currentfill}{rgb}{0.000000,0.000000,0.000000}%
\pgfsetfillcolor{currentfill}%
\pgfsetlinewidth{0.803000pt}%
\definecolor{currentstroke}{rgb}{0.000000,0.000000,0.000000}%
\pgfsetstrokecolor{currentstroke}%
\pgfsetdash{}{0pt}%
\pgfsys@defobject{currentmarker}{\pgfqpoint{-0.048611in}{0.000000in}}{\pgfqpoint{-0.000000in}{0.000000in}}{%
\pgfpathmoveto{\pgfqpoint{-0.000000in}{0.000000in}}%
\pgfpathlineto{\pgfqpoint{-0.048611in}{0.000000in}}%
\pgfusepath{stroke,fill}%
}%
\begin{pgfscope}%
\pgfsys@transformshift{0.548324in}{1.733334in}%
\pgfsys@useobject{currentmarker}{}%
\end{pgfscope}%
\end{pgfscope}%
\begin{pgfscope}%
\definecolor{textcolor}{rgb}{0.000000,0.000000,0.000000}%
\pgfsetstrokecolor{textcolor}%
\pgfsetfillcolor{textcolor}%
\pgftext[x=0.399017in, y=1.694754in, left, base]{\color{textcolor}{\rmfamily\fontsize{8.330000}{9.996000}\selectfont\catcode`\^=\active\def^{\ifmmode\sp\else\^{}\fi}\catcode`\%=\active\def%{\%}$\mathdefault{6}$}}%
\end{pgfscope}%
\begin{pgfscope}%
\definecolor{textcolor}{rgb}{0.000000,0.000000,0.000000}%
\pgfsetstrokecolor{textcolor}%
\pgfsetfillcolor{textcolor}%
\pgftext[x=0.343462in,y=1.185642in,,bottom,rotate=90.000000]{\color{textcolor}{\rmfamily\fontsize{10.000000}{12.000000}\selectfont\catcode`\^=\active\def^{\ifmmode\sp\else\^{}\fi}\catcode`\%=\active\def%{\%}$\delta_{\ell^1}^{(q)} \ [\%]$}}%
\end{pgfscope}%
\begin{pgfscope}%
\pgfsetrectcap%
\pgfsetmiterjoin%
\pgfsetlinewidth{0.803000pt}%
\definecolor{currentstroke}{rgb}{0.000000,0.000000,0.000000}%
\pgfsetstrokecolor{currentstroke}%
\pgfsetdash{}{0pt}%
\pgfpathmoveto{\pgfqpoint{0.548324in}{0.521284in}}%
\pgfpathlineto{\pgfqpoint{0.548324in}{1.850000in}}%
\pgfusepath{stroke}%
\end{pgfscope}%
\begin{pgfscope}%
\pgfsetrectcap%
\pgfsetmiterjoin%
\pgfsetlinewidth{0.803000pt}%
\definecolor{currentstroke}{rgb}{0.000000,0.000000,0.000000}%
\pgfsetstrokecolor{currentstroke}%
\pgfsetdash{}{0pt}%
\pgfpathmoveto{\pgfqpoint{2.850000in}{0.521284in}}%
\pgfpathlineto{\pgfqpoint{2.850000in}{1.850000in}}%
\pgfusepath{stroke}%
\end{pgfscope}%
\begin{pgfscope}%
\pgfsetrectcap%
\pgfsetmiterjoin%
\pgfsetlinewidth{0.803000pt}%
\definecolor{currentstroke}{rgb}{0.000000,0.000000,0.000000}%
\pgfsetstrokecolor{currentstroke}%
\pgfsetdash{}{0pt}%
\pgfpathmoveto{\pgfqpoint{0.548324in}{0.521284in}}%
\pgfpathlineto{\pgfqpoint{2.850000in}{0.521284in}}%
\pgfusepath{stroke}%
\end{pgfscope}%
\begin{pgfscope}%
\pgfsetrectcap%
\pgfsetmiterjoin%
\pgfsetlinewidth{0.803000pt}%
\definecolor{currentstroke}{rgb}{0.000000,0.000000,0.000000}%
\pgfsetstrokecolor{currentstroke}%
\pgfsetdash{}{0pt}%
\pgfpathmoveto{\pgfqpoint{0.548324in}{1.850000in}}%
\pgfpathlineto{\pgfqpoint{2.850000in}{1.850000in}}%
\pgfusepath{stroke}%
\end{pgfscope}%
\begin{pgfscope}%
\pgfsetbuttcap%
\pgfsetmiterjoin%
\definecolor{currentfill}{rgb}{1.000000,1.000000,1.000000}%
\pgfsetfillcolor{currentfill}%
\pgfsetfillopacity{0.800000}%
\pgfsetlinewidth{1.003750pt}%
\definecolor{currentstroke}{rgb}{0.800000,0.800000,0.800000}%
\pgfsetstrokecolor{currentstroke}%
\pgfsetstrokeopacity{0.800000}%
\pgfsetdash{}{0pt}%
\pgfpathmoveto{\pgfqpoint{2.056703in}{1.347583in}}%
\pgfpathlineto{\pgfqpoint{2.769014in}{1.347583in}}%
\pgfpathquadraticcurveto{\pgfqpoint{2.792153in}{1.347583in}}{\pgfqpoint{2.792153in}{1.370722in}}%
\pgfpathlineto{\pgfqpoint{2.792153in}{1.769014in}}%
\pgfpathquadraticcurveto{\pgfqpoint{2.792153in}{1.792153in}}{\pgfqpoint{2.769014in}{1.792153in}}%
\pgfpathlineto{\pgfqpoint{2.056703in}{1.792153in}}%
\pgfpathquadraticcurveto{\pgfqpoint{2.033564in}{1.792153in}}{\pgfqpoint{2.033564in}{1.769014in}}%
\pgfpathlineto{\pgfqpoint{2.033564in}{1.370722in}}%
\pgfpathquadraticcurveto{\pgfqpoint{2.033564in}{1.347583in}}{\pgfqpoint{2.056703in}{1.347583in}}%
\pgfpathlineto{\pgfqpoint{2.056703in}{1.347583in}}%
\pgfpathclose%
\pgfusepath{stroke,fill}%
\end{pgfscope}%
\begin{pgfscope}%
\pgfsetbuttcap%
\pgfsetroundjoin%
\definecolor{currentfill}{rgb}{0.172549,0.482353,0.713725}%
\pgfsetfillcolor{currentfill}%
\pgfsetlinewidth{0.501875pt}%
\definecolor{currentstroke}{rgb}{0.000000,0.000000,0.000000}%
\pgfsetstrokecolor{currentstroke}%
\pgfsetdash{}{0pt}%
\pgfsys@defobject{currentmarker}{\pgfqpoint{-0.026896in}{-0.026896in}}{\pgfqpoint{0.026896in}{0.026896in}}{%
\pgfpathmoveto{\pgfqpoint{0.000000in}{0.026896in}}%
\pgfpathlineto{\pgfqpoint{-0.026896in}{-0.026896in}}%
\pgfpathlineto{\pgfqpoint{0.026896in}{-0.026896in}}%
\pgfpathlineto{\pgfqpoint{0.000000in}{0.026896in}}%
\pgfpathclose%
\pgfusepath{stroke,fill}%
}%
\begin{pgfscope}%
\pgfsys@transformshift{2.195537in}{1.673143in}%
\pgfsys@useobject{currentmarker}{}%
\end{pgfscope}%
\end{pgfscope}%
\begin{pgfscope}%
\definecolor{textcolor}{rgb}{0.000000,0.000000,0.000000}%
\pgfsetstrokecolor{textcolor}%
\pgfsetfillcolor{textcolor}%
\pgftext[x=2.403787in,y=1.642773in,left,base]{\color{textcolor}{\rmfamily\fontsize{8.330000}{9.996000}\selectfont\catcode`\^=\active\def^{\ifmmode\sp\else\^{}\fi}\catcode`\%=\active\def%{\%}$\bm{v}^\text{PINN}_\text{target}$}}%
\end{pgfscope}%
\begin{pgfscope}%
\pgfsetbuttcap%
\pgfsetroundjoin%
\definecolor{currentfill}{rgb}{0.172549,0.482353,0.713725}%
\pgfsetfillcolor{currentfill}%
\pgfsetlinewidth{0.501875pt}%
\definecolor{currentstroke}{rgb}{0.000000,0.000000,0.000000}%
\pgfsetstrokecolor{currentstroke}%
\pgfsetdash{}{0pt}%
\pgfsys@defobject{currentmarker}{\pgfqpoint{-0.038036in}{-0.038036in}}{\pgfqpoint{0.038036in}{0.038036in}}{%
\pgfpathmoveto{\pgfqpoint{-0.000000in}{-0.038036in}}%
\pgfpathlineto{\pgfqpoint{0.038036in}{0.000000in}}%
\pgfpathlineto{\pgfqpoint{0.000000in}{0.038036in}}%
\pgfpathlineto{\pgfqpoint{-0.038036in}{0.000000in}}%
\pgfpathlineto{\pgfqpoint{-0.000000in}{-0.038036in}}%
\pgfpathclose%
\pgfusepath{stroke,fill}%
}%
\begin{pgfscope}%
\pgfsys@transformshift{2.195537in}{1.468212in}%
\pgfsys@useobject{currentmarker}{}%
\end{pgfscope}%
\end{pgfscope}%
\begin{pgfscope}%
\definecolor{textcolor}{rgb}{0.000000,0.000000,0.000000}%
\pgfsetstrokecolor{textcolor}%
\pgfsetfillcolor{textcolor}%
\pgftext[x=2.403787in,y=1.437842in,left,base]{\color{textcolor}{\rmfamily\fontsize{8.330000}{9.996000}\selectfont\catcode`\^=\active\def^{\ifmmode\sp\else\^{}\fi}\catcode`\%=\active\def%{\%}$p^\text{PINN}_\text{target}$}}%
\end{pgfscope}%
\end{pgfpicture}%
\makeatother%
\endgroup%

%% file: figures/section5/targeted/re_1000/err_vmag.pgf
%% Creator: Matplotlib, PGF backend
%%
%% To include the figure in your LaTeX document, write
%%   \input{<filename>.pgf}
%%
%% Make sure the required packages are loaded in your preamble
%%   \usepackage{pgf}
%%
%% Also ensure that all the required font packages are loaded; for instance,
%% the lmodern package is sometimes necessary when using math font.
%%   \usepackage{lmodern}
%%
%% Figures using additional raster images can only be included by \input if
%% they are in the same directory as the main LaTeX file. For loading figures
%% from other directories you can use the `import` package
%% and then include the figures with
%%   \import{<path to file>}{<filename>.pgf}
%%
%% Matplotlib used the following preamble
%%   \def\mathdefault#1{#1}
%%   \everymath=\expandafter{\the\everymath\displaystyle}
%%   \usepackage{amsmath}\usepackage{bm}
%%   \makeatletter\@ifpackageloaded{underscore}{}{\usepackage[strings]{underscore}}\makeatother
%%
\begingroup%
\makeatletter%
\begin{pgfpicture}%
\pgfpathrectangle{\pgfpointorigin}{\pgfqpoint{2.800000in}{2.800000in}}%
\pgfusepath{use as bounding box, clip}%
\begin{pgfscope}%
\pgfsetbuttcap%
\pgfsetmiterjoin%
\definecolor{currentfill}{rgb}{1.000000,1.000000,1.000000}%
\pgfsetfillcolor{currentfill}%
\pgfsetlinewidth{0.000000pt}%
\definecolor{currentstroke}{rgb}{1.000000,1.000000,1.000000}%
\pgfsetstrokecolor{currentstroke}%
\pgfsetdash{}{0pt}%
\pgfpathmoveto{\pgfqpoint{0.000000in}{0.000000in}}%
\pgfpathlineto{\pgfqpoint{2.800000in}{0.000000in}}%
\pgfpathlineto{\pgfqpoint{2.800000in}{2.800000in}}%
\pgfpathlineto{\pgfqpoint{0.000000in}{2.800000in}}%
\pgfpathlineto{\pgfqpoint{0.000000in}{0.000000in}}%
\pgfpathclose%
\pgfusepath{fill}%
\end{pgfscope}%
\begin{pgfscope}%
\pgfsys@transformshift{0.816250in}{0.955000in}%
\pgftext[left,bottom]{\includegraphics[interpolate=true,width=1.213750in,height=1.213750in]{figures/./section5/targeted/re_1000//err_vmag-img0.png}}%
\end{pgfscope}%
\begin{pgfscope}%
\pgfsys@transformshift{0.870000in}{1.008750in}%
\pgftext[left,bottom]{\includegraphics[interpolate=true,width=1.105000in,height=1.105000in]{figures/./section5/targeted/re_1000//err_vmag-img1.png}}%
\end{pgfscope}%
\begin{pgfscope}%
\pgfsys@transformshift{0.423750in}{0.396250in}%
\pgftext[left,bottom]{\includegraphics[interpolate=true,width=1.998750in,height=1.777500in]{figures/./section5/targeted/re_1000//err_vmag-img2.png}}%
\end{pgfscope}%
\begin{pgfscope}%
\pgfsetbuttcap%
\pgfsetmiterjoin%
\definecolor{currentfill}{rgb}{1.000000,1.000000,1.000000}%
\pgfsetfillcolor{currentfill}%
\pgfsetlinewidth{0.000000pt}%
\definecolor{currentstroke}{rgb}{0.000000,0.000000,0.000000}%
\pgfsetstrokecolor{currentstroke}%
\pgfsetstrokeopacity{0.000000}%
\pgfsetdash{}{0pt}%
\pgfpathmoveto{\pgfqpoint{0.334911in}{2.244461in}}%
\pgfpathlineto{\pgfqpoint{2.510628in}{2.244461in}}%
\pgfpathlineto{\pgfqpoint{2.510628in}{2.353247in}}%
\pgfpathlineto{\pgfqpoint{0.334911in}{2.353247in}}%
\pgfpathlineto{\pgfqpoint{0.334911in}{2.244461in}}%
\pgfpathclose%
\pgfusepath{fill}%
\end{pgfscope}%
\begin{pgfscope}%
\pgfsys@transformshift{0.335000in}{2.245000in}%
\pgftext[left,bottom]{\includegraphics[interpolate=true,width=2.176250in,height=0.108750in]{figures/./section5/targeted/re_1000//err_vmag-img3.png}}%
\end{pgfscope}%
\begin{pgfscope}%
\pgfsetbuttcap%
\pgfsetroundjoin%
\definecolor{currentfill}{rgb}{0.000000,0.000000,0.000000}%
\pgfsetfillcolor{currentfill}%
\pgfsetlinewidth{0.803000pt}%
\definecolor{currentstroke}{rgb}{0.000000,0.000000,0.000000}%
\pgfsetstrokecolor{currentstroke}%
\pgfsetdash{}{0pt}%
\pgfsys@defobject{currentmarker}{\pgfqpoint{0.000000in}{0.000000in}}{\pgfqpoint{0.000000in}{0.048611in}}{%
\pgfpathmoveto{\pgfqpoint{0.000000in}{0.000000in}}%
\pgfpathlineto{\pgfqpoint{0.000000in}{0.048611in}}%
\pgfusepath{stroke,fill}%
}%
\begin{pgfscope}%
\pgfsys@transformshift{0.337109in}{2.353247in}%
\pgfsys@useobject{currentmarker}{}%
\end{pgfscope}%
\end{pgfscope}%
\begin{pgfscope}%
\definecolor{textcolor}{rgb}{0.000000,0.000000,0.000000}%
\pgfsetstrokecolor{textcolor}%
\pgfsetfillcolor{textcolor}%
\pgftext[x=0.337109in,y=2.443525in,,bottom]{\color{textcolor}{\rmfamily\fontsize{8.330000}{9.996000}\selectfont\catcode`\^=\active\def^{\ifmmode\sp\else\^{}\fi}\catcode`\%=\active\def%{\%}$\mathdefault{\ensuremath{-}0.865}$}}%
\end{pgfscope}%
\begin{pgfscope}%
\pgfsetbuttcap%
\pgfsetroundjoin%
\definecolor{currentfill}{rgb}{0.000000,0.000000,0.000000}%
\pgfsetfillcolor{currentfill}%
\pgfsetlinewidth{0.803000pt}%
\definecolor{currentstroke}{rgb}{0.000000,0.000000,0.000000}%
\pgfsetstrokecolor{currentstroke}%
\pgfsetdash{}{0pt}%
\pgfsys@defobject{currentmarker}{\pgfqpoint{0.000000in}{0.000000in}}{\pgfqpoint{0.000000in}{0.048611in}}{%
\pgfpathmoveto{\pgfqpoint{0.000000in}{0.000000in}}%
\pgfpathlineto{\pgfqpoint{0.000000in}{0.048611in}}%
\pgfusepath{stroke,fill}%
}%
\begin{pgfscope}%
\pgfsys@transformshift{1.427185in}{2.353247in}%
\pgfsys@useobject{currentmarker}{}%
\end{pgfscope}%
\end{pgfscope}%
\begin{pgfscope}%
\definecolor{textcolor}{rgb}{0.000000,0.000000,0.000000}%
\pgfsetstrokecolor{textcolor}%
\pgfsetfillcolor{textcolor}%
\pgftext[x=1.427185in,y=2.443525in,,bottom]{\color{textcolor}{\rmfamily\fontsize{8.330000}{9.996000}\selectfont\catcode`\^=\active\def^{\ifmmode\sp\else\^{}\fi}\catcode`\%=\active\def%{\%}$\mathdefault{0.007}$}}%
\end{pgfscope}%
\begin{pgfscope}%
\pgfsetbuttcap%
\pgfsetroundjoin%
\definecolor{currentfill}{rgb}{0.000000,0.000000,0.000000}%
\pgfsetfillcolor{currentfill}%
\pgfsetlinewidth{0.803000pt}%
\definecolor{currentstroke}{rgb}{0.000000,0.000000,0.000000}%
\pgfsetstrokecolor{currentstroke}%
\pgfsetdash{}{0pt}%
\pgfsys@defobject{currentmarker}{\pgfqpoint{0.000000in}{0.000000in}}{\pgfqpoint{0.000000in}{0.048611in}}{%
\pgfpathmoveto{\pgfqpoint{0.000000in}{0.000000in}}%
\pgfpathlineto{\pgfqpoint{0.000000in}{0.048611in}}%
\pgfusepath{stroke,fill}%
}%
\begin{pgfscope}%
\pgfsys@transformshift{1.968854in}{2.353247in}%
\pgfsys@useobject{currentmarker}{}%
\end{pgfscope}%
\end{pgfscope}%
\begin{pgfscope}%
\definecolor{textcolor}{rgb}{0.000000,0.000000,0.000000}%
\pgfsetstrokecolor{textcolor}%
\pgfsetfillcolor{textcolor}%
\pgftext[x=1.968854in,y=2.443525in,,bottom]{\color{textcolor}{\rmfamily\fontsize{8.330000}{9.996000}\selectfont\catcode`\^=\active\def^{\ifmmode\sp\else\^{}\fi}\catcode`\%=\active\def%{\%}$\mathdefault{0.879}$}}%
\end{pgfscope}%
\begin{pgfscope}%
\pgfsetbuttcap%
\pgfsetroundjoin%
\definecolor{currentfill}{rgb}{0.000000,0.000000,0.000000}%
\pgfsetfillcolor{currentfill}%
\pgfsetlinewidth{0.803000pt}%
\definecolor{currentstroke}{rgb}{0.000000,0.000000,0.000000}%
\pgfsetstrokecolor{currentstroke}%
\pgfsetdash{}{0pt}%
\pgfsys@defobject{currentmarker}{\pgfqpoint{0.000000in}{0.000000in}}{\pgfqpoint{0.000000in}{0.048611in}}{%
\pgfpathmoveto{\pgfqpoint{0.000000in}{0.000000in}}%
\pgfpathlineto{\pgfqpoint{0.000000in}{0.048611in}}%
\pgfusepath{stroke,fill}%
}%
\begin{pgfscope}%
\pgfsys@transformshift{2.510523in}{2.353247in}%
\pgfsys@useobject{currentmarker}{}%
\end{pgfscope}%
\end{pgfscope}%
\begin{pgfscope}%
\definecolor{textcolor}{rgb}{0.000000,0.000000,0.000000}%
\pgfsetstrokecolor{textcolor}%
\pgfsetfillcolor{textcolor}%
\pgftext[x=2.510523in,y=2.443525in,,bottom]{\color{textcolor}{\rmfamily\fontsize{8.330000}{9.996000}\selectfont\catcode`\^=\active\def^{\ifmmode\sp\else\^{}\fi}\catcode`\%=\active\def%{\%}$\mathdefault{1.752}$}}%
\end{pgfscope}%
\begin{pgfscope}%
\definecolor{textcolor}{rgb}{0.000000,0.000000,0.000000}%
\pgfsetstrokecolor{textcolor}%
\pgfsetfillcolor{textcolor}%
\pgftext[x=1.422769in,y=2.601535in,,base]{\color{textcolor}{\rmfamily\fontsize{10.000000}{12.000000}\selectfont\catcode`\^=\active\def^{\ifmmode\sp\else\^{}\fi}\catcode`\%=\active\def%{\%}$f^{(\bm{v})}_\text{err} \ [\%]$}}%
\end{pgfscope}%
\begin{pgfscope}%
\definecolor{textcolor}{rgb}{0.000000,0.000000,0.000000}%
\pgfsetstrokecolor{textcolor}%
\pgfsetfillcolor{textcolor}%
\pgftext[x=2.510628in,y=2.587647in,right,bottom]{\color{textcolor}{\rmfamily\fontsize{8.330000}{9.996000}\selectfont\catcode`\^=\active\def^{\ifmmode\sp\else\^{}\fi}\catcode`\%=\active\def%{\%}$\times\mathdefault{10^{1}}\mathdefault{}$}}%
\end{pgfscope}%
\begin{pgfscope}%
\pgfsetrectcap%
\pgfsetmiterjoin%
\pgfsetlinewidth{0.803000pt}%
\definecolor{currentstroke}{rgb}{0.000000,0.000000,0.000000}%
\pgfsetstrokecolor{currentstroke}%
\pgfsetdash{}{0pt}%
\pgfpathmoveto{\pgfqpoint{0.334911in}{2.244461in}}%
\pgfpathlineto{\pgfqpoint{0.334911in}{2.298854in}}%
\pgfpathlineto{\pgfqpoint{0.334911in}{2.353247in}}%
\pgfpathlineto{\pgfqpoint{2.510628in}{2.353247in}}%
\pgfpathlineto{\pgfqpoint{2.510628in}{2.298854in}}%
\pgfpathlineto{\pgfqpoint{2.510628in}{2.244461in}}%
\pgfpathlineto{\pgfqpoint{0.334911in}{2.244461in}}%
\pgfpathclose%
\pgfusepath{stroke}%
\end{pgfscope}%
\end{pgfpicture}%
\makeatother%
\endgroup%

%% file: figures/section5/random/re_1000/err_vmag.pgf
%% Creator: Matplotlib, PGF backend
%%
%% To include the figure in your LaTeX document, write
%%   \input{<filename>.pgf}
%%
%% Make sure the required packages are loaded in your preamble
%%   \usepackage{pgf}
%%
%% Also ensure that all the required font packages are loaded; for instance,
%% the lmodern package is sometimes necessary when using math font.
%%   \usepackage{lmodern}
%%
%% Figures using additional raster images can only be included by \input if
%% they are in the same directory as the main LaTeX file. For loading figures
%% from other directories you can use the `import` package
%% and then include the figures with
%%   \import{<path to file>}{<filename>.pgf}
%%
%% Matplotlib used the following preamble
%%   \def\mathdefault#1{#1}
%%   \everymath=\expandafter{\the\everymath\displaystyle}
%%   \usepackage{amsmath}\usepackage{bm}
%%   \makeatletter\@ifpackageloaded{underscore}{}{\usepackage[strings]{underscore}}\makeatother
%%
\begingroup%
\makeatletter%
\begin{pgfpicture}%
\pgfpathrectangle{\pgfpointorigin}{\pgfqpoint{2.800000in}{2.800000in}}%
\pgfusepath{use as bounding box, clip}%
\begin{pgfscope}%
\pgfsetbuttcap%
\pgfsetmiterjoin%
\definecolor{currentfill}{rgb}{1.000000,1.000000,1.000000}%
\pgfsetfillcolor{currentfill}%
\pgfsetlinewidth{0.000000pt}%
\definecolor{currentstroke}{rgb}{1.000000,1.000000,1.000000}%
\pgfsetstrokecolor{currentstroke}%
\pgfsetdash{}{0pt}%
\pgfpathmoveto{\pgfqpoint{0.000000in}{0.000000in}}%
\pgfpathlineto{\pgfqpoint{2.800000in}{0.000000in}}%
\pgfpathlineto{\pgfqpoint{2.800000in}{2.800000in}}%
\pgfpathlineto{\pgfqpoint{0.000000in}{2.800000in}}%
\pgfpathlineto{\pgfqpoint{0.000000in}{0.000000in}}%
\pgfpathclose%
\pgfusepath{fill}%
\end{pgfscope}%
\begin{pgfscope}%
\pgfsys@transformshift{0.816250in}{0.955000in}%
\pgftext[left,bottom]{\includegraphics[interpolate=true,width=1.215000in,height=1.213750in]{figures/./section5/random/re_1000//err_vmag-img0.png}}%
\end{pgfscope}%
\begin{pgfscope}%
\pgfsys@transformshift{0.871250in}{1.008750in}%
\pgftext[left,bottom]{\includegraphics[interpolate=true,width=1.105000in,height=1.105000in]{figures/./section5/random/re_1000//err_vmag-img1.png}}%
\end{pgfscope}%
\begin{pgfscope}%
\pgfsys@transformshift{0.425000in}{0.396250in}%
\pgftext[left,bottom]{\includegraphics[interpolate=true,width=1.998750in,height=1.777500in]{figures/./section5/random/re_1000//err_vmag-img2.png}}%
\end{pgfscope}%
\begin{pgfscope}%
\pgfsetbuttcap%
\pgfsetmiterjoin%
\definecolor{currentfill}{rgb}{1.000000,1.000000,1.000000}%
\pgfsetfillcolor{currentfill}%
\pgfsetlinewidth{0.000000pt}%
\definecolor{currentstroke}{rgb}{0.000000,0.000000,0.000000}%
\pgfsetstrokecolor{currentstroke}%
\pgfsetstrokeopacity{0.000000}%
\pgfsetdash{}{0pt}%
\pgfpathmoveto{\pgfqpoint{0.335864in}{2.244461in}}%
\pgfpathlineto{\pgfqpoint{2.511369in}{2.244461in}}%
\pgfpathlineto{\pgfqpoint{2.511369in}{2.353236in}}%
\pgfpathlineto{\pgfqpoint{0.335864in}{2.353236in}}%
\pgfpathlineto{\pgfqpoint{0.335864in}{2.244461in}}%
\pgfpathclose%
\pgfusepath{fill}%
\end{pgfscope}%
\begin{pgfscope}%
\pgfsys@transformshift{0.336250in}{2.245000in}%
\pgftext[left,bottom]{\includegraphics[interpolate=true,width=2.175000in,height=0.108750in]{figures/./section5/random/re_1000//err_vmag-img3.png}}%
\end{pgfscope}%
\begin{pgfscope}%
\pgfsetbuttcap%
\pgfsetroundjoin%
\definecolor{currentfill}{rgb}{0.000000,0.000000,0.000000}%
\pgfsetfillcolor{currentfill}%
\pgfsetlinewidth{0.803000pt}%
\definecolor{currentstroke}{rgb}{0.000000,0.000000,0.000000}%
\pgfsetstrokecolor{currentstroke}%
\pgfsetdash{}{0pt}%
\pgfsys@defobject{currentmarker}{\pgfqpoint{0.000000in}{0.000000in}}{\pgfqpoint{0.000000in}{0.048611in}}{%
\pgfpathmoveto{\pgfqpoint{0.000000in}{0.000000in}}%
\pgfpathlineto{\pgfqpoint{0.000000in}{0.048611in}}%
\pgfusepath{stroke,fill}%
}%
\begin{pgfscope}%
\pgfsys@transformshift{0.337107in}{2.353236in}%
\pgfsys@useobject{currentmarker}{}%
\end{pgfscope}%
\end{pgfscope}%
\begin{pgfscope}%
\definecolor{textcolor}{rgb}{0.000000,0.000000,0.000000}%
\pgfsetstrokecolor{textcolor}%
\pgfsetfillcolor{textcolor}%
\pgftext[x=0.337107in,y=2.443514in,,bottom]{\color{textcolor}{\rmfamily\fontsize{8.330000}{9.996000}\selectfont\catcode`\^=\active\def^{\ifmmode\sp\else\^{}\fi}\catcode`\%=\active\def%{\%}$\mathdefault{\ensuremath{-}1.227}$}}%
\end{pgfscope}%
\begin{pgfscope}%
\pgfsetbuttcap%
\pgfsetroundjoin%
\definecolor{currentfill}{rgb}{0.000000,0.000000,0.000000}%
\pgfsetfillcolor{currentfill}%
\pgfsetlinewidth{0.803000pt}%
\definecolor{currentstroke}{rgb}{0.000000,0.000000,0.000000}%
\pgfsetstrokecolor{currentstroke}%
\pgfsetdash{}{0pt}%
\pgfsys@defobject{currentmarker}{\pgfqpoint{0.000000in}{0.000000in}}{\pgfqpoint{0.000000in}{0.048611in}}{%
\pgfpathmoveto{\pgfqpoint{0.000000in}{0.000000in}}%
\pgfpathlineto{\pgfqpoint{0.000000in}{0.048611in}}%
\pgfusepath{stroke,fill}%
}%
\begin{pgfscope}%
\pgfsys@transformshift{1.430122in}{2.353236in}%
\pgfsys@useobject{currentmarker}{}%
\end{pgfscope}%
\end{pgfscope}%
\begin{pgfscope}%
\definecolor{textcolor}{rgb}{0.000000,0.000000,0.000000}%
\pgfsetstrokecolor{textcolor}%
\pgfsetfillcolor{textcolor}%
\pgftext[x=1.430122in,y=2.443514in,,bottom]{\color{textcolor}{\rmfamily\fontsize{8.330000}{9.996000}\selectfont\catcode`\^=\active\def^{\ifmmode\sp\else\^{}\fi}\catcode`\%=\active\def%{\%}$\mathdefault{0.015}$}}%
\end{pgfscope}%
\begin{pgfscope}%
\pgfsetbuttcap%
\pgfsetroundjoin%
\definecolor{currentfill}{rgb}{0.000000,0.000000,0.000000}%
\pgfsetfillcolor{currentfill}%
\pgfsetlinewidth{0.803000pt}%
\definecolor{currentstroke}{rgb}{0.000000,0.000000,0.000000}%
\pgfsetstrokecolor{currentstroke}%
\pgfsetdash{}{0pt}%
\pgfsys@defobject{currentmarker}{\pgfqpoint{0.000000in}{0.000000in}}{\pgfqpoint{0.000000in}{0.048611in}}{%
\pgfpathmoveto{\pgfqpoint{0.000000in}{0.000000in}}%
\pgfpathlineto{\pgfqpoint{0.000000in}{0.048611in}}%
\pgfusepath{stroke,fill}%
}%
\begin{pgfscope}%
\pgfsys@transformshift{1.970321in}{2.353236in}%
\pgfsys@useobject{currentmarker}{}%
\end{pgfscope}%
\end{pgfscope}%
\begin{pgfscope}%
\definecolor{textcolor}{rgb}{0.000000,0.000000,0.000000}%
\pgfsetstrokecolor{textcolor}%
\pgfsetfillcolor{textcolor}%
\pgftext[x=1.970321in,y=2.443514in,,bottom]{\color{textcolor}{\rmfamily\fontsize{8.330000}{9.996000}\selectfont\catcode`\^=\active\def^{\ifmmode\sp\else\^{}\fi}\catcode`\%=\active\def%{\%}$\mathdefault{1.257}$}}%
\end{pgfscope}%
\begin{pgfscope}%
\pgfsetbuttcap%
\pgfsetroundjoin%
\definecolor{currentfill}{rgb}{0.000000,0.000000,0.000000}%
\pgfsetfillcolor{currentfill}%
\pgfsetlinewidth{0.803000pt}%
\definecolor{currentstroke}{rgb}{0.000000,0.000000,0.000000}%
\pgfsetstrokecolor{currentstroke}%
\pgfsetdash{}{0pt}%
\pgfsys@defobject{currentmarker}{\pgfqpoint{0.000000in}{0.000000in}}{\pgfqpoint{0.000000in}{0.048611in}}{%
\pgfpathmoveto{\pgfqpoint{0.000000in}{0.000000in}}%
\pgfpathlineto{\pgfqpoint{0.000000in}{0.048611in}}%
\pgfusepath{stroke,fill}%
}%
\begin{pgfscope}%
\pgfsys@transformshift{2.510521in}{2.353236in}%
\pgfsys@useobject{currentmarker}{}%
\end{pgfscope}%
\end{pgfscope}%
\begin{pgfscope}%
\definecolor{textcolor}{rgb}{0.000000,0.000000,0.000000}%
\pgfsetstrokecolor{textcolor}%
\pgfsetfillcolor{textcolor}%
\pgftext[x=2.510521in,y=2.443514in,,bottom]{\color{textcolor}{\rmfamily\fontsize{8.330000}{9.996000}\selectfont\catcode`\^=\active\def^{\ifmmode\sp\else\^{}\fi}\catcode`\%=\active\def%{\%}$\mathdefault{2.498}$}}%
\end{pgfscope}%
\begin{pgfscope}%
\definecolor{textcolor}{rgb}{0.000000,0.000000,0.000000}%
\pgfsetstrokecolor{textcolor}%
\pgfsetfillcolor{textcolor}%
\pgftext[x=1.423616in,y=2.601525in,,base]{\color{textcolor}{\rmfamily\fontsize{10.000000}{12.000000}\selectfont\catcode`\^=\active\def^{\ifmmode\sp\else\^{}\fi}\catcode`\%=\active\def%{\%}$f^{(\bm{v})}_\text{err} \ [\%]$}}%
\end{pgfscope}%
\begin{pgfscope}%
\definecolor{textcolor}{rgb}{0.000000,0.000000,0.000000}%
\pgfsetstrokecolor{textcolor}%
\pgfsetfillcolor{textcolor}%
\pgftext[x=2.511369in,y=2.587636in,right,bottom]{\color{textcolor}{\rmfamily\fontsize{8.330000}{9.996000}\selectfont\catcode`\^=\active\def^{\ifmmode\sp\else\^{}\fi}\catcode`\%=\active\def%{\%}$\times\mathdefault{10^{1}}\mathdefault{}$}}%
\end{pgfscope}%
\begin{pgfscope}%
\pgfsetrectcap%
\pgfsetmiterjoin%
\pgfsetlinewidth{0.803000pt}%
\definecolor{currentstroke}{rgb}{0.000000,0.000000,0.000000}%
\pgfsetstrokecolor{currentstroke}%
\pgfsetdash{}{0pt}%
\pgfpathmoveto{\pgfqpoint{0.335864in}{2.244461in}}%
\pgfpathlineto{\pgfqpoint{0.335864in}{2.298849in}}%
\pgfpathlineto{\pgfqpoint{0.335864in}{2.353236in}}%
\pgfpathlineto{\pgfqpoint{2.511369in}{2.353236in}}%
\pgfpathlineto{\pgfqpoint{2.511369in}{2.298849in}}%
\pgfpathlineto{\pgfqpoint{2.511369in}{2.244461in}}%
\pgfpathlineto{\pgfqpoint{0.335864in}{2.244461in}}%
\pgfpathclose%
\pgfusepath{stroke}%
\end{pgfscope}%
\end{pgfpicture}%
\makeatother%
\endgroup%

%% file: figures/section5/vel_error_Re_targeted_vs_random.pgf
%% Creator: Matplotlib, PGF backend
%%
%% To include the figure in your LaTeX document, write
%%   \input{<filename>.pgf}
%%
%% Make sure the required packages are loaded in your preamble
%%   \usepackage{pgf}
%%
%% Also ensure that all the required font packages are loaded; for instance,
%% the lmodern package is sometimes necessary when using math font.
%%   \usepackage{lmodern}
%%
%% Figures using additional raster images can only be included by \input if
%% they are in the same directory as the main LaTeX file. For loading figures
%% from other directories you can use the `import` package
%% and then include the figures with
%%   \import{<path to file>}{<filename>.pgf}
%%
%% Matplotlib used the following preamble
%%   \def\mathdefault#1{#1}
%%   \everymath=\expandafter{\the\everymath\displaystyle}
%%   \usepackage{amsmath}\usepackage{bm}
%%   \makeatletter\@ifpackageloaded{underscore}{}{\usepackage[strings]{underscore}}\makeatother
%%
\begingroup%
\makeatletter%
\begin{pgfpicture}%
\pgfpathrectangle{\pgfpointorigin}{\pgfqpoint{3.000000in}{2.000000in}}%
\pgfusepath{use as bounding box, clip}%
\begin{pgfscope}%
\pgfsetbuttcap%
\pgfsetmiterjoin%
\definecolor{currentfill}{rgb}{1.000000,1.000000,1.000000}%
\pgfsetfillcolor{currentfill}%
\pgfsetlinewidth{0.000000pt}%
\definecolor{currentstroke}{rgb}{1.000000,1.000000,1.000000}%
\pgfsetstrokecolor{currentstroke}%
\pgfsetdash{}{0pt}%
\pgfpathmoveto{\pgfqpoint{0.000000in}{0.000000in}}%
\pgfpathlineto{\pgfqpoint{3.000000in}{0.000000in}}%
\pgfpathlineto{\pgfqpoint{3.000000in}{2.000000in}}%
\pgfpathlineto{\pgfqpoint{0.000000in}{2.000000in}}%
\pgfpathlineto{\pgfqpoint{0.000000in}{0.000000in}}%
\pgfpathclose%
\pgfusepath{fill}%
\end{pgfscope}%
\begin{pgfscope}%
\pgfsetbuttcap%
\pgfsetmiterjoin%
\definecolor{currentfill}{rgb}{1.000000,1.000000,1.000000}%
\pgfsetfillcolor{currentfill}%
\pgfsetlinewidth{0.000000pt}%
\definecolor{currentstroke}{rgb}{0.000000,0.000000,0.000000}%
\pgfsetstrokecolor{currentstroke}%
\pgfsetstrokeopacity{0.000000}%
\pgfsetdash{}{0pt}%
\pgfpathmoveto{\pgfqpoint{0.597016in}{0.498776in}}%
\pgfpathlineto{\pgfqpoint{2.845756in}{0.498776in}}%
\pgfpathlineto{\pgfqpoint{2.845756in}{1.850000in}}%
\pgfpathlineto{\pgfqpoint{0.597016in}{1.850000in}}%
\pgfpathlineto{\pgfqpoint{0.597016in}{0.498776in}}%
\pgfpathclose%
\pgfusepath{fill}%
\end{pgfscope}%
\begin{pgfscope}%
\pgfpathrectangle{\pgfqpoint{0.597016in}{0.498776in}}{\pgfqpoint{2.248740in}{1.351224in}}%
\pgfusepath{clip}%
\pgfsetbuttcap%
\pgfsetroundjoin%
\definecolor{currentfill}{rgb}{0.172549,0.482353,0.713725}%
\pgfsetfillcolor{currentfill}%
\pgfsetlinewidth{0.501875pt}%
\definecolor{currentstroke}{rgb}{0.000000,0.000000,0.000000}%
\pgfsetstrokecolor{currentstroke}%
\pgfsetdash{}{0pt}%
\pgfsys@defobject{currentmarker}{\pgfqpoint{-0.026896in}{-0.026896in}}{\pgfqpoint{0.026896in}{0.026896in}}{%
\pgfpathmoveto{\pgfqpoint{0.000000in}{0.026896in}}%
\pgfpathlineto{\pgfqpoint{-0.026896in}{-0.026896in}}%
\pgfpathlineto{\pgfqpoint{0.026896in}{-0.026896in}}%
\pgfpathlineto{\pgfqpoint{0.000000in}{0.026896in}}%
\pgfpathclose%
\pgfusepath{stroke,fill}%
}%
\begin{pgfscope}%
\pgfsys@transformshift{0.699231in}{1.652495in}%
\pgfsys@useobject{currentmarker}{}%
\end{pgfscope}%
\begin{pgfscope}%
\pgfsys@transformshift{0.926377in}{1.495936in}%
\pgfsys@useobject{currentmarker}{}%
\end{pgfscope}%
\begin{pgfscope}%
\pgfsys@transformshift{1.153522in}{0.701098in}%
\pgfsys@useobject{currentmarker}{}%
\end{pgfscope}%
\begin{pgfscope}%
\pgfsys@transformshift{1.380668in}{0.560195in}%
\pgfsys@useobject{currentmarker}{}%
\end{pgfscope}%
\begin{pgfscope}%
\pgfsys@transformshift{1.607813in}{0.614389in}%
\pgfsys@useobject{currentmarker}{}%
\end{pgfscope}%
\begin{pgfscope}%
\pgfsys@transformshift{1.834959in}{0.614389in}%
\pgfsys@useobject{currentmarker}{}%
\end{pgfscope}%
\begin{pgfscope}%
\pgfsys@transformshift{2.062104in}{0.590303in}%
\pgfsys@useobject{currentmarker}{}%
\end{pgfscope}%
\begin{pgfscope}%
\pgfsys@transformshift{2.289250in}{0.592711in}%
\pgfsys@useobject{currentmarker}{}%
\end{pgfscope}%
\begin{pgfscope}%
\pgfsys@transformshift{2.516395in}{0.618002in}%
\pgfsys@useobject{currentmarker}{}%
\end{pgfscope}%
\begin{pgfscope}%
\pgfsys@transformshift{2.743541in}{0.719163in}%
\pgfsys@useobject{currentmarker}{}%
\end{pgfscope}%
\end{pgfscope}%
\begin{pgfscope}%
\pgfpathrectangle{\pgfqpoint{0.597016in}{0.498776in}}{\pgfqpoint{2.248740in}{1.351224in}}%
\pgfusepath{clip}%
\pgfsetbuttcap%
\pgfsetroundjoin%
\definecolor{currentfill}{rgb}{0.843137,0.098039,0.109804}%
\pgfsetfillcolor{currentfill}%
\pgfsetlinewidth{0.501875pt}%
\definecolor{currentstroke}{rgb}{0.000000,0.000000,0.000000}%
\pgfsetstrokecolor{currentstroke}%
\pgfsetdash{}{0pt}%
\pgfsys@defobject{currentmarker}{\pgfqpoint{-0.026896in}{-0.026896in}}{\pgfqpoint{0.026896in}{0.026896in}}{%
\pgfpathmoveto{\pgfqpoint{0.000000in}{-0.026896in}}%
\pgfpathcurveto{\pgfqpoint{0.007133in}{-0.026896in}}{\pgfqpoint{0.013974in}{-0.024062in}}{\pgfqpoint{0.019018in}{-0.019018in}}%
\pgfpathcurveto{\pgfqpoint{0.024062in}{-0.013974in}}{\pgfqpoint{0.026896in}{-0.007133in}}{\pgfqpoint{0.026896in}{0.000000in}}%
\pgfpathcurveto{\pgfqpoint{0.026896in}{0.007133in}}{\pgfqpoint{0.024062in}{0.013974in}}{\pgfqpoint{0.019018in}{0.019018in}}%
\pgfpathcurveto{\pgfqpoint{0.013974in}{0.024062in}}{\pgfqpoint{0.007133in}{0.026896in}}{\pgfqpoint{0.000000in}{0.026896in}}%
\pgfpathcurveto{\pgfqpoint{-0.007133in}{0.026896in}}{\pgfqpoint{-0.013974in}{0.024062in}}{\pgfqpoint{-0.019018in}{0.019018in}}%
\pgfpathcurveto{\pgfqpoint{-0.024062in}{0.013974in}}{\pgfqpoint{-0.026896in}{0.007133in}}{\pgfqpoint{-0.026896in}{0.000000in}}%
\pgfpathcurveto{\pgfqpoint{-0.026896in}{-0.007133in}}{\pgfqpoint{-0.024062in}{-0.013974in}}{\pgfqpoint{-0.019018in}{-0.019018in}}%
\pgfpathcurveto{\pgfqpoint{-0.013974in}{-0.024062in}}{\pgfqpoint{-0.007133in}{-0.026896in}}{\pgfqpoint{0.000000in}{-0.026896in}}%
\pgfpathlineto{\pgfqpoint{0.000000in}{-0.026896in}}%
\pgfpathclose%
\pgfusepath{stroke,fill}%
}%
\begin{pgfscope}%
\pgfsys@transformshift{0.699231in}{1.788581in}%
\pgfsys@useobject{currentmarker}{}%
\end{pgfscope}%
\begin{pgfscope}%
\pgfsys@transformshift{0.926377in}{1.651291in}%
\pgfsys@useobject{currentmarker}{}%
\end{pgfscope}%
\begin{pgfscope}%
\pgfsys@transformshift{1.153522in}{0.952797in}%
\pgfsys@useobject{currentmarker}{}%
\end{pgfscope}%
\begin{pgfscope}%
\pgfsys@transformshift{1.380668in}{0.856453in}%
\pgfsys@useobject{currentmarker}{}%
\end{pgfscope}%
\begin{pgfscope}%
\pgfsys@transformshift{1.607813in}{0.990130in}%
\pgfsys@useobject{currentmarker}{}%
\end{pgfscope}%
\begin{pgfscope}%
\pgfsys@transformshift{1.834959in}{1.055162in}%
\pgfsys@useobject{currentmarker}{}%
\end{pgfscope}%
\begin{pgfscope}%
\pgfsys@transformshift{2.062104in}{1.063592in}%
\pgfsys@useobject{currentmarker}{}%
\end{pgfscope}%
\begin{pgfscope}%
\pgfsys@transformshift{2.289250in}{1.051549in}%
\pgfsys@useobject{currentmarker}{}%
\end{pgfscope}%
\begin{pgfscope}%
\pgfsys@transformshift{2.516395in}{1.017829in}%
\pgfsys@useobject{currentmarker}{}%
\end{pgfscope}%
\begin{pgfscope}%
\pgfsys@transformshift{2.743541in}{1.098517in}%
\pgfsys@useobject{currentmarker}{}%
\end{pgfscope}%
\end{pgfscope}%
\begin{pgfscope}%
\pgfsetbuttcap%
\pgfsetroundjoin%
\definecolor{currentfill}{rgb}{0.000000,0.000000,0.000000}%
\pgfsetfillcolor{currentfill}%
\pgfsetlinewidth{0.803000pt}%
\definecolor{currentstroke}{rgb}{0.000000,0.000000,0.000000}%
\pgfsetstrokecolor{currentstroke}%
\pgfsetdash{}{0pt}%
\pgfsys@defobject{currentmarker}{\pgfqpoint{0.000000in}{-0.048611in}}{\pgfqpoint{0.000000in}{0.000000in}}{%
\pgfpathmoveto{\pgfqpoint{0.000000in}{0.000000in}}%
\pgfpathlineto{\pgfqpoint{0.000000in}{-0.048611in}}%
\pgfusepath{stroke,fill}%
}%
\begin{pgfscope}%
\pgfsys@transformshift{0.699231in}{0.498776in}%
\pgfsys@useobject{currentmarker}{}%
\end{pgfscope}%
\end{pgfscope}%
\begin{pgfscope}%
\definecolor{textcolor}{rgb}{0.000000,0.000000,0.000000}%
\pgfsetstrokecolor{textcolor}%
\pgfsetfillcolor{textcolor}%
\pgftext[x=0.699231in,y=0.408498in,,top]{\color{textcolor}{\rmfamily\fontsize{6.500000}{7.800000}\selectfont\catcode`\^=\active\def^{\ifmmode\sp\else\^{}\fi}\catcode`\%=\active\def%{\%}30}}%
\end{pgfscope}%
\begin{pgfscope}%
\pgfsetbuttcap%
\pgfsetroundjoin%
\definecolor{currentfill}{rgb}{0.000000,0.000000,0.000000}%
\pgfsetfillcolor{currentfill}%
\pgfsetlinewidth{0.803000pt}%
\definecolor{currentstroke}{rgb}{0.000000,0.000000,0.000000}%
\pgfsetstrokecolor{currentstroke}%
\pgfsetdash{}{0pt}%
\pgfsys@defobject{currentmarker}{\pgfqpoint{0.000000in}{-0.048611in}}{\pgfqpoint{0.000000in}{0.000000in}}{%
\pgfpathmoveto{\pgfqpoint{0.000000in}{0.000000in}}%
\pgfpathlineto{\pgfqpoint{0.000000in}{-0.048611in}}%
\pgfusepath{stroke,fill}%
}%
\begin{pgfscope}%
\pgfsys@transformshift{0.926377in}{0.498776in}%
\pgfsys@useobject{currentmarker}{}%
\end{pgfscope}%
\end{pgfscope}%
\begin{pgfscope}%
\definecolor{textcolor}{rgb}{0.000000,0.000000,0.000000}%
\pgfsetstrokecolor{textcolor}%
\pgfsetfillcolor{textcolor}%
\pgftext[x=0.926377in,y=0.408498in,,top]{\color{textcolor}{\rmfamily\fontsize{6.500000}{7.800000}\selectfont\catcode`\^=\active\def^{\ifmmode\sp\else\^{}\fi}\catcode`\%=\active\def%{\%}50}}%
\end{pgfscope}%
\begin{pgfscope}%
\pgfsetbuttcap%
\pgfsetroundjoin%
\definecolor{currentfill}{rgb}{0.000000,0.000000,0.000000}%
\pgfsetfillcolor{currentfill}%
\pgfsetlinewidth{0.803000pt}%
\definecolor{currentstroke}{rgb}{0.000000,0.000000,0.000000}%
\pgfsetstrokecolor{currentstroke}%
\pgfsetdash{}{0pt}%
\pgfsys@defobject{currentmarker}{\pgfqpoint{0.000000in}{-0.048611in}}{\pgfqpoint{0.000000in}{0.000000in}}{%
\pgfpathmoveto{\pgfqpoint{0.000000in}{0.000000in}}%
\pgfpathlineto{\pgfqpoint{0.000000in}{-0.048611in}}%
\pgfusepath{stroke,fill}%
}%
\begin{pgfscope}%
\pgfsys@transformshift{1.153522in}{0.498776in}%
\pgfsys@useobject{currentmarker}{}%
\end{pgfscope}%
\end{pgfscope}%
\begin{pgfscope}%
\definecolor{textcolor}{rgb}{0.000000,0.000000,0.000000}%
\pgfsetstrokecolor{textcolor}%
\pgfsetfillcolor{textcolor}%
\pgftext[x=1.153522in,y=0.408498in,,top]{\color{textcolor}{\rmfamily\fontsize{6.500000}{7.800000}\selectfont\catcode`\^=\active\def^{\ifmmode\sp\else\^{}\fi}\catcode`\%=\active\def%{\%}100}}%
\end{pgfscope}%
\begin{pgfscope}%
\pgfsetbuttcap%
\pgfsetroundjoin%
\definecolor{currentfill}{rgb}{0.000000,0.000000,0.000000}%
\pgfsetfillcolor{currentfill}%
\pgfsetlinewidth{0.803000pt}%
\definecolor{currentstroke}{rgb}{0.000000,0.000000,0.000000}%
\pgfsetstrokecolor{currentstroke}%
\pgfsetdash{}{0pt}%
\pgfsys@defobject{currentmarker}{\pgfqpoint{0.000000in}{-0.048611in}}{\pgfqpoint{0.000000in}{0.000000in}}{%
\pgfpathmoveto{\pgfqpoint{0.000000in}{0.000000in}}%
\pgfpathlineto{\pgfqpoint{0.000000in}{-0.048611in}}%
\pgfusepath{stroke,fill}%
}%
\begin{pgfscope}%
\pgfsys@transformshift{1.380668in}{0.498776in}%
\pgfsys@useobject{currentmarker}{}%
\end{pgfscope}%
\end{pgfscope}%
\begin{pgfscope}%
\definecolor{textcolor}{rgb}{0.000000,0.000000,0.000000}%
\pgfsetstrokecolor{textcolor}%
\pgfsetfillcolor{textcolor}%
\pgftext[x=1.380668in,y=0.408498in,,top]{\color{textcolor}{\rmfamily\fontsize{6.500000}{7.800000}\selectfont\catcode`\^=\active\def^{\ifmmode\sp\else\^{}\fi}\catcode`\%=\active\def%{\%}500}}%
\end{pgfscope}%
\begin{pgfscope}%
\pgfsetbuttcap%
\pgfsetroundjoin%
\definecolor{currentfill}{rgb}{0.000000,0.000000,0.000000}%
\pgfsetfillcolor{currentfill}%
\pgfsetlinewidth{0.803000pt}%
\definecolor{currentstroke}{rgb}{0.000000,0.000000,0.000000}%
\pgfsetstrokecolor{currentstroke}%
\pgfsetdash{}{0pt}%
\pgfsys@defobject{currentmarker}{\pgfqpoint{0.000000in}{-0.048611in}}{\pgfqpoint{0.000000in}{0.000000in}}{%
\pgfpathmoveto{\pgfqpoint{0.000000in}{0.000000in}}%
\pgfpathlineto{\pgfqpoint{0.000000in}{-0.048611in}}%
\pgfusepath{stroke,fill}%
}%
\begin{pgfscope}%
\pgfsys@transformshift{1.607813in}{0.498776in}%
\pgfsys@useobject{currentmarker}{}%
\end{pgfscope}%
\end{pgfscope}%
\begin{pgfscope}%
\definecolor{textcolor}{rgb}{0.000000,0.000000,0.000000}%
\pgfsetstrokecolor{textcolor}%
\pgfsetfillcolor{textcolor}%
\pgftext[x=1.607813in,y=0.408498in,,top]{\color{textcolor}{\rmfamily\fontsize{6.500000}{7.800000}\selectfont\catcode`\^=\active\def^{\ifmmode\sp\else\^{}\fi}\catcode`\%=\active\def%{\%}1000}}%
\end{pgfscope}%
\begin{pgfscope}%
\pgfsetbuttcap%
\pgfsetroundjoin%
\definecolor{currentfill}{rgb}{0.000000,0.000000,0.000000}%
\pgfsetfillcolor{currentfill}%
\pgfsetlinewidth{0.803000pt}%
\definecolor{currentstroke}{rgb}{0.000000,0.000000,0.000000}%
\pgfsetstrokecolor{currentstroke}%
\pgfsetdash{}{0pt}%
\pgfsys@defobject{currentmarker}{\pgfqpoint{0.000000in}{-0.048611in}}{\pgfqpoint{0.000000in}{0.000000in}}{%
\pgfpathmoveto{\pgfqpoint{0.000000in}{0.000000in}}%
\pgfpathlineto{\pgfqpoint{0.000000in}{-0.048611in}}%
\pgfusepath{stroke,fill}%
}%
\begin{pgfscope}%
\pgfsys@transformshift{1.834959in}{0.498776in}%
\pgfsys@useobject{currentmarker}{}%
\end{pgfscope}%
\end{pgfscope}%
\begin{pgfscope}%
\definecolor{textcolor}{rgb}{0.000000,0.000000,0.000000}%
\pgfsetstrokecolor{textcolor}%
\pgfsetfillcolor{textcolor}%
\pgftext[x=1.834959in,y=0.408498in,,top]{\color{textcolor}{\rmfamily\fontsize{6.500000}{7.800000}\selectfont\catcode`\^=\active\def^{\ifmmode\sp\else\^{}\fi}\catcode`\%=\active\def%{\%}2000}}%
\end{pgfscope}%
\begin{pgfscope}%
\pgfsetbuttcap%
\pgfsetroundjoin%
\definecolor{currentfill}{rgb}{0.000000,0.000000,0.000000}%
\pgfsetfillcolor{currentfill}%
\pgfsetlinewidth{0.803000pt}%
\definecolor{currentstroke}{rgb}{0.000000,0.000000,0.000000}%
\pgfsetstrokecolor{currentstroke}%
\pgfsetdash{}{0pt}%
\pgfsys@defobject{currentmarker}{\pgfqpoint{0.000000in}{-0.048611in}}{\pgfqpoint{0.000000in}{0.000000in}}{%
\pgfpathmoveto{\pgfqpoint{0.000000in}{0.000000in}}%
\pgfpathlineto{\pgfqpoint{0.000000in}{-0.048611in}}%
\pgfusepath{stroke,fill}%
}%
\begin{pgfscope}%
\pgfsys@transformshift{2.062104in}{0.498776in}%
\pgfsys@useobject{currentmarker}{}%
\end{pgfscope}%
\end{pgfscope}%
\begin{pgfscope}%
\definecolor{textcolor}{rgb}{0.000000,0.000000,0.000000}%
\pgfsetstrokecolor{textcolor}%
\pgfsetfillcolor{textcolor}%
\pgftext[x=2.062104in,y=0.408498in,,top]{\color{textcolor}{\rmfamily\fontsize{6.500000}{7.800000}\selectfont\catcode`\^=\active\def^{\ifmmode\sp\else\^{}\fi}\catcode`\%=\active\def%{\%}3000}}%
\end{pgfscope}%
\begin{pgfscope}%
\pgfsetbuttcap%
\pgfsetroundjoin%
\definecolor{currentfill}{rgb}{0.000000,0.000000,0.000000}%
\pgfsetfillcolor{currentfill}%
\pgfsetlinewidth{0.803000pt}%
\definecolor{currentstroke}{rgb}{0.000000,0.000000,0.000000}%
\pgfsetstrokecolor{currentstroke}%
\pgfsetdash{}{0pt}%
\pgfsys@defobject{currentmarker}{\pgfqpoint{0.000000in}{-0.048611in}}{\pgfqpoint{0.000000in}{0.000000in}}{%
\pgfpathmoveto{\pgfqpoint{0.000000in}{0.000000in}}%
\pgfpathlineto{\pgfqpoint{0.000000in}{-0.048611in}}%
\pgfusepath{stroke,fill}%
}%
\begin{pgfscope}%
\pgfsys@transformshift{2.289250in}{0.498776in}%
\pgfsys@useobject{currentmarker}{}%
\end{pgfscope}%
\end{pgfscope}%
\begin{pgfscope}%
\definecolor{textcolor}{rgb}{0.000000,0.000000,0.000000}%
\pgfsetstrokecolor{textcolor}%
\pgfsetfillcolor{textcolor}%
\pgftext[x=2.289250in,y=0.408498in,,top]{\color{textcolor}{\rmfamily\fontsize{6.500000}{7.800000}\selectfont\catcode`\^=\active\def^{\ifmmode\sp\else\^{}\fi}\catcode`\%=\active\def%{\%}4000}}%
\end{pgfscope}%
\begin{pgfscope}%
\pgfsetbuttcap%
\pgfsetroundjoin%
\definecolor{currentfill}{rgb}{0.000000,0.000000,0.000000}%
\pgfsetfillcolor{currentfill}%
\pgfsetlinewidth{0.803000pt}%
\definecolor{currentstroke}{rgb}{0.000000,0.000000,0.000000}%
\pgfsetstrokecolor{currentstroke}%
\pgfsetdash{}{0pt}%
\pgfsys@defobject{currentmarker}{\pgfqpoint{0.000000in}{-0.048611in}}{\pgfqpoint{0.000000in}{0.000000in}}{%
\pgfpathmoveto{\pgfqpoint{0.000000in}{0.000000in}}%
\pgfpathlineto{\pgfqpoint{0.000000in}{-0.048611in}}%
\pgfusepath{stroke,fill}%
}%
\begin{pgfscope}%
\pgfsys@transformshift{2.516395in}{0.498776in}%
\pgfsys@useobject{currentmarker}{}%
\end{pgfscope}%
\end{pgfscope}%
\begin{pgfscope}%
\definecolor{textcolor}{rgb}{0.000000,0.000000,0.000000}%
\pgfsetstrokecolor{textcolor}%
\pgfsetfillcolor{textcolor}%
\pgftext[x=2.516395in,y=0.408498in,,top]{\color{textcolor}{\rmfamily\fontsize{6.500000}{7.800000}\selectfont\catcode`\^=\active\def^{\ifmmode\sp\else\^{}\fi}\catcode`\%=\active\def%{\%}5000}}%
\end{pgfscope}%
\begin{pgfscope}%
\pgfsetbuttcap%
\pgfsetroundjoin%
\definecolor{currentfill}{rgb}{0.000000,0.000000,0.000000}%
\pgfsetfillcolor{currentfill}%
\pgfsetlinewidth{0.803000pt}%
\definecolor{currentstroke}{rgb}{0.000000,0.000000,0.000000}%
\pgfsetstrokecolor{currentstroke}%
\pgfsetdash{}{0pt}%
\pgfsys@defobject{currentmarker}{\pgfqpoint{0.000000in}{-0.048611in}}{\pgfqpoint{0.000000in}{0.000000in}}{%
\pgfpathmoveto{\pgfqpoint{0.000000in}{0.000000in}}%
\pgfpathlineto{\pgfqpoint{0.000000in}{-0.048611in}}%
\pgfusepath{stroke,fill}%
}%
\begin{pgfscope}%
\pgfsys@transformshift{2.743541in}{0.498776in}%
\pgfsys@useobject{currentmarker}{}%
\end{pgfscope}%
\end{pgfscope}%
\begin{pgfscope}%
\definecolor{textcolor}{rgb}{0.000000,0.000000,0.000000}%
\pgfsetstrokecolor{textcolor}%
\pgfsetfillcolor{textcolor}%
\pgftext[x=2.743541in,y=0.408498in,,top]{\color{textcolor}{\rmfamily\fontsize{6.500000}{7.800000}\selectfont\catcode`\^=\active\def^{\ifmmode\sp\else\^{}\fi}\catcode`\%=\active\def%{\%}6000}}%
\end{pgfscope}%
\begin{pgfscope}%
\definecolor{textcolor}{rgb}{0.000000,0.000000,0.000000}%
\pgfsetstrokecolor{textcolor}%
\pgfsetfillcolor{textcolor}%
\pgftext[x=1.721386in,y=0.278868in,,top]{\color{textcolor}{\rmfamily\fontsize{10.000000}{12.000000}\selectfont\catcode`\^=\active\def^{\ifmmode\sp\else\^{}\fi}\catcode`\%=\active\def%{\%}$\text{Re}$}}%
\end{pgfscope}%
\begin{pgfscope}%
\pgfsetbuttcap%
\pgfsetroundjoin%
\definecolor{currentfill}{rgb}{0.000000,0.000000,0.000000}%
\pgfsetfillcolor{currentfill}%
\pgfsetlinewidth{0.803000pt}%
\definecolor{currentstroke}{rgb}{0.000000,0.000000,0.000000}%
\pgfsetstrokecolor{currentstroke}%
\pgfsetdash{}{0pt}%
\pgfsys@defobject{currentmarker}{\pgfqpoint{-0.048611in}{0.000000in}}{\pgfqpoint{-0.000000in}{0.000000in}}{%
\pgfpathmoveto{\pgfqpoint{-0.000000in}{0.000000in}}%
\pgfpathlineto{\pgfqpoint{-0.048611in}{0.000000in}}%
\pgfusepath{stroke,fill}%
}%
\begin{pgfscope}%
\pgfsys@transformshift{0.597016in}{0.536109in}%
\pgfsys@useobject{currentmarker}{}%
\end{pgfscope}%
\end{pgfscope}%
\begin{pgfscope}%
\definecolor{textcolor}{rgb}{0.000000,0.000000,0.000000}%
\pgfsetstrokecolor{textcolor}%
\pgfsetfillcolor{textcolor}%
\pgftext[x=0.373252in, y=0.507174in, left, base]{\color{textcolor}{\rmfamily\fontsize{6.500000}{7.800000}\selectfont\catcode`\^=\active\def^{\ifmmode\sp\else\^{}\fi}\catcode`\%=\active\def%{\%}$\mathdefault{2.5}$}}%
\end{pgfscope}%
\begin{pgfscope}%
\pgfsetbuttcap%
\pgfsetroundjoin%
\definecolor{currentfill}{rgb}{0.000000,0.000000,0.000000}%
\pgfsetfillcolor{currentfill}%
\pgfsetlinewidth{0.803000pt}%
\definecolor{currentstroke}{rgb}{0.000000,0.000000,0.000000}%
\pgfsetstrokecolor{currentstroke}%
\pgfsetdash{}{0pt}%
\pgfsys@defobject{currentmarker}{\pgfqpoint{-0.048611in}{0.000000in}}{\pgfqpoint{-0.000000in}{0.000000in}}{%
\pgfpathmoveto{\pgfqpoint{-0.000000in}{0.000000in}}%
\pgfpathlineto{\pgfqpoint{-0.048611in}{0.000000in}}%
\pgfusepath{stroke,fill}%
}%
\begin{pgfscope}%
\pgfsys@transformshift{0.597016in}{0.837184in}%
\pgfsys@useobject{currentmarker}{}%
\end{pgfscope}%
\end{pgfscope}%
\begin{pgfscope}%
\definecolor{textcolor}{rgb}{0.000000,0.000000,0.000000}%
\pgfsetstrokecolor{textcolor}%
\pgfsetfillcolor{textcolor}%
\pgftext[x=0.373252in, y=0.808249in, left, base]{\color{textcolor}{\rmfamily\fontsize{6.500000}{7.800000}\selectfont\catcode`\^=\active\def^{\ifmmode\sp\else\^{}\fi}\catcode`\%=\active\def%{\%}$\mathdefault{5.0}$}}%
\end{pgfscope}%
\begin{pgfscope}%
\pgfsetbuttcap%
\pgfsetroundjoin%
\definecolor{currentfill}{rgb}{0.000000,0.000000,0.000000}%
\pgfsetfillcolor{currentfill}%
\pgfsetlinewidth{0.803000pt}%
\definecolor{currentstroke}{rgb}{0.000000,0.000000,0.000000}%
\pgfsetstrokecolor{currentstroke}%
\pgfsetdash{}{0pt}%
\pgfsys@defobject{currentmarker}{\pgfqpoint{-0.048611in}{0.000000in}}{\pgfqpoint{-0.000000in}{0.000000in}}{%
\pgfpathmoveto{\pgfqpoint{-0.000000in}{0.000000in}}%
\pgfpathlineto{\pgfqpoint{-0.048611in}{0.000000in}}%
\pgfusepath{stroke,fill}%
}%
\begin{pgfscope}%
\pgfsys@transformshift{0.597016in}{1.138259in}%
\pgfsys@useobject{currentmarker}{}%
\end{pgfscope}%
\end{pgfscope}%
\begin{pgfscope}%
\definecolor{textcolor}{rgb}{0.000000,0.000000,0.000000}%
\pgfsetstrokecolor{textcolor}%
\pgfsetfillcolor{textcolor}%
\pgftext[x=0.373252in, y=1.109324in, left, base]{\color{textcolor}{\rmfamily\fontsize{6.500000}{7.800000}\selectfont\catcode`\^=\active\def^{\ifmmode\sp\else\^{}\fi}\catcode`\%=\active\def%{\%}$\mathdefault{7.5}$}}%
\end{pgfscope}%
\begin{pgfscope}%
\pgfsetbuttcap%
\pgfsetroundjoin%
\definecolor{currentfill}{rgb}{0.000000,0.000000,0.000000}%
\pgfsetfillcolor{currentfill}%
\pgfsetlinewidth{0.803000pt}%
\definecolor{currentstroke}{rgb}{0.000000,0.000000,0.000000}%
\pgfsetstrokecolor{currentstroke}%
\pgfsetdash{}{0pt}%
\pgfsys@defobject{currentmarker}{\pgfqpoint{-0.048611in}{0.000000in}}{\pgfqpoint{-0.000000in}{0.000000in}}{%
\pgfpathmoveto{\pgfqpoint{-0.000000in}{0.000000in}}%
\pgfpathlineto{\pgfqpoint{-0.048611in}{0.000000in}}%
\pgfusepath{stroke,fill}%
}%
\begin{pgfscope}%
\pgfsys@transformshift{0.597016in}{1.439334in}%
\pgfsys@useobject{currentmarker}{}%
\end{pgfscope}%
\end{pgfscope}%
\begin{pgfscope}%
\definecolor{textcolor}{rgb}{0.000000,0.000000,0.000000}%
\pgfsetstrokecolor{textcolor}%
\pgfsetfillcolor{textcolor}%
\pgftext[x=0.322327in, y=1.410399in, left, base]{\color{textcolor}{\rmfamily\fontsize{6.500000}{7.800000}\selectfont\catcode`\^=\active\def^{\ifmmode\sp\else\^{}\fi}\catcode`\%=\active\def%{\%}$\mathdefault{10.0}$}}%
\end{pgfscope}%
\begin{pgfscope}%
\pgfsetbuttcap%
\pgfsetroundjoin%
\definecolor{currentfill}{rgb}{0.000000,0.000000,0.000000}%
\pgfsetfillcolor{currentfill}%
\pgfsetlinewidth{0.803000pt}%
\definecolor{currentstroke}{rgb}{0.000000,0.000000,0.000000}%
\pgfsetstrokecolor{currentstroke}%
\pgfsetdash{}{0pt}%
\pgfsys@defobject{currentmarker}{\pgfqpoint{-0.048611in}{0.000000in}}{\pgfqpoint{-0.000000in}{0.000000in}}{%
\pgfpathmoveto{\pgfqpoint{-0.000000in}{0.000000in}}%
\pgfpathlineto{\pgfqpoint{-0.048611in}{0.000000in}}%
\pgfusepath{stroke,fill}%
}%
\begin{pgfscope}%
\pgfsys@transformshift{0.597016in}{1.740409in}%
\pgfsys@useobject{currentmarker}{}%
\end{pgfscope}%
\end{pgfscope}%
\begin{pgfscope}%
\definecolor{textcolor}{rgb}{0.000000,0.000000,0.000000}%
\pgfsetstrokecolor{textcolor}%
\pgfsetfillcolor{textcolor}%
\pgftext[x=0.322327in, y=1.711474in, left, base]{\color{textcolor}{\rmfamily\fontsize{6.500000}{7.800000}\selectfont\catcode`\^=\active\def^{\ifmmode\sp\else\^{}\fi}\catcode`\%=\active\def%{\%}$\mathdefault{12.5}$}}%
\end{pgfscope}%
\begin{pgfscope}%
\definecolor{textcolor}{rgb}{0.000000,0.000000,0.000000}%
\pgfsetstrokecolor{textcolor}%
\pgfsetfillcolor{textcolor}%
\pgftext[x=0.266771in,y=1.174388in,,bottom,rotate=90.000000]{\color{textcolor}{\rmfamily\fontsize{10.000000}{12.000000}\selectfont\catcode`\^=\active\def^{\ifmmode\sp\else\^{}\fi}\catcode`\%=\active\def%{\%}$\delta_{\ell^1}^{(\bm{v})} \ [\%]$}}%
\end{pgfscope}%
\begin{pgfscope}%
\pgfsetrectcap%
\pgfsetmiterjoin%
\pgfsetlinewidth{0.803000pt}%
\definecolor{currentstroke}{rgb}{0.000000,0.000000,0.000000}%
\pgfsetstrokecolor{currentstroke}%
\pgfsetdash{}{0pt}%
\pgfpathmoveto{\pgfqpoint{0.597016in}{0.498776in}}%
\pgfpathlineto{\pgfqpoint{0.597016in}{1.850000in}}%
\pgfusepath{stroke}%
\end{pgfscope}%
\begin{pgfscope}%
\pgfsetrectcap%
\pgfsetmiterjoin%
\pgfsetlinewidth{0.803000pt}%
\definecolor{currentstroke}{rgb}{0.000000,0.000000,0.000000}%
\pgfsetstrokecolor{currentstroke}%
\pgfsetdash{}{0pt}%
\pgfpathmoveto{\pgfqpoint{2.845756in}{0.498776in}}%
\pgfpathlineto{\pgfqpoint{2.845756in}{1.850000in}}%
\pgfusepath{stroke}%
\end{pgfscope}%
\begin{pgfscope}%
\pgfsetrectcap%
\pgfsetmiterjoin%
\pgfsetlinewidth{0.803000pt}%
\definecolor{currentstroke}{rgb}{0.000000,0.000000,0.000000}%
\pgfsetstrokecolor{currentstroke}%
\pgfsetdash{}{0pt}%
\pgfpathmoveto{\pgfqpoint{0.597016in}{0.498776in}}%
\pgfpathlineto{\pgfqpoint{2.845756in}{0.498776in}}%
\pgfusepath{stroke}%
\end{pgfscope}%
\begin{pgfscope}%
\pgfsetrectcap%
\pgfsetmiterjoin%
\pgfsetlinewidth{0.803000pt}%
\definecolor{currentstroke}{rgb}{0.000000,0.000000,0.000000}%
\pgfsetstrokecolor{currentstroke}%
\pgfsetdash{}{0pt}%
\pgfpathmoveto{\pgfqpoint{0.597016in}{1.850000in}}%
\pgfpathlineto{\pgfqpoint{2.845756in}{1.850000in}}%
\pgfusepath{stroke}%
\end{pgfscope}%
\begin{pgfscope}%
\pgfsetbuttcap%
\pgfsetmiterjoin%
\definecolor{currentfill}{rgb}{1.000000,1.000000,1.000000}%
\pgfsetfillcolor{currentfill}%
\pgfsetfillopacity{0.800000}%
\pgfsetlinewidth{1.003750pt}%
\definecolor{currentstroke}{rgb}{0.800000,0.800000,0.800000}%
\pgfsetstrokecolor{currentstroke}%
\pgfsetstrokeopacity{0.800000}%
\pgfsetdash{}{0pt}%
\pgfpathmoveto{\pgfqpoint{2.052459in}{1.359852in}}%
\pgfpathlineto{\pgfqpoint{2.764770in}{1.359852in}}%
\pgfpathquadraticcurveto{\pgfqpoint{2.787909in}{1.359852in}}{\pgfqpoint{2.787909in}{1.382990in}}%
\pgfpathlineto{\pgfqpoint{2.787909in}{1.769014in}}%
\pgfpathquadraticcurveto{\pgfqpoint{2.787909in}{1.792153in}}{\pgfqpoint{2.764770in}{1.792153in}}%
\pgfpathlineto{\pgfqpoint{2.052459in}{1.792153in}}%
\pgfpathquadraticcurveto{\pgfqpoint{2.029320in}{1.792153in}}{\pgfqpoint{2.029320in}{1.769014in}}%
\pgfpathlineto{\pgfqpoint{2.029320in}{1.382990in}}%
\pgfpathquadraticcurveto{\pgfqpoint{2.029320in}{1.359852in}}{\pgfqpoint{2.052459in}{1.359852in}}%
\pgfpathlineto{\pgfqpoint{2.052459in}{1.359852in}}%
\pgfpathclose%
\pgfusepath{stroke,fill}%
\end{pgfscope}%
\begin{pgfscope}%
\pgfsetbuttcap%
\pgfsetroundjoin%
\definecolor{currentfill}{rgb}{0.172549,0.482353,0.713725}%
\pgfsetfillcolor{currentfill}%
\pgfsetlinewidth{0.501875pt}%
\definecolor{currentstroke}{rgb}{0.000000,0.000000,0.000000}%
\pgfsetstrokecolor{currentstroke}%
\pgfsetdash{}{0pt}%
\pgfsys@defobject{currentmarker}{\pgfqpoint{-0.026896in}{-0.026896in}}{\pgfqpoint{0.026896in}{0.026896in}}{%
\pgfpathmoveto{\pgfqpoint{0.000000in}{0.026896in}}%
\pgfpathlineto{\pgfqpoint{-0.026896in}{-0.026896in}}%
\pgfpathlineto{\pgfqpoint{0.026896in}{-0.026896in}}%
\pgfpathlineto{\pgfqpoint{0.000000in}{0.026896in}}%
\pgfpathclose%
\pgfusepath{stroke,fill}%
}%
\begin{pgfscope}%
\pgfsys@transformshift{2.191293in}{1.673143in}%
\pgfsys@useobject{currentmarker}{}%
\end{pgfscope}%
\end{pgfscope}%
\begin{pgfscope}%
\definecolor{textcolor}{rgb}{0.000000,0.000000,0.000000}%
\pgfsetstrokecolor{textcolor}%
\pgfsetfillcolor{textcolor}%
\pgftext[x=2.399543in,y=1.642773in,left,base]{\color{textcolor}{\rmfamily\fontsize{8.330000}{9.996000}\selectfont\catcode`\^=\active\def^{\ifmmode\sp\else\^{}\fi}\catcode`\%=\active\def%{\%}$\bm{v}^\text{PINN}_\text{target}$}}%
\end{pgfscope}%
\begin{pgfscope}%
\pgfsetbuttcap%
\pgfsetroundjoin%
\definecolor{currentfill}{rgb}{0.843137,0.098039,0.109804}%
\pgfsetfillcolor{currentfill}%
\pgfsetlinewidth{0.501875pt}%
\definecolor{currentstroke}{rgb}{0.000000,0.000000,0.000000}%
\pgfsetstrokecolor{currentstroke}%
\pgfsetdash{}{0pt}%
\pgfsys@defobject{currentmarker}{\pgfqpoint{-0.026896in}{-0.026896in}}{\pgfqpoint{0.026896in}{0.026896in}}{%
\pgfpathmoveto{\pgfqpoint{0.000000in}{-0.026896in}}%
\pgfpathcurveto{\pgfqpoint{0.007133in}{-0.026896in}}{\pgfqpoint{0.013974in}{-0.024062in}}{\pgfqpoint{0.019018in}{-0.019018in}}%
\pgfpathcurveto{\pgfqpoint{0.024062in}{-0.013974in}}{\pgfqpoint{0.026896in}{-0.007133in}}{\pgfqpoint{0.026896in}{0.000000in}}%
\pgfpathcurveto{\pgfqpoint{0.026896in}{0.007133in}}{\pgfqpoint{0.024062in}{0.013974in}}{\pgfqpoint{0.019018in}{0.019018in}}%
\pgfpathcurveto{\pgfqpoint{0.013974in}{0.024062in}}{\pgfqpoint{0.007133in}{0.026896in}}{\pgfqpoint{0.000000in}{0.026896in}}%
\pgfpathcurveto{\pgfqpoint{-0.007133in}{0.026896in}}{\pgfqpoint{-0.013974in}{0.024062in}}{\pgfqpoint{-0.019018in}{0.019018in}}%
\pgfpathcurveto{\pgfqpoint{-0.024062in}{0.013974in}}{\pgfqpoint{-0.026896in}{0.007133in}}{\pgfqpoint{-0.026896in}{0.000000in}}%
\pgfpathcurveto{\pgfqpoint{-0.026896in}{-0.007133in}}{\pgfqpoint{-0.024062in}{-0.013974in}}{\pgfqpoint{-0.019018in}{-0.019018in}}%
\pgfpathcurveto{\pgfqpoint{-0.013974in}{-0.024062in}}{\pgfqpoint{-0.007133in}{-0.026896in}}{\pgfqpoint{0.000000in}{-0.026896in}}%
\pgfpathlineto{\pgfqpoint{0.000000in}{-0.026896in}}%
\pgfpathclose%
\pgfusepath{stroke,fill}%
}%
\begin{pgfscope}%
\pgfsys@transformshift{2.191293in}{1.468212in}%
\pgfsys@useobject{currentmarker}{}%
\end{pgfscope}%
\end{pgfscope}%
\begin{pgfscope}%
\definecolor{textcolor}{rgb}{0.000000,0.000000,0.000000}%
\pgfsetstrokecolor{textcolor}%
\pgfsetfillcolor{textcolor}%
\pgftext[x=2.399543in,y=1.437842in,left,base]{\color{textcolor}{\rmfamily\fontsize{8.330000}{9.996000}\selectfont\catcode`\^=\active\def^{\ifmmode\sp\else\^{}\fi}\catcode`\%=\active\def%{\%}$\bm{v}^\text{PINN}_\text{rand}$}}%
\end{pgfscope}%
\end{pgfpicture}%
\makeatother%
\endgroup%

%% file: figures/section5/press_error_Re_targeted_vs_random.pgf
%% Creator: Matplotlib, PGF backend
%%
%% To include the figure in your LaTeX document, write
%%   \input{<filename>.pgf}
%%
%% Make sure the required packages are loaded in your preamble
%%   \usepackage{pgf}
%%
%% Also ensure that all the required font packages are loaded; for instance,
%% the lmodern package is sometimes necessary when using math font.
%%   \usepackage{lmodern}
%%
%% Figures using additional raster images can only be included by \input if
%% they are in the same directory as the main LaTeX file. For loading figures
%% from other directories you can use the `import` package
%% and then include the figures with
%%   \import{<path to file>}{<filename>.pgf}
%%
%% Matplotlib used the following preamble
%%   \def\mathdefault#1{#1}
%%   \everymath=\expandafter{\the\everymath\displaystyle}
%%   \usepackage{amsmath}\usepackage{bm}
%%   \makeatletter\@ifpackageloaded{underscore}{}{\usepackage[strings]{underscore}}\makeatother
%%
\begingroup%
\makeatletter%
\begin{pgfpicture}%
\pgfpathrectangle{\pgfpointorigin}{\pgfqpoint{3.000000in}{2.000000in}}%
\pgfusepath{use as bounding box, clip}%
\begin{pgfscope}%
\pgfsetbuttcap%
\pgfsetmiterjoin%
\definecolor{currentfill}{rgb}{1.000000,1.000000,1.000000}%
\pgfsetfillcolor{currentfill}%
\pgfsetlinewidth{0.000000pt}%
\definecolor{currentstroke}{rgb}{1.000000,1.000000,1.000000}%
\pgfsetstrokecolor{currentstroke}%
\pgfsetdash{}{0pt}%
\pgfpathmoveto{\pgfqpoint{0.000000in}{0.000000in}}%
\pgfpathlineto{\pgfqpoint{3.000000in}{0.000000in}}%
\pgfpathlineto{\pgfqpoint{3.000000in}{2.000000in}}%
\pgfpathlineto{\pgfqpoint{0.000000in}{2.000000in}}%
\pgfpathlineto{\pgfqpoint{0.000000in}{0.000000in}}%
\pgfpathclose%
\pgfusepath{fill}%
\end{pgfscope}%
\begin{pgfscope}%
\pgfsetbuttcap%
\pgfsetmiterjoin%
\definecolor{currentfill}{rgb}{1.000000,1.000000,1.000000}%
\pgfsetfillcolor{currentfill}%
\pgfsetlinewidth{0.000000pt}%
\definecolor{currentstroke}{rgb}{0.000000,0.000000,0.000000}%
\pgfsetstrokecolor{currentstroke}%
\pgfsetstrokeopacity{0.000000}%
\pgfsetdash{}{0pt}%
\pgfpathmoveto{\pgfqpoint{0.597016in}{0.498776in}}%
\pgfpathlineto{\pgfqpoint{2.845756in}{0.498776in}}%
\pgfpathlineto{\pgfqpoint{2.845756in}{1.850000in}}%
\pgfpathlineto{\pgfqpoint{0.597016in}{1.850000in}}%
\pgfpathlineto{\pgfqpoint{0.597016in}{0.498776in}}%
\pgfpathclose%
\pgfusepath{fill}%
\end{pgfscope}%
\begin{pgfscope}%
\pgfpathrectangle{\pgfqpoint{0.597016in}{0.498776in}}{\pgfqpoint{2.248740in}{1.351224in}}%
\pgfusepath{clip}%
\pgfsetbuttcap%
\pgfsetroundjoin%
\definecolor{currentfill}{rgb}{0.843137,0.098039,0.109804}%
\pgfsetfillcolor{currentfill}%
\pgfsetlinewidth{0.501875pt}%
\definecolor{currentstroke}{rgb}{0.000000,0.000000,0.000000}%
\pgfsetstrokecolor{currentstroke}%
\pgfsetdash{}{0pt}%
\pgfsys@defobject{currentmarker}{\pgfqpoint{-0.026896in}{-0.026896in}}{\pgfqpoint{0.026896in}{0.026896in}}{%
\pgfpathmoveto{\pgfqpoint{-0.026896in}{-0.026896in}}%
\pgfpathlineto{\pgfqpoint{0.026896in}{-0.026896in}}%
\pgfpathlineto{\pgfqpoint{0.026896in}{0.026896in}}%
\pgfpathlineto{\pgfqpoint{-0.026896in}{0.026896in}}%
\pgfpathlineto{\pgfqpoint{-0.026896in}{-0.026896in}}%
\pgfpathclose%
\pgfusepath{stroke,fill}%
}%
\begin{pgfscope}%
\pgfsys@transformshift{0.699231in}{1.788581in}%
\pgfsys@useobject{currentmarker}{}%
\end{pgfscope}%
\begin{pgfscope}%
\pgfsys@transformshift{0.926377in}{1.545521in}%
\pgfsys@useobject{currentmarker}{}%
\end{pgfscope}%
\begin{pgfscope}%
\pgfsys@transformshift{1.153522in}{0.661782in}%
\pgfsys@useobject{currentmarker}{}%
\end{pgfscope}%
\begin{pgfscope}%
\pgfsys@transformshift{1.380668in}{0.590733in}%
\pgfsys@useobject{currentmarker}{}%
\end{pgfscope}%
\begin{pgfscope}%
\pgfsys@transformshift{1.607813in}{0.711017in}%
\pgfsys@useobject{currentmarker}{}%
\end{pgfscope}%
\begin{pgfscope}%
\pgfsys@transformshift{1.834959in}{0.796399in}%
\pgfsys@useobject{currentmarker}{}%
\end{pgfscope}%
\begin{pgfscope}%
\pgfsys@transformshift{2.062104in}{0.836909in}%
\pgfsys@useobject{currentmarker}{}%
\end{pgfscope}%
\begin{pgfscope}%
\pgfsys@transformshift{2.289250in}{0.848750in}%
\pgfsys@useobject{currentmarker}{}%
\end{pgfscope}%
\begin{pgfscope}%
\pgfsys@transformshift{2.516395in}{0.821328in}%
\pgfsys@useobject{currentmarker}{}%
\end{pgfscope}%
\begin{pgfscope}%
\pgfsys@transformshift{2.743541in}{0.811980in}%
\pgfsys@useobject{currentmarker}{}%
\end{pgfscope}%
\end{pgfscope}%
\begin{pgfscope}%
\pgfpathrectangle{\pgfqpoint{0.597016in}{0.498776in}}{\pgfqpoint{2.248740in}{1.351224in}}%
\pgfusepath{clip}%
\pgfsetbuttcap%
\pgfsetroundjoin%
\definecolor{currentfill}{rgb}{0.172549,0.482353,0.713725}%
\pgfsetfillcolor{currentfill}%
\pgfsetlinewidth{0.501875pt}%
\definecolor{currentstroke}{rgb}{0.000000,0.000000,0.000000}%
\pgfsetstrokecolor{currentstroke}%
\pgfsetdash{}{0pt}%
\pgfsys@defobject{currentmarker}{\pgfqpoint{-0.038036in}{-0.038036in}}{\pgfqpoint{0.038036in}{0.038036in}}{%
\pgfpathmoveto{\pgfqpoint{-0.000000in}{-0.038036in}}%
\pgfpathlineto{\pgfqpoint{0.038036in}{0.000000in}}%
\pgfpathlineto{\pgfqpoint{0.000000in}{0.038036in}}%
\pgfpathlineto{\pgfqpoint{-0.038036in}{0.000000in}}%
\pgfpathlineto{\pgfqpoint{-0.000000in}{-0.038036in}}%
\pgfpathclose%
\pgfusepath{stroke,fill}%
}%
\begin{pgfscope}%
\pgfsys@transformshift{0.699231in}{1.787957in}%
\pgfsys@useobject{currentmarker}{}%
\end{pgfscope}%
\begin{pgfscope}%
\pgfsys@transformshift{0.926377in}{1.564218in}%
\pgfsys@useobject{currentmarker}{}%
\end{pgfscope}%
\begin{pgfscope}%
\pgfsys@transformshift{1.153522in}{0.669884in}%
\pgfsys@useobject{currentmarker}{}%
\end{pgfscope}%
\begin{pgfscope}%
\pgfsys@transformshift{1.380668in}{0.560195in}%
\pgfsys@useobject{currentmarker}{}%
\end{pgfscope}%
\begin{pgfscope}%
\pgfsys@transformshift{1.607813in}{0.619402in}%
\pgfsys@useobject{currentmarker}{}%
\end{pgfscope}%
\begin{pgfscope}%
\pgfsys@transformshift{1.834959in}{0.613793in}%
\pgfsys@useobject{currentmarker}{}%
\end{pgfscope}%
\begin{pgfscope}%
\pgfsys@transformshift{2.062104in}{0.588240in}%
\pgfsys@useobject{currentmarker}{}%
\end{pgfscope}%
\begin{pgfscope}%
\pgfsys@transformshift{2.289250in}{0.592603in}%
\pgfsys@useobject{currentmarker}{}%
\end{pgfscope}%
\begin{pgfscope}%
\pgfsys@transformshift{2.516395in}{0.616286in}%
\pgfsys@useobject{currentmarker}{}%
\end{pgfscope}%
\begin{pgfscope}%
\pgfsys@transformshift{2.743541in}{0.695436in}%
\pgfsys@useobject{currentmarker}{}%
\end{pgfscope}%
\end{pgfscope}%
\begin{pgfscope}%
\pgfsetbuttcap%
\pgfsetroundjoin%
\definecolor{currentfill}{rgb}{0.000000,0.000000,0.000000}%
\pgfsetfillcolor{currentfill}%
\pgfsetlinewidth{0.803000pt}%
\definecolor{currentstroke}{rgb}{0.000000,0.000000,0.000000}%
\pgfsetstrokecolor{currentstroke}%
\pgfsetdash{}{0pt}%
\pgfsys@defobject{currentmarker}{\pgfqpoint{0.000000in}{-0.048611in}}{\pgfqpoint{0.000000in}{0.000000in}}{%
\pgfpathmoveto{\pgfqpoint{0.000000in}{0.000000in}}%
\pgfpathlineto{\pgfqpoint{0.000000in}{-0.048611in}}%
\pgfusepath{stroke,fill}%
}%
\begin{pgfscope}%
\pgfsys@transformshift{0.699231in}{0.498776in}%
\pgfsys@useobject{currentmarker}{}%
\end{pgfscope}%
\end{pgfscope}%
\begin{pgfscope}%
\definecolor{textcolor}{rgb}{0.000000,0.000000,0.000000}%
\pgfsetstrokecolor{textcolor}%
\pgfsetfillcolor{textcolor}%
\pgftext[x=0.699231in,y=0.408498in,,top]{\color{textcolor}{\rmfamily\fontsize{6.500000}{7.800000}\selectfont\catcode`\^=\active\def^{\ifmmode\sp\else\^{}\fi}\catcode`\%=\active\def%{\%}30}}%
\end{pgfscope}%
\begin{pgfscope}%
\pgfsetbuttcap%
\pgfsetroundjoin%
\definecolor{currentfill}{rgb}{0.000000,0.000000,0.000000}%
\pgfsetfillcolor{currentfill}%
\pgfsetlinewidth{0.803000pt}%
\definecolor{currentstroke}{rgb}{0.000000,0.000000,0.000000}%
\pgfsetstrokecolor{currentstroke}%
\pgfsetdash{}{0pt}%
\pgfsys@defobject{currentmarker}{\pgfqpoint{0.000000in}{-0.048611in}}{\pgfqpoint{0.000000in}{0.000000in}}{%
\pgfpathmoveto{\pgfqpoint{0.000000in}{0.000000in}}%
\pgfpathlineto{\pgfqpoint{0.000000in}{-0.048611in}}%
\pgfusepath{stroke,fill}%
}%
\begin{pgfscope}%
\pgfsys@transformshift{0.926377in}{0.498776in}%
\pgfsys@useobject{currentmarker}{}%
\end{pgfscope}%
\end{pgfscope}%
\begin{pgfscope}%
\definecolor{textcolor}{rgb}{0.000000,0.000000,0.000000}%
\pgfsetstrokecolor{textcolor}%
\pgfsetfillcolor{textcolor}%
\pgftext[x=0.926377in,y=0.408498in,,top]{\color{textcolor}{\rmfamily\fontsize{6.500000}{7.800000}\selectfont\catcode`\^=\active\def^{\ifmmode\sp\else\^{}\fi}\catcode`\%=\active\def%{\%}50}}%
\end{pgfscope}%
\begin{pgfscope}%
\pgfsetbuttcap%
\pgfsetroundjoin%
\definecolor{currentfill}{rgb}{0.000000,0.000000,0.000000}%
\pgfsetfillcolor{currentfill}%
\pgfsetlinewidth{0.803000pt}%
\definecolor{currentstroke}{rgb}{0.000000,0.000000,0.000000}%
\pgfsetstrokecolor{currentstroke}%
\pgfsetdash{}{0pt}%
\pgfsys@defobject{currentmarker}{\pgfqpoint{0.000000in}{-0.048611in}}{\pgfqpoint{0.000000in}{0.000000in}}{%
\pgfpathmoveto{\pgfqpoint{0.000000in}{0.000000in}}%
\pgfpathlineto{\pgfqpoint{0.000000in}{-0.048611in}}%
\pgfusepath{stroke,fill}%
}%
\begin{pgfscope}%
\pgfsys@transformshift{1.153522in}{0.498776in}%
\pgfsys@useobject{currentmarker}{}%
\end{pgfscope}%
\end{pgfscope}%
\begin{pgfscope}%
\definecolor{textcolor}{rgb}{0.000000,0.000000,0.000000}%
\pgfsetstrokecolor{textcolor}%
\pgfsetfillcolor{textcolor}%
\pgftext[x=1.153522in,y=0.408498in,,top]{\color{textcolor}{\rmfamily\fontsize{6.500000}{7.800000}\selectfont\catcode`\^=\active\def^{\ifmmode\sp\else\^{}\fi}\catcode`\%=\active\def%{\%}100}}%
\end{pgfscope}%
\begin{pgfscope}%
\pgfsetbuttcap%
\pgfsetroundjoin%
\definecolor{currentfill}{rgb}{0.000000,0.000000,0.000000}%
\pgfsetfillcolor{currentfill}%
\pgfsetlinewidth{0.803000pt}%
\definecolor{currentstroke}{rgb}{0.000000,0.000000,0.000000}%
\pgfsetstrokecolor{currentstroke}%
\pgfsetdash{}{0pt}%
\pgfsys@defobject{currentmarker}{\pgfqpoint{0.000000in}{-0.048611in}}{\pgfqpoint{0.000000in}{0.000000in}}{%
\pgfpathmoveto{\pgfqpoint{0.000000in}{0.000000in}}%
\pgfpathlineto{\pgfqpoint{0.000000in}{-0.048611in}}%
\pgfusepath{stroke,fill}%
}%
\begin{pgfscope}%
\pgfsys@transformshift{1.380668in}{0.498776in}%
\pgfsys@useobject{currentmarker}{}%
\end{pgfscope}%
\end{pgfscope}%
\begin{pgfscope}%
\definecolor{textcolor}{rgb}{0.000000,0.000000,0.000000}%
\pgfsetstrokecolor{textcolor}%
\pgfsetfillcolor{textcolor}%
\pgftext[x=1.380668in,y=0.408498in,,top]{\color{textcolor}{\rmfamily\fontsize{6.500000}{7.800000}\selectfont\catcode`\^=\active\def^{\ifmmode\sp\else\^{}\fi}\catcode`\%=\active\def%{\%}500}}%
\end{pgfscope}%
\begin{pgfscope}%
\pgfsetbuttcap%
\pgfsetroundjoin%
\definecolor{currentfill}{rgb}{0.000000,0.000000,0.000000}%
\pgfsetfillcolor{currentfill}%
\pgfsetlinewidth{0.803000pt}%
\definecolor{currentstroke}{rgb}{0.000000,0.000000,0.000000}%
\pgfsetstrokecolor{currentstroke}%
\pgfsetdash{}{0pt}%
\pgfsys@defobject{currentmarker}{\pgfqpoint{0.000000in}{-0.048611in}}{\pgfqpoint{0.000000in}{0.000000in}}{%
\pgfpathmoveto{\pgfqpoint{0.000000in}{0.000000in}}%
\pgfpathlineto{\pgfqpoint{0.000000in}{-0.048611in}}%
\pgfusepath{stroke,fill}%
}%
\begin{pgfscope}%
\pgfsys@transformshift{1.607813in}{0.498776in}%
\pgfsys@useobject{currentmarker}{}%
\end{pgfscope}%
\end{pgfscope}%
\begin{pgfscope}%
\definecolor{textcolor}{rgb}{0.000000,0.000000,0.000000}%
\pgfsetstrokecolor{textcolor}%
\pgfsetfillcolor{textcolor}%
\pgftext[x=1.607813in,y=0.408498in,,top]{\color{textcolor}{\rmfamily\fontsize{6.500000}{7.800000}\selectfont\catcode`\^=\active\def^{\ifmmode\sp\else\^{}\fi}\catcode`\%=\active\def%{\%}1000}}%
\end{pgfscope}%
\begin{pgfscope}%
\pgfsetbuttcap%
\pgfsetroundjoin%
\definecolor{currentfill}{rgb}{0.000000,0.000000,0.000000}%
\pgfsetfillcolor{currentfill}%
\pgfsetlinewidth{0.803000pt}%
\definecolor{currentstroke}{rgb}{0.000000,0.000000,0.000000}%
\pgfsetstrokecolor{currentstroke}%
\pgfsetdash{}{0pt}%
\pgfsys@defobject{currentmarker}{\pgfqpoint{0.000000in}{-0.048611in}}{\pgfqpoint{0.000000in}{0.000000in}}{%
\pgfpathmoveto{\pgfqpoint{0.000000in}{0.000000in}}%
\pgfpathlineto{\pgfqpoint{0.000000in}{-0.048611in}}%
\pgfusepath{stroke,fill}%
}%
\begin{pgfscope}%
\pgfsys@transformshift{1.834959in}{0.498776in}%
\pgfsys@useobject{currentmarker}{}%
\end{pgfscope}%
\end{pgfscope}%
\begin{pgfscope}%
\definecolor{textcolor}{rgb}{0.000000,0.000000,0.000000}%
\pgfsetstrokecolor{textcolor}%
\pgfsetfillcolor{textcolor}%
\pgftext[x=1.834959in,y=0.408498in,,top]{\color{textcolor}{\rmfamily\fontsize{6.500000}{7.800000}\selectfont\catcode`\^=\active\def^{\ifmmode\sp\else\^{}\fi}\catcode`\%=\active\def%{\%}2000}}%
\end{pgfscope}%
\begin{pgfscope}%
\pgfsetbuttcap%
\pgfsetroundjoin%
\definecolor{currentfill}{rgb}{0.000000,0.000000,0.000000}%
\pgfsetfillcolor{currentfill}%
\pgfsetlinewidth{0.803000pt}%
\definecolor{currentstroke}{rgb}{0.000000,0.000000,0.000000}%
\pgfsetstrokecolor{currentstroke}%
\pgfsetdash{}{0pt}%
\pgfsys@defobject{currentmarker}{\pgfqpoint{0.000000in}{-0.048611in}}{\pgfqpoint{0.000000in}{0.000000in}}{%
\pgfpathmoveto{\pgfqpoint{0.000000in}{0.000000in}}%
\pgfpathlineto{\pgfqpoint{0.000000in}{-0.048611in}}%
\pgfusepath{stroke,fill}%
}%
\begin{pgfscope}%
\pgfsys@transformshift{2.062104in}{0.498776in}%
\pgfsys@useobject{currentmarker}{}%
\end{pgfscope}%
\end{pgfscope}%
\begin{pgfscope}%
\definecolor{textcolor}{rgb}{0.000000,0.000000,0.000000}%
\pgfsetstrokecolor{textcolor}%
\pgfsetfillcolor{textcolor}%
\pgftext[x=2.062104in,y=0.408498in,,top]{\color{textcolor}{\rmfamily\fontsize{6.500000}{7.800000}\selectfont\catcode`\^=\active\def^{\ifmmode\sp\else\^{}\fi}\catcode`\%=\active\def%{\%}3000}}%
\end{pgfscope}%
\begin{pgfscope}%
\pgfsetbuttcap%
\pgfsetroundjoin%
\definecolor{currentfill}{rgb}{0.000000,0.000000,0.000000}%
\pgfsetfillcolor{currentfill}%
\pgfsetlinewidth{0.803000pt}%
\definecolor{currentstroke}{rgb}{0.000000,0.000000,0.000000}%
\pgfsetstrokecolor{currentstroke}%
\pgfsetdash{}{0pt}%
\pgfsys@defobject{currentmarker}{\pgfqpoint{0.000000in}{-0.048611in}}{\pgfqpoint{0.000000in}{0.000000in}}{%
\pgfpathmoveto{\pgfqpoint{0.000000in}{0.000000in}}%
\pgfpathlineto{\pgfqpoint{0.000000in}{-0.048611in}}%
\pgfusepath{stroke,fill}%
}%
\begin{pgfscope}%
\pgfsys@transformshift{2.289250in}{0.498776in}%
\pgfsys@useobject{currentmarker}{}%
\end{pgfscope}%
\end{pgfscope}%
\begin{pgfscope}%
\definecolor{textcolor}{rgb}{0.000000,0.000000,0.000000}%
\pgfsetstrokecolor{textcolor}%
\pgfsetfillcolor{textcolor}%
\pgftext[x=2.289250in,y=0.408498in,,top]{\color{textcolor}{\rmfamily\fontsize{6.500000}{7.800000}\selectfont\catcode`\^=\active\def^{\ifmmode\sp\else\^{}\fi}\catcode`\%=\active\def%{\%}4000}}%
\end{pgfscope}%
\begin{pgfscope}%
\pgfsetbuttcap%
\pgfsetroundjoin%
\definecolor{currentfill}{rgb}{0.000000,0.000000,0.000000}%
\pgfsetfillcolor{currentfill}%
\pgfsetlinewidth{0.803000pt}%
\definecolor{currentstroke}{rgb}{0.000000,0.000000,0.000000}%
\pgfsetstrokecolor{currentstroke}%
\pgfsetdash{}{0pt}%
\pgfsys@defobject{currentmarker}{\pgfqpoint{0.000000in}{-0.048611in}}{\pgfqpoint{0.000000in}{0.000000in}}{%
\pgfpathmoveto{\pgfqpoint{0.000000in}{0.000000in}}%
\pgfpathlineto{\pgfqpoint{0.000000in}{-0.048611in}}%
\pgfusepath{stroke,fill}%
}%
\begin{pgfscope}%
\pgfsys@transformshift{2.516395in}{0.498776in}%
\pgfsys@useobject{currentmarker}{}%
\end{pgfscope}%
\end{pgfscope}%
\begin{pgfscope}%
\definecolor{textcolor}{rgb}{0.000000,0.000000,0.000000}%
\pgfsetstrokecolor{textcolor}%
\pgfsetfillcolor{textcolor}%
\pgftext[x=2.516395in,y=0.408498in,,top]{\color{textcolor}{\rmfamily\fontsize{6.500000}{7.800000}\selectfont\catcode`\^=\active\def^{\ifmmode\sp\else\^{}\fi}\catcode`\%=\active\def%{\%}5000}}%
\end{pgfscope}%
\begin{pgfscope}%
\pgfsetbuttcap%
\pgfsetroundjoin%
\definecolor{currentfill}{rgb}{0.000000,0.000000,0.000000}%
\pgfsetfillcolor{currentfill}%
\pgfsetlinewidth{0.803000pt}%
\definecolor{currentstroke}{rgb}{0.000000,0.000000,0.000000}%
\pgfsetstrokecolor{currentstroke}%
\pgfsetdash{}{0pt}%
\pgfsys@defobject{currentmarker}{\pgfqpoint{0.000000in}{-0.048611in}}{\pgfqpoint{0.000000in}{0.000000in}}{%
\pgfpathmoveto{\pgfqpoint{0.000000in}{0.000000in}}%
\pgfpathlineto{\pgfqpoint{0.000000in}{-0.048611in}}%
\pgfusepath{stroke,fill}%
}%
\begin{pgfscope}%
\pgfsys@transformshift{2.743541in}{0.498776in}%
\pgfsys@useobject{currentmarker}{}%
\end{pgfscope}%
\end{pgfscope}%
\begin{pgfscope}%
\definecolor{textcolor}{rgb}{0.000000,0.000000,0.000000}%
\pgfsetstrokecolor{textcolor}%
\pgfsetfillcolor{textcolor}%
\pgftext[x=2.743541in,y=0.408498in,,top]{\color{textcolor}{\rmfamily\fontsize{6.500000}{7.800000}\selectfont\catcode`\^=\active\def^{\ifmmode\sp\else\^{}\fi}\catcode`\%=\active\def%{\%}6000}}%
\end{pgfscope}%
\begin{pgfscope}%
\definecolor{textcolor}{rgb}{0.000000,0.000000,0.000000}%
\pgfsetstrokecolor{textcolor}%
\pgfsetfillcolor{textcolor}%
\pgftext[x=1.721386in,y=0.278868in,,top]{\color{textcolor}{\rmfamily\fontsize{10.000000}{12.000000}\selectfont\catcode`\^=\active\def^{\ifmmode\sp\else\^{}\fi}\catcode`\%=\active\def%{\%}$\text{Re}$}}%
\end{pgfscope}%
\begin{pgfscope}%
\pgfsetbuttcap%
\pgfsetroundjoin%
\definecolor{currentfill}{rgb}{0.000000,0.000000,0.000000}%
\pgfsetfillcolor{currentfill}%
\pgfsetlinewidth{0.803000pt}%
\definecolor{currentstroke}{rgb}{0.000000,0.000000,0.000000}%
\pgfsetstrokecolor{currentstroke}%
\pgfsetdash{}{0pt}%
\pgfsys@defobject{currentmarker}{\pgfqpoint{-0.048611in}{0.000000in}}{\pgfqpoint{-0.000000in}{0.000000in}}{%
\pgfpathmoveto{\pgfqpoint{-0.000000in}{0.000000in}}%
\pgfpathlineto{\pgfqpoint{-0.048611in}{0.000000in}}%
\pgfusepath{stroke,fill}%
}%
\begin{pgfscope}%
\pgfsys@transformshift{0.597016in}{0.748410in}%
\pgfsys@useobject{currentmarker}{}%
\end{pgfscope}%
\end{pgfscope}%
\begin{pgfscope}%
\definecolor{textcolor}{rgb}{0.000000,0.000000,0.000000}%
\pgfsetstrokecolor{textcolor}%
\pgfsetfillcolor{textcolor}%
\pgftext[x=0.455813in, y=0.719475in, left, base]{\color{textcolor}{\rmfamily\fontsize{6.500000}{7.800000}\selectfont\catcode`\^=\active\def^{\ifmmode\sp\else\^{}\fi}\catcode`\%=\active\def%{\%}$\mathdefault{5}$}}%
\end{pgfscope}%
\begin{pgfscope}%
\pgfsetbuttcap%
\pgfsetroundjoin%
\definecolor{currentfill}{rgb}{0.000000,0.000000,0.000000}%
\pgfsetfillcolor{currentfill}%
\pgfsetlinewidth{0.803000pt}%
\definecolor{currentstroke}{rgb}{0.000000,0.000000,0.000000}%
\pgfsetstrokecolor{currentstroke}%
\pgfsetdash{}{0pt}%
\pgfsys@defobject{currentmarker}{\pgfqpoint{-0.048611in}{0.000000in}}{\pgfqpoint{-0.000000in}{0.000000in}}{%
\pgfpathmoveto{\pgfqpoint{-0.000000in}{0.000000in}}%
\pgfpathlineto{\pgfqpoint{-0.048611in}{0.000000in}}%
\pgfusepath{stroke,fill}%
}%
\begin{pgfscope}%
\pgfsys@transformshift{0.597016in}{1.060025in}%
\pgfsys@useobject{currentmarker}{}%
\end{pgfscope}%
\end{pgfscope}%
\begin{pgfscope}%
\definecolor{textcolor}{rgb}{0.000000,0.000000,0.000000}%
\pgfsetstrokecolor{textcolor}%
\pgfsetfillcolor{textcolor}%
\pgftext[x=0.404888in, y=1.031090in, left, base]{\color{textcolor}{\rmfamily\fontsize{6.500000}{7.800000}\selectfont\catcode`\^=\active\def^{\ifmmode\sp\else\^{}\fi}\catcode`\%=\active\def%{\%}$\mathdefault{10}$}}%
\end{pgfscope}%
\begin{pgfscope}%
\pgfsetbuttcap%
\pgfsetroundjoin%
\definecolor{currentfill}{rgb}{0.000000,0.000000,0.000000}%
\pgfsetfillcolor{currentfill}%
\pgfsetlinewidth{0.803000pt}%
\definecolor{currentstroke}{rgb}{0.000000,0.000000,0.000000}%
\pgfsetstrokecolor{currentstroke}%
\pgfsetdash{}{0pt}%
\pgfsys@defobject{currentmarker}{\pgfqpoint{-0.048611in}{0.000000in}}{\pgfqpoint{-0.000000in}{0.000000in}}{%
\pgfpathmoveto{\pgfqpoint{-0.000000in}{0.000000in}}%
\pgfpathlineto{\pgfqpoint{-0.048611in}{0.000000in}}%
\pgfusepath{stroke,fill}%
}%
\begin{pgfscope}%
\pgfsys@transformshift{0.597016in}{1.371640in}%
\pgfsys@useobject{currentmarker}{}%
\end{pgfscope}%
\end{pgfscope}%
\begin{pgfscope}%
\definecolor{textcolor}{rgb}{0.000000,0.000000,0.000000}%
\pgfsetstrokecolor{textcolor}%
\pgfsetfillcolor{textcolor}%
\pgftext[x=0.404888in, y=1.342705in, left, base]{\color{textcolor}{\rmfamily\fontsize{6.500000}{7.800000}\selectfont\catcode`\^=\active\def^{\ifmmode\sp\else\^{}\fi}\catcode`\%=\active\def%{\%}$\mathdefault{15}$}}%
\end{pgfscope}%
\begin{pgfscope}%
\pgfsetbuttcap%
\pgfsetroundjoin%
\definecolor{currentfill}{rgb}{0.000000,0.000000,0.000000}%
\pgfsetfillcolor{currentfill}%
\pgfsetlinewidth{0.803000pt}%
\definecolor{currentstroke}{rgb}{0.000000,0.000000,0.000000}%
\pgfsetstrokecolor{currentstroke}%
\pgfsetdash{}{0pt}%
\pgfsys@defobject{currentmarker}{\pgfqpoint{-0.048611in}{0.000000in}}{\pgfqpoint{-0.000000in}{0.000000in}}{%
\pgfpathmoveto{\pgfqpoint{-0.000000in}{0.000000in}}%
\pgfpathlineto{\pgfqpoint{-0.048611in}{0.000000in}}%
\pgfusepath{stroke,fill}%
}%
\begin{pgfscope}%
\pgfsys@transformshift{0.597016in}{1.683255in}%
\pgfsys@useobject{currentmarker}{}%
\end{pgfscope}%
\end{pgfscope}%
\begin{pgfscope}%
\definecolor{textcolor}{rgb}{0.000000,0.000000,0.000000}%
\pgfsetstrokecolor{textcolor}%
\pgfsetfillcolor{textcolor}%
\pgftext[x=0.404888in, y=1.654320in, left, base]{\color{textcolor}{\rmfamily\fontsize{6.500000}{7.800000}\selectfont\catcode`\^=\active\def^{\ifmmode\sp\else\^{}\fi}\catcode`\%=\active\def%{\%}$\mathdefault{20}$}}%
\end{pgfscope}%
\begin{pgfscope}%
\definecolor{textcolor}{rgb}{0.000000,0.000000,0.000000}%
\pgfsetstrokecolor{textcolor}%
\pgfsetfillcolor{textcolor}%
\pgftext[x=0.349332in,y=1.174388in,,bottom,rotate=90.000000]{\color{textcolor}{\rmfamily\fontsize{10.000000}{12.000000}\selectfont\catcode`\^=\active\def^{\ifmmode\sp\else\^{}\fi}\catcode`\%=\active\def%{\%}$\delta_{\ell^1}^{(p)} \ [\%]$}}%
\end{pgfscope}%
\begin{pgfscope}%
\pgfsetrectcap%
\pgfsetmiterjoin%
\pgfsetlinewidth{0.803000pt}%
\definecolor{currentstroke}{rgb}{0.000000,0.000000,0.000000}%
\pgfsetstrokecolor{currentstroke}%
\pgfsetdash{}{0pt}%
\pgfpathmoveto{\pgfqpoint{0.597016in}{0.498776in}}%
\pgfpathlineto{\pgfqpoint{0.597016in}{1.850000in}}%
\pgfusepath{stroke}%
\end{pgfscope}%
\begin{pgfscope}%
\pgfsetrectcap%
\pgfsetmiterjoin%
\pgfsetlinewidth{0.803000pt}%
\definecolor{currentstroke}{rgb}{0.000000,0.000000,0.000000}%
\pgfsetstrokecolor{currentstroke}%
\pgfsetdash{}{0pt}%
\pgfpathmoveto{\pgfqpoint{2.845756in}{0.498776in}}%
\pgfpathlineto{\pgfqpoint{2.845756in}{1.850000in}}%
\pgfusepath{stroke}%
\end{pgfscope}%
\begin{pgfscope}%
\pgfsetrectcap%
\pgfsetmiterjoin%
\pgfsetlinewidth{0.803000pt}%
\definecolor{currentstroke}{rgb}{0.000000,0.000000,0.000000}%
\pgfsetstrokecolor{currentstroke}%
\pgfsetdash{}{0pt}%
\pgfpathmoveto{\pgfqpoint{0.597016in}{0.498776in}}%
\pgfpathlineto{\pgfqpoint{2.845756in}{0.498776in}}%
\pgfusepath{stroke}%
\end{pgfscope}%
\begin{pgfscope}%
\pgfsetrectcap%
\pgfsetmiterjoin%
\pgfsetlinewidth{0.803000pt}%
\definecolor{currentstroke}{rgb}{0.000000,0.000000,0.000000}%
\pgfsetstrokecolor{currentstroke}%
\pgfsetdash{}{0pt}%
\pgfpathmoveto{\pgfqpoint{0.597016in}{1.850000in}}%
\pgfpathlineto{\pgfqpoint{2.845756in}{1.850000in}}%
\pgfusepath{stroke}%
\end{pgfscope}%
\begin{pgfscope}%
\pgfsetbuttcap%
\pgfsetmiterjoin%
\definecolor{currentfill}{rgb}{1.000000,1.000000,1.000000}%
\pgfsetfillcolor{currentfill}%
\pgfsetfillopacity{0.800000}%
\pgfsetlinewidth{1.003750pt}%
\definecolor{currentstroke}{rgb}{0.800000,0.800000,0.800000}%
\pgfsetstrokecolor{currentstroke}%
\pgfsetstrokeopacity{0.800000}%
\pgfsetdash{}{0pt}%
\pgfpathmoveto{\pgfqpoint{2.060433in}{1.359852in}}%
\pgfpathlineto{\pgfqpoint{2.764770in}{1.359852in}}%
\pgfpathquadraticcurveto{\pgfqpoint{2.787909in}{1.359852in}}{\pgfqpoint{2.787909in}{1.382990in}}%
\pgfpathlineto{\pgfqpoint{2.787909in}{1.769014in}}%
\pgfpathquadraticcurveto{\pgfqpoint{2.787909in}{1.792153in}}{\pgfqpoint{2.764770in}{1.792153in}}%
\pgfpathlineto{\pgfqpoint{2.060433in}{1.792153in}}%
\pgfpathquadraticcurveto{\pgfqpoint{2.037294in}{1.792153in}}{\pgfqpoint{2.037294in}{1.769014in}}%
\pgfpathlineto{\pgfqpoint{2.037294in}{1.382990in}}%
\pgfpathquadraticcurveto{\pgfqpoint{2.037294in}{1.359852in}}{\pgfqpoint{2.060433in}{1.359852in}}%
\pgfpathlineto{\pgfqpoint{2.060433in}{1.359852in}}%
\pgfpathclose%
\pgfusepath{stroke,fill}%
\end{pgfscope}%
\begin{pgfscope}%
\pgfsetbuttcap%
\pgfsetroundjoin%
\definecolor{currentfill}{rgb}{0.843137,0.098039,0.109804}%
\pgfsetfillcolor{currentfill}%
\pgfsetlinewidth{0.501875pt}%
\definecolor{currentstroke}{rgb}{0.000000,0.000000,0.000000}%
\pgfsetstrokecolor{currentstroke}%
\pgfsetdash{}{0pt}%
\pgfsys@defobject{currentmarker}{\pgfqpoint{-0.026896in}{-0.026896in}}{\pgfqpoint{0.026896in}{0.026896in}}{%
\pgfpathmoveto{\pgfqpoint{-0.026896in}{-0.026896in}}%
\pgfpathlineto{\pgfqpoint{0.026896in}{-0.026896in}}%
\pgfpathlineto{\pgfqpoint{0.026896in}{0.026896in}}%
\pgfpathlineto{\pgfqpoint{-0.026896in}{0.026896in}}%
\pgfpathlineto{\pgfqpoint{-0.026896in}{-0.026896in}}%
\pgfpathclose%
\pgfusepath{stroke,fill}%
}%
\begin{pgfscope}%
\pgfsys@transformshift{2.199266in}{1.673143in}%
\pgfsys@useobject{currentmarker}{}%
\end{pgfscope}%
\end{pgfscope}%
\begin{pgfscope}%
\definecolor{textcolor}{rgb}{0.000000,0.000000,0.000000}%
\pgfsetstrokecolor{textcolor}%
\pgfsetfillcolor{textcolor}%
\pgftext[x=2.407516in,y=1.642773in,left,base]{\color{textcolor}{\rmfamily\fontsize{8.330000}{9.996000}\selectfont\catcode`\^=\active\def^{\ifmmode\sp\else\^{}\fi}\catcode`\%=\active\def%{\%}$p^\text{PINN}_\text{rand}$}}%
\end{pgfscope}%
\begin{pgfscope}%
\pgfsetbuttcap%
\pgfsetroundjoin%
\definecolor{currentfill}{rgb}{0.172549,0.482353,0.713725}%
\pgfsetfillcolor{currentfill}%
\pgfsetlinewidth{0.501875pt}%
\definecolor{currentstroke}{rgb}{0.000000,0.000000,0.000000}%
\pgfsetstrokecolor{currentstroke}%
\pgfsetdash{}{0pt}%
\pgfsys@defobject{currentmarker}{\pgfqpoint{-0.038036in}{-0.038036in}}{\pgfqpoint{0.038036in}{0.038036in}}{%
\pgfpathmoveto{\pgfqpoint{-0.000000in}{-0.038036in}}%
\pgfpathlineto{\pgfqpoint{0.038036in}{0.000000in}}%
\pgfpathlineto{\pgfqpoint{0.000000in}{0.038036in}}%
\pgfpathlineto{\pgfqpoint{-0.038036in}{0.000000in}}%
\pgfpathlineto{\pgfqpoint{-0.000000in}{-0.038036in}}%
\pgfpathclose%
\pgfusepath{stroke,fill}%
}%
\begin{pgfscope}%
\pgfsys@transformshift{2.199266in}{1.480481in}%
\pgfsys@useobject{currentmarker}{}%
\end{pgfscope}%
\end{pgfscope}%
\begin{pgfscope}%
\definecolor{textcolor}{rgb}{0.000000,0.000000,0.000000}%
\pgfsetstrokecolor{textcolor}%
\pgfsetfillcolor{textcolor}%
\pgftext[x=2.407516in,y=1.450111in,left,base]{\color{textcolor}{\rmfamily\fontsize{8.330000}{9.996000}\selectfont\catcode`\^=\active\def^{\ifmmode\sp\else\^{}\fi}\catcode`\%=\active\def%{\%}$p^\text{PINN}_\text{target}$}}%
\end{pgfscope}%
\end{pgfpicture}%
\makeatother%
\endgroup%

%% file: figures/section6/vel_profile_phi0.pgf
%% Creator: Matplotlib, PGF backend
%%
%% To include the figure in your LaTeX document, write
%%   \input{<filename>.pgf}
%%
%% Make sure the required packages are loaded in your preamble
%%   \usepackage{pgf}
%%
%% Also ensure that all the required font packages are loaded; for instance,
%% the lmodern package is sometimes necessary when using math font.
%%   \usepackage{lmodern}
%%
%% Figures using additional raster images can only be included by \input if
%% they are in the same directory as the main LaTeX file. For loading figures
%% from other directories you can use the `import` package
%% and then include the figures with
%%   \import{<path to file>}{<filename>.pgf}
%%
%% Matplotlib used the following preamble
%%   \def\mathdefault#1{#1}
%%   \everymath=\expandafter{\the\everymath\displaystyle}
%%   \usepackage{amsmath}\usepackage{bm}
%%   \makeatletter\@ifpackageloaded{underscore}{}{\usepackage[strings]{underscore}}\makeatother
%%
\begingroup%
\makeatletter%
\begin{pgfpicture}%
\pgfpathrectangle{\pgfpointorigin}{\pgfqpoint{3.000000in}{2.500000in}}%
\pgfusepath{use as bounding box, clip}%
\begin{pgfscope}%
\pgfsetbuttcap%
\pgfsetmiterjoin%
\definecolor{currentfill}{rgb}{1.000000,1.000000,1.000000}%
\pgfsetfillcolor{currentfill}%
\pgfsetlinewidth{0.000000pt}%
\definecolor{currentstroke}{rgb}{1.000000,1.000000,1.000000}%
\pgfsetstrokecolor{currentstroke}%
\pgfsetdash{}{0pt}%
\pgfpathmoveto{\pgfqpoint{0.000000in}{0.000000in}}%
\pgfpathlineto{\pgfqpoint{3.000000in}{0.000000in}}%
\pgfpathlineto{\pgfqpoint{3.000000in}{2.500000in}}%
\pgfpathlineto{\pgfqpoint{0.000000in}{2.500000in}}%
\pgfpathlineto{\pgfqpoint{0.000000in}{0.000000in}}%
\pgfpathclose%
\pgfusepath{fill}%
\end{pgfscope}%
\begin{pgfscope}%
\pgfsetbuttcap%
\pgfsetmiterjoin%
\definecolor{currentfill}{rgb}{1.000000,1.000000,1.000000}%
\pgfsetfillcolor{currentfill}%
\pgfsetlinewidth{0.000000pt}%
\definecolor{currentstroke}{rgb}{0.000000,0.000000,0.000000}%
\pgfsetstrokecolor{currentstroke}%
\pgfsetstrokeopacity{0.000000}%
\pgfsetdash{}{0pt}%
\pgfpathmoveto{\pgfqpoint{0.651925in}{0.536658in}}%
\pgfpathlineto{\pgfqpoint{2.846822in}{0.536658in}}%
\pgfpathlineto{\pgfqpoint{2.846822in}{1.717454in}}%
\pgfpathlineto{\pgfqpoint{0.651925in}{1.717454in}}%
\pgfpathlineto{\pgfqpoint{0.651925in}{0.536658in}}%
\pgfpathclose%
\pgfusepath{fill}%
\end{pgfscope}%
\begin{pgfscope}%
\pgfsetbuttcap%
\pgfsetroundjoin%
\definecolor{currentfill}{rgb}{0.000000,0.000000,0.000000}%
\pgfsetfillcolor{currentfill}%
\pgfsetlinewidth{0.803000pt}%
\definecolor{currentstroke}{rgb}{0.000000,0.000000,0.000000}%
\pgfsetstrokecolor{currentstroke}%
\pgfsetdash{}{0pt}%
\pgfsys@defobject{currentmarker}{\pgfqpoint{0.000000in}{-0.048611in}}{\pgfqpoint{0.000000in}{0.000000in}}{%
\pgfpathmoveto{\pgfqpoint{0.000000in}{0.000000in}}%
\pgfpathlineto{\pgfqpoint{0.000000in}{-0.048611in}}%
\pgfusepath{stroke,fill}%
}%
\begin{pgfscope}%
\pgfsys@transformshift{0.668553in}{0.536658in}%
\pgfsys@useobject{currentmarker}{}%
\end{pgfscope}%
\end{pgfscope}%
\begin{pgfscope}%
\definecolor{textcolor}{rgb}{0.000000,0.000000,0.000000}%
\pgfsetstrokecolor{textcolor}%
\pgfsetfillcolor{textcolor}%
\pgftext[x=0.668553in,y=0.446381in,,top]{\color{textcolor}{\rmfamily\fontsize{8.330000}{9.996000}\selectfont\catcode`\^=\active\def^{\ifmmode\sp\else\^{}\fi}\catcode`\%=\active\def%{\%}$\mathdefault{0.00}$}}%
\end{pgfscope}%
\begin{pgfscope}%
\pgfsetbuttcap%
\pgfsetroundjoin%
\definecolor{currentfill}{rgb}{0.000000,0.000000,0.000000}%
\pgfsetfillcolor{currentfill}%
\pgfsetlinewidth{0.803000pt}%
\definecolor{currentstroke}{rgb}{0.000000,0.000000,0.000000}%
\pgfsetstrokecolor{currentstroke}%
\pgfsetdash{}{0pt}%
\pgfsys@defobject{currentmarker}{\pgfqpoint{0.000000in}{-0.048611in}}{\pgfqpoint{0.000000in}{0.000000in}}{%
\pgfpathmoveto{\pgfqpoint{0.000000in}{0.000000in}}%
\pgfpathlineto{\pgfqpoint{0.000000in}{-0.048611in}}%
\pgfusepath{stroke,fill}%
}%
\begin{pgfscope}%
\pgfsys@transformshift{1.084253in}{0.536658in}%
\pgfsys@useobject{currentmarker}{}%
\end{pgfscope}%
\end{pgfscope}%
\begin{pgfscope}%
\definecolor{textcolor}{rgb}{0.000000,0.000000,0.000000}%
\pgfsetstrokecolor{textcolor}%
\pgfsetfillcolor{textcolor}%
\pgftext[x=1.084253in,y=0.446381in,,top]{\color{textcolor}{\rmfamily\fontsize{8.330000}{9.996000}\selectfont\catcode`\^=\active\def^{\ifmmode\sp\else\^{}\fi}\catcode`\%=\active\def%{\%}$\mathdefault{0.02}$}}%
\end{pgfscope}%
\begin{pgfscope}%
\pgfsetbuttcap%
\pgfsetroundjoin%
\definecolor{currentfill}{rgb}{0.000000,0.000000,0.000000}%
\pgfsetfillcolor{currentfill}%
\pgfsetlinewidth{0.803000pt}%
\definecolor{currentstroke}{rgb}{0.000000,0.000000,0.000000}%
\pgfsetstrokecolor{currentstroke}%
\pgfsetdash{}{0pt}%
\pgfsys@defobject{currentmarker}{\pgfqpoint{0.000000in}{-0.048611in}}{\pgfqpoint{0.000000in}{0.000000in}}{%
\pgfpathmoveto{\pgfqpoint{0.000000in}{0.000000in}}%
\pgfpathlineto{\pgfqpoint{0.000000in}{-0.048611in}}%
\pgfusepath{stroke,fill}%
}%
\begin{pgfscope}%
\pgfsys@transformshift{1.499953in}{0.536658in}%
\pgfsys@useobject{currentmarker}{}%
\end{pgfscope}%
\end{pgfscope}%
\begin{pgfscope}%
\definecolor{textcolor}{rgb}{0.000000,0.000000,0.000000}%
\pgfsetstrokecolor{textcolor}%
\pgfsetfillcolor{textcolor}%
\pgftext[x=1.499953in,y=0.446381in,,top]{\color{textcolor}{\rmfamily\fontsize{8.330000}{9.996000}\selectfont\catcode`\^=\active\def^{\ifmmode\sp\else\^{}\fi}\catcode`\%=\active\def%{\%}$\mathdefault{0.04}$}}%
\end{pgfscope}%
\begin{pgfscope}%
\pgfsetbuttcap%
\pgfsetroundjoin%
\definecolor{currentfill}{rgb}{0.000000,0.000000,0.000000}%
\pgfsetfillcolor{currentfill}%
\pgfsetlinewidth{0.803000pt}%
\definecolor{currentstroke}{rgb}{0.000000,0.000000,0.000000}%
\pgfsetstrokecolor{currentstroke}%
\pgfsetdash{}{0pt}%
\pgfsys@defobject{currentmarker}{\pgfqpoint{0.000000in}{-0.048611in}}{\pgfqpoint{0.000000in}{0.000000in}}{%
\pgfpathmoveto{\pgfqpoint{0.000000in}{0.000000in}}%
\pgfpathlineto{\pgfqpoint{0.000000in}{-0.048611in}}%
\pgfusepath{stroke,fill}%
}%
\begin{pgfscope}%
\pgfsys@transformshift{1.915654in}{0.536658in}%
\pgfsys@useobject{currentmarker}{}%
\end{pgfscope}%
\end{pgfscope}%
\begin{pgfscope}%
\definecolor{textcolor}{rgb}{0.000000,0.000000,0.000000}%
\pgfsetstrokecolor{textcolor}%
\pgfsetfillcolor{textcolor}%
\pgftext[x=1.915654in,y=0.446381in,,top]{\color{textcolor}{\rmfamily\fontsize{8.330000}{9.996000}\selectfont\catcode`\^=\active\def^{\ifmmode\sp\else\^{}\fi}\catcode`\%=\active\def%{\%}$\mathdefault{0.06}$}}%
\end{pgfscope}%
\begin{pgfscope}%
\pgfsetbuttcap%
\pgfsetroundjoin%
\definecolor{currentfill}{rgb}{0.000000,0.000000,0.000000}%
\pgfsetfillcolor{currentfill}%
\pgfsetlinewidth{0.803000pt}%
\definecolor{currentstroke}{rgb}{0.000000,0.000000,0.000000}%
\pgfsetstrokecolor{currentstroke}%
\pgfsetdash{}{0pt}%
\pgfsys@defobject{currentmarker}{\pgfqpoint{0.000000in}{-0.048611in}}{\pgfqpoint{0.000000in}{0.000000in}}{%
\pgfpathmoveto{\pgfqpoint{0.000000in}{0.000000in}}%
\pgfpathlineto{\pgfqpoint{0.000000in}{-0.048611in}}%
\pgfusepath{stroke,fill}%
}%
\begin{pgfscope}%
\pgfsys@transformshift{2.331354in}{0.536658in}%
\pgfsys@useobject{currentmarker}{}%
\end{pgfscope}%
\end{pgfscope}%
\begin{pgfscope}%
\definecolor{textcolor}{rgb}{0.000000,0.000000,0.000000}%
\pgfsetstrokecolor{textcolor}%
\pgfsetfillcolor{textcolor}%
\pgftext[x=2.331354in,y=0.446381in,,top]{\color{textcolor}{\rmfamily\fontsize{8.330000}{9.996000}\selectfont\catcode`\^=\active\def^{\ifmmode\sp\else\^{}\fi}\catcode`\%=\active\def%{\%}$\mathdefault{0.08}$}}%
\end{pgfscope}%
\begin{pgfscope}%
\pgfsetbuttcap%
\pgfsetroundjoin%
\definecolor{currentfill}{rgb}{0.000000,0.000000,0.000000}%
\pgfsetfillcolor{currentfill}%
\pgfsetlinewidth{0.803000pt}%
\definecolor{currentstroke}{rgb}{0.000000,0.000000,0.000000}%
\pgfsetstrokecolor{currentstroke}%
\pgfsetdash{}{0pt}%
\pgfsys@defobject{currentmarker}{\pgfqpoint{0.000000in}{-0.048611in}}{\pgfqpoint{0.000000in}{0.000000in}}{%
\pgfpathmoveto{\pgfqpoint{0.000000in}{0.000000in}}%
\pgfpathlineto{\pgfqpoint{0.000000in}{-0.048611in}}%
\pgfusepath{stroke,fill}%
}%
\begin{pgfscope}%
\pgfsys@transformshift{2.747054in}{0.536658in}%
\pgfsys@useobject{currentmarker}{}%
\end{pgfscope}%
\end{pgfscope}%
\begin{pgfscope}%
\definecolor{textcolor}{rgb}{0.000000,0.000000,0.000000}%
\pgfsetstrokecolor{textcolor}%
\pgfsetfillcolor{textcolor}%
\pgftext[x=2.747054in,y=0.446381in,,top]{\color{textcolor}{\rmfamily\fontsize{8.330000}{9.996000}\selectfont\catcode`\^=\active\def^{\ifmmode\sp\else\^{}\fi}\catcode`\%=\active\def%{\%}$\mathdefault{0.10}$}}%
\end{pgfscope}%
\begin{pgfscope}%
\definecolor{textcolor}{rgb}{0.000000,0.000000,0.000000}%
\pgfsetstrokecolor{textcolor}%
\pgfsetfillcolor{textcolor}%
\pgftext[x=1.749374in,y=0.292060in,,top]{\color{textcolor}{\rmfamily\fontsize{10.000000}{12.000000}\selectfont\catcode`\^=\active\def^{\ifmmode\sp\else\^{}\fi}\catcode`\%=\active\def%{\%}$x\;[\text{m}]$}}%
\end{pgfscope}%
\begin{pgfscope}%
\pgfsetbuttcap%
\pgfsetroundjoin%
\definecolor{currentfill}{rgb}{0.000000,0.000000,0.000000}%
\pgfsetfillcolor{currentfill}%
\pgfsetlinewidth{0.803000pt}%
\definecolor{currentstroke}{rgb}{0.000000,0.000000,0.000000}%
\pgfsetstrokecolor{currentstroke}%
\pgfsetdash{}{0pt}%
\pgfsys@defobject{currentmarker}{\pgfqpoint{-0.048611in}{0.000000in}}{\pgfqpoint{-0.000000in}{0.000000in}}{%
\pgfpathmoveto{\pgfqpoint{-0.000000in}{0.000000in}}%
\pgfpathlineto{\pgfqpoint{-0.048611in}{0.000000in}}%
\pgfusepath{stroke,fill}%
}%
\begin{pgfscope}%
\pgfsys@transformshift{0.651925in}{0.590331in}%
\pgfsys@useobject{currentmarker}{}%
\end{pgfscope}%
\end{pgfscope}%
\begin{pgfscope}%
\definecolor{textcolor}{rgb}{0.000000,0.000000,0.000000}%
\pgfsetstrokecolor{textcolor}%
\pgfsetfillcolor{textcolor}%
\pgftext[x=0.351768in, y=0.551751in, left, base]{\color{textcolor}{\rmfamily\fontsize{8.330000}{9.996000}\selectfont\catcode`\^=\active\def^{\ifmmode\sp\else\^{}\fi}\catcode`\%=\active\def%{\%}$\mathdefault{0.00}$}}%
\end{pgfscope}%
\begin{pgfscope}%
\pgfsetbuttcap%
\pgfsetroundjoin%
\definecolor{currentfill}{rgb}{0.000000,0.000000,0.000000}%
\pgfsetfillcolor{currentfill}%
\pgfsetlinewidth{0.803000pt}%
\definecolor{currentstroke}{rgb}{0.000000,0.000000,0.000000}%
\pgfsetstrokecolor{currentstroke}%
\pgfsetdash{}{0pt}%
\pgfsys@defobject{currentmarker}{\pgfqpoint{-0.048611in}{0.000000in}}{\pgfqpoint{-0.000000in}{0.000000in}}{%
\pgfpathmoveto{\pgfqpoint{-0.000000in}{0.000000in}}%
\pgfpathlineto{\pgfqpoint{-0.048611in}{0.000000in}}%
\pgfusepath{stroke,fill}%
}%
\begin{pgfscope}%
\pgfsys@transformshift{0.651925in}{0.859065in}%
\pgfsys@useobject{currentmarker}{}%
\end{pgfscope}%
\end{pgfscope}%
\begin{pgfscope}%
\definecolor{textcolor}{rgb}{0.000000,0.000000,0.000000}%
\pgfsetstrokecolor{textcolor}%
\pgfsetfillcolor{textcolor}%
\pgftext[x=0.351768in, y=0.820485in, left, base]{\color{textcolor}{\rmfamily\fontsize{8.330000}{9.996000}\selectfont\catcode`\^=\active\def^{\ifmmode\sp\else\^{}\fi}\catcode`\%=\active\def%{\%}$\mathdefault{0.01}$}}%
\end{pgfscope}%
\begin{pgfscope}%
\pgfsetbuttcap%
\pgfsetroundjoin%
\definecolor{currentfill}{rgb}{0.000000,0.000000,0.000000}%
\pgfsetfillcolor{currentfill}%
\pgfsetlinewidth{0.803000pt}%
\definecolor{currentstroke}{rgb}{0.000000,0.000000,0.000000}%
\pgfsetstrokecolor{currentstroke}%
\pgfsetdash{}{0pt}%
\pgfsys@defobject{currentmarker}{\pgfqpoint{-0.048611in}{0.000000in}}{\pgfqpoint{-0.000000in}{0.000000in}}{%
\pgfpathmoveto{\pgfqpoint{-0.000000in}{0.000000in}}%
\pgfpathlineto{\pgfqpoint{-0.048611in}{0.000000in}}%
\pgfusepath{stroke,fill}%
}%
\begin{pgfscope}%
\pgfsys@transformshift{0.651925in}{1.127799in}%
\pgfsys@useobject{currentmarker}{}%
\end{pgfscope}%
\end{pgfscope}%
\begin{pgfscope}%
\definecolor{textcolor}{rgb}{0.000000,0.000000,0.000000}%
\pgfsetstrokecolor{textcolor}%
\pgfsetfillcolor{textcolor}%
\pgftext[x=0.351768in, y=1.089219in, left, base]{\color{textcolor}{\rmfamily\fontsize{8.330000}{9.996000}\selectfont\catcode`\^=\active\def^{\ifmmode\sp\else\^{}\fi}\catcode`\%=\active\def%{\%}$\mathdefault{0.02}$}}%
\end{pgfscope}%
\begin{pgfscope}%
\pgfsetbuttcap%
\pgfsetroundjoin%
\definecolor{currentfill}{rgb}{0.000000,0.000000,0.000000}%
\pgfsetfillcolor{currentfill}%
\pgfsetlinewidth{0.803000pt}%
\definecolor{currentstroke}{rgb}{0.000000,0.000000,0.000000}%
\pgfsetstrokecolor{currentstroke}%
\pgfsetdash{}{0pt}%
\pgfsys@defobject{currentmarker}{\pgfqpoint{-0.048611in}{0.000000in}}{\pgfqpoint{-0.000000in}{0.000000in}}{%
\pgfpathmoveto{\pgfqpoint{-0.000000in}{0.000000in}}%
\pgfpathlineto{\pgfqpoint{-0.048611in}{0.000000in}}%
\pgfusepath{stroke,fill}%
}%
\begin{pgfscope}%
\pgfsys@transformshift{0.651925in}{1.396533in}%
\pgfsys@useobject{currentmarker}{}%
\end{pgfscope}%
\end{pgfscope}%
\begin{pgfscope}%
\definecolor{textcolor}{rgb}{0.000000,0.000000,0.000000}%
\pgfsetstrokecolor{textcolor}%
\pgfsetfillcolor{textcolor}%
\pgftext[x=0.351768in, y=1.357953in, left, base]{\color{textcolor}{\rmfamily\fontsize{8.330000}{9.996000}\selectfont\catcode`\^=\active\def^{\ifmmode\sp\else\^{}\fi}\catcode`\%=\active\def%{\%}$\mathdefault{0.03}$}}%
\end{pgfscope}%
\begin{pgfscope}%
\pgfsetbuttcap%
\pgfsetroundjoin%
\definecolor{currentfill}{rgb}{0.000000,0.000000,0.000000}%
\pgfsetfillcolor{currentfill}%
\pgfsetlinewidth{0.803000pt}%
\definecolor{currentstroke}{rgb}{0.000000,0.000000,0.000000}%
\pgfsetstrokecolor{currentstroke}%
\pgfsetdash{}{0pt}%
\pgfsys@defobject{currentmarker}{\pgfqpoint{-0.048611in}{0.000000in}}{\pgfqpoint{-0.000000in}{0.000000in}}{%
\pgfpathmoveto{\pgfqpoint{-0.000000in}{0.000000in}}%
\pgfpathlineto{\pgfqpoint{-0.048611in}{0.000000in}}%
\pgfusepath{stroke,fill}%
}%
\begin{pgfscope}%
\pgfsys@transformshift{0.651925in}{1.665268in}%
\pgfsys@useobject{currentmarker}{}%
\end{pgfscope}%
\end{pgfscope}%
\begin{pgfscope}%
\definecolor{textcolor}{rgb}{0.000000,0.000000,0.000000}%
\pgfsetstrokecolor{textcolor}%
\pgfsetfillcolor{textcolor}%
\pgftext[x=0.351768in, y=1.626687in, left, base]{\color{textcolor}{\rmfamily\fontsize{8.330000}{9.996000}\selectfont\catcode`\^=\active\def^{\ifmmode\sp\else\^{}\fi}\catcode`\%=\active\def%{\%}$\mathdefault{0.04}$}}%
\end{pgfscope}%
\begin{pgfscope}%
\definecolor{textcolor}{rgb}{0.000000,0.000000,0.000000}%
\pgfsetstrokecolor{textcolor}%
\pgfsetfillcolor{textcolor}%
\pgftext[x=0.296212in,y=1.127056in,,bottom,rotate=90.000000]{\color{textcolor}{\rmfamily\fontsize{10.000000}{12.000000}\selectfont\catcode`\^=\active\def^{\ifmmode\sp\else\^{}\fi}\catcode`\%=\active\def%{\%}$\nicefrac{\|\bm{v}\|_2}{\omega}\;[\nicefrac{\text{m}}{rad}]$}}%
\end{pgfscope}%
\begin{pgfscope}%
\pgfpathrectangle{\pgfqpoint{0.651925in}{0.536658in}}{\pgfqpoint{2.194896in}{1.180795in}}%
\pgfusepath{clip}%
\pgfsetrectcap%
\pgfsetroundjoin%
\pgfsetlinewidth{1.505625pt}%
\definecolor{currentstroke}{rgb}{0.843137,0.098039,0.109804}%
\pgfsetstrokecolor{currentstroke}%
\pgfsetdash{}{0pt}%
\pgfpathmoveto{\pgfqpoint{0.751693in}{0.697825in}}%
\pgfpathlineto{\pgfqpoint{1.498705in}{1.663653in}}%
\pgfpathlineto{\pgfqpoint{1.500702in}{1.663781in}}%
\pgfpathlineto{\pgfqpoint{1.512687in}{1.640294in}}%
\pgfpathlineto{\pgfqpoint{1.524671in}{1.617347in}}%
\pgfpathlineto{\pgfqpoint{1.536655in}{1.594920in}}%
\pgfpathlineto{\pgfqpoint{1.550636in}{1.569382in}}%
\pgfpathlineto{\pgfqpoint{1.564618in}{1.544491in}}%
\pgfpathlineto{\pgfqpoint{1.578599in}{1.520215in}}%
\pgfpathlineto{\pgfqpoint{1.592581in}{1.496527in}}%
\pgfpathlineto{\pgfqpoint{1.606562in}{1.473400in}}%
\pgfpathlineto{\pgfqpoint{1.620544in}{1.450811in}}%
\pgfpathlineto{\pgfqpoint{1.634525in}{1.428736in}}%
\pgfpathlineto{\pgfqpoint{1.648507in}{1.407152in}}%
\pgfpathlineto{\pgfqpoint{1.662488in}{1.386039in}}%
\pgfpathlineto{\pgfqpoint{1.676470in}{1.365377in}}%
\pgfpathlineto{\pgfqpoint{1.692449in}{1.342292in}}%
\pgfpathlineto{\pgfqpoint{1.708428in}{1.319747in}}%
\pgfpathlineto{\pgfqpoint{1.724407in}{1.297717in}}%
\pgfpathlineto{\pgfqpoint{1.740385in}{1.276178in}}%
\pgfpathlineto{\pgfqpoint{1.756364in}{1.255109in}}%
\pgfpathlineto{\pgfqpoint{1.772343in}{1.234490in}}%
\pgfpathlineto{\pgfqpoint{1.788322in}{1.214302in}}%
\pgfpathlineto{\pgfqpoint{1.804301in}{1.194525in}}%
\pgfpathlineto{\pgfqpoint{1.822277in}{1.172748in}}%
\pgfpathlineto{\pgfqpoint{1.840253in}{1.151448in}}%
\pgfpathlineto{\pgfqpoint{1.858229in}{1.130604in}}%
\pgfpathlineto{\pgfqpoint{1.876206in}{1.110195in}}%
\pgfpathlineto{\pgfqpoint{1.894182in}{1.090201in}}%
\pgfpathlineto{\pgfqpoint{1.912158in}{1.070605in}}%
\pgfpathlineto{\pgfqpoint{1.930134in}{1.051391in}}%
\pgfpathlineto{\pgfqpoint{1.950108in}{1.030468in}}%
\pgfpathlineto{\pgfqpoint{1.970082in}{1.009976in}}%
\pgfpathlineto{\pgfqpoint{1.990055in}{0.989893in}}%
\pgfpathlineto{\pgfqpoint{2.010029in}{0.970203in}}%
\pgfpathlineto{\pgfqpoint{2.030002in}{0.950887in}}%
\pgfpathlineto{\pgfqpoint{2.049976in}{0.931929in}}%
\pgfpathlineto{\pgfqpoint{2.071947in}{0.911471in}}%
\pgfpathlineto{\pgfqpoint{2.093918in}{0.891410in}}%
\pgfpathlineto{\pgfqpoint{2.115889in}{0.871726in}}%
\pgfpathlineto{\pgfqpoint{2.137860in}{0.852403in}}%
\pgfpathlineto{\pgfqpoint{2.159830in}{0.833425in}}%
\pgfpathlineto{\pgfqpoint{2.183799in}{0.813098in}}%
\pgfpathlineto{\pgfqpoint{2.207767in}{0.793145in}}%
\pgfpathlineto{\pgfqpoint{2.231735in}{0.773549in}}%
\pgfpathlineto{\pgfqpoint{2.255704in}{0.754294in}}%
\pgfpathlineto{\pgfqpoint{2.281669in}{0.733802in}}%
\pgfpathlineto{\pgfqpoint{2.307635in}{0.713674in}}%
\pgfpathlineto{\pgfqpoint{2.333601in}{0.693893in}}%
\pgfpathlineto{\pgfqpoint{2.359566in}{0.674443in}}%
\pgfpathlineto{\pgfqpoint{2.387529in}{0.653850in}}%
\pgfpathlineto{\pgfqpoint{2.415492in}{0.633606in}}%
\pgfpathlineto{\pgfqpoint{2.443455in}{0.613695in}}%
\pgfpathlineto{\pgfqpoint{2.473416in}{0.592713in}}%
\pgfpathlineto{\pgfqpoint{2.475413in}{0.591326in}}%
\pgfpathlineto{\pgfqpoint{2.477410in}{0.590720in}}%
\pgfpathlineto{\pgfqpoint{2.507371in}{0.611312in}}%
\pgfpathlineto{\pgfqpoint{2.537331in}{0.631575in}}%
\pgfpathlineto{\pgfqpoint{2.569289in}{0.652847in}}%
\pgfpathlineto{\pgfqpoint{2.601247in}{0.673781in}}%
\pgfpathlineto{\pgfqpoint{2.633204in}{0.694395in}}%
\pgfpathlineto{\pgfqpoint{2.667159in}{0.715963in}}%
\pgfpathlineto{\pgfqpoint{2.701114in}{0.737203in}}%
\pgfpathlineto{\pgfqpoint{2.737067in}{0.759354in}}%
\pgfpathlineto{\pgfqpoint{2.747054in}{0.765447in}}%
\pgfpathlineto{\pgfqpoint{2.747054in}{0.765447in}}%
\pgfusepath{stroke}%
\end{pgfscope}%
\begin{pgfscope}%
\pgfpathrectangle{\pgfqpoint{0.651925in}{0.536658in}}{\pgfqpoint{2.194896in}{1.180795in}}%
\pgfusepath{clip}%
\pgfsetbuttcap%
\pgfsetroundjoin%
\pgfsetlinewidth{1.505625pt}%
\definecolor{currentstroke}{rgb}{0.172549,0.482353,0.713725}%
\pgfsetstrokecolor{currentstroke}%
\pgfsetdash{{1.500000pt}{2.475000pt}}{0.000000pt}%
\pgfpathmoveto{\pgfqpoint{0.751693in}{0.697819in}}%
\pgfpathlineto{\pgfqpoint{1.498705in}{1.663601in}}%
\pgfpathlineto{\pgfqpoint{1.500702in}{1.588604in}}%
\pgfpathlineto{\pgfqpoint{1.502700in}{1.406226in}}%
\pgfpathlineto{\pgfqpoint{1.504697in}{1.279790in}}%
\pgfpathlineto{\pgfqpoint{1.506695in}{1.194513in}}%
\pgfpathlineto{\pgfqpoint{1.508692in}{1.130430in}}%
\pgfpathlineto{\pgfqpoint{1.510689in}{1.084878in}}%
\pgfpathlineto{\pgfqpoint{1.512687in}{1.057828in}}%
\pgfpathlineto{\pgfqpoint{1.514684in}{1.038645in}}%
\pgfpathlineto{\pgfqpoint{1.516681in}{1.024289in}}%
\pgfpathlineto{\pgfqpoint{1.518679in}{1.020154in}}%
\pgfpathlineto{\pgfqpoint{1.520676in}{1.016351in}}%
\pgfpathlineto{\pgfqpoint{1.522673in}{1.015385in}}%
\pgfpathlineto{\pgfqpoint{1.524671in}{1.017304in}}%
\pgfpathlineto{\pgfqpoint{1.528665in}{1.023308in}}%
\pgfpathlineto{\pgfqpoint{1.532660in}{1.028676in}}%
\pgfpathlineto{\pgfqpoint{1.534658in}{1.030455in}}%
\pgfpathlineto{\pgfqpoint{1.536655in}{1.031416in}}%
\pgfpathlineto{\pgfqpoint{1.540650in}{1.031951in}}%
\pgfpathlineto{\pgfqpoint{1.542647in}{1.033275in}}%
\pgfpathlineto{\pgfqpoint{1.544644in}{1.034838in}}%
\pgfpathlineto{\pgfqpoint{1.548639in}{1.036082in}}%
\pgfpathlineto{\pgfqpoint{1.550636in}{1.036411in}}%
\pgfpathlineto{\pgfqpoint{1.552634in}{1.035749in}}%
\pgfpathlineto{\pgfqpoint{1.556628in}{1.034846in}}%
\pgfpathlineto{\pgfqpoint{1.562621in}{1.032428in}}%
\pgfpathlineto{\pgfqpoint{1.564618in}{1.031339in}}%
\pgfpathlineto{\pgfqpoint{1.568613in}{1.028667in}}%
\pgfpathlineto{\pgfqpoint{1.574605in}{1.024784in}}%
\pgfpathlineto{\pgfqpoint{1.578599in}{1.021429in}}%
\pgfpathlineto{\pgfqpoint{1.582594in}{1.017594in}}%
\pgfpathlineto{\pgfqpoint{1.592581in}{1.009258in}}%
\pgfpathlineto{\pgfqpoint{1.596576in}{1.005446in}}%
\pgfpathlineto{\pgfqpoint{1.602568in}{1.000526in}}%
\pgfpathlineto{\pgfqpoint{1.606562in}{0.997576in}}%
\pgfpathlineto{\pgfqpoint{1.612555in}{0.992794in}}%
\pgfpathlineto{\pgfqpoint{1.618547in}{0.988666in}}%
\pgfpathlineto{\pgfqpoint{1.622541in}{0.986060in}}%
\pgfpathlineto{\pgfqpoint{1.628533in}{0.982605in}}%
\pgfpathlineto{\pgfqpoint{1.634525in}{0.978447in}}%
\pgfpathlineto{\pgfqpoint{1.644512in}{0.972340in}}%
\pgfpathlineto{\pgfqpoint{1.648507in}{0.970039in}}%
\pgfpathlineto{\pgfqpoint{1.654499in}{0.966676in}}%
\pgfpathlineto{\pgfqpoint{1.662488in}{0.961990in}}%
\pgfpathlineto{\pgfqpoint{1.668481in}{0.958875in}}%
\pgfpathlineto{\pgfqpoint{1.676470in}{0.955586in}}%
\pgfpathlineto{\pgfqpoint{1.686457in}{0.951821in}}%
\pgfpathlineto{\pgfqpoint{1.692449in}{0.949590in}}%
\pgfpathlineto{\pgfqpoint{1.700438in}{0.947058in}}%
\pgfpathlineto{\pgfqpoint{1.722409in}{0.941157in}}%
\pgfpathlineto{\pgfqpoint{1.736391in}{0.937885in}}%
\pgfpathlineto{\pgfqpoint{1.750372in}{0.935012in}}%
\pgfpathlineto{\pgfqpoint{1.766351in}{0.932086in}}%
\pgfpathlineto{\pgfqpoint{1.796311in}{0.927527in}}%
\pgfpathlineto{\pgfqpoint{1.816285in}{0.924286in}}%
\pgfpathlineto{\pgfqpoint{1.836259in}{0.920659in}}%
\pgfpathlineto{\pgfqpoint{1.844248in}{0.918851in}}%
\pgfpathlineto{\pgfqpoint{1.866219in}{0.913637in}}%
\pgfpathlineto{\pgfqpoint{1.876206in}{0.911070in}}%
\pgfpathlineto{\pgfqpoint{1.890187in}{0.906654in}}%
\pgfpathlineto{\pgfqpoint{1.916153in}{0.897724in}}%
\pgfpathlineto{\pgfqpoint{1.926140in}{0.894019in}}%
\pgfpathlineto{\pgfqpoint{1.942119in}{0.887162in}}%
\pgfpathlineto{\pgfqpoint{1.958097in}{0.879952in}}%
\pgfpathlineto{\pgfqpoint{1.978071in}{0.870670in}}%
\pgfpathlineto{\pgfqpoint{1.994050in}{0.862940in}}%
\pgfpathlineto{\pgfqpoint{2.002039in}{0.859059in}}%
\pgfpathlineto{\pgfqpoint{2.014023in}{0.852895in}}%
\pgfpathlineto{\pgfqpoint{2.041986in}{0.837659in}}%
\pgfpathlineto{\pgfqpoint{2.051973in}{0.832335in}}%
\pgfpathlineto{\pgfqpoint{2.069949in}{0.822413in}}%
\pgfpathlineto{\pgfqpoint{2.077939in}{0.817952in}}%
\pgfpathlineto{\pgfqpoint{2.089923in}{0.810985in}}%
\pgfpathlineto{\pgfqpoint{2.101907in}{0.803881in}}%
\pgfpathlineto{\pgfqpoint{2.113891in}{0.796918in}}%
\pgfpathlineto{\pgfqpoint{2.155836in}{0.771022in}}%
\pgfpathlineto{\pgfqpoint{2.175809in}{0.758964in}}%
\pgfpathlineto{\pgfqpoint{2.195783in}{0.747009in}}%
\pgfpathlineto{\pgfqpoint{2.227741in}{0.728295in}}%
\pgfpathlineto{\pgfqpoint{2.241722in}{0.720516in}}%
\pgfpathlineto{\pgfqpoint{2.261696in}{0.709267in}}%
\pgfpathlineto{\pgfqpoint{2.297648in}{0.690082in}}%
\pgfpathlineto{\pgfqpoint{2.335598in}{0.671360in}}%
\pgfpathlineto{\pgfqpoint{2.359566in}{0.660519in}}%
\pgfpathlineto{\pgfqpoint{2.375545in}{0.653750in}}%
\pgfpathlineto{\pgfqpoint{2.391524in}{0.647205in}}%
\pgfpathlineto{\pgfqpoint{2.399513in}{0.644138in}}%
\pgfpathlineto{\pgfqpoint{2.423482in}{0.636065in}}%
\pgfpathlineto{\pgfqpoint{2.437463in}{0.631663in}}%
\pgfpathlineto{\pgfqpoint{2.457437in}{0.626265in}}%
\pgfpathlineto{\pgfqpoint{2.469421in}{0.623242in}}%
\pgfpathlineto{\pgfqpoint{2.485400in}{0.619382in}}%
\pgfpathlineto{\pgfqpoint{2.529342in}{0.610525in}}%
\pgfpathlineto{\pgfqpoint{2.549315in}{0.606970in}}%
\pgfpathlineto{\pgfqpoint{2.581273in}{0.601939in}}%
\pgfpathlineto{\pgfqpoint{2.599249in}{0.599592in}}%
\pgfpathlineto{\pgfqpoint{2.615228in}{0.597713in}}%
\pgfpathlineto{\pgfqpoint{2.635202in}{0.595694in}}%
\pgfpathlineto{\pgfqpoint{2.659170in}{0.593631in}}%
\pgfpathlineto{\pgfqpoint{2.677146in}{0.592452in}}%
\pgfpathlineto{\pgfqpoint{2.693125in}{0.591576in}}%
\pgfpathlineto{\pgfqpoint{2.709104in}{0.590980in}}%
\pgfpathlineto{\pgfqpoint{2.723085in}{0.590747in}}%
\pgfpathlineto{\pgfqpoint{2.743059in}{0.590553in}}%
\pgfpathlineto{\pgfqpoint{2.747054in}{0.590331in}}%
\pgfpathlineto{\pgfqpoint{2.747054in}{0.590331in}}%
\pgfusepath{stroke}%
\end{pgfscope}%
\begin{pgfscope}%
\pgfpathrectangle{\pgfqpoint{0.651925in}{0.536658in}}{\pgfqpoint{2.194896in}{1.180795in}}%
\pgfusepath{clip}%
\pgfsetbuttcap%
\pgfsetroundjoin%
\pgfsetlinewidth{1.505625pt}%
\definecolor{currentstroke}{rgb}{0.992157,0.682353,0.380392}%
\pgfsetstrokecolor{currentstroke}%
\pgfsetdash{{5.550000pt}{2.400000pt}}{0.000000pt}%
\pgfpathmoveto{\pgfqpoint{0.751693in}{0.697819in}}%
\pgfpathlineto{\pgfqpoint{1.498705in}{1.663601in}}%
\pgfpathlineto{\pgfqpoint{1.500702in}{1.646225in}}%
\pgfpathlineto{\pgfqpoint{1.502700in}{1.604746in}}%
\pgfpathlineto{\pgfqpoint{1.504697in}{1.584288in}}%
\pgfpathlineto{\pgfqpoint{1.506695in}{1.575807in}}%
\pgfpathlineto{\pgfqpoint{1.508692in}{1.572375in}}%
\pgfpathlineto{\pgfqpoint{1.510689in}{1.570287in}}%
\pgfpathlineto{\pgfqpoint{1.512687in}{1.568442in}}%
\pgfpathlineto{\pgfqpoint{1.514684in}{1.566375in}}%
\pgfpathlineto{\pgfqpoint{1.516681in}{1.563931in}}%
\pgfpathlineto{\pgfqpoint{1.518679in}{1.561070in}}%
\pgfpathlineto{\pgfqpoint{1.522673in}{1.554803in}}%
\pgfpathlineto{\pgfqpoint{1.526668in}{1.547948in}}%
\pgfpathlineto{\pgfqpoint{1.532660in}{1.537219in}}%
\pgfpathlineto{\pgfqpoint{1.540650in}{1.522076in}}%
\pgfpathlineto{\pgfqpoint{1.554631in}{1.494657in}}%
\pgfpathlineto{\pgfqpoint{1.574605in}{1.455494in}}%
\pgfpathlineto{\pgfqpoint{1.586589in}{1.432517in}}%
\pgfpathlineto{\pgfqpoint{1.598573in}{1.410133in}}%
\pgfpathlineto{\pgfqpoint{1.612555in}{1.384638in}}%
\pgfpathlineto{\pgfqpoint{1.626536in}{1.359836in}}%
\pgfpathlineto{\pgfqpoint{1.638520in}{1.339111in}}%
\pgfpathlineto{\pgfqpoint{1.652502in}{1.315397in}}%
\pgfpathlineto{\pgfqpoint{1.664486in}{1.295649in}}%
\pgfpathlineto{\pgfqpoint{1.676470in}{1.276283in}}%
\pgfpathlineto{\pgfqpoint{1.690451in}{1.254224in}}%
\pgfpathlineto{\pgfqpoint{1.706430in}{1.229670in}}%
\pgfpathlineto{\pgfqpoint{1.722409in}{1.205846in}}%
\pgfpathlineto{\pgfqpoint{1.732396in}{1.191276in}}%
\pgfpathlineto{\pgfqpoint{1.740385in}{1.179875in}}%
\pgfpathlineto{\pgfqpoint{1.756364in}{1.157411in}}%
\pgfpathlineto{\pgfqpoint{1.770346in}{1.138244in}}%
\pgfpathlineto{\pgfqpoint{1.784327in}{1.119639in}}%
\pgfpathlineto{\pgfqpoint{1.798309in}{1.101398in}}%
\pgfpathlineto{\pgfqpoint{1.808296in}{1.088682in}}%
\pgfpathlineto{\pgfqpoint{1.822277in}{1.071138in}}%
\pgfpathlineto{\pgfqpoint{1.830266in}{1.061348in}}%
\pgfpathlineto{\pgfqpoint{1.848243in}{1.039681in}}%
\pgfpathlineto{\pgfqpoint{1.860227in}{1.025646in}}%
\pgfpathlineto{\pgfqpoint{1.872211in}{1.011863in}}%
\pgfpathlineto{\pgfqpoint{1.884195in}{0.998318in}}%
\pgfpathlineto{\pgfqpoint{1.896179in}{0.985024in}}%
\pgfpathlineto{\pgfqpoint{1.904169in}{0.976380in}}%
\pgfpathlineto{\pgfqpoint{1.918150in}{0.961623in}}%
\pgfpathlineto{\pgfqpoint{1.936126in}{0.943066in}}%
\pgfpathlineto{\pgfqpoint{1.952105in}{0.927110in}}%
\pgfpathlineto{\pgfqpoint{1.968084in}{0.911614in}}%
\pgfpathlineto{\pgfqpoint{1.986060in}{0.894609in}}%
\pgfpathlineto{\pgfqpoint{2.002039in}{0.880128in}}%
\pgfpathlineto{\pgfqpoint{2.026008in}{0.858864in}}%
\pgfpathlineto{\pgfqpoint{2.047978in}{0.840245in}}%
\pgfpathlineto{\pgfqpoint{2.063957in}{0.827118in}}%
\pgfpathlineto{\pgfqpoint{2.087926in}{0.808007in}}%
\pgfpathlineto{\pgfqpoint{2.103904in}{0.795808in}}%
\pgfpathlineto{\pgfqpoint{2.127873in}{0.778145in}}%
\pgfpathlineto{\pgfqpoint{2.141854in}{0.768165in}}%
\pgfpathlineto{\pgfqpoint{2.155836in}{0.758528in}}%
\pgfpathlineto{\pgfqpoint{2.169817in}{0.749176in}}%
\pgfpathlineto{\pgfqpoint{2.187794in}{0.737393in}}%
\pgfpathlineto{\pgfqpoint{2.207767in}{0.724830in}}%
\pgfpathlineto{\pgfqpoint{2.217754in}{0.718796in}}%
\pgfpathlineto{\pgfqpoint{2.235730in}{0.708216in}}%
\pgfpathlineto{\pgfqpoint{2.261696in}{0.693614in}}%
\pgfpathlineto{\pgfqpoint{2.285664in}{0.680931in}}%
\pgfpathlineto{\pgfqpoint{2.301643in}{0.672792in}}%
\pgfpathlineto{\pgfqpoint{2.335598in}{0.656555in}}%
\pgfpathlineto{\pgfqpoint{2.357569in}{0.647022in}}%
\pgfpathlineto{\pgfqpoint{2.367556in}{0.642910in}}%
\pgfpathlineto{\pgfqpoint{2.381537in}{0.637353in}}%
\pgfpathlineto{\pgfqpoint{2.397516in}{0.631178in}}%
\pgfpathlineto{\pgfqpoint{2.421484in}{0.622898in}}%
\pgfpathlineto{\pgfqpoint{2.435466in}{0.618294in}}%
\pgfpathlineto{\pgfqpoint{2.453442in}{0.613116in}}%
\pgfpathlineto{\pgfqpoint{2.475413in}{0.607270in}}%
\pgfpathlineto{\pgfqpoint{2.487397in}{0.604320in}}%
\pgfpathlineto{\pgfqpoint{2.511365in}{0.599199in}}%
\pgfpathlineto{\pgfqpoint{2.527344in}{0.596237in}}%
\pgfpathlineto{\pgfqpoint{2.549315in}{0.592651in}}%
\pgfpathlineto{\pgfqpoint{2.559302in}{0.591378in}}%
\pgfpathlineto{\pgfqpoint{2.569289in}{0.590791in}}%
\pgfpathlineto{\pgfqpoint{2.573284in}{0.591073in}}%
\pgfpathlineto{\pgfqpoint{2.595254in}{0.593075in}}%
\pgfpathlineto{\pgfqpoint{2.609236in}{0.594092in}}%
\pgfpathlineto{\pgfqpoint{2.625215in}{0.594949in}}%
\pgfpathlineto{\pgfqpoint{2.641194in}{0.595382in}}%
\pgfpathlineto{\pgfqpoint{2.657173in}{0.595504in}}%
\pgfpathlineto{\pgfqpoint{2.669157in}{0.595390in}}%
\pgfpathlineto{\pgfqpoint{2.683138in}{0.595035in}}%
\pgfpathlineto{\pgfqpoint{2.699117in}{0.594333in}}%
\pgfpathlineto{\pgfqpoint{2.713099in}{0.593459in}}%
\pgfpathlineto{\pgfqpoint{2.729077in}{0.592164in}}%
\pgfpathlineto{\pgfqpoint{2.745056in}{0.590552in}}%
\pgfpathlineto{\pgfqpoint{2.747054in}{0.590331in}}%
\pgfpathlineto{\pgfqpoint{2.747054in}{0.590331in}}%
\pgfusepath{stroke}%
\end{pgfscope}%
\begin{pgfscope}%
\pgfpathrectangle{\pgfqpoint{0.651925in}{0.536658in}}{\pgfqpoint{2.194896in}{1.180795in}}%
\pgfusepath{clip}%
\pgfsetbuttcap%
\pgfsetroundjoin%
\pgfsetlinewidth{1.505625pt}%
\definecolor{currentstroke}{rgb}{0.670588,0.850980,0.913725}%
\pgfsetstrokecolor{currentstroke}%
\pgfsetdash{{9.600000pt}{2.400000pt}{1.500000pt}{2.400000pt}}{0.000000pt}%
\pgfpathmoveto{\pgfqpoint{0.751693in}{0.697823in}}%
\pgfpathlineto{\pgfqpoint{1.498705in}{1.663639in}}%
\pgfpathlineto{\pgfqpoint{1.500702in}{1.657573in}}%
\pgfpathlineto{\pgfqpoint{1.502700in}{1.639133in}}%
\pgfpathlineto{\pgfqpoint{1.504697in}{1.626930in}}%
\pgfpathlineto{\pgfqpoint{1.506695in}{1.620776in}}%
\pgfpathlineto{\pgfqpoint{1.508692in}{1.617058in}}%
\pgfpathlineto{\pgfqpoint{1.512687in}{1.610453in}}%
\pgfpathlineto{\pgfqpoint{1.516681in}{1.602999in}}%
\pgfpathlineto{\pgfqpoint{1.522673in}{1.590756in}}%
\pgfpathlineto{\pgfqpoint{1.548639in}{1.536626in}}%
\pgfpathlineto{\pgfqpoint{1.558626in}{1.516679in}}%
\pgfpathlineto{\pgfqpoint{1.568613in}{1.497246in}}%
\pgfpathlineto{\pgfqpoint{1.578599in}{1.478335in}}%
\pgfpathlineto{\pgfqpoint{1.590584in}{1.456343in}}%
\pgfpathlineto{\pgfqpoint{1.600570in}{1.438535in}}%
\pgfpathlineto{\pgfqpoint{1.612555in}{1.417768in}}%
\pgfpathlineto{\pgfqpoint{1.624539in}{1.397601in}}%
\pgfpathlineto{\pgfqpoint{1.636523in}{1.378001in}}%
\pgfpathlineto{\pgfqpoint{1.648507in}{1.358914in}}%
\pgfpathlineto{\pgfqpoint{1.660491in}{1.340313in}}%
\pgfpathlineto{\pgfqpoint{1.674473in}{1.319231in}}%
\pgfpathlineto{\pgfqpoint{1.690451in}{1.295887in}}%
\pgfpathlineto{\pgfqpoint{1.706430in}{1.273268in}}%
\pgfpathlineto{\pgfqpoint{1.720412in}{1.254079in}}%
\pgfpathlineto{\pgfqpoint{1.734393in}{1.235337in}}%
\pgfpathlineto{\pgfqpoint{1.750372in}{1.214562in}}%
\pgfpathlineto{\pgfqpoint{1.764354in}{1.196772in}}%
\pgfpathlineto{\pgfqpoint{1.778335in}{1.179423in}}%
\pgfpathlineto{\pgfqpoint{1.796311in}{1.157677in}}%
\pgfpathlineto{\pgfqpoint{1.808296in}{1.143511in}}%
\pgfpathlineto{\pgfqpoint{1.822277in}{1.127270in}}%
\pgfpathlineto{\pgfqpoint{1.832264in}{1.115907in}}%
\pgfpathlineto{\pgfqpoint{1.846245in}{1.100260in}}%
\pgfpathlineto{\pgfqpoint{1.860227in}{1.084902in}}%
\pgfpathlineto{\pgfqpoint{1.874208in}{1.069808in}}%
\pgfpathlineto{\pgfqpoint{1.896179in}{1.046649in}}%
\pgfpathlineto{\pgfqpoint{1.908163in}{1.034302in}}%
\pgfpathlineto{\pgfqpoint{1.922145in}{1.020146in}}%
\pgfpathlineto{\pgfqpoint{1.936126in}{1.006218in}}%
\pgfpathlineto{\pgfqpoint{1.956100in}{0.986747in}}%
\pgfpathlineto{\pgfqpoint{1.974076in}{0.969575in}}%
\pgfpathlineto{\pgfqpoint{1.998045in}{0.947277in}}%
\pgfpathlineto{\pgfqpoint{2.018018in}{0.929131in}}%
\pgfpathlineto{\pgfqpoint{2.035994in}{0.913117in}}%
\pgfpathlineto{\pgfqpoint{2.059963in}{0.892255in}}%
\pgfpathlineto{\pgfqpoint{2.077939in}{0.876941in}}%
\pgfpathlineto{\pgfqpoint{2.099910in}{0.858574in}}%
\pgfpathlineto{\pgfqpoint{2.125875in}{0.837413in}}%
\pgfpathlineto{\pgfqpoint{2.149844in}{0.818319in}}%
\pgfpathlineto{\pgfqpoint{2.175809in}{0.798121in}}%
\pgfpathlineto{\pgfqpoint{2.201775in}{0.778382in}}%
\pgfpathlineto{\pgfqpoint{2.223746in}{0.762080in}}%
\pgfpathlineto{\pgfqpoint{2.251709in}{0.741821in}}%
\pgfpathlineto{\pgfqpoint{2.275677in}{0.724828in}}%
\pgfpathlineto{\pgfqpoint{2.307635in}{0.702600in}}%
\pgfpathlineto{\pgfqpoint{2.337595in}{0.682114in}}%
\pgfpathlineto{\pgfqpoint{2.359566in}{0.667527in}}%
\pgfpathlineto{\pgfqpoint{2.375545in}{0.657298in}}%
\pgfpathlineto{\pgfqpoint{2.393521in}{0.646198in}}%
\pgfpathlineto{\pgfqpoint{2.399513in}{0.642644in}}%
\pgfpathlineto{\pgfqpoint{2.417490in}{0.632745in}}%
\pgfpathlineto{\pgfqpoint{2.429474in}{0.626420in}}%
\pgfpathlineto{\pgfqpoint{2.437463in}{0.622415in}}%
\pgfpathlineto{\pgfqpoint{2.451445in}{0.616133in}}%
\pgfpathlineto{\pgfqpoint{2.465426in}{0.610375in}}%
\pgfpathlineto{\pgfqpoint{2.479408in}{0.604956in}}%
\pgfpathlineto{\pgfqpoint{2.485400in}{0.602725in}}%
\pgfpathlineto{\pgfqpoint{2.507371in}{0.596049in}}%
\pgfpathlineto{\pgfqpoint{2.513363in}{0.594398in}}%
\pgfpathlineto{\pgfqpoint{2.521352in}{0.592322in}}%
\pgfpathlineto{\pgfqpoint{2.523350in}{0.592015in}}%
\pgfpathlineto{\pgfqpoint{2.533336in}{0.591507in}}%
\pgfpathlineto{\pgfqpoint{2.535334in}{0.591548in}}%
\pgfpathlineto{\pgfqpoint{2.549315in}{0.594416in}}%
\pgfpathlineto{\pgfqpoint{2.565294in}{0.596884in}}%
\pgfpathlineto{\pgfqpoint{2.575281in}{0.598180in}}%
\pgfpathlineto{\pgfqpoint{2.587265in}{0.599477in}}%
\pgfpathlineto{\pgfqpoint{2.597252in}{0.600343in}}%
\pgfpathlineto{\pgfqpoint{2.607239in}{0.601029in}}%
\pgfpathlineto{\pgfqpoint{2.617225in}{0.601538in}}%
\pgfpathlineto{\pgfqpoint{2.631207in}{0.601860in}}%
\pgfpathlineto{\pgfqpoint{2.641194in}{0.601804in}}%
\pgfpathlineto{\pgfqpoint{2.655175in}{0.601424in}}%
\pgfpathlineto{\pgfqpoint{2.665162in}{0.600957in}}%
\pgfpathlineto{\pgfqpoint{2.681141in}{0.599810in}}%
\pgfpathlineto{\pgfqpoint{2.697120in}{0.598193in}}%
\pgfpathlineto{\pgfqpoint{2.711101in}{0.596417in}}%
\pgfpathlineto{\pgfqpoint{2.725083in}{0.594298in}}%
\pgfpathlineto{\pgfqpoint{2.737067in}{0.592225in}}%
\pgfpathlineto{\pgfqpoint{2.747054in}{0.590331in}}%
\pgfpathlineto{\pgfqpoint{2.747054in}{0.590331in}}%
\pgfusepath{stroke}%
\end{pgfscope}%
\begin{pgfscope}%
\pgfsetrectcap%
\pgfsetmiterjoin%
\pgfsetlinewidth{0.803000pt}%
\definecolor{currentstroke}{rgb}{0.000000,0.000000,0.000000}%
\pgfsetstrokecolor{currentstroke}%
\pgfsetdash{}{0pt}%
\pgfpathmoveto{\pgfqpoint{0.651925in}{0.536658in}}%
\pgfpathlineto{\pgfqpoint{0.651925in}{1.717454in}}%
\pgfusepath{stroke}%
\end{pgfscope}%
\begin{pgfscope}%
\pgfsetrectcap%
\pgfsetmiterjoin%
\pgfsetlinewidth{0.803000pt}%
\definecolor{currentstroke}{rgb}{0.000000,0.000000,0.000000}%
\pgfsetstrokecolor{currentstroke}%
\pgfsetdash{}{0pt}%
\pgfpathmoveto{\pgfqpoint{2.846822in}{0.536658in}}%
\pgfpathlineto{\pgfqpoint{2.846822in}{1.717454in}}%
\pgfusepath{stroke}%
\end{pgfscope}%
\begin{pgfscope}%
\pgfsetrectcap%
\pgfsetmiterjoin%
\pgfsetlinewidth{0.803000pt}%
\definecolor{currentstroke}{rgb}{0.000000,0.000000,0.000000}%
\pgfsetstrokecolor{currentstroke}%
\pgfsetdash{}{0pt}%
\pgfpathmoveto{\pgfqpoint{0.651925in}{0.536658in}}%
\pgfpathlineto{\pgfqpoint{2.846822in}{0.536658in}}%
\pgfusepath{stroke}%
\end{pgfscope}%
\begin{pgfscope}%
\pgfsetrectcap%
\pgfsetmiterjoin%
\pgfsetlinewidth{0.803000pt}%
\definecolor{currentstroke}{rgb}{0.000000,0.000000,0.000000}%
\pgfsetstrokecolor{currentstroke}%
\pgfsetdash{}{0pt}%
\pgfpathmoveto{\pgfqpoint{0.651925in}{1.717454in}}%
\pgfpathlineto{\pgfqpoint{2.846822in}{1.717454in}}%
\pgfusepath{stroke}%
\end{pgfscope}%
\begin{pgfscope}%
\pgfsetbuttcap%
\pgfsetmiterjoin%
\definecolor{currentfill}{rgb}{1.000000,1.000000,1.000000}%
\pgfsetfillcolor{currentfill}%
\pgfsetfillopacity{0.800000}%
\pgfsetlinewidth{1.003750pt}%
\definecolor{currentstroke}{rgb}{0.800000,0.800000,0.800000}%
\pgfsetstrokecolor{currentstroke}%
\pgfsetstrokeopacity{0.800000}%
\pgfsetdash{}{0pt}%
\pgfpathmoveto{\pgfqpoint{0.254695in}{1.881572in}}%
\pgfpathlineto{\pgfqpoint{2.717277in}{1.881572in}}%
\pgfpathquadraticcurveto{\pgfqpoint{2.740416in}{1.881572in}}{\pgfqpoint{2.740416in}{1.904711in}}%
\pgfpathlineto{\pgfqpoint{2.740416in}{2.272447in}}%
\pgfpathquadraticcurveto{\pgfqpoint{2.740416in}{2.295586in}}{\pgfqpoint{2.717277in}{2.295586in}}%
\pgfpathlineto{\pgfqpoint{0.254695in}{2.295586in}}%
\pgfpathquadraticcurveto{\pgfqpoint{0.231556in}{2.295586in}}{\pgfqpoint{0.231556in}{2.272447in}}%
\pgfpathlineto{\pgfqpoint{0.231556in}{1.904711in}}%
\pgfpathquadraticcurveto{\pgfqpoint{0.231556in}{1.881572in}}{\pgfqpoint{0.254695in}{1.881572in}}%
\pgfpathlineto{\pgfqpoint{0.254695in}{1.881572in}}%
\pgfpathclose%
\pgfusepath{stroke,fill}%
\end{pgfscope}%
\begin{pgfscope}%
\pgfsetrectcap%
\pgfsetroundjoin%
\pgfsetlinewidth{1.505625pt}%
\definecolor{currentstroke}{rgb}{0.843137,0.098039,0.109804}%
\pgfsetstrokecolor{currentstroke}%
\pgfsetdash{}{0pt}%
\pgfpathmoveto{\pgfqpoint{0.277834in}{2.185774in}}%
\pgfpathlineto{\pgfqpoint{0.393529in}{2.185774in}}%
\pgfpathlineto{\pgfqpoint{0.509223in}{2.185774in}}%
\pgfusepath{stroke}%
\end{pgfscope}%
\begin{pgfscope}%
\definecolor{textcolor}{rgb}{0.000000,0.000000,0.000000}%
\pgfsetstrokecolor{textcolor}%
\pgfsetfillcolor{textcolor}%
\pgftext[x=0.601779in,y=2.145281in,left,base]{\color{textcolor}{\rmfamily\fontsize{8.330000}{9.996000}\selectfont\catcode`\^=\active\def^{\ifmmode\sp\else\^{}\fi}\catcode`\%=\active\def%{\%}$\tilde{\bm{v}}$}}%
\end{pgfscope}%
\begin{pgfscope}%
\pgfsetbuttcap%
\pgfsetroundjoin%
\pgfsetlinewidth{1.505625pt}%
\definecolor{currentstroke}{rgb}{0.172549,0.482353,0.713725}%
\pgfsetstrokecolor{currentstroke}%
\pgfsetdash{{1.500000pt}{2.475000pt}}{0.000000pt}%
\pgfpathmoveto{\pgfqpoint{0.277834in}{2.002294in}}%
\pgfpathlineto{\pgfqpoint{0.393529in}{2.002294in}}%
\pgfpathlineto{\pgfqpoint{0.509223in}{2.002294in}}%
\pgfusepath{stroke}%
\end{pgfscope}%
\begin{pgfscope}%
\definecolor{textcolor}{rgb}{0.000000,0.000000,0.000000}%
\pgfsetstrokecolor{textcolor}%
\pgfsetfillcolor{textcolor}%
\pgftext[x=0.601779in,y=1.961801in,left,base]{\color{textcolor}{\rmfamily\fontsize{8.330000}{9.996000}\selectfont\catcode`\^=\active\def^{\ifmmode\sp\else\^{}\fi}\catcode`\%=\active\def%{\%}$\bm{v}^{\mathrm{ref}}$$(\mathrm{Re}=50)$}}%
\end{pgfscope}%
\begin{pgfscope}%
\pgfsetbuttcap%
\pgfsetroundjoin%
\pgfsetlinewidth{1.505625pt}%
\definecolor{currentstroke}{rgb}{0.992157,0.682353,0.380392}%
\pgfsetstrokecolor{currentstroke}%
\pgfsetdash{{5.550000pt}{2.400000pt}}{0.000000pt}%
\pgfpathmoveto{\pgfqpoint{1.542652in}{2.185774in}}%
\pgfpathlineto{\pgfqpoint{1.658346in}{2.185774in}}%
\pgfpathlineto{\pgfqpoint{1.774041in}{2.185774in}}%
\pgfusepath{stroke}%
\end{pgfscope}%
\begin{pgfscope}%
\definecolor{textcolor}{rgb}{0.000000,0.000000,0.000000}%
\pgfsetstrokecolor{textcolor}%
\pgfsetfillcolor{textcolor}%
\pgftext[x=1.866596in,y=2.145281in,left,base]{\color{textcolor}{\rmfamily\fontsize{8.330000}{9.996000}\selectfont\catcode`\^=\active\def^{\ifmmode\sp\else\^{}\fi}\catcode`\%=\active\def%{\%}$\bm{v}^{\mathrm{ref}}$$(\mathrm{Re}=1000)$}}%
\end{pgfscope}%
\begin{pgfscope}%
\pgfsetbuttcap%
\pgfsetroundjoin%
\pgfsetlinewidth{1.505625pt}%
\definecolor{currentstroke}{rgb}{0.670588,0.850980,0.913725}%
\pgfsetstrokecolor{currentstroke}%
\pgfsetdash{{9.600000pt}{2.400000pt}{1.500000pt}{2.400000pt}}{0.000000pt}%
\pgfpathmoveto{\pgfqpoint{1.542652in}{1.996121in}}%
\pgfpathlineto{\pgfqpoint{1.658346in}{1.996121in}}%
\pgfpathlineto{\pgfqpoint{1.774041in}{1.996121in}}%
\pgfusepath{stroke}%
\end{pgfscope}%
\begin{pgfscope}%
\definecolor{textcolor}{rgb}{0.000000,0.000000,0.000000}%
\pgfsetstrokecolor{textcolor}%
\pgfsetfillcolor{textcolor}%
\pgftext[x=1.866596in,y=1.955628in,left,base]{\color{textcolor}{\rmfamily\fontsize{8.330000}{9.996000}\selectfont\catcode`\^=\active\def^{\ifmmode\sp\else\^{}\fi}\catcode`\%=\active\def%{\%}$\bm{v}^{\mathrm{ref}}$$(\mathrm{Re}=5000)$}}%
\end{pgfscope}%
\end{pgfpicture}%
\makeatother%
\endgroup%